\pdfoutput=1
\documentclass[a4paper,11pt]{article}

\newcommand{\changelocaltocdepth}[1]{%
  \addtocontents{toc}{\protect\setcounter{tocdepth}{#1}}%
  \setcounter{tocdepth}{#1}%
}

\usepackage{fullpage}

\usepackage[utf8]{inputenc}
\usepackage[T1]{fontenc}
\usepackage{microtype}
\usepackage{lmodern}

\usepackage{lmodern}
\usepackage{newunicodechar}
\newunicodechar{−}{-}

\usepackage[english]{babel}
\usepackage[autostyle, english = american]{csquotes}
\MakeOuterQuote{"}
\usepackage{xspace}

\usepackage{mathtools}
\usepackage{amssymb}
\usepackage{bm}
\makeatletter\g@addto@macro\bfseries{\boldmath}\makeatother
\usepackage{nicefrac}

\usepackage{physics}
\usepackage{slashed}

\usepackage{subcaption}
\usepackage{wrapfig}
\usepackage{graphicx}

\usepackage{booktabs}
\usepackage{makecell}
\usepackage{float}

\usepackage{subfiles}

\usepackage{hyperref}
\hypersetup{colorlinks=true,linkcolor=blue,urlcolor=black,linktocpage=true,citecolor=blue}
\usepackage[capitalise,noabbrev]{cleveref}

\newcommand{\bea}{\begin{eqnarray}}
\newcommand{\eea}{\end{eqnarray}}
\newcommand{\Y}{\mathcal{Y}}
\newcommand{\M}{\mathcal{M}}
\newcommand{\Mbar}{\bar{\mathcal{M}}}
\newcommand\psiB{\psi_{\mathcal{B}}}
\newcommand\psiBbar{\bar{\psi}_{\mathcal{B}}}
\newcommand{\Bsm}{\mathcal{B}_{\rm SM}}
\newcommand{\Bbarsm}{\bar{\mathcal{B}}_{\rm SM}}
\newcommand{\phiB}{\phi_{\mathcal{B}}}
\newcommand{\phiBbar}{\phi^*_{\mathcal{B}}}
\newcommand{\YBAU}{Y_{\mathcal{B}}^{\rm meas}}
\newcommand{\dcp}{\text{Dark}_{\rm CP}}
\newcommand{\cO}{\mathcal{O}}
\newcommand{\Bri}{\text{Br}\left(B_i^0\rightarrow \psiBbar \, \mathcal{B}_{\rm SM}\right)}
\newcommand{\B}{\mathcal{B}}
\newcommand{\Bp}{{\B^+}}
\newcommand{\ACP}{A_{\rm CP}}
\newcommand{\acpf}{a^{f_\M}_{\rm CP}}
\newcommand{\ACPB}{{\tilde{A}_{\rm CP}}}
\newcommand{\prn}[1]{ \left(  #1 \right) }
\newcommand{\email}[1]{\normalfont{\href{mailto:#1}{\nolinkurl{#1}}}}

\begin{document}

\begin{titlepage}
\pagenumbering{roman}

\vspace*{-1cm}
\noindent\begin{tabular*}{\linewidth}{@{}l@{\extracolsep{\fill}}r@{}}
&
 \includegraphics[width=.25\linewidth]{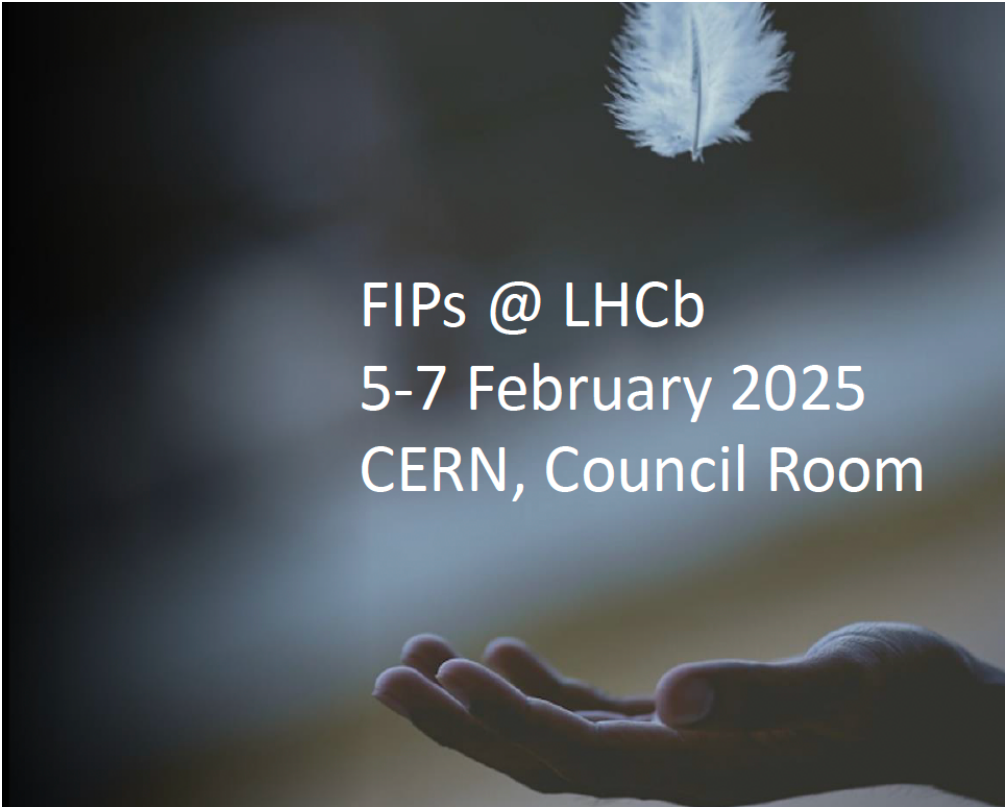} \\
\hline
\end{tabular*}

\vspace*{1cm}

\begin{center}\bf\LARGE
 Feebly-Interacting Particles: \\
 FIPs at LHCb --- Workshop Report \\
 2025 Edition
\end{center}

\vspace*{.5cm}

\begin{center}
J.~Alimena$^{1}$,  
J.~Boyd$^{2}$,   
G.~Cacciapaglia$^{3}$, 
A.~Casais~Vidal$^{4,*}$,
X.~Cid~Vidal$^{5}$, 
S.~Collaviti$^{6}$, 
A.~De~Oyanguren~Campos$^{7}$, 
G.~Dalla~Valle~Garcia$^{8}$, 
G.~Elor$^{9}$, 
G.~Ferretti$^{10}$, 
D.~Gorbunov$^{11}$,
E.~Goudzovski$^{12}$,
J.~Hajer$^{13}$,
J.~Jerhot$^{14}$,
B.~Kishor~Jashal$^{7}$,
V.~Kholoimov$^{7}$,
J.~Klaric$^{15}$,
F.~Kling$^{1,27}$,
E.~Kriukova$^{11,29}$,
Y.~Kyselov$^{16}$,
G.~Lanfranchi$^{17,*}$,
C.~Langenbruch$^{18}$,
S.~Libralon$^{7}$, 
F.~Martinez~Vidal$^{7}$, 
A.~Merli$^{6}$,
M.~Ovchynnikov$^{2}$,
J.~Pfaller$^{19}$,
G.~Perez$^{20}$,
P.~Reimitz$^{21,28}$,
I.~Sanderswood$^{7}$,
L.~Shchutska$^{6}$,
V.~Svintozelskyi$^{7}$, 
E.~Torro~Pastor$^{22}$,
Y.~Tsai$^{23}$
A.~Usachov$^{24,*}$,
L.~Vale~Silva$^{25}$,
C.~Vázquez~Sierra$^{26,*}$,
F.~Volle$^{12}$,
J.~Zhuo$^{7}$,
J.~Zurita$^{7}$
\end{center}

\vspace*{.5cm}

\begin{abstract}\noindent
With the establishment and maturation of the experimental programs
searching for new physics with sizeable couplings at the LHC, there is an increasing interest
in the broader particle and astrophysics community for exploring the physics of light and
feebly-interacting particles as a paradigm complementary to a New Physics sector at the TeV
scale and beyond. FIPs@LHCb continues the successful series of the FIPs workshops, FIPs 2020 and FIPs 2022. The main focus of the workshop was to explore the LHCb potential to search for FIPs thanks to the new software trigger deployed during the recent upgrade. Equally important goals of the workshop were to update the available parameter space in the commonly used FIPs benchmarks by including recent results from the high energy physics community and to discuss recent theory progress necessary for a more accurate definition of observables related to FIP benchmarks.
This document presents the summary of the talks presented at the workshops and the outcome of subsequent
discussions. 

\end{abstract}

\vskip 2cm
\begin{center} 
{CC BY 4.0 license}
\end{center}

\vfill
\noindent
\begin{footnotesize}
$^{*}$Corresponding authors:
Adrian Casais Vidal (\email{adrian.casais.vidal@cern.ch}),
Gaia Lanfranchi (\email{Gaia.Lanfranchi@lnf.infn.it}),
Andrii Usachov (\email{Andrii.Usachov@cern.ch}.),
Carlos Vázquez Sierra (\email{carlos.vazquez@cern.ch})
\end{footnotesize}
\end{titlepage}

\clearpage

\begin{center}\itshape\footnotesize
$^{1}$Deutsches Elektronen-Synchrotron DESY, Notkestr. 85, 22607 Hamburg, Germany \\
$^{2}$European Organization for Nuclear Research (CERN), Geneva, Switzerland \\
$^{3}$Laboratoire de Physique Th\'eorique et Hautes \'Energies (LPTHE), UMR 7589, Sorbonne Universit\'e \& CNRS/INP, 4 place Jussieu, 75252 Paris Cedex 05, France. \\
$^{4}$ Laboratory for Nuclear Science. 
Massachusetts Institute of Technology. Cambridge, MA. United States of America. \\
$^{5}$Instituto Galego de F\'isica de Altas Enerx\'ias, Universidade de Santiago de Compostela, Santiago, Spain \\
$^{6}$Institute of Physics, Ecole Polytechnique Fédérale de Lausanne (EPFL), Lausanne, Switzerland.\\
$^{7}$Instituto de F\'{\i}sica Corpuscular, CSIC-Universitat de Val\`encia, E-46980, Paterna, Valencia, Spain. \\
$^{8}$Karlsruher Institut für Technologie (KIT), Hermann-von-Helmholtz-Platz 1,  76344 Eggenstein-Leopoldshafen, Germany.\\
$^{9}$Theory Group, The Weinberg Institute for Theoretical Physics, University of Texas at Austin, Austin, TX 78712, United States. \\
$^{10}$Department of Physics, Chalmers University of Technology, Fysikg{\aa}rden 1, 41296 G\"oteborg, Sweden.\\
$^{11}$Institute for Nuclear Research of the Russian Academy of Sciences, 60th October Anniversary prospect, 7a, Moscow 117312, Russia. \\
$^{12}$School of Physics and Astronomy, University of Birmingham, Edgbaston, Birmingham, B15~2TT, United Kingdom.\\
$^{13}$Departamento de Física, Instituto Superior Técnico (IST), Universidade de Lisboa, Av. Rovisco Pais 1, 1049-001 Lisboa, Portugal.\\
$^{14}$Max-Planck-Institut f\"ur Physik (Werner-Heisenberg-Institut), Boltzmannstr. 8, 85748 Garching bei M\"unchen, Germany.\\
$^{15}$University of Amsterdam, Science Park 904, 1098 XH Amsterdam, The Netherlands. \\
$^{16}$Taras Shevchenko National University of Kyiv, Kyiv, Ukraine. \\
$^{17}$Laboratori Nazionali di Frascati dell’INFN, via E. Fermi 56, 00044 Frascati (Rome), Italy. \\
$^{18}$Physikalisches Institut, Heidelberg University, Im Neuenheimer Feld 226, 69120 Heidelberg, Germany. \\
$^{19}$Department of Physics, University of Cincinnati, OH 45220, USA. \\
$^{20}$Department of Particle Physics and Astrophysics, Weizmann Institute of Science, Rehovot 7610001, Israel. \\
$^{21}$Instituto de F\'{i}sica, Universidade de S\~{a}o Paulo, 05508-090 S\~{a}o Paulo, SP, Brasil.\\
$^{22}$Instituto de Física Corpuscular (IFIC), Centro Mixto Universidad de Valencia - CSIC, Valencia; Spain. \\
$^{23}$Department of Physics and Astronomy, University of Notre Dame, IN 46556, USA. \\
$^{24}$Nikhef National Institute for Subatomic Physics, Amsterdam, Netherlands.\\
$^{25}$Departamento de Matem\'{a}ticas, F\'{i}sica y Ciencias Tecnol\'{o}gicas, Universidad Cardenal Herrera-CEU, CEU Universities, 46115 Alfara del Patriarca, Val\`{e}ncia, Spain. \\
$^{26}$Departmento de Física y Ciencias de la Tierra, Universidade da Coruña, A Coruña, Spain \\
$^{27}$Department of Physics and Astronomy, University of California, Irvine, CA 92697-4575, USA \\
$^{28}$Department of Physics and Astronomy, University of Victoria, Victoria BC V8P 5C2, Canada \\
$^{29}$Institute of Theoretical and Mathematical Physics, Lomonosov Moscow State University, Moscow 119991, Russia
\end{center}

\clearpage

\tableofcontents

\cleardoublepage

\setcounter{page}{1}
\pagenumbering{arabic}

\vskip 5mm
\noindent

\clearpage

\section{Introduction}
\label{sec:intro}
This document serves as both the handbook and proceedings of the \textit{FIPs@LHCb} workshop, held at CERN from 5 to 7 February 2025. It summarizes the main contributions and discussions from the workshop, which brought together theorists and experimentalists working on searches for Feebly Interacting Particles (FIPs) at LHCb and other experiments. Intended as both a scientific reference and a practical guide, this handbook provides a comprehensive overview of the current status, key challenges, and future directions in the rapidly developing field.

FIPs are motivated by several open questions in fundamental particle physics, such as the strong CP problem, the nature of dark matter, the origin of the matter--antimatter asymmetry, the existence of neutrino masses and oscillations, and the hierarchy problem. The characteristic GeV-scale invariant masses and long lifetimes of many FIP scenarios make the LHCb experiment particularly well-suited for their study.
The structure of this document follows the scientific program of the workshop and is organized as follows.

Section~\ref{sec:ckm} explores connections between FIPs and the Cabibbo--Kobayashi--Maskawa (CKM) matrix. It includes theoretical models predicting time-dependent modifications to flavor observables via ultralight dark matter couplings, along with current and projected precision on CKM parameters from global fits, and implications for new physics in $B$-meson mixing.

Section~\ref{sec:portal} presents benchmark models and their phenomenology, including dark photons, dark scalars, axion-like particles (ALPs), heavy neutral leptons (HNLs), and $B$ mesogenesis models. It discusses theoretical motivations, production and decay channels, and the role of hadronic uncertainties, with a view to supporting reinterpretation and sensitivity comparisons across experiments.

Section~\ref{sec:fips_experiments} reviews the experimental landscape, summarizing recent results and future prospects. It highlights LHCb’s performance following Upgrade~I and its enhanced capabilities in Upgrade~II, with a focus on FIP reconstruction. Results and plans from ATLAS and CMS are discussed, including searches covering both low- and high-mass regimes using increasingly sophisticated dedicated triggers. Contributions from Belle~II, NA62, other kaon experiments, FASER, and CODEX-b emphasize the complementarity of experimental approaches in geometry, timing, and sensitivity to different lifetimes and couplings.

Section~\ref{sec:current-status} presents the current global picture of the FIP searches, including up-to-date results and sensitivity projections for the benchmark models introduced earlier. It provides a comparative view across experiments and highlights target regions for the future. Finally, Section~\ref{sec:conclusions} offers an outlook on collaborative priorities and directions for the coming years.

This report is intended to serve as a lasting record of the workshop and a resource for the broader community. It reflects the collaborative spirit of the field and supports ongoing efforts to expand the discovery potential of the LHC and related experiments in the search for FIPs.

\clearpage
\section{FIPs and CKM}
\label{sec:ckm}
\changelocaltocdepth{2}

\subsection{Ultra-light Dark Matter searches and CKM --- \textit{G.~Perez}}
\label{ssec:perez}
\textit{Author: Gilad Perez, \email{gilad.perez@weizmann.ac.il}}  \\
\subsubsection{\label{sec:introduction}Introduction}

From a modern effective field theory (EFT) perspective of quantum field theories (QFTs), the strong CP problem can be understood in four different levels of sophistication (to match with the "Four Sons" of the Passover's Haggadah). The first would say that the strong CP problem is associated with the smallness of the strong CP phase (Is it a problem?).
The second would say that it is associated with the hierarchy between the CKM phase (being order one) and the strong CP one (being smaller than $10^{-10}$, how different this is from the flavor puzzle, dealing with the hierarchical structure of the masses and mixing?).
The third one would say that it is associated with the hierarchy between the CKM phase (being order one) and the strong CP one and noting that the strong CP phase, at 7-loop order~\cite{Ellis:1978hq}, is expected to suffer from a logarithmic divergent proportional to the CKM phase (albeit with a tiny coefficient, thus giving a negligible additive contribution for any appreciable UV cutoff).
It is possible that once adding gravitational interaction, this picture is changed as has been emphasized in~\cite{Dvali:2005an}, but this goes beyond the scope of EFT and this review.

Within the EFT framework are three widely discussed solutions to the strong CP problem.  These are the possibility of a massless $u$ quark, for some time ruled out by lattice gauge theory, the Peccei-Quinn solution~\cite{Peccei:1977hh}, and the possibility of spontaneous breaking of CP, which we will refer to as the Nelson-Barr solution~\cite{Nelson:1983zb,Nelson:1984hg,Barr:1984qx} (for earlier attempts see~\cite{Mohapatra:1978fy,Georgi:1978xz}, see also~\cite{Hiller:2001qg, Harnik:2004su} for a different but related approach).  In this work, following our paper~\cite{Dine:2024bxv}, we focus on aspects of the Nelson-Barr class of models.  In particular, the minimal realization of the Nelson-Barr solution involves an additional vector-like fermion, as well as a single complex scalar that spontaneously breaks the CP symmetry~\cite{Bento:1991ez}.
Some special structure is required to obtain a working model, which can be the result of a discrete or (non-anomalous) approximate continuous symmetry. As we show below, the symmetry could render the phase of the scalar field comparatively light since it is only broken by the standard model (SM) quark flavor-mixing. In this paper, we explore the possibility that this boson is {\it extremely} light and can be a viable ultralight dark matter (UDM) candidate.
The same symmetry
suppresses otherwise dangerous contributions to the strong CP phase. Unlike the case of the QCD axion, the light field has linear scalar couplings to quarks, which arise as a result of the fact that the CKM angles and phase depend linearly on the UDM field.
The requirement that it does not lead to unobserved long-range forces places a lower bound on the scale of
symmetry breaking, but at the same time opens up the possibility to an observable time-dependent signal in flavor factories or future atomic and nuclear clock experiments.

\subsubsection{\label{sec:model}A minimal Nelson Barr model}
In this paper, we study the minimal model of Ref.~\cite{Bento:1991ez}. It introduces an additional vector-like quark pair $q$ ($\bar q$) that carries the same (opposite) SM charge as the right-handed up-quark, in addition to a neutral complex scalar, $\Phi=(f+\rho)\exp(i\theta)/\sqrt2$\,. The Lagrangian contains the following couplings
\begin{equation}
    {\cal L} \supset \mu \bar q q + (g_{i} \Phi+\tilde g_{i}\Phi^*) \bar u_{i} q+ y^{u}_{ij} \tilde{H}Q_i \bar u_{j}+ y^{d}_{ij} H Q_i \bar d_{j}+ \dots \label{eq:lagrangian}
\end{equation}
It is further assumed that the theory is CP-conserving, and CP is only spontaneously broken by the expectation value of $\Phi$\,, such that $\mu,y^u$ and $y^d$ are real. 
In the presence of the Higgs vacuum expectation value, $v$\,, the up-quark mass is given by the following $4\times 4$ matrix:
\begin{equation}
    {\cal M}_u = \left ( \begin{matrix} \mu & B  \cr 0 & m_u \end{matrix} \right ),\ \ m_u = y^uv,~B_i=(g_{i} \Phi+\tilde g_{i}\Phi^*)\,.
    \label{eq: 4times4 quark matrix}
\end{equation}
Due to the absence of the bottom left entry, and the fact that $\Phi$ only appears in the off-diagonal entry of the above matrix, $\arg(\det({\cal M}_u))=0$ holds and no QCD phase is introduced (while the CKM phase is unconstrained).  This can be ensured by introducing an additional ${Z}_2$ symmetry under which $q,~\overline{q}$ and $\Phi$ are odd while the SM fields are even. An approximate U(1) flavor symmetry can also enforce this structure instead, which also protects the mass of the phase of $\Phi$ (see more details below).

To identify the quark masses and mixings, we focus on the structure of  ${\cal M}_u{\cal M}_u^\dagger$\,.
Assuming that the vector-like quark is heavy $\mu,|B|\gg m_u$\,, we can integrate it out and are left with an effective up-quark mass matrix $\tilde{m}_u$ satisfying
\begin{equation}
    (\tilde m_u \tilde m_u^\dagger )_{ij}=\left (  (m_u m_u^T )_{ij} - \frac{(m_u)_{ik} B^\dagger_k B_\ell (m_u^T)_{\ell j}} {\mu^2 + B_fB_f^\dagger} \right ) \,.\label{eq: effective quark mass matrix}
\end{equation}
The CKM matrix is the product of the $SU(3)$ matrix required to diagonalize $\tilde m_u \tilde m_u^\dagger$ and the left $SO(3)$ rotation required to diagonalize $y^{d}$\,. Assuming $\langle\theta\rangle={\cal O}(1)$\,, $\mu\lesssim |B|$\,, and the vectors $g$ and $\tilde g$ are of comparable magnitude and not parallel in flavor space (see discussion in~\cite{Davidi:2017gir}), the resulting CKM matrix has an $\mathcal{O}(1)$ CP violating phase.

In the rest of the paper, we investigate the possibility that the potential generating the VEV of $\theta$ is shallow (again, here we refer to~\cite{Davidi:2017gir}, which effectively already used this property within the relaxion framework). This corresponds to the presence of a global symmetry holding to a very good
approximation.  In this case, one finds at low energies a remaining light scalar field. The couplings of this light scalar $\phi$ are found by replacing the VEV $\theta$ with $\theta_0+\phi/f$\,. We discuss them in detail in the next sections, focusing on the following set of couplings. We take $g\propto(1,0,0),\ \tilde{g}\propto (0,1,0)$\,, with $m^u$ diagonal and $m^d=V^{d\dagger}\mathrm{diag}(m_d,m_s,m_b)$\,, where $V^{d}\in SO(3)$ is the real valued CKM matrix of the original Lagrangian. We will see below that this choice gives the best prospects for detecting the model through flavor physics.

These parameters can arise naturally, for example if a global shift symmetry in $\theta$ rotates $\Phi \rightarrow e^{i \theta} \Phi$\,, and $\bar u_1 \rightarrow e^{-i \theta} \bar u_1, ~\bar u_2 \rightarrow e^{i \theta} \bar u_2$\,. This symmetry is only broken by the off-diagonal entries of the Yukawa matrices, turning the angular variable $\theta$ into a pseudo-Nambu-Goldstone boson. In particular, we show in \cref{sec:challenges} that loop corrections to the mass are suppressed by the Yukawas being small as well as the CKM matrix being mostly diagonal. 
The symmetries we consider are free of anomalies, so they differ from a Peccei-Quinn symmetry.
They further account for the lightness of $\phi$ as well as suppressing potentially dangerous contributions to the strong CP phase discussed in~\cite{Dine:2015jga}.

\subsubsection{\label{sec:osc_CKM}Oscillating flavour observables}
With the parameters given above one can then readily check that $\tilde m_u \tilde m_u^\dagger$ of \cref{eq: effective quark mass matrix} is diagonalized by a rotation of the first two types of up-quarks $O_{12}$ and a phase transformation $P$ removing the phase induced by the complex scalar $\Phi$
\begin{equation}
P=\mathrm{diag}\left(1,\exp\left[-2i\left(\theta_0+\frac{\phi}{f}\right)\right],1\right)\,.
\end{equation}
The resulting effective CKM matrix is given by
\begin{equation}
    V= O_{12}PV^{d}\,.\label{eq:CKM matrix}
\end{equation}
The CKM phase is directly related to the phase $\theta$ of the complex scalar, which consists of a background value $\theta_0$ and a light scalar $\phi$. If $\phi$ is light enough and constitutes dark matter, the CKM phase oscillates in time. The amplitude and period of the oscillation depend on the local dark matter density, $\rho_\mathrm{DM}\approx2.5\times10^{-6}~\mathrm{eV}^4$\,, and on the mass of the scalar, $m_\phi$:
\begin{equation}
    \frac{\phi(t)}{f}=\frac{\sqrt{2\rho_\mathrm{DM}}}{f m_\phi}\cos(m_\phi t)\,.\label{eq:uldm_osc}
\end{equation}
 Note that if $O_{12}= \mathbb{1}$ the phase can be removed. In fact the Jarlskog invariant, which all CP-violating observables are proportional to, vanishes if the rotation angle $\theta_{12}$ goes to zero
\begin{align}
    J &=  \text{Im}\left(V_{ud}V^*_{ub}V_{tb}V^*_{td}\right)\nonumber\\
    &= \frac{1}{2}|V_{tb}||V_{td}| \sin{2(\theta_0+\phi/f)}\sin{2{\theta}_{12}}\left(V^{d}_{cd}V^{d}_{ub}-V_{cb}^yV^{d}_{ud}\right)\,.
\end{align}
The time independent part given by $\theta_0$  should match the observed value of $J\approx 3\times 10^{-5}$\,.
Assuming $|V|\approx |V^{d}|$\,, this leads to $\kappa\equiv\sin{2\theta_0}\sin{2{\theta}_{12}}\approx 0.2$\,. To linear order, the $\phi$ dependence of CP-violating observables is therefore given by
\begin{equation}
    J = J^{0}\left(1+\frac{2}{\tan 2\theta_0}\frac{\phi}{f}\right)\,,
\end{equation}
where $J^0$ is the time independent part obtained by setting $\theta=\theta_0$\,.
In our following sensitivity estimates we assume besides $|V|\approx |V^{d}|$ that $\theta_0$ is a random phase such that $\sin{2\theta_0}\sim\cos{2\theta_0}\sim1$\,. 

It turns out that with our choice of parameters, the absolute value of the CKM entries involving the up and charm quarks vary in time.
\begin{equation}
|V_{u/c\,j}|^2\approx|V^{\mathrm{0}}_{u/c\,j}|^2\left(1\mp 2\frac{V^{d}_{c/u\, j}}{V^{d}_{u/c\, j}}\kappa \frac{\phi}{f}\right)\,,\label{eq: CKM matrix dependence phi}
\end{equation}
where $|V^{\mathrm{0}}_{ij}|$ is the time-independent part of the CKM matrix, with $i=u,c,t\,,$ and $j=d,s,b$\,.
Note that, with our particular choice of $g$ and $\tilde g$\,, the CKM elements involving the top quark are unaffected, and are thus real and have absolute value fixed to $|V_{tj}|=|V^{d}_{tj}|$\,. This choice leads to the weakest bounds from other searches, as we show in the next section.

We conclude that the parameters most susceptible to the oscillations of $\phi$ are $|V_{us}|$\,, $|V_{ub}|$\,, $|V_{cd}|$, and the CKM phase. For these, the relative change due to the oscillation is approximately given by the amplitude of $\phi/f$ in Eq.~\eqref{eq:uldm_osc}\,. Any observable $O$ proportional to these parameters, e.g. a meson decay width, therefore also oscillates with the same relative amplitude.
The amplitude of the oscillation in these observables $\Delta O$ relative to the background value $\overline{O}$ is given as
\begin{align}
    \frac{\Delta O}{\overline{O}}&=\mathcal{O}(1)\frac{\sqrt{2\rho_\mathrm{DM}}}{f m_\phi} \nonumber\\
    &\sim10^{-5}\times\frac{10^{14}~\mathrm{GeV}}{f}\times\frac{10^{-21}~\mathrm{eV}}{m_{\phi}}\,,\label{eq: amplitude observables}
\end{align}
where we assumed that the light scalar is all of the dark matter. The value of $f\sim10^{14}$ GeV saturates the bounds discussed in the next section, while $m_\phi\sim10^{-21}$ eV is at the lower limit set by cosmological and astrophysical observations (see Ref.~\cite{Hui:2021tkt} for a recent review).

To get a sense of the achievable experimental sensitivities, we would like to estimate the error that can be achieved in a given experiment. While most experiments currently report their results under the assumption of a time-independent observable, if the time-stamped data is made available, it could be possible to search for time variations through a Fourier-like analysis. However, statistical fluctuations will hinder the experimental ability to observe oscillations at a certain frequency. The spectral power of such statistical noise is frequency-independent (for the non-zero frequencies). For an experiment of length $T$ and time resolution $\Delta t$, one can show that if the phase of the oscillations is unknown, the relative amplitude of oscillations with period $T>\tau>2\Delta t$ can be constrained to the level of $\sim2/\sqrt{N_0}$, where $N_0$ is the expected total number of events in the time-independent case. For a more detailed discussion of how to perform a measurement of this kind, see Refs.~\cite{Dev:2020kgz,Losada:2021bxx,Losada:2023zap}.

As a conservative benchmark, one may quote an existing study aimed at the determination of the observable, which was calculated under the assumption that it is constant in time. Such studies, however, impose stringent cuts on a given dataset in order to keep systematic errors under control. The number of used events is then much smaller than the total number of recorded events. Assuming that systematic errors don't vary over time, there is, however, no need for such cuts when searching for oscillations. As an optimistic benchmark, we may therefore assume that a relative statistical error $\approx2/\sqrt{N_{0}}$ can be achieved, where $N_{0}$ is the number of events an experiment has collected. Here we also note that while in many flavor observables theoretical uncertainties (such as on the hadronic parameters) are important, they are irrelevant for the uncertainty on the oscillations at a non-zero frequency.

To get a sense for the achievable experimental sensitivities, we consider the following simple approach to measure these oscillations. Suppose we have a data set that allows us to measure one of these observables with a statistical error $\Delta O$\,. We further assume that this data set was acquired over a period $T$\,, and the events are evenly distributed over this time. We can then split this period into $N$ time bins, to obtain $N$ independent measurements with statistical error $\approx \sqrt{N}\Delta O$\,. With the data split in this way, the amplitude of oscillations in the observable with period $\tau$ satisfying $T>\tau>2T/N$ can be constrained at the level of $\Delta O$\,. For masses of $m_\phi\gtrsim 10^{-21}\text{ eV}$\,, the oscillation period is a couple of months and therefore matches the time over which accelerator experiments typically run. For a more detailed discussion of how to perform a measurement of this kind, see Refs.~\cite{Dev:2020kgz,Losada:2021bxx,Losada:2023zap}.

Lastly, we need to estimate the statistical error that can be achieved in a given experiment. As a conservative benchmark, one may quote an existing study aimed at the determination of the absolute value of the observable. Such studies, however, impose stringent cuts on a given dataset in order to keep systematic errors under control. The number of used events is then much smaller than the total number of recorded events. Assuming that systematic errors don't vary over time, there is, however, no need for such cuts when searching for oscillations. As an optimistic benchmark, we may therefore assume that a relative statistical error $\approx2/\sqrt{N_0}$ can be achieved, where $N_0$ is the number of events an experiment has collected.

\begin{figure}
    \centering\hspace*{-2cm}
    \includegraphics[width=.57 \textwidth]{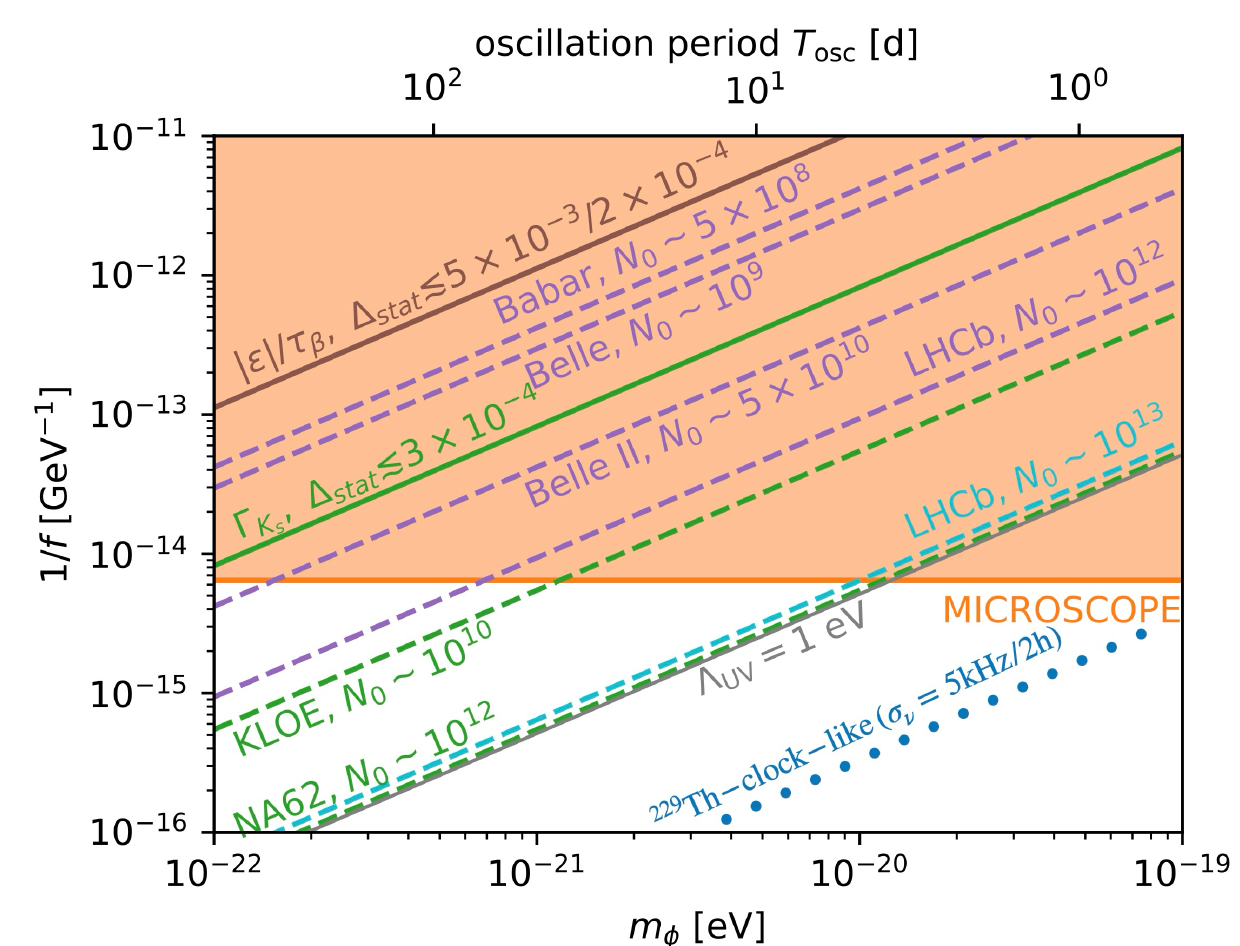}
    \caption{Potential reach of various collider searches for oscillations of the CKM matrix elements. Straight lines indicate quoted sensitivities, while dashed lines assume that for a relative measurement, an $\mathcal{O}(1)$ fraction of the events that an experiment has on tape $N_0$ can be used. In green we show measurements of $|V_{us}|$ obtained from Kaons, in purple of $|V_{ub}|$ from B-Mesons, and in blue of $|V_{cd}|$ from D-Mesons. The brown line refers to both oscillation of CP-violation in the Kaon system as well as lifetimes in $\beta$-decays, which happen to be of the same strength. The orange region is excluded from fifth-force searches, while the gray line refers to the potential fine-tuning of the scalar mass. }
    \label{fig:zoomed-limits}
\end{figure}

Given the large number of Kaons produced in experiments like KLOE, NA48, and NA62, searching for variations in $|V_{us}|$, setting the decay width of Kaons, is perhaps one of the most promising candidates. For the determination of $|V_{us}|$ e.g. the absolute branching ratio of $K^+\rightarrow \mu^+ \nu(\gamma)$ is used. It has been determined with a relative statistical error of $\sim 2\times 10^{-3}$  using $N_0\approx 10^{6}$ Kaons by the KLOE experiment \cite{ KLOE:2005xes}. For our purposes, however, every observable proportional to $|V_{us}|$ will work independently of whether it is suited to extract the matrix element. This makes the measurement of the $K_s$ lifetime possibly the leading candidate with a relative statistical error of $\sim 3\times 10^{-4}$ \cite{KLOE:2010yit}. To be optimistic, one might consider such a study involving the $N_{0}\sim 10^{10}$ Kaons KLOE recorded, or even the $N_{0}\sim 10^{13}$ recorded by NA62. The Kaons produced by NA62 are, however, strongly boosted, which could further complicate the determination of absolute widths. The bounds one might obtain from these observables are shown in green in Figure~\ref{fig:zoomed-limits}.

Similarly, $|V_{ub}|^2$ can be determined from the decay of B-mesons, although such decays are rare compared to the ones involving $V_{cb}$\,. The number of decays involving $|V_{ub}|^2$ is suppressed by $\approx|V_{ub}|^2/|V_{cb}|^2\approx 0.006$\,. However, given that Babar, Belle, Belle II, and LHCb have produced $N_0\sim 5\times 10^8\,, 10^9\,, 5\times 10^{10}$ and $10^{12}$ $b\overline{b}$-pairs sensitivity to unconstrained parameter space might still be obtained. This can be seen from Figure~\cref{fig:zoomed-limits}, where we show the resulting reach in purple. Additionally, these experiments have in principle an even larger number of D-mesons on tape, given that the production cross-section is about an order of magnitude larger than for the B-mesons. These allow the determination of $|V_{cd}|^2$ although only a fraction $\approx|V_{cd}|^2/|V_{cs}|^2\approx 0.05$ of D-mesons decays in an appropriate way. The expected reach is shown in blue in Figure~\cref{fig:zoomed-limits} for LHCb only in order to avoid clutter. It becomes clear that via this channel, b-factories should be able to rival the sensitivity of NA62\footnote{We thank Frédéric Blanc and Maria Vieites Diaz for pointing this out to us.}.

Further, one may consider CP-violating observables. The Kaon system proves to be the most precise, with e.g. the decay rate of $K_L\rightarrow\pi^+\pi^-$ relative to the rate of $K_L\rightarrow\pi^\pm e^\mp\nu_e$ determined with a relative statistical error of $\sim 5\times 10^{-3}$ \cite{NA48:2006jeq}. The resulting bound is shown in brown in Figure~\ref{fig:zoomed-limits}. Due to CP violation being small, one has to expect this sensitivity to scale poorer than $|V_{us}|$ when considering a sample of Kaons with a fixed size $N_0$\,.

Lastly, we also want to mention the possibility of observing variations in $|V_{ud}|$ via oscillations in the lifetime of $\beta$-decays. Since $|V_{ud}|^2\approx 1-|V_{us}|^2$ the relative oscillations are suppressed by $\approx|V_{us}|^2/|V_{ud}|^2\approx\lambda_\mathrm{Cabbibo}^2$ compared to the observables discussed above. Nevertheless, radioactive nuclei are abundantly available, and one might therefore hope to accomplish a competitive sensitivity. The measurements used to determine $|V_{ud}|$ currently achieve relative uncertainties of $\approx 2\times 10^{-4}$ \cite{Towner:2010zz}, which taking into account the reduced sensitivity leads to a bound with about the same strength as the search for CP violation we just discussed and is shown by the same brown line in Figure~\ref{fig:zoomed-limits}. For these measurements, it is, however, imperative that the sample only consists of one isotope of a given element. In order to observe oscillations, this is not required, but in principle, any highly radioactive sample should work.


\subsubsection{\label{sec:challenges}Challenges and prospects }

The variation of the absolute values of the CKM matrix in \cref{eq: CKM matrix dependence phi} leads to a dependence on $\phi$ of the quark masses through quantum corrections. Such couplings of $\phi$ to the quark masses are strongly constrained by searches for violations of the equivalence principle. In the near future, one can expect even stronger bounds from nuclear clocks. The coupling to the quark masses further generates a mass correction for $\phi$ itself. This is used to judge whether the small masses considered in the previous section are fine-tuned.

The strongest current bounds on such couplings stem from the MICROSCOPE mission \cite{Touboul:2017grn,MICROSCOPE:2022doy} searching for violations of the equivalence principle \cite{Banerjee:2022sqg}.
MICROSCOPE constraints the differential acceleration between a platinum and a titanium test mass relative to the common gravitational acceleration caused by Earth to be less than $\eta=(-1.5\pm2.7)\times 10^{-15}$\,\cite{MICROSCOPE:2022doy}.
We follow \cite{Damour:2010rp} to estimate the resulting variations of atomic masses, where for the strange quark we only take into account the variation of the nucleon masses \cite{Junnarkar:2013ac,Shifman:1978zn}.
In this way we arrive at a bound $1/f<9\times 10^{-17}$~GeV$^{-1}$ for masses $m_{\phi}<10^{-13}$~eV, shown in orange in \cref{fig:zoomed-limits}.

In the future, nuclear clocks are expected to drastically improve the sensitivity to variations in the quark masses~\cite{Campbell:2012zzb,Peik:2020cwm,Elwell:2024qyh}. Through the coupling discussed above, our model would introduce periodic oscillations in the quark masses, just like in the CKM entries, if the scalar constitutes dark matter. In \cref{fig:zoomed-limits} we have indicated the expected current reach of the laser excitation of Th-229 using the analysis of~\cite{Fuchs:2024edo}, with the relevant assumptions discussed in~\cite{Caputo:2024doz}.
 (for a full operational power of future nuclear clock, see for instance~\cite{Banerjee:2022sqg,Banerjee:2020kww,Kim:2022ype}).

A coupling of the type $\mathcal{L}\supset \Delta m_q(\phi)\overline{q}q$ additionally introduces a correction to the scalar potential of $\phi$ through a tadpole,
\begin{equation}
    V(\phi)\supset \Delta m_q(\phi)\frac{m_q\Lambda^2_{\text{UV}}}{16\pi^2}\,,
\end{equation}
where $\Lambda_{UV}$ is a UV cutoff. The minima of this potential are CP conserving, as one can see from the exact expressions for $\Delta m_q(\phi)\propto |V_{UD}(\phi)|^2$, where all $\phi$-dependence is $\propto \cos{2(\theta_0+\phi/f)}$\,. This is expected since our effective Lagrangian \cref{eq:lagrangian} is CP-conserving \cite{Vafa:1984xg}. The part of the potential that generates the CP-violating VEV $\theta_0$ is therefore necessarily due to additional new physics. The term above gives a correction to the scalar mass, $\Delta m_{\phi}$\,, which is dominated by the charm and bottom quark  contributions:
\begin{equation}
    \Delta m_{\phi}\simeq \frac{\sqrt{12}}{16\pi^2}|V_{cb}||V_{ub}|\sin{2\theta_{12}}\cos{2\theta_0}\ y_c y_b\frac{v \Lambda_{\text{UV}}}{f}\,.
\end{equation}
A model where $m_\phi\ll\Delta m_\phi$ would be considered fine-tuned. Saturating $m_\phi\sim \Delta m_\phi$ we can in fact work out the amplitude of oscillation for a given UV cut-off as in \cref{eq: amplitude observables}
\begin{equation}
    \frac{\Delta O}{\overline{O}}\sim 1.1 \times 10^{-6}\frac{\text{eV}}{\Lambda_{\mathrm{UV}}}\,.
\end{equation}
We conclude that, while corrections to the scalar mass are suppressed by the smallness of the Yukawa couplings, by the smallness of the off-diagonal entries of the CKM matrix, and by only arising at two loops, the amplitudes that we highlighted as detectable in the previous section still require a decent amount of fine-tuning.
Nuclear clocks will be able to probe both natural regions and regions consistent with a minimal misalignment scenario for UDM production \cite{Preskill:1982cy}, satisfying the relation $f\gtrsim 10^{18}\rm\, GeV\,\left({10^{-27} \,eV\over m_\phi}\right)^{1/4},$ indicated by the black line.

\clearpage
\subsection{Current and Expected experimental precision in the determination of CKM triangle  --- \textit{L.~Vale~Silva}}
\label{ssec:vale-silva}
\textit{Author: Luiz Vale Silva, \email{luiz.valesilva@uchceu.es}}  \\
Under the Standard Model (SM) framework, we discuss the extraction of the Cabibbo-Kobayashi-Maskawa (CKM) matrix elements from a global fit, which combines observables that satisfy the double requirement of experimental and theoretical precision. We employ the \textsf{\textit{CKM}fitter} package, consisting of a frequentist framework that applies the \textit{Range} fit scheme to handle theoretical uncertainties.
In light of the planned LHCb Upgrades, Belle~II and its possible upgrade, and moreover the broad interest in flavour physics in the tera-$Z$ phase of the proposed FCC-$e e$ program, we also discuss the expected future precision.
Finally, we briefly discuss constraints from a global fit on New Physics (NP) contributions to $B_d$ and $B_s$ mixing when considering these benchmark, future scenarios.

\subsubsection{Introduction}


Flavour changing processes test the SM, and consist in an avenue to look for its extension.
The breaking of flavour symmetries in the SM originates from the Yukawa couplings of the SM Higgs.
This originates the spectrum of fermion masses and the CKM matrix in the quark sector, which
collects the fundamental parameters of the SM describing quark flavour mixing.
The CKM matrix in the SM is a 3-by-3 unitary matrix, parameterized by three mixing angles and a single source of CP violation \cite{Kobayashi:1973fv}.

The CKM matrix turns out to be highly hierarchical, with elements closer to the diagonal larger in modulus. A useful, rephasing invariant, parameterization showing clearly this feature is the Wolfenstein parameterization, which consists of four real parameters $ A $, $ \lambda $, $ \bar\rho $ and $ \bar\eta $:

\begin{equation}
	{\color{black} \lambda} = \frac{\vert V_{us} \vert}{(\vert V_{ud} \vert^2 + \vert V_{us} \vert^2)^{1/2}} \, , \quad {\color{black} A \lambda^2} = \frac{\vert V_{cb} \vert}{(\vert V_{ud} \vert^2 + \vert V_{us} \vert^2)^{1/2}} \, , \quad {\color{black} \bar\rho+i \bar\eta} = - \frac{V^{}_{ud} V^{*}_{ub}}{V^{}_{cd} V^{*}_{cb}} \,,
\end{equation}
where no assumption is made about the size of $ \lambda $. The small size of $ \lambda $ accommodates the hierarchical structure just evoked.
The unitarity of the CKM matrix leads to a useful graphical representation: in the left panel of Fig.~\ref{fig:rhoetaBd_pulls}, the so-called $B_d$ unitarity triangle is displayed.
It is of utmost importance to appreciate that the single parameter $\bar\eta$ must be at the origin of CP violation across distinct flavour sectors. As seen in Fig.~\ref{fig:rhoetaBd_pulls}, both bottom and strange sectors are represented, while it is still unclear if and how the charm sector fits in this picture \cite{Khodjamirian:2017zdu,Pich:2023kim}.
In the SM, these parameters are constant in time; their variation due to the presence of NP is discussed elsewhere in this document.


A rich variety of observables is used to probe the structure of the CKM matrix, including:

\begin{itemize}
    \item processes dominated by SM tree-level contributions,
    \item processes for which in the SM there is no tree-level contribution,
    \item processes that individually rule out a vanishing CP-violating phase,
    \item or yet processes that are individually compatible with the absence of CP violation.
\end{itemize}

\noindent
On a more technical level, some observables are presently dominated by experimental uncertainties (e.g., $\alpha$, $\beta$, $\gamma$), while others are dominated by theoretical sources of uncertainty.
The latter category of uncertainties is due mainly to hadronic effects, ubiquitous to quark flavour physics.
We heavily rely on Lattice QCD calculations to provide hadronic inputs, more importantly: bag parameters, decay constants, and form factors,
which are relevant for meson-mixing observables, and leptonic and semi-leptonic decays of hadrons.
In the SM or in the presence of NP as in Ref.~\cite{Dine:2024bxv}, these hadronic effects are constant in time. This is used in the latter reference in order to extract any signal pointing at the modulation in time of the CKM matrix elements, for which the goal is not the extraction of the CKM matrix itself.

Over the last two decades, exquisite accuracy has been reached in the extraction of the CKM matrix elements,
thanks to experimental (e.g., $ B $-factories, LHCb, etc.) and theoretical (e.g., Lattice QCD) developments. In the following, we briefly discuss the preliminary 2023 update of the combination and extraction in the SM of the CKM matrix elements by the \textsf{\textit{CKM}fitter} Collaboration~\cite{Hocker:2001xe,Charles:2004jd,Charles:2011va,Charles:2015gya,website}.
Further discussion is found in Ref.~\cite{ValeSilva:2024jml}, while the 2025 update is currently being prepared.
The primary goal is testing the SM, and possibly pointing out tensions in its picture of flavour transitions, very likely to occur in the presence of physics beyond the SM.

\subsubsection{Current and expected precision}

The \textsf{\textit{CKM}fitter} Collaboration employs a frequentist approach based on a $ \chi^2 $ analysis. The scheme used to incorporate theoretical uncertainties is called \textit{Range} fit (\textit{R}fit). It means that one varies freely, without any penalty from the $ \chi^2 $, the true value of the fixed (and thus not stochastic), but unknown, theoretical effect $ \delta $ inside the quoted uncertainty $ \pm \Delta $, i.e., $ \delta \in [- \Delta, \Delta] $.
The minimum of $ \chi^2 $ sets the goodness of the fit.
Subsequently, Confidence Level (C.L.) intervals are built when studying the $ \chi^2 $ around the best-fit ``point'' determined by the argument of $\chi^2_{min}$.
Important theoretical uncertainties are usually present in the quark flavour sector. Further details are found in the references already quoted.
While here we adopt the \textit{R}fit scheme introduced above, other frequentist methodologies are discussed in Ref.~\cite{Charles:2016qtt}.

\begin{table}[h]
    \centering
    \begin{tabular}{ccccc}
        Quantity & Current (\textit{R}fit) & Unc. Phase I & Unc. Phase II & Unc. Phase III \\
        $ \sin (2 \beta)_{[c \bar{c}]} $ & 0.011 & 0.005 & 0.002 & 0.0008 \\
        $ \gamma $ & 3.4$^{\circ}$ & 1$^{\circ}$ & 0.25$^{\circ}$ & 0.20$^{\circ}$ \\
        $ | V_{ub} | / | V_{cb} | $ & 10 \% & 3 \% & 1 \% & id \\
        $ (\phi_s)_{[b \to c \bar{c} s]} $ & 21 mrad & 14 mrad & 4 mrad & 2 mrad \\
    \end{tabular}
    \caption{Current and future (relative) $ 68 \% $ uncertainties of some of the quantities used in the global fit analysis. ``id'' stands for the same uncertainty as in the previous entry of the same row. LHC will play a crucial role in improving the extraction of these quantities in Phases~I and II.}
    \label{tab:EXP_projections_1}
\end{table}

\begin{table}[h]
    \centering
    \begin{tabular}{ccccc}
        Quantity & Current (\textit{R}fit) & Unc. Phase I & Unc. Phase II & Unc. Phase III \\
        $ | V_{ub} |_{\rm SL} $ & 0.19 & 0.042 & 0.032 & id \\
        $ | V_{cb} | $ & 0.61 & 0.60 & 0.44 & 0.17 \\ 
        $ \alpha $ & 3.7$^{\circ}$ & 0.6$^{\circ}$ & id & id \\
        $ \mathcal{B} (B \to \tau \nu) $ & 22 \% & 5 \% & 2 \% & 1 \% \\
    \end{tabular}
    \caption{Some of the quantities used in the global fit analysis. Belle~II will play a leading role in improving the extraction of these quantities in Phases~I and II. See the caption of Tab.~\ref{tab:EXP_projections_1}.}
    \label{tab:EXP_projections_2}
\end{table}

\begin{table}[h]
    \centering
    \begin{tabular}{cccc}
        Quantity & Current (\textit{R}fit) & Unc. Phase I & Unc. Phases II \& III \\
        $ f_K $ & 0.5 \% & 0.3 \% & id \\
        $ f_+^{K \to \pi} (0) $ & 0.4 \% & 0.1 \% & id \\
        $ f_{B_s} $ & 1.1 \% & 0.5 \% & id \\
        $ B_{B_s} $ & 3.2 \% & 0.8 \% & 0.6 \% \\
        $ f_{B_s} / f_{B_d} $ & 0.6 \% & 0.4 \% & id \\
        $ B_{B_s} / B_{B_d} $ & 2.4 \% & 0.5 \% & 0.3 \% \\
    \end{tabular}
    \caption{Current and future relative uncertainties of some of the theoretical inputs from Lattice QCD used in the global fit analysis.}
    \label{tab:QCD_projections}
\end{table}

\begin{figure}[t]
	\centering
    \includegraphics[scale=0.4,trim={0.5cm 0.5cm 0.3cm 1.8cm},clip]{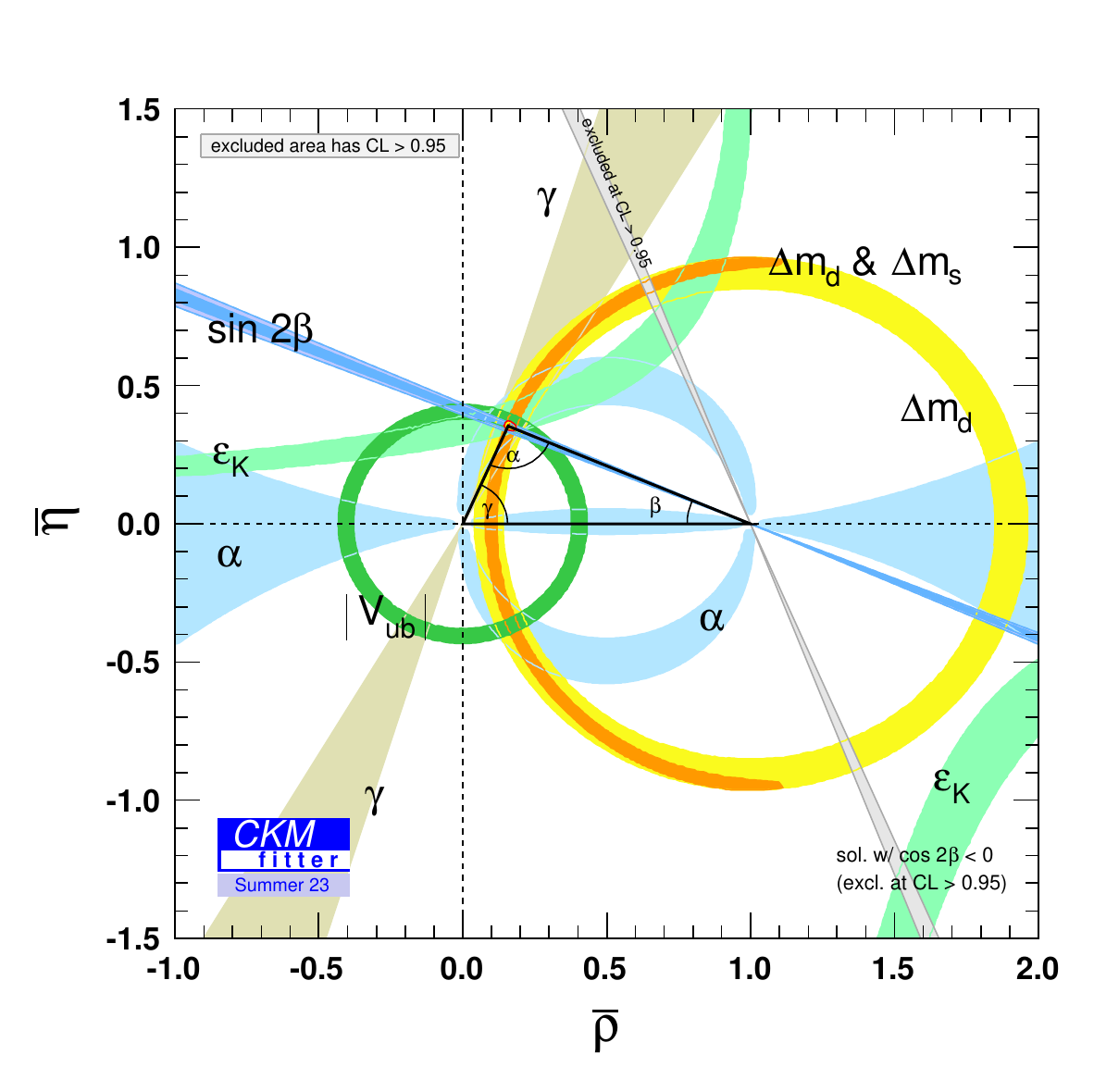} \hspace{8mm}
	\includegraphics[scale=0.415,trim={3cm 0.5cm 0.5cm 0.5cm},clip]{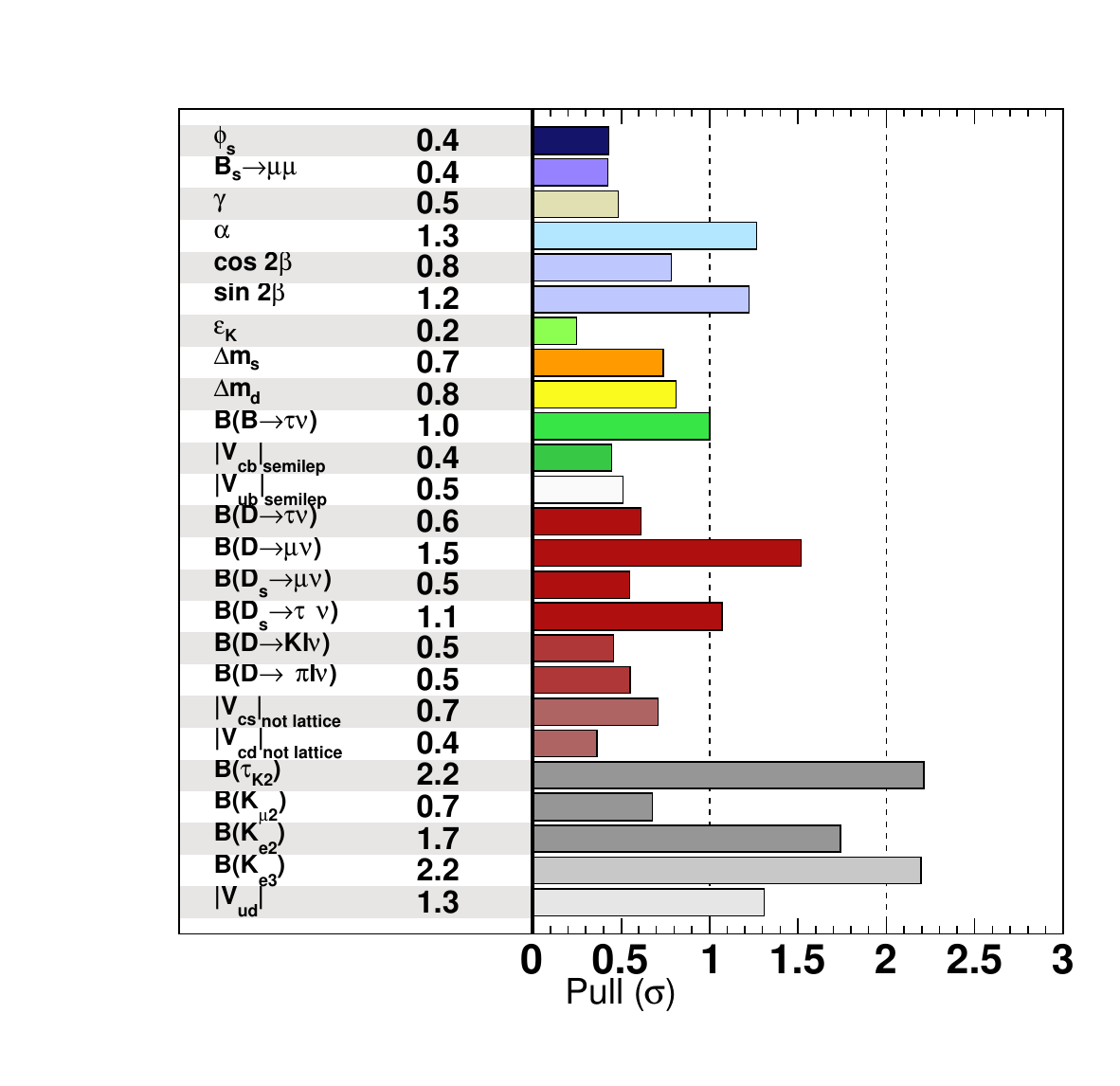}
	\caption{{\it Left panel.} Combination of constraints used in the extraction of the Wolfenstein parameters in the $ (\bar\rho \,, \bar\eta) $ plane. The apex of the triangle, at $ \bar{\rho} \simeq 0.16 $ and $ \bar{\eta} \simeq 0.36 $, corresponds to the best-fit point. Also shown are the $ 68 \% $ (hashed red) and $ 95 \% $ (yellow with red contour) C.L. regions of $ \bar{\rho} $ and $ \bar{\eta} $. {\it Right panel.} Pulls for individual observables in units of $ \sigma $.}\label{fig:rhoetaBd_pulls}
\end{figure}

Experimental and theoretical inputs, together with the pertinent references, are found in Ref.~\cite{website}.
Relative uncertainties are given in Tabs.~\ref{tab:EXP_projections_1}, \ref{tab:EXP_projections_2} and \ref{tab:QCD_projections} for some of them.
Using the SM as the null hypothesis, the results of the global fit are excellent: the p-value is $ 67\% $, which corresponds to $ 0.4\sigma $, if all uncertainties are treated as Gaussian.
This consistent overall picture, as measured by the $\chi^2_{min}$, allows for a meaningful interpretation of the flavour data in terms of the SM CKM matrix. Preliminary results for the relative uncertainties of the extracted Wolfenstein parameters are given in Tab.~\ref{tab:CKM_relative_uncertainties}.
In the global fit analysis, $ \lambda $ is precisely determined from  $ | V_{ud} | $ (superallowed nuclear transitions) and $ | V_{us} | $ (semi-leptonic and leptonic kaon decays), while a precise determination of $ A $ comes from a combination of $ | V_{cb} | $, $ | V_{ub} | $, and further information from the global fit. The extraction of $ \bar{\rho} $ and $ \bar{\eta} $ is illustrated in the left panel of Fig.~\ref{fig:rhoetaBd_pulls}, where it is clear the dominant role played by $ \sin (2 \beta) $ and $ \Delta m_{d} / \Delta m_{s} $, but also $ | V_{ub} |/| V_{cb} | $, in the determination of the CP violating phase.
In this respect, we highlight an important experimental progress in the extraction of the angle $\beta$, with a substantial decrease of the uncertainty that reflects in a more precise extraction of the $(\bar\rho , \, \bar\eta)$ apex. We employ the value $ \sin (2 \beta) = 0.708 \pm 0.011 $ (i.e., an uncertainty on $\beta$ of $0.45^\text{o}$) from a \textsf{HFLAV} combination, to be compared with our previous input $ \sin (2 \beta) = 0.699 \pm 0.017 $ used in the 2021 global fit.

\begin{table}[]
    \centering
    \begin{tabular}{cccccc}
        Quantity & Current (\textit{R}fit) & Fully agree & Unc. Phase I & Unc. Phase II & Unc. Phase III \\
        $A$ & 1.8 \% & 1.9 \% & 1.4 \% & 1.2 \% & 0.9 \% \\
        $\lambda$ & 0.2 \% & 0.3 \% & 0.3 \% & 0.3 \% & 0.3 \% \\
        $\bar{\rho}$ & 10 \% & 9.3 \% & 3.9 \% & 1.8 \% & 1.5 \% \\
        $\bar{\eta}$ & 3.7 \% & 3.6 \% & 1.4 \% & 0.7 \% & 0.4 \% \\
    \end{tabular}
    \caption{Relative uncertainties derived from the sizes of the extracted $ 95 \% $ C.L. intervals with respect to their best-fit values. The column marked as ``Fully agree'' consists of relative uncertainties derived once shifting the central values of the current inputs such that the SM is perfectly consistent with the data (i.e., p-value = 1), when employing Gaussian uncertainties for most quantities. Preliminary results.}
    \label{tab:CKM_relative_uncertainties}
\end{table}

Each observable considered in the fit provides a test of the SM, as measured by the pulls:

\begin{equation}
	Pull_{\mathcal{O}_{}} = \sqrt{\chi^2_{min} - \chi^2_{min, !\mathcal{O}_{}}} \,,
\end{equation}
where $ !\mathcal{O}_{} $ means that the $ \chi^2 $ is built without considering $ \mathcal{O}_{} $.
The values of the pulls given in the right panel of Fig.~\ref{fig:rhoetaBd_pulls} imply that the SM predictions are in good agreement with the individual measurements. Since the observables carry correlations, and since theoretical uncertainties are not distributed normally (indeed, they are not distributed at all), the distribution of the pulls is not normal, and the number of observables which have a pull larger than $ n \times \sigma $, for a certain $ n $, is not properly a meaningful information.
In a similar way, global fits based on selected subsets of inputs (for instance, including only observables which are not induced at the tree level in the SM) show consistency.


Looking towards the future,
flavour will remain an important physics case for collider and non-collider experimental proposals.
Conversely, future data will shape the field, further testing the SM and presenting anomalies, while possibly revealing new tensions.
Presently, we are on the verge of a new era in flavour physics:
LHCb and Belle~II will collect large data sets,
and there are proposals for extending their operations \cite{Cerri:2018ypt,Belle-II:2018jsg}; see Ref.~\cite{Ablikim:2019hff} for the BESIII future physics program.

To discuss projections, we define the benchmark phases:

\begin{itemize}
    \item Phase~I consists of LHCb Upgrade~I with a luminosity of $50/$fb, together with Belle~II with a luminosity of $50/$ab;
    \item in Phase~II, these luminosities are multiplied respectively by 6 and 5, following future upgrade proposals;
    \item and a more speculative Phase~III, which consists of Phase~II together with a promising candidate for the next generation of colliders, namely, the Future Circular Collider (FCC) \cite{FCC:2018byv}, more specifically its electron-positron phase.
\end{itemize}
An important caveat is in order: the projected luminosities of Belle~II and its upgrade may not be achieved in the short term, given the latest updated expectations.
On the other hand, FCC-$e e$ can contribute substantially to flavour physics due to the large number of $Z$ and $W$-pairs produced, which decay into all possible heavy flavours, in a clean experimental environment \cite{Monteil:2021ith,Grossman:2021xfq}.

Projections for the uncertainties of the specific observables included in our analysis are found in Ref.~\cite{Charles:2020dfl} and the references quoted therein; Tabs.~\ref{tab:EXP_projections_1}, \ref{tab:EXP_projections_2} and \ref{tab:QCD_projections} summarize some key observables and theoretical inputs.
Many of these projections are still simplistic, being first attempts to estimate future uncertainties.
We highlight the following:
\begin{itemize}
    \item Two key quantities in probing NP in $B$ meson mixing are $|V_{ub}|$ and $|V_{cb}|$, which are going to be precisely extracted in exclusive semi-leptonic decays at Belle~II \cite{Belle-II:2018jsg}.
    FCC-$e e$ can produce a qualitatively new measurement of $|V_{cb}|$, based on $W^+ \to c \bar{b}$ thanks to the tagging of heavy flavours; see Refs.~\cite{Marzocca:2024mkc,Liang:2024hox} for recent estimates. See Tabs.~\ref{tab:EXP_projections_1}, \ref{tab:EXP_projections_2}.
    \item Starting from Phase~I, uncertainties on the determinations of the angles $\alpha, \gamma$ ($\beta_s$) will be better than $1^\text{o}$ (10~mrad, respectively) \cite{Belle-II:2018jsg,FCC:2018byv,Cerri:2018ypt}. Claiming such precision requires dealing with effects that at the present moment may be sub-leading, but will increase in importance once statistical uncertainties decrease. Two notable examples consist of penguin pollution (affecting notably $ \beta $), and isospin breaking (affecting $ \alpha $). However, these theoretical uncertainties are not included in this current study. See Tabs.~\ref{tab:EXP_projections_1}, \ref{tab:EXP_projections_2}.
    \item Bag parameters from Lattice QCD extractions will be known at a level better than $1 \%$ \cite{Belle-II:2018jsg,Cerri:2018ypt}. For these, projections in the literature go as far as Phase~II, and in the case of Phase~III we take the same inputs as in Phase~II. See Tab.~\ref{tab:QCD_projections}.
\end{itemize}

The projections displayed in Tab.~\ref{tab:CKM_relative_uncertainties}, which updates the analysis shown in Ref.~\cite{Cerri:2018ypt}, were obtained with \textsf{\textit{CKM}live} \cite{ckmlive}, which is an open-access platform, and can therefore be reproduced by any external user. We note in particular that in Phase~I (the less speculative projection), the determination of the apex of the $B_d$ unitarity triangle will improve by a factor of about $ 2.6 $.


\subsubsection{New physics in $ B $ meson mixing}



\begin{figure}
    \centering
    \includegraphics[scale=0.38,clip,bb=15 15 550 470]{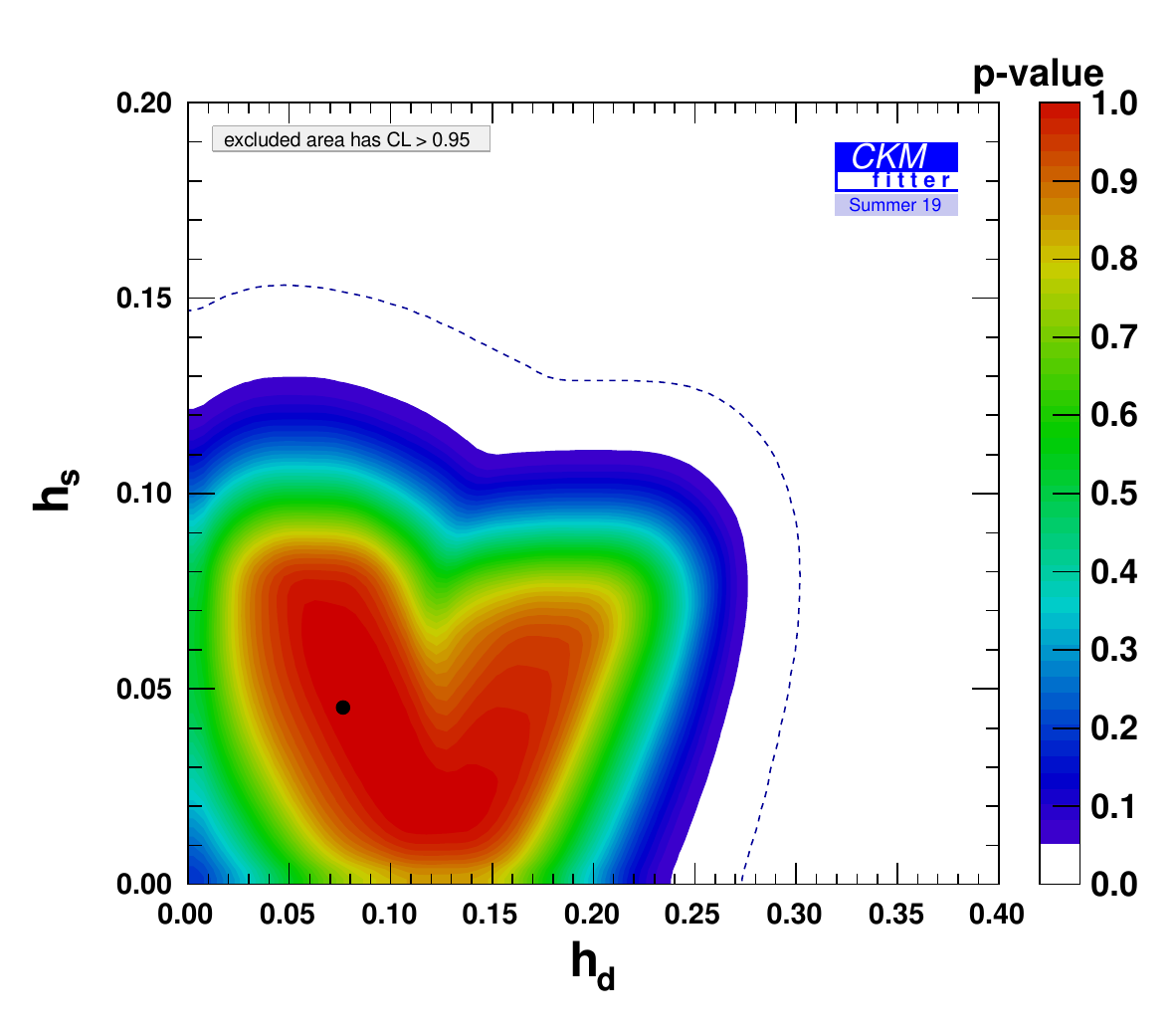}
    \hspace{4mm}
    \includegraphics[scale=0.38,clip,bb=15 15 550 470]{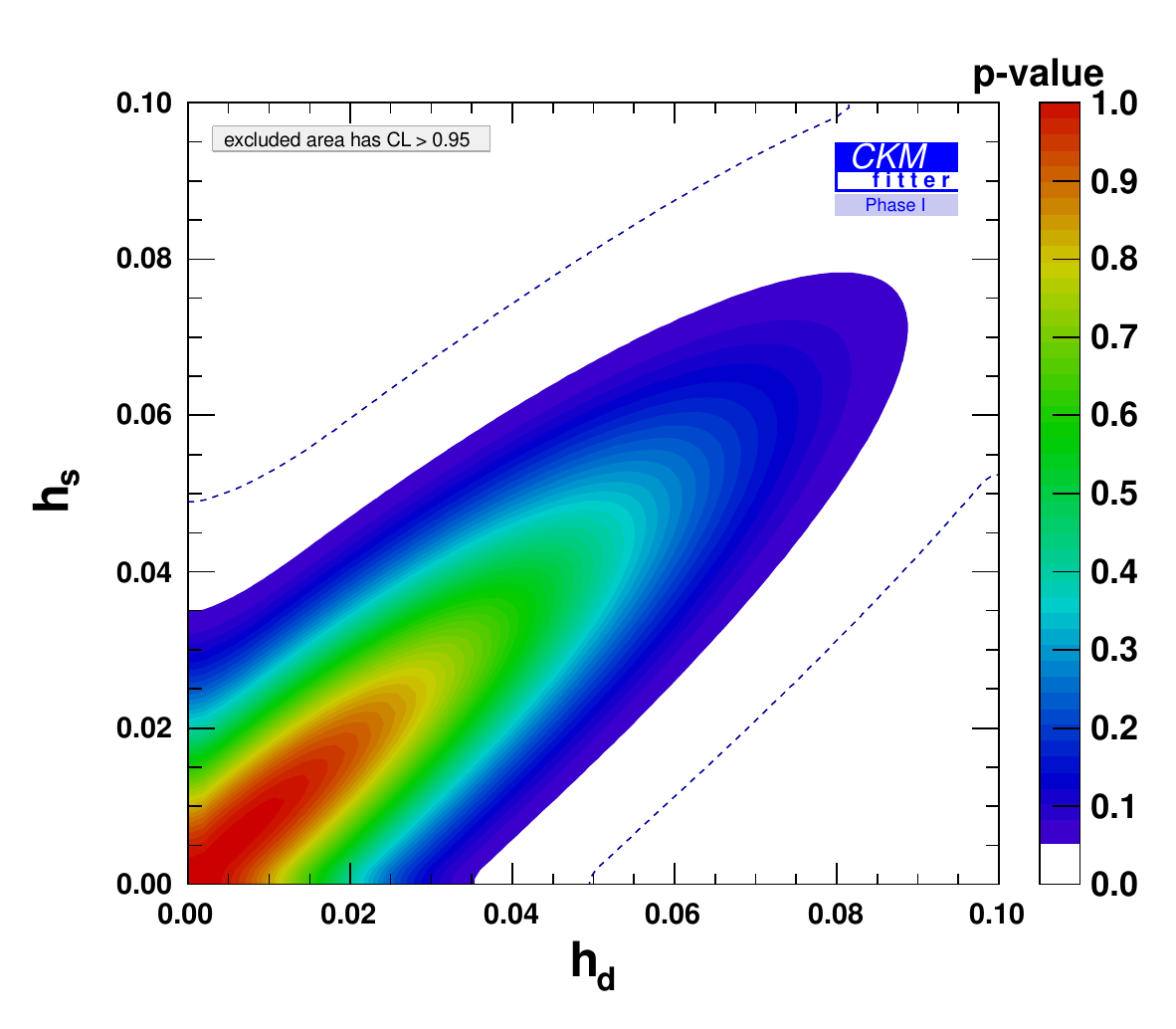}
    \caption{{\it Left panel.} Current sensitivities to $h_d-h_s$ in $B_{d}$ and $B_{s}$ mixings as of Summer 2019 \cite{website}. The black dot indicates the best-fit point. {\it Right panel.} Phase~I sensitivity to $h_d-h_s$ in $B_{d}$ and $B_{s}$ mixings. In both panels, the SM point corresponds to the origin, and the dotted curves show the 99.7\%\,C.L. ($3\sigma$) contours.}
    \label{fig:hdhs}
\end{figure}

The mixing of neutral mesons has been playing a key role in the formulation and testing of the SM. It is also sensitive to some of the highest beyond-the-SM scales probed in laboratory experiments.
The left panel of Fig.~\ref{fig:hdhs} shows the present status of the allowed size of NP \cite{Charles:2020dfl}, where $ h_d $ and $ h_s $ parametrize the sizes of NP contributions relative to the SM, coming from a sector heavy compared to the bottom mass scale.
We observe that the SM point is favored at about $1 \sigma$, but that NP can be as large as $20-30\%$ of the SM.
In presence of NP, the accuracy in the extraction of the Wolfenstein parameters $ \bar{\rho} $ and $ \bar{\eta} $ of the CKM matrix degrade by a factor $3$, illustrating the role of the observables affected by NP of the kind $|\Delta B| = 2$ in extracting these parameters.
Previous analyses are found in Refs.~\cite{Lenz:2010gu,Lenz:2012az,Charles:2013aka}; see also Refs.~\cite{UTfit:2006onp,UTfit:2007eik}.

The planned LHCb Upgrades, Belle~II and its possible upgrade, and the tera-$Z$ phase of the proposed FCC-$e e$ program have a huge potential of unveiling NP contributions affecting flavour observables \cite{Cerri:2018ypt,Belle-II:2018jsg,FCC:2018byv},
neutral meson mixing being only a particular physics case; see also the analyses of Refs.~\cite{DeBruyn:2022zhw,Miro:2024fid,Descotes-Genon:2018foz,Biswas:2021pic}. As seen in the right panel of Fig.~\ref{fig:hdhs}, near future bounds will push these NP contributions below the 10\% level (at 95\% C.L.).
Phase~II is expected to lead to a less impressive improvement of these constraints, by a factor 1.5 compared to Phase~I.
To increase the latter factor, progress is needed in key quantities beyond current expectations, namely, the determinations of hadronic inputs (decay constants and bag parameters) and perturbative QCD corrections, and the extraction of the CKM matrix element $ |V_{cb}| $. Improving the latter by a sizable factor (namely, 20) has an effect similar to the one achievable by adding FCC-$e e$ to Phase~II, which, however, results from a first look into FCC-$e e$ flavour physics capabilities.
These constraints on $h_d-h_s$ translate into sensitivities to tree-level NP contributions to meson mixing at the scale of hundreds ($B_s$ mixing) to thousands ($B_d$ mixing) of TeV.
Therefore, even double insertions of dimension-6 operators at one loop can be used to set strong constraints on their Wilson coefficients \cite{ValeSilva:2022tph}.

\subsubsection{Conclusions}

The SM provides a consistent picture of flavour transitions and CP violation in the quark sector.
Tab.~\ref{tab:CKM_relative_uncertainties} shows the current high level of precision in the extraction of fundamental parameters of the SM, achieved after decades of experimental and theoretical efforts.
Given the large set of precisely measured and predicted observables, we can also consider the extraction of bounds on possible NP contributions.
The size of the allowed NP in $ |\Delta B| = 2 $ processes is still large, of the order of $20-30\%$ of the SM, and has a large impact on the extraction of CKM parameters.
In view of the ongoing and future flavour physics program, we also consider prospects concerning the extraction of CKM-related quantities, and searches for NP in $ |\Delta B| = 2 $.
The extraction of the parameters $\bar\rho$ and $\bar\eta$ describing the unitarity triangle is expected to improve by a factor close to 3 at the beginning of the next decade, while the sensitivity to NP in $ |\Delta B| = 2 $ will increase by a similar factor.

\clearpage
\section{FIP benchmarks: Results and prospects}
\label{sec:portal}

\subsection{FIP phenomenology: Status and challenges --- \textit{M.~Ovchynnikov}}
\label{ssec:fips-pheno}
\textit{Author: Maksym Ovchynnikov, \email{maksym.ovchynnikov@cern.ch}}  \\
\label{sssec:ovchynnikov}
\subsubsection{Introduction}

Searches for FIPs are a very actively developing area of experimental physics~\cite{Beacham:2019nyx,Antel:2023hkf,PBC:2025sny}. Various experiments that may look for FIPs, including main LHC detectors with new FIP-oriented triggers~\cite{CMS:2015sjc,ATLAS:2014fka,LHCb:2015nkv}, LHC-based experiments~\cite{ANUBIS:Satterthwaite:2839063,CODEX-b:2025rck,MATHUSLA:2025eth,FASER:2018bac,SNDLHC:2022ihg,FASER:2020gpr,Gorkavenko:2023nbk}, beam dumps~\cite{NA64:2017vtt,NA62:2017rwk,SHiP:2025ows,Apyan:2022tsd,DUNE:2015lol}, and future colliders~\cite{Blondel:2022qqo,Boyarsky:2022epg}, are currently running, have been recently approved, or are in the stage of actively developing proposals.

These experiments may potentially probe orders of magnitude in the FIP parameter space, in the plane of FIP mass-FIP coupling to the Standard Model. If nothing is found, they will exclude this parameter space. Otherwise, if events with FIPs are observed, they may be utilized to reveal information about FIPs, such as their coupling pattern, spin, or relation to the resolution of BSM problems~\cite{Tastet:2019nqj,Tastet:2021vwp,Mikulenko:2023iqq,Cao:2024rzb,DallaValleGarcia:2025aeq}.

The central point in these efforts is to relate the observed events to the given FIP, which requires understanding its phenomenology at accelerator experiments, i.e., how it may be produced and how it subsequently interacts with detectors (i.e., decays, scatters, etc.). Once the phenomenology is known, it may be incorporated in FIP event generators~\cite{Kling:2021fwx,Jerhot:2022chi,Ovchynnikov:2023cry}, which can be used to obtain ``lightweight'' sensitivities or be interfaced to the full simulation framework of the given experiment.

{\small
    \begin{table}[h!]
        \centering
        \begin{tabular}{|c|c|c|c|c|}
           \hline Model &(Effective) Lagrangian & What it looks like  \\ \hline
            HNL $\bm{N}$ & $\bm{Y\bar{L}\tilde{H}N+\text{\textbf{h.c.}}}$ & \makecell{Heavy neutrino with \\ interaction suppressed by $\bm{U \sim Yv_{h}/m_{N}\ll 1}$} \\ \hline
            Higgs-like scalar $\bm{S}$ & $\bm{c_{1}H^{\dagger}HS^{2}+c_{2}H^{\dagger}HS}$ & \makecell{A light Higgs boson with \\ interaction suppressed by $\bm{\theta \sim c_{2}v_{h}/m_{h}}$} \\ \hline
            Vector mediator $\bm{V}$ & $\bm{-\frac{\epsilon}{2}B_{\mu\nu}V^{\mu\nu}+gV^{\mu}J_{\mu,B}}$ & \makecell{A massive photon/vector meson with \\ interaction suppressed by $\bm{\epsilon/g}$}  \\ \hline
            ALP $\bm{a}$ &$\bm{ag_{a}G^{\mu\nu}\tilde{G}_{\mu\nu}+}\dots$ & \makecell{A $\bm{\pi^{0}/\eta/\eta'}$-like particle with the interaction \\ suppressed by $\bm{f_{\pi}g_{a}}$} \\ \hline
        \end{tabular}
        \caption{Various simplest models adding FIPs: Heavy Neutral Leptons (HNL), Higgs-like scalars, vector mediators, and axion-like particles (ALPs). Here, $\bar{L},H,B,J_{\mu,B},G$ stay for the $SU(2)_{L}$ lepton doublet, the Higgs doublet, the $U(1)_{Y}$ field, the baryon current, and the gluon field.}
        \label{tab:FIPs-characterization}
    \end{table}
}

FIPs with masses in the GeV range are special. On the one hand, they may be copiously produced at accelerator experiments (e.g., in decays of various mesons), which enables the possibility to observe a lot of events with them and extract important insights. In addition, the laboratory and cosmological/astrophysical constraints complementarily define the target FIP parameter space from above and below to be probed by future experiments. The main challenge is the complexity of the description of FIP interactions with hadrons: the mass scale is close to $\Lambda_{\text{QCD}}$. As a result, at some FIP mass scale, we have to match the description of the phenomenology in terms of perturbative QCD with the one based on interactions with various hadronic bound states, such as mesons and nucleons.

In this proceeding, I summarize the status of the phenomenology of the various FIP models. I will concentrate on the simplest models, adding gauge-invariant interactions with the SM fields, either renormalizable or having a simple UV completion~\cite{Alekhin:2015byh,Beacham:2019nyx}. Those include, e.g., dark photons and mediators coupled to the combination of the baryon and lepton currents, Higgs-like scalars, axion-like particles, and heavy neutral leptons.

\subsubsection{FIP phenomenology: how mixing with mesons leads to uncertainties}

The interaction Lagrangian and qualitative characterization of the mentioned FIPs are summarized in Table~\ref{tab:FIPs-characterization}.

For bosonic FIPs, one of the main features making hadronic interactions complicated is their possible linear coupling to quark or gluon bilinear operators. These operators have a non-zero matrix element between the vacuum and a one-particle hadronic bound state $h$, e.g., a meson such as $\pi^{0}$. As a result, there is a mixing between the FIP and $h$, parametrized by the mixing angle $\theta_{Xh}$. Being typically tiny, $\theta_{Xh}$ gets resonantly enhanced around the FIP masses $m_{X} = m_{h}$; thus, mixing dominates the phenomenology of the FIPs. It makes it essential to understand the interactions of the $h$-like particle, which, though, has a different mass than the actual hadron.

The problem is that realistic LLPs mix not with a single meson, but with families of mesons defined by the spin and quantum numbers, such as strangeness. Thus, apart the ``ground state'' $h$, the FIP has mixing with its excitations $h',h''$, etc.; for the example of the mixing with $\pi^{0}$, the excitations are $\pi^{0}(1300)$ and $\pi^{0}(1800)$. On top of that, in realistic models, the mixing with $\pi^{0}$ implies the existence of the mixing with the heavier $\eta,\eta'$ (and their excited states).

Understanding the properties of heavy excitations is far from being complete~\cite{Giacosa:2024epf}. In particular, for particular states, measurements of masses and decay widths have huge experimental error bars. The nature of some of them (for instance, whether they are a 2-quark or a 4-quark bound state) is not fully clarified. Existing frameworks incorporating the interactions of the mesons, such as the extended linear sigma model, do not include all the excitations.

\begin{figure}[t!]
    \centering
    \includegraphics[width=\linewidth]{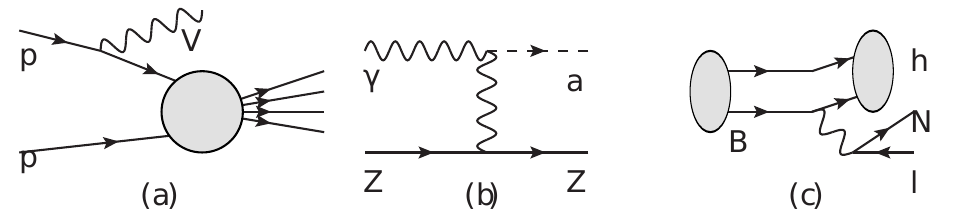}
    \caption{Examples of the production mechanisms of FIPs: from left to right, the proton bremsstrahlung producing a vector FIP $V$, the Primakov scattering of photons off a nucleus $Z$ (producing the ALP $a$), and semileptonic decays of a $B$ meson into an HNL $N$.}
    \label{fig:fip-production}
\end{figure}

Depending on the FIP, the problem may be partially avoided by using experimental data on some processes that share the same interactions as the target new physics particle. However, in most cases, it is impossible or is only possible indirectly. Below, I briefly discuss the situation for the FIPs from Table~\ref{tab:FIPs-characterization}:

\begin{figure}[t!]
    \centering
    \includegraphics[width=0.5\linewidth]{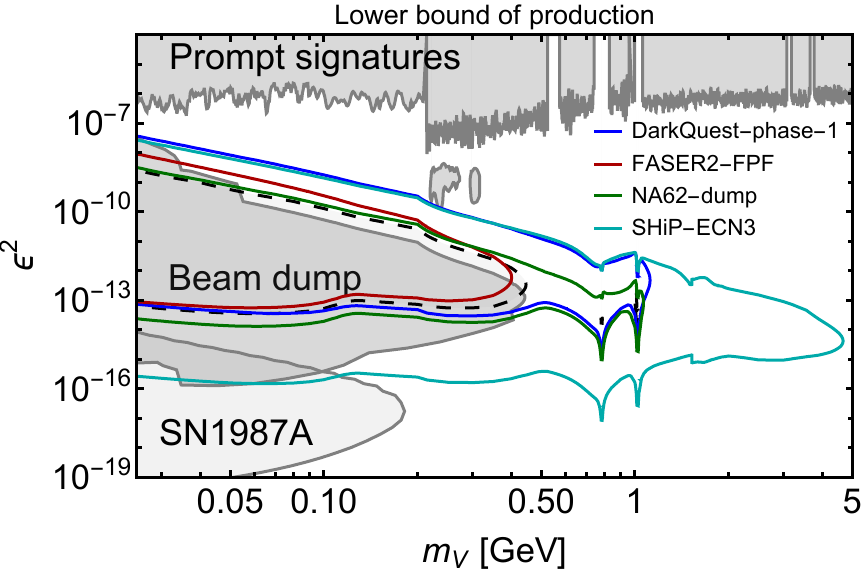}~ \includegraphics[width=0.5\linewidth]{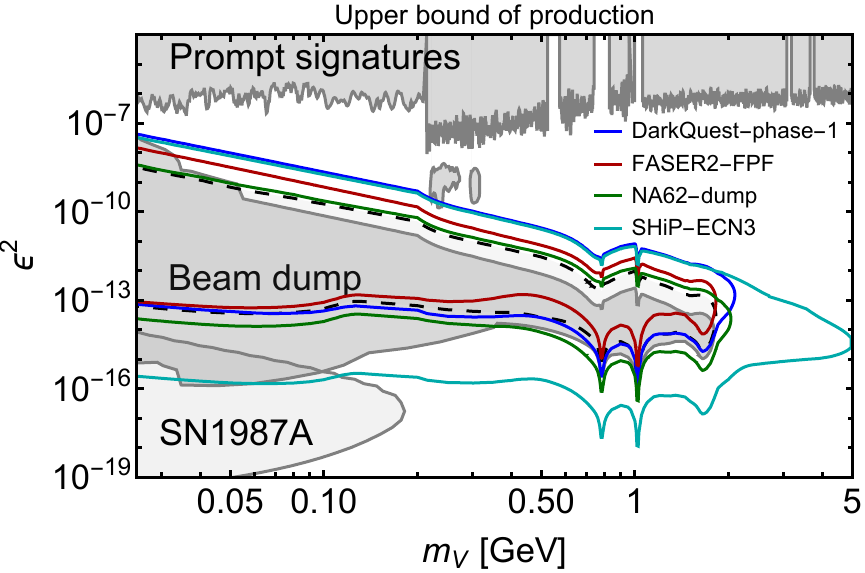}
    \caption{Parameter space of dark photons, including constraints from past searches (shown in gray) and sensitivities of future experiments (shown by colored curves). The left plot shows the parameter space assuming the minimal possible dark photon flux within the theoretical uncertainty, whereas the right plot corresponds to the maximal possible flux. The description of the uncertainty of dark photons and the figures (generated by the \texttt{SensCalc} package~\cite{Ovchynnikov:2023cry}) are taken from Ref.~\cite{Kyselov:2024dmi}.}
    \label{fig:dark-photon-uncertainty}
\end{figure}

\begin{itemize}
    \item \textbf{HNLs.} These particles are fermions behaving like heavy neutrinos; their interaction is described by 4-body Fermi operators, and thus they do not mix with hadrons. Their phenomenology, including production and decays, is well-understood and may be summarized in terms of the semileptonic decays of various mesons and electroweak bosons driven by weak interactions (see Fig.~\ref{fig:fip-production} as an example of this and other production mechanisms of FIPs), and subsequent decays into leptons or a combination of leptons and hadrons~\cite{Bondarenko:2018ptm}. The uncertainty in the HNL decays is within $\mathcal{O}(1)$ and comes from the complexity in merging exclusive decays and perturbative decays into quarks.
    \item \textbf{ALPs without hadronic coupling.} Such particles have relatively clean production and decay patterns. For example, at accelerator experiments, ALPs coupled to the $U(1)_{Y}$ SM field are mainly produced by the Primakov scattering of photons off target or the surrounding infrastructure~\cite{Dobrich:2019dxc}, whereas those coupled to the $SU(2)_{L}$ field are produced by flavor-changing neutral current (FCNC) decays $B\to Y_{s/d}+\text{ALP}$, where $Y_{s/d}$ is a hadronic state including an $s/d$ quark. Decays are mainly purely electromagnetic: $a\to \gamma\gamma$.
    \item \textbf{Dark photons.} Dark photons have mixing with $\rho^{0},\omega,\phi$ mesons and their excitations. Their decay palette may be extracted from the electromagnetic data $e^{+}e^{-}\to \text{hadrons}$: using the framework of Hidden Local Symmetry of vector meson dominance, the scattering events may be decomposed into $\rho,\omega,\phi$-like components and then related to the dark photon decay width via the so-called $R$-ratio~\cite{Ilten:2018crw}. Such an opportunity does not exist for dark photon production at photon accelerators. Despite recent progress~\cite{Foroughi-Abari:2024xlj,Kyselov:2024dmi,Kyselov:2025uez}, the uncertainties in the production may reach 1-2 orders of magnitude. They mainly come from the proton bremsstrahlung (where the elastic proton form-factor includes the contributions from vector meson excitations) and deep inelastic scattering (which probes the domain of low parton's energy fractions $x$ and scales $Q \simeq m_{\text{FIP}}$).
    \item \textbf{Vector mediators coupled to the baryon current.} These particles have mixing with $\omega,\phi$ mesons. The status of their phenomenology is similar to the dark photon case, with one more complexity: for the proton bremsstrahlung production channel, there is no experimental data to fit the elastic proton form-factor.
    \item \textbf{ALPs coupled to hadrons.} These ALPs have mixing with $\pi^{0},\eta,\eta'$ and their excitations. Additional complexity is introduced by the coupling to gluons: to translate it to the interactions with these mesons, one has to perform the ambiguous chiral rotation of the light quark fields. The observables must be independent of the chiral rotation parameter, which is reached by combining the contributions from the mixing of ALPs with different mesons and the multi-field operator from the initial Lagrangian~\cite{Bauer:2021wjo,Ovchynnikov:2025gpx}. Unlike the dark photons, no direct extraction of the decay widths is possible, and one has to build the effective ALP interaction Lagrangian with mesons~\cite{Aloni:2018vki,DallaValleGarcia:2023xhh,Ovchynnikov:2025gpx}. The resulting phenomenology description is very sensitive to the addition of various interaction operators of heavy pseudoscalar mesons, which has to be done in a self-consistent way to reproduce the observables, such as partial decay widths of the mesons; it is subject to further investigation. The production of these ALPs heavily depends on the coupling pattern. For example, if the dominant coupling is to gluons, the main channels are decays of light mesons, proton bremsstrahlung, partonic fragmentation, and the Drell-Yan process, with uncertainties similar to the dark photon case~\cite{DallaValleGarcia:2023xhh,Kyselov:2025uez}. If there is sizeable interaction with top quarks, then, as far as there are many $B$ mesons at the given experiment, the dominant production mode is FCNC decay $B\to X_{s/d}+\text{ALP}$~\cite{DallaValleGarcia:2023xhh}.
    \item \textbf{Higgs-like scalars.} The scalars have mixing with the meson $f_{0}(980)$ and its excitations. The main production mode is the FCNC decay $B\to X_{s}+\text{scalar}$~\cite{Boiarska:2019jym}, unless the experiments have too low yields of $B$\!s. Then, the main channel becomes the proton bremsstrahlung. Similarly to the ALP case, the scalar decay widths cannot be directly extracted from the experimental data. In the literature, they are calculated using the information from the scattering processes $\pi\pi\to \pi \pi/KK$, which provide the input in the framework using the method of dispersion relations~\cite{Monin:2018lee,Blackstone:2024ouf}. The decay uncertainty on the decays into a pair of mesons is within an order of magnitude, whereas the widths of more complicated decays (potentially dominating the decay width of GeV-scale scalars), such as $\text{scalar}\to 4\pi$, have not even been computed.
\end{itemize}

\subsubsection{Impact of uncertainties: dark photons and Higgs-like scalars as examples}

To illustrate the impact of theoretical uncertainties in the FIP phenomenology, let us consider two examples: dark photons and Higgs-like scalars.

The parameter space of dark photons in the GeV mass range is shown in Fig.~\ref{fig:dark-photon-uncertainty}. The dominant constraints arise from prompt searches at LHCb, beam dump experiments such as CHARM and E137, and supernovae~\cite{Antel:2023hkf}. As mentioned above, the main uncertainty in the dark photon phenomenology comes from the dark photon production in the proton bremsstrahlung and the Drell-Yan process. Depending on the position within the uncertainty band, the parameter space that was excluded or will be probed by future experiments may vary by a factor of a few in terms of masses and by 1 order of magnitude in terms of couplings.

\begin{figure}[t!]
    \centering
    \includegraphics[width=0.5\linewidth]{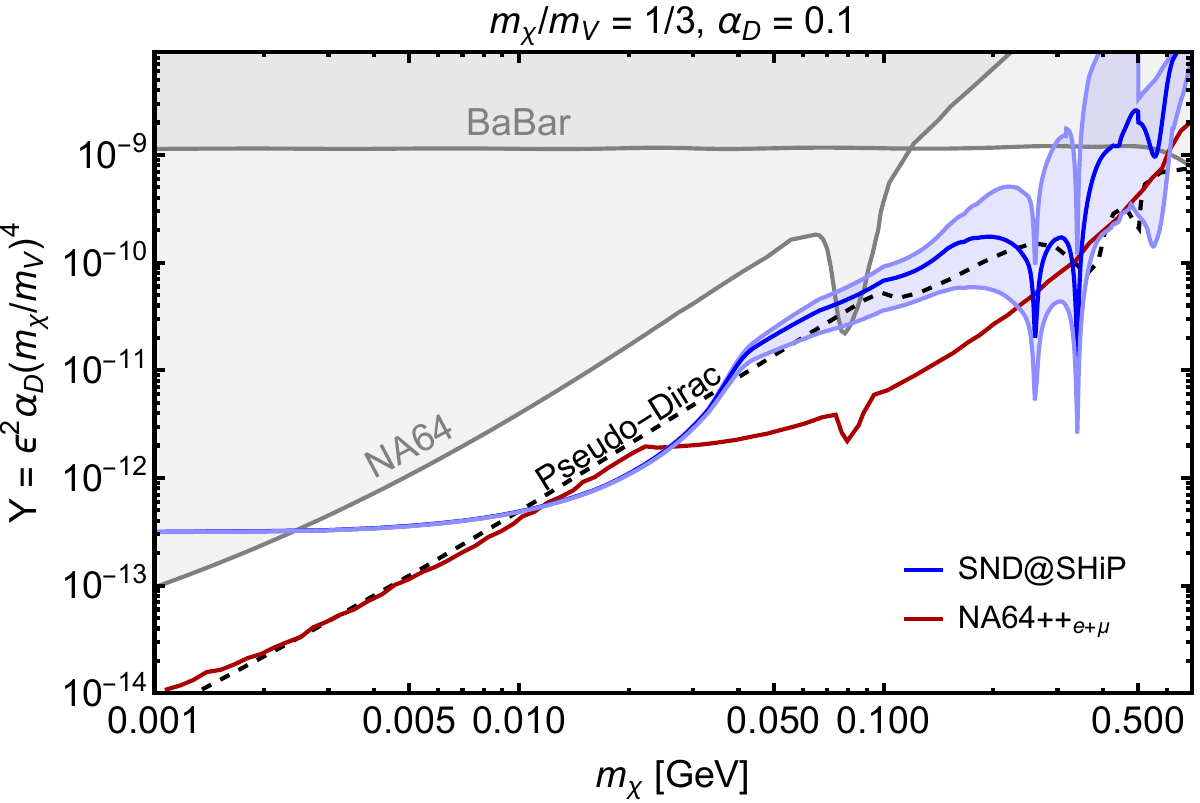}
    \caption{The parameter space of the model of quasi-elastic light dark matter coupled to dark photons (also known as the BC2 model~\cite{Beacham:2019nyx,Antel:2023hkf}). The uncertainty in the dark photon production at proton accelerator experiments translates to the uncertainty of the sensitivity of the SHiP experiment. The figure is obtained using the \texttt{SensCalc} package~\cite{Ovchynnikov:2023cry} and is taken from Ref.~\cite{Kyselov:2024dmi}.}
    \label{fig:LDM-uncertainty}
\end{figure}

The same uncertainty translates to the models of dark sectors coupled via dark photons. As an illustration, consider the model of quasi-elastic light dark matter (LDM) coupled via dark photons (known as the BC2 model~\cite{Beacham:2019nyx}). At laboratory experiments, these particles may be produced in decays of dark photons and then manifest themselves via missing energy or by scattering off electrons or nucleons. At proton accelerator experiments, such as SHiP, the uncertainty in dark photon production leads to comparable uncertainty in the sensitivity to the LDM, see Fig.~\ref{fig:LDM-uncertainty}. Similarly, some other LDM models, including inelastic dark matter or dark matter coupled to different vector mediators, suffer from these uncertainties as well.

\begin{figure}[t!]
    \centering
    \includegraphics[width=0.5\linewidth]{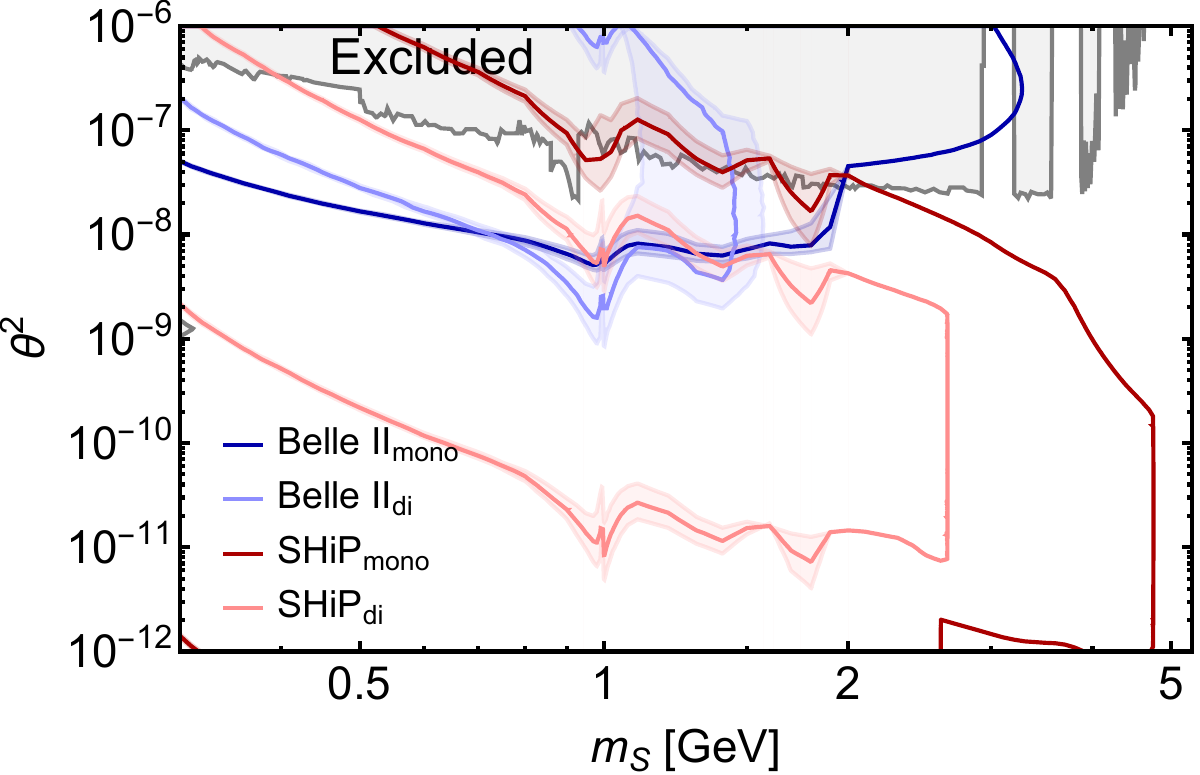}~\includegraphics[width=0.5\linewidth]{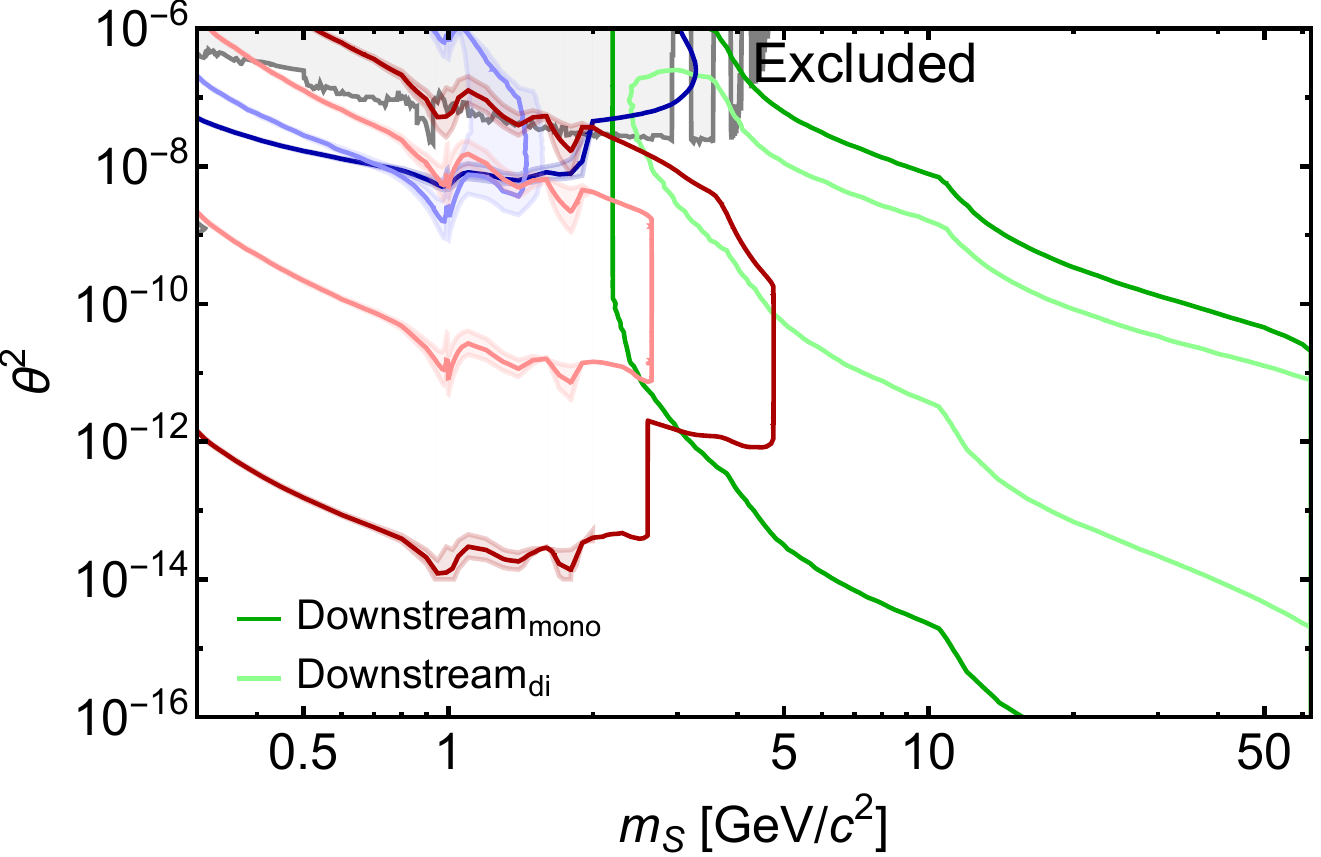}
    \caption{The parameter space of Higgs-like scalars (see Table~\ref{tab:FIPs-characterization}) assuming the branching ratio $\text{Br}(h\to SS) = 0.1$, close to the maximal one allowed by model-independent measurements of Higgs signal strength. The gray domain shows the excluded parameter space, whereas the colored curves denote the sensitivities of different experiments to different signatures with decaying scalars: a single decaying scalar per event (``mono''), and a pair of scalars (``di''). The band shows the theoretical uncertainty in scalar decays from~\cite{Blackstone:2024ouf}. The left panel shows the zoom-in sensitivity of SHiP and Belle II, whereas the right panel also includes the sensitivity of the \texttt{Downstream} algorithm at LHCb~\cite{Gorkavenko:2023nbk,Kholoimov:2025cqe} obtained under conservative selection of the decay events. The figures are obtained using the modification of the \texttt{SensCalc} package~\cite{Ovchynnikov:2023cry} and are taken from Ref.~\cite{DallaValleGarcia:2025aeq}.}
    \label{fig:scalar-uncertainty}
\end{figure}

Now, consider the model of Higgs-like scalars. In the presence of the tri-linear $hSS$ coupling (remind Table~\ref{tab:FIPs-characterization}), long-lived scalars may be searched for by events with one or two displaced decays. Its parameter space is shown in Fig.~\ref{fig:scalar-uncertainty}. Depending on the signature, the theoretical uncertainties may affect the sensitivity by a factor of two in terms of mass and a factor of a few in terms of coupling.

\subsubsection{Conclusions}

Ongoing and future experiments have a powerful potential to explore the parameter space of FIPs. However, there are sizeable theoretical uncertainties in the description of the phenomenology of FIPs with mass in the GeV range. The most significant contribution to the uncertainties comes from hadronic interactions, namely the mixing of FIPs with hadronic bound states such as mesons. While some FIPs, like heavy neutral leptons and ALPs with non-hadronic couplings, do not have mixing, others exhibit it with scalar, pseudoscalar, and vector mesons.

It leads to sizeable uncertainties in both the production and decay of the FIPs. They may span by orders of magnitude and heavily affect the parameter space of FIPs, for example, potentially reopening the domain of masses and couplings that may have been considered as excluded. The status of these uncertainties is tightly connected to the maturity of the spectroscopy of mesons with mass $M<2\text{ GeV}$. Fortunately, the ongoing progress in spectroscopy and improving measurements of various processes that may be used to extract information about the FIP phenomenology may improve the situation in the future.

In the meantime, it is necessary to refine the parameter space of FIPs by incorporating the uncertainties, considering both the domain of excluded masses and couplings and the sensitivities of future experiments. It may be done in a unified and systematic fashion by incorporating the FIP phenomenology together with uncertainties in the available FIP event generators and integrating them into the full simulation frameworks.

\subsection{Dark photon}
\label{ssec:DP}
\changelocaltocdepth{3}
\subsubsection{Dark photon production in proton bremsstrahlung --- \textit{P. Reimitz and E. Kriukova}}
\label{sssec:DP-proton-bremss}
\textit{Author: Peter Reimitz, \email{peter@if.usp.br} and Ekaterina Kriukova, \email{kriukovaea@my.msu.ru}}

\paragraph*{Introduction}

A large number of experiments have already searched for visible decays of dark photons, as promising candidates for new physics~\cite{Lanfranchi:2020crw}. There are also several projects, such as LHCb~\cite{Ilten:2016tkc}, DarkQuest~\cite{Berlin:2018pwi}, T2K~\cite{Araki:2023xgb}, DUNE~\cite{Berryman:2019dme}, and SHiP~\cite{SHiP:2020vbd}, that are going to further perform analogous searches for dark photons produced in proton-proton collisions in the next decade. The feature that they have in common is the focus on the range of dark photon masses around 1 GeV, which was rarely studied earlier. In this region, the known theoretical predictions for dark photon production rates have significant uncertainties~\cite{Foroughi-Abari:2024xlj}. Moreover, some theoretical predictions seriously differ from each other~\cite{Foroughi-Abari:2021zbm,Gorbunov:2023jnx}. This proceeding aims at clarifying these differences and presenting the modern state of the art in proton bremsstrahlung calculations.

The portal formalism serves as a convenient framework to extend the Standard Model (SM) Lagrangian $\mathcal{L}_\text{SM}$ by adding to it terms that are composed of the products of the SM and dark sector operators~\cite{Agrawal:2021dbo}. In this work, we consider the vector portal and its mediator, dark photon $\tilde{A}^\prime_\mu$, in the context of the minimal dark photon model (BC1 in Physics Beyond Colliders classification~\cite{Beacham:2019nyx}), that does not include any other particles of the dark sector. The model Lagrangian is as follows
\begin{equation}
    \mathcal{L}=\mathcal{L}_\text{SM}-\frac{1}{4}\tilde{F}^\prime_{\mu \nu}\tilde{F}^{\prime \mu \nu}-\frac{\epsilon}{2\cos\theta_W}\tilde{F}^\prime_{\mu\nu}B^{\mu \nu}+\frac{m_{\gamma'}^2}{2}\tilde{A}^\prime_\mu \tilde{A}^{\prime \mu},
\end{equation}
where $\tilde{F}^\prime_{\mu\nu}$ and $B_{\mu \nu}$ are the dark photon and the SM hypercharge field strength tensors, and the model parameters are the kinetic mixing parameter $\epsilon$ and dark photon mass $m_{\gamma'}$. By the simultaneous rotation of the SM gauge fields with the dark photon field, one can diagonalize the kinetic terms and obtain the interaction of the dark photon with the SM electromagnetic current $J_\text{em}^\mu$ in the form $-\epsilon e J_\text{em}^\mu A^{\prime}_\mu$.  

For dark photon masses around 1 GeV, the most popular production mechanisms are neutral meson decays ($m_{\gamma^\prime} < 0.4$\,GeV), proton bremsstrahlung ($0.4\text{\,GeV} < m_{\gamma^\prime} < 1.8$\,GeV), and the Drell-Yan process ($m_{\gamma^\prime} > 1.8$\,GeV). Next, we will focus on the predictions for proton bremsstrahlung.

\paragraph*{Elastic bremsstrahlung}

An important, although not the dominant type of proton bremsstrahlung, is the elastic process $pp\rightarrow \gamma^\prime pp$. Some key steps towards the calculation of its cross section were previously made in~\cite{Kim:1973he}, where the so-called generalized Weizsacker-Williams (WW) approximation was obtained for electron bremsstrahlung. There, the electron bremsstrahlung cross section was connected with the cross section of the $2\rightarrow 2$ subprocess containing the virtual photon approximated as being on-shell (i.e. the square of its 4-momentum $q^2\simeq 0$). 

This result has been recently applied to proton bremsstrahlung, giving its differential cross section in the WW approximation~\cite{Foroughi-Abari:2021zbm,Gorbunov:2023jnx}
\begin{equation} \label{eq:WW}
\left[\frac{\dd^2 \sigma(pp\rightarrow \gamma^\prime pp)}{\dd z \dd k^2_\perp}\right]_\text{WW}=\frac{\epsilon^2\alpha z(1-z)}{16\pi^2H^2} A_{2\rightarrow 2} \chi_\mathbb{P} \abs{F_{\text{VMD}}}^2 F_{\text{virt}}^2\,,
\end{equation}
where $z$ is the ratio of 3-momentum taken by the dark photon from the incident proton in the beam direction in the laboratory frame, $k_\perp$ is the transverse momentum of the dark photon, $\alpha$ is the fine-structure constant, and $H$ is the frequently occurring combination, characterizing the off-shellness of the virtual proton with mass $m_p$
\begin{equation}
H\equiv k^2_\perp+(1-z)m^2_{\gamma '}+z^2m_p^2.
\end{equation}
The off-shellness of the virtual proton with 4-momentum $p-k$ is taken into account by the phenomenological hadronic form factor 
\begin{equation} \label{eq:off-shell}
F_{\text{virt}}\equiv \frac{\Lambda_p^4}{\Lambda_p^4+\left((p-k)^2-m_p^2\right)^2}
\end{equation}
with the scale $\Lambda_p=1.5\text{ GeV}$~\cite{Feuster:1998cj}.
The expressions for the $2\rightarrow 2$ subprocess matrix element $A_{2\rightarrow 2}$, the hypothetical vector pomeron flux emitted by the target proton $\chi_\mathbb{P}$ (analogous to photon flux in the case of electron bremsstrahlung) and the vector meson dominance form factor $F_{\text{VMD}}$ can be found in~\cite{Foroughi-Abari:2021zbm,Gorbunov:2023jnx}. 

The validity of the WW approximation for the production of various mediators in electron bremsstrahlung was checked in Refs.~\cite{Liu:2017htz,Voronchikhin:2024vfu,Voronchikhin:2024ygo}. For elastic proton bremsstrahlung, the differential hypothetical pomeron flux peaks at a non-negligible scale $\sqrt{t}$ of about 350 MeV (see Fig.\,1 in~\cite{Gorbunov:2023jnx}), which motivated the further study of its applicability in~\cite{Gorbunov:2023jnx,Kriukova:2024wsi}, which we discuss below.   

The Blumlein-Brunner (BB) approximation that has been widely used in dark photon phenomenology earlier assumes that protons exchange a hypothetical virtual vector boson with zero 4-momentum \cite{Blumlein:2013cua}. Despite the attractiveness and simplicity of the result obtained therein, we caution against using the latter approximation due to insufficient justification of its applicability. It has been shown that the BB approximation significantly overestimates the elastic bremsstrahlung cross section~\cite{Gorbunov:2023jnx}. 

The elastic proton bremsstrahlung as the sum of two contributions with the initial and final state radiation is studied explicitly in the recent works \cite{Foroughi-Abari:2021zbm,Gorbunov:2023jnx,Kriukova:2024wsi}. The $2\rightarrow 3$ elastic bremsstrahlung process with virtual vector Donnachie-Landshoff pomeron exchange \cite{Donnachie:1983hf} is examined numerically in \cite{Foroughi-Abari:2021zbm}. Fig.~7 of~\cite{Foroughi-Abari:2021zbm} shows its good agreement with the WW approximation~\eqref{eq:WW}. In \cite{Gorbunov:2023jnx,Kriukova:2024wsi} the cross section of the same process is calculated analytically for protons exchanging a massless hypothetical vector particle with a photon propagator. In contrast to \cite{Kim:1973he,Blumlein:2013cua}, the non-zero momentum transfer between the protons $t\equiv -q^2$ is taken into account there. The resulting differential cross section explicitly depends on $t$ (see (3.38) in~\cite{Gorbunov:2023jnx}), but at the same time agrees with the WW approximation~\eqref{eq:WW} within the accuracy of 3--9\,\%.

\begin{figure}[t]
    \centering
    \includegraphics[width=0.7\linewidth]{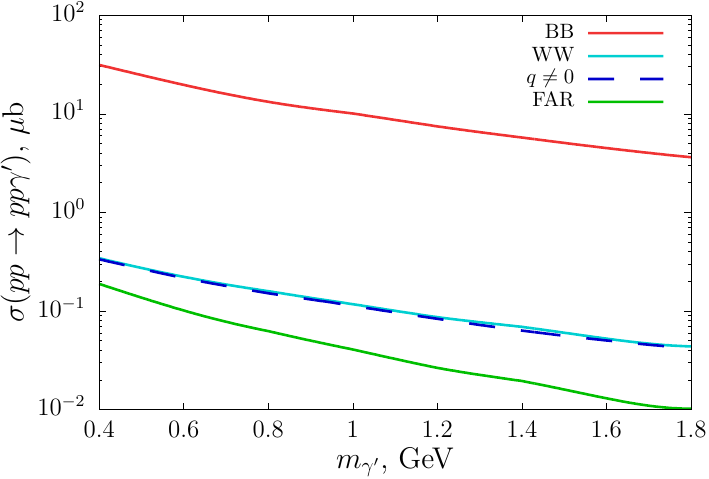}~
    \caption{Full elastic bremsstrahlung cross section as function of dark photon mass without form factor $F_\text{VMD}$ obtained in BB approximation~\cite{Blumlein:2013cua} (red line), WW approximation~\cite{Kim:1973he} (light blue line), direct $2\rightarrow3$ process calculation with vector pomeron~\cite{Foroughi-Abari:2021zbm} (green line) and by taking into account the non-zero momentum transfer between the protons~\cite{Gorbunov:2023jnx} (dark blue dached line). The figure is taken from~\cite{Gorbunov:2023jnx}.}
    \label{fig:comp-elastic}
\end{figure}

The results obtained in \cite{Foroughi-Abari:2021zbm,Gorbunov:2023jnx,Kim:1973he,Blumlein:2013cua} for the full elastic bremsstrahlung cross section are compared in figure~\ref{fig:comp-elastic}
(see also figures 5,\,6 of \cite{Gorbunov:2023jnx} and figures 1,\,2 in~\cite{Foroughi-Abari:2021zbm} for the comparison of results for differential cross section). All of them, except the BB approximation~\cite{Blumlein:2013cua}, are consistent with each other. The apparent difference between the result of~\cite{Foroughi-Abari:2021zbm} and the WW approximation is due to the application of the phenomenological hadronic form factor~\eqref{eq:off-shell} in the former. To summarize, the WW approximation~\eqref{eq:WW} can be used to calculate the elastic proton bremsstrahlung with good accuracy. It is also important to note that the BB answer~\cite{Blumlein:2013cua} is very different from all the others, and there are difficulties in reproducing it, so it should be used with caution for estimating the cross section of both elastic and inelastic proton bremsstrahlung. 

\paragraph*{Inelastic bremsstrahlung}
The dominant contribution to proton bremsstrahlung comes from the inelastic process $pp\rightarrow\gamma^\prime X$, which we consider in this paragraph. Following the tradition~\cite{Foroughi-Abari:2021zbm}, we study the initial state radiation (ISR) of the dark photon and neglect the input of the final state radiation, during which the dark photon is produced by the products of proton dissociation. Thus, the dark photon is produced by the vertex $pp\gamma^\prime$, and it is important to accurately include the input of proton electromagnetic form factors. 

\begin{figure}[t]
    \centering
    \includegraphics[width=0.7\linewidth]{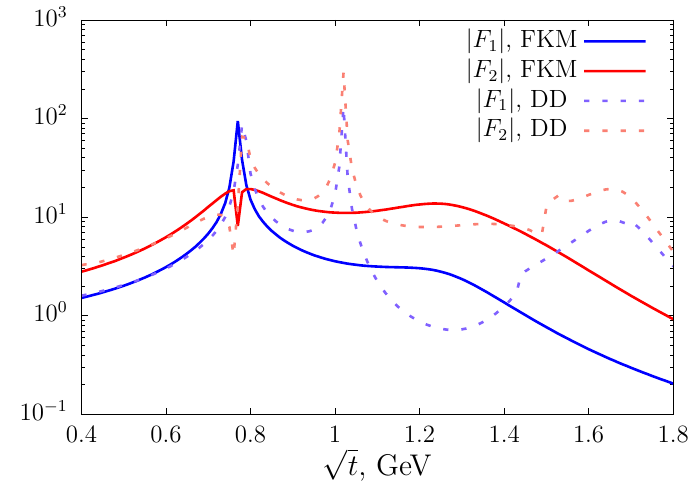}~
    \caption{Absolute values of Dirac $F_1(t)$ and Pauli $F_2(t)$ proton electromagnetic form factors according to the extended vector meson dominance model~\cite{Faessler:2009tn} (solid lines) and to the unitary and analytic model~\cite{Dubnickova:2020heq} (dashed lines).}
    \label{fig:absFFs}
\end{figure}

If both protons are on-shell, the matrix elements of the electromagnetic current $j^\mu_\text{em}\equiv\sum_i Q_i \bar{q_i} \gamma^\mu q_i$ can be expressed through the Dirac $F_1(t)$ and Pauli $F_2(t)$ electromagnetic form factors of the proton
\begin{equation}
    \bra{p(p_2)}j^\mu_\text{em}\ket{p(p_1)}\equiv \overline{u}(p_2) \left[F_1(t)\gamma^\mu+i\frac{F_2(t)}{2m_p}\sigma^{\mu\nu}(p_2-p_1)_\nu\right] u(p_1),	
\end{equation}
where $\sigma_{\mu \nu}\equiv i\left[\gamma_\mu, \gamma_\nu\right]/2$ and the (dark) photon momentum squared $t\equiv (p_2-p_1)^2$ in our case is equal to $m^2_{\gamma^\prime}$. The electromagnetic form factors of the proton can be measured in two kinds of experiments, depending on the value of $t$. For space-like momentum transfer ($t<0$), they are real and are measured in the electron-proton scattering, while for time-like $t$ they are complex-valued and can be measured with the help of electron-positron annihilation for $t$ above the threshold, $t>4m_p^2$. The intermediate region $0<t<4m_p^2$ is called \textit{unphysical} and is not accessible for measurements with $2\rightarrow 2$ processes. Thus, the value of proton electromagnetic form factors there, that is relevant for our study of dark photons with masses 0.4--1.8\,GeV, can be obtained only as a result of an interpolation between data in space-like and time-like regions. Figure~\ref{fig:absFFs}
shows the absolute values of the Dirac $F_1(t)$ and Pauli $F_2(t)$ proton electromagnetic form factors in the unphysical region according to two sets of fits, based on the well-known extended vector meson dominance model with artificial resonances~\cite{Faessler:2009tn} (for the recent update of this fit see~\cite{Kuzmin:2024ozz}) and on the unitary and analytic model containing the impacts of experimentally confirmed neutral vector mesons~\cite{Dubnickova:2020heq}. As can be seen from figure~\ref{fig:absFFs}, the electromagnetic form factors in the unphysical region significantly depend on the fit used, and this inevitably introduces uncertainty in the prediction for the inelastic proton bremsstrahlung cross section (see the discussion of possible uncertainties below). For a recent study presenting updated nucleon form factors applicable
to a broader class of BSM scenarios, and addressing both proton and
neutron bremsstrahlung production, see Ref.~\cite{Kling:2025udr}

\begin{figure}[t]
    \centering
    \includegraphics[width=0.4\linewidth]{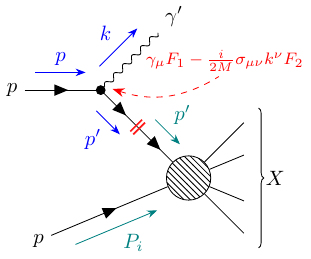}
    \caption{Feynman diagram for dark photon radiation in the initial state as the most important contribution to inelastic proton bremsstrahlung.}
    \label{fig:inel-brem}
\end{figure}

To estimate the dark photon production during the inelastic bremsstrahlung cross section, we consider the dominant contribution of the initial state radiation from the incident proton in the laboratory frame shown in Figure~\ref {fig:inel-brem}.
The momenta of the incident proton and dark photon read
\begin{align}
    p &= \{E_p, 0, 0, P\},\\
    k &= \{E_k, k_\perp \cos \varphi, k_\perp \sin \varphi, zP\}.
\end{align}
Now we briefly describe the factorization procedure that has been implemented in recent works~\cite{Foroughi-Abari:2024xlj,Boiarska:2019jym,Foroughi-Abari:2021zbm,Gorbunov:2024iyu,Gorbunov:2024vrc}. We consider the intermediate protons closest to the mass surface and thus giving the largest contribution to the bremsstrahlung cross section. For such protons, the diagram in figure~\ref{fig:inel-brem} is cut along the double red line. Thus, it is possible to extract in the cross section the contribution of the inelastic process $pp\rightarrow X$ and the master splitting function $w_\text{mas}(z,k^2_\perp)$, characterizing the probability that the incident proton emits a dark photon. As a result, the differential cross section of inelastic bremsstrahlung factorizes into the form
\begin{equation}
    \frac{\dd^2 \sigma(pp\rightarrow \gamma^\prime X)}{\dd z \dd k^2_\perp}\simeq w_\text{mas}(z, k^2_\perp) F^2_\text{virt}(z, k^2_\perp) \sigma(pp\rightarrow X),
\end{equation}
where we use the phenomenological hadronic form factor~\eqref{eq:off-shell} to keep only the intermediate protons that are the closest to the mass shell. 
The master splitting function $w_\text{mas}(z, k^2_\perp)$ is expressed in terms of quadratic combinations of the form factors~\cite{Foroughi-Abari:2021zbm,Gorbunov:2024vrc}
\begin{equation} 
    w_\text{mas}(z, k^2_\perp)\equiv w_1(z, k^2_\perp)|F_1|^2+w_2(z, k^2_\perp)|F_2|^2+w_{12}(z, k^2_\perp)\Re\left(F_1F^*_2\right),
\end{equation}
where three auxiliary splitting functions read
\begin{align} 
    w_1(z, k^2_\perp)&\equiv \frac{\epsilon^2 \alpha}{2\pi H} \left(z-\frac{z\left(1-z\right)}{H}\left(2m_p^2+m^2_{\gamma^\prime}\right)+\frac{H}{2zm^2_{\gamma^\prime}}\right),\label{eq:w1}\\
    w_2(z, k^2_\perp)&\equiv \frac{\epsilon^2 \alpha}{2\pi H} \frac{m^2_{\gamma^\prime}}{8m_p^2} \left(z-\frac{z\left(1-z\right)}{H}\left(8m_p^2+m^2_{\gamma^\prime}\right)+\frac{2H}{zm^2_{\gamma^\prime}}\right),\label{eq:w2}\\ 
    w_{12}(z, k^2_\perp)&\equiv \frac{\epsilon^2 \alpha}{2\pi H} \left(\frac{3z}{2}-\frac{3m^2_{\gamma^\prime}z\left(1-z\right)}{H}\right).\label{eq:w12}
\end{align}
The detailed derivation of the Dirac form factor contribution, $w_1(z, k^2_\perp)$, can be found in~\cite{Foroughi-Abari:2021zbm}. Two other splitting functions, $w_2(z, k^2_\perp)$ and $w_{12}(z, k^2_\perp)$, also related to the Pauli form factor, were obtained in the recent works~\cite{Gorbunov:2024iyu,Gorbunov:2024vrc}. At the same time, it was suggested to use the effective auxiliary splitting functions~\cite{Foroughi-Abari:2024xlj}. The Dawson correction, which was applied in Ref.~\cite{Foroughi-Abari:2024xlj} to the auxiliary splitting functions in order to eliminate potential $1/m^2_{\gamma^\prime}$ singularity arising for small dark photon masses, will be discussed in the next paragraph.

In the end, it is important to note that the absolute value of the Pauli form factor $F_2(t)$ in a wide range of masses exceeds the absolute value of the Dirac form factor $F_1(t)$. Therefore, to estimate the inelastic bremsstrahlung cross section, it is necessary to take into account \textit{three} auxiliary splitting functions \eqref{eq:w1}--\eqref{eq:w12} or their effective analogues \eqref{eq:w2},\,\eqref{eq:w1eff}--\eqref{eq:w12eff} instead of only the Dirac contribution, as it was previously accepted in the literature.

\paragraph*{Dawson Approach}

Direct calculations of the splitting functions using momenta in the infinite momentum frame lead to a splitting function $w_1(z, k^2_\perp)$ with an unphysical $1/m_{\gamma^\prime}^2$ singularity, arising from the longitudinal polarization of the vector boson. A similar unphysical divergence for vector boson mass $m_V\to 0$ appears at the quark-level in $q\to q' V$, as observed in~\cite{Masouminia:2021kne}. To resolve this, Ref.~\cite{Masouminia:2021kne} employs the Dawson approach~\cite{Dawson:1984gx} which removes the longitudinal polarization component proportional to $k^\mu$. This is justified by the relation $\bar{u}(p')\slashed{k} u(p)=0$, implying that the longitudinal polarization term proportional to $k^\mu$ vanishes when sandwiched between spinors of the incoming and outgoing fermions with four momenta $p$ and $p'=p-k$. Note, however, that $p'=p-k$ does not hold in the infinite momentum frame where $E_{p'}\neq E_p-E_k$. 

The effective $w_1(z, k^2_\perp)$ splitting function then takes the form
\begin{align} \label{eq:w1eff}
    w^{\rm eff}_{1}(z,k^2_\perp)=\frac{\alpha \epsilon^2}{2\pi H}\bigg[ \frac{1+(1-z)^2}{z} -z(1-z)\bigg(\frac{2m_p^2+m_{\gamma^\prime}^2}{H}\bigg)
\bigg],
\end{align}
which remains well-behaved in the massless limit and resembles the Altarelli-Parisi splitting kernel for the SM photon for $m_{\gamma^\prime}\to 0$. This correction also slightly modifies $w_{12}(z, k^2_\perp)$ to
\begin{align} \label{eq:w12eff}
    w^{\rm eff}_{12}(z, k^2_\perp) = \frac{\alpha\epsilon^2}{2\pi H}\bigg[ z-z(1-z)\frac{3m_{\gamma^\prime}^2}{H}\bigg].
\end{align}
The splitting function $w_2(z, k^2_\perp)$, however, remains unaffected if $k^\mu$ is used instead of $(p'-p)^\mu$ in the calculation of the splitting function involving the $F_2$ component of the nucleon couplings. Specifically, in the vertex term $\sigma^{\mu\nu}k_\mu \varepsilon_{\nu,L}$ the component of the longitudinal polarization $\varepsilon_L$ proportional to $k^\mu$ cancels anyway since $\sigma^{\mu\nu}k_\mu k_\nu=0$. 

In Ref.~\cite{Foroughi-Abari:2024xlj}, $(p'-p)_\mu$ has been used, leading to deviations in the computed $w_{12},w_2$ splitting kernels compared to the results presented here. The fact that $p\neq p'+k$  appears to influence the results, potentially introducing an unphysical divergence while also impacting $w_{12}$ and $w_2$ depending on the chosen momentum definitions in the tensor part of the splitting function calculation. This issue requires further investigation in future studies.

\paragraph*{Uncertainties}

We identify two major sources of uncertainties for dark photon production in proton bremsstrahlung:
\begin{itemize}
    \item electromagnetic form factor uncertainties
    \item off-shell form factor uncertainties
\end{itemize}
The dark photon production rate is highly sensitive to the choice of the electromagnetic form factors. Refs.~\cite{Foroughi-Abari:2024xlj,Gorbunov:2024iyu,Gorbunov:2024vrc} discuss two common approaches to describe and parametrize the form factors: the Dispersion-relation (DR) model~\cite{Lin:2021xrc}, and the Unitarity-Analytic (UA) model~\cite{Adamuscin:2016rer}. Both models yield similar behaviour for the real-valued Sachs form factors $G_M(t), G_E(t)$, and $G_{\rm eff}(t)$ in the physical regions for $t<0$ and $t>4m_p^2$ as well as below the $\rho,\omega,\phi$ resonances. However, in the crucial GeV-scale dark photon production region, $1~{\rm GeV}^2< t<4m_p^2$, they exhibit significant differences. 

While the DR model provides an excellent fit to both proton and neutron data, it suffers from the presence of unphysical, zero-width resonances above the $\phi$ mass, making it less suitable for describing dark photon bremsstrahlung in that region. As an alternative,~\cite{Foroughi-Abari:2024xlj,Gorbunov:2024iyu,Gorbunov:2024vrc} adopt the UA model as a working solution, which, while slightly less precise in fitting data, avoids unphysical resonances in the key region of interest. All in all, the models primarily differ in the region $m_\phi^2 < t < 4m_p^2$. To assess the intrinsic uncertainty in choosing the fit function parametrization, we vary the resonance masses around their documented experimental values~\cite{ParticleDataGroup:2022pth}. The results can be found in Fig.\,2 of~\cite{Foroughi-Abari:2024xlj}. Further varying the widths would significantly inflate the uncertainty band significantly, reducing confidence in the fit in this region as indicated by dashed lines in the final production rates of Ref.~\cite{Foroughi-Abari:2024xlj}.

The second major source of uncertainty arises from the off-shell form factor~\eqref{eq:off-shell}. In the quasi-real approximation (QRA), the intermediate momentum $p'$ should not be too far off-shell, though the precise definition of "too far" remains ambiguous. The form factor $\mathcal{K}$ is designed to filter out potentially unphysical kinematic contributions, with the degree of suppression controlled by the scale $\Lambda_p$. Since $\Lambda_p$ is not theoretically determined, its choice is somewhat arbitrary. However, as we will see, benchmark scenarios can provide mild constraints on its maximal value. To estimate the associated uncertainty, we vary $\Lambda_p$ within the range $1.0< \Lambda_p/{\rm GeV}<2.0$ around the default value $\Lambda_p=1.5$~GeV. 

\paragraph*{Benchmarks}

To assess the impact of the Dawson approach and the off-shell form factor, we consider two benchmark scenarios: (i) differential production distribution comparisons with quasi-elastic scattering, and (ii) inclusive $\rho$-meson production.

First, we compare the QRA-based description to the kinematics of the well-defined quasi-elastic scattering, which correctly scales in the soft photon limit and suppresses large off-shellness. As shown in the lower left and center panels of Fig.\,3 in~\cite{Foroughi-Abari:2024xlj}, the expected suppression of large off-shell contributions is only restored when including the off-shell form factor. Additionally, the unphysical features at small angles and large momenta present in the QRA distribution are mitigated by the Dawson approach. Overall, incorporating both corrections enhances confidence in using the QRA approach by aligning its consistency of kinematic distribution with that derived for the well-defined quasi-elastic scattering.

Second, we benchmark the QRA description by using the ISR differential rates to compare them with data for inclusive $\rho(770)$ meson production. As a reference for an efficient description of $\rho$ meson production data, we use the parametrization of~\cite{Bonesini:2001iz}, which was fit to N27 data at $\sqrt{s}=27.5~$ GeV. We describe the Bremsstrahlung-like emissions of $\rho$ meson by the QRA approach by using an effective $\rho$-nucleon vertex $\mathcal{L}_{\rm eff} = g_{\rho NN}\overline{N}\gamma^\mu \Vec{\rho}_\mu\cdot\Vec{\tau}N$ with an average experimental value of $g_{\rho NN}=2.9 \pm 0.3$~\cite{Oset:1983mvn,Downum:2006re,Riska:2000gd} where we neglected the small tensor coupling contribution. Since the Bremsstrahlung-like processes represent only a fraction of the total production of $\rho$ mesons in the $p p \to \rho^0+X$ production rate, this comparison serves as an upper limit on the role of QRA, in particular for low momentum where the description is expected to be valid. 

We observe in Fig.\,6 of~\cite{Foroughi-Abari:2024xlj} that the comparison, in particular, sets a limit on the $\Lambda_p$ scale of the off-shell form factor. It is apparent that values of $\Lambda_p$ of a few GeVs fail to pass the benchmark test, suggesting that $\Lambda_p$ must be smaller. Additionally, we observe that the FWW approach overproduces $\rho$ mesons across the entire data range.

\paragraph*{Conclusions}
In this contribution, we have collected recent theoretical results on the production of dark photons with masses about 1 GeV in proton-proton collisions and have commented on the status of the available answers in the literature. In particular, for the process of elastic proton bremsstrahlung, it has been shown that the WW approximation works with good accuracy. For the inelastic proton bremsstrahlung, the role of the accuracy in determining the proton electromagnetic form factors in the unphysical region was emphasized. In addition, it was shown that the inelastic bremsstrahlung cross section always depends on three auxiliary splitting functions and each of them should be taken into account in the answer. There were also given the effective splitting functions obtained in the Dawson approach, which do not contain singularities for small dark photon masses. Finally, we discussed possible sources of uncertainties in the bremsstrahlung cross section and the benchmarks of quasi-real approximation.

\subsection{Dark Scalar}
\label{ssec:DS}

\subsubsection{Hadronic decay rate of a scalar with GDA approach --- \textit{D. Gorbunov}}
\label{sssec:gorbunov}
\textit{Author: Dmitry Gorbunov, \email{Dmitry.Gorbunov@cern.ch}}\\
\paragraph*{Introduction}
To explain neutrino oscillations, dark matter phenomena, baryon asymmetry of the Universe, and many other issues, we definitely need some new
physics. There are many suggestions in literature on what might be
responsible for some (or all) of these issues, but the present
understanding of quantum field theory makes the models with light
particles theoretically stronger motivated than those with the heavy
stuff. The problem is that the heavy particles induce quantum
corrections to the mass of the Standard Model (SM) Higgs boson, which lift
it up to the scale of new physics, thus destabilizing the electroweak energy
scale. The light new particles circumvent this problem, but must be only
feebly coupled to the SM particles in order to demystify the absence
of any direct hint of them in particle physics experiments. Naturally,
they must be singlets with respect to the SM gauge group, which is an
alternative to Grand Unified Theory extensions of the SM.

The tiny couplings are not a novelty in the SM, where the values of the
Higgs Yukawa couplings to the first-generation fermions are of the
order of $10^{-6}-10^{-5}$ and hardly to be experimentally tested in
the foreseeable future. The couplings between the SM particles and the new
SM singlets are generally called {\it portals}, and the renormalizable
Interactions between the hidden and visible sectors are highlighted
from both the experimental and theoretical sides. Indeed,
Non-renormalizable couplings are suppressed by some energy scale,
which makes the low-energy experiments less sensitive to the new
physics, though it is light, as compared to the high-energy
experiments. Likewise, such models are effective, and the energy scale entering the interaction terms indicates where it must be
completed. The renormalizable portals---scalar, fermion, vector---can
be tested at low energies without penalties, and may form a theory
which works properly up to the Planck scale as a complete model
of particle physics.

We are interested here in the scalar portal, which couples the SM
Higgs doublet $H$ to a scalar $X$ from a hidden sector via potential terms  
\begin{equation}
\label{scalar-portal}
V=\beta H^\dagger H X^2 + \mu H^\dagger H X\,,
\end{equation}
where $\beta$ is dimensionless coupling and $\mu$ has a dimension of mass. Both terms are renormalizable, the second term is absent if the scalar $X$ carries some charge under the hidden sector symmetry
group.

While the choice of interaction of the form \eqref{scalar-portal}
is generally attractive due to renormalizability, it is also very strongly motivated by cosmology,
where typically the inflation in the early Universe is attributed
to the dynamics of the singlet scalar field. The inflationary stage
makes the Universe spatially flat and very homogeneous, but empty, except for 
the homogeneous inflaton field (and its fluctuations), see e.g.\,\cite{Gorbunov:2011zzc}. When inflation
terminates, the energy from the inflaton field must be transferred to the SM particles, which thermalise and form the primordial plasma. The most economical way to do it is by invoking the interaction \eqref{scalar-portal} between the SM Higgs and the inflaton $X$. The scalar portal produces the SM Higgs bosons, which scatter off each other and populate the early Universe with SM particles, thus reheating the Universe and solving the entropy problem of the Hot Big
Bang theory. 

While the inflaton as a scalar field is the most natural choice, and a reheating stage must exist, so some type of coupling between the inflaton and other fields must exist, it may not be of the scalar portal form. Moreover, the inflaton field may be (much) heavier
than the electroweak scale. Maybe, but certainly not
necessary. This statement can be illustrated with
a simple model\,\cite{Bezrukov:2009yw,Bezrukov:2013fca}
of viable inflation and reheating, which indeed exploits the scalar portal to reheat the Universe, and where the inflaton field is light, in the GeV mass
range. The model Lagrangian reads
\begin{align*}
  S_{X \mathrm{SM}} = &\
    \!\!\int\!\! \sqrt{-g}\, d^4x\left( {\cal L}_\mathrm{SM} 
      + {\cal L}_{\text{ext}} + {\cal L}_\text{grav} \right)
  , \nonumber \\
  {\cal L}_{\text{ext}} =          &\ 
     \frac{1}{2} \partial_\mu X\partial^\mu X +\frac{1}{2} m_X^2 X^2-\frac{ \beta}{4} X^4
     - \lambda \left( H^\dagger H - \frac{\alpha}{\lambda} X^2 \right)^2
  , \label{4*} \\
  {\cal L}_\text{grav} =    &\ -\frac{M_{Pl}^2/(8\pi)+{ \xi} X^2}{2}R\,,
\end{align*}
with SM Lagrangian ${\cal L}_\mathrm{SM}$ and Hilbert--Einstein
Lagrangian ${\cal L}_\text{grav}$ for gravity, $M_{Pl}$ is the Planck mass and dimensionless
parameter $\xi$ stands in front of non-minimal gravitational coupling of the inflaton $X$. At a high value of $X$, the inflaton goes due to the inflaton self-coupling $\beta$, while both the inflaton and Higgs fields are rolling down side by side along the value $\lambda H^\dagger H =\alpha X^2$. A combination of dimensionless parameters $\beta$ and $\xi$ are tuned \cite{Bezrukov:2014nza} to fit the observed amplitude of matter perturbations originated
from the quantum fluctuations of the inflaton field.

The inflaton mass term in the Lagrangian above enters with the wrong sign, and after inflation ensures the non-zero vacuum expectation value of the inflaton field
$\langle X\rangle \neq0$. The latter induces the mass parameter $v$ for the electroweak vacuum via the scalar portal coupling, so $v^2/2=\alpha\langle X\rangle ^2/\lambda$, which fixes the value of the dimensionless parameter $\alpha$. Upon the spontaneous breaking of the electroweak symmetry, the SM Higgs mixes with the inflaton field, the
mixing angle $\theta$ (considered to be small) is
related to the model parameters as
\[
\theta^2=\frac{2\beta v^2}{m_\chi^2}=\frac{2\alpha}{\lambda},
\]
and the inflaton mass $m_\chi$ is related to the SM Higgs boson mass
$m_h$ as
\[
 m_\chi = m_h \sqrt{\frac{\beta}{2\alpha}}
 = \sqrt{\frac{\beta}{\lambda\theta^2}}.
 \]
 The stronger the mixing, the more efficient the energy transfer from the inflaton to the Higgs boson, and the higher the reheating temperature. The requirement of getting the reheating temperature above the electroweak scale bounds the mixing from below and consequently places the inflaton into GeV mass region.  
This mixing angle $\theta$ and (to some extent) the non-minimal coupling $\xi$ entirely define the phenomenology of the light inflaton. There were direct searches for the light inflaton, including searches performed at LHCb, see e.g.\,\cite{LHCb:2015nkv}. 

In what follows, we investigate the phenomenology of a general light scalar $X$ singlet with respect to the SM gauge group, whose coupling to the SM fields is limited by mixing with the SM boson. Hence, there are two free parameters in the model: the scalar mass $m_X$ and the mixing angle $\theta$. 

\paragraph*{Production and decay modes of a hidden scalar}

\begin{figure}[!htb]
\centerline{
  \includegraphics[width=0.45\textwidth]{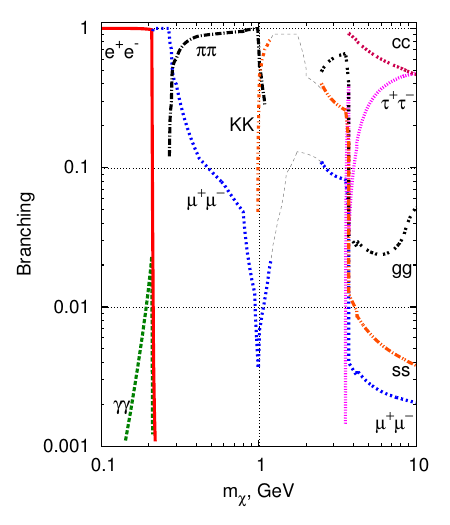}
\includegraphics[width=0.45\textwidth]{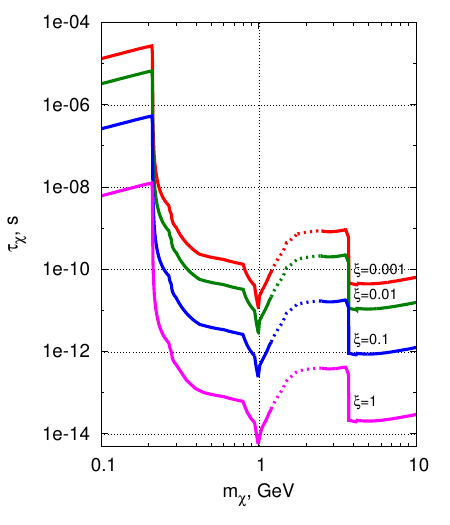}
}
\caption{Scalar decay branching ratios (left panel) and scalar
  lifetime (right panel) in the model of light inflaton\,\cite{Bezrukov:2013fca}.}
\label{fig:bran-and-life}
\end{figure}

Since all couplings of the light scalar to the SM fields are due to the mixing with the SM Higgs bosons, all the scalar decay modes and the corresponding branching ratios are those exhibited by the SM Higgs boson, would it be light, of the same mass as the light scalar. Likewise, the production modes of the light scalar are the
same as those for the light SM Higgs boson.
The phenomenology of the light SM Higgs boson had been thoroughly studied in the seventies  \cite{Ellis:1975ap,Shifman:1979eb}, when it was expected to be much lighter than it is in reality. These results can be adopted for the light inflaton. Naturally, the branching ratios of the scalar depend only on its mass, while its lifetime depends both on mass and mixing angle. At a given mass, the mixing angle is a free parameter, but it may be fixed in a particular
extension of the SM. To illustrate typical values of the scalar lifetime, we chose those favoured by the light inflaton model discussed above and present in Fig.\,\ref{fig:bran-and-life}  
both branching ratios (which do not depend
on mixing) and lifetime (which does depend on the mixing angle) adopted
from Ref.\,\cite{Bezrukov:2013fca}.

One immediately observes on these
plots the main subject of this discussion: there are solid predictions
for scalar of mass well below 1\,GeV and well above 1\,GeV, but not in
between. The reason is in hadronic decay modes: light scalars decay
into light mesons, heavy scalars decay into quark and gluons which are
subsequently fragmented into light hadrons. The first case can be
described within Chiral Perturbation Theory (ChPT), while the second case can  be properly addressed by
perturbative Quantum Chromodynamics. The situation with 1\,GeV
scalar is much more complicated. The authors of
Ref.\,\cite{Donoghue:1990xh}  considered the
decay of 1\,GeV mass SM Higgs boson into pions and estimated the pion
interaction in the final state by making use of the dispersion
relation. We explain the procedure below, but the conclusion was that due
to very strong pion interaction in the final state the scalar decay
rate into pions, calculated within the ChPT, gets amplified
by a huge numerical factor, which depends on mass and varies from several to
hundred. The problem had not been resolved, since right after
the publication of Ref.\,\cite{Donoghue:1990xh} the new experimental 
results were announced and clearly indicated that the SM Higgs boson is heavier than
half of the $Z$-boson. The problem has raised up
nowadays \cite{Bezrukov:2009yw}, when the
scalar portal coupling and light scalars in general became
popular.

\begin{figure}[!htb]
  \centerline{
  \includegraphics[width=0.45\textwidth]{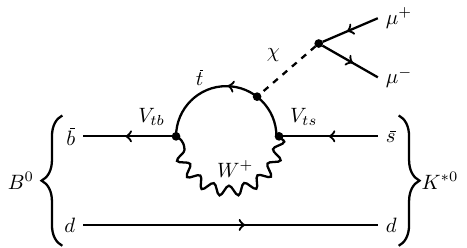}
\includegraphics[width=0.45\textwidth]{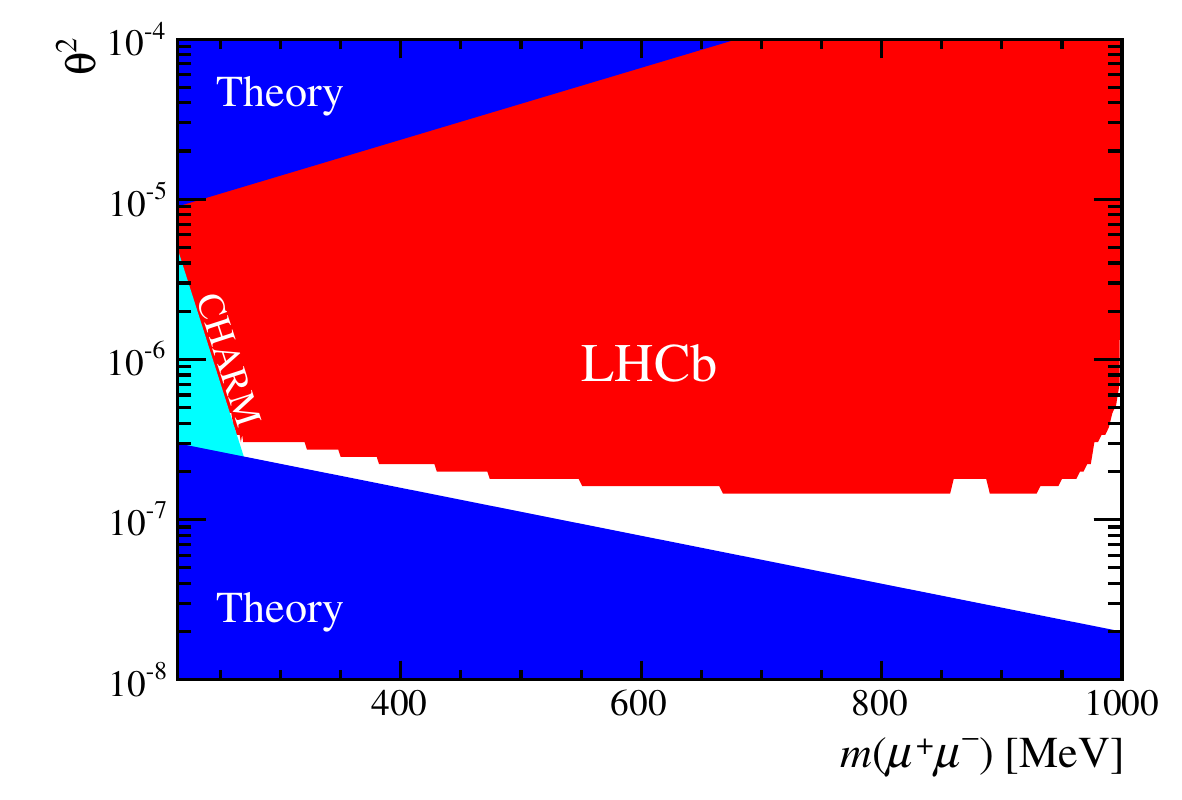}
}
\caption{Feynman diagram of scalar production in $B$-meson decay (left
  panel). Experimental limits on the model parameter space of the
  light inflaton (right panel). The plots are adopted from
  Ref.\,\cite{LHCb:2015nkv}.}
\label{fig:LHCb-inflaton}
\end{figure}

Remarkably, calculations of the light scalar production do not
suffer from huge uncertainties due to strong interactions. The main
production channel is via meson decays, the meson decay rates are
dominated by 1-loop Feynman diagrams involving the top-quark
contribution. The Feynman diagram of light inflaton production in the decay of  
$B^0$-meson with subsequent decay of the inflaton into muon pair
is presented in Fig.\,\ref{fig:LHCb-inflaton}
along with results taken from the LHCb paper devoted to
searches for the light inflaton
\cite{Bezrukov:2009yw,Bezrukov:2013fca} in this channel.

However, uncertainties in the estimates of the scalar decay rates into
hadrons generically prevent one from placing direct limits on the model
parameters, because the hadronic modes dominate the scalar decay
and hence define the scalar lifetime, see
Fig.\,\ref{fig:bran-and-life}.

\paragraph*{QCD uncertainties for the hadronic modes}

\begin{figure}[!htb]
 \centerline{
  \includegraphics[width=0.5\textwidth]{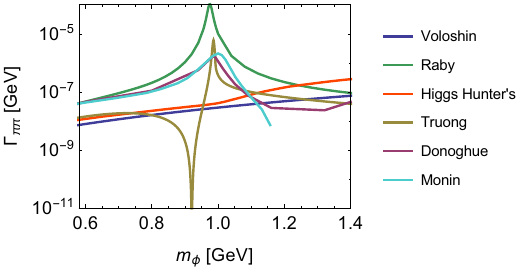}
\includegraphics[width=0.5\textwidth]{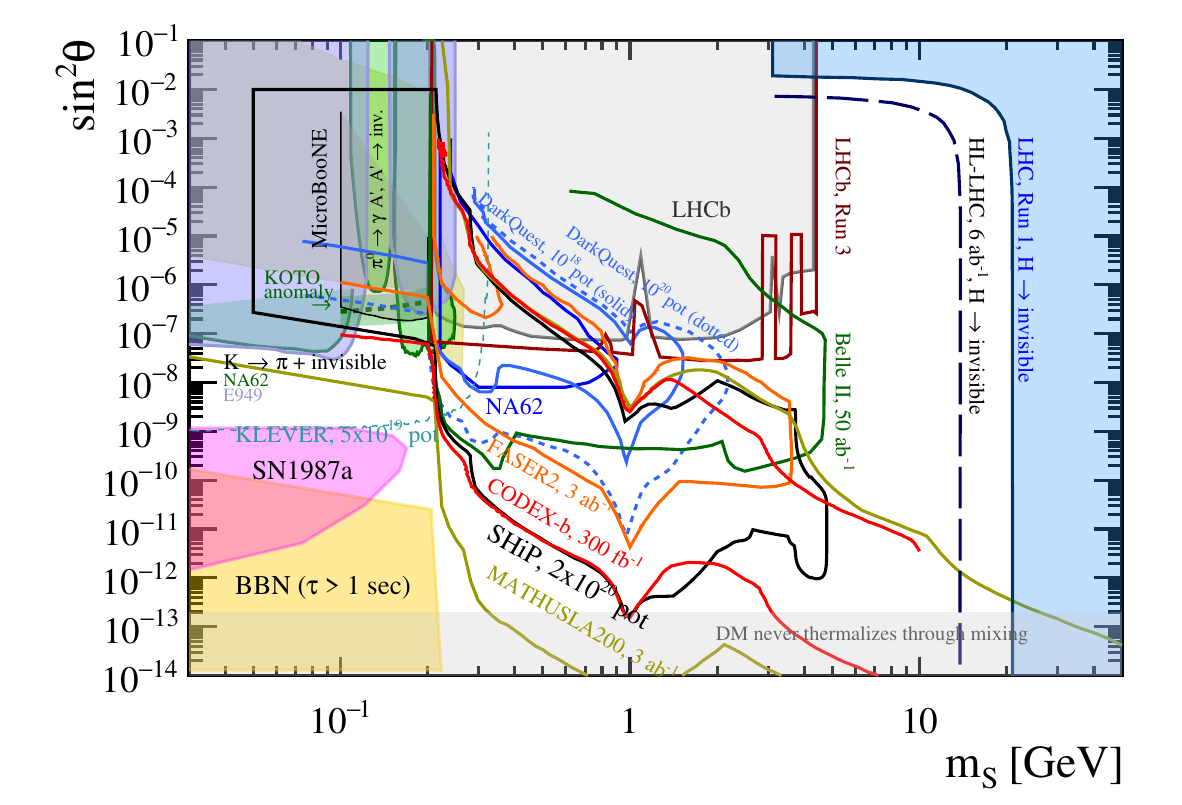}
}
  \caption{{\it Left panel:}
    various predictions of scalar decay rate into pions as a
  function of scalar mass $m_\phi$ (left
  panel), for the references see Ref.\,\cite{Winkler:2018qyg}.
  {\it Right panel:} the estimates of sensitivity
  to scalar model parameters (mixing angle $\theta$ and scalar mass
  $m_S$) obtained for various experiments and projects, for the
  references see Ref.\,\cite{Agrawal:2021dbo}.}
\label{fig:dispersion}
\end{figure}

There were several attempts in the literature to address the issue of
hadronic decays of a light singlet scalar, based on some interpolations between the
ChPT and QCD and application of the dispersion relation
approach. However, until very recent \cite{Blackstone:2024ouf}, there were no any (not to mention
reliable) 
estimates of the uncertainties of the performed calculations. The
results for the scalar decay rate into pions $\Gamma_{\pi\pi}$
are nicely summarized on left panel of Fig.\,\ref{fig:dispersion}
adopted from Ref.\,\cite{Winkler:2018qyg}. One concludes, that various
attempts reveal different results, though typically they predict a huge
and sufficiently broad peak in the region of 1\,GeV. Numerically, the 
answers deviate  from each other by factors of a dozen, a hundred, or
even a thousand.

The absence of any reliable estimates of the uncertainties
in each of this calculations certainly must warn everyone: the
matter is certainly not settled. However, for new experimental
projects some estimates of their sensitivities to the model parameters
must be performed. Presently, the experimental community follows the
paper\,\cite{Winkler:2018qyg}, where the dispersion relation approach
of Ref.\,\cite{Donoghue:1990xh} was repeated with updated experimental
data on pion scatterings. The decay rate is depicted in Fig.\,\ref{fig:unc} 
with blue line. The high peak in the decay rate leads to 
the pronounced dips
of the sensitivity curves in the 1\,GeV region of the scalar mass, see
an example on the right panel of Fig.\,\ref{fig:dispersion}.  

Two different cases must be recognized with respect to the
uncertainties in the hadronic decay rates. The first is the situation where the decay length of the scalar $X$ is much shorter
than the detector size, $l_{dec}=c\tau_X\gamma_X \ll
L_{detector}$. In this case, all the produced scalars decay inside the
detector, and the {\it number of signal events} in a particular channel (the
different channels are typically considered separately due to
different backgrounds) {\it is proportional to the corresponding
  branching ratio.} Since the hadronic modes, mostly pions in case of
light scalar, dominate, one expects to see a peak in GeV range for the
sensitivity lines obtained for the muon pair as a scalar signature and
no peaks for the pion pairs. The second is the opposite situation, when the scalar decay length much exceeds the detector size and the
distance from the target to the detector (otherwise, the scalar flux in the
detector is exponentially suppressed), $l_{dec}=c\tau_X\gamma_X \gg
L_{detector}$. Then the number of signal events in a particular
channel is proportional to the product of the branching ratio and the
number of scalar decays inside the detector. The latter is
suppressed by the ratio of the detector length to the decay
length. Hence, {\it the number of signal events depends only on the
decay rate to a given final state.} Therefore, the
sensitivity lines are expected to exhibit peaks for the final pion state and be
smooth for muons.

The scalar $S$ couples to the SM fermions through mixing with the
SM Higgs boson,
\[
{\cal L} = - \theta\sum_f { \frac{m_f}{v}\bar{\psi}_f
    \psi_f}\,S 
\]
In fact, these couplings also induce through the quantum corrections the scalar
coupling to gluons, as they do for the SM Higgs. It is convenient then
to describe the scalar interactions at low energies via Yukawa couplings to
the light quarks $u$, $d$, $s$ and hadronic energy momentum tensor
$T_{\mu\nu}$. The amplitude of the scalar decay into pions is
proportional to the invariant quantity (for the decay $s\equiv M_S^2$)
\cite{Voloshin:1985tc}
\[
G_\pi(s) = \frac{2}{9} \Theta_\pi(s)+ \frac{7}{9}
\left( \Gamma_\pi(s) + \Delta_\pi(s) \right). 
\]
governed by the
three form factors defined as
\begin{align*}
\langle {\pi^i(p)\pi^k(p')} \left| {T_\mu^\mu}\right|0\rangle  &\equiv \Theta_\pi(s) \delta^{i k},
\\
\langle {\pi^i(p)\pi^k(p')} \left|{ m_u \bar{u}u + m_d \bar{d}d} \right| 0 \rangle &\equiv 
\Gamma_\pi(s) \delta^{i k}, \\
\langle {\pi^i(p)\pi^k(p')} \left| {m_s \bar{s}s} \right| 0\rangle &\equiv \Delta_\pi(s) \delta^{i k}.
\end{align*}
Similar expressions can be introduced for kaons. 
At small $s=(p+p')^2=M_S^2$ we can compute the form factors within the ChPT, the
leading order gives 
\begin{align*}
\Theta_\pi(s)  &= s + 2m_\pi^2, \qquad & \Theta_K(s)   = s + 2m_K^2\,,
\\
\Gamma_\pi(s)  &= m_\pi^2, \qquad & \Gamma_K(s)   = \frac{12}{ m_\pi^2}\,,
\\
\Delta_\pi(s)  &= 0, \qquad & \Delta_K(s)  = m_K^2 - \frac{12}{ m_\pi^2}\,.
\end{align*}

The outgoing pions can re-scatter right after production, which
impacts on the scalar decay rate. Naively, for a sufficiently light
scalar, the latter process, $\pi\pi\to\pi\pi$, can be described within
the one-channel approximation to the $S$-matrix of the hadronic
states. Namely: only vacuum and two-pion states are accounted for. 
Then the $S$-matrix element is just a complex phase,
$e^{i\delta(s)}$. Since any form factor $A(s)$ is an analytic function of
$s$, it can be generically expressed via its imaginary part as  
\[
A(s)=\frac{1}{\pi}\int _0^\infty ds'\frac{\text{Im} A(s')}{s'-s-i\epsilon}\,.
\]
Applying this formula for any of the scalar decay amplitudes above
and exploiting the
unitarity of the one-channel $S$-matrix, one obtains for each of the
scalar form factors
\[
F(z)=F(0)\exp{\frac{z}{\pi}\int_{4m_\pi^2}^\infty ds
  \frac{\delta(s)}{s(s-z)}}.
  \]
The function $\delta(s)$ can be inferred by fitting to the
experimental data on scatterings
$\pi\pi\to\pi\pi$, treated within the one-channel
approximation. In this way, the results presented on the left panel of
Fig.\,\ref{fig:dispersion} were obtained. 

There are several issues with application of the dispersion relation
to the calculation of the scalar decay rates into
hadrons, which partly explain such different
results depicted on left panel of
Fig.\,\ref{fig:dispersion}. We briefly mention some of them here, see
Ref.\,\cite{Bezrukov:2018yvd} for more details. First, the leading order
expressions for the form factors $\Theta(s)$ and the next-to-leading order 
expressions for the other form factors grow with energy, while unitarity
requires them to vanish at $s\to\infty$. This problem has been
acknowledged but ignored in the original paper
\cite{Donoghue:1990xh}. Setting $\Theta(s)=0$ at $s\to\infty$ by hand
(what one typically sees in the literature)
may or may not change the result for the light scalar but definitely
changes it for the scalar mass in the GeV range. Second, the true hadronic
$S$-matrix includes many channels, while typically only one-channel
(two pions)  or two-channel (two pions and two kaons) approximations
are utilized. Naturally, this, so called truncated approach, deals with
a non-unitary matrix, and to improve the approximation we need to add
other channels, $\eta\eta$, $4\pi$,
etc.\,\cite{Moussallam:1999aq}. Third, with the new channels accounted for the 
new heavy
resonances enter the game. On this way one observes, that
the heavier resonances change the answer for lighter
scalars\,\cite{Monin:2018lee}. Fourth, the new heavy channels are
strongly coupled, as indicated by large branching ratios into
multipion states. Hence, the result for the scalar form factor
explicitly depends on the way one adds the new
channels\,\cite{Ropertz:2018stk}.

Given the situation above, it is natural to ask for alternative
approaches to the problem with scalar hadronic decay modes. Below we
describe the method based on the Generalized Distribution Amplitudes
(GDA, for a review see \cite{Diehl:2003ny}). The calculations of the
scalar decay form factors to pions are presented in Ref.\,
\cite{Gorbunov:2023lga}.

\paragraph*{An alternative approach with GDA}

The starting point of calculations is the $u$- and $d$-quark
contributions to the gravitational form factor of pions, defined as
\begin{equation}
\label{Grav}
\langle \pi^a(p)\pi^b(p')|T^{\mu\nu}_q(0)|0\rangle 
\equiv \frac{\delta^{ab}}{2} \left( \left( s \,\eta^{\mu\nu}-P^\mu P^\nu\right)
{\Theta_{1,q}(s)}+\Delta ^\mu \Delta^\nu {\Theta_{2,q}(s)}\right), 
\end{equation}
where $P\equiv p+p'$ and $\Delta\equiv p-p'$, and $T^{\mu\nu}_q(x)$ is
the operator of quark energy-momentum tensor. One  sums up over the
$u$- and $d$-quarks and straightforwardly
obtains from this formula the relation between the form factors
$\Theta(s)_{1(2)}\equiv \Theta(s)_{1(2),u}+ \Theta(s)_{1(2),d}$ and
formula which defines the quark contribution to the scalar decay,
\[
{\Gamma_\pi(s)}=s \left( \frac{3}{2}{\Theta_{1}(s)}-\frac{1}{2}
{\Theta_{2}(s)}\right) +2 m^2_\pi {\Theta_{2}(s)}\,.
\]

\begin{figure}[!htb]
\centerline{  \includegraphics[width=0.5\textwidth]{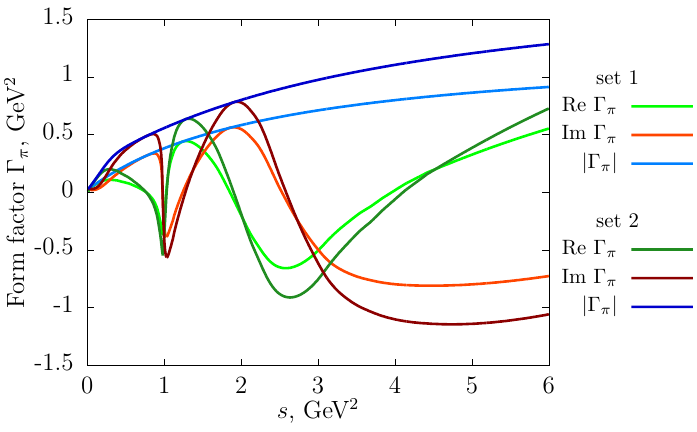}
\hskip 0.05\textwidth
\includegraphics[width=0.4\textwidth]{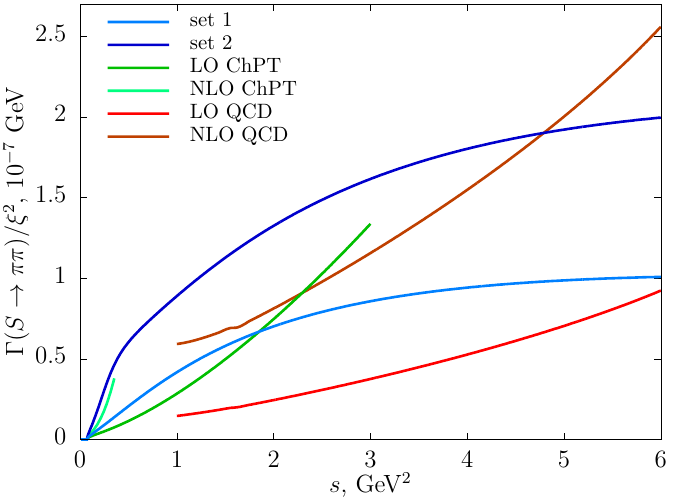}
}
  \caption{{\it Left panel:}
    real, imaginary and absolute values of $\Gamma_\pi(s)$ obtained
    within the GDA approach on the base of a fit to experimental data
    on $\gamma^*\gamma\to\pi^0\pi^0$ scattering, see
    Ref.\,\cite{Gorbunov:2023lga} for details.  
  {\it Right panel:} comparison of the obtained with GDA decay rates based on
  two different fits with predictions of ChPT and QCD, see
    Ref.\,\cite{Gorbunov:2023lga} for details.}
\label{fig:GDA}
\end{figure}

The form factors $\Theta_{1(2),q}(s)$ have been inferred in
Ref.\,\cite{Kumano:2017lhr} from numerical fit to Belle data on 
$\gamma^*\gamma\to\pi^0\pi^0$ scattering. The fit was
obtained with the help of technique, developed for Generalized 
Distribution Amplitudes \cite{Diehl:2003ny}. The fit includes the
regular and resonance contributions, accounts for the kaon pair
threshold and utilizes the phase shifts for the $s$- and $d$-waves (both
contribute to the $\gamma\gamma$ scattering) taken from analysis of
Ref.\,\cite{Bydzovsky:2016vdx}. 
The results for the real and imaginary parts of $\Gamma_\pi(s)$ as well as
the absolute value of this form factor are presented on right panel of
Fig. \,\ref{fig:GDA}
There are two sets of parameters used when fitting in
Ref.\,\cite{Kumano:2017lhr}, but quantitatively the behaviours are the
same. One observes, though both real and imaginary parts of the form factor
oscillate, the absolute value does not, exhibiting monotonic but
rather moderate growth with squared energy $s$. While hadronic
resonances were taken into account at the fitting, their contributions
turned out to be unrecognizable in the final result. This fact may be
attributed to the nature of $f_0(980)$ resonance apparently dominating
the peak observed at 1\,GeV in the dispersion
relation approach in Fig.\,\ref{fig:dispersion}:
this hadronic state is most probably a four-quark, 
$qq\bar q\bar q$,  
rather than a 2-quark, $q\bar q$,
state \cite{ParticleDataGroup:2024cfk}.
Therefore, it does not
contribute to the gravitational form factor defined by the 2-quark $q\bar q$
operator \eqref{Grav} as the fits of Ref.\,\cite{Kumano:2017lhr}
reveal. Likewise, it does not contribute to the scalar decay form
factor $\Gamma_\pi(s)$ defined by the similar two-quark operator. The
obtained results\,\cite{Gorbunov:2023lga}, based on GDA,
nicely match at large $s$ to the next-to-leading order
QCD predictions, as the right panel of Fig.\,\ref{fig:GDA}
demonstrates. The GDA-based results at small $s$ are consistent with the leading order
ChPT predictions. The numerical fit to the form factor, inferred from
the fitting with set 1, reads\,\cite{Gorbunov:2023lga},
\[
    |\Gamma_\pi(x)|=(0.0687 + 0.0270 x^{1/2} + 0.697 x^{3/4} 
    -0.283 x)\text{ GeV}^2\,,
\]
where $x\equiv s/\text{GeV}^2$. 

\begin{figure}[!htb]
 \centerline{ \includegraphics[width=0.48\textwidth]{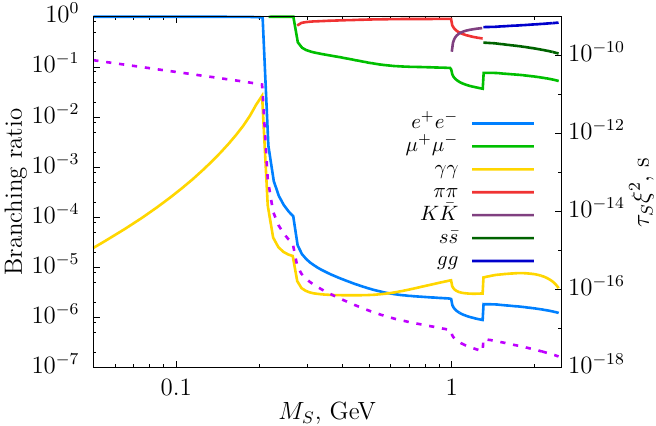}
\hskip 0.05\textwidth
\includegraphics[width=0.42\textwidth]{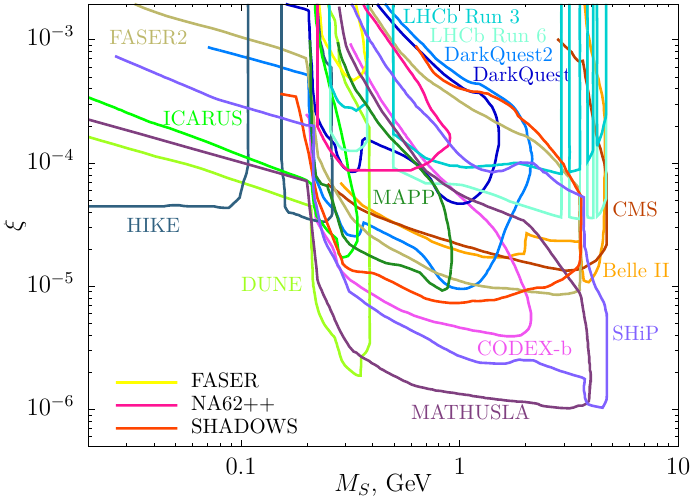}
}
  \caption{{\it Left panel:}
    Scalar decay branching ratios with hadronic rates estimated by making use of 
  the GDA method, adopted from Ref.\,\cite{Gorbunov:2023lga}.  
  {\it Right panel:} Refined sensitivity curves from
  Ref.\,\cite{Gorbunov:2023lga}.}
\label{fig:results}
\end{figure}

\begin{figure}[!htb]
\centerline{\includegraphics[width=0.9\textwidth]{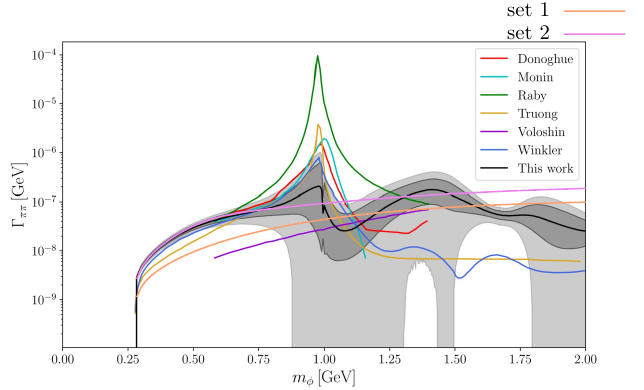}}
\caption{Various predictions based on the dispersion relations for the
scalar decay rate into pions and the associated uncertainties (grey areas), adopted
from Ref.\cite{Blackstone:2024ouf}.
We added here the DGA predictions for the two different fitting
sets to illustrate the uncertainties of that method (a factor of two) and deviations
between the two predictions: based on the GDA and based on the dispersion relations.}
\label{fig:unc}
\end{figure}

We conclude that the GDA technique allows us to obtain the relevant
predictions for the hadronic form factor. The uncertainties may be
estimated by comparing the two different fits to the form factors
$Q_{1(2),q}$, that is numerically about a factor of two. 
Assuming that other hadronic form factors exhibit similar monotonic
behaviour, one can obtain, with a factor of two uncertainty, the predictions
for the scalar decay branching ratios and correspondingly refine the
sensitivity plots of experiments and projects to models with the light
scalars, see Fig.\,\ref{fig:results}. 

\paragraph*{Discussion}

The application of GDA technique to the problem of hadronic decay
rates of a light scalar seems rather promising. 
However, the GDA calculations
have been performed only for up- and down-quark operators, and only
for the two-pion final states, so in the end we know only one form factor
$\Gamma_{\pi}(s)$. To fully address the issue of the hadronic decays we
need much more. There may be kaons and $\eta$-mesons in the final
states, and strange quark operators and gluonic operators must be
evaluated. Some of the tasks may be accomplished with measurements of
the two photon scatterings (one photon is off-shell)
\[
\gamma^*\gamma\to KK\,,\;\;\;\;\;\gamma^*\gamma\to \eta\eta\,,
\]
which can be undertaken at the operating electron-positron colliders,
experiments BESIII and BelleII.

Remarkably, the very recent study \cite{Blackstone:2024ouf}
presents not only the rates but also
the estimate of the
uncertainty in the prediction for the scalar decay rate into pions,
obtained within the dispersion relation technique.
The results are shown in Fig.\,\ref{fig:unc}
by grey colours. One
observes, that the uncertainty factor varies from several to several
hundreds indeed. {\it The very important note concerns
the blue line, which is typically adopted by
experimentalists to use in the evaluations of project
sensitivities. At large mass it deviates from the QCD estimates by two
orders of magnitude, see right panel of Fig.\,\ref{fig:GDA}.}

We plot in Fig.\,\ref{fig:unc} the results obtained within the
GDA technique (set 1 and set 2). The new
results \cite{Blackstone:2024ouf}, obtained within the dispersion relations, 
are much closer to that obtained within the GDA technique, as compared to the
old results of Ref.\,\cite{Winkler:2018qyg}, given the uncertainties
of the order of hundred.

\subsubsection{Exclusive displaced hadronic decays of light scalar at LHCb --- \textit{X.~Cid Vidal, Y.~Tsai, J.~F.~Zurita}}
\label{sssec:zurita}
\textit{Author: Xabier Cid Vidal}

\noindent\textit{Author: Yuhsin Tsai}

\noindent\textit{Author: Jose Francisco Zurita, \email{josefranciscozurita@yahoo.com}}\\
{\it As the main results of this contribution are published~\cite{CidVidal:2019urm}, we will give a succinct version here, referring the interested reader to that publication for details. The only new result here is the impact of the new calculation of the light scalar decays into exclusive hadronic (pions and kaons) states~\cite{Blackstone:2024ouf}, which has a non-negligible impact with respect to the previous result by Winkler~\cite{Winkler:2018qyg}.\\}

\paragraph{A light new scalar} A new scalar $S$ can be added to the Standard Model in an economical manner through a single dimension 4 operator, aka ``the Higgs Portal'' (see e.g.~\cite{Patt:2006fw, Wells:2004di}, and also the discussion in Section~\ref{ssec:fips-pheno} in this volume). A key phenomenological feature of this setup is that, when kinematically allowed, the portal coupling enables a new (non-Standard Model) decay of the 125 GeV scalar (from now on, it will be called the Higgs boson for simplicity).  Indeed, these Exotic Higgs Decays (see e.g.~\cite{Curtin:2013fra, Cepeda:2021rql} for reviews) have enlarged the initial LHC search programme. Due to size of the indirect constraints on the $H \to SS$ branching ratio from a global analysis of all Higgs data (around 12 \% currently~\cite{ATLAS:2024fkg, CMS:2022dwd}, which will be reduced to 4 \% at the HL-LHC~\cite{deBlas:2019rxi}), the portal coupling is small enough as to for the $S$ be a Long-Lived Particle (see e.g.~\cite{Alimena:2019zri} for a review on LLPs), but given the expected ${\cal O} (10^{8})$ Higgs bosons to be produced at the LHC, it is clear that in principle direct tests of branching fractions in the $10^{-4}$ or $10^{-5}$ are statistically possible. Hence one can adopt a simplified approach and parametrize this scalar in terms of its mass $m_S$ and its lifetime $c \tau$ -- in concrete models, one would have $c \tau (m_S)$ (as it is normally parametrized in the simplest portal model, with just a mixing angle $\theta$, see e.g.~\cite{Boiarska:2019jym}), but for a flexible reinterpretation we consider these two variables as independent. Indeed, the S decays (and hence its lifetime) for a given mass depend also on the assumed flavour structure; for instance one could have a leptophobic (hadrophilic) scalar, which could moreover couple only to up-type or down-type quarks, as done in e.g.~\cite{Batell:2018fqo}, or one could consider a larger model inspired by other considerations, e.g. relaxion models~\cite{Fuchs:2020cmm}, or the scalar S could also be serving as a portal to a dark sector, hence adding new final states to S, as happens in dark Higgs models~\cite{Ferber:2023iso}. In a nutshell, we would be presenting model-independent results in the $m_S-c \tau$ plane, and also interpreting these results for concrete incarnations of scalar portals.

When $m_S \gtrsim 2 m_b \sim 10$ GeV, the parameter space is well covered by displaced jet searches from ATLAS, CMS and LHCb. When $c \tau \gtrsim 1$ m, it is clear that the number of decays in the main LHC detectors would be small, hence being more prone to be probed by a new transverse dedicated LLP detector such as MATHUSLA or CODEX-b~\footnote{For a longitudinal detector like FASER this is more difficult, as Higgs boson production in the forward region is severely suppressed}. Hence, the region below 10 GeV (see Ref.~\cite{Boiarska:2019vid} for a review of the GeV scale Higgs portal) and with lifetimes in the 1mm–1m range, is a daunting task for ATLAS and CMS, but it is a fertile territory to leverage the unique capabilities of the LHCb detector (see e.g.~\cite{Borsato:2021aum}). Indeed, for masses below a few GeV, the scalar can dominantly decay into pairs of hadrons ($DD, KK, \pi \pi$), which the LHCb detector can identify, while for ATLAS and CMS, there is no dedicated hadronic PID.

\begin{figure}[t]
\includegraphics[width=1.0\textwidth]{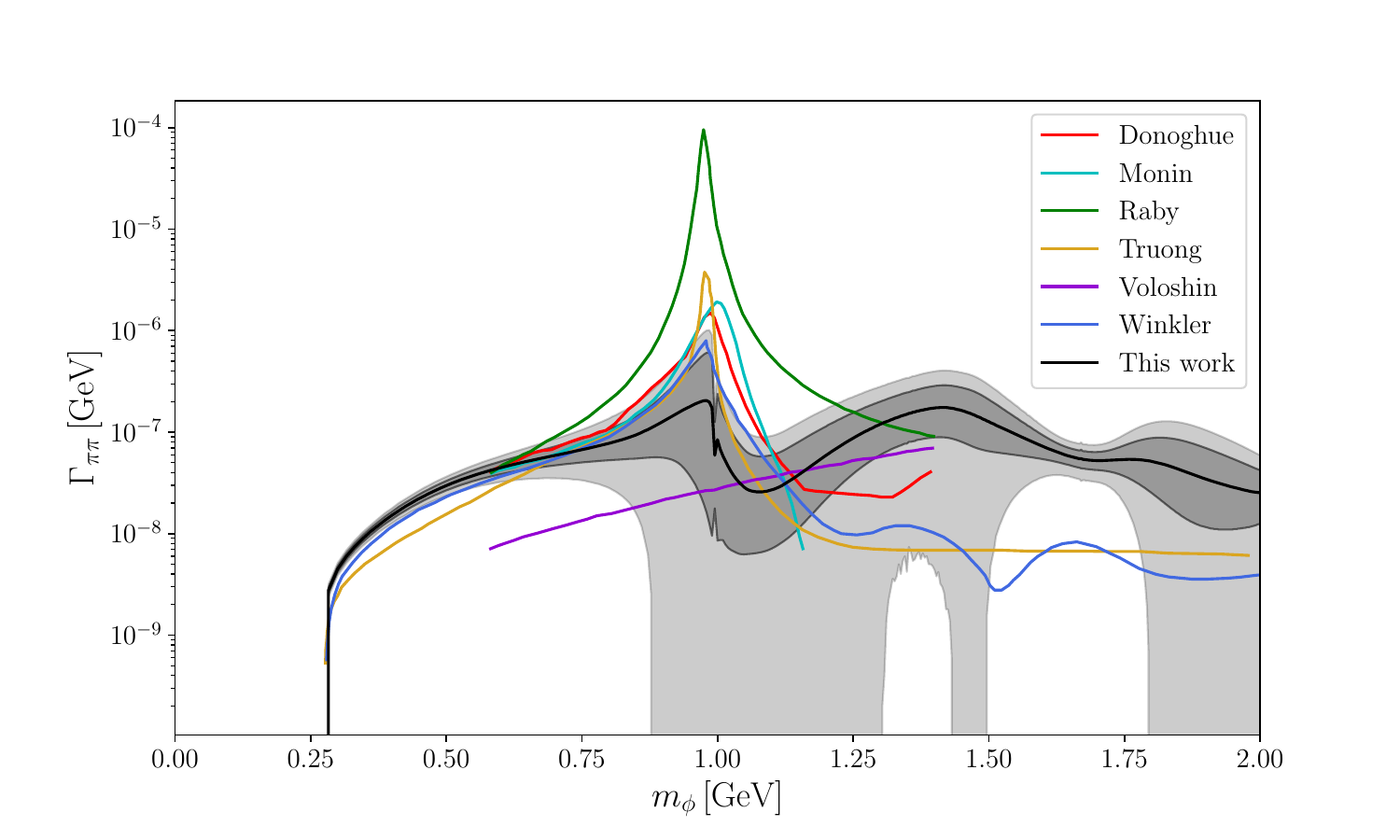}
\caption{Estimations for $\Gamma_{\pi \pi} = \Gamma (S \to \pi \pi)$ using different methods. The main result from~\cite{Blackstone:2024ouf} is the solid black line. Reproduced verbatim from ~\cite{Blackstone:2024ouf}.}
\label{f.comparisons}
\end{figure}

The calculation of the decay widths for $m_S$ > 2 GeV can be done by considering $1 \to 2$ decays into individual quarks, and it is rather straightforward; the results can be found in e.g.~\cite{Boiarska:2019jym}. For $m_S < 2$ GeV, the quark picture is no longer applicable --- the scalar decays into hadrons, and non-perturbative techniques are required to make a proper estimation of decay widths such as $\Gamma (S \to \pi \pi), \Gamma (S \to K K)$. Last year, the work of Blackstone et al.~\cite {Blackstone:2024ouf} largely improved the previous estimations made by Winkler~\cite{Winkler:2018qyg}. The latter were employed in our publication~\cite{CidVidal:2019urm}. To show the impact of this result, we show in Figure~\ref{f.comparisons} the partial width of $S$ into two pions ($\Gamma_{\pi \pi}$), taken from Figure 7 of~\cite{Blackstone:2024ouf}.  We can immediately see that $\Gamma_{\pi \pi}$ can be one order of magnitude larger than what was previously estimated for the 1-2 GeV mass range, while from the kinematic threshold up to one GeV this result yields about a factor of 2-3 lower than the expectation. A similar behaviour is obtained for the Kaon case, and moreover, these results are readily available in the Python public code HipsofCobra~\footnote{\url{https://github.com/blackstonep/hipsofcobra} .}.

\paragraph{Search strategy} Signal and background events for $p p \to H \to S S \to K^+ K^- K^+ K^-$ are generated using Pythia 8.1~\cite{Sjostrand:2007gs}. Our search strategy selects $K^{\pm}$ candidate satisfying $p_T > 0.5$ GeV and $2 < \eta < 5$, which are used to reconstruct S candidates requiring the kaons to be close to each other ($d(K,K) < 0.1$) and to have some transverse momentum, $p_T(K K) = p_T(S) > 10$ GeV. Furthermore, the position (vertex) of S has to fullfill $2 < \rho / {\rm cm} < 25 $ and $z < 400$ mm, with $\rho$ (z) being the radial (perpendicular to the beam axis z) distance, and it must point to the primary vertex with impact parameter larger than 0.1 mm. We apply isolation criteria for muons and veto specific regions in $m_{KK}$ to account for misidentification of other hadrons. Our events are classified into signal regions, depending on their decay position with respect to the VELO, whether 1 or 2 S candidates are present, and on whether the candidate kaons are isolated or not. Among the eight possible signal regions, for each point in the $m_S - c \tau$ plane, we select the most sensitive.

\begin{figure}[t]
\centering
\includegraphics[width=1.0\textwidth]{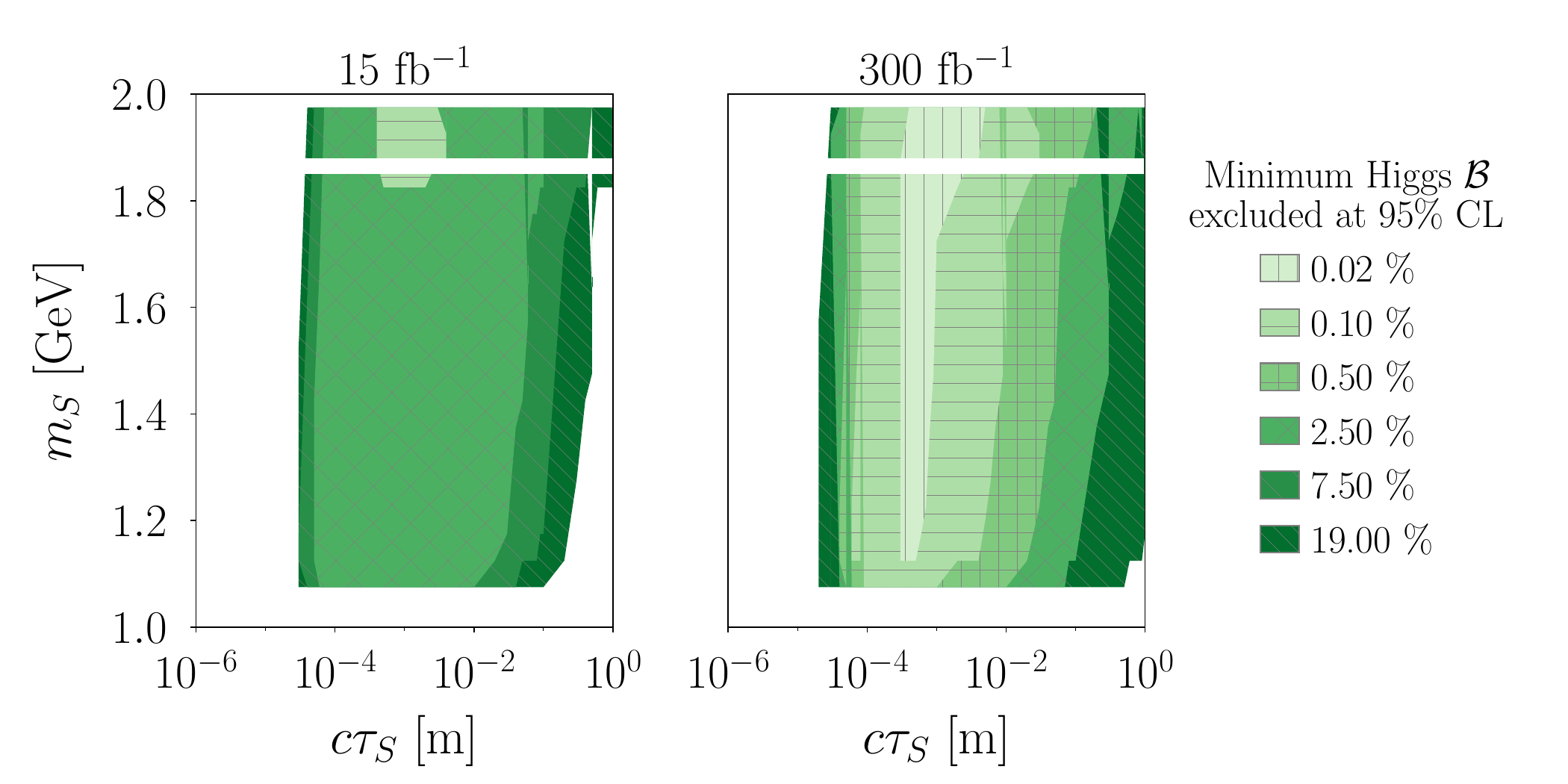}
\caption{Expected exclusions on the exotic branching fraction $H \to S S, S -> K^+ K^-$ in the $m_S - c \tau$ plane. We consider total integrated luminosities of 15 and 300 fb$^{-1}$. Taken from ~\cite{CidVidal:2019urm}.}
\label{f.modelindep}
\end{figure}

\paragraph{Results} The model-independent results in this plane are shown in Figure~\ref{f.modelindep}. We consider total integrated luminosities of 15 and 300 fb$^{-1}$, which correspond to the planned Run 3 and the ongoing discussions for Run 4, respectively.  Note that our study can probe exotic branching fractions down to about $2 \times 10^{-4}$, which is better than the expected 4 \% achievable at the HL-LHC.

\begin{figure}[t]
\centering
\includegraphics[width=0.9\textwidth]{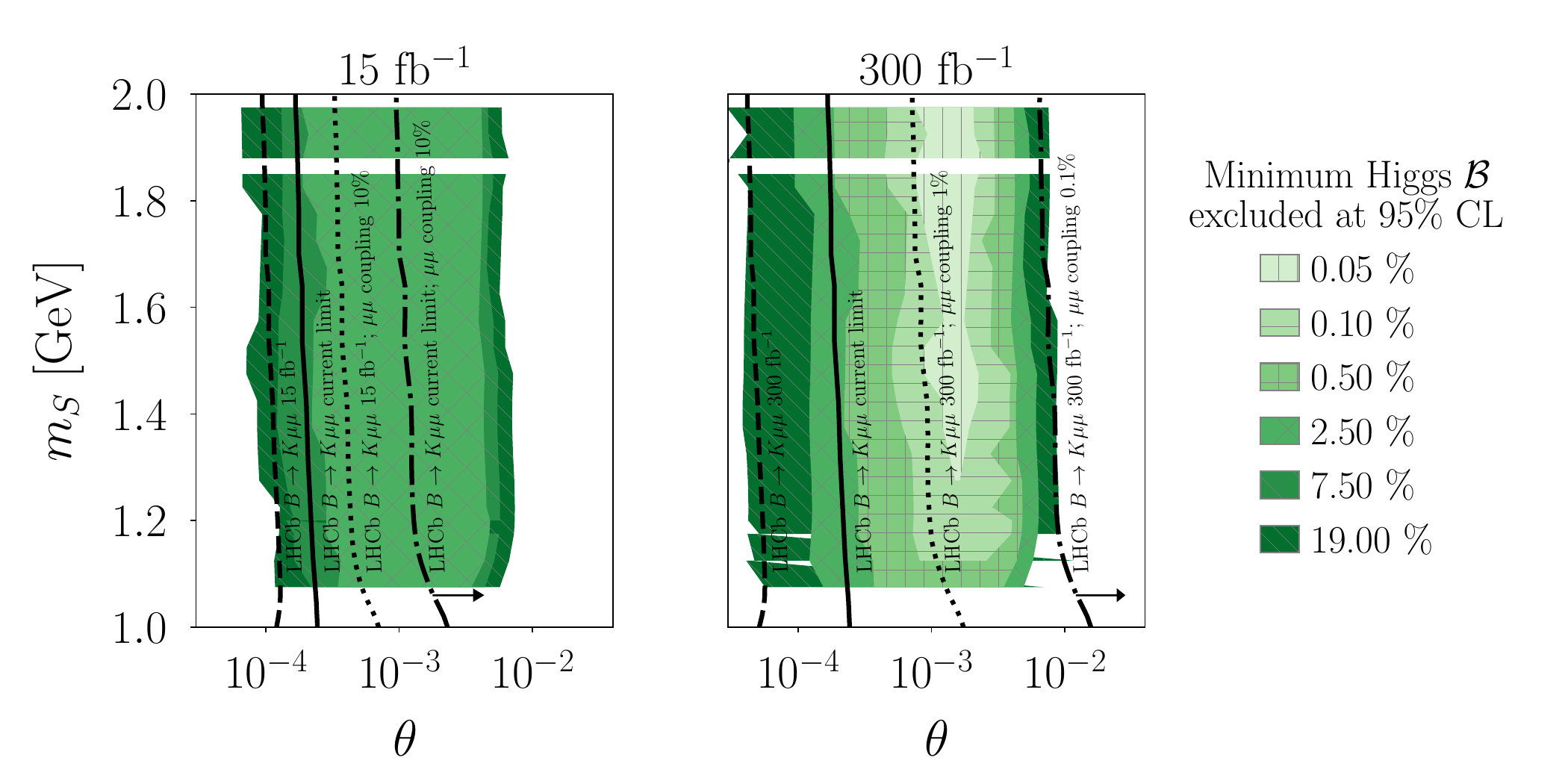}
\includegraphics[width=0.9\textwidth]{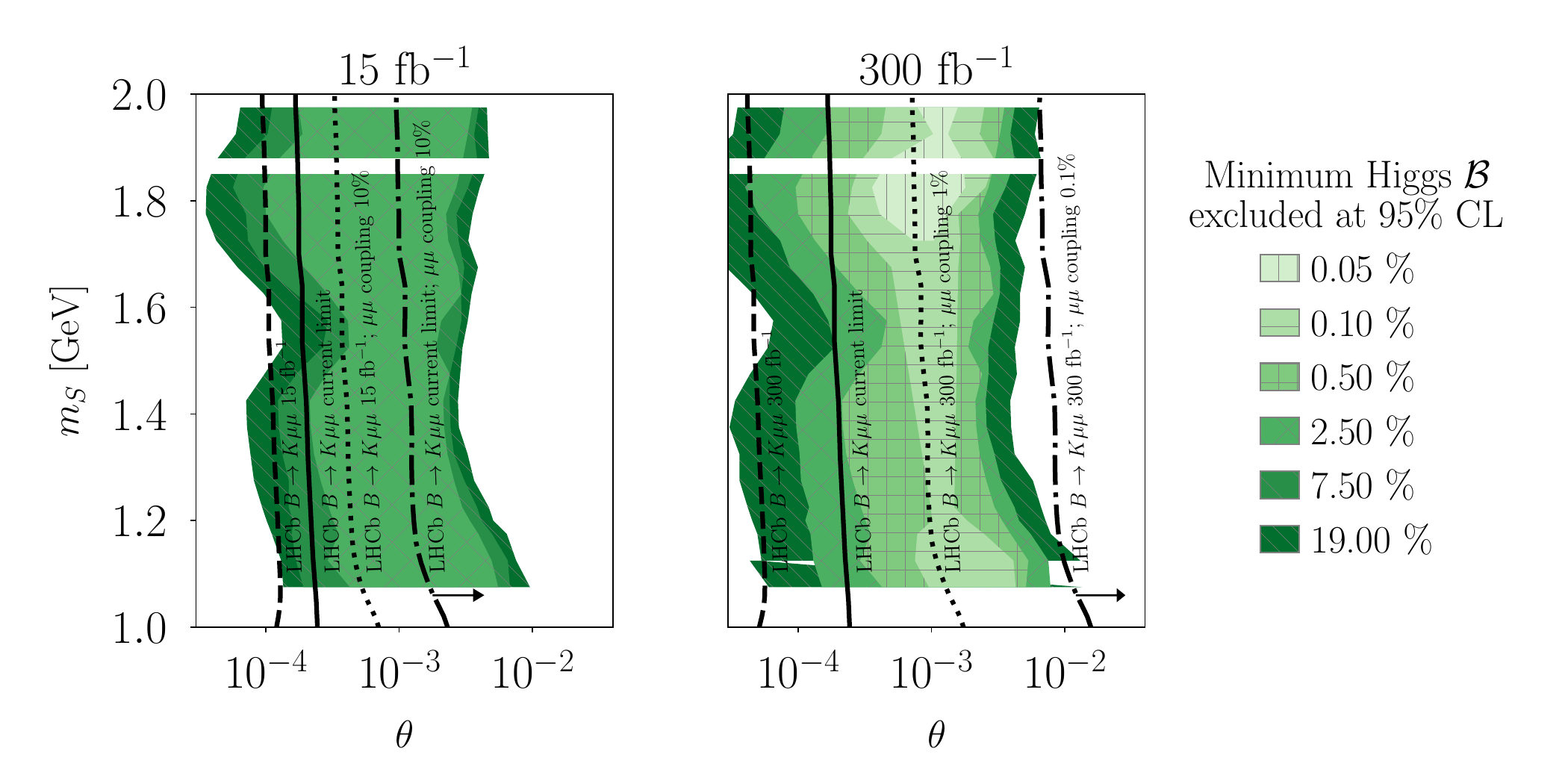}
\caption{Expected exclusions on the exotic branching fraction $H \to S S, S \to K^+ K^-$ in the $m_S - \theta$ plane for the Higgs portal through mixing model. We consider total integrated luminosities of 15 and 300 fb$^{-1}$. The upper row is taken from~\cite{CidVidal:2019urm} and estimates the $S$ decays using the Winkler results~\cite{Winkler:2018qyg}, while the lower row (unpublished) has been rescaled using the new results from Blackstone et al.~\cite {Blackstone:2024ouf}.}
\label{f.limitsHP}
\end{figure}

In the above figure, there is no need for any theoretical input, besides the calculation of the Higgs production cross section, for which we employed state-of-the-art values reported by the Higgs Cross Section Working Group. For specific BSM scenarios, one can set constraints on the fundamental parameters of the theory. To this extent, we consider two theoretical scenarios: the Higgs Portal (through mixing) and the Hadrophilic Higgs Portal. For the Higgs portal, the relevant Lagrangian, omitting kinetic and mass terms, reads
\begin{equation}
\label{eq:higgsportal}
{\cal L} \supset - \theta \Bigl( \frac{m_f}{v} S \bar{f} f - 2 \frac{m_W^2}{v} S W^+ W^- -  \frac{m_Z^2}{v} S Z^2  \Bigr) + \frac{\epsilon}{2} S^2 (h^2 + 2 h v) \, ,
\end{equation}
where there is an implicit sum over all SM fermions $f$ and where we are working in the approximation that the scalar mixing angle $\theta$ is small, hence $s_{\theta} \approx \theta$. In this model, it is customary to display results using the scalar mass and the mixing angle $\theta$, instead of our variable $c \tau$. We then show the limits interpreted in this model in Figure~\ref{f.limitsHP} for our benchmark luminosities. The upper row shows our published results, using the partial widths from Winkler~\cite{Winkler:2018qyg}, while the lower row uses the partial width calculation of Blackstone et al.~\cite {Blackstone:2024ouf}. We note that due to the inherited interactions with leptons, this model is better constrained by the $B \to K \mu \mu$ searches than from our proposed search.

\begin{figure}[t]
\includegraphics[width=0.9\textwidth]{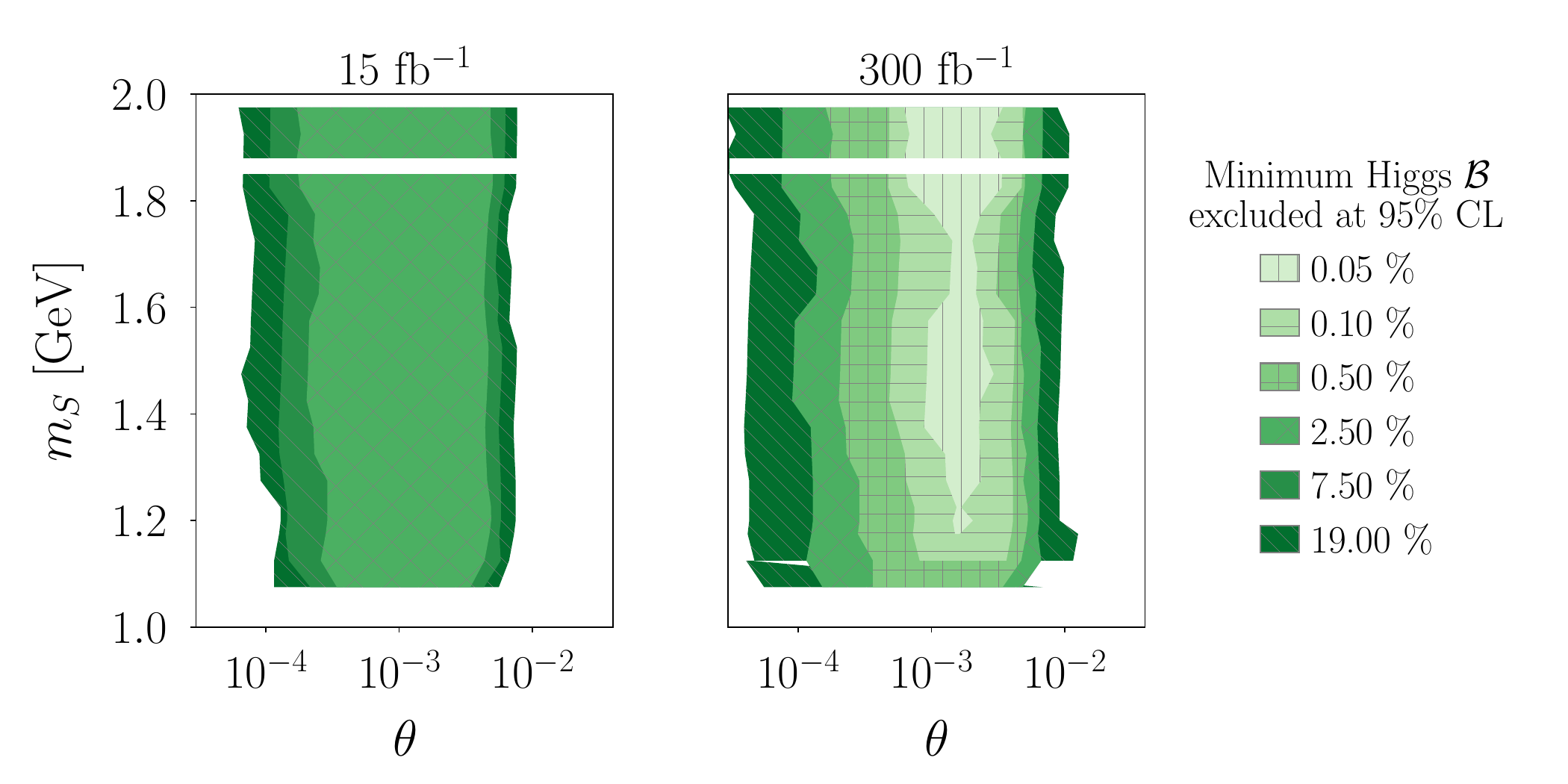}
\includegraphics[width=0.9\textwidth]{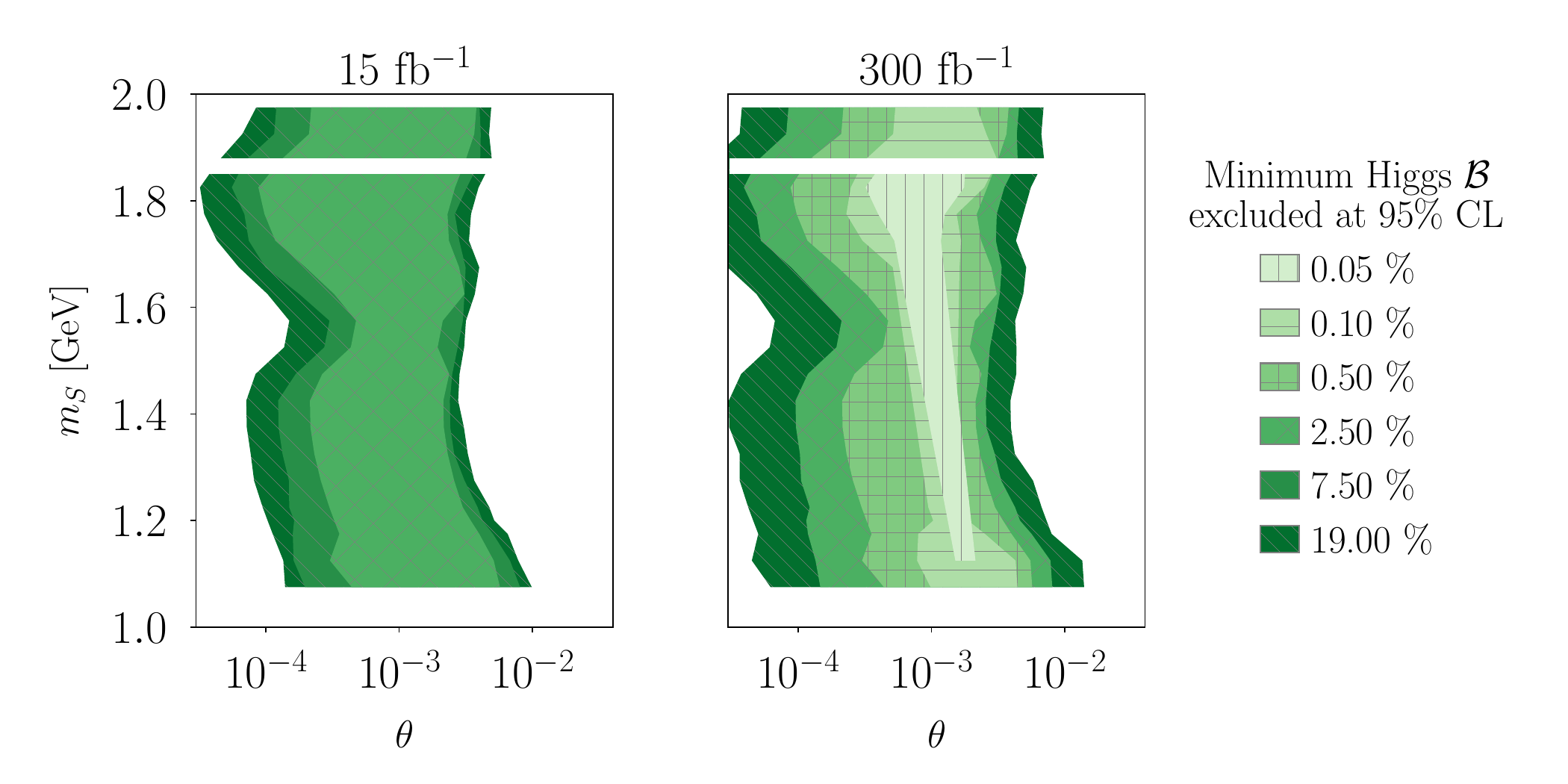}
\caption{Expected exclusions on the exotic branching fraction $H \to S S, S \to K^+ K^-$ in the $m_S - \theta$ plane for the Hadrophilic Higgs portal model. We consider total integrated luminosities of 15 and 300 fb$^{-1}$. The upper row is taken from ~\cite{CidVidal:2019urm} and estimates the $S$ decays using the Winkler results~\cite{Winkler:2018qyg}, while the lower row (unpublished) has been rescaled using the new results from Blackstone et al.~\cite {Blackstone:2024ouf}.}
\label{f.limitsHHP}
\end{figure}

For the Hadrophilic Higgs Portal, the Lagrangian reads (ignoring again mass and kinetic terms)
\begin{equation}
{\cal L} \supset \frac{m_{q_i}}{M} S \bar{q_i} q_i + \frac{\epsilon}{2}  S^2 (h^2 + 2 h v) \,
\end{equation}
where $M$ is a new Physics scale that generates the S-q-q interactions. We see that we can identify $\theta = v/M$ and match our results into the Higgs Portal Lagrangian Eq.~\ref{eq:higgsportal}, with the notable exception that there are no couplings of the scalar to leptons, and hence the strong bounds from $B \to K S, S \to \mu \mu$ do not apply. We show the expected bounds in Figure~\ref{f.limitsHHP}.

\paragraph{Conclusions and Outlook}
The LHCb detector is able to probe New Physics that is ``stealth" to ATLAS and CMS, for instance, light mass resonances, hadronic final states, and displaced particles. In this contribution we have examined the LHCb prospects to test exotic Higgs decays for masses of few GeV, complementing the existing efforts from ATLAS and CMS using displaced jets and also the ``untagged'' Higgs decays limits, as well of those of dedicated detectors such as MATHUSLA or CODEX-b, which can cover the large lifetime regime, $c \tau \gtrsim 1$ m. It is worth noting that in the case of ``untagged'' Higgs decays, even if an excess due to a 2 GeV hadronically decaying long-lived scalar were to be found, the LHCb detector would be a necessary tool to characterise this New Physics, measure the mass, lifetime, and branching fractions of the new scalar. This consideration fully supports a dedicated program for displaced hadronic decays of light resonances at LHCb. While we have focused on kaons, our strategy can be implemented for other final states (e.g. $\pi \pi$ or $DD$), other production modes (e.g. from rare B-decays) as well as consider novel signatures, such as a Higgs decaying into dark showers, giving rise to multiple displaced vertices in a single event.

\subsection{Axion-like Particles}
\label{ssec:ALPs}

\subsubsection{New paths to ALPs discovery --- \textit{G.~Cacciapaglia, G.~Ferretti, C.~Vázquez~Sierra, X.~Cid~Vidal}}
\label{sssec:ferretti}
\textit{Author: Giacomo Cacciapaglia}

\noindent \textit{Author: Xabier Cid Vidal}

\noindent \textit{Author: Gabriele Ferretti, \email{gabriele.ferretti@chalmers.se}}

\noindent \textit{Author: Carlos Vázquez Sierra}\\
%

\newcommand{\beq}{\begin{equation}}
\newcommand{\eeq}{\end{equation}}
\newcommand{\beqs}{\begin{eqnarray}}
\newcommand{\eeqs}{\end{eqnarray}}
\newcommand{\eit}{\end{itemize}}
\newcommand{\bce}{\begin{center}}
\newcommand{\ece}{\end{center}}
\newcommand{\ben}{\begin{enumerate}}
\newcommand{\een}{\end{enumerate}}
\newcommand{\hc}{\mathrm{h.c.}}
\newcommand{\diag}{\mathrm{diag}}
\newcommand{\nn}{\nonumber}
\newcommand{\sw}{{\mathrm s}_{\mathrm w}}
\newcommand{\cw}{{\mathrm c}_{\mathrm w}}
\newcommand{\cc}{{\mathcal{C}}}
\newcommand{\Lcal}{{\mathcal{L}}}
\newcommand{\Ocal}{{\mathcal{O}}}

Many models of physics Beyond the Standard Model (BSM) give rise to an Axion-Like Particle (ALP) as an additional light degree of freedom. ALPs characteristically emerge as pseudo-Goldstone bosons from spontaneously broken Abelian global symmetries. Often, they are the only BSM particles left below the Electro-Weak (EW) scale. 
A general effective Lagrangian describing the interaction of the ALP $a$ with the light SM degrees of freedom, after integrating out the top, the Higgs and the heavy vector bosons,
is given by
\begin{align}
    {\mathcal{L}}_{\rm ALP} =&\frac{1}{2}(\partial_\mu a)^2 -\frac{1}{2}m_a^2 a^2 -i\frac{a}{f}\sum_{\psi\not=t,\nu} \eta_{\psi\psi} m_\psi \bar{\psi}\gamma^5\psi + \frac{a}{4\pi f}\left(\eta_{GG} \alpha_s G_{\mu\nu}^a\tilde{G}^{\mu\nu a} + \eta_{\gamma \gamma} \alpha F_{\mu\nu}\tilde{F}^{\mu\nu} \right) \nn\\
    & +i\frac{a}{f}\left(\eta_{ds} m_s \bar{d}P_R s + \eta_{db} m_b \bar{d} P_R b +\eta_{sb} m_b \bar{s} P_R b + \hc\right)\nn\\
    &-i\frac{a}{f}\left(\eta_{ds} m_d \bar{d}P_L s + \eta_{db} m_d \bar{d} P_L b +\eta_{sb} m_s \bar{s} P_L b + \hc\right),
\label{lagrangian}
\end{align}
where $\eta_{\dots}$ are dimensionless couplings controlling the interaction of the ALP with the SM fields, $m_a$ is the ALP mass and $f$ is the ALP decay constant. Without loss of generality, we can fix $f=1\mbox{ TeV}$ by appropriately rescaling the couplings $\eta_{\dots}$. The remaining symbols denote SM fields and parameters in the standard notation.

For the flavor-conserving couplings to fermions in the first line, the sum runs over all SM fermions except the top, which has been integrated out, and neutrinos, which can be ignored in view of their tiny masses. 
We also included in (\ref{lagrangian}) the unavoidable Flavor-Violating (FV) interactions in the down-type quark sector, arising from the Renormalization Group (RG) analysis \cite{Bauer:2017ris, Bauer:2020jbp, Bauer:2021mvw} of the Lagrangian \cite{Georgi:1986df}. In principle, FV interactions could be added in the up-type quark and lepton sectors; however, they are less relevant for our discussion and, therefore, we omit them.

The most notorious coupling in (\ref{lagrangian}) is the one to photons, which leads to a plethora of possible experimental searches. However, in models with small $\eta_{\gamma\gamma}$, other decays open up that could be of interest at collider experiments.
In \cite{BuarqueFranzosi:2021kky}, motivated by Composite Higgs Models (CHM), we proposed two novel discovery channels of interest for LHCb: $a\to D^+ D^-$ and $a\to \tau^+ \tau^-$, with main production via gluon fusion. 
They are relevant from masses right above the $DD$ mass threshold up to tens of GeV. 

In the above mass region, the ALP decays promptly. As an example, in models with negligible decays into photons, at threshold $m_a = 2 m_D$ the dominant partial decay width in $\tau$ leptons is given by $\Gamma_{a\to \tau\tau} = 1.5 \times 10^{-7}$~GeV for $\eta_{\tau\tau} = 1$ and $f = 1$~TeV. Above threshold, $a \to DD$ quickly catches up, giving a similar contribution to the total width. We also recall that, assuming lepton flavor universality, $\eta_{ee} =\eta_{\mu\mu} = \eta_{\tau\tau}$, the branching ratios into leptons admit the bound $\mathcal{B}(a\to e^+ e^-) \leq 2.7 \times 10^{-7}$ and $\mathcal{B}(a\to \mu^+ \mu^-) \leq 0.011$ for an ALP in the relevant mass region.
We first present model-independent projections for the two searches in $a\to D^+ D^-$ and $a\to \tau^+ \tau^-$, before commenting on model-dependent issues. 

\begin{figure}
\centering
\includegraphics[width=0.9\textwidth]{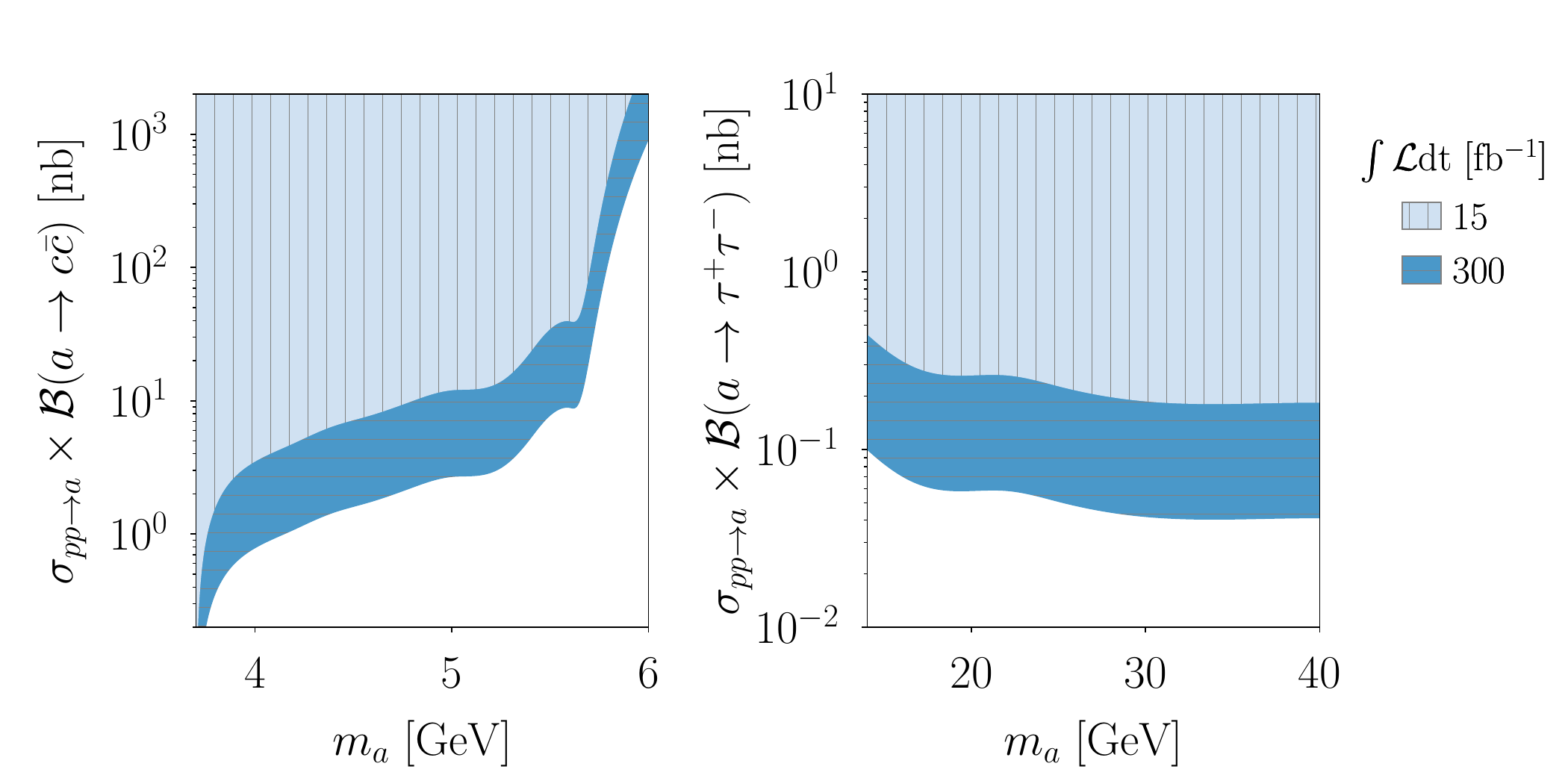}
\caption{Left: Model independent projected bounds on $\sigma(pp\to a)\times \mathcal{B}(a\to c\bar{c})$ for the reference integrated luminosities from the channel $a \to D^+ D^- \to (K^{-} \pi^+\pi^+) (K^+ \pi^- \pi^-)$.  Right: Model independent projected bounds on $\sigma(pp\to a) \times \mathcal{B}(a\to \tau^+\tau^-)$ for the reference integrated luminosities.}
\label{fig:bounds_exp}
\end{figure}

The first channel ($a\to D^+ D^-$) is of interest just above the threshold $2 m_D \lesssim m_a$ and it is based on the possibility of fully reconstructing the $a\to D^+ D^-$ decay via the $D^\pm\to K^\mp \pi^\pm \pi^\pm$ mode, which has a branching ratio of 9.38\%. We showed that LHCb is sensitive to this signal in a window of a few GeV for the ALP mass above threshold \cite{BuarqueFranzosi:2021kky}. 
In Fig.~\ref{fig:bounds_exp} (left), we show the projected bounds for this channel in a model-independent way.  

By far, the main kinematic cut leading to a large increase in the signal over background ratio $S/B$  is the one that selects $D^+ D^-$ pairs with an invariant mass in a window of $m_a \pm 20$~MeV. This captures essentially all of the signal while rejecting most of the background,  which follows a continuous distribution. 
The signal degrades very quickly for $m_a > 5.5\mbox{ GeV}$, due to the opening up of more inclusive decay modes containing multi-pions and kaons, so this channel is of potential interest only for the discovery of ALPs with a mass range between $3.8$ and $5.5$~GeV. It would be interesting to update the study in light of the recent LHCb upgrades.

The second channel ($a\to \tau^+\tau^-$) corresponds to one of the dominant decay channels for ALP masses below the $Z$ mass, and it has also not yet been searched for at LHCb.
In the previous study \cite{BuarqueFranzosi:2021kky}, the most sensitive decay channel was found to be the fully leptonic opposite flavor and opposite sign channel $a \to \tau^+\,\tau^-\to \mu^\pm\, e^\mp\, + 4\nu $, in spite of the small branching ratio of this channel: $2\times  \mathcal{B}(\tau\to e + 2\nu) \times \mathcal{B}(\tau\to \mu + 2\nu) = 6.2\%$.

We imposed various isolation requirements on the $\ell=e, \mu$, as well as geometric conditions on the tracks to ensure prompt decay. In addition, we set the following kinematic cuts: $p_T(\ell_1)>7.5\mbox{ GeV}$, ~~$p_T(\ell_2)>5\mbox{ GeV}$, $p_T(2\ell)>15\mbox{ GeV}$, $2 < \eta(2\ell)< 4.5$, $m_a-\Delta_{L, m_a} < m(2\ell) < m_a+\Delta_{R, m_a}$
for appropriate values of $\Delta_{L/R, m_a}$ that optimize the efficiency. 

Contrary to the $D^+ D^-$ case, the presence of neutrinos in $\tau$ decay does not allow for a proper bump-hunt, and thus the optimal window size $\Delta = \Delta_{L, m_a}+ \Delta_{R, m_a}$ is much larger, having values ranging between $7.5$ and $23$~GeV, depending on the ALP mass hypothesis.

The main sources of background are $b\bar b$ production and DY di-tau production, as shown in Table~\ref{tab:ditau}. We considered only these two and ignored multi-jet, di-boson, and DY production of fermions other than the $\tau$.
\begin{table}[bth]
    \centering
    \begin{tabular}{l|c|c|}
Process  & Cross section 14 TeV & Efficiency $14< m_a[\mbox{GeV}]< 40$ \\
  \hline\hline
$ b\bar b$ &  $ (5.62\pm 0.82)\times 10^{11}\mbox{fb}$  &   $[1.75\times 10^{-9}, \,\, 4.46\times 10^{-9}]$ \\
  \hline
 DY &   $ (4.49\pm 0.24)\times 10^{6}\mbox{fb}$ &  $[6.76\times 10^{-5}, \,\, 7.45\times 10^{-4}]$  \\
  \hline
 Signal & $10^6 \sim 10^9 \times (v/f)^2\,\mbox{fb}$  &  $[5.23 \times 10^{-4}, \,\, 2.06\times 10^{-3}]$  \\
 \hline
\end{tabular}
    \caption{Summary of the total cross sections of the two main backgrounds ($b \bar b$ and DY) and the expected signal from the models \cite{BuarqueFranzosi:2021kky}, together with the range in efficiency after the cuts, as the ALP mass varies from $14$ to $40$~GeV.}
    \label{tab:ditau}
\end{table}
The results of the analysis for this channel are presented in Fig.~\ref{fig:bounds_exp} (right) in a model-independent way. 

\begin{figure}[h]
\begin{center}
\includegraphics[width=0.45\textwidth]{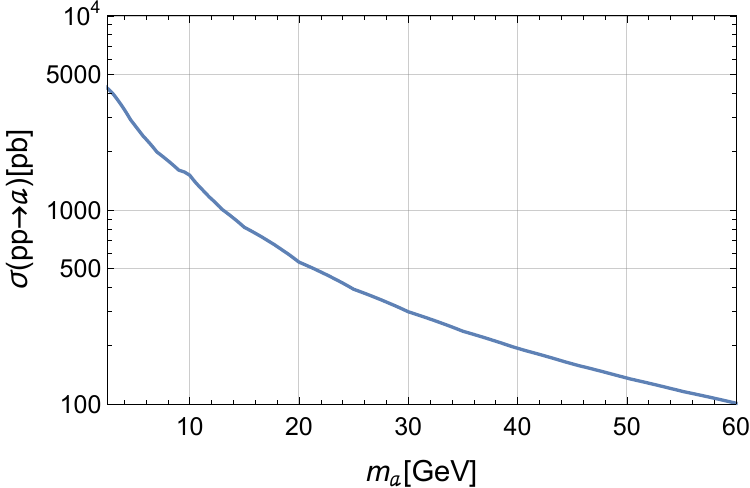} \; \;
\includegraphics[width=0.45\textwidth]{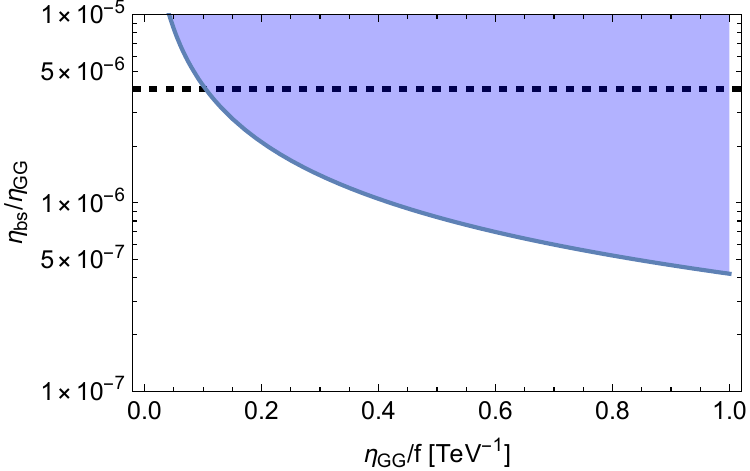}
\end{center}
\caption{Left: Gluon fusion production cross section at LHC for $\sqrt s=14\mbox{ TeV}$, setting $f=1\mbox{ TeV}$ and $\eta_{GG}=1$. Right: Region in $\eta_{bs}$ vs. $\eta_{GG}$ where gluon fusion dominates (below the dashed line). The blue region is excluded by rare $B$ meson decays for the maximal $\mathcal{B}(a\to \mu\mu) = 1.1\%$. Note that at low masses, a large-scale dependence is present. We only show the central value obtained for $\mu_F=\mu_R=m_a$ and refer to~\cite{Haisch:2016hzu} for a treatment of the error estimated by varying these parameters.}
\label{fig:xs14TeV}
\end{figure}

We suggest that, \emph{for promptly decaying ALPs}, plots like those in Fig.~\ref{fig:bounds_exp} are the most useful way to present the experimental results, lacking a signal. The model builders can then compare these bounds with their model predictions by overlapping the appropriate curve, as has been done in \cite{BuarqueFranzosi:2021kky} for the models in \cite{Ferretti:2016upr,Belyaev:2016ftv}.
The gluon fusion production cross section of the ALP, for $\sqrt s=14\mbox{ TeV}$ at LHC, setting $f=1\mbox{ TeV}$ and $\eta_{GG}=1$, can be computed with HIGLU \cite{Spira:1995rr} and is shown in Fig.~\ref{fig:xs14TeV} (left).

One can question how generic the dominance of gluon fusion production is for the relevant mass ranges. We checked that, for $\eta_{GG} \sim \eta_{\psi\psi}$, production via DY, $q\bar{q} \to a$, is always subdominant. In any case, we would not expect dramatic changes in the kinematic distribution from the DY channel; hence, similar bounds should be obtained. However, for masses below the $B$ meson mass, as relevant for the $DD$ channel, the ALP may also be produced in $B$ decays via the FV couplings, $pp\to B\bar{B}\ (B\to aX)$. We estimate the latter from the total production of $B$ mesons via 
\begin{align}
\sigma_{B\bar{B}\ (B\to aX)} = \sigma_{b\bar b}\times \frac{\Gamma(B\to aX)}{\Gamma_B},
\end{align} 
where $\sigma_{b\bar b}=5.\times 10^5\mbox{ nb}$ is the total $b\bar b$ cross section at LHC for $\sqrt s=14\mbox{ TeV}$, $\Gamma_B = \hbar/\tau_B= 4.01\times 10^{-13}\mbox{ GeV}$ is the total width of the $B$-meson, and we estimate the $B$ meson FV decay width via the quark formula
\begin{align}
   \Gamma(B\to aX)\approx\Gamma(b\to a s) = \frac{1}{32 \pi f^2}|\eta_{sb}|^2 m_b^3 \left(1-\frac{m_a^2}{m_B^2}\right)^2,
\end{align}
where we have considered as an example the effect of the FV coupling $\eta_{bs}$.

For $m_a=2 m_D$, the gluon fusion cross section is $\sigma_{ggF}= |\eta_{GG}|^2 (\mbox{TeV}/f)^2\, \bar\sigma$, where $\bar\sigma=3470~\mbox{ pb}$ from Fig.~\ref{fig:xs14TeV}. Comparing the two, we find that gluon fusion dominates as long as $\eta_{bs}/\eta_{GG} \leq 4\times 10^{-6}$, as shown by the dashed line in Fig.~\ref{fig:xs14TeV} (right). We also reported in blue the region excluded by rare $B$ decays, in this case dominated by $B^- \to K^- \mu \mu$, leading to $\eta_{bs}/f \times \sqrt{\mathcal{B}(a\to \mu\mu)} \leq 4.4 \times 10^{-8}~\mbox{TeV}^{-1}$ \cite{Bauer:2021mvw}, where the region in the plot corresponds to the worst case scenario $\mathcal{B}(a\to \mu \mu)  = 1.1\%$ (see also \cite{Ferretti:2025zsq}). Similar results can be obtained for FV coupling with the down, $\eta_{bd}$. We see that gluon fusion, indeed, dominates for sizable couplings to gluons in the allowed parameter space and for a decay constant smaller than considered in studies on Long Lived Particles \cite{DallaValleGarcia:2023xhh}.

We conclude with some general comments on the FV interactions, which are subject to very strong experimental bounds \cite{Bauer:2021mvw}. In the CHM context, and many other realizations, the FV coupling in the down sector are generated by the top couplings, hence they are proportional to $V^*_{ti} V_{tj}$, with $i,j=d,s,b$ ($i\not =j$) and $V$ the CKM matrix: $\eta_{ij} = \kappa \times V^*_{ti} V_{tj}$. The proportionality coefficient $\kappa$ can have various origins and varying orders of magnitude, depending on the high-energy model. This is discussed in general in \cite{Bauer:2020jbp,Bauer:2021mvw}, and their results are used in the CHM context in \cite{Ferretti:2025zsq}. 

There can be a $\Ocal(1)$ contribution to $\kappa$ if the high-energy Lagrangian from which (\ref{lagrangian}) arises does not preserve flavor. In this case, an ALP mass $m_a<m_B-m_K$ is fully excluded. If the UV Lagrangian preserves flavor, FV interactions still arise from RG running to the EW scale and from integrating out the heavy SM fields, also leading to strong constraints in the low mass region.

For $m_a> m_B$, $B_{d,s}\to\mu\mu$ decay via an off-shell ALP sets constraints on the quantities $\eta_{\mu\mu}\eta_{db}/V^*_{td} V_{tb}$ or $\eta_{\mu\mu}\eta_{sb}/V^*_{ts} V_{tb}$. However, in this mass range, $\eta_{ij}/V^*_{ti} V_{tj}$ can be much larger, even of order one, as long as one stays away from the pole in the $m_a\approx m_B$ region. The relevant discussion is found in \cite{Bauer:2021mvw}, using the SM computation of \cite{Beneke:2019slt}. In \cite{Ferretti:2025zsq}, the bounds are given for a generic CHM using the experimental values \cite{ParticleDataGroup:2024cfk}.

%



\subsection{Heavy Neutral Leptons}
\label{ssec:HNL}

\subsubsection{New benchmark models for HNLs searches --- \textit{J.~Klaric}}
\label{sssec:klaric}
\textit{Author: Juraj Klaric, \email{jklaric@nikhef.nl}}\\
\textit{
    This contribution provides a brief summary of the motivation for the new HNL benchmark models that were introduced as a part of the efforts of the FIPs centre HNL working group.
    Previous discussions can be found in the 2022 FIPs workshop report~\cite{Antel:2023hkf}, as well as a detailed discussion in the paper~\cite{Drewes:2022akb}.
}
\paragraph{Introduction}
Heavy Neutral Leptons (HNLs) are a thoroughly studied extension of the Standard Model (SM) of particle physics that could simultaneously generate neutrino masses through the seesaw mechanism~\cite{Minkowski:1977sc,Glashow:1979nm,Gell-Mann:1979vob,Mohapatra:1979ia,Yanagida:1980xy,Schechter:1980gr},
create the observed Baryon Asymmetry of the Universe (BAU) via leptogenesis~\cite{Fukugita:1986hr},
and even constitute Dark Matter (DM)~\cite{Dodelson:1993je}.

Crucially, HNLs can address these shortcomings while within the reach of existing and near-future experiments.
To interpret the results of experimental searches and their projected sensitivity,
it is crucial to understand how the realistic models map onto simplified phenomenological benchmarks that are typically assumed to generate experimental bounds.

The commonly studied simplified phenomenological model introduces one HNL field $N$ that interacts with the SM $W$, $Z$ and Higgs ($h$) fields through the \emph{neutrino portal} according to the following Lagrangian
\begin{equation}
 \mathcal L
\supset
- \frac{m_W}{v} \overline N \theta^*_\alpha \gamma^\mu e_{L \alpha} W^+_\mu
- \frac{m_Z}{\sqrt 2 v} \overline N \theta^*_\alpha \gamma^\mu \nu_{L \alpha} Z_\mu
- \frac{M}{v} \theta_\alpha h \overline{\nu_L}_\alpha N
+ \text{h.c.}
\ ,\label{PhenoModelLagrangian}
\end{equation}
where $e_{L \alpha}$ and $\nu_{L \alpha}$ are the charged leptons and the SM neutrinos, respectively, and $v \approx 174$ GeV is the vacuum expectation value of the Higgs field.
This model has four parameters: the HNL mass $M$,
and the three mixing angles $\theta_\alpha$ that determine the strength of the HNL interactions $U_\alpha^2 = |\theta_\alpha|^2$ with the SM flavours $\alpha = e, \mu,\tau$.
The overall coupling strength is characterized by summing over the flavoured mixing angles $U^2 = \sum_\alpha U_\alpha^2$.
In this simplified scenario with one HNL,
the phases in $\theta_\alpha$ are unphysical, so it is sufficient to characterize the modulus of the mixing angle $U_\alpha^2$.

Finally, the HNL field $N$, can either be a lepton number conserving (LNC) \emph{Dirac} or
lepton number violating (LNV) \emph{Majorana} field.
In realistic scenarios, HNLs do not necessarily correspond to the purely LNV or purely LNC case, but can allow for both types of processes.
The amount of lepton number violation is therefore often characterized by the ratio of LNV and LNC decays $R_{\ell\ell}$,
that takes values $\in [0,1]$ in realistic scenarios and leads to very rich phenomenology~\cite{Atre:2009rg,Drewes:2013gca,Deppisch:2015qwa,Anamiati:2016uxp,Antusch:2016ejd,Chun:2017spz,Cai:2017mow,Abdullahi:2022jlv},
with the limiting Dirac case corresponding to $R_{\ell\ell} = 0$, and Majorana to $R_{\ell\ell} = 1$.

These five parameters $(M, U_e^2, U_\mu^2, U_\tau^2, R_{\ell\ell})$ can still be too complex to be exhaustively explored in experimental searches.
In particular, for a fixed HNL mass $M$, mixing $U^2$ and amount of LNV $R_{\ell\ell}$,
the detection prospects highly depend on the flavour ratios $U_e^2: U_\mu^2: U_\tau^2$ (see e.g.~\cite{SHiP:2018xqw,Drewes:2018gkc,Tastet:2021vwp,CMS:2023ovx,CMS:2024xdq,CMS:2024ita}).

Most searches, therefore, focus on specific HNL benchmarks - assuming Dirac or Majorana HNLs that only couple to a single lepton flavour. This corresponds to the three benchmarks
\ref{BC6}--\ref{BC8} defined in~\cite{Beacham:2019nyx},
\begin{subequations}
	\begin{align}
        \label{BC6}
		U_e^2 : U_\mu^2 : U_\tau^2 = 1:0:0, \tag{BC6}\\
        \label{BC7}
		U_e^2 : U_\mu^2 : U_\tau^2 = 0:1:0, \tag{BC7}\\
        \label{BC8}
		U_e^2 : U_\mu^2 : U_\tau^2 = 0:0:1, \tag{BC8}
	\end{align}
\end{subequations}
with $R_{\ell\ell}=0$ for Dirac or $R_{\ell\ell}=1$ for Majorana HNLs.

These single-flavour benchmarks come with significant limitations:
1. they can neither generate the observed neutrino masses and oscillations, nor produce the observed BAU
2. they lack lepton flavour violation (LFV), which is a distinct signature for HNLs searches where it is difficult to observe a macroscopic displacement, and LNV signals are suppressed (see e.g.~\cite{Tastet:2021vwp}),
3. they cannot be probed by dedicated searches for charged LFV~\cite{Alonso:2012ji,deGouvea:2015euy,Abada:2015oba,Fernandez-Martinez:2016lgt,Abada:2018nio,Granelli:2022eru,Urquia-Calderon:2022ufc},
4. they are highly sensitive to small deviations from the single flavour limit, as even percent-level deviations (in particular for a pure-$\tau$ mixing as in \ref{BC8}) can significantly change the experimental sensitivities~\cite{Drewes:2018gkc}.

\begin{figure}
    \centering
    \includegraphics[height=0.25\textheight]{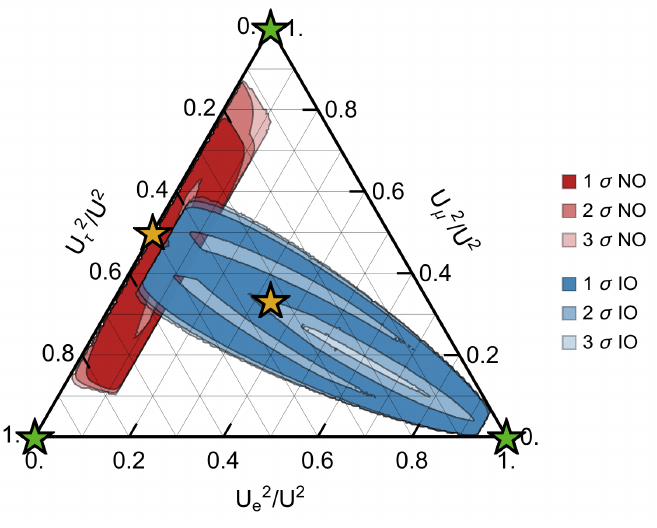}
    \includegraphics[height=0.25\textheight]{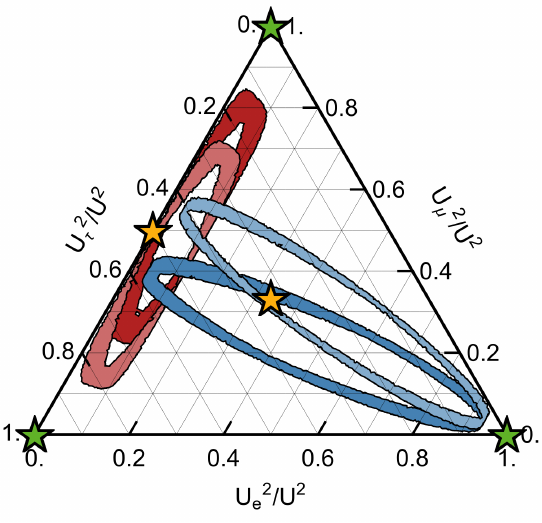}
    \caption{
        The range of mixing ratios $U_e^2 : U_\mu^2 : U_\tau^2$ compatible with the measured light neutrino properties in the minimal seesaw model for Normal Ordering (NO, red) and Inverted Ordering (IO, blue). The benchmark points from~\ref{NewBenchmarks} are indicated by yellow stars, and those from~\ref{BC6}--\ref{BC8} by green stars.
        \textbf{Left:} Contours showing the range allowed by current neutrino oscillation data~\cite{Esteban:2020cvm}.
        \textbf{Right:} Future projection based on 15 years of data taking at DUNE~\cite{DUNE:2020jqi}, assuming a true value of $\delta=-\pi/2$ and two benchmark values for the light neutrino mixing angle: $s_{23}^2\equiv \sin^2\theta_{23}=0.58$ (darker regions) and $s_{23}^2=0.42$ (lighter regions).
        Similar sensitivity can be expected at Hyper-Kamiokande~\cite{Hyper-Kamiokande:2018ofw,Ballett:2016daj}. Figures from~\cite{Drewes:2022akb}.
    }
    \label{fig:ternary}
\end{figure}

To make a closer connection to realistic neutrino mass models and as a part of the efforts of the FIPs physics centre HNL working group, we proposed two additional benchmarks in~\cite{Drewes:2022akb}:
\begin{subequations}
	\label{NewBenchmarks}
	\begin{align}
		\label{NObenchmark}
		U_e^2 : U_\mu^2 : U_\tau^2 &= 0:1:1\,,\\
		\label{IObenchmark}
		U_e^2 : U_\mu^2 : U_\tau^2 &= 1:1:1\,,
	\end{align}
\end{subequations}
which can be used with both $R_{\ell\ell}=0,1$.
Together with the benchmarks in~\ref{BC6}--\ref{BC8}, these capture a more complete physical picture of realistic models in accelerator-based experiments (see \cref{fig:ternary}).
Below we will summarize the criteria that were used to choose the new benchmarks and how they relate to realistic models of HNLs.

\paragraph{Criteria for benchmark selection}
Realistic HNL models based on the type-I seesaw mechanism differ from the phenomenological model in~\eqref{PhenoModelLagrangian} in several important ways.
The most important difference comes from the very number of HNLs that are present in the theory,
where each of the massive light neutrinos requires one corresponding HNL.
Since current neutrino oscillation data only provides evidence for the existence of two massive
light neutrinos, the lightest neutrino in the minimal type-I seesaw model is massless
($m_\text{lightest} = 0$), corresponding to $n=2$ HNL species.
However, if the lightest neutrino is found to be massive ($m_\text{lightest} \neq 0$),
a third HNL immediately becomes necessary.

This is also where the major difference and difficulty compared to the phenomenological model comes from: with more particles, we immediately get many more parameters.
After removing all the unphysical phases, one is left with $7 n -3$ physical parameters for $n$ HNLs.

The major challenge in selecting good benchmarks, therefore, corresponds to finding an adequate mapping from the realistic type-I seesaw model to the model in~\eqref{PhenoModelLagrangian}.

Fortunately, many of the parameters in the realistic seesaw model are already constrained by the requirement to reproduce the light neutrino masses and oscillations: if we take into account the measured values of the light neutrino mass differences, the three angles and one CP violating phase of the Pontecorvo--Maki--Nakagawa--Sakata (PMNS) matrix $V_\nu$, 6 of the aforementioned parameters become constrained, leaving us with only $7 n - 9$ free parameters.
The biggest impact of these constraints is on the mixing ratios $U^2_e : U^2_\mu : U^2_\tau$,
which are almost completely determined by the matrix $V_\nu$ in the minimal seesaw scenario. To evaluate the possible benchmark models, we considered the following criteria:

\begin{enumerate}
	\item \textbf{Consistency with neutrino oscillation data.} \label{crit1}
		One of the primary motivations for the existence of HNLs is that they can explain the origin of the light neutrino masses and their properties.
        When choosing benchmark HNL models, one needs to take into account that the range of allowed
        ratios $U_e^2:U_\mu^2:U_\tau^2$ is closely connected to the properties of the light neutrinos in the minimal type-I seesaw model.
        To identify this range, in~\cite{Drewes:2022akb}  we used an updated global fit to light neutrino oscillation data~\cite{Esteban:2020cvm}, compared to previous studies~\cite{SHiP:2018xqw, Drewes:2018gkc,Tastet:2021vwp}.
        We also estimated how measurements of the CP phase $\delta_{CP}$, and other neutrino properties at future neutrino oscillation experiments could modify the allowed range of HNL flavour ratios, as shown in \cref{fig:ternary}.
	\item \textbf{Added value.} \label{crit2}
        Introducing new benchmark models is only warranted if they can lead to significantly different predictions compared to the existing single-flavour benchmarks in \ref{BC6}--\ref{BC8}.
        This favours benchmark points that have a sizeable mixture of more than one lepton flavour.
        While already percent-level changes of the single-flavour benchmarks can significantly affect the expected experimental sensitivity
        as they open up new production and decay channels (for instance, for \ref{BC8}),
        sensitivities near benchmarks with more than one lepton flavour can be quite robust to such changes.
        Additionally, benchmarks with sizeable couplings to more than one lepton flavour can also lead to lepton-flavour violating signatures, which are not possible in the case of single-flavour benchmarks.
	\item \textbf{Symmetry considerations.}
        Given that our benchmark selection follows a ``bottom-up'' approach, criteria \ref{crit1} and \ref{crit2} took precedence in our decision-making process.
        However, when these primary criteria yield multiple viable options without a clear preference, theoretical model-building constraints provide valuable guidance. In particular, discrete symmetry frameworks applied to fermion mixing matrices~(cf.~\cite{Xing:2015fdg,King:2017guk,Xing:2020ijf}) offer additional compelling theoretical motivation for specific flavour ratios.
    \item \textbf{Simplicity.}
        Preference is given to $U_e^2:U_\mu^2:U_\tau^2$ ratios that can be expressed in simple terms, facilitating clear communication within the research community. When multiple options are similarly justified by other criteria, we favour benchmarks with straightforward numerical representations that are easy to implement, discuss, and remember.
	\item \textbf{Leptogenesis.}
        Another strong motivation for the existence of HNLs is baryogenesis through leptogenesis, specifically the low-scale leptogenesis mechanisms~\cite{Akhmedov:1998qx,Pilaftsis:2003gt,Asaka:2005pn}.
        Existing studies of the parameter space indicate that a significant part of this parameter space is already within reach of existing experiments (cf.~\cite{Klaric:2020phc,Klaric:2021cpi,Hernandez:2022ivz})
		Therefore, we favoured flavour ratios that can allow for leptogenesis within the experimentally testable part of the parameter space.
\end{enumerate}

\paragraph{The minimal type-I seesaw model} In this minimal scenario, the SM is extended by two HNLs that are required to reproduce the observed light neutrino masses.
The HNLs are characterized by their masses $M_I$ and mixing angles $\theta_{alpha I}$, through which they generate the masses of the light neutrinos, which can be expressed as:
\begin{align}
    \label{eq:seesaw_formula}
    (m_\nu)_{\alpha \beta} \approx - \theta_{\alpha I} M_I \theta^T_{I \beta}\,.
\end{align}
This relation between the light neutrino mass matrix, the HNL masses and mixing angles strongly constrains the allowed mixing ratios.
Once the bounds on the neutrino masses are taken into account, this model has 11 free parameters, with 6 of them that are already measured or constrained by the light neutrino oscillation experiments.

The region of parameter space that is accessible to most accelerator-based experiments typically corresponds to mixing angles much larger than one would expect from the seesaw formula
\begin{align}
    \label{eq:LargeMixings}
    |\theta_{\alpha I}|^2 \gg \frac{\sum_i m_i}{M_I}\,,
\end{align}
where $m_i$ are the masses of the light neutrinos.
To achieve this limit without causing large radiative corrections to the light neutrino masses, the two HNLs need to respect a generalised global $U(1)_{B-L}$ symmetry~\cite{Shaposhnikov:2006nn,Kersten:2007vk}.
Such a symmetry is a common feature of a number of low-scale seesaw models, such as
the \emph{linear}~\cite{Akhmedov:1995ip,Akhmedov:1995vm,Gavela:2009cd}
and \emph{inverse}~\cite{Wyler:1982dd,Mohapatra:1986aw,Mohapatra:1986bd,Bernabeu:1987gr,Branco:1988ex} seesaw models, as well as in the $\nu$MSM~\cite{Asaka:2005pn,Asaka:2005an}.
In these scenarios, the smallness of the light neutrino masses does not come from the suppression by a large HNL mass $M_I$, but is associated with small $B-L$ breaking parameters, allowing for mixing angles that are $\mathcal{O}(1)$ without spoiling the stability of the light neutrino masses under radiative corrections.
In such a scenario, the HNL masses are close to degenerate with $M_1 \approx M_2 \equiv M$, and their mixing angles are approximately equal up to a phase $\theta_{\alpha 1} \approx i \theta_{\alpha 2}$, which ensures an approximate cancellation in~\eqref{eq:seesaw_formula}.
These constraints bring the realistic scenario much closer to the phenomenological model in~\eqref{PhenoModelLagrangian}, albeit with constraints on the mixing angles $\theta_{\alpha I}$ coming from the neutrino masses $m_i$ and the PMNS matrix $V_\nu$.

Existing neutrino oscillation experiments have successfully measured two of the light neutrino mass splittings, three angles of the matrix $V_\nu$, and constrained one phase, $\delta_{CP}$~\cite{Esteban:2020cvm}.
The remaining \emph{Majorana} phase cannot be directly constrained by neutrino oscillation experiments, but could instead be probed in processes such as neutrinoless double beta $(0\nu\beta\beta)$ decay.
Since only the two light neutrino mass differences are known, there are two possible realizations of the light neutrino spectrum currently allowed:
\emph{normal} (NO) and \emph{inverted} (IO) ordering.
To determine the allowed range of the mixing ratios
$U_e^2 : U_\mu^2 : U_\tau^2$, we varied over the neutrino mass ordering,
the unknown Majorana phase,
as well as two of the least constrained neutrino oscillation parameters, phase $\delta_{CP}$, and the angle $\theta_{23}$,
by taking the 2-d $\chi^2$ projections from~\cite{Esteban:2020cvm}.

The compatibility with neutrino oscillation data is the primary criterion for selecting benchmark scenarios~\ref{crit1}, making it essential to compare the allowed mixing ratio ranges shown in Fig.~\ref{fig:ternary} with our proposed benchmarks.
While specific parameter choices can result in mixing patterns dominated by a single flavour, it is important to note that none of the single-flavour benchmarks (\ref{BC6}--\ref{BC8}) can be strictly realized within the minimal seesaw model.

Experimental sensitivities can vary by several orders of magnitude even with small deviations from single-flavour mixing limits.
This observation from existing data strongly motivates the introduction of the two benchmarks in \ref{NewBenchmarks}.
The benchmark in \ref{IObenchmark} is achievable only for the IO case; however, it can effectively represent NO scenarios with substantial electron mixing ($U_e^2/U^2 \sim 0.1$), as experimental sensitivities remain relatively consistent when flavour mixings vary within the same order of magnitude.
The benchmark in \ref{NObenchmark} is not strictly realized in either mass ordering, yet it serves as a good approximation for the NO case with minimal electron mixing ($U_e^2/U^2 \sim 10^{-3}$).

\paragraph{Leptogenesis}
One of the major features of the type-I seesaw mechanism is that it also provides a way to generate the observed BAU via low-scale leptogenesis~\cite{Akhmedov:1998qx,Pilaftsis:2003gt,Asaka:2005pn}.
In the minimal scenario, this is possible within a wide range of HNL masses and mixing angles~\cite{Klaric:2020phc,Klaric:2021cpi,Hernandez:2022ivz}.
While low-scale leptogenesis is compatible with the full range of mixing patterns that are allowed by the type-I seesaw mechanism, specific mixing patterns can be more favourable and allow for a wider range of the remaining parameters~\cite{Antusch:2017pkq,Hernandez:2022ivz}.
Situations where the HNL approximately decouples from one of the lepton flavours can lead to a highly \emph{flavour asymmetric washout},
which can allow for both larger overall mixing angles and larger HNL mass splittings than for a generic point in the parameter space.
This motivates the choice \cref{NObenchmark}, where the electron approximately decouples, as well as \ref{BC6},
for which both the $\tau$ and $\mu$ mixings can be suppressed.

\paragraph{Dirac or Majorana HNLs}
Individually, each of the HNLs in the minimal type-I seesaw mechanism is a Majorana field,
and should therefore violate lepton number.
However, in the symmetry-protected scenario, the contributions from the two HNLs can interfere destructively, which leads to an overall suppression of LNV.
In practice, this means that depending on the exact parameters of the theory, the amount of LNV can vary between
$R_{\ell\ell}=1$ and $R_{\ell\ell}=0$.
Therefore, when mapping the minimal type-I seesaw model onto the phenomenological Lagrangian \cref{PhenoModelLagrangian}, one should consider both Dirac $R_{\ell\ell}=0$ and Majorana $R_{\ell\ell}=1$ HNLs, as both can appear as limiting cases of the realistic type-I seesaw scenario.
This choice also sets the relation between the mixing angle and decay width of HNLs, where Majorana HNLs have a decay width that is twice as large compared to the Dirac HNL case.

\paragraph{The model with 3 HNLs}
The minimal model with two HNLs can only generate two of the light neutrino masses.
If it turns out that all three light neutrinos are massive, a third HNL immediately becomes necessary.
In contrast to the minimal scenario, the case with 3 HNLs has 18 free parameters.
The mixing patterns in this scenario are no longer only sensitive to $V_\nu$ (which comes with one additional Majorana phase), but also depend on the additional HNL parameters (two at leading order).
This greatly extends the range of mixing patterns that are allowed.
While the allowed range is quite similar to \cref{fig:ternary} for $m_\text{lightest}=0$, already
$m_\mathrm{lightest} \gtrsim 10^{-5}$ eV can cause a significant deviation from this pattern,
with $m_\mathrm{lightest} \sim 10^{-2}$ eV allowing the full range of flavour ratios~\cite{Chrzaszcz:2019inj}.
Another advantage of this scenario is that leptogenesis becomes possible with significantly larger
mixing angles compared to the minimal model~\cite{Abada:2018oly,Drewes:2021nqr}.
While this does not necessarily constrain the choice of new benchmark scenarios, this model can further support the benchmarks~\ref{BC6}--\ref{BC8} which are not possible in the minimal case.

\paragraph{Connection to indirect probes}
New benchmarks can also be motivated by their connection to indirect probes of HNLs such as neutrinoless double beta decay and searches for rare processes such as charged lepton flavour violation.

\paragraph{Neutrinoless double beta decay}
The Majorana nature of the light neutrinos and the associated LNV is one of the key predictions of the minimal type-I seesaw model.
This type of LNV could be probed in experimental searches for neutrinoless double beta decay.
The rate of this process depends on a CP-violating \emph{Majorana} phase of $V_\nu$, which is inaccessible through neutrino oscillation experiments.

Interestingly, the same phase determines the mixing ratios in this scenario, which allows us to correlate the rate of the $0\nu\beta\beta$ decay with the ratio
$U_e^2/U^2$~\cite{Drewes:2022akb}.
The prospects of further constraining the allowed mixing ratios unfortunately remain slim due to the large theoretical uncertainties in the computed $0\nu\beta\beta$ decay rate.

Interestingly, HNLs could also modify the $0\nu\beta\beta$ decay rate directly~\cite{Bezrukov:2005mx,Blennow:2010th,Asaka:2011pb,Lopez-Pavon:2012yda,Drewes:2016lqo,Hernandez:2016kel,Asaka:2016zib}.
The simple relation between the mixing $U_e^2$ and the $0\nu\beta\beta$ rate only holds in the cases where we can neglect the HNL contribution to this process.
HNLs could also lead to an enhancement or a suppression of this rate compared to the naive expectation.
Together with other bounds on the HNL properties, this could lead to clear target regions for HNL searches~\cite{deVries:2024rfh}.

\paragraph{Charged Lepton Flavour Violation}
HNLs generically allow for processes that lead to charged lepton flavour violation. While such processes are not possible for the benchmarks ~\ref{BC6}--\ref{BC8}, the benchmarks in \eqref{NewBenchmarks} would allow for such processes.
This also opens up a complementary probe of HNLs in processes such as $\mu^\pm \rightarrow e^\pm \gamma$, $\mu^\pm \rightarrow e^\pm e^+ e^-$.
These can probe HNLs with masses beyond the collider-testable region, up to a few TeV; however, they are limited to large mixing angles $U^2\gtrsim 10^{-7}$.
While such mixings are typically incompatible with leptogenesis in the minimal model, they could be realized in the scenario with 3 HNLs.

\paragraph{Summary}
HNLs are an exciting extension of the SM that could simultaneously offer a solution to several shortcomings of the SM.
HNLs in the GeV-mass range could be produced at several existing and upcoming experiments.
The bounds on HNL mixing angles $U_\alpha^2$ fundamentally depend on what we assume about ratios.
Most searches rely on single-flavour benchmarks \ref{BC6}--\ref{BC8},
which are in tension with realistic models of neutrino masses -- one of the primary motivations for the existence of HNLs.
Another drawback of the single-flavour benchmarks is that they do not allow for LFV processes,
which could play a crucial role when other smoking-gun signals of HNLs (such as displaced vertices or LNV) are absent.
To address these shortcomings, in~\cite{Drewes:2022akb} we introduced two additional benchmarks~\ref{NewBenchmarks}.
Together, these five benchmark scenarios can cover a wide range of phenomena and signals that exist in realistic models of neutrino masses.

\subsubsection{Phenomenology of GeV-scale HNLs --- \textit{F.~Kling}}
\label{sssec:kling}
\textit{Author: Felix Kling, \email{felix.kling@desy.de}}\\
\paragraph{Introduction and Motivation} The SM is a remarkably successful theory of particle physics and includes all of the observed particles in nature. The neutrino in the SM is special for several reasons: i) Unlike all other fermions, which have both left- and right-handed fields, only left-handed neutrino fields exist. ii) The masses of neutrinos are much smaller than the weak scale, $m_\nu / v \sim 10^{-13}$, and also much smaller than the electron mass, which is the next lightest fermion, $m_\nu/m_e \sim 10^{-7}$. This can be interpreted as a hint that our understanding of neutrinos is incomplete.

Indeed, right-handed neutrinos may exist but may have simply escaped our attention. Such a right-handed neutrino $\nu_R$ would need to be a singlet under all SM gauge groups. It would not interact via the electromagnetic, strong, or weak forces but would only couple to the SM via the (very feeble) Yukawa interaction $\sim \frac{m_{\nu}}{v} \bar L \tilde H \nu_R$. In addition, since it is a singlet under all gauge groups, it could have a (potentially very heavy) Majorana mass $\sim M \bar \nu_R^c \nu_R$. Both its feeble interaction and potentially heavy mass would explain the non-observation of this right-handed neutrino state.

At the same time, the Majorana mass could explain the smallness of the neutrino masses. This is most easily seen in the Type~I seesaw model with one additional right-handed neutrino $\nu_R$. The corresponding Lagrangian is
\begin{equation}
\mathcal{L} = i \bar{\nu}_R \partial \!\!\! / \, \nu_R - y \bar L \tilde H \nu_R - m_M \bar{\nu}^c_R \nu_R + h.c. \ ,
\end{equation}
where the second term gives a Dirac mass $m_D = y v / \sqrt{2}$. In the limit $m_M \gg m_D$, diagonalizing the mass matrix yields two mass eigenstates: the active neutrino $\nu \approx \nu_L - \theta \nu_R$ with $m_\nu \approx m_D^2/m_M$ and the sterile neutrino $N \approx \nu_R + \theta \nu_L$ with $m_N \approx m_M$, where the mixing angle is $\theta \approx m_D/m_M$. We see that a small neutrino mass of $m_\nu \sim 0.1~\text{eV}$ naturally arises if the sterile neutrino has a Majorana mass $m_N \sim m_M \sim y^2 \cdot 10^{14}~\text{GeV}$. Due to the mixing, the sterile neutrino also obtains a coupling to weak gauge bosons, which is suppressed by $\theta^2 \approx m_D^2/m_M^2 \approx m_\nu / m_N$. A sterile neutrino $N$ resulting from the simple Type-I seesaw mechanism with only one additional state is therefore either too heavy or too weakly coupled to be probed by accelerator or collider experiments. However, in practice, we expect several generations of sterile neutrinos. In this case, a non-trivial structure of the Majorana mass matrix could lead to a broader spectrum of masses and couplings.

In addition, there are other variants of the seesaw mechanism that can yield lighter sterile neutrinos with sizable couplings. One example is the inverse seesaw mechanism, which contains two fermions, $\nu_R$ and $\nu_s$, that are singlets under all gauge groups and have opposite fermion numbers. The corresponding Lagrangian is
\begin{equation}
\mathcal{L} = y \bar L \tilde H \nu_R + m_M \bar \nu_R^c \nu_S + \mu \bar{\nu}_S^c \nu_S + h.c.
\end{equation}
Assuming a mass ordering $\mu \ll y v \ll m_M$, the active neutrino mass is $m_\nu \sim \mu y^2 v^2 / m_M^2$, while the sterile neutrino mass is $m_N \sim m_M$. In this case, a light active neutrino with mass $m_\nu \sim 0.1~\text{eV}$ can be achieved by choosing Yukawa couplings $y \sim 10^{-4}$, a Majorana mass $m_M \sim \text{GeV}$, and $\mu \sim 1~\text{keV}$. The resulting mixing between left- and right-handed neutrinos, $\theta \sim m_D/m_M \sim 10^{-2}$, can then be sizable. This scenario can therefore lead to a sterile neutrino with GeV-scale mass and sizable couplings, which are within reach of upcoming experiments. \medskip

\paragraph{Effective Phenomenological Model of HNLs} In the literature, a variety of models for neutrino mass generation have been proposed that could lead to observable sterile neutrinos, also called heavy neutral leptons (HNLs). In practice, realistic models typically contain at least three generations and many free parameters that affect the spectrum of masses and mixings.

For phenomenological studies, it is often practical to consider an effective model in which the phenomenology is dominated by a single HNL $N$, while others either contribute subdominantly or are outside the mass range of interest. Barring rather special scenarios, for example, where the HNLs are almost mass-degenerate and can oscillate into each other on relevant length scales~\cite{Tastet:2019nqj, Antel:2023hkf}, the signal rates for HNLs in models with two or more HNLs with significant mixings can be determined simply by adding together the signal rates for each HNL considered separately. In this case, the neutrino flavor eigenstates can be expressed in terms of the mass eigenstates as
\begin{equation}
\nu_\alpha = \sum_{i = 1}^{3} V_{\alpha i} \,\nu_i + U_\alpha N^c ,
\end{equation}
where $V_{\alpha i }$ is the PMNS matrix parameterizing the active neutrino content, and $U_\alpha$ for $\alpha = e$, $\mu$, and $\tau$ parameterizes the active-sterile neutrino mixing. Through these mixings $U_\alpha$, the HNL obtains a small coupling to SM gauge bosons. The charged-current (CC) and neutral-current (NC) interaction terms are
\begin{equation}
\mathcal{L}^{\text{CC}} = -\frac{g}{\sqrt{2}}\sum_{\alpha} U^*_\alpha  W_\mu^+  \,\overline{N^c} \, \gamma^\mu l_\alpha + \text{h.c.}
\qquad \text{and} \qquad
\mathcal{L}^{\text{NC}} = - \frac{g}{2\cos\theta_W}\sum_{\alpha} U^*_\alpha Z_\mu  \overline{N^c}\, \gamma^\mu \nu_\alpha+ \text{h.c.}
\end{equation}
Previous studies performed in the context of the CERN Physics Beyond Colliders (PBC) Initiative~\cite{Beacham:2019nyx, Alemany:2019vsk} mainly focused on benchmarks where one of the mixings dominates while the others vanish. However, realistic models typically feature mixed couplings. Some motivated benchmarks have, for example, been proposed in Ref.~\cite{Drewes:2022akb}. \medskip

\paragraph{HNL Phenomenology in HNLCalc} Compared to other models of long-lived particles, such as the dark photon or dark Higgs, HNLs have a large number of production and decay channels. Moreover, these decays are typically described by complicated expressions for differential branching fractions. As a result, simulating HNLs is more challenging than for other models. To address this, the authors of Ref.~\cite{Feng:2024zfe} developed the \texttt{HNLCalc} Python library, which calculates HNL production rates, lifetimes, and decay branching fractions for arbitrary mixings $U_\alpha$. The tool also includes an interface to the \texttt{FORESEE} event generator~\cite{Kling:2021fwx}. \medskip

\paragraph{HNL Production} At the LHC, high-energy proton-proton collisions produce a large number of hadrons and tau leptons, which can decay into HNLs via CC and NC interactions. HNLs with masses in the GeV range are therefore primarily generated in the decays of these heavy mesons and tau leptons.

\begin{table}[tbp]
\footnotesize
\setlength{\tabcolsep}{2pt}
\centering
\begin{tabular}{ c |c c c c c c}
\hline\hline
$P \to l N$
& $\pi^+ \to l^+ N$                 & $K^+ \to l^+ N$
& $D^+ \to l^+ N$                   & $D_s^+ \to l^+ N$
& $B^+ \to l^+ N$                   & $B_c^+ \to l^+ N$
\\
\hline
$P \to P^\prime l N$
& $K^+ \to \pi^0 l^+ N$             &  $K_S \to \pi^{+} l^{-} N$
& $K_L \to \pi^{+} l^{-} N$         & $D^0 \to K^- l^+ N$
& $\bar{D}^0 \to \pi^+ l^- N$       & $D^+ \to \pi^0 l^+ N$\\
&$D^+ \to \eta l^+ N$              & $D^+ \to \eta^\prime l^+ N$
& $D^+ \to \bar{K}^0 l^+ N$         & $D_s^+ \to \bar{K^0} l^+ N$
& $D_s^+ \to \eta l^+ N$            & $D_s^+ \to \eta^\prime l^+ N$\\
&$B^+ \to \pi^0 l^+ N$             & $B^+ \to \eta l^+ N$
& $B^+ \to \eta^\prime l^+ N$       & $B^+ \to \bar{D}^0 l^+ N$
& $B^0 \to \pi^- l^+ N$             & $B^0 \to D^- l^+ N$\\
&$B^0_s \to K^- l^+ N$             & $B^0_s \to D^-_s l^+ N$
& $B^+_c \to D^0 l^+ N$             & $B^+_c \to \eta_c l^+ N$
& $B^+_c \to B^0 l^+ N$             & $B^+_c \to B^0_s l^+ N$\\
\hline
$P \to V l N$
& $D^0 \to \rho^- l^+ N$            & $D^0 \to K^{*-} l^+ N$
& $D^+ \to \rho^0 l^+ N$            & $D^+ \to \omega l^+ N$
& $D^+ \to \bar{K}^{*0} l^+ N$      & $D^+_s \to K^{*0} l^+ N$\\
&$D^+_s \to \phi l^+ N$            & $B^+ \to \rho^0 l^+ N$
& $B^+ \to \omega l^+ N$            & $B^+ \to \bar{D}^{*0} l^+ N$
& $B^0 \to \rho^- l^+ N$            & $B^0 \to D^{*-} l^+ N$\\
&$B^0_s \to K^{*-} l^+ N$          & $B^0_s \to D^{*-}_s l^+ N$
& $B_c^+ \to D^{*0} l^+ N$          & $B_c^+ \to J/\psi \; l^+ N$
& $B^+_c \to B^{*0} l^+ N$          & $B^+_c \to B^{*0}_s l^+ N$ \\
\hline
$\tau \to H N$
& $\tau^+ \to \pi^+ N$              & $\tau^+ \to K^+ N$
& $\tau^+ \to \rho^+ N$             & $\tau^+ \to K^{*+} N$ &&\\
\hline
$\tau^+ \to l^+ \nu N$
& $\tau^+ \to l^+ \bar{\nu}_\tau N$ & $\tau^+ \to l^+ \nu_l N$
&&&& \\
\hline \hline
\end{tabular}
\caption{HNL production processes included in \texttt{HNLCalc}. The processes are ordered by increasing parent particle mass; $P$, $V$, and $H$ denote pseudoscalar mesons, vector mesons, and hadrons, respectively; $l=e,\mu,\tau$; and $N$ is the HNL.  Charge-conjugate processes are also implemented, but are not explicitly listed here.}
\label{tab:production modes}
\end{table}

The \texttt{HNLCalc} library includes over 100 production channels, summarized in Table~\ref{tab:production modes}. These encompass both two-body and three-body decays of mesons and tau leptons. Notably, the parent hadrons considered are pseudoscalar mesons. Decays of vector mesons do not contribute significantly to HNL production, because the decays to HNLs compete with decays mediated by the strong interactions and so are highly suppressed, and HNL production in baryon decays is also not included, since they are subdominant. The relevant decay rate expressions are taken from the literature, most notably Ref.~\cite{Gorbunov:2007ak}, with various typos and errors in existing sources identified and corrected.

\begin{figure}[t]
\includegraphics[width=0.44\textwidth]{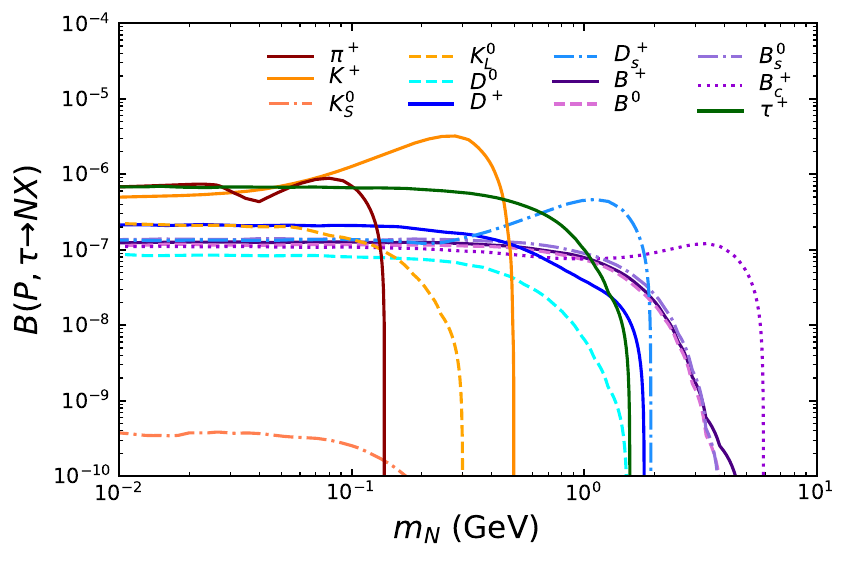}
\includegraphics[width=0.26\textwidth]{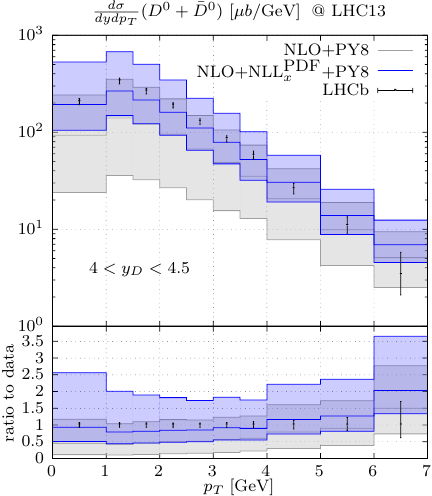}
\includegraphics[width=0.26\textwidth]{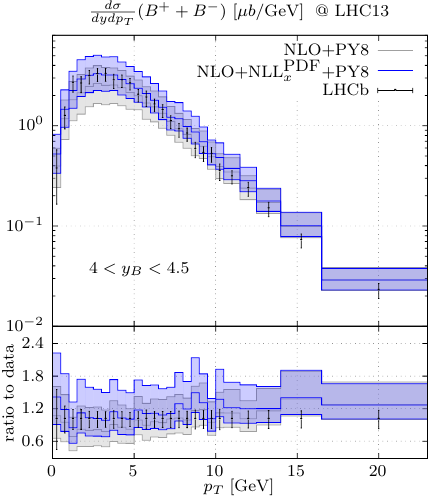}
\caption{Left: Branching fractions of mesons and tau leptons into HNLs as functions of the HNL mass $m_N$ for $U_e = U_\mu = U_\tau$ where $\sum U_\alpha^2 = 10^{-6}$ as obtained in Ref.~\cite{Feng:2024zfe}. Center and Right: State-of-the-art predictions for the production of D-meson and B-meson in comparison with LHCb as obtained in Ref.~\cite{Buonocore:2023kna}.}
\label{fig:HNL-BR}
\end{figure}

The branching fractions for meson and tau lepton decays into HNLs are presented in the left panel of Fig.~\ref{fig:HNL-BR} as functions of the HNL mass. For illustrative purposes, we show results for the $U_e = U_\mu = U_\tau$ benchmark model. In this figure, we have fixed $\sum U_\alpha^2 = 10^{-6}$, a value that is approximately at the limit of current constraints. We see that branching fractions of order $10^{-6}$ are allowed.

Let us further note a few general features: i) The branching fractions are typically larger for lighter parent mesons, as their competing SM decay modes have smaller widths. ii) The branching fractions vanish when the decay modes become kinematically inaccessible. iii) For parent mesons where the dominant production mode is chirality suppressed -- such as $K^+$ mesons, where the primary HNL decay mode is $K^+ \to \ell^+ N$ -- the branching fraction vanishes at $m_N = 0$ and increases with $m_N$. In such cases, the branching fraction can be large and even maximal near the kinematic threshold. iv) Since more light mesons than heavy mesons are produced at the LHC, the dominant HNL production mechanism is the decay of the lightest parent mesons for which the decay is kinematically allowed.

In addition to the decay rates of hadrons into HNLs, one also needs to know the parent hadron fluxes, particularly for heavy charm- and bottom-flavored hadrons. Fortunately, their production rates have been constrained by LHCb open charm and bottom data~\cite{LHCb:2015swx, LHCb:2017vec} and have been used to validate and improve simulations. A new state-of-the-art QCD prediction for heavy hadron production has been presented in Ref.~\cite{Buonocore:2023kna}. It includes radiative corrections at next-to-leading order using \texttt{POWHEG}~\cite{Nason:2004rx, Frixione:2007vw, Alioli:2010xd}, employs the \texttt{NNPDF3.1sx+LHCb} parton distribution function including small-x resummation at next-to-leading logarithmic accuracy that was fitted to LHCb data~\cite{Ball:2017otu, Bertone:2018dse} and uses \texttt{Pythia}~\cite{Sjostrand:2014zea} to match the predictions to parton showers and to provide a realistic description of hadronization effects. We emphasize the importance of the latter, as fragmentation functions are known to become invalid for low-transverse-momentum particles at hadron colliders~\cite{Bhattacharya:2023zei}, since they do not, for example, capture effects of hadronization with remnants or the observed enhanced baryon production~\cite{ALICE:2021rzj}, both of which can substantially alter predictions. \medskip

\paragraph{HNL Decays} Once produced, HNLs decay to SM final states through the CC and NC interactions. \texttt{HNLCalc} includes over 100 individual decay channels, including leptonic 3-body decays such as $N \to 3\nu$ and $\ell^-_\alpha \ell^+_\beta \nu$, semi-leptonic 2-body decays such as $N \to \nu \pi^{0}$ and $\ell^+ \pi^-$, and semi-leptonic 3-body decays. The corresponding HNL decay branching fractions are computed using formulas derived in Refs.~\cite{Coloma:2020lgy, Bondarenko:2018ptm}.

To properly simulate the detector's response to an HNL decay, it is important to accurately know and represent its final states. For example, it can make a significant difference experimentally whether an HNL decays to states with charged tracks, e.g., $N \to \nu \rho \to \nu \pi^+ \pi^-$, or to states with only final state photons, e.g., $N \to \nu \pi^0 \to \nu \gamma \gamma$. For this reason, it is generally insufficient to specify HNL decays into quarks, as this would rely on hadronization tools to obtain hadronic final states. These tools are known to perform poorly when the invariant mass of the hadronic final state is close to or below the QCD confinement scale, because, for example, this treatment fails to model hadronic resonances and kinematic thresholds. More generally, in this regime, the factorization theorem loses validity, meaning that one cannot factorize the HNL decay into quarks and use quark hadronization into hadrons anymore. On the other hand, for $m_N > 1$~GeV, decays into single mesons are insufficient for computing the total HNL decay width, as multi-meson decays become important in this region. In this case, we follow the approach taken in Refs.~\cite{Coloma:2020lgy, Bondarenko:2018ptm} and calculate the total hadronic width by summing up HNL decay widths into the quark-level final states $\ell q \bar{q}'$ and $\nu q \bar{q}$, including appropriate loop correction and phase space suppression factors to account for thresholds.

\begin{figure}[t]
\includegraphics[width=1.0\textwidth]{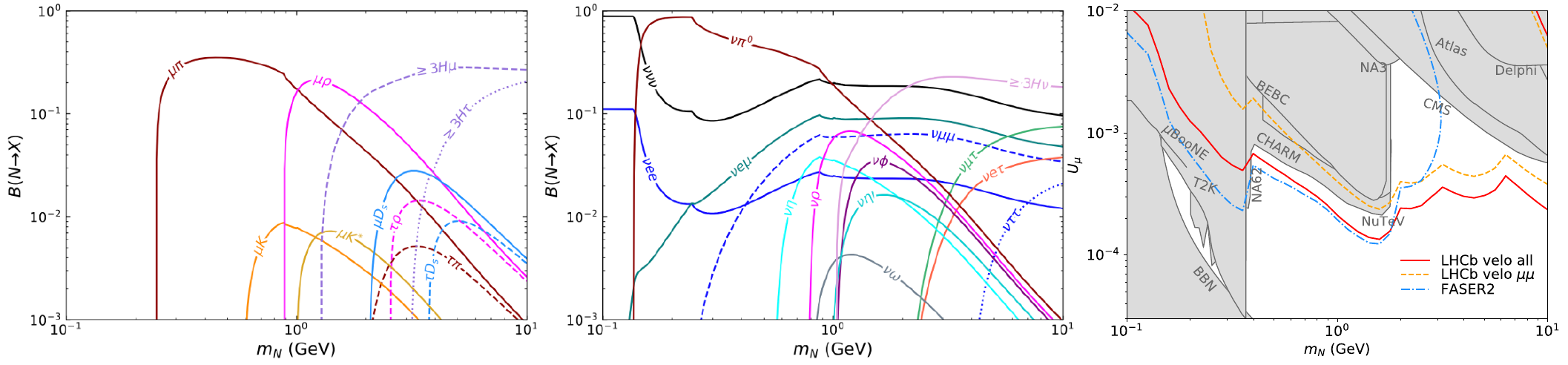}
\caption{Left and Center: HNL branching fractions in the $U_e = U_\mu = U_\tau$ benchmark model for the dominant decays via CC and NC interactions as functions of $m_N$. Right: Sensitivity of LHCb and FASER2 to HNLs with $U_\mu \neq 0$ and $U_e = U_\tau = 0$.}
\label{fig:HNL-BR-decay}
\end{figure}

Included in our single meson decays are decays into $\rho$ mesons. Ref.~\cite{Bondarenko:2018ptm} shows that the dominant source of two-pion final states is decays into $\rho$ mesons followed by $\rho \to \pi \pi$. To estimate the decay width to even higher multiplicity final states with $\geq 3$ hadrons, we take the difference between the quark-level decay width and the single-meson total width. In the left and central panels of Fig.~\ref{fig:HNL-BR-decay}, we plot the branching fractions for all relevant modes in the $U_e = U_\mu = U_\tau$ benchmark case. For $m_N < m_{\pi}$, HNL decays are dominated by NC decays into the 3-body final states $\nu \bar{\nu} \nu$ and $\nu e^+ e^-$. For $m_N > m_{\pi}$, the 2-body decays $N \to l^\pm \pi^\mp$ and $N \to \nu \pi^0$ become kinematically accessible, leading to a sharp drop in the lifetime. For masses $m_N > m_{\pi}$, the hadronic decay modes become dominant, and the invisible decay branching fraction drops below 20\%.\medskip

\paragraph{HNL Sensitivities} Let us now use the previously discussed HNL phenomenology to estimate the sensitivity of LHCb for HNL searches in the mixing versus mass plane. For concreteness, we will consider the scenario in which $U_\mu \neq 0$ and $U_e = U_\tau = 0$. However, we note that the existing constraints and sensitivities are quite similar for most other benchmark models.

As an illustration, we consider HNLs decaying within the LHCb VELO, assuming an integrated luminosity of $300~\text{fb}^{-1}$. In particular, we require the decay vertex to be within LHCb's pseudorapidity acceptance of $2 < \eta < 5$ and within $10~\text{cm} < z < 60~\text{cm}$ of the interaction point along the longitudinal direction. The minimal distance requirement is intended to reject possible backgrounds associated with displaced vertices from heavy hadron or tau lepton decays. We further require the HNL energy to exceed 10~GeV to avoid low-energy backgrounds. We estimate the corresponding event rate using the \texttt{FORESEE} generator~\cite{Kling:2021fwx}. For simplicity, we assume 100\% efficiency for all visible final states and no backgrounds. We note, however, that additional cuts to suppress backgrounds may need to be applied in a realistic analysis. The corresponding sensitivity curve is shown in the right panel of Fig.~\ref{fig:HNL-BR-decay} and compared to that of FASER2~\cite{Kling:2018wct, FASER:2018eoc, Feng:2024zfe}. This indicates the great potential of LHCb to search for HNLs.

\subsubsection{Heavy neutrino anti-neutrino oscillations --- \textit{J.~Hajer}}
\label{sssec:hajer}
\textit{Author: Jan Hajer, \email{jan@hajer.com}}\\
%
%


\begin{figure}
\small
\begin{subcaptionblock}{.55\textwidth}\centering
\includegraphics[width=\linewidth]{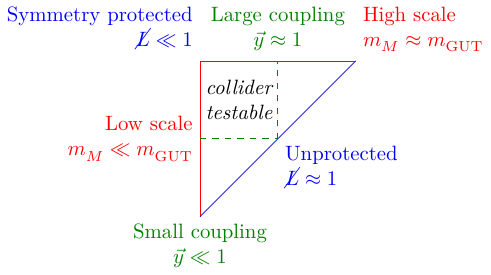}
\caption{Parameter regions of type I seesaw models} \label{hajer:triangle}
\end{subcaptionblock}%
\hfil
\begin{subcaptionblock}{.45\textwidth}\centering
\includegraphics[width=\linewidth]{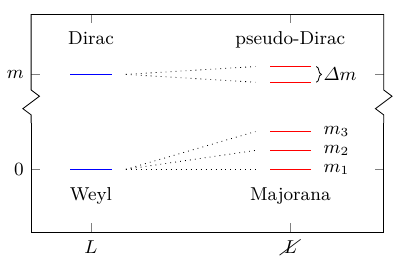}
\caption{Symmetry breaking induced mass splitting} \label{hajer:mass-splitting}
\end{subcaptionblock}%
\caption{
Panel \subref{hajer:triangle}:
In order to ensure tiny neutrino masses, the parameters of the type I seesaw are constrained.
For a single right-handed neutrino, either its mass must be very heavy or its coupling very small.
In the case of two right-handed neutrinos, the two contributions to the neutrino masses can also cancel each other, as is the case in the symmetry-protected seesaw scenarios.
Panel \subref{hajer:mass-splitting}:
A small breaking of the $U(1)_L$ not only generates tiny and protected neutrino masses but also induces a small mass splitting to the Dirac HNL, producing a pseudo-Dirac pair of two Majorana neutrinos.
} \label{hajer:depictions}
\end{figure}

In order to explain the observed neutrino oscillation data within the type I seesaw extension of the Standard Model (SM), at least two right-handed Majorana neutrinos are necessary.
Furthermore, they must be either so heavy or feebly interacting that a discovery at current or future experiments is barely possible.
The exceptions are models that ensure a large cancellation between the neutrino mass contributions of these two right-handed neutrinos.
This cancellation can be protected by a \emph{lepton number}-like $U(1)_L$ symmetry (LNLS), leading to symmetry-protected seesaw scenarios \cite{Antusch:2022ceb}, see \cref{hajer:triangle}.
Choosing lepton charges such that
\begin{equation}
L(\vec \nu) = -L(N_1) = L(N_2) = 1 ,
\end{equation}
allows only one Dirac and one Majorana mass, $m_D$ and $m_M$, respectively, and leads to a degenerate mass for the two right-handed neutrinos and corresponds to a model that introduces a single Dirac HNL.
The LNLS ensures that the SM neutrinos remain exactly massless.
Breaking this symmetry generates the additional masses $\mu_D$, $\mu_M$, and $\mu_M^\prime$, which are thus protected and can be ensured to remain small.
The resulting mass contribution to the Lagrangian reads
\begin{equation}
\mathcal L_m =
\begin{pmatrix} \vec \nu \\ N_1 \\ N_2 \end{pmatrix}^{\!\!\!t}
\begin{pmatrix} 0 & \vec m_D^{} & \vec \mu_D^{} \\ \vec m_D^T & \mu_M^\prime & m_M^{} \\ \vec \mu_D^T & m_M^{} & \mu_M^{} \end{pmatrix}
\begin{pmatrix} \vec \nu \\ N_1 \\ N_2 \end{pmatrix} .
\end{equation}
The new masses generate not only tiny masses for the SM neutrinos but also lead to a splitting between the degenerate masses of the two right-handed neutrinos, resulting in a pseudo-Dirac pair, see \cref{hajer:mass-splitting}.
If $\mu_D$ is the only small mass, the mass matrix corresponds to the linear seesaw, while $\mu_M$ leads to the inverse seesaw.

\begin{figure}
\begin{subcaptionblock}{.332\textwidth}
\includegraphics[width=\linewidth]{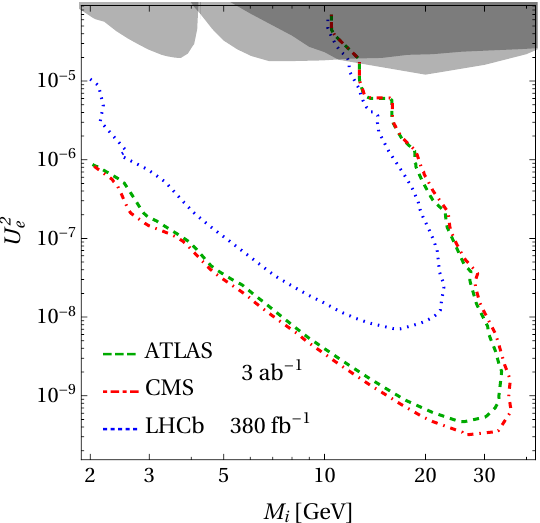}
\caption{Electron coupling} \label{hajer:e}
\end{subcaptionblock}%
\hfil
\begin{subcaptionblock}{.332\textwidth}
\includegraphics[width=\linewidth]{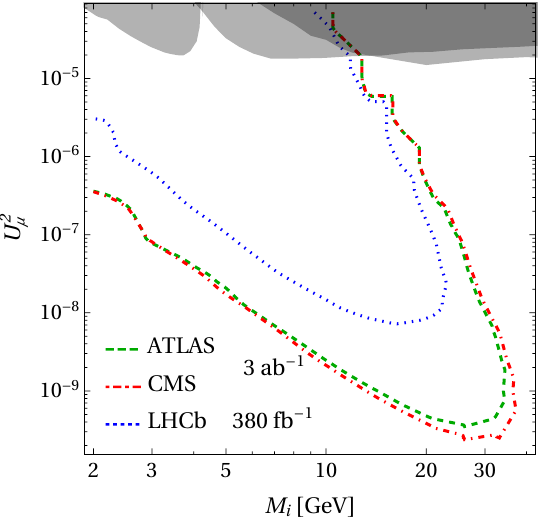}
\caption{Muon coupling} \label{hajer:mu}
\end{subcaptionblock}%
\hfil
\begin{subcaptionblock}{.332\textwidth}
\includegraphics[width=\linewidth]{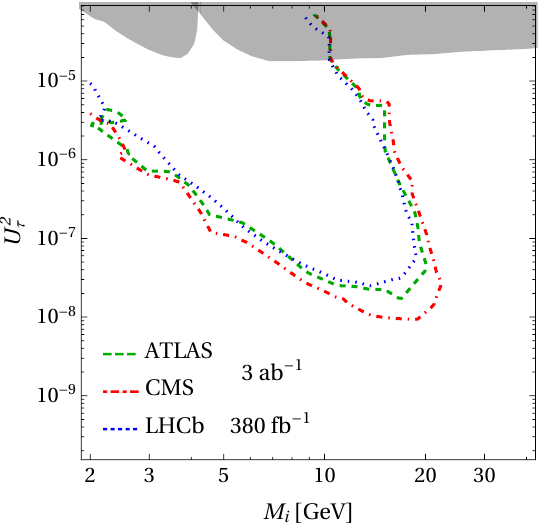}
\caption{$\tau$-lepton coupling} \label{hajer:tau}
\end{subcaptionblock}%
\caption{
Comparison of the sensitivities of the ATLAS, CMS, and LHCb detectors after adding searches of all final states for HNLs that couple solely to electrons, muons, and $\tau$-leptons in panels \subref{hajer:e}, \subref{hajer:mu}, and \subref{hajer:tau}, respectively \cite{Drewes:2019fou}.
While CMS and ATLAS outperform LHCb for HNLs that couple purely to electrons and muons, lower trigger requirements allow LHCb to be competitive for HNLs that couple only to $\tau$-leptons.
} \label{hajer:sensitivitiy}
\end{figure}

As discussed above, the usual benchmark models that add a single collider detectable Majorana or Dirac HNL to the SM fail to produce the correct neutrino mass scale.
However, they suffice as simplified limiting cases that capture some of the predicted HNL properties.
On the one hand, the decay width of the Dirac HNL agrees with the decay width of the pseudo-Dirac HNL.
However, the absence of LNV in this model makes it unsuitable for searches that depend on it.
On the other hand, the Majorana HNL has a decay width that differs from the decay width of the pseudo-Dirac HNL and generically produces too many lepton number violating events, when compared the the pseudo-Dirac HNL.
The discovery potential of such HNLs at current and future colliders has been discussed in great detail \cite{Alimena:2019zri}.
Here we present a phenomenological comparison of the potential for the ATLAS, CMS, and LHCb detectors to discover HNLs in long-lived particle searches during the HL-LHC in \cref{hajer:sensitivitiy} \cite{Drewes:2019fou}.
The reach corresponds to the completion of the HL-LHC and assumes a combination of searches over all final states.
The smaller integrated luminosity of LHCb results in a smaller discovery reach for HNLs that interact only with electrons and muons.
In the case of HNLs that couple only to $\tau$-leptons, this is compensated by the lower trigger requirements of LHCb, leading to a competitive reach of $U_\tau^2 \approx 2 \cdot 10^{-8}$ for HNLs with a mass of $m_N \approx 20$ GeV.

\begin{figure}
\begin{subcaptionblock}{.5\textwidth}
\includegraphics[width=\linewidth]{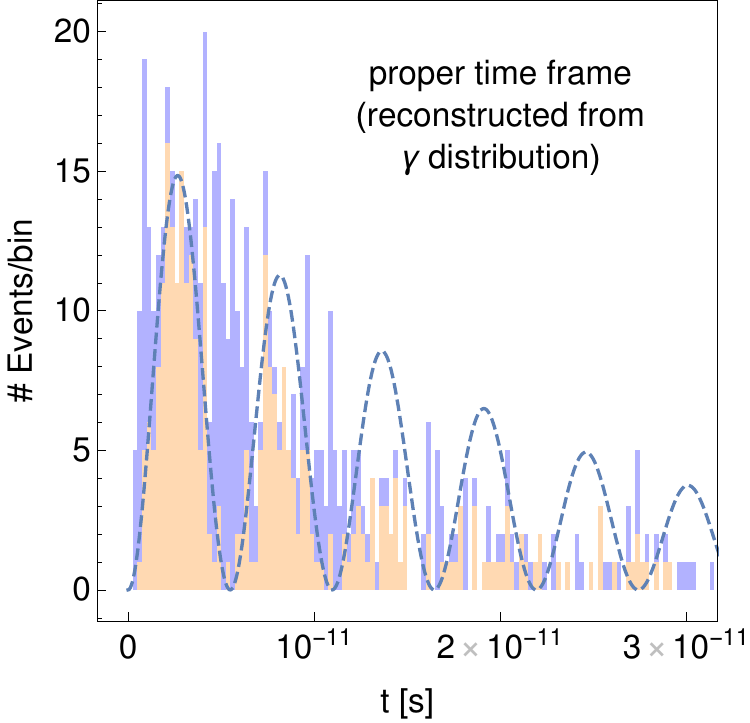}
\caption{LHCb} \label{hajer:LHCb}
\end{subcaptionblock}%
\hfil
\begin{subcaptionblock}{.5\textwidth}
\includegraphics[width=\linewidth]{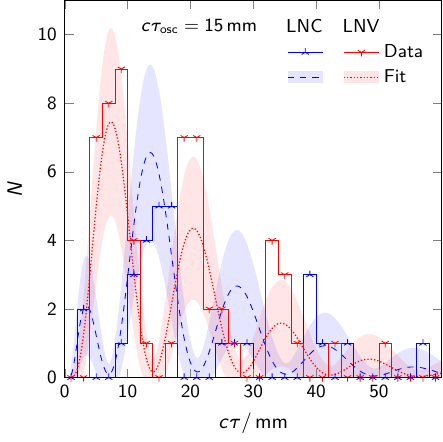}
\caption{CMS} \label{hajer:CMS}
\end{subcaptionblock}%
\caption{
The neutrino-antineutrino oscillations of pseudo-Dirac HNLs can be observed in the rest frame of the decaying HNL as an oscillation between lepton-number-conserving and lepton-number-violating events.
That the reconstruction is possible has been demonstrated for the LHCb detector with a focus on events with a very large Lorentz boost \cite{Antusch:2017ebe} as shown in panel \subref{hajer:LHCb}.
For the CMS detector, this study has been generalised and augmented with a statistical framework that allows the quantification of the discoverability of oscillations \cite{Antusch:2022hhh}, see panel \subref{hajer:CMS}.
} \label{hajer:oscillations}
\end{figure}

As it is the case with neutral mesons, the tiny mass splitting $\Delta m$ within pseudo-Dirac pairs of heavy neutrinos leads to heavy neutrino-antineutrino oscillations.
These oscillations are observable as oscillations between lepton number-conserving (LNC) and lepton number-violating (LNV) processes.
The oscillation length in the rest frame of the HNL is given by the mass splitting, and the oscillation probability as a function of the proper time $\tau$ reads
\begin{equation}
P^{\nicefrac{\text{LNC}}{\text{LNV}}}_\text{osc}(\tau) = \frac{1 \pm \cos (\Delta m \tau )}{2}
\end{equation}
The probability of discovering lepton-number-conserving and violating processes is further influenced by the usual decay probability density function of the decaying HNL
\begin{align}
P^{\nicefrac{\text{LNC}}{\text{LNV}}}(\tau) &= P^{\nicefrac{\text{LNC}}{\text{LNV}}}_\text{osc}(\tau) P_\text{decay}(\tau)
, &
P_\text{decay}(\tau) &= \Gamma \exp\left(- \Gamma \tau\right).
\end{align}
This allows to define the ratio of the integrated probabilities
\begin{equation}
R_{ll}
= \frac{P^{\text{LNV}}}{P^{\text{LNC}}}
= \frac{\Delta m^2}{\Delta m^2 + 2 \Gamma^2}
\end{equation}
which can be measured as a ratio between the total number of observed events with opposite and same-sign leptons.
However, for oscillation lengths of the order of millimetres and given a sufficient number of measured HNL decays, the oscillations can also be resolved, allowing for a direct measurement of LNV.
For the LHCb and CMS detectors, it has been shown that such oscillations are reconstructable with a sufficiently high significance and can be used to measure the amount of LNV induced by the seesaw, see \cref{hajer:oscillations}.

\begin{figure}
\includegraphics[width=\linewidth]{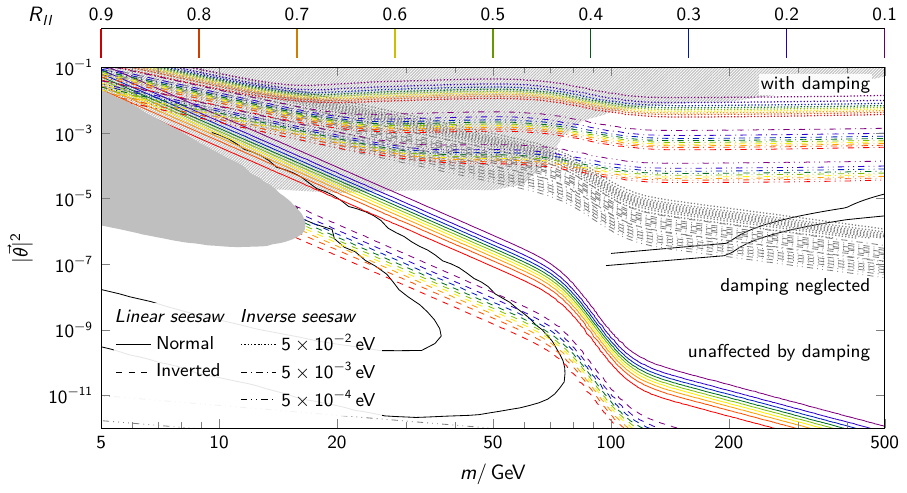}
\caption{
The transition from mostly lepton number-conserving to mostly lepton number-violating events happens in fairly narrow bands in the parameter space, as demonstrated for five different benchmark models of the linear and inverse seesaw.
The damping due to decoherence does not affect the benchmark models that a relevant for long-lived particle searches, but can become significant for short-lived benchmark models, leading to increased lepton number violation \cite{Antusch:2023nqd}.
} \label{hajer:damping}
\end{figure}

Since particle-antiparticle oscillations are inherently quantum mechanical, they are affected by decoherence, leading to damped oscillations \cite{Antusch:2023nqd}.
The damping is mostly affected by an increased mass splitting.
Therefore, it pushes bands of constant LNV towards smaller decay widths and therefore smaller interaction strengths as shown in \cref{hajer:damping}



\changelocaltocdepth{2}
\subsection{New directions in inelastic dark matter: \textnormal{Not-so-inelastic Dark Matter} --- \textit{G. Dalla Valle Garcia}}
\label{sssec:DP-not-so-inelastic}
\textit{Author: Giovani Dalla Valle Garcia, \email{giovani.garcia@student.kit.edu}}  \\

\subsubsection{Introduction}
The nature of dark matter (DM) remains one of the most profound mysteries in physics. Direct detection (DD) experiments have set stringent constraints on heavy DM candidates with masses above a few tenths of a GeV~\cite{LZCollaboration:2024lux}. Combined with indirect detection (ID) probes~\cite{Slatyer:2015jla,Gaskins:2016cha}, these constraints significantly restrict the parameter space of models in which DM is produced via thermal decoupling, particularly the traditional Weakly Interacting Massive Particle (WIMP) paradigm~\cite{Arcadi:2017kky}. Nevertheless, the freeze-out mechanism -- where particles are produced via thermal decoupling-- remains a compelling DM production scenario due to its independence from the universe’s initial conditions and its strong predictive power.

Recently, a wide range of alternatives to traditional WIMPs have been proposed~\cite{Cirelli:2024ssz}. To preserve the viability of thermal freeze-out, two primary approaches can be followed: \footnote{A third approach would involve considering DM annihilating (only into new particles that subsequently decay) only into neutrinos,  which evade ID searches~\cite{Arguelles:2019ouk}. However, since SM neutrinos belong to lepton doublets, avoiding annihilation into charged leptons requires a non-trivial extension of the theory~\cite{Blennow:2019fhy}.}
\begin{enumerate}
    \item \textbf{Heavy DM models}, which automatically evade ID constraints while employing suppressed scattering rates to (model-dependently) escape DD constraints.
    \item \textbf{Light DM models}, which naturally avoid DD limits while suppressing annihilation via velocity-dependent cross-sections\footnote{This includes $p$-wave annihilation cross-sections $\langle \sigma v \rangle \propto v^{2}$ (or even higher velocity suppressions) and scenarios with fine-tuned masses enabling resonant annihilation or kinematically forbidden channels.} or an asymmetry between the two (co)annihilating particles, thereby evading ID constraints.\footnote{As discussed below, not-so-inelastic DM (niDM)~\cite{Garcia:2024uwf} exhibits features of all these approaches.}
\end{enumerate}

This second class of models provides well-motivated DM candidates that can be probed at collider experiments, as their masses fall within experimentally accessible ranges, with the only suppression coming from their inherently feeble couplings. Moreover, these models often feature multi-particle dark sectors with long-lived particles, leading to novel signatures at high-energy facilities.

A key example is inelastic DM (iDM)~\cite{Tucker-Smith:2001myb}, which consists of two DM states: a ground state $\chi$ and an excited state $\chi^\ast$ differing only by their masses $m_{\chi^\ast} > m_\chi$. The defining feature of iDM models is the absence of diagonal couplings of the form $X\chi\chi$, where $X$ represents any mediator other than the DM states. Instead, only off-diagonal couplings of the type $X\chi\chi^\ast$ are allowed. As a consequence, tree-level DM scattering can only occur inelastically, $\chi X \to \chi^\ast X$, and is suppressed when the kinetic energy is insufficient to convert $\chi$ into $\chi^\ast$, thereby relaxing DD constraints. Furthermore, the same off-diagonal interaction renders $\chi^\ast$ unstable, leading to an asymmetric DM scenario: after freeze-out, ground states no longer encounter excited states to annihilate with, suppressing late-time annihilation and avoiding ID constraints.

A well-studied realization of iDM is the pseudo-Dirac iDM model, in which DM consists of two Majorana fermions coupled to the Standard Model (SM) via a dark photon (DP) portal. These models often assume an accidental parity symmetry in the dark sector to suppress diagonal couplings. However, this symmetry is not strictly necessary, as diagonal couplings in Majorana DM models are naturally suppressed due to the $p$-wave nature of their annihilations. Indeed, Majorana DM coupled via a DP portal remains a viable candidate for sub-GeV thermal DM~\cite{Berlin:2018bsc}.

In this context, we explore an ultraviolet-complete generalization of the standard pseudo-Dirac iDM scenario, allowing for generic parity violation in the dark sector. This significantly alters the DM relic density constraints by introducing additional annihilation channels. As a result, a new viable parameter space emerges, where DM is primarily produced via standard $p$-wave annihilation through diagonal couplings, while its main experimental interactions remain dominated by the off-diagonal interaction. We refer to this regime as not-so-inelastic DM (niDM). Our analysis identifies two viable DM mass ranges for niDM:
\begin{itemize}
    \item \text{$m_\chi \sim 500$~MeV}, which will be fully tested by upcoming analyses at NA64 and Belle II.
    \item \text{$m_\chi \sim 5$~GeV}, which remains challenging to probe even with future experiments.
\end{itemize}

\begin{figure}[h]
     \centering
     \includegraphics[width=0.75\textwidth]{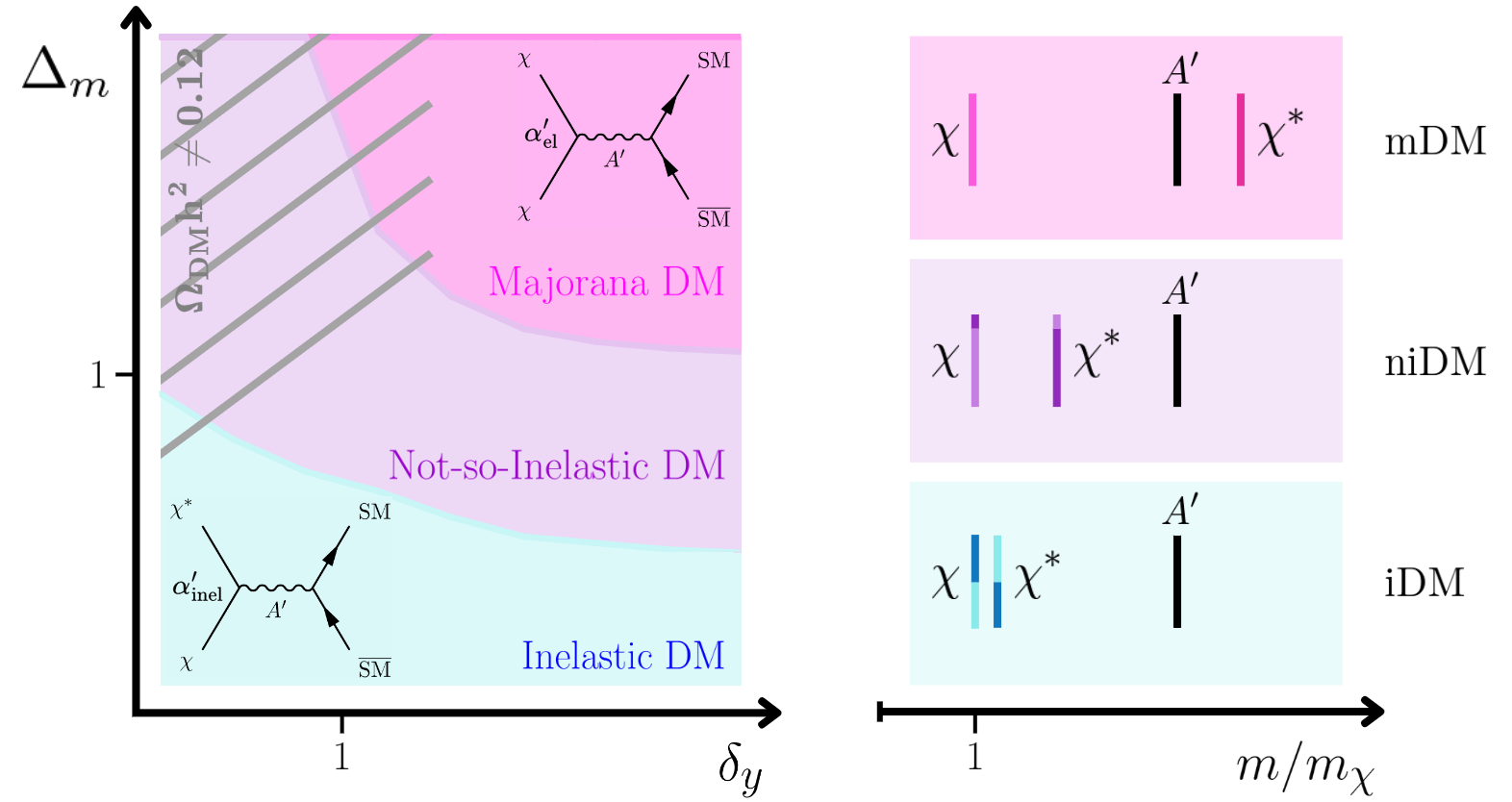}
        \caption{Schematic illustration of the particle spectrum and the continuous transition from inelastic DM (iDM) over not-so-inelastic DM (niDM) to Majorana DM (mDM). The left panel shows the effect of the normalized mass splitting $\Delta_m$ and Yukawa asymmetry $\delta_y$, see \cref{eq:deltas}, where we also indicate the relevance of the elastic (diagonal) and inelastic (off-diagonal) interactions. The right panel shows the typical relative mass spectrum of $\chi,\chi^\ast$ and $A'$ for different regimes of the normalized mass splitting $\Delta_m$.}
        \label{fig:spectrum}
\end{figure}

Notably, our model provides a continuous transition between standard Majorana DM, pseudo-Dirac iDM, and even Dirac DM (in the limit $m_{\chi^\ast} = m_\chi$). This generic framework accommodates a broad class of fermionic DM models coupled via a DP portal,\footnote{This holds for DM coupled via the dark Higgs (DH) portal as well~\cite{DallaValleGarcia:2024zva}.} enabling a comprehensive characterization and potential exclusion of these various scenarios, see \cref{fig:spectrum}.

\subsubsection{The Model}

We consider a dark sector with a new $U(1)'$ gauge symmetry, associated with a gauge boson $A'$ and a gauge coupling $e'$ (with fine-structure constant $\alpha' \equiv e'^2/4\pi$). The dark sector contains a Dirac fermion $\chi_D = \chi_L + \chi_R$, which is a singlet under the SM gauge group but carries a $U(1)'$ charge $Q_{\chi}' \equiv 1$, ensuring anomaly cancellation~\cite{Costa:2020krs}.\footnote{Adopting $Q_{\chi}' \equiv 1$ entails no loss of generality, as any charge choice can be absorbed into the gauge coupling $e'$.} Additionally, we introduce a SM-singlet scalar field $H'$ with charge $Q_{H'}' = -2 Q_{\chi}'$ under $U(1)'$. This field acquires a vacuum expectation value (vev), $\langle H' \rangle = w/\sqrt{2}$, spontaneously breaking the gauge symmetry and generating a mass for $A'$ as well as Majorana mass terms for $\chi_L$ and $\chi_R$.

The new physics Lagrangian is given by:
\begin{equation} \label{eq:NewPhysicsLagrangian}
    \mathcal{L}_{\text{NP}} =  \mathcal{L}_{\chi} + \mathcal{L}_{A'} + \mathcal{L}_{H'}  \, ,
\end{equation}
where
\begin{align}
    \mathcal{L}_{\chi}  = & i \bar{\chi}_L \slashed{D} \chi_L + i \bar{\chi}_R \slashed{D} \chi_R - m_D^{\ast} \bar\chi_L\chi_R - \sqrt{2}  H' ( y_L\bar{\chi}^c_L \chi_L +  y_R  \bar{\chi}^c_R \chi_R) + \text{h.c.} \,, \\
    \mathcal{L}_{A'} = & -\dfrac{1}{4} A'^{\mu \nu} A'_{\mu\nu} - \dfrac{1}{2} \dfrac{\epsilon}{\cos{\theta_w}} B^{\mu \nu} A'_{\mu\nu} \, , \\
    \mathcal{L}_{H'} = & (D^{\mu} H')^*(D_{\mu} H') + \mu_s^2 \abs{H'}^2 - \lambda_s \abs{H'}^4 -  \lambda_{hs} \abs{H'}^2 \abs{H}^2 \, ,
    \label{eq:Lscalar}
\end{align}
where $H$ denotes the SM Higgs field, $B$ the SM hypercharge gauge boson, and the covariant derivative is given by $i D_{\mu} \phi = i \partial_{\mu} \phi - Q_{\phi}' e' A'_{\mu} \phi$.
The charge-conjugated field is defined as $\psi^c = C \gamma_0^T \psi^*$, where $C$ is the charge conjugation matrix. Vector fields $V^{\mu}$ with two Lorentz indices $V^{\mu\nu}$ denote their respective field strength tensors. For simplicity, we assume $m_D = m_D^\ast$ and that the scalar sector plays a negligible role in the phenomenology. The most general case is discussed in ref.~\cite{Garcia:2024uwf}.

After $U(1)'$ symmetry breaking, the gauge boson acquires a mass $m_{A'} = 2 e' w$. The fermion mass terms must then be diagonalized via a unitary transformation, yielding two physical Majorana states, $\chi$ and $\chi^\ast$, with a mass hierarchy $m_{\chi^\ast} > m_\chi$. This transformation also modifies the interaction Lagrangian, introducing both diagonal and off-diagonal interactions between $\chi$, $\chi^\ast$, and the gauge boson $A'$. These interactions are characterized by the diagonal and off-diagonal  fine-structure constants, respectively:
\begin{align}\label{eq:alpha}
    \alpha'_\text{el} = \alpha' \cos^2{2\theta} && \text{and}&& \alpha'_\text{inel} = \alpha' \sin^2{2\theta} \, ,
\end{align}
where the mixing angle $\theta$ is given by:
\begin{equation}\label{eq:rotationAngleTheta}
    \cos{2\theta} =  - \dfrac{\delta_y \Delta_m}{(2+\delta_y)(2+\Delta_m)} \, .
\end{equation}
Here, we define the normalized dark left-right Yukawa asymmetry and the relative mass splitting between the two Majorana mass eigenstates as: \footnote{Without loss of generality, we take $\delta_y\geq0$, which can always be ensured by appropriately relabeling $\chi_L \leftrightarrow \chi_R^c$.}
\begin{align}\label{eq:deltas}
    \delta_y \equiv \frac{y_R - y_L}{y_L} && \text{and} && \Delta_m \equiv \frac{m_{\chi^{\ast}}-m_{\chi}}{m_{\chi}} \, .
\end{align}

To conclude, we note that Big Bang Nucleosynthesis imposes a robust lower bound of approximately $m_\chi \gtrsim 10 \, \mathrm{MeV}$ on any thermally produced DM particle~\cite{Sabti:2019mhn}. In the not-so-inelastic DM (niDM) regime, where $\Delta_m \sim 1$, the decay channel $\chi^\ast \to \chi e^+ e^-$ is always kinematically allowed. This ensures that the excited state $\chi^\ast$ is short-lived on cosmological timescales, rendering its abundance negligible. Further details can be found in ref.~\cite{Garcia:2024uwf}.

\paragraph{Relic abundance}

We compute the DM relic abundance using the public tool \texttt{MicrOmegas v5.3.41}~\cite{Alguero:2022inz,Pukhov:2004ca,Alloul:2013bka}. To account for non-perturbative QCD effects, we adopt the standard approach of using the measured cross-section ratio  $
    R(s) \equiv {\sigma(e^- e^+ \to \textit{hadrons})}/{\sigma(e^- e^+ \to \mu^- \mu^+)} $
to rescale the muonic annihilation cross-sections into the hadronic ones as~\cite{Izaguirre:2015yja,Berlin:2018bsc}:
\begin{equation}
    \sigma(\chi^{(\ast)} \chi^{(\ast)}  \to \textit{hadrons}) = R(s) \, \sigma(\chi^{(\ast)} \chi^{(\ast)} \to \mu^- \mu^+) \, .
\end{equation}
See Appendix A of ref.~\cite{Duerr:2020muu} for further details.\footnote{We consider only the regime where $m_{A'} > 2 m_\chi$, ensuring that pure dark sector annihilations $\chi\chi \to A'A'$ are kinematically forbidden.}

For sizable mass splittings, $\Delta_m \gtrsim 0.4$, the Boltzmann suppression of $\chi^\ast$ relative to $\chi$ is already too strong for coannihilations $\chi\chi^\ast\to XX$ to compete with the annihilation channel $\chi\chi \to XX$, even when diagonal couplings are significantly smaller than off-diagonal ones, $\alpha'_\text{el}/\alpha'_\text{inel} \sim 10^{-2}$.

\subsubsection{Experimental Searches}

\paragraph{(In)direct Detection}

For the mass splittings considered in this work, inelastic scatterings in DD experiments are kinematically suppressed and, therefore, irrelevant. However, in the niDM regime, sizable diagonal couplings also arise, leading to suppressed operators in the effective Lagrangian for direct detection, specifically, the operators $\mathcal{O}_8^N \propto \mathbf{v}_{\perp}$ and $\mathcal{O}_9^N \propto \mathbf{q}$~\cite{Fitzpatrick:2012ix,Bishara:2017nnn}. Using the public code DDCalc~\cite{GAMBIT:2018eea}, we estimate the constraints and sensitivities from DD experiments for these operators.

For indirect detection, off-diagonal interactions are again negligible, as the excited states $\chi^{\ast}$ are too heavy to be produced in DM scattering and too short-lived to be efficiently produced in astrophysical systems~\cite{Baryakhtar:2020rwy,Emken:2021vmf}. Additionally, there is no significant cosmological abundance of the excited states. The diagonal annihilation processes are strongly velocity-suppressed due to their $p$-wave nature. As a result, there are no relevant constraints on the (co)annihilation cross section.

Finally, elastic scattering via the diagonal couplings is found to be below current observational constraints for $m_\text{DM} = m_{A'}/3 > 10 \, \mathrm{MeV}$ and $\alpha_\text{el}' < 0.5$.

\paragraph{Colliders}

The values of the kinetic mixing parameter $\epsilon$ required to reproduce the correct DM relic abundance $\Omega_{\rm DM}h^2=0.11$ via freeze-out suggest that DM particles could be detected in high-energy experiments, such as colliders and beam-dump experiments. Excited state production is extremely relevant since $\chi^\ast$ decays into $\chi$ plus visible SM particles for mass splittings $\Delta_m < 2$. The decay length of $\chi^\ast$ plays a crucial role in determining which experiments are most suitable for niDM searches. For $m_\chi \sim 100\,\mathrm{MeV}$, the decay length varies across different mass splitting regimes: for $\Delta_m \lesssim 0.2$, the decays occur outside the detector, leading to missing energy signatures in both beam dump and electron-positron collider experiments. For $\Delta_m \gtrsim 0.6$, the decays are prompt but suffer from high background noise. The intermediate region, $0.3 \lesssim \Delta_m \lesssim 0.5$, may give rise to displaced vertices signatures. Longer decay lengths are characteristic of lighter DM masses, which may be observable in beam-dump experiments. Additional collider probes include DM scattering events in downstream detectors, which probe DM independently of its state or elastic/inelastic couplings ratio.

\subsubsection{Results}

We begin by considering a DM mass of 200 MeV and quantitatively explore the $(\delta_y, \Delta_m)$ parameter plane, previously presented in \cref{fig:spectrum}.

\begin{figure}[h]
     \includegraphics[width=0.95\textwidth]{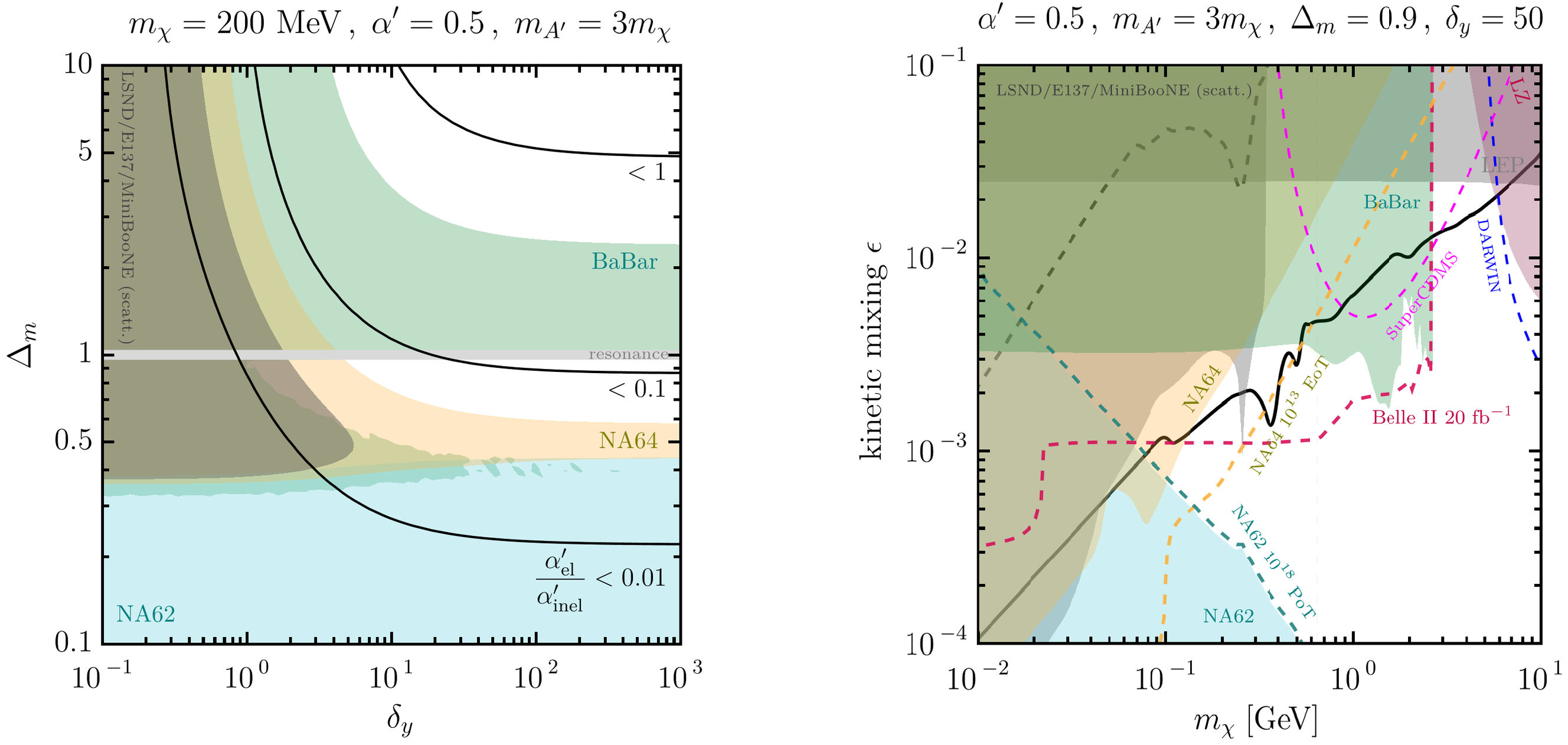}
        \caption{{\bf{Left:}} Region in the $(\delta_y,\Delta_m)$ parameter plane where the kinetic mixing required to reproduce the observed DM relic abundance is experimentally excluded. Black lines indicate the ratio of diagonal $\alpha'_{\text{el}}$ to off-diagonal $\alpha'_{\text{inel}}$ couplings, see \cref{eq:alpha}. {\bf{Right:}} Current bounds and projected sensitivity to the kinetic mixing $\epsilon$ as a function of the DM mass in the niDM model. The solid black curve indicates the DM relic density curve for niDM, while the dashed one is for the corresponding iDM ($\delta_y=0$).}
        \label{fig:allBeampDumps}
\end{figure}

In the left panel of \cref{fig:allBeampDumps}, we show the constraints on the niDM parameters. For each point in the $(\delta_y, \Delta_m)$ plane, we compute the kinetic mixing parameter required to reproduce the observed DM abundance and check for its experimental exclusions. Excluded regions are shown in blue ($\chi^\ast$ decay at NA62 -- proton beam-dump~\cite{NA62:2023nhs}), yellow (missing energy at NA64 -- electron active-beam-dump~\cite{NA64:2023wbi}), green (mono-$\gamma$ plus missing energy at BaBar -- electron-positron collider~\cite{BaBar:2017tiz}) and gray (DM scattering at far detectors~\cite{Berlin:2018pwi} -- LSND~\cite{deNiverville:2011it}, E137~\cite{Batell:2014mga}, and MiniBooNE~\cite{MiniBooNE:2017nqe}). The light gray band indicates the resonance condition $m_{\chi} + m_{\chi^\ast} = m_{A'} \pm 1\%$. Black lines represent constant ratios of diagonal to off-diagonal couplings.

For $\Delta_m < 1$, BaBar's constraints are suppressed due to kinematic allowed decays $A' \to \chi \chi^\ast\to2\chi+{\rm SM}$ which whenever detected are vetoed from the analysis, turning DM to be viable in a large region of parameter space, $\Delta_m \sim 0.6$--0.9 and $\delta_y \gtrsim 1$. This corresponds to the niDM regime, where diagonal interactions dominate the relic density, and off-diagonal signatures dominate in experiments.

All allowed regions for niDM at $m_\chi = 200$ MeV in \cref{fig:allBeampDumps} are expected to be probed in the near future by the Belle II experiment~\cite{Belle-II:2018jsg} with only a luminosity of 20 fb$^{-1}$ -- Belle II has collected already more than 424 fb$^{-1}$~\cite{Tenchini:2023wow}. NA64 is also expected to provide significant progress with an ultimate goal of $10^{13}$  electrons-on-target (EoT)~\cite{Na64SPSC}. The right panel of \cref{fig:allBeampDumps} illustrates the impact of these near-future improvements, which can be clearly visualized in the ($m_\chi,\epsilon$) plane. The solid black curve shows the kinetic mixing required to match the observed DM relic abundance for the niDM case -- to be compared to the dashed curve for its respective iDM equivalent ($\delta_y=0$). Solid colorful lines represent current experimental bounds, while dashed lines indicate projected sensitivities. Remarkably, Belle II is expected to completely probe DM masses below $\lesssim 3$ GeV.

In addition to collider constraints, we present bounds from the LZ DD experiment~\cite{LZCollaboration:2024lux}, which already probes the upper mass range we considered ($m_\chi \sim 10$ GeV). Thus, these experiments clearly provide complementary
information on niDM models. Future DD experiments such as SuperCDMS~\cite{SuperCDMS:2016wui} and DARWIN~\cite{Schumann:2015cpa} are expected to probe values of $\epsilon \alpha_{\rm el}'$ as small as $\sim 10^{-4}$ over a broader mass range. However, a gap remains for $m_\chi \sim 5$ GeV, highlighting an exciting avenue for future research.

\subsubsection{Conclusion}

Not-so-inelastic Dark Matter is an ultraviolet-complete extension of Majorana and pseudo-Dirac dark matter models, coupled to the Standard Model through a dark photon and/or dark Higgs boson. By allowing parity to be generically broken in the dark sector, we identified two viable DM mass regions for the dark photon portal scenario where "inelastic" dark matter with large mass splittings is allowed: a few hundred MeV and a few GeV. The former will be fully probed by the NA64 and Belle II experiments in the near future, while the latter remains challenging to explore, even with future experiments. Additionally, we found that the model is viable only for large dark fine-structure constants, $\alpha' \gtrsim 0.1$, implying that the dark Higgs mass must be similar to the masses of other dark sector particles in order to preserve perturbative unitarity. Consequently, the not-so-inelastic Dark Matter phenomenology at colliders can be richer than currently explored, potentially including light dark Higgs bosons that can be searched for in current and future detectors.\footnote{This is already attracting attention from both the particle phenomenology theory and Belle II communities, in the particular case of pseudo-Dirac inelastic Dark Matter~\cite{Duerr:2020muu,belleIIpresentation}.}

\subsection{Production of new physics particles via mixing with neutral mesons --- \textit{Y.~Kyselov}}
\label{ssec:kyselov}
\textit{Author: Yehor Kyselov, \email{kiselev883@gmail.com}}  \\
\subsubsection{Introduction}

Various Standard Model extensions add FIPs, which have mass and/or kinetic mixing with mesons. Namely, the interaction eigenstate of a meson, labeled $m^0_{\text{int}}$, is not a pure mass eigenstate but instead contains a small admixture of the LLP (further denoted by $X$):
\begin{equation}
    m^0_{\text{int}} \approx m^0 + \theta_{m^0X}\,X,
\end{equation}
where $\theta_{m^0X}$ is the mixing angle that parametrizes the strength of the coupling. This mixing angle exhibits a pole-like structure resonantly enhanced when the mass of the LLP is close to the mass of the meson ($m_X \approx m^0$).

Mixing plays a crucial role in high-energy accelerator experiments, as it substantially contributes to the flux of LLPs in the GeV mass range through production processes such as meson decays, parton hadronization, and proton bremsstrahlung. This production mechanism is especially important for ongoing and future searches at intensity frontier experiments, including the recently approved Downstream algorithm at LHCb~\cite{Kholoimov:2025cqe} and the SHiP experiment.

However, existing descriptions of mixing-induced LLP production often rely on simplified assumptions and approximations, leading to inconsistencies in predicting production rates and kinematics. We systematically investigate the production modes and develop a refined description that overcomes these limitations. The proceedings are based on our work Ref.~\cite{Kyselov:2025uez}.

\subsubsection{Models where mixing is present}
Several Standard Model extensions predict long-lived particles (LLPs) that mix with mesons.

One such example involves Higgs-like scalars $S$, described by the following Lagrangian~\cite{Beacham:2019nyx}:
\begin{equation}
    \mathcal{L} = \alpha_{1} S H^{\dagger}H + \alpha_{2} S^{2}H^{\dagger}H,
\end{equation}
which, below the electroweak scale, yields effective Yukawa interactions:
\begin{equation}
    \mathcal{L}_{\text{eff}} \subset \theta S \sum_{f} y_f\, \bar{f}f.
\end{equation}
Here, $ H $ is the SM Higgs doublet, and $ \alpha_1, \alpha_2 $ are coupling constants. The term $ \theta $ is the mixing angle between the scalar and the Higgs boson, while $ y_f $ represents the Yukawa coupling of fermion $ f $. These interactions introduce mixing between scalars and scalar mesons, such as $ f_{0} $ and their excitations.

Another class of models considers pseudoscalar axion-like particles, of ALPs $a$, whose Lagrangian takes the form~\cite{Bauer:2020jbp}
\begin{equation}
    \mathcal{L} = c_{G}\frac{a}{f_{a}}\frac{\alpha_{S}}{4\pi}G^{a}_{\mu\nu}\tilde{G}^{a,\mu\nu} + c_{q}\frac{\partial_{\mu}a}{f_{a}}\,\bar{q}\gamma^{\mu}\gamma_{5}q + \dots.
\end{equation}
In this expression, $f_a$ is the ALP decay constant, $ G^a_{\mu\nu} $ is the gluon field strength tensor, and $ \tilde{G}^{a,\mu\nu} $ is its dual. The coefficients $ c_G $ and $ c_q $ parametrize the ALP-gluon and ALP-quark couplings, respectively. This Lagrangian introduces both mass and kinetic mixing between ALPs and pseudoscalar mesons, such as $ \pi^0, \eta, \eta' $, and their excitations.

Vector mediators have the Lagrangian given by~\cite{Alekhin:2015byh,Beacham:2019nyx,Ilten:2018crw}
\begin{equation}
    \mathcal{L} = V_{\mu}\left(e\epsilon\, J^{\mu}_{\text{EM}}+\sqrt{4\pi\alpha_{B}}\Big(J^{\mu}_{B}-\sum_{\alpha}c_{\alpha}J^{\mu}_{L,\alpha}\Big)\right).
\end{equation}
Here, $ V_{\mu} $ is the vector mediator field. The terms $ J^{\mu}_{\text{EM}}, J^{\mu}_{B} $, and $ J^{\mu}_{L,\alpha} $ correspond to the electromagnetic, baryonic, and lepton currents, respectively, while $ \alpha_B $ and $ c_{\alpha} $ are model-dependent couplings. The Standard Model vector currents have non-zero vacuum expectation values between the vacuum and vector meson states, such as $ \rho^{0}, \phi, \omega $. This fact is known as vector meson dominance~\cite{Fujiwara:1984mp}.

These particles may also serve as mediators between the Standard Model particles and a hypothetical dark sector, such as dark matter~\cite{Berlin:2018jbm,Beacham:2019nyx,Antel:2023hkf,Fitzpatrick:2023xks,Foguel:2024lca}. Thus, the mixing may also significantly affect the production of the latter.

\subsubsection{Production channels via mixing}

\begin{figure}[t]
    \centering
    \includegraphics[width=0.9\textwidth]{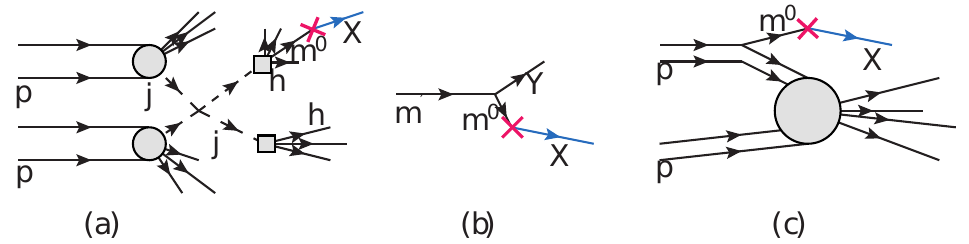}
    \caption{Diagrams of various production channels of a LLP particle $X$ via the mixing with a neutral meson $m^{0}$: the parton fragmentation in the deep-inelastic collisions (the diagram (a)), decays of heavy mesons $m^{'} \to X + Y$, with $Y$ being some Standard Model state (the diagram (b)), and proton bremsstrahlung (the diagram (c)). The red cross indicates the mixing vertex.}
    \label{fig:production-channels}
\end{figure}

There are several production processes of an LLP $X$ with mass in the GeV range to which the mixing directly contributes (see Fig.~\ref{fig:production-channels}).

In deep-inelastic proton collisions, partons may fragment into a meson $ m^{0} $. Due to mixing, there is a small probability that instead of an LLP $ X $ appears. We refer to this production mode as production in \textbf{fragmentation} (see diagram (a) at Fig.~\ref{fig:production-channels}).

Next, \textbf{decays of heavy mesons} (see diagram (b) at Fig.~\ref{fig:production-channels}). Mesons $ m^{\prime} $ can decay into $ X $ plus other particles: $m^{\prime} \to X + \text{other}.$ For example, if $ X $ mixes with the neutral pion $ \pi^0$, a decay process such as
\begin{equation}
    \rho^{+} \to \pi^{+} \pi^{0}
\end{equation}
automatically implies that another channel will be present,
\begin{equation}
    \rho^{+} \to \pi^{+} X,
\end{equation}
with the branching ratio suppressed by the square of the mixing angle.

\textbf{Proton bremsstrahlung} (see diagram (c) at Fig.~\ref{fig:production-channels}), overall is an inelastic process, since the final state includes additional particles from the subsequent collision. However, using old-fashioned perturbation theory, this process can be modeled as the emission of the LLP $X$ from an incoming proton as initial state radiation—essentially occurring "before" the inelastic collision takes place. In this picture, the mixing induces a resonant enhancement of the bremsstrahlung amplitude via the elastic $ppX$ form factor around the masses of $m^0$ (see, e.g.,~\cite{Faessler:2009tn,Foroughi-Abari:2021zbm}).

\subsubsection{Common Approximation for LLP Flux}
A common approximation adopted in the literature~\cite{Berlin:2018jbm,Jerhot:2022chi} is to estimate the LLP flux from some of the production channels where mixing contributes by scaling
\begin{equation}
\frac{d^{2}N_{X}}{d\theta_{X}dE_{X}} = \sum_{m^{0}}|\theta_{m^{0}X}|^{2}\frac{d^{2}N_{m^{0}}}{d\theta_{X}dE_{X}}.
\label{eq:approximation}
\end{equation}
I.e., the LLP flux is approximated by the flux of the corresponding meson times the square of the mixing angle.

Different sub-processes contributing to the production of mesons in~\eqref{eq:approximation} would have a different dependence on LLP mass. Neglecting this can lead to significant inaccuracies in predicting LLP production rates and kinematics. For instance, LLPs mixing with $ \pi^{0} $ may be produced by fragmentation and meson decays at low masses, but once the LLP mass exceeds the kinematic threshold of meson decays, the contribution from the decay channels vanishes, reducing the production flux. Another issue is that for ALPs, the squared mixing angles depend on unphysical chiral rotation parameters~\cite{Ovchynnikov:2025gpx}.

Despite these drawbacks, the approach is still widely used in LLP event rate calculators~\cite{Kling:2021fwx,Jerhot:2022chi,Ovchynnikov:2023cry}. These issues also apply to extensions of minimal models, where LLPs act as mediators for hypothetical dark sectors, including light dark matter.

\subsubsection{Revised approach}

\begin{figure}[t!]
    \centering
    \includegraphics[width=0.5\linewidth]{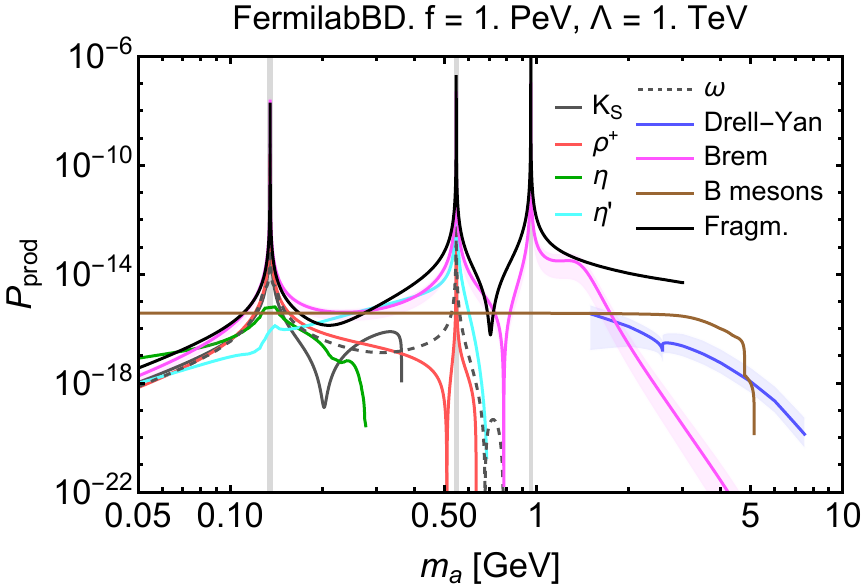}~\includegraphics[width=0.5\linewidth]{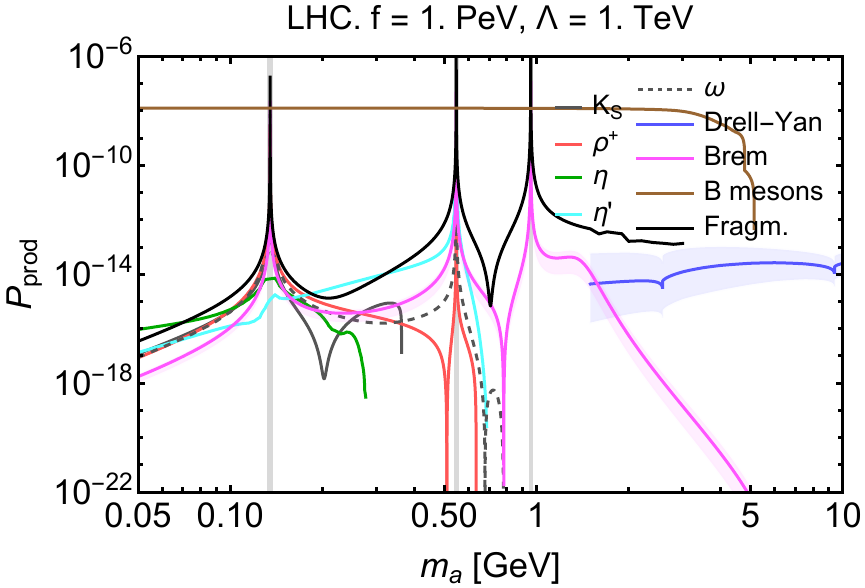}
    \\     \includegraphics[width=0.5\linewidth]{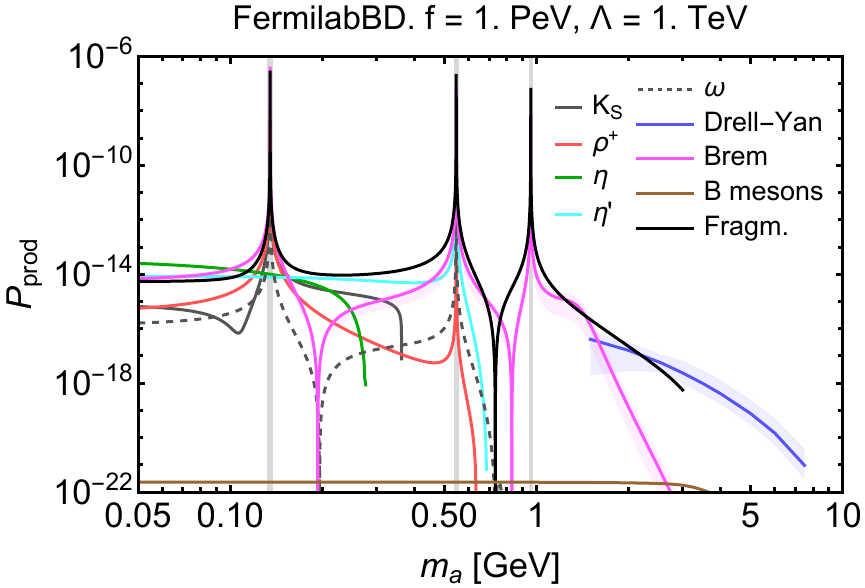}~\includegraphics[width=0.5\linewidth]{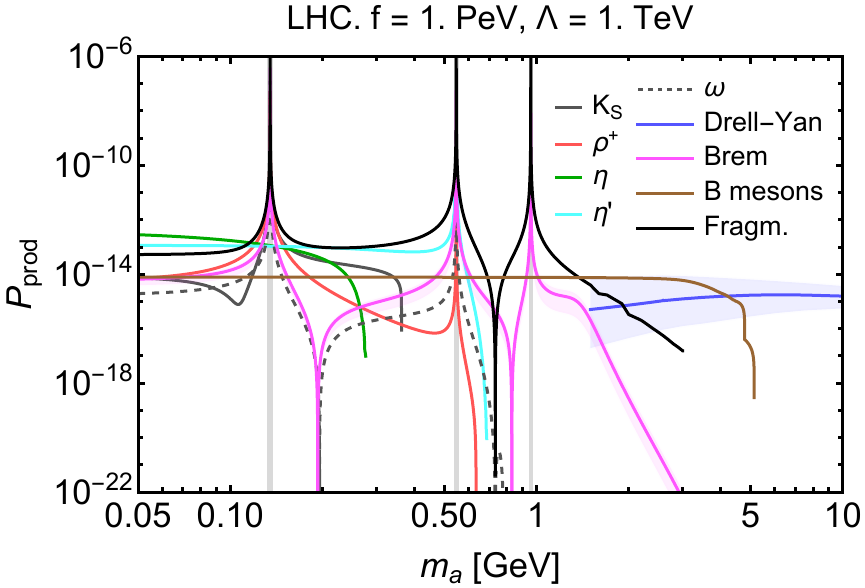}
    \caption{Production probabilities of the axion-like particles at FermilabBD and LHC facilities at the scale $\Lambda = 1\text{ TeV}$. Top panel: ALPs with the universal coupling to quarks ($c_{q} = 1, c_{G} = 0$ ). Bottom panel: ALPs coupled to gluons ($c_{q} = 0, c_{G} = 1$). }
    \label{fig:ALP-production}
\end{figure}

We have developed a refined approach to describing the LLP production channels where the mixing contributes. It is based on the dedicated treatment of all these modes and their incorporation in event generators \texttt{PYTHIA8}~\cite{Bierlich:2022pfr} and \texttt{SensCalc}~\cite{Ovchynnikov:2023cry}. We have applied to several models -- ALPs with coupling to gluons and universal coupling to fermions, dark photons, and mediators coupled to the baryon current.

The calculations of the mixing angles are described in detail in our work~\cite{Kyselov:2025uez}. The calculation of meson decays into LLPs follows the studies in Refs.~\cite{Ilten:2018crw, Ovchynnikov:2025gpx}. We also implement updated bremsstrahlung descriptions from Refs.~\cite{Foroughi-Abari:2024xlj, Blinov:2021say,Ovchynnikov:2025gpx}, taking into account the theoretical uncertainties as discussed in Ref.~\cite{Kyselov:2024dmi}.

In our modified version of \texttt{PYTHIA8}\footnote{Modification of the \texttt{PYTHIA8} code is public and available at \url{https://gitlab.com/YehorKyselyov/pythia-mixing/-/tree/dev}}, we change the fragmentation chain of mesons $ m^{0} $ by replacing them with LLPs $ X $ at a rate given by the mixing angle. For ALPs, however, we use a “generalized mixing parameter” that includes both the mixing contribution and the direct interactions of ALPs with mesons, defined such that it is independent of unphysical chiral rotation parameters. We also discussed earlier decays of hard mesons using built-in Pythia methods.

Our programs in this framework are designed for various experimental facilities, including the CERN LHC, Fermilab, CERN SPS, the Serpukhov Accelerator, and the Future Circular Collider. For selected experiments, we also incorporated corresponding tunes that approximately reproduce the light meson production experimental data, see Refs.~\cite{Fieg:2023kld,Dobrich:2019dxc}
Within this framework, our results show that for vector mediators, meson decays dominate their production for masses below the $\eta$ meson mass, while at higher masses, fragmentation and bremsstrahlung become increasingly relevant. The uncertainties associated with bremsstrahlung remain significant due to the sensitivity to the proton form factor parameterization and intermediate proton virtuality, while uncertainties of the fragmentation channel remain relatively small and stable, around 30\%.

For ALPs, in the universal quark couplings case, decays of $B$ mesons provide a dominant production mode as far as the amount of $B$ mesons is large. In the case of the dominant coupling to gluons, fragmentation dominates the yield (see figure \ref{fig:ALP-production}).

\subsubsection{Conclusion}

We have developed a refined approach to describe the production of LLPs that mix with neutral mesons, addressing inconsistencies in traditional methods and providing a more precise theoretical framework. By implementing a separate treatment for each production mechanism, incorporating a modified fragmentation model in \texttt{PYTHIA8}, and refining decay and bremsstrahlung descriptions, we have ensured a more accurate prediction of LLP flux and kinematics. This improved treatment has direct implications for interpreting previous, current, and future LLP searches at experiments such as LHCb, SHiP, FASER, and others.

\subsection{ New mesogenesis discovery opportunities at LHCb --- \textit{G.~Elor}}
\label{ssec:elor}
\textit{Author: Gilly Elor, \email{gilly.elor@austin.utexas.edu}}
\begin{table}[t!]
\renewcommand{\arraystretch}{1}
 \setlength{\arrayrulewidth}{.15mm}
\centering
\footnotesize
\setlength{\tabcolsep}{0.18 em}
\begin{tabular}{ | c | c | c | c | c  |  c|}
    \hline
  Mechansim & CPV & Dark Sector & Observables  &  Relevant Experiments \\
    \hline \hline
    $B^0$  Mesogenesis
    &  $B_s^0 \,\, \& \,\, B_d^0$
    &  dark baryons
    & $A^{s,d}_{sl}$
    & LHCb     \\
   See Ref. \cite{Elor:2018twp}
    &   oscillations
    &
    & $\text{Br} (B^0\rightarrow \mathcal{B}_{\rm SM}+ X)$
    &  $B$ Factories, LHCb    \\
    \hline

    &
    &
    & $A_{CP}^D$
    & $B$ Factories, LHCb \\
    $D^+$  Mesogenesis
    &  $D^\pm$ decays
    &   dark leptons  & $\text{Br}_{D^+}$
    &  $B$ Factories, LHCb \\
     See Ref.  \cite{Elor:2020tkc}
    &
    &    and dark baryons
    &  $\text{Br} (D^+ \rightarrow \ell^+ + X)$
    &  peak searches e.g. PSI, PIENU \\
    \hline
     &
     &
     & $A_{CP}^B$
     & $B$ Factories, LHCb\\
    $B^+$  Mesogenesis
     &  $B^\pm$ decays
     & dark leptons
     & $\text{Br}_{B^+}$
     &$B$ Factories, LHCb   \\
       See Ref. \cite{Elahi:2021jia}
     &
     & and dark baryons
     &  $\text{Br} (B^+ \rightarrow \ell^+ + X)$
     & peak searches e.g. PSI, PIENU  \\
    \hline
     &
     &
     & $A_{CP}^{B_c}$
     &  LHCb, FCC \\
     $B^+_c$  Mesogenesis
     &  $B^\pm_c$ decays & dark baryons
     & $\text{Br}_{B_c^+}$
     & LHCb, FCC \\
      See Ref.  \cite{Elahi:2021jia}
     &
     &
     &  $\text{Br} (B^+\rightarrow \mathcal{B}_{\rm SM}^++ X)$
     &  $B$ Factories, LHCb \\
     \hline
      \hline
        $\dcp$ Mesogenesis
     &   either $B^0_d$, $B^0_s$,
     & dark baryons
     &  $A_{\rm CP}^{\rm dark}$
     & EDMs, Flavor Observables  \\
  See Ref. \cite{darkmeso}
     &  $B^\pm$, $B_c^\pm$ decays
     & and dark CP phase
     & $\text{Br} (\mathcal{M} \rightarrow \mathcal{B}_{\rm SM} + X)$
     &  $B$ Factories, LHCb\\
     \hline
        Mesogenesis
     &   $B_s^0 \,\, \& \,\, B_d^0$
     & dark baryons and
     &  $A_{\rm sl, \, \rm SM}^{\rm s,d}$
     & LHCb  \\
  with Morphing
  &  oscillations
     & dark first order
     & $\text{Br} (B^0 \rightarrow \mathcal{B}_{\rm SM} + X)$
     &   $B$ Factories, LHCb\\
     See Ref. \cite{Elor:2024cea,stealthmeso}
     &
     &phase transition
     & Gravitational Waves
     & Pulsar Timing Arrays, CMB \\
     \hline

\end{tabular}
\caption{Summary of the various proposed flavours (mechanisms) of Mesogenesis. Also listed are sources of CP Violation, the details of the dark sector, observables directly related to the baryon asymmetry (additional indirect or model-dependent signals are not shown), and the relevant experiments for each observable. X denotes missing energy. }
\label{tab:theoryspace}
\end{table}

This subsection summarizes updates to the discovery potential of mesogenesis (late-time baryogenesis and dark matter production) at LHCb.

\subsubsection{The Mesogenesis Framework}
Mesogenesis \cite{Elor:2018twp,Elor:2020tkc,Elahi:2021jia,Elor:2024cea, darkmeso, stealthmeso} refers to a class of \emph{testable} baryogenesis mechanisms in which the observed baryon asymmetry of the Universe (BAU) is directly linked to Standard Model (SM) meson systems $\mathcal{M} \supset \{D^\pm, B^\pm, B_{d,s}^0, B_c^\pm \}$. In these constructions, SM mesons $\M$ decay into SM baryons $\mathcal{B}_{\rm SM}$ and dark matter states, thereby generating both the dark matter abundance and the BAU. The BAU has been determined from measurements of the Cosmic Microwave Background (CMB) \cite{Ade:2015xua,Aghanim:2018eyx} and light element abundances after Big Bang Nucleosynthesis (BBN) \cite{Cyburt:2015mya,pdg} to be (in co-moving yield units):  $\YBAU \equiv (n_{\mathcal{B}}-n_{\bar{\mathcal{B}}})/s = \left( 8.718 \pm 0.004 \right) \times 10^{-11}$, where $n_\B$ are particle number densities and $s$ is the entropy density. Mesogenesis operates at ``late times" when the temperature of the Universe $T_R$ was of order MeV scales, specifically $\Lambda_{\rm QCD} > T_R > T_{\rm BBN}  \sim 5\, \text{MeV}$, i.e. after the strong force confines but before the epoch of BBN.  This is achieved through the existence of a late-time matter-dominated era in which the energy density of the Universe was dominated by a scalar field $\Phi$. The exact UV nature of $\Phi$ has no impact on the phenomenology and so we simply assume that at around $T_R$ the $\Phi$ field decays, at a rate $\Gamma_\Phi \sim H(T_R) \sim 10^{-21} \text{GeV}^{-1}$, to equal numbers of quarks and anti-quarks which then quickly hadronize to form SM charged and neutral mesons and anti-mesons. Since the BAU is generated at MeV scale temperatures by SM mesons with GeV scale masses, Sakharov's \cite{sakharov} out-of-thermal-equilibrium condition is satisfied by construction. The produced mesons subsequently undergo a CP-violating process, parameterized by the charge asymmetry $ A_{CP}^{\rm}$ (an experimental observable), thus satisfying Sakharov's C and CP violation condition \cite{sakharov}. The final Sakharov condition, baryon number ``violation", is satisfied through \emph{baryon number conserving decays} of SM mesons into dark sector states carrying SM baryon (and possibly lepton) number. The product of this observable branching fraction and the CP violation is then directly linked to the BAU: $ \YBAU \propto  \prod \text{Br}  \times A_{\rm CP}$, where  $\prod \text{Br}$ is taken to be the product of branching fractions for every decay involved in the process (in mesogenesis variations involving dark leptons \cite{Elor:2020tkc,Elahi:2021jia} the dark sector is populated by a cascade of decays).

Table~\ref{tab:theoryspace} summarizes all currently proposed mechanisms of mesogenesis. Existing mechanisms differ in the meson system that sources the CP violation, as well as the content of the dark sector. However, generic to all scenarios is a colored mediator $\Y$ which is required to connect the visible and dark sectors to facilitates the decay(s) $\mathcal{M} \rightarrow \mathcal{B}_{\rm SM} + \rm{invisible}$, through an effective operator $ \cO_{d_k,u_i d_j} = \mathcal{C}_{d_k,u_i d_j} \epsilon_{\alpha\beta\gamma} \left(\bar{\psi}_{\mathcal{B}} d^\alpha_k\right) \left(\bar{d^c}^\beta_j u^\gamma_i\right)$,
where $(i,j)$ are flavour and $(\alpha,\beta,\gamma)$ are color indices, and $\mathcal{C} \propto 1/M_{\Y}^2$ is the Wilson coefficient. This operator can be generated in a UV model with a heavy colored mediator $\mathcal{Y}$, and a baryon-number charged dark fermion $\psiB$. There are two possible hyper-charge assignments to the TeV-scale colored scalar mediator $\Y$:
\begin{align}
\label{eq:L}
\mathcal{L}_{\Y_{2/3}}&=-\sum_{i,j}y_{d_{i}d_{j}}\,\epsilon_{abc}\,\Y^{a}d_{Ri}^{b}d_{Rj}^{c}-\sum_{k}y_{\psi u_{k}}\Y_{a}^{*}u_{Rk}^{a}\psi_{R}+\text{h.c}\\\mathcal{L}_{\Y_{-1/3}}&=-\sum_{i,j}y_{u_{i}d_{j}}\,\epsilon_{abc}\,\Y^{a}u_{Ri}^{b}d_{Rj}^{c}-\sum_{ij}y_{Q_{i}Q_{j}}\,\epsilon_{abc}\,\epsilon_{\alpha\beta}Y^{a}Q_{Li}^{b\alpha}Q_{Lj}^{c\beta}-\sum_{k}y_{\psi d_{k}}\Y_{a}^{*}d_{Rk}^{a}\psi_{R}+\text{h.c.}\,,
\end{align}
where $a,b,c$ are color, $i,j,k$ are flavor and $\alpha \,, \beta$  are $SU(2)_L$ indices.  Alternatively, the mediator could be a vector with SM charge assignment $(3,2)_{1/6}$, in which case the following interactions are allowed by all symmetries:
\begin{align}
    \label{eq:L2}
    \mathcal{L}_{1/6} = - \sum_{ij} y_{Q_i d_j} \epsilon_{abc} \epsilon_{\alpha \beta} \Y_\mu^{a \alpha} Q_{L i}^{b \beta} \sigma^\mu d_{R j}^c
    - \sum_k y_{\psi Q_k} \Y_\mu^{\dag a \alpha} Q_{Lk}^{a \alpha} \sigma^\mu \psi_R + \text{h.c.}\,.
\end{align}
The mass of $\Y$ in all models is roughly constrained to be at the TeV scale due to limits from ATLAS and CMS searches for colored scalars (e.g. squarks in supersymmetric models \cite{Alonso-Alvarez:2019fym}). For detailed bounds on $\Y$ see \cite{Alonso-Alvarez:2021qfd,Alonso-Alvarez:2021oaj,stealthmeso}.

In all cases $\psiB$ is not stable and must further decays into the a dark scalar baryon $\phiB$ and a Majorana fermion $\xi$ compose the dark matter (which can be stabilized by a $\mathbb{Z}_2$ symmetry): $\mathcal{L}_d = y_d \bar{\psi} \phi \xi$. ($\phiB$ and $\xi$ are both present in the dark matter halo today with relative densities fixed by the requirement that $\psiB$ carries baryon number; specifically, $\rho_\xi / \rho_{\phiB} = 5 m_p/m_{\phiB} - 1$ with $\rho_{\rm tot} = \rho_{\phiB} + \rho_\xi = 0.5 \text{GeV}/\text{cm}^3$ and the masses of the dark particles must satisfy $m_{\M}- m_{\Bsm} > m_{\psiB} > m_p + m_e  = 938.78 \, \text{MeV}$, as discussed in \cite{Berger:2023ccd,Elor:2018twp,Elor:2020tkc,Elahi:2021jia,Elor:2024cea, darkmeso, stealthmeso}). Critically, the signatures of various ``flavors" of mesogenesis populate a web of experimental observables where signals directly linked to the BAU production of one mechanism are indirect signals of another (and therefore populate a different region of parameter space).  This remarkable degree of testability is arguably unique to the mesogenesis framework. In all cases, mesogenesis predicts that product of experimental observables $\prod \text{Br}  \times A_{\rm CP} \gtrsim 10^{-8}$ i.e. mesogenesis can be fully tested (up to some weak caveats) by a measurement the Wilson coefficient down to a sensitivity corresponding to a branching fraction of  (see \cite{darkmeso}):
\bea
\text{Br} \left(\mathcal{M}\rightarrow \Bsm + \text{invisible} \right) \gtrsim 10^{-8}.
\label{eq:masterbound}
\eea
We hope this serves as a benchmark to help guide experimental searches towards the discovery (or complete exclusion) of the mesogenesis framework.
We now summarize the specific signatures of the mechanisms outlined in Table \ref{tab:theoryspace} and highlight LHCb-specific opportunities.  Mesogenesis can be further subdivided into mechanisms involving decays to dark baryons and those involving decays to dark leptons (where a lepton asymmetry is transferred into a baryon asymmetry via dark sector scatterings).

\subsubsection{$B^0$ Mesogenesis}
\label{subsubsec:neutralBmeso}
\begin{figure}[t!]
\centering
\includegraphics[width=1\textwidth]{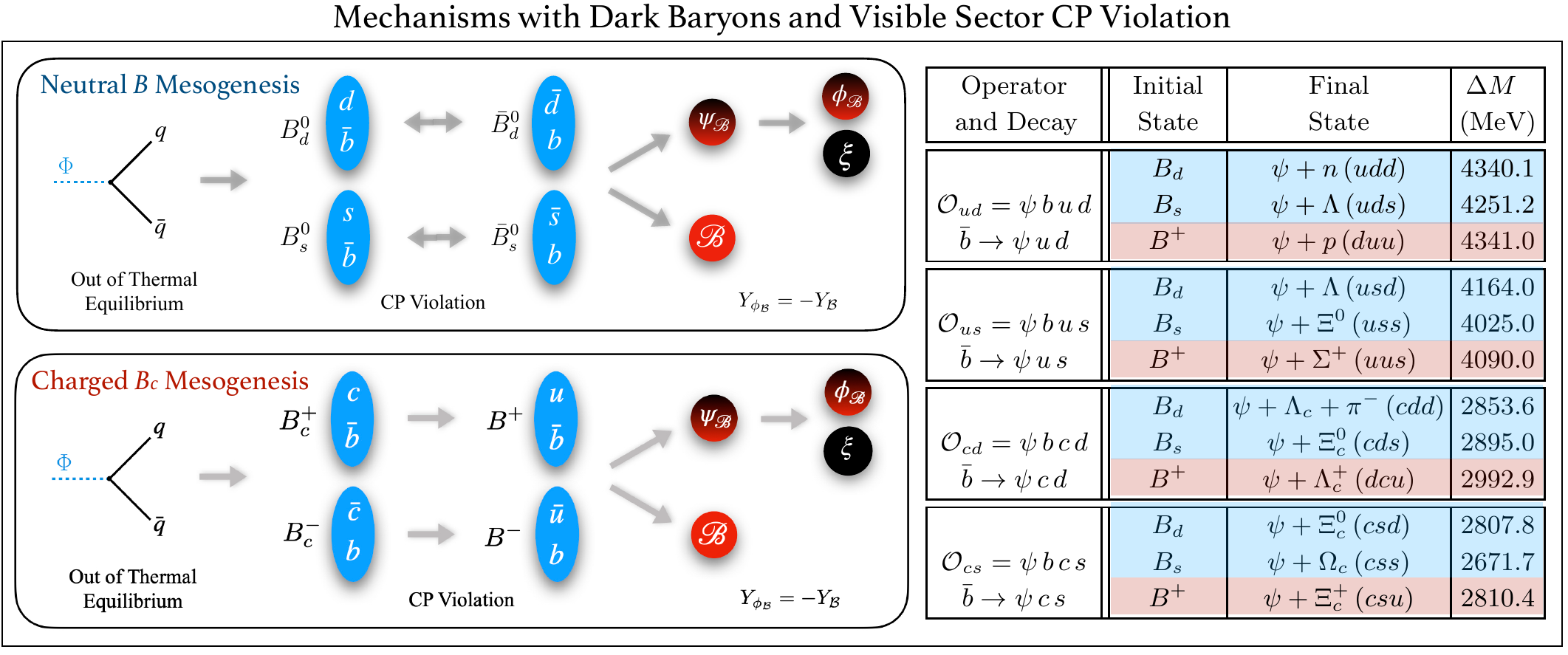}
\vspace{-.2cm}
\caption{Depiction of mesogenesis mechanisms involving dark sector baryons and visible sector sources of CP violation. \emph{Left:} Cartoon depictions of the the Neutral $B$ (top) and Charged $B_c$ (bottom) Mesogenesis mechanisms. \emph{Right}: Channels for which Neutral $B$ (blue) and Charged $B_c$ (red) Mesogenesis can proceed.  In both cases, all decays will be present (one mechanism's direct signal is the other's indirect signal). For theoretical calculations of these branching fractions, see \cite{Elor:2022jxy}. The Belle-II collaboration (using Belle-I data) \cite{Belle:2021gmc} and BaBar \cite{BaBar:2023rer} have set limits on $B_{d}^0 \rightarrow \Lambda^0 \psiB$. Additionally,  BaBar mesogenesis searches for $B^+ \rightarrow p \psiB$ \cite{BaBar:2023dtq} and $B^+ \rightarrow \Lambda_c^+ \psiB$ \cite{BaBar:2024qqx}.  LHCb has developed a search \cite{Rodriguez:2021urv} for some of these channels as well as additional indirect signals.}
\label{fig:DarkBvisible}
\end{figure}

In Neutral $B$ Mesogenesis \cite{Elor:2018twp,Elor:2024cea,stealthmeso} the CP violation in $B_{s,d}^0$ oscillations is leveraged to generate the BAU:
\bea
Y_{\mathcal{B}} &\simeq 5 \times 10^{-5} \, \sum_{i=d,s} \bigl[ \Bri  A_{sl}^i \bigr]  \alpha_i (T_R)\,, \\ \nonumber
&\quad \text{where} \quad A_{sl}^i \equiv \frac{\Gamma \left( \bar{B}_i^0 \rightarrow B_i^0 \rightarrow f\right) - \Gamma\left({B}_i^0  \rightarrow \bar{B}_i^0 \rightarrow \bar{f}\right) }{\Gamma\left( \bar{B}_i^0  \rightarrow B_i^0\rightarrow f\right) + \Gamma \left( B_i^0  \rightarrow \bar{B}_i^0 \rightarrow \bar{f}\right) }\,.
\eea
Here the inclusive branching fraction $\text{Br}$ is over all possible final states baryons, $A_{ sl}^{i=s,d}$ is the semi-leptonic asymmetry which is an \emph{observable} CP violating parameter in  $B^0_{s,d}$ meson oscillations (which decay to a measured final state $f$). The $T_R$ dependent functions $\alpha_i \in \left[0,1 \right]$ capture additional numerics and higher temperature decoherence effects that spoil $B^0$-$\bar{B}^0$ oscillations (and thus the generated BAU) \cite{Elor:2018twp}. Results are weakly dependent on $\Phi$ parameters.  Generically, $B^0_s$ mesons dominate production at higher temperatures $T_R > 35$ MeV while $B_d$ mesons are more relevant at lower $T_R$. Assuming $T_R \sim 20 MeV$ producing the observed BAU requires $\Bri  \times A_{sl}^i \gtrsim 10^{-6}$. Given limits on $A_{sl}^i$ this leads to the prediction: $\Bri \gtrsim 10^{-5}-10^{-4}$.

Figure \ref{fig:DarkBvisible} depicts a cartoon of the Neutral $B_{s,d}^0$-Mesogenesis mechanism in the top left. On the right side of that figure, we list $B_{s,d}^0$  decay channels to various final state SM baryons that can generate the BAU for a given flavorful choice of operator  $\cO_{d_k,u_i d_j}$. For theoretical calculations of these branching fractions, see \cite{Elor:2022jxy}. Improved measurements of $A_{sl}^{s,d}$ are projected to be made by LHCb and other experiments. Meanwhile, dedicated neutral $B$-Mesogenesis searches at $B$factories are setting limits on the branching fraction of exotic meson decays to SM baryons and missing energy. In particular, Belle-I and BaBar data was used to constrain the branching fraction $B_d^0 \rightarrow \Lambda_0 \psiB$ \cite{Belle:2021gmc, BaBar:2023rer} to $10^{-5}$-$10^{-4}$. Additionally, BaBar set similar limits on indirect signals of this mechanism: $B^+ \rightarrow p \psiB$ \cite{BaBar:2023dtq} and $B^+ \rightarrow \Lambda_c^+ \psiB$ \cite{BaBar:2024qqx}. While these limits can be evaded with additional dark sector dynamics (namely a dark sector phase transition), this comes at the cost of non-trivial dark sector model building \cite{Elor:2024cea,stealthmeso}. LHCb has developed a search \cite{Rodriguez:2021urv} for some of these channels as well as additional indirect signals discussed below.

\subsubsection{$B_c^+$ Mesogenesis}
\begin{figure}[t!]
\centering
\includegraphics[width=0.6\textwidth]{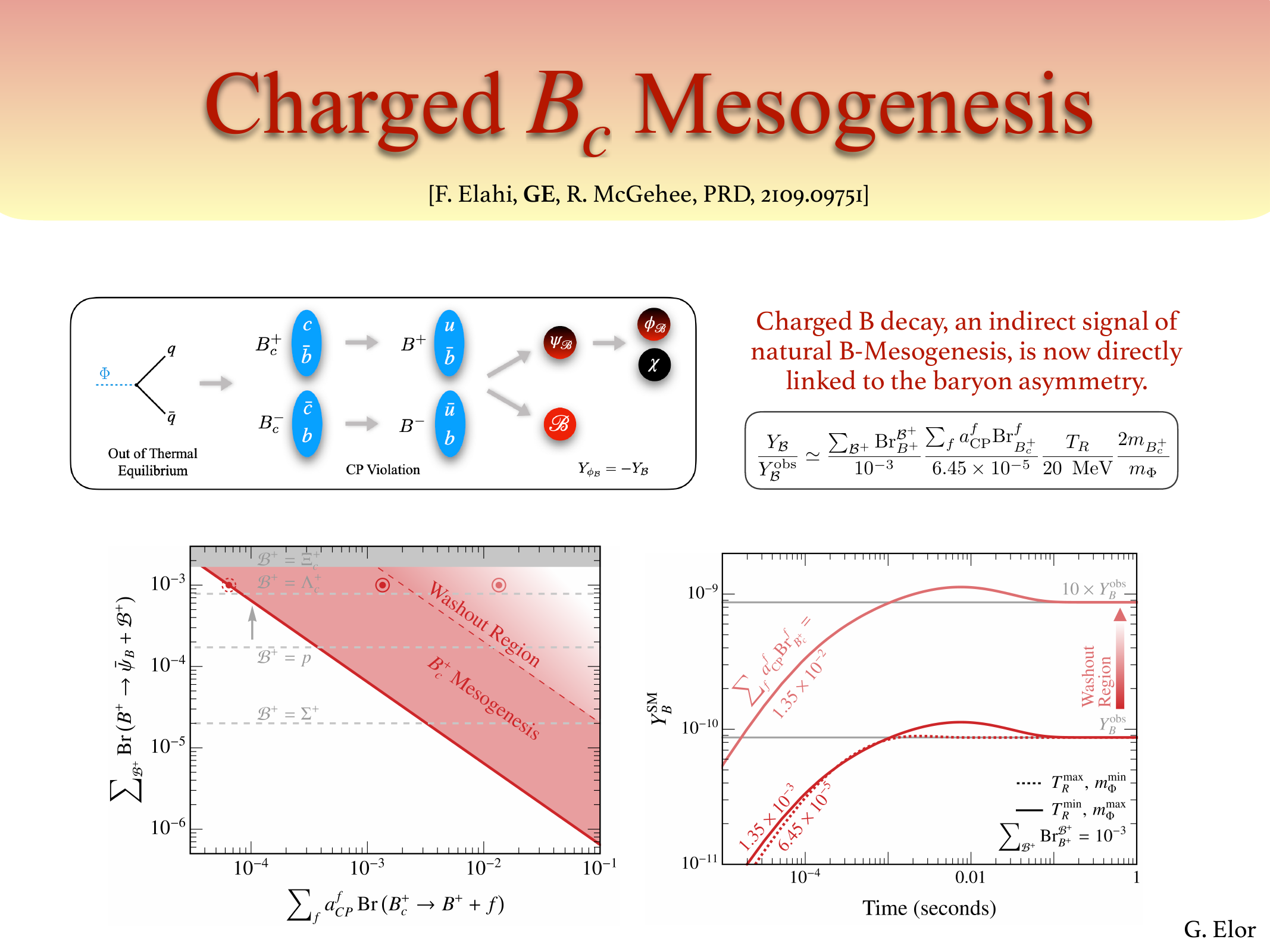}
\vspace{-.2cm}
\caption{Viable parameter space for $B_c^+$ Mesogenesis in red as presented in \cite{Elahi:2021jia}. Recast constraints for different final-state $\Bp$ are shown in gray \cite{Alonso-Alvarez:2021qfd}.}
\label{fig:BcParamPlot}
\end{figure}

In Charged $B_c^+$ Mesogenesis \cite{Elahi:2021jia}, the BAU is generated from the decays: $B_c^+ \to \, B^+  f_\M$ where  $f_\M = (D,\, K,\, \pi)$ is a light neutral SM meson.  For a given $f_\M$, CP violation is parameterized by the charge asymmetry observable:
\bea
\ACP^{f_\M} = \frac{\Gamma(B^+_c \to f_\M) - \Gamma(B^-_c \to \bar{f_\M})}{\Gamma(B^+_c \to f_\M) + \Gamma(B^-_c \to \bar{f_\M})}\,.
\eea
The BAU generation process completes when the daughter $B^+$ meson quickly decays into a charged SM baryon and dark baryon  $B^+ \to \, \psiBbar  \mathcal{B}^+$. The SM baryon asymmetry will depend the experimental observables $\ACP^{f_\M}$ and branching fractions of both $B_c^+$ and $B^+$ decays:  $Y_{\B} \propto \sum_{f_\M} \acpf  \text{Br}(B_c^+ \rightarrow B^+  f_\M) \times \sum_{\Bp} \text{Br}(B^+ \rightarrow \Bp  \psiBbar)$, where $\acpf \equiv \ACP^{f_\M} /\prn{1+\ACP^{f_\M}}$. Unlike Neutral $B$ Mesogenesis, $ACP^{f_\M}$ can be negative and still generate the BAU by simply swapping the baryon number of $\psiB$ and $\psiBbar$. The bottom left of Figure \ref{fig:DarkBvisible} depicts this mechanism. As discussed above, the possible $B^+$ decay channels summarized on the right-hand side of this figure correspond to decays that can generate the BAU through Charged $B_c^+$ Mesogenesis, while the neutral $B^0_{s,d}$ are now an indirect signal. Numerically integrating the Boltzmann equations  \cite{Elahi:2021jia}, the BAU is given by
\bea
\frac{Y_{\mathcal{B}}}{Y_{\mathcal{B}}^{\rm meas}}   \simeq \frac{\sum_{\Bp} \text{Br}(B^+ \rightarrow \Bp  \psiBbar)}{10^{-3}}   \frac{\sum_{f_\M} \acpf   \text{Br}(B_c^+ \rightarrow B^+  f_\M) }{ 6.45 \times 10^{-5}} \frac{T_R}{20 \, \text{ MeV}} \frac{2 m_{B_c^+}}{m_\Phi}\,.
\eea
Figure \ref{fig:BcParamPlot} shows the parameter space for Charged $B_c$ Mesogenesis in terms of the experimental observables discussed above.

ATLAS, CMS, and LHCb have all measured the $B_c^+$ mass and lifetime \cite{Anderlini:2014dha}. LHCb, in particular, is well suited for conducting searches of $B_c^+$ decays \cite{Yuan:2014mya,Tuning:2013nio}. There have been numerous measurements at LHCb of the branching fraction of fully hadronic $B_c^+$ decays which involve weak transitions of a $b$ to a $c$-quark \emph{e.g.} \cite{LHCb:2012ihf,LHCb:2012ag,LHCb:2013kwl,LHCb:2017lpu}. The decays relevant for $B_c^+$ Mesogenesis involve a $B^+$ meson in the final state i.e. the $b$-quark acts as a spectator. LHCb has made such a measurement, of $B_c^+ \to B^0_s \, \pi^+$ \cite{LHCb:2013xlg}.  But to date, no such searches exist for similar modes more directly relevant for Charged $B_c^+$ Mesogenesis, and both $\acpf$ and $\text{Br}(B_c^+ \rightarrow B^+  f_\M)$ have not yet been measured. Hopefully, these will become possible with increased luminosity at LHCb \cite{Gouz:2002kk}. $B_c^+$ Mesogenesis directly links the generation of the BAU to these observables and thus strongly motivates looking for these yet undiscovered CP-violating $B_c^+$ decays. Regarding $\ACP^{f_\M}$, most studies of the CPV within SM decays consider decays to $D^+$ and $K^+$, which are expected to have sizable CPV \cite{PhysRevD.62.057503}, rather than $B^+$. There are currently no measurements of $\ACP^{f_\M}$ for the decay modes relevant for $B_c^+$ Mesogenesis.

\subsubsection{$\dcp$ Mesogenesis}
\begin{figure}[t!]
\centering
\includegraphics[width=1\textwidth]{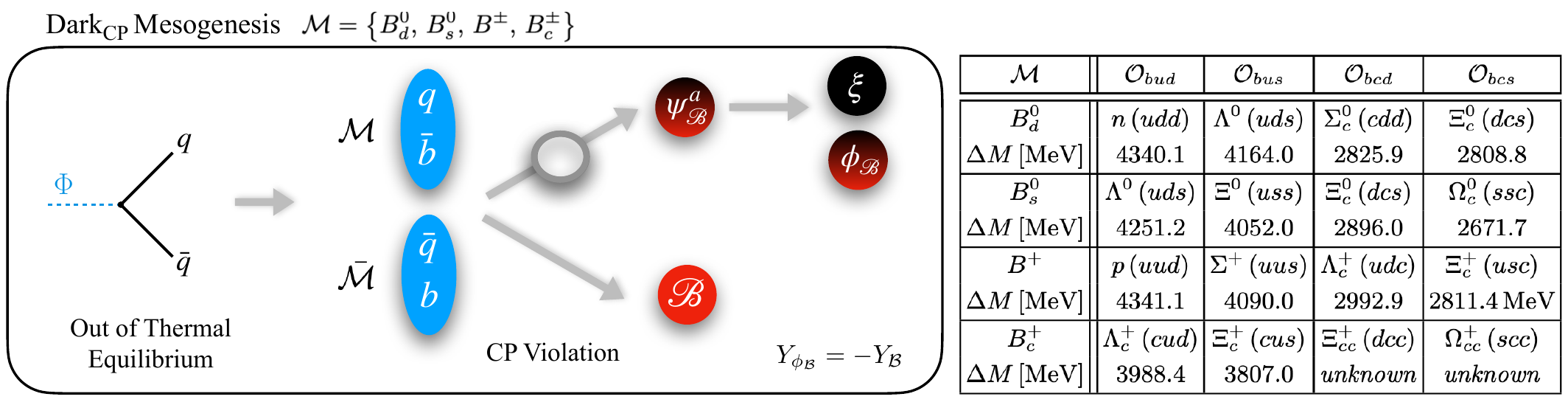}
\vspace{-.2cm}
\caption{\emph{Left:} Cartoon depiction of $\dcp$ $\M$ Mesogenesis. CP violation arises in the $\M$ meson decay itself through 1-loop and tree-level interference from diagrams with additional dark sector flavors. \emph{Right:}  For each initial meson $\mathcal{M}$ we list the daughter SM baryon arising from the decay of the $\bar{b}$ via the four possible flavorful operators, $\mathcal{O}_{b,u_i d_j} \equiv i \epsilon_{\alpha \beta \gamma} b^\alpha (\bar{d}_j^{c \beta} u_i^\gamma)$, resulting in the decay  $\mathcal{M} \rightarrow \mathcal{B}_{\rm SM} \, \psiB$. For each channel the the released energy $\Delta M = M_{\mathcal{M}} - M_{\mathcal{B}_{\rm SM}} \equiv m_{\psiB}^{\rm max}$ is quoted. }
\label{fig:cartoonDarkCPV}
\end{figure}
In $\dcp$ Mesogenesis \cite{darkmeso} the CP violation is entirely sequestered into the dark sector, and is currently unconstrained (e.g. by direct measurements at $B$ factories or from measurements of electric dipole moments). As such, this mesogenesis variant allows for the derivation of the limit stated above in Eq.~\eqref{eq:masterbound} on the minimum meson branching fraction through which any variant of mesogenesis can generate the BAU (up to the caveats discussed above).

$\dcp$ $\M$ Mesogenesis is arguably the simplest version of mesogenesis. CP violation in decays of SM mesons
\bea
\M = \left\{ B_d^0, \, B_s^0, \, B^\pm, \, B_c^\pm \right\},
\label{eq:M}
\eea
directly into the GeV scale dark baryon $\psiB$ and SM baryon,
$\Gamma \left( \M \rightarrow \psiB \mathcal{B}_{\rm SM} \right) \neq \Gamma \left( \Mbar \rightarrow \bar{\psiB}  \, \bar{\mathcal{B}}_{\rm SM} \right)$,
arises from the interactions in the dark sector, thus satisfying the C and CP violation Sakharov condition (see the model in \cite{darkmeso}).
As with all other variants of mesogenesis, since baryon number is extended to the dark sector, it remains a good symmetry, leading to the production of equal and opposite baryon asymmetries in the dark and visible sectors.  The left side of Figure ~\ref{fig:cartoonDarkCPV} illustrates this mechanism. The right side of that figure lists all possible final state daughter baryons for a given choice of decaying $\M$ in Eq.~\eqref{eq:M} and for each flavorful operator mediating the decay.

\begin{table*}[t]
\renewcommand{\arraystretch}{1.05}
  \setlength{\arrayrulewidth}{.2mm}
\centering
\small
\setlength{\tabcolsep}{0.18 em}
\begin{tabular}{ c  c }
\begin{tabular}{ |c || c | c | c  |}
    \hline
    Operator 			&  \,\, Initial \,\,  &  Final 	&     \,\,\, $\Delta M$     \,\,\,  \\
    and Decay   			&  Baryon &  Mesons				& (MeV)       \\ \hline \hline
& $\Lambda_b^0 (udb)$ & $\pi^0, \,\pi^+ \, \pi^-$ & 5484, 5340 \\
& $\Sigma_b^+ (uub) $ & $\pi^0 \pi^+ $ & 5536 \\
& $\Sigma_b^- (ddb)$ &  $\pi^0 \pi^- $ & 5541 \\
 & $\Sigma_b^0 (udb)$ & $\pi^0, \,\pi^+ \pi^-$ & \emph{unknown} \\
$\mathcal{O}_{bud}$  & $\Xi_b^0 (usb)$ & \emph{none} & \emph{n.a}  \\
$b \rightarrow \bar{u} \bar{d} \psiBbar $ & $\Xi_b^- (dsb)$ & $\pi^- \bar{K}^0 ,\, \pi^0 K^-$ &  5160, 5168\\
& $\Xi^0_{bb} (ubb)$& $\pi^0 \bar{B}_d^0\,, \pi^+ B^-$ &  \emph{unknown}\\
& $\Xi_{bb}^- (dbb)$ & $\pi^- B^0_d\,, \pi^0 B^-$ & \emph{unknown}\\
& $\Xi^+_{cb} (ucb)$ & $\pi^0 D^+, \, \pi^+ D^0$ & \emph{unknown} \\
& $\Xi^0_{cb} (dcb)$ & $\pi^0 D^0, \, \pi^- D^+$ & \emph{unknown}\\
& $\Omega_b^- (ssb)$ & $K^- \bar{K}^0$ & 5055 \\
\hline \hline
& $\Lambda_b^0 (udb)$ & $\bar{D}^0 \pi^0,\, \pi^+ D^-$ & 3620, 3610 \\
& $\Sigma_b^+ (uub) $ & $\bar{D}^0 \pi^+$ & 3811 \\
& $\Sigma_b^- (ddb)$ &  $D^- \pi^0$ & 3806  \\
 & $\Sigma_b^0 (udb)$ & $\bar{D}^0 \pi^0 ,\, \pi^+ D^-$ & \emph{unknown} \\
$\mathcal{O}_{bcd}$& $\Xi_b^0 (usb)$ &  $\bar{D}^0 \bar{K}^0 ,\, \pi^+ D_s^-$ & 3430, 3684  \\
$b \rightarrow \bar{c} \bar{d} \psiBbar $ & $\Xi_b^- (dsb)$ & $D^- \bar{K}^0,\, \pi^0 D_s^-$ & 3430, 3694 \\
 & $\Xi^0_{bb} (ubb)$ & $\bar{D}^0 \bar{B}_d^0 ,\, \pi^+ B_c^-$ &  \emph{unknown}\\
& $\Xi_{bb}^- (dbb)$ & $D^- B_d^0,\,\pi^0 B_c^-$ & \emph{unknown}\\
& $\Xi^+_{cb} (ucb)$ & $\bar{D}^0 D^+,\, \pi^+ \eta_c$ & \emph{unknown} \\
& $\Xi^0_{cb} (dcb)$ & $D^- D^+,\, \pi^0 \eta_c$ & \emph{unknown}\\
& $\Omega_b^- (ssb)$ & $D_s^- \bar{K}^0$ & 3580 \\
\hline
\end{tabular}
&
\begin{tabular}{ |c || c | c | c  |}
    \hline
    Operator 			&  \,\, Initial \,\,  &  Final 	&     \,\,\, $\Delta M$     \,\,\,  \\
    and Decay   			&  Baryon &  Mesons				& (MeV)       \\ \hline \hline
& $\Lambda_b^0 (udb)$ & $\pi^9 K^0, \, K^+ \pi^-$ & 4986 \\
& $\Sigma_b^+ (uub) $ & $ \pi^0 K^+ $ & 5172 \\
& $\Sigma_b^- (ddb)$ &  $\pi^- K^0$ & 5181 \\
 & $\Sigma_b^0 (udb)$ & $K^0 \pi^0 \,, K^+ \pi^-$ & \emph{unknown} \\
$\mathcal{O}_{bus}$ & $\Xi_b^0 (usb)$ &  $\eta', K^+K^-$ & 4834, 4804  \\
$b \rightarrow \bar{u} \bar{s} \psiBbar $ & $\Xi_b^- (dsb)$ & $\pi^- \eta',\, K^0 K^-$ & 4699, 4805  \\
& $\Xi^0_{bb} (ubb)$ & $\pi^0 B_s^0,\, K^+ B^-$ &  \emph{unknown}\\
& $\Xi_{bb}^- (dbb)$ & $\pi^- \bar{B}_s^0 ,\, K^0 B^-$ & \emph{unknown}\\
& $\Xi^+_{cb} (ucb)$ & $\pi^0 D_s^+, \, K^+ D^0$ & \emph{unknown} \\
& $\Xi^0_{cb} (dcb)$ & $\pi^- D_s^+,\, K^0 D^0$ & \emph{unknown}\\
& $\Omega_b^- (ssb)$ & $\eta K^-$ &  5005 \\
\hline \hline
& $\Lambda_b^0 (udb)$ & $\bar{D}^0 K^0, K^+ D^-$ & 3257, 3256 \\
& $\Sigma_b^+ (uub) $ & $\bar{D}^0 K^+$ & 3452 \\
& $\Sigma_b^- (ddb)$ &  $D^- D^0$ & 2082  \\
 & $\Sigma_b^0 (udb)$ & $\bar{D}^0  K^0 ,\, K^+ D^-$ & \emph{unknown} \\
$\mathcal{O}_{bcs}$ & $\Xi_b^0 (usb)$ &  $\bar{D}^0 \bar{B}_s^0$ & 1351 \\
 $b \rightarrow \bar{c} \bar{s} \psiBbar $ & $\Xi_b^- (dsb)$ & $D^- \eta, \, K^0 D_s^-$ &3386, 3331  \\
& $\Xi^0_{bb} (ubb)$ & $\bar{D}^0 \bar{B}_s^0,\, K^+ \bar{B}_s^0$ &  \emph{unknown}\\
& $\Xi_{bb}^- (dbb)$ & $D^- \bar{B}_s^0,\, K^0 B_c^-$ & \emph{unknown}\\
& $\Xi^+_{cb} (ucb)$ & $\bar{D}^0 D_s^+,\, K^+ \eta_c$ & \emph{unknown} \\
& $\Xi^0_{cb} (dcb)$ & $D^- D_s^+, \, K^0 \eta_c$ & \emph{unknown}\\
& $\Omega_b^- (ssb)$ & $\eta D_s^-$ & 3531  \\
\hline
\end{tabular}
\end{tabular}
\caption{Summary of all possible heavy baryon decays into mesons and missing energy that Eq.~\eqref{eq:L}--\eqref{eq:L2} can induce.}
\label{tab:bBaryonDecays}
\end{table*}

The SM baryon asymmetry is computed by solving (see  \cite{darkmeso}):
\bea
\frac{d (n_{\phiBbar} - n_{\phiB})}{dt}+ 3 H (n_{\phiBbar} - n_{\phiB}) &=&  2\, \Gamma_\Phi \text{Br}_\Phi^{\M} n_\Phi
\, A_{CP}^{\rm dark}
\text{Br}_{\M}, \\ \nonumber
\text{where} \,\,A_{CP}^{\rm dark} &\equiv&
 \frac{
 \Gamma (\Mbar\rightarrow \psiBbar \Bbarsm) - \Gamma (\M \rightarrow \psiB \Bsm)}{ \Gamma (\Mbar\rightarrow \psiBbar \Bbarsm) + \Gamma (\M \rightarrow \psiB \Bsm)}\,.
\label{eq:asymBE}
\eea
The \emph{dark CP asymmetry}  $A_{CP}^{\rm dark}$ is defined in analogy to the visible sector CP-violating observables. $\text{Br}_\Phi^{\M}$ is the branching fraction of $\Phi$ into the meson $\M$ under consideration and incorporates the quark fragmentation function. Here $A_{CP}^{\rm dark}$ is chosen such that a positive value generates the BAU. However, a negative value can also generate the BAU by swapping the baryon number charge assignments of $\psiB$ and $\psiBbar$ (and therefore also $\phiB$ and $\phiBbar$). Numerically integrating the Boltzmann equations yields the following prediction for the parameter space \cite{darkmeso}:
\bea
\hspace{-0.25in}
\frac{Y_{\mathcal{B}}}{Y_{\mathcal{B}}^{obs}} \simeq \frac{ \text{Br}_{\M} }{1.3 \times 10^{-8}} A_{CP}^{dark} \frac{T_R}{60 \, \text{MeV}}
\frac{2m_{\M}}{m_\Phi}\,.
\label{eq:BAU}
\eea
Since $A_{CP}^{\rm dark}$ may be order one \cite{darkmeso}, Eq.~\eqref{eq:BAU} indicates that generating the BAU from with $\dcp$ $\M$-Mesogenesis yields the lower limit quoted above in Eq.~\eqref{eq:masterbound} on the branching fraction of mesons (in Eq.~\eqref{eq:M}) into the baryons (listed in Figure ~\ref{fig:cartoonDarkCPV}) and missing energy.  Below this limit, mesogenesis (in any variation) is no longer well motivated. On the flip side, a measurement of $\text{Br}_{\M} \equiv \text{Br} (\M \rightarrow \mathcal{B}_{\rm SM} \psiB)$ in the range of $10^{-8}$-$10^{-5}$, while excluding visible sector CP violation variants of Mesogenesis, would be a signal of $\dcp$ Mesogenesis. Coupled with complementary probes (e.g. for the dark matter \cite{Berger:2023ccd}) and indirect signals $\dcp$, Mesogenesis can be entirely reconstructible.

$B$-factories are well-suited to search for such decays of the lighter charged and neutral $B$ mesons, while LHCb would be needed to set limits on the decays through which $\dcp$ $B_c^+$-Mesogenesis can proceed.  However, it would also be interesting to explore opportunities at LHCb for $B^+$ and $B^0$ decays in analogy to the search proposed in \cite{Rodriguez:2021urv}. While searches discussed in Section~\ref{subsubsec:neutralBmeso} begin to disfavor certain channels through which the original Neutral $B$-Mesogenesis mechanism \cite{Elor:2018twp} can proceed, there are still 3-4 orders of magnitude in sensitivity through which $\dcp$ $B_{d,s}^0$ and $\dcp$ $B^+$-Mesogenesis can be evoked to generate the BAU. The Belle-II collaboration has planned to set limits on the $\text{Br}$ observables for all four channels using Belle-I data, and hopefully also with new data coming in from Belle-II. Belle-II should have the capabilities to reach sensitivities down to the $10^{-7}$ level --- approaching the lower limit on the branching fraction for which $\dcp$ $B^0_{s,d}$ and $\dcp B^+$ Mesogenesis can generate the entire BAU (we expect upcoming electric dipole moment measurements to probe the remaining parameter space \cite{darkmeso}). Note that given a detection of the above-mentioned decays, we expect an experiment to be able to reconstruct the conjugate decay and therefore also measure $A_{CP}^{\rm dark}$. Thus, both observables related to the BAU can, in principle, be reconstructed. On the flip side, current $B$-Mesogenesis searches \cite{Belle:2021gmc, BaBar:2023rer} trigger on the final state baryon but not the anti-baryon. As such, an asymmetry is not measured and $A_{\rm CP}^{\rm dark}$ is currently entirely unconstrained by direct searches. For the heavier $\dcp \, B_c^+$ Mesogenesis, LHCb or a $B_c$ physics program at a future collider may be required to search for the $B_c^\pm$ decays in Figure ~\ref{fig:cartoonDarkCPV}.

\subsubsection{Indirect Signals at LHCb}
Hadron colliders, and LHCb in particular, can also indirectly probe mesogenesis. The colored mediator in Eq.~\eqref{eq:L}-\eqref{eq:L2} will also induce heavy baryon decays to mesons and missing energy.  LHCb has already designed a search for some of these channels \cite{Rodriguez:2021urv}. Table~\ref{tab:bBaryonDecays} comprehensively lists possible decays which are kinematically allowed i.e. where $\Delta M > m_{\psiB} \gtrsim 1 \, \text{GeV}$. The mass difference for heavy resonant baryon decays is not shown, but these can be particularly interesting triggers for LHCb. Not included in Table~\ref{tab:bBaryonDecays} are decays of doubly beautiful and other exotic heavy baryons of unknown masses.

Note that the dark matter ($\xi$ and the symmetric component of $\phiB$) in mechanisms involving dark baryons can be further probed via searches for induced nucleon decay   \cite{Berger:2023ccd}, which complements collider searches and increases reconstructability.

\subsubsection{$D^+$ and $B^+$ Mesogenesis}
\begin{figure}[t!]
\centering
\includegraphics[width=0.85\textwidth]{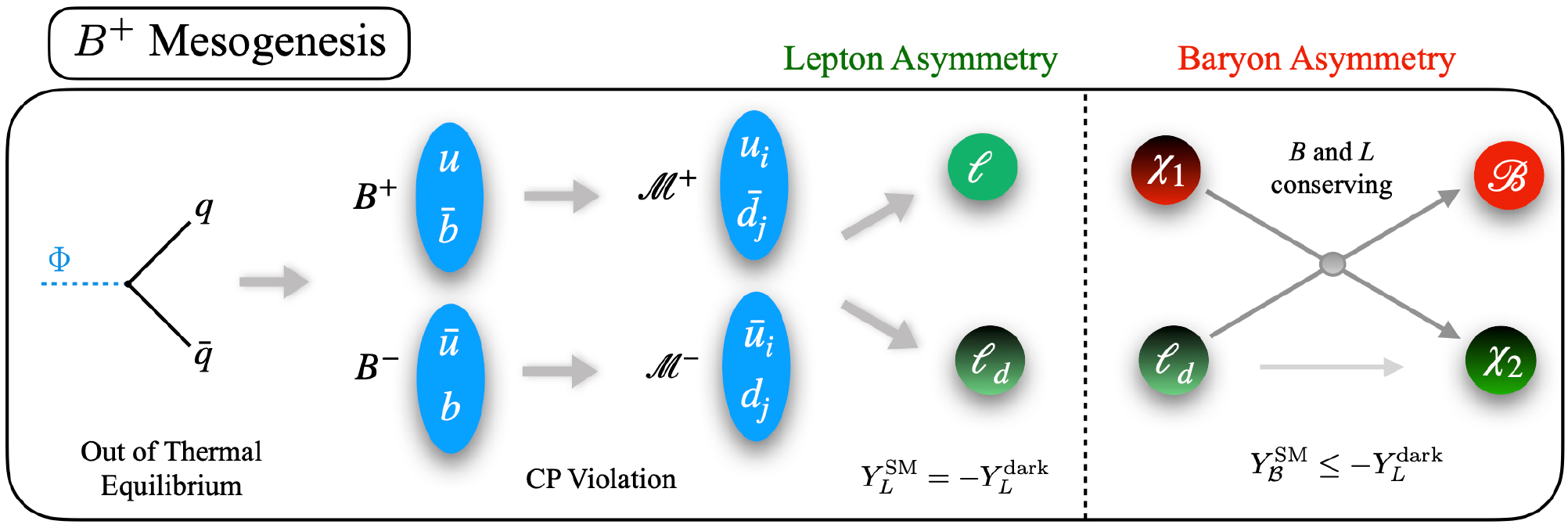}
\vspace{-.2cm}
\caption{An illustration adapted from \cite{Elahi:2021jia} of the way in which $B^+$ Mesogenesis realizes the Sakharov conditions. At MeV scales, $B^\pm$ mesons are produced and undergo CP-violating SM decays to charged mesons $\M^\pm = \left\{\pi^\pm, K^\pm, D^\pm, D^\pm_s, K^{\ast +} \right\}$. The charged mesons subsequently decay into a dark lepton, generating an equal and opposite dark and visible lepton asymmetry. Dark sector scatterings involving dark states carrying lepton and baryon number then transfer the lepton asymmetry into an equal and opposite dark and SM baryon asymmetry. }
\label{fig:ChargedBMesoCartoon}
\end{figure}
\begin{figure}[t!]
\centering
\includegraphics[width=0.9\textwidth]{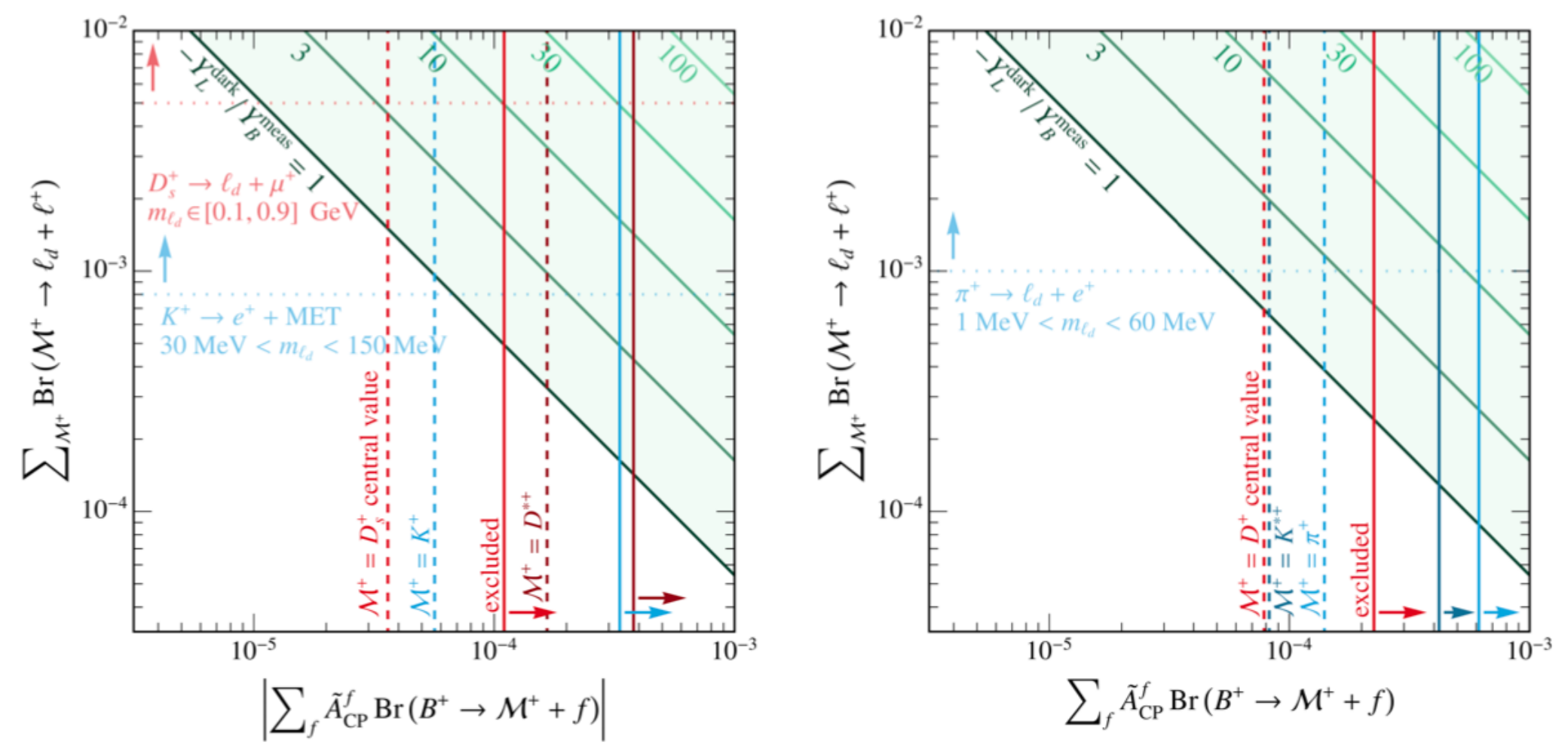}
\vspace{-.2cm}
\caption{Parameter space plot for $B^+$-Mesogenesis from \cite{Elahi:2021jia} for various daughter $\M^+$ and resonances of $\M^+$ (which should be particularly interesting for an LHCb trigger). The region of parameter space where a lepton asymmetry of equal to or greater than the observed baryon asymmerty is satisfied is shown in green, along with central values and bounds on $\sum_f \ACPB^f \text{Br}_{B^+}^f$ and a bound on $\text{Br} \prn{\pi^+ \to \ell_d + e^+}$. Channels shown here have a net positive central value for the summed CPV. A similar plot for $D^+$-Mesogenesis can be found in \cite{Elor:2020tkc}.}
\label{fig:ChargedDMesoPlots}
\end{figure}

In $D^+$ Mesogenesis \cite{Elor:2020tkc} and $B^+$ Mesogenesis \cite{Elahi:2021jia}, the CP violation in charged $D$ and $B$ meson decays is leveraged, respectively. These mesons are too light to decay directly into dark sector baryons $\psiB$ whose mass must be at the GeV scale to kinematically forbid proton decay and ensure the stability of matter. To circumvent this,  a lepton asymmetry is first produced via the decay:
\bea
D^+ \,\, \text{or} \,\, B^+ \, \to \, \M^+ \,+\,  \M \,, \quad \text{followed by} \quad \M^+  \, \to \, \ell_d \,+\, \ell^+ \,,
\eea
where $\M^+$ is a charged SM meson: $\pi^+$, $K^+$, or $D^+$, $D_s^+$ (in the case of $B^+$ decays) or a resonant meson $K^{\ast +}$, $D^{\ast +}$; $\ell_d$ is a dark lepton with SM lepton number $L = 1$ and mass $m_{\ell_d} < m_{\M^+} - m_\ell$; and the SM charged lepton $\ell^+$ can be a positron, antimuon, or antitau (in the case of $D^+$ and $D_s^+$ decays). The lepton asymmetry is then directly related to the CP asymmetry (which can be positive or negative), which for e.g. $B^+$ mesogenesis is defined as
\bea
\ACPB^f = \frac{\Gamma \prn{B^+ \to f } - \Gamma \prn{ B^- \to f }}{\Gamma \prn{B^+ \to f } + \Gamma \prn{ B^- \to f }} \,,
\eea
for the final state $f$, and the branching fractions of the two decays. Figure~\ref{fig:ChargedDMesoPlots} shows the results of solving the Boltzmann equations for the lepton asymmetry (see \cite{Elor:2020tkc,Elahi:2021jia}). The scenario in which $B^+$ first decays into a charged resonance is well suited for searches at LHCb. Furthermore, searches at hadron colliders and $B$ factories are projected to improve the measurements of the branching fraction and CPV in charged $B^+$ decays to final state pions, kaons, and $D$ mesons, and likewise for $D$ meson decays to pions and kaons in $D^+$ mesogenesis. The lepton asymmetry is then efficiently transferred into a baryon asymmetry through dark sector scatterings mediated by the possible interactions listed in Eq.~\eqref{eq:L}-\eqref{eq:L2}  --- which searches for the previous (dark baryon only) variants of mesogenesis will probe. Figure~\ref{fig:ChargedBMesoCartoon} depicts a cartoon of this scenario.

\clearpage
\changelocaltocdepth{3}
\section{FIPs experimental results and prospects}
\label{sec:fips_experiments}

\subsection{FIPs at LHCb in Upgrade 1 and Upgrade 2: Challenges and opportunities}
\label{ssec:fips_at_lhcb}

\subsubsection{LHCb Upgrade 1 detector and the new software trigger --- \textit{A.~Casais~Vidal}}
\label{sssec:casais-vidal}

\textit{Author: Adrian~Casais~Vidal, \email{adrian.casais.vidal@cern.ch}}  \\
The LHCb detector is a spectrometer instrumented in the forward region \cite{LHCb:2008vvz}. Although originally designed to study heavy flavour physics at the LHC, numerous results outside this domain have established LHCb as a versatile, general-purpose detector. During Runs 1–2, LHCb collected about 9 fb$^{-1}$ of data and produced many precision measurements. However, the upgraded LHCb Run 3 is designed to extend the physics reach by operating at five times higher instantaneous luminosity ($\simeq 2\times10^{33}\mathrm{cm}^{-2}\mathrm{s}^{-1}$) and reading out the entire detector at the 40 MHz LHC bunch-crossing rate \cite{LHCb:2023hlw}.

This demanded a comprehensive overhaul of LHCb’s detectors and trigger architecture. The upgrade, completed during the LHC’s Long Shutdown 2 (2019–2022), enables LHCb to accumulate at least 50 $\mathrm{fb}^{-1}$ over Run 3 and beyond. Key features include a revamped tracking system \cite{Bediaga:2013tje, LHCb:2014uqj}, improved particle identification (PID) sub-detectors \cite{LHCb:2013urp}, 30 MHz full-detector readout with no hardware trigger, and a fully software-based real-time event selection.

\paragraph{Tracking system upgrade}

LHCb’s silicon vertex detector was replaced with a new high-granularity pixel VELO that can withstand the higher luminosity and output 40 MHz of data. The upgraded VELO consists of 52 modules of hybrid silicon pixel sensors $(55\times55 \mu \mathrm{m}^2$ pixels, totalling $\sim41$ million channels) arranged around the beamline \cite{Bediaga:2013tje}.

Replacing the former TT station in front of the dipole magnet, the UT is a new silicon micro-strip detector. It comprises four planes of high-granularity sensors covering a larger area, ensuring efficient track seeding before the magnet. The UT’s design improves acceptance and compensates for the removal of the TT, using modern front-end electronics for 40 MHz readout. Together with the VELO, it provides precise momentum measurements for low-momentum tracks and seeds tracks entering the downstream regions.

Downstream of the magnet, the legacy Inner and Outer Trackers were replaced by the SciFi tracker. This detector is built from 2.5 m-long scintillating fibre mats read out by arrays of silicon photomultipliers (SiPMs) at the ends. The SciFi tracker covers the full acceptance ($x \simeq 6$ m width) with three stations of four layers each. Its design provides robust pattern recognition and hit resolution of $\sim70 \mu\mathrm{m}$, comparable to the former straw-tube Outer Tracker but with much finer granularity. The SciFi and UT together enable efficient tracking of charged particles after the magnet at 30 MHz \cite{LHCb:2014uqj}.

\paragraph{Particle Identification and Calorimeter Improvements}

LHCb’s Ring-Imaging Cherenkov (RICH1 and RICH2) system, crucial for hadron PID, was upgraded to sustain the 40 MHz readout and high particle flux. The overall optical design remained the same, but the photodetectors were replaced: the original hybrid photon detectors (HPDs) were swapped for faster pixelated multi-anode photomultiplier tubes (MaPMTs) with new readout electronics. This upgrade increased the RICH readout rate from the prior 1 MHz limit to the full 40 MHz, in line with the trigger-less readout scheme \cite{LHCb:2013urp}.

The electromagnetic (ECAL) and hadronic (HCAL) calorimeters from Runs 1–2 were mostly retained in terms of absorber and scintillator structure. However, the upstream Scintillating Pad Detector (SPD) and Preshower (PS) planes were removed in the upgrade.

The muon detector consists of four stations (M2–M5) of multi-wire proportional chambers (MWPCs) placed downstream of the calorimeters and interleaved with iron absorbers \cite{Barbosa-Marinho:504326}. In the upgrade, the first station, M1 (situated in front of the ECAL) was removed. Station M1 had been used in the old Level-0 trigger to enhance momentum resolution for muons, but at higher luminosity, it would suffer intense radiation and occupancy with no hardware trigger to justify it.

A key characteristic the Upgrade I is the removal of the hardware L0 trigger and the adoption of a trigger-less readout: every LHC bunch-crossing ($\sim30$ MHz of inelastic collisions in LHCb) is read out from all subdetectors into a high-bandwidth data acquisition system. Event builder nodes assemble complete events at 30 MHz (with an input rate of $\sim4$ TB/s of raw data). This strategy maximizes trigger efficiency by giving the software High-Level Trigger (HLT) access to full event information for every collision. An online farm of thousands of CPUs and GPUs is used to process this data in real time. Two sequential software trigger stages are employed, HLT1 and HLT2, both running event reconstruction and selection algorithms \cite{CERN-LHCC-2018-014}.
\paragraph{HLT1}

The HLT1 is executed on a GPU farm (using the custom Allen framework \cite{Allen-paper})  to exploit parallel processing. It performs a fast partial event reconstruction of each collision, including tracking and vertexing, to select interesting events at the full 30 MHz input. The primary tasks of HLT1 are to reconstruct all charged particle tracks, find primary and secondary vertices, identify muons, and apply inclusive selections (e.g. high-$p_T$ two-track topological triggers) \cite{CERN-LHCC-2020-006}. This reduces the rate by roughly a factor of 20–30, outputting $\mathcal{O}(1 \mathrm{MHz})$ of triggered events. LHCb’s is the first GPU-based trigger in HEP, and HLT1 on GPUs has been running successfully since 2022. In addition to the baseline charged-particle reconstruction, HLT1 has been augmented with algorithms for neutral object reconstruction (photons, $\pi^0$) and electron identification with bremsstrahlung. It also incorporates downstream tracking to recover tracks that do not have VELO hits \cite{Kholoimov:2025cqe}, thereby allowing triggers on long-lived particles decaying outside the vertex detector. A notable innovation in HLT1 is the use of neural-network-based algorithms for improved selection and data quality: for example, dedicated Lipschitz-constrained neural networks \cite{Kitouni:2021fkh} have been integrated to enhance the inclusive topological triggers \cite{Delaney:2023swp} and perform robust particle identification (see Fig.~\ref{fig:leptonID}). These Lipschitz neural networks are designed with bounded network response, providing stability against detector fluctuations and enabling reliable real-time selections. As a result, HLT1 achieves high efficiency for heavy-flavour decays and even potential long-lived exotic signals while keeping the output rate manageable. The HLT1 stage also includes a novel 30 MHz real-time monitoring strategy: for instance, the “BuSca” buffer scanner algorithm continuously scans the HLT1 output buffer for displaced-track signatures that might indicate long-lived particle decays, without requiring a specific trigger line \cite{LHCB-FIGURE-2024-018}. This provides a blind search capability for feebly-interacting particles that complements the standard trigger lines.

\paragraph{HLT2}

Events accepted by HLT1 are buffered to disk (up to $\sim 40$ PB buffer capacity) such that HLT2 can be executed asynchronously, effectively acting as a “semi-offline” processing stage. HLT2 runs on a CPU farm and performs full event reconstruction, including refined tracking, particle identification (using RICH, CALO, muon information), and exclusive/semi-inclusive selections \cite{CERN-LHCC-2018-014, CERN-LHCC-2018-007}. Thanks to an extensive real-time calibration and alignment framework, HLT2 reconstruction achieves offline-quality performance. Calibration constants (e.g. RICH refractive index, tracker alignments, calorimeter calibrations) are updated in real time during data-taking – typically at the per-fill level – and applied promptly in the HLT processing. This ensures that quantities like particle momenta and PID variables are as accurate as in an offline reprocessing. HLT2 implements hundreds of trigger lines, ranging from inclusive streams (e.g. multi-track topological selections) to specific exclusive decays, and even specialized lines for exotic signatures. Because HLT2 is buffered, its output throughput can be flexibly managed, and large latency algorithms (such as high-precision PID or complex neural network classifiers) can be used without impacting data taking. The full HLT2 output rate is on the order of tens of kHz, limited by bandwidth and storage considerations.

\paragraph{Dataflow and persistency}
An important aspect of the Run 3 trigger is the Turbo model for data storage, building on LHCb’s Run 2 “triggered real-time analysis” concept. Rather than writing out every accepted event in raw format for offline reconstruction, a majority of events are persisted in a reduced format containing only the information necessary for physics analysis (e.g. reconstructed particle candidates and a subset of detector data). This is referred to as the Turbo stream (or selective persistency) \cite{Benson:2015yzo,CERN-LHCC-2018-014}. Events in the Turbo stream are fully reconstructed and stripped of low-level detector data; analysts can directly use these for measurements without needing a separate offline processing. On the other hand, a subset of triggers (typically those for which complex offline recalibration is needed or for which signal purity is critical) are designated to the Full stream, in which the raw detector data is also saved to tape. For Run 3, thanks to improved online reconstruction, the Turbo stream constitutes the bulk of the data output, on the order of $70\%$ of the trigger yield. Only $\sim20\%$ of events are saved in the raw Full format, and a small fraction (a few percent) are kept for exclusive calibration samples. This approach dramatically reduces the offline storage and CPU requirements, effectively shifting the reconstruction burden to the trigger farm. The two-stream model (Turbo vs. Full) is essential for handling the high output rates while preserving the ability to do full offline analysis for critical channels \cite{CERN-LHCC-2018-014,CERN-LHCC-2018-007}. In practice, the Turbo data format and real-time analysis paradigm proved extremely successful: already in 2024 the LHCb trigger recorded a dataset (in equivalent integrated luminosity) about as large as Run 1 + Run 2 combined, demonstrating the experiment’s new capability.

The LHCb Upgrade I has transformed the experiment for the high-luminosity era. The detector underwent extensive upgrades: a new pixel VELO and tracking system to cope with 5× luminosity, improved RICH, calorimeters, and muon detectors to handle 40 MHz readout and provide enhanced PID. The hardware trigger was completely removed and replaced with a two-level software trigger that fully reconstructs events in real time. HLT1 on GPUs and HLT2 on CPUs, supplemented by real-time alignment and calibration, achieve offline-quality physics performance online. Advanced techniques, such as GPU parallelization, machine learning inference within trigger algorithms, and clever buffering and data scouting (e.g. BuSca for LLP signals), have been deployed for the first time in a hadron collider experiment. The combination of a trigger-less readout and selective persistence model enables LHCb to record unprecedented volumes of data – the 2024 run alone yielded approximately the combined amount of data of Runs 1 and 2 – while retaining the flexibility to search for both precise flavour physics signals and long-lived new physics signatures. With Run 3 underway, the upgraded LHCb detector and its novel real-time processing are expected to deliver a rich harvest of results and have set a new paradigm for trigger and data acquisition in collider experiments.

\subsubsection{LHCb dark photon results and prospects for LHCb Upgrade 1 --- \textit{F.~Volle}}
\label{sssec:volle}
\textit{Author: Felicia Volle, \email{felicia.carolin.volle@cern.ch}}\\
Dark photons (DPs) can but do not have to be produced by kinetic mixing with the Standard Model (SM) photon. Thanks to the kinetic mixing, massive DPs possess a direct coupling to the SM currents and the non-zero invariant mass, $m_{A^\prime}$. Experimentally, the non-zero mass is helpful in distinguishing between the dark and the ordinary photon. In this contribution, the focus is set on DP decays to visible final states, because they are more accessible experimentally.

In the FIPs community, DPs receive a lot of attention thanks to the natural suppression of the coupling strength of the dark photon to electromagnetically charged Standard Model (SM) fermions, $\varepsilon$, which can be achieved by the kinetic mixing between the new dark $U(1)_{D}$ with the SM $U(1)_{Q/Y}$ field as a correction at loop level of some UV completion~\cite{Fabbrichesi:2020wbt}. \\

LHCb proved its sensitivity to DPs by analysing promptly produced dimuon pairs~\cite{Aaij:2017rft,LHCb:2019vmc}. The kinetic mixing of the DP with the SM photon is assumed. A minimal model is assumed, where the DP is the only addition to the SM. It is named the ``minimal DP model''. It holds, as well, if energy conservation prohibits the decay of the DP to dark-sector particles. However, the search can be re-interpreted for a non-negligible partial decay width of dark photons decaying to invisible final states.

\begin{figure}[htb]
  \begin{center}
    \includegraphics[width=0.79\linewidth]{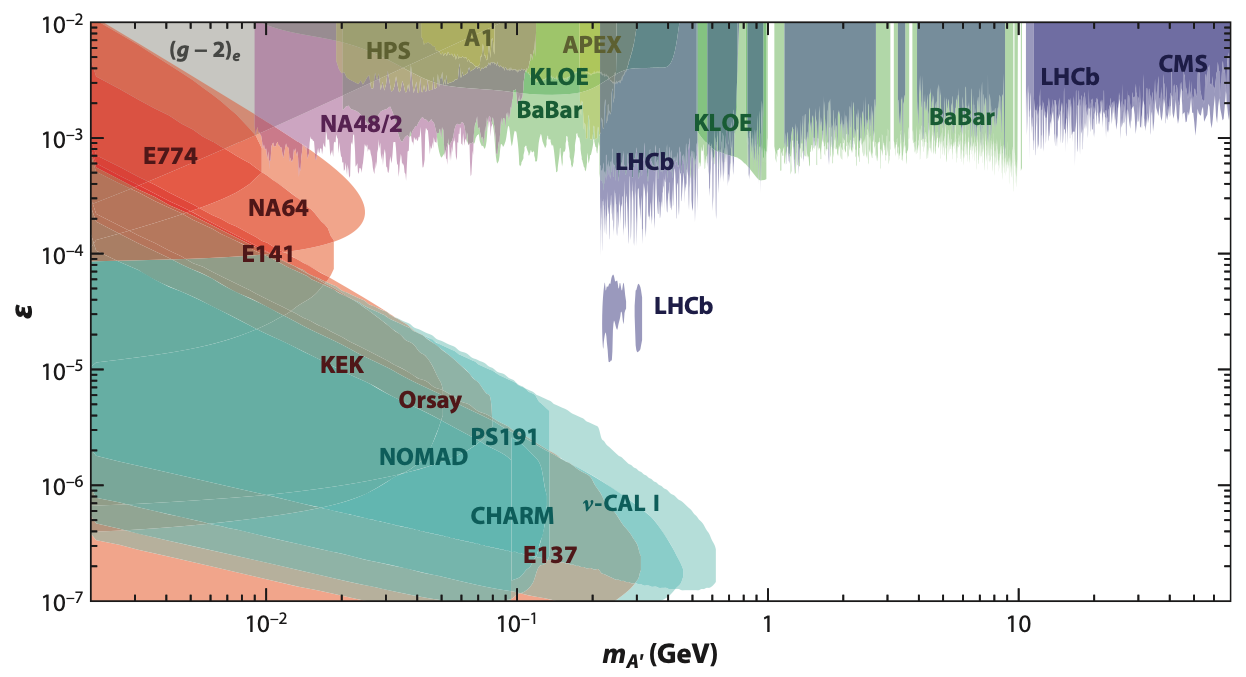}
    \vspace*{-0.5cm}
  \end{center}
  \caption{Exclusion limits of the DP mass, $m_{A^\prime}$, and kinetic-mixing strength, $\varepsilon$, in the ``Minimal dark photon model''~\cite{LHCb:2019vmc}. }
  \label{fig:DP2mumu_exclusion}
\end{figure}

The most recent analysis, Ref.~\cite{LHCb:2019vmc}, exploits 5.5\,fb$^{-1}$ of the Run~2 dataset collected by the LHCb detector. The DP yield is normalized to the off-shell photon yield in order to cancel most systematic uncertainties. Promptly decaying DPs have been searched for in a wide mass range from the dimuon mass threshold up to 70\,GeV. The mass region around prominent resonances has been vetoed, and a narrow resonance over the continuum background has been searched in the dimuon invariant mass spectrum. The search is conducted in steps of half of the dimuon mass resolution. Kinetic-mixing strengths, $\varepsilon$, of up to $10^{-4}$ could be exploited, but no excess has been found. Exclusion limits at 90\% confidence level have been established, which are shown in Fig.~\ref{fig:DP2mumu_exclusion}. The newest results from NA62~\cite{NA62:2023nhs} and CMS~\cite{CMS:2023hwl} are not shown in this figure, but LHCb still places the most stringent limits below 0.5\,GeV, and above 10\,GeV.

For accessing lower kinetic mixing strengths, larger DP lifetimes have to be tested~\cite{Curtin:2014cca}. In LHCb, testing larger DP lifetimes is challenging because of the presence of the vertex locator material, in which photon conversions to fermion-antifermion states may happen. Therefore, the detector material was mapped~\cite{Alexander:2018png}. The origin vertices of the dimuon pair consistent with the material location have been vetoed. The search for displaced DP decay signatures covered a reduced DP mass range from 214 to 350\,MeV, and was sensitive to $\varepsilon$ values in between $10^{-4}$ and $10^{-5}$. Again, no excess could be stated, and limits have been placed in the unexplored parameter space (see fig.~\ref{fig:DP2mumu_exclusion}).

A non-minimal search of a narrow resonance in the dimuon invariant mass spectrum has been published using 5.1\,fb$^{-1}$ of LHCb's Run~2 data~\cite{LHCb:2020ysn}. This model-independent search explores four different DP decay configurations without requiring kinetic mixing. These configurations include inclusive DP searches in the prompt dimuon invariant mass spectrum with and without an associated $b$-jet. Moreover, a displaced search has been conducted in the dimuon mass spectrum, which is, in the first case, pointing back to the primary vertex, and not in the second case. This diverse search could place limits on several models containing low-mass dimuon resonances. \\

\begin{figure}[tb]
  \begin{center}
    \includegraphics[width=0.65\linewidth]{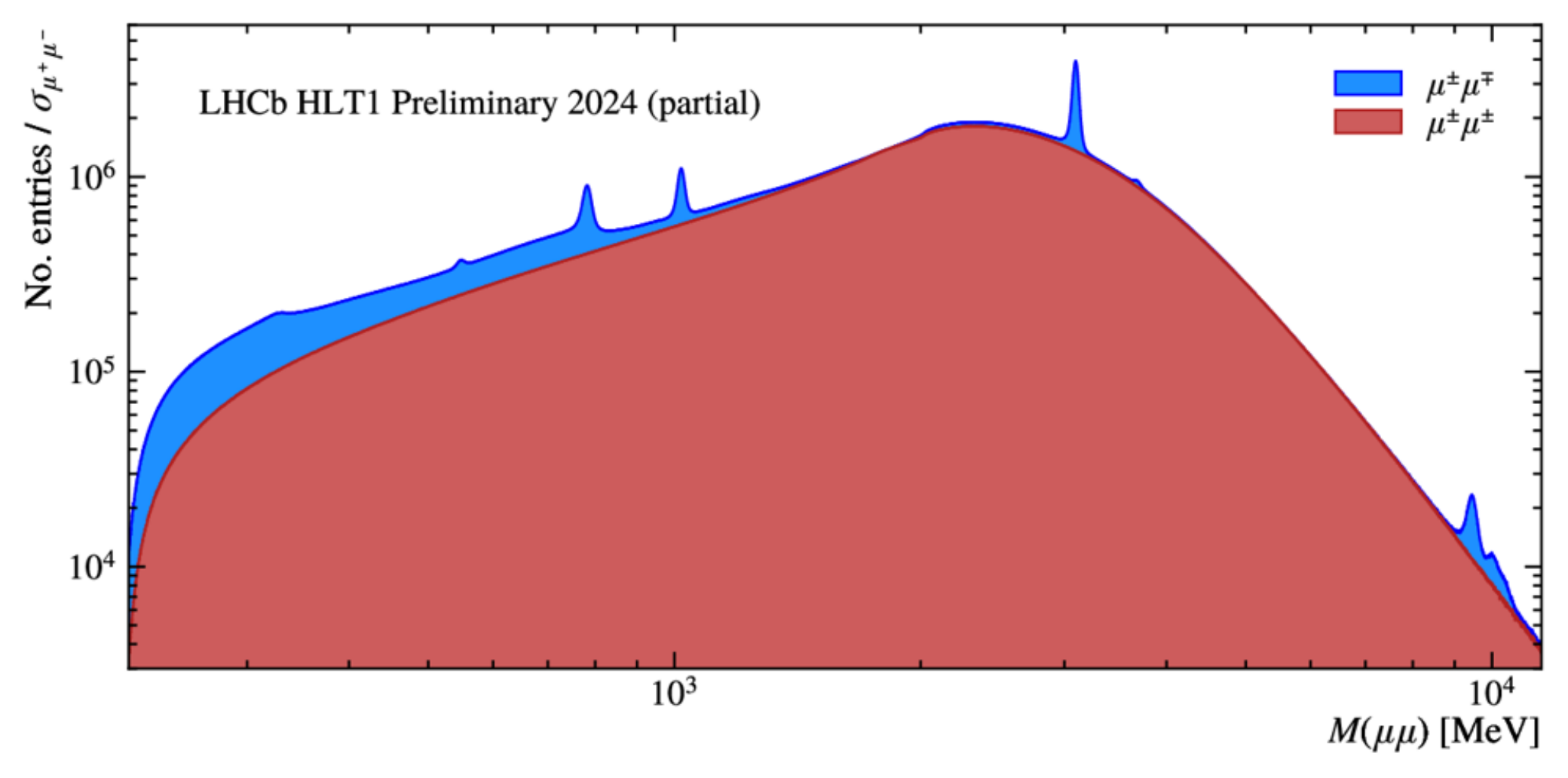}
    \vspace*{-0.5cm}
  \end{center}
  \caption{ The dimuon invariant mass spectrum after the first stage of the fully software-based trigger is shown in a range from 211\,MeV to 11.5\,GeV~\cite{LHCB-FIGURE-2024-029}. }
  \label{fig:dimuon}
\end{figure}

\begin{figure}[tb]
  \begin{center}
    \includegraphics[width=0.4\linewidth]{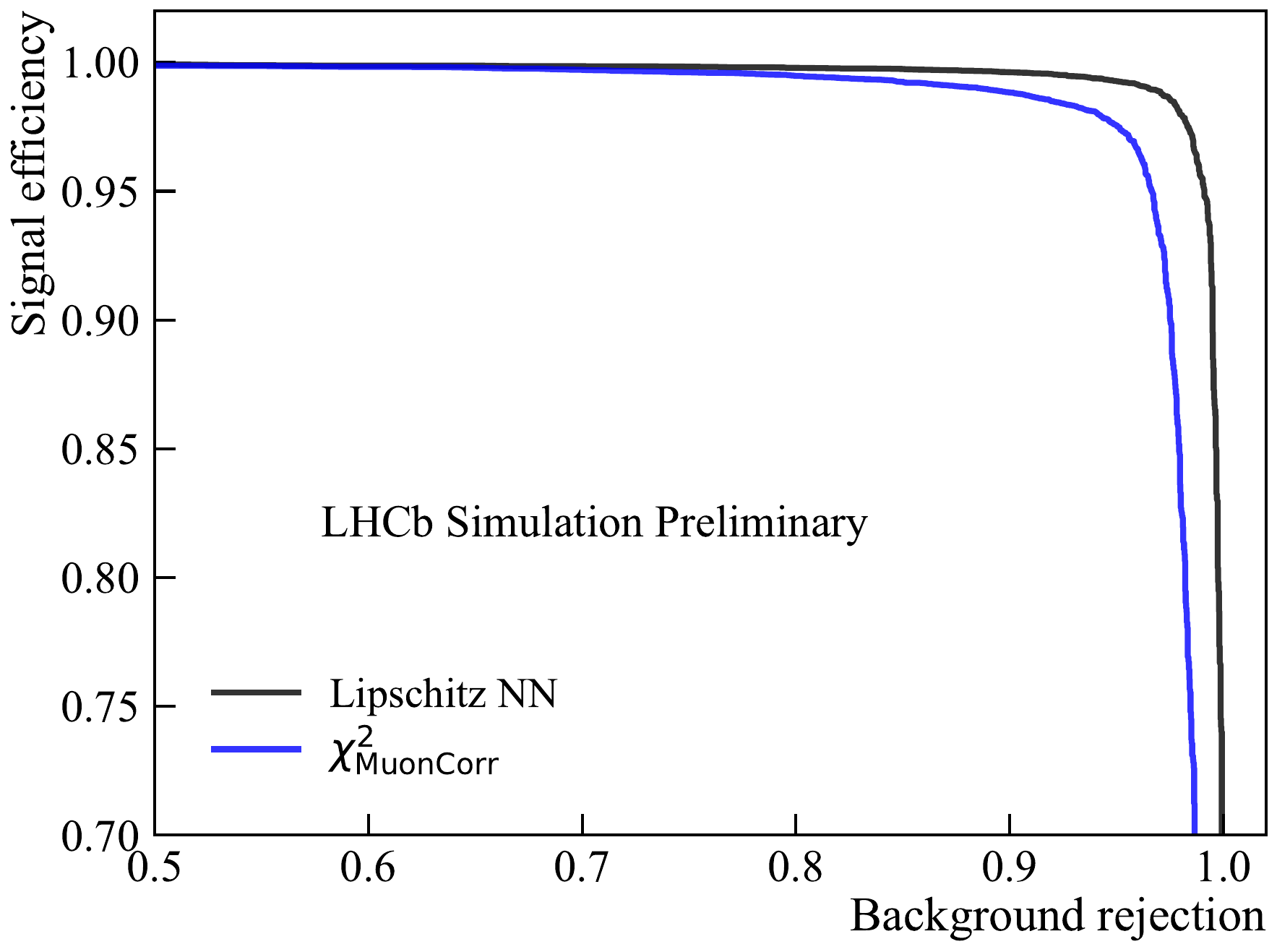}
    \includegraphics[width=0.4\linewidth]{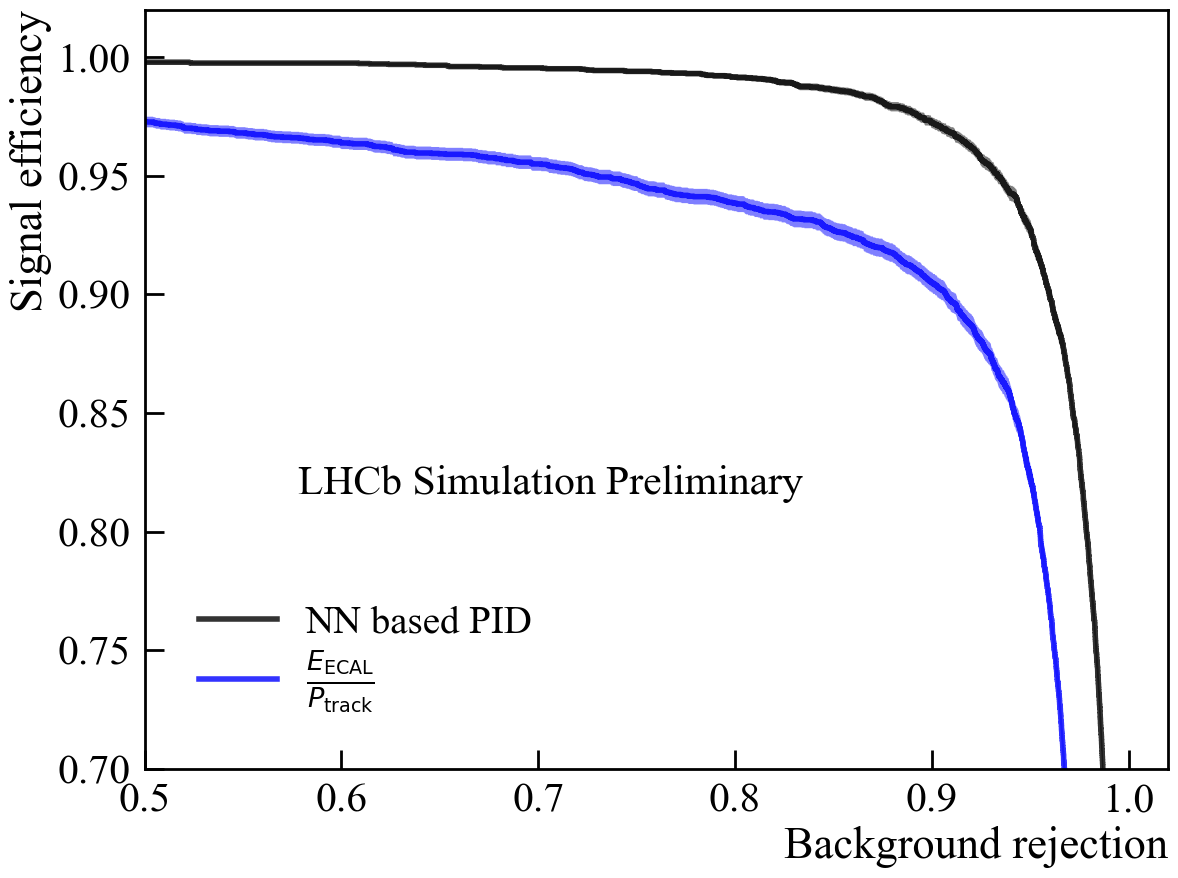}
    \vspace*{-0.5cm}
  \end{center}
  \caption{ Performance of the muon and electron identification neural network employed in the first trigger stage of the LHCb's fully software-based trigger in Run~3~\cite{LHCB-FIGURE-2024-029,LHCB-FIGURE-2024-003}. }
  \label{fig:leptonID}
\end{figure}

In 2024, the LHCb experiment has collected 9.56\,fb$^{-1}$ of $pp$-collisions at 13.6\,TeV, which exceeds the size of the combined Run~1 and 2 dataset. The large dataset permits the DP search using $A^\prime \to \mu\mu$ decays, the exploration of a larger area in the $\varepsilon$-$m_{A^\prime}$ plane. An additional enhancement is achieved by employing muon identification neural networks that are already in the first trigger stage. Its performance is evaluated in Fig.~\ref{fig:leptonID}. The clean dimuon invariant mass spectrum out of the first trigger stage is shown in Fig.~\ref{fig:dimuon}.

In order to solve the puzzle of the nature of Dark Matter, it is crucial to explore the fully accessible parameter space in colliders. An underexplored area is that of the dark photon mass below the dimuon mass threshold. This motivates the search for dark photons decaying into dielectron pairs.

Electrons are more challenging in LHCb than muons because of their lower efficiency and poorer identification. In Run~3, the software-based trigger enables LHCb to counteract those two difficulties.

A better electron identification is achieved by employing an electron identification at the first trigger stage, which performance is depicted in Fig.~\ref{fig:leptonID}.
An increase in electron efficiencies has been achieved in Run~3 by triggering on the tracks, which allows looser kinematic requirements than previously possible by triggering on the transverse energy deposit in the electromagnetic calorimeters. The improvement has been evaluated in $B^+ \to K^+ \ell^+\ell^-$ decays in Ref.~\cite{LHCB-FIGURE-2024-030}. While for the muon case, an improvement is achieved predominantly for low $B^+$ transverse momenta, an improvement in the full $B^+$ transverse momentum range of up to a factor four has been realised. The DP searches in LHCb will profit from the expected larger efficiencies of low-momentum leptons. \\

\begin{figure}[tb]
  \begin{center}
    \includegraphics[width=0.49\linewidth]{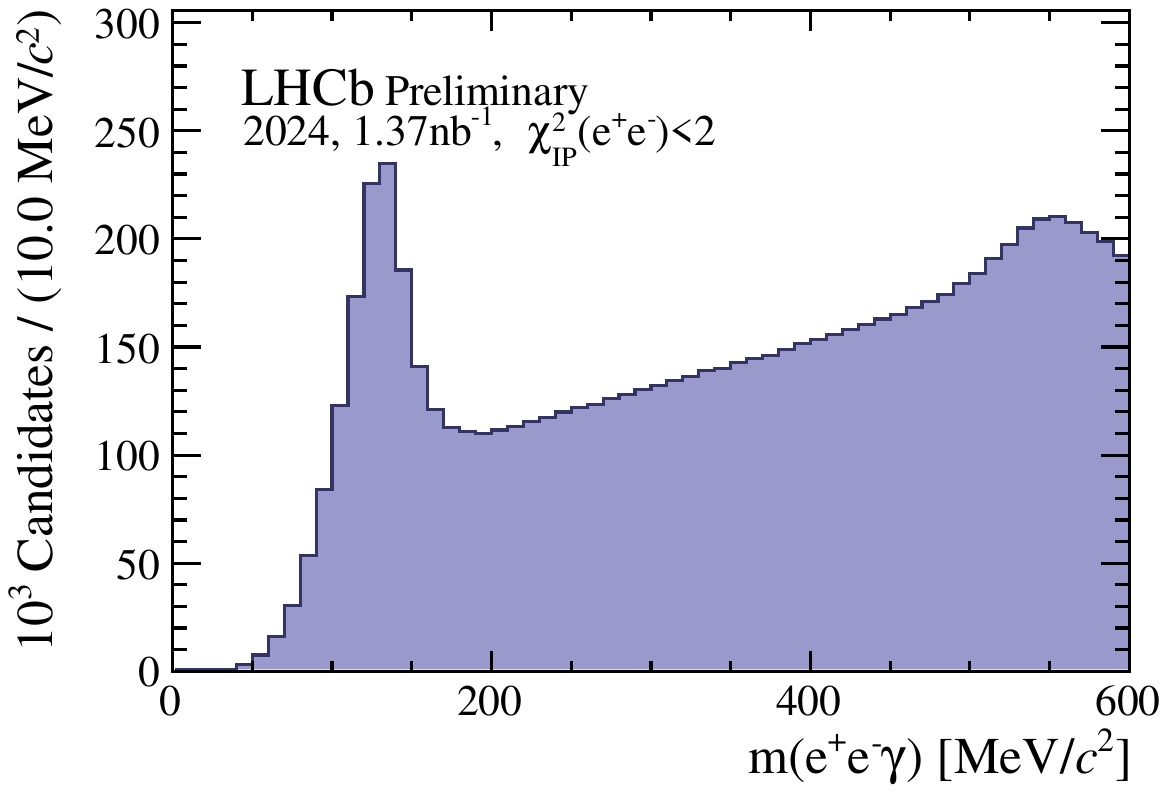}
    \vspace*{-0.5cm}
  \end{center}
  \caption{ The $e^+e^-\gamma$ invariant mass is shown for the case where the di-electron pair decays promptly~\cite{LHCB-FIGURE-2025-003}. }
  \label{fig:pi0massplot}
\end{figure}

The dimuon search analyses all dimuon pairs produced in the $pp$-collision. Such a search is challenging for electrons because of the more abundant backgrounds and the worse resolution. One idea of the DP search is to use the decay of $\pi^0$ and $\eta$ meson decays to $e^+e^-\gamma$. Using $\pi^0$ and $\eta$ mesons in the initial state gives a handle on the backgrounds. The $e^+e^-\gamma$ invariant mass spectrum in 2024 data is shown for prompt $e^+e^-$ decays in Fig.~\ref{fig:pi0massplot}. The $\pi^0$ and $\eta$ mesons are visible just after the preselection.

\begin{figure}[tb]
  \begin{center}
    \includegraphics[width=0.49\linewidth]{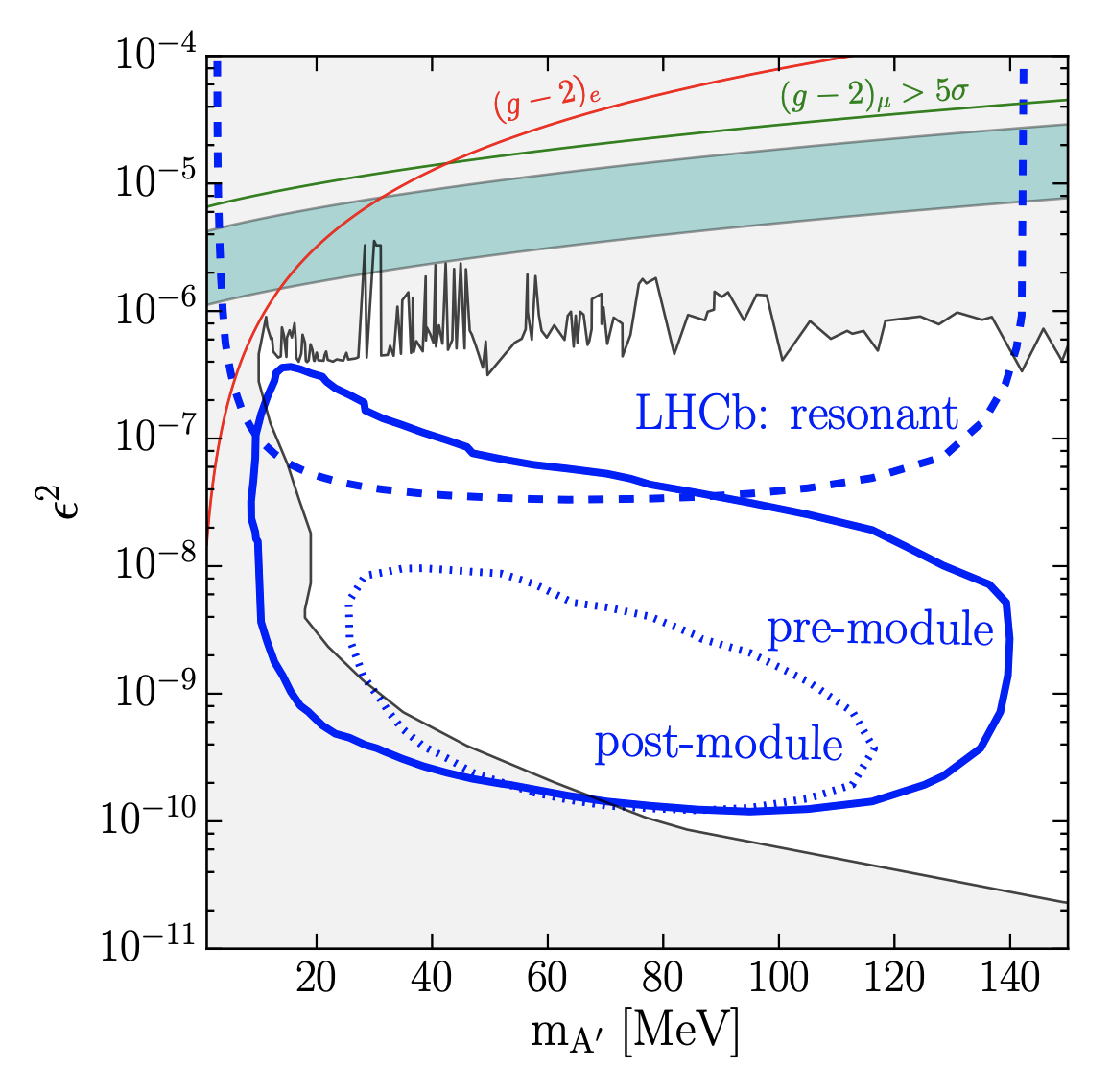}
    \vspace*{-0.5cm}
  \end{center}
  \caption{ Sensitivity projections of the potential bounds on the dark photon mass, $m_{A^\prime}$, and kinetic-mixing strength, $\varepsilon$, with 15\,fb$^{-1}$ of LHCb's Run~3 data~\cite{Ilten:2015hya}. }
  \label{fig:sensitivity_charm}
\end{figure}

Another idea is to search for DPs in $D^{*0} \to D^0 A^\prime(\to e^+e^-)$ decays in Run~3. Thanks to the narrow width of the $D^{*0}$, a mass constraint on its mass permits the recalculation of the dielectron invariant mass spectrum; this enables a better resolution of the dielectron final state, which is crucial because of the frequent Bremsstrahlung emission of electrons, worsening their mass resolution. The sensitivity study in Ref.~\cite{Ilten:2015hya} promises the exclusion of a large uncovered area in the DP parameter space, as shown in Fig.~\ref{fig:sensitivity_charm}. The study only requires 15\,fb$^{-1}$, which is supposed to be collected at the latest end of 2025.

\subsubsection{LHCb Upgrade 1: results and prospects for triggers with displaced vertices --- \textit{J.~Zhuo}}
\label{sssec:zhuo}
\textit{Author: Brij Kishor Jashal, \email{brij.kishor.jashal@cern.ch}}  \\
\textit{Author: Valerii Kholoimov, \email{valerii.kholoimov@cern.ch}}  \\
\textit{Author: Arantza De Oyanguren Campos, \email{arantza.de.oyanguren.campos@cern.ch}}  \\
\textit{Author: Volodymyr Svintozelskyi, \email{volodymyr.svintozelskyi@cern.ch}}  \\
\textit{Author: Jiahui Zhuo, \email{jiahui.zhuo@cern.ch}}  \\
\paragraph{Introduction}

In Upgrade I, the LHCb experiment completely renewed the tracking sub-detectors and introduced a new trigger system to handle an instantaneous luminosity five times higher than that of previous running periods. Since the hardware trigger (L0) reaches saturation at such high luminosities, it was entirely removed in this upgrade and all detector readouts are processed directly by software trigger system (HLT1 and HLT2). This system performs Real-Time Analysis to reconstruct events and make trigger decisions based on the reconstructed objects. In particular, HLT1 operates at 30 MHz, directly handling the full detector readout. This high-throughput requirement is achieved using GPU acceleration (Allen \cite{Allen-paper}).

In the LHCb tracking system, downstream tracks are reconstructed using only the UT and SciFi sub-detectors. Since the UT is located approximately 2 meters from the primary vertices, these tracks can be used to reconstruct the decay products of LLPs if the decay occurs within that distance. In previous running periods, downstream track reconstruction was only possible in HLT2 and was considered infeasible in HLT1 due to strict timing constraints. However, thanks to the upgraded trigger system, downstream track reconstruction is now successfully implemented in HLT1 in Run 3. With the removal of the L0 trigger, we are now able to reconstruct and trigger on LLP decays using downstream tracks directly at 30 MHz.

\paragraph{Downstream tracking}

\begin{figure}[t!]
    \centering
    \includegraphics[width=0.45\textwidth]{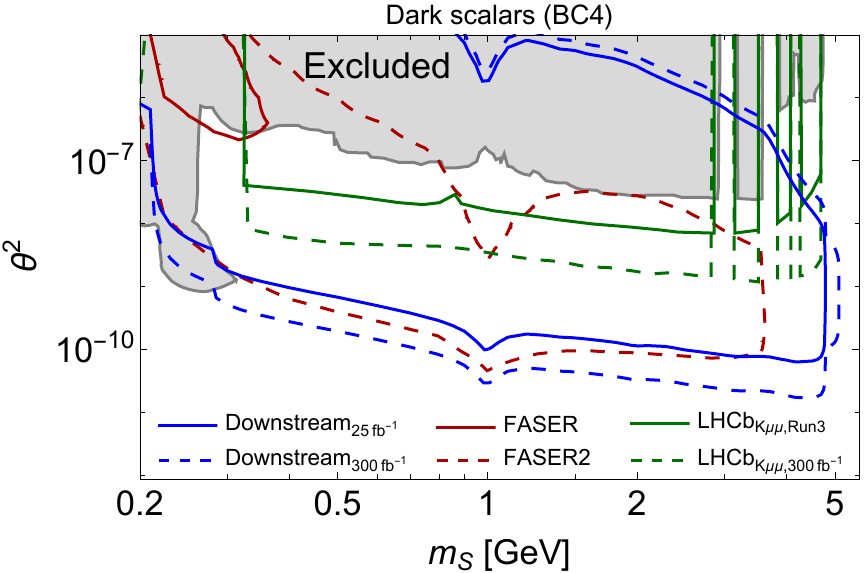}~\includegraphics[width=0.45\textwidth]{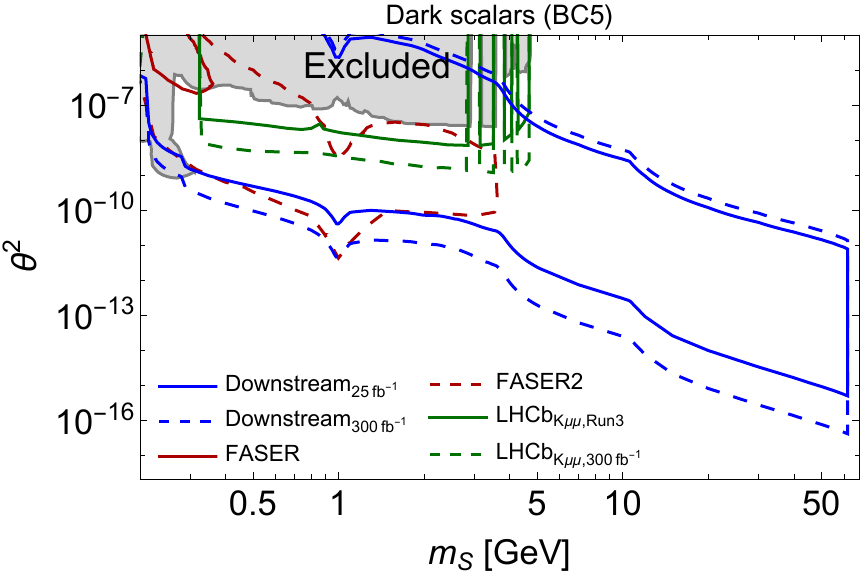}
    \caption{Sensitivity to Higgs-like scalars, models \textit{BC4} (left) and \textit{BC5} (right). The excluded domain, as well as sensitivities of FASER, FASER2, and the search of $B\to KS(\to \mu\mu)$ are taken from~\cite{Antel:2023hkf}, where S is the long-lived Higgs-like scalar.}
    \label{fig:downstream_sensitivities}
\end{figure}

Recent sensitivity studies \cite{downstream_sensitivity_study} and \cite{Gorkavenko:2023nbk} have shown that downstream tracks can significantly enhance the sensitivity of the LHCb detector for BSM LLP searches, increasing it by more than two orders of magnitude, as illustrated in Fig. \ref{fig:downstream_sensitivities}.

The HLT1 downstream tracking algorithm \cite{downstream_tracking_paper} was developed in 2022, integrated into the HLT1 system in 2023, and became operational for the first time in 2024 at the nominal Run 3 luminosity. It ran successfully during the last two weeks of proton-proton data-taking in 2024, collecting approximately 1~fb\(^{-1}\) of data. It is expected to continue operating in the rest of Run 3.

\begin{figure}[t!]
    \centering
    \includegraphics[width=0.8\textwidth]{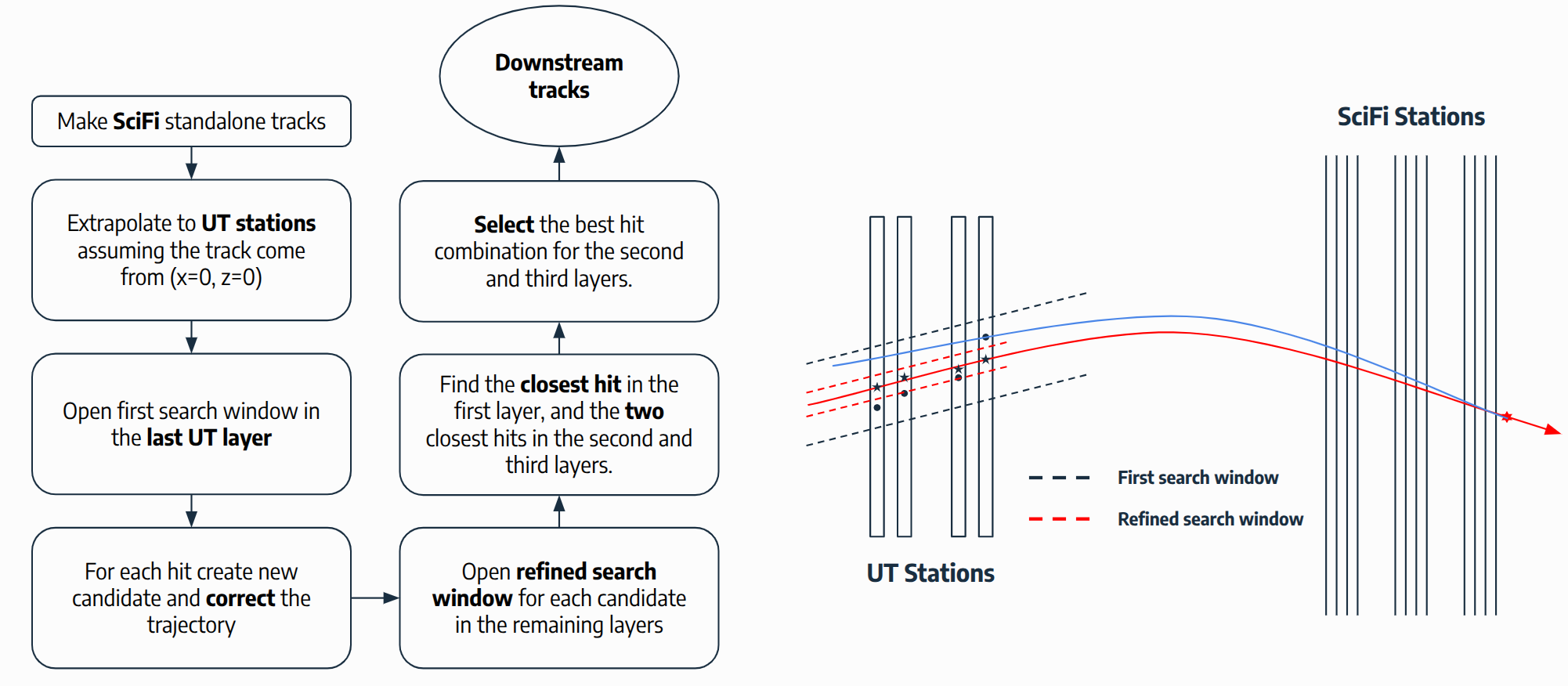}
    \caption{The flowchart of HLT1 downstream tracking (left) and an illustration of the tracking design (right)~\cite{downstream_tracking_paper}.}
    \label{fig:downstream_tracking_flowchart}
\end{figure}

Fig. \ref{fig:downstream_tracking_flowchart} shown the design of the HLT1 downstream tracking. The algorithm uses SciFi standalone tracks as input, extrapolates the track state to the UT stations assuming the track originates from the point \((x=0, z=0)\), and opens a wide search window in the last UT layer. For each hit within the acceptance, it creates a downstream candidate and corrects the track trajectory using the hit position. It then continues extrapolating to the remaining UT layers, opening refined search windows to add hits closest to the track trajectory. In this step, only one closest hit is considered in the first UT layer, and up to two closest hits are considered in the second and third layers. Finally, the algorithm selects the best hit combination and performs a fit using all selected UT hits to obtain an updated track state.

\begin{figure}[t!]
\centering
\includegraphics[width=0.45\textwidth]{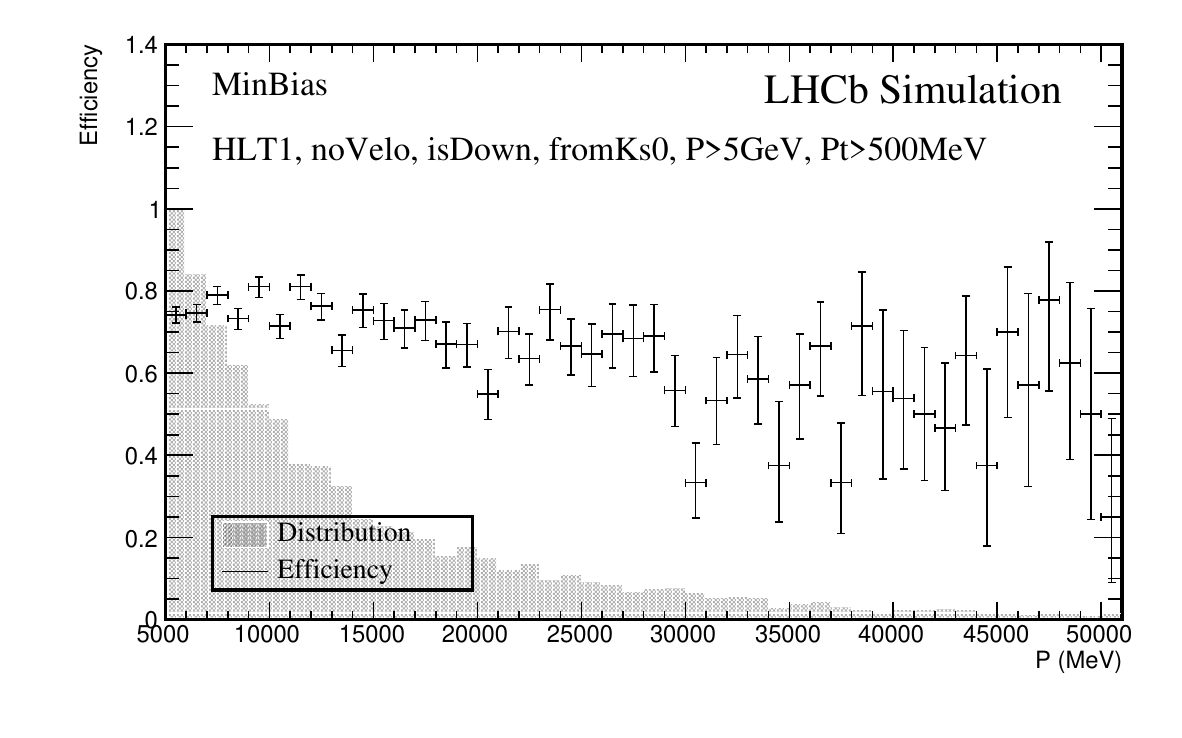}
\includegraphics[width=0.45\textwidth]{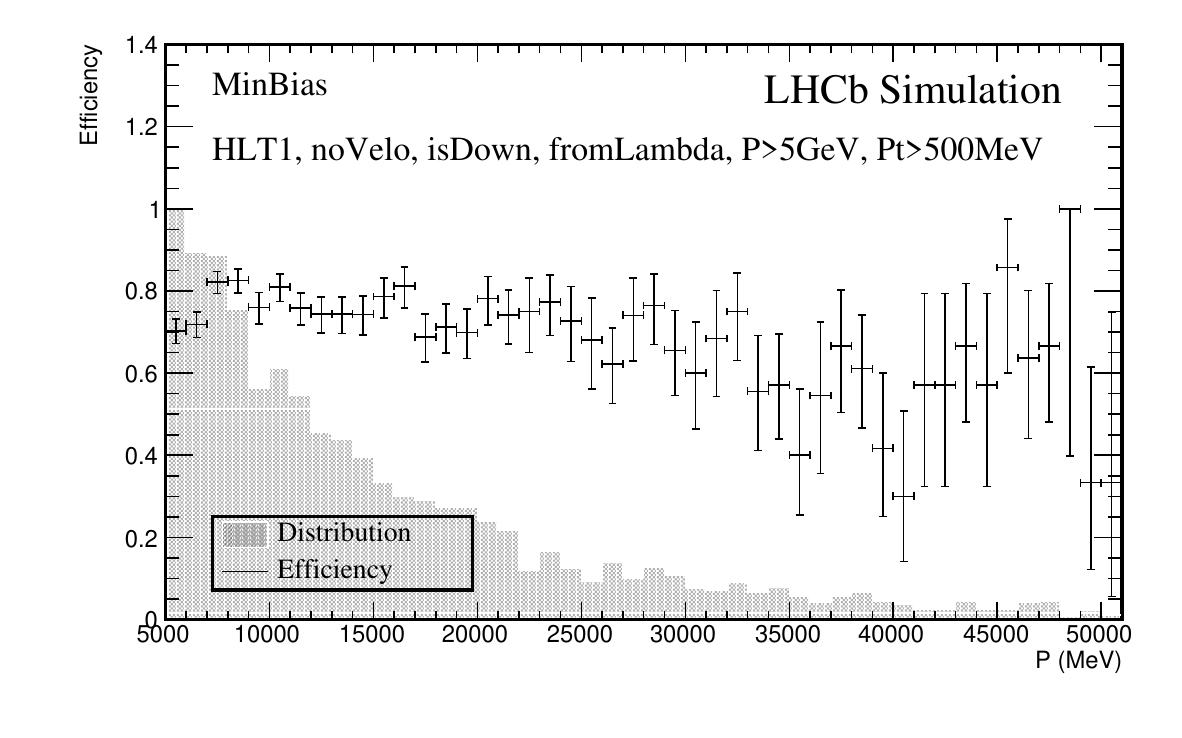}
\caption{Efficiency of the downstream tracking algorithm as function of the momentum to reconstruct prompt $K_S^0 \to \pi^+ \pi^-$ decays (top) and $\Lambda \to p \pi^-$ decays (bottom)~ \cite{downstream_tracking_paper}.}
\label{fig:downstream_tracking_efficiency}
\end{figure}

\begin{figure}[t!]
\centering
\includegraphics[width=0.45\textwidth]{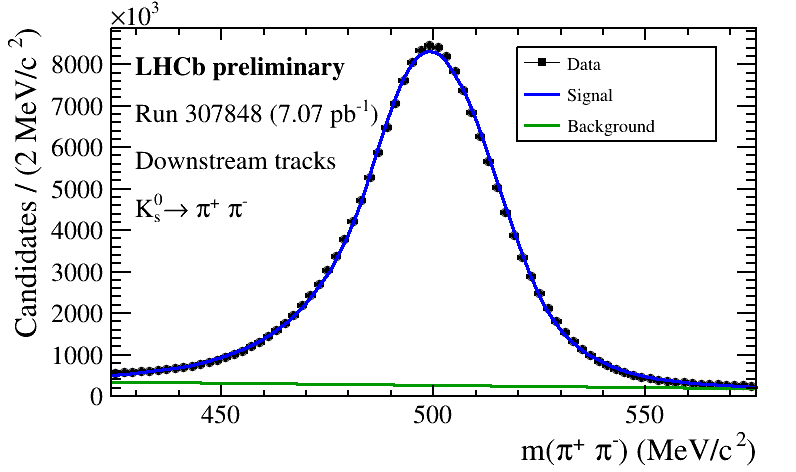}
\includegraphics[width=0.45\textwidth]{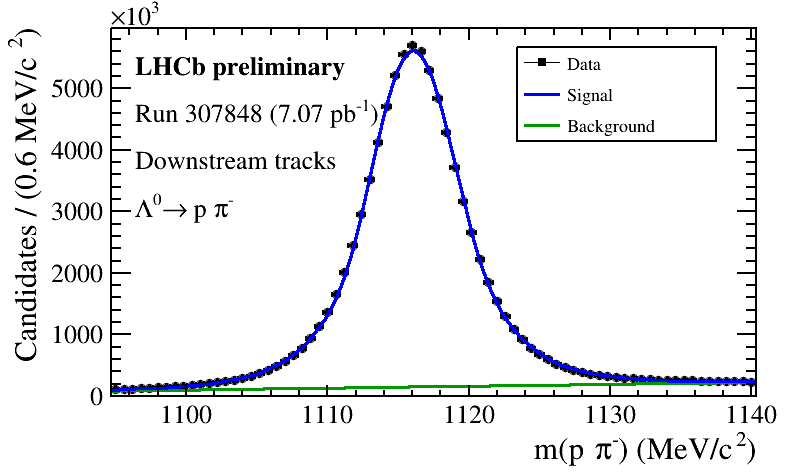}
\caption{The mass resolution of prompt $K_S^0 \to \pi^+ \pi^-$ decays (left) and $\Lambda \to p \pi^-$ decays (right) with downstream tracks~\cite{LHCB-FIGURE-2024-035}.}
\label{fig:downstream_tracking_resolution}
\end{figure}

The main contributions of downstream tracks come from $\Lambda \to p \pi^-$ and $K_S^0 \to \pi^+ \pi^-$ decays, as these particles have longer lifetimes than typical B mesons. As shown in Fig \ref{fig:downstream_tracking_efficiency}, the new HLT1 downstream tracking achieves approximately 75\% tracking efficiency for high-momentum downstream tracks from $\Lambda$ and $K_S^0$. The mass resolutions of $K_S^0$ and $\Lambda$ reconstructed from two downstream tracks in HLT1 are 15.3~MeV/$c^2$ and 3.2~MeV/$c^2$, respectively, as illustrated in Fig \ref{fig:downstream_tracking_resolution}.

\begin{figure}[t!]
    \centering
    \includegraphics[width=0.6\textwidth]{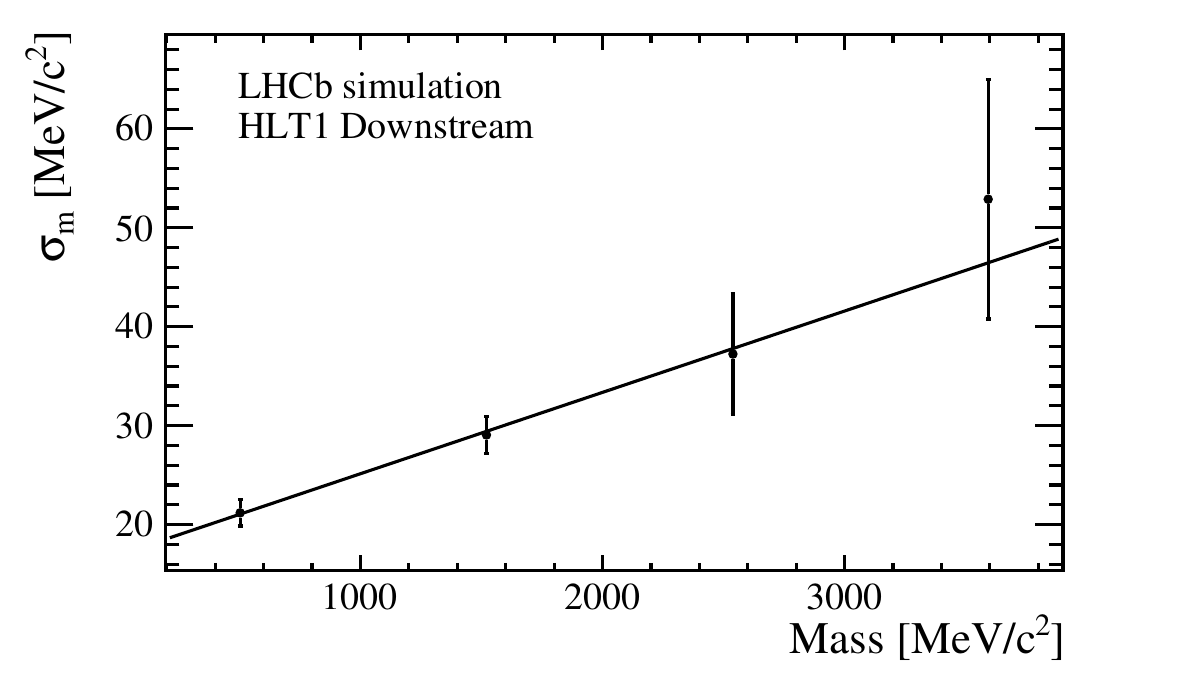}
    \caption{The mass resolution of general two-body decays of BSM LLPs reconstructed with two downstream tracks~\cite{LHCB-FIGURE-2024-018}.}
    \label{fig:downstream_llp_resolution_as_function_of_mass}
\end{figure}

Although the downstream tracking was originally designed for reconstructing $K_S^0$ and $\Lambda$ decays, it can also be used for BSM LLP decays. Based on LHCb simulations, general two-body decays of BSM LLPs show good mass resolution when reconstructed with the HLT1 downstream tracking algorithm, as shown in Fig. \ref{fig:downstream_llp_resolution_as_function_of_mass}. However, for very heavy BSM LLPs decaying into light particles (e.g., two pions), the tracking efficiency is lower compared to $K_S^0$ and $\Lambda$. This inefficiency is mainly due to the assumption that tracks originate from $(x=0, z=0)$, which becomes inaccurate for such decays.

This efficiency issue was identified at the end of 2024 and was quickly addressed in early 2025 by redesigning the tracking algorithm. The new approach removes the $(x=0, z=0)$ constraint and instead performs a general triplet search: instead of opening a search window only in the last UT layer, the algorithm scans all possible combinations of SciFi standalone tracks with UT hits within the acceptance in the $y$-direction (because the LHCb magnetic field has very small components in the $x$ and $z$ directions, the track extrapolation in the $y$-$z$ plane is nearly a straight line and independent of momentum). This new algorithm was integrated into the HLT1 in the beginning of 2025 and is expected to be used for the rest of Run 3. An HLT2 implementation is currently under development. \\

\paragraph{BuSca Project}

After the removal of the L0 trigger, all detector readouts are temporarily stored in a buffer while waiting for HLT1 processing. The BuSca project, received its name from ``Buffer Scanning for BSM LLP searches'' was developed to perform LLP searches at 30~MHz by combining both HLT1 and HLT2. BuSca consists of two main components:

\begin{figure}[t!]
\centering
\includegraphics[width=0.45\textwidth]{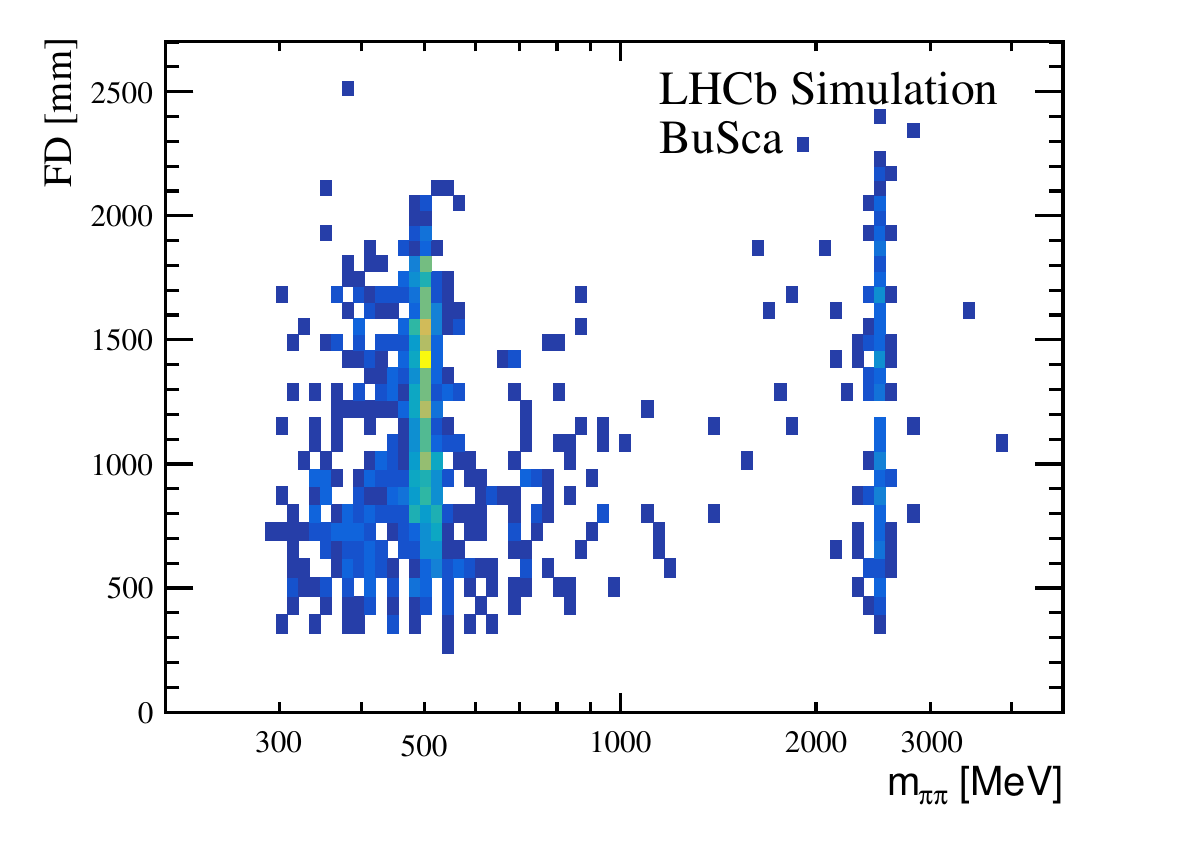}
\includegraphics[width=0.45\textwidth]{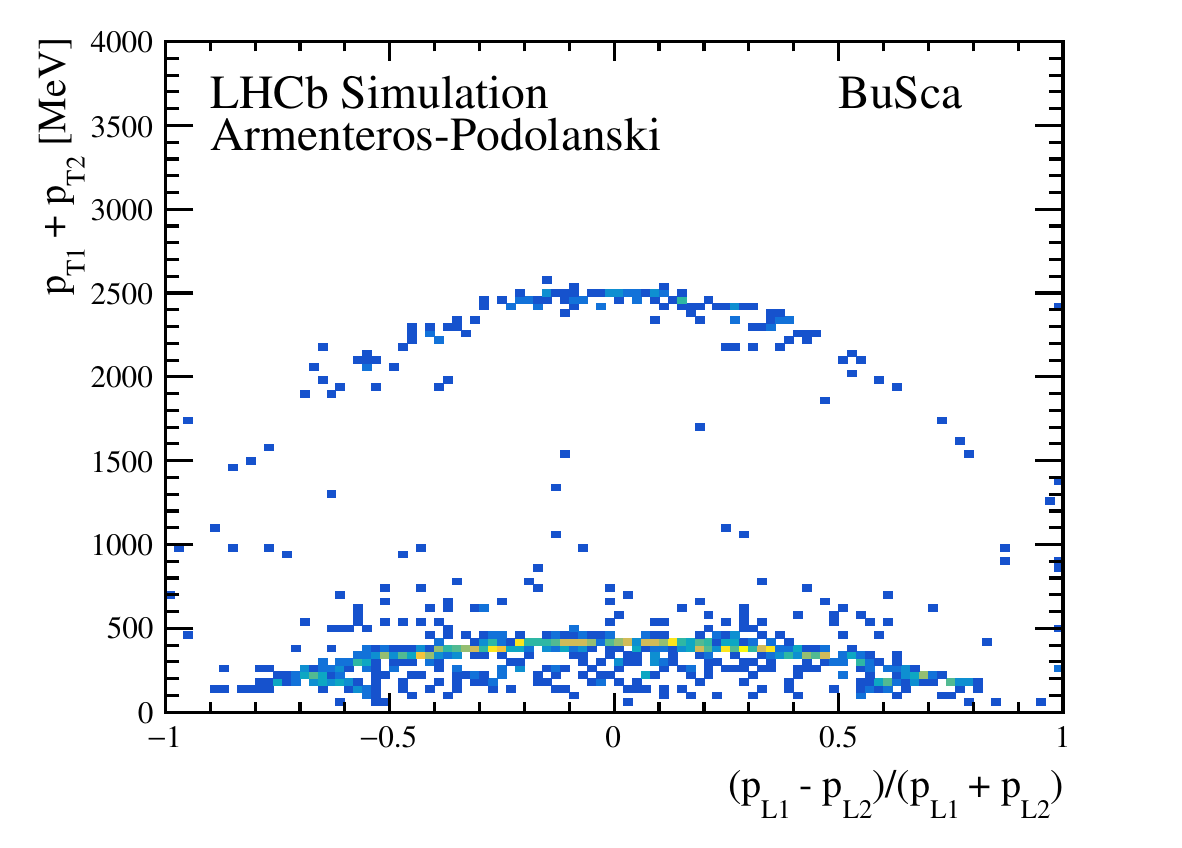}
\caption{The real-time histogram-based monitoring for mass and flight distance (left) and Armenteros-Podolanski plot (right) with downstream tracks in BuSca project~\cite{LHCB-FIGURE-2024-018}.}
\label{fig:busca_monitoring}
\end{figure}

\begin{itemize}
    \item A real-time, histogram-based monitoring system that fills trigger-less histograms at 30~MHz for fast anomaly detection and to provide hints for trigger line development, as shown in Fig. \ref{fig:busca_monitoring}.
    \item A set of HLT1+HLT2 trigger lines designed to select general, model-independent BSM displaced vertices. Currently, the system uses only downstream tracks and targets two-body decays, but it is expected to be extended to other track types and multi-body decays.
\end{itemize}

The HLT1 and HLT2 trigger lines have been integrated into the LHCb trigger system and were tested during the final two weeks of 2024 proton-proton collisions. The triggered data is now being used to study potential background contributions to the search:

\begin{figure}[t!]
    \centering
    \includegraphics[width=0.6\textwidth]{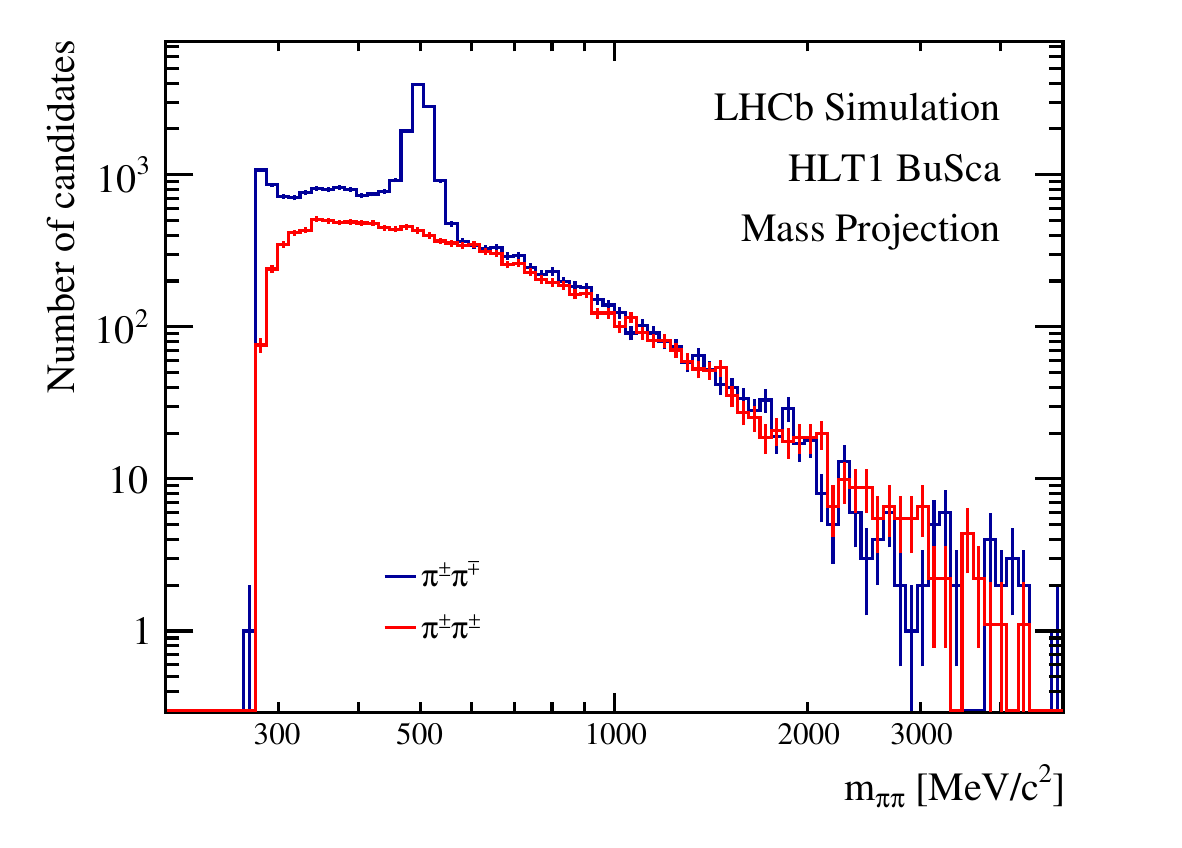}
    \caption{Two-track mass candidates, from MC simulation, assuming di-pion mass, selected by BuSca trigger lines~\cite{LHCB-FIGURE-2024-036}.}
    \label{fig:busca_same_sign}
\end{figure}

\begin{itemize}
    \item \textbf{Physics backgrounds}: Most Standard Model hadronic resonances (e.g., $J/\psi$, $\phi$, etc.) have short lifetimes and can be suppressed using flight distance cuts. The dominant contributions to this background come from $K_S^0$, $K_L^0$, and $\Lambda$ decays due to their longer lifetimes, but these decays are well understood and can be rejected using kinematic and mass window cuts.

    \item \textbf{Material interactions}: Light resonances such as $\rho$, $f_2$, etc., can be produced via interactions with detector material. These typically have short lifetimes and located in specific spatial regions like the beam pipe or tracker layers, making them suppressible using geometric vetoes.

    \item \textbf{Combinatorial backgrounds}: The dominant background arises from random combinations of two downstream tracks. As shown in Fig \ref{fig:busca_same_sign}, this background decreases with mass and can be modeled using same-sign downstream vertex candidates. We aim to suppress this background by training an MVA model on same-sign downstream vertices from real data. The background level can then be controlled by applying a tight cut on the MVA response.
\end{itemize}

The BuSca project plan to update the HLT1 and HLT2 trigger lines for 2025 data-taking, with the goal of performing an analysis with Run 3 data.

\paragraph{Retina DWT} \cite{LHCB-TDR-025}

\begin{figure}[t!]
    \centering
    \includegraphics[width=0.8\textwidth]{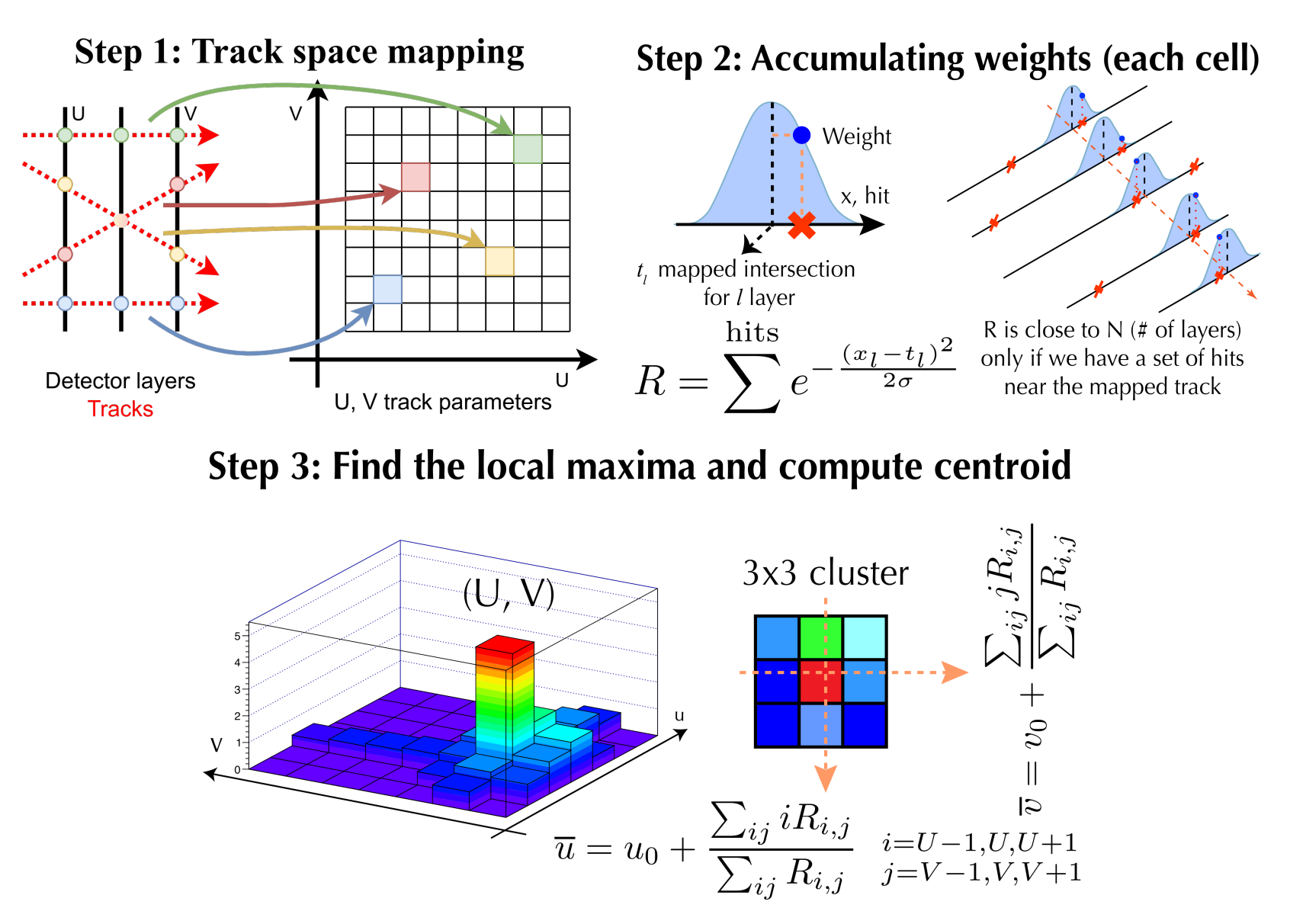}
    \caption{Track reconstruction steps with the Retina DWT~\cite{LHCB-TDR-025}.}
    \label{fig:retina_dwt_reconstruction}
\end{figure}

\begin{figure}[t!]
    \centering
    \includegraphics[width=0.8\textwidth]{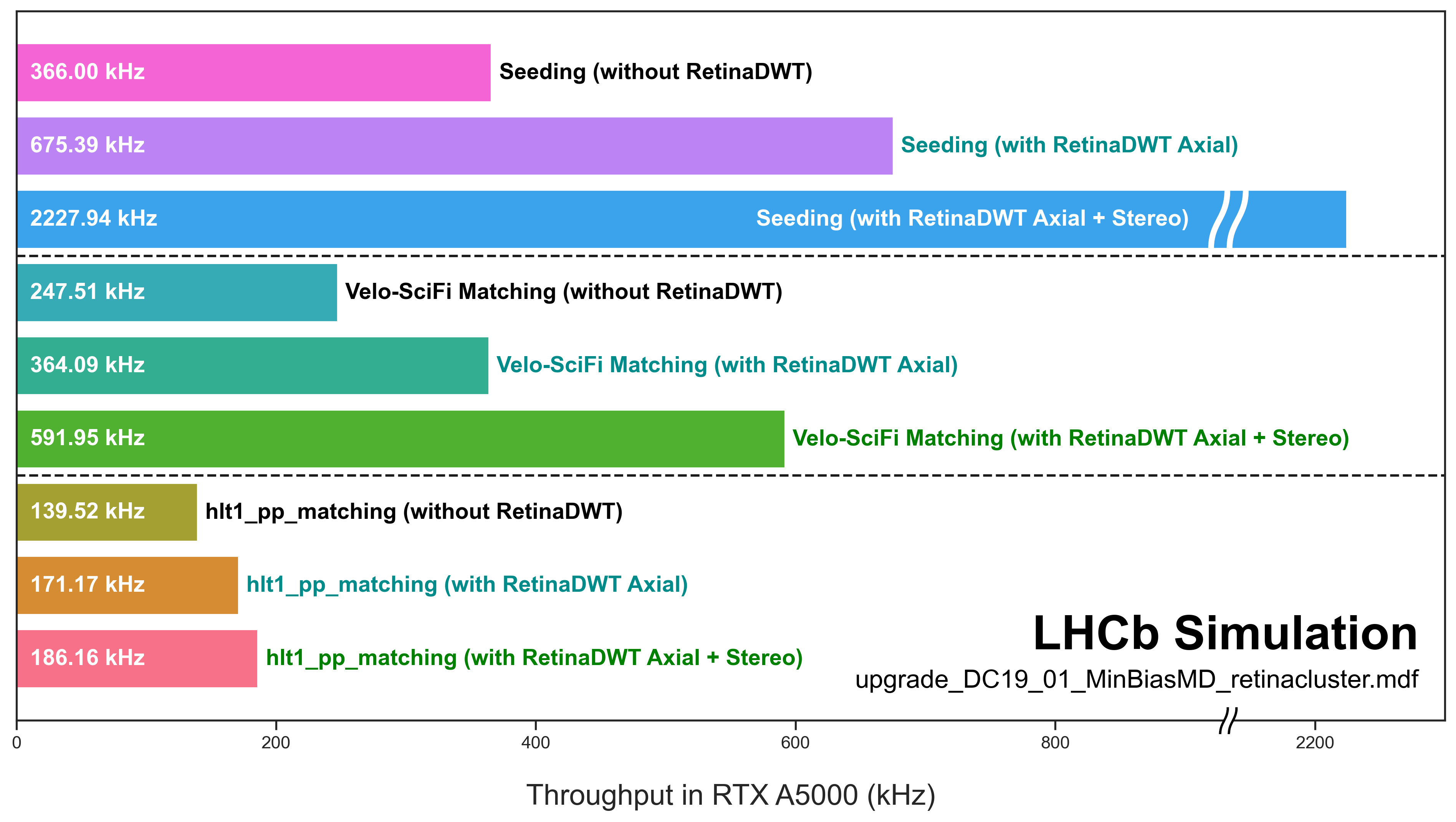}
    \caption{Throughput of various Allen reconstruction sequences, comparing the baseline fully GPU-based (black text) and the same sequences using Retina axial T-track (teal text) or Retina axial+stereo T-track (green text, white when super-imposed to the bar)~\cite{LHCB-TDR-025}.}
    \label{fig:retina_dwt_throughput}
\end{figure}

The LHCb experiment is expected to increase luminosity again in Upgrade II, the complexity of tracking algorithms, particularly triplet search algorithms, scales as \(\mathcal{O}(N^3)\). Although GPUs can reduce this to \(\mathcal{O}(N^2)\) through parallelization, it remains a significant challenge to perform track reconstruction at such high luminosities. The RetinaDWT (Retina DoWnstream Tracker) is one proposed solution to address this challenge. The idea behind RetinaDWT is to offload part of the tracking task from the GPU to FPGA. Specifically, it proposes to reconstruct SciFi standalone tracks directly on the FPGA with very low latency, immediately after detector readout, and pass them to the HLT (both HLT1 and HLT2) as additional input.

RetinaDWT is approved to be deployed for Run 4 data-taking to explore the feasibility of FPGA acceleration in the trigger system for Upgrade II. It aims to free up resources in the HLT as shown in Fig. \ref{fig:retina_dwt_throughput} (it is expected to free up 15\% of HLT1 computing time), thereby improving the quality of track reconstruction. The name ``Retina'' is inspired by the "Artificial Retina" concept, which refers to a generalized numerical application of the Hough transform. As shown in Fig \ref{fig:retina_dwt_reconstruction}, each cell in the track parameter space corresponds to a potential SciFi standalone track. Each cell computes a weighted sum of nearby hits consistent with that track hypothesis. The reconstructed tracks are identified as local maxima in the cell matrix. The FPGA is capable of process all cells in parallel, providing tracking results with high throughput and low latency. The resources it frees up in HLT1 and HLT2 are expected to be used to improve the downstream track reconstruction during Run 4. \\

\paragraph{Summary}

In this section, we reviewed the recent developments in downstream track reconstruction within the HLT system of LHCb experiment. These tracks have the potential to significantly enhance the sensitivity of the detector to BSM LLP searches. We introduced the BuSca project, designed for general displaced LLP decay searches at 30~MHz by combining both HLT1 and HLT2, targeting a analyses using Run 3 data. Additionally, we presented the RetinaDWT project, which leverages FPGA technology to help event reconstruction in the HLT by offloading part of the tracking task to the FPGA, RetinaDWT helps free up HLT resources and is expected to improve track reconstruction quality in Upgrade Ib (Run 4).

\subsubsection{LHCb Upgrade 1: results and prospects for triggers with very displaced vertices with focus on Dark scalars --- \textit{I.~Sanderswood}}
\label{sssec:sanderswood}
\textit{Author: Izaac Sanderswood, \email{izaac.sanderswood@cern.ch}}  \\
The search for FIPs decaying with very displaced vertices is motivated by scenarios in the dark sector and other physics beyond the Standard Model~\cite{Alimena:2019zri}. Recent studies have investigated the potential of the LHCb detector for triggering on and reconstructing these unique signatures. In this note, emphasis is placed on the reconstruction of decays that occur in the magnet region, which spans from approximately 2.5 m to 7.6 m downstream of the interaction point. The unique forward geometry and discrete tracking system of LHCb are exploited to extend the sensitivity to LLPs, even in regions where other detectors may lack coverage.

The LHCb detector~\cite{LHCb:2008vvz,LHCb:2023hlw} features a segmented tracking system designed to provide high-precision measurements over a large acceptance in the forward region. The tracking system is comprised of several key components:
\begin{itemize}
    \item \textbf{Vertex Locator (VELO):} Surrounding the interaction point, the VELO offers high-resolution vertex reconstruction.
    \item \textbf{Upstream Tracker (UT):} Positioned just before the magnet, the UT provides additional tracking information, especially for particles originating near the interaction point.
    \item \textbf{SciFi Stations:} Located downstream of the magnet, the SciFi tracking stations are critical for reconstructing the momentum of particles as they traverse the magnetic field.
\end{itemize}

\paragraph{The LHCb Detector and Tracking System}

The magnet region, placed between the UT and the SciFi stations, is crucial for momentum measurement. The design of the detector allows for the reconstruction of different types of tracks:
\begin{itemize}
    \item \textbf{Long tracks:} Reconstructed using hits in at least the VELO and SciFi stations. These tracks provide the highest precision but are typically limited to decays close to the interaction point. The inclusion of UT information helps to reduce the ``ghost track'' rate from mis-matched VELO and SciFi segments.
    \item \textbf{Downstream tracks:} Reconstructed from hits in the UT and SciFi stations, allowing for the analysis of decays occurring further downstream, up to about 2.5~m.
    \item \textbf{T tracks:} Reconstructed solely with SciFi hits, these tracks enable the reconstruction of LLP decays that occur far from the primary interaction point.
\end{itemize}

\begin{figure}
    \centering
    \includegraphics[width=0.6\textwidth]{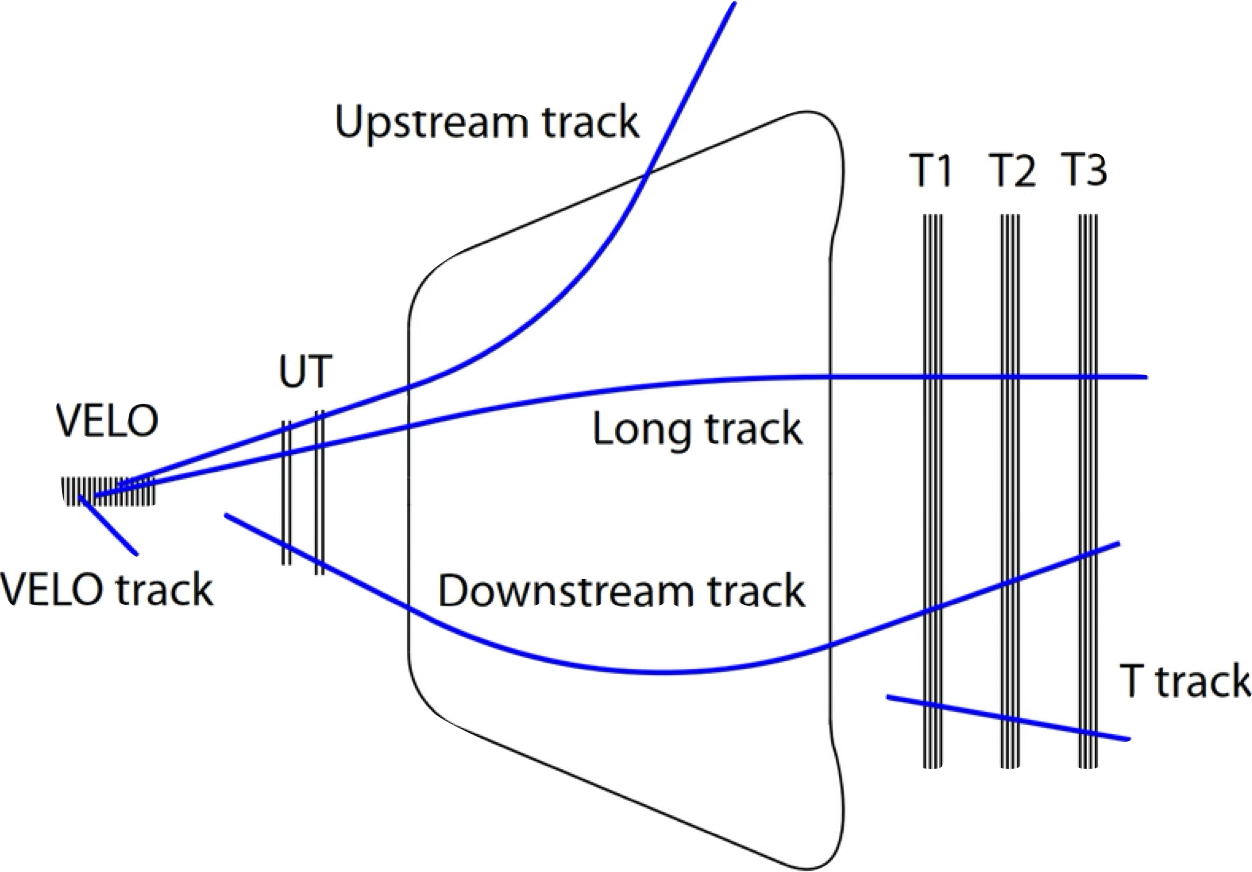}
    \caption{Schematic diagram of the LHCb tracking system, indicating the locations of the VELO, UT, and SciFi stations.}
    \label{fig:tracking}
\end{figure}

Figure~\ref{fig:tracking} provides a schematic overview of the tracking system, highlighting the placement of the VELO, UT, and SciFi stations relative to the magnet region.

\paragraph{Feasibility Studies and Reconstruction of Displaced Decays}

\begin{figure}
    \centering
    \includegraphics[width=0.8\textwidth]{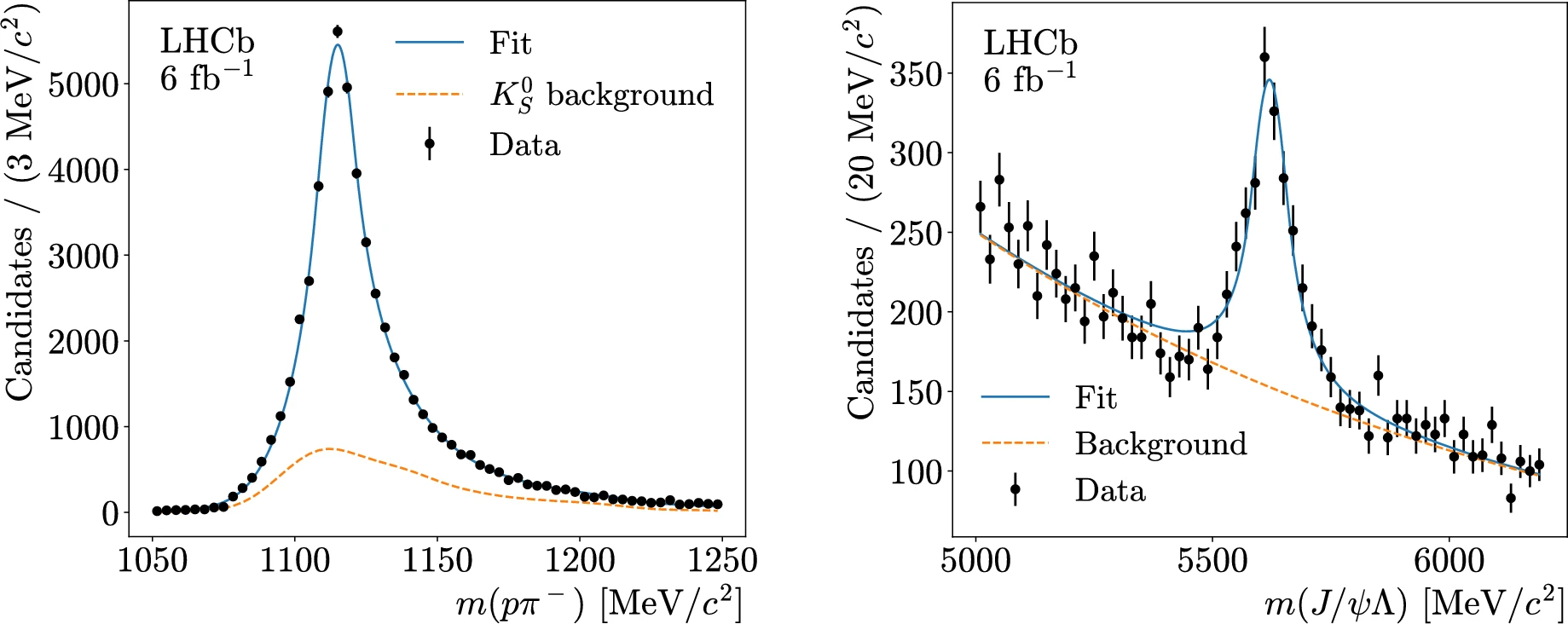}\\
    \includegraphics[width=0.8\textwidth]{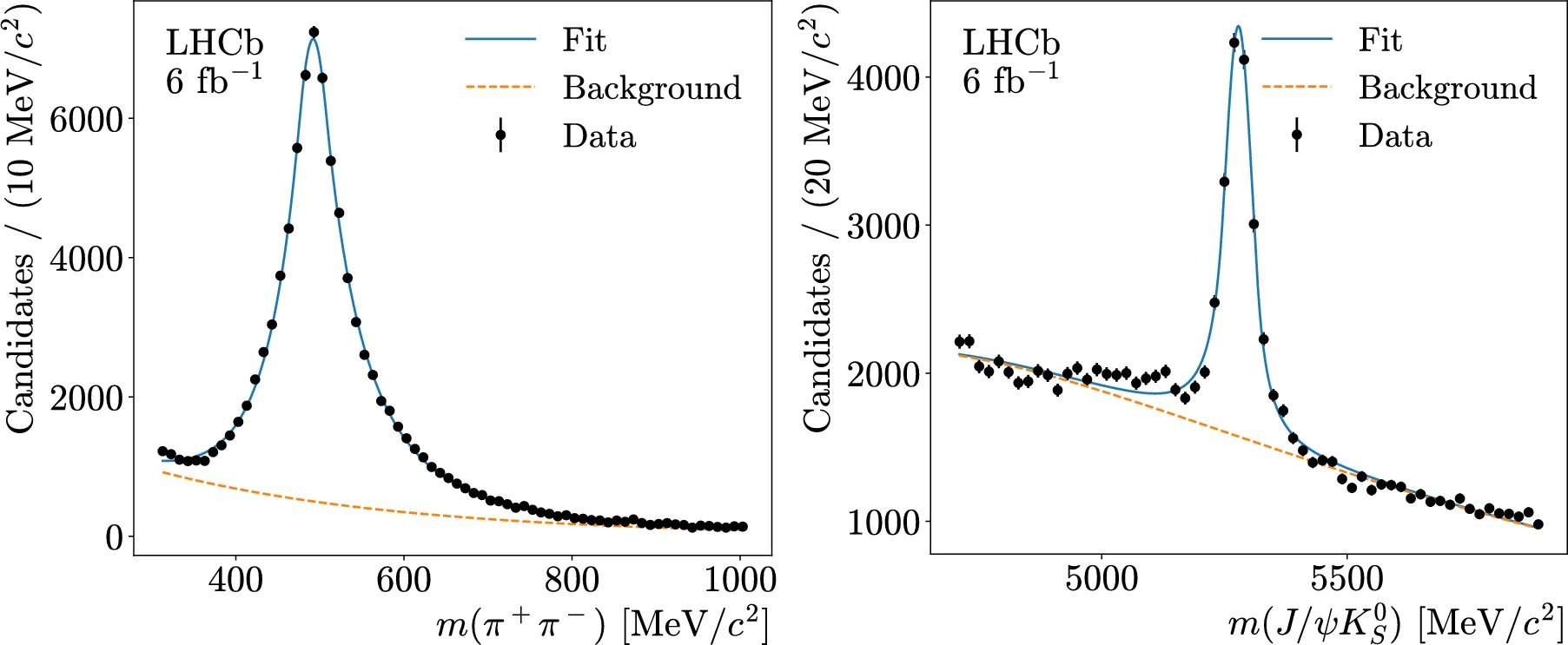}\\
    \caption{Invariant mass distributions of $\Lambda$ baryons (top) and $K_S^0$ mesons (bottom) reconstructed from decays occurring in the magnet region. As shown in Ref.~\cite{LHCb:2022kwc}.}
    \label{fig:invariant_mass}
\end{figure}

\begin{figure}
    \centering
    \includegraphics[width=0.8\textwidth]{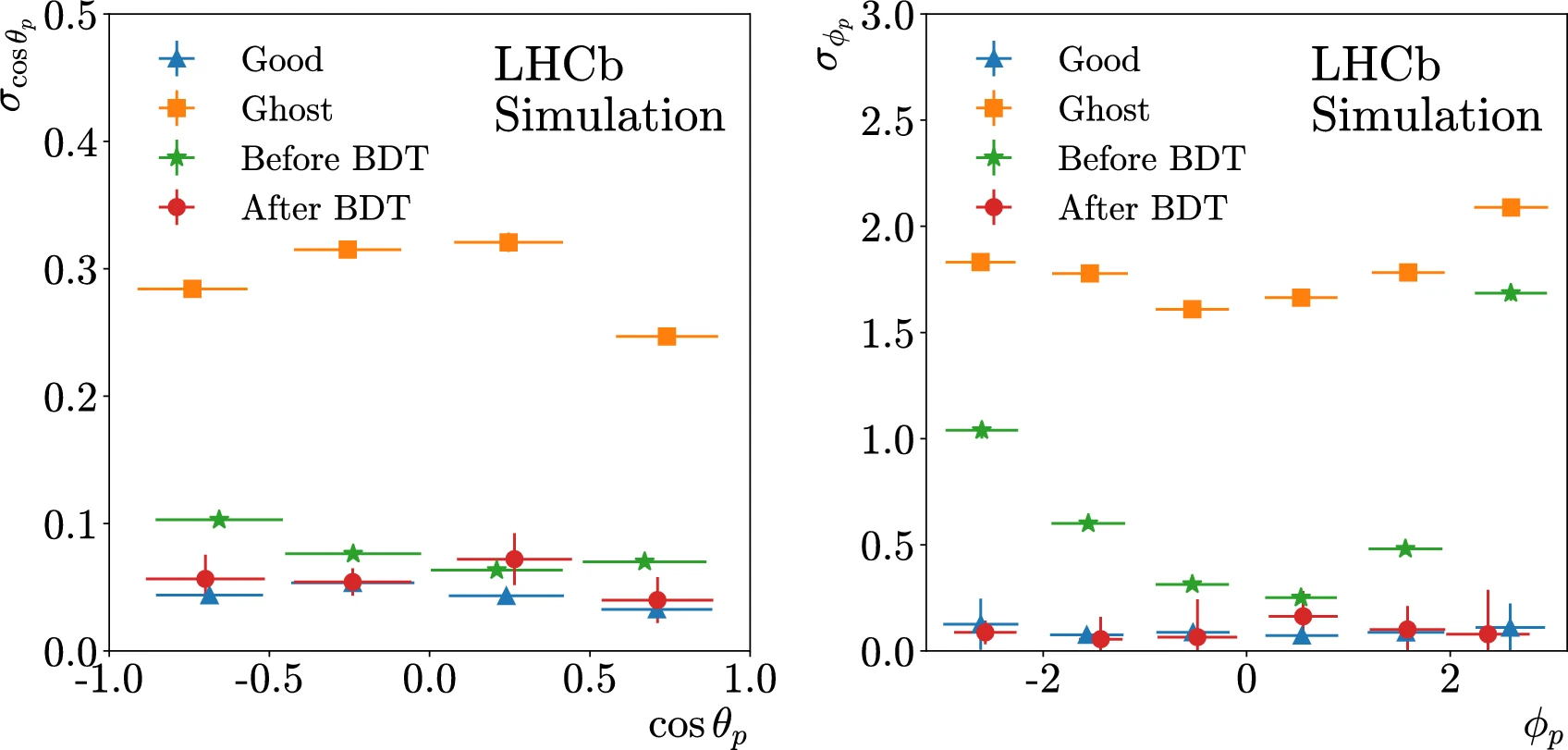}\\
    \includegraphics[width=0.8\textwidth]{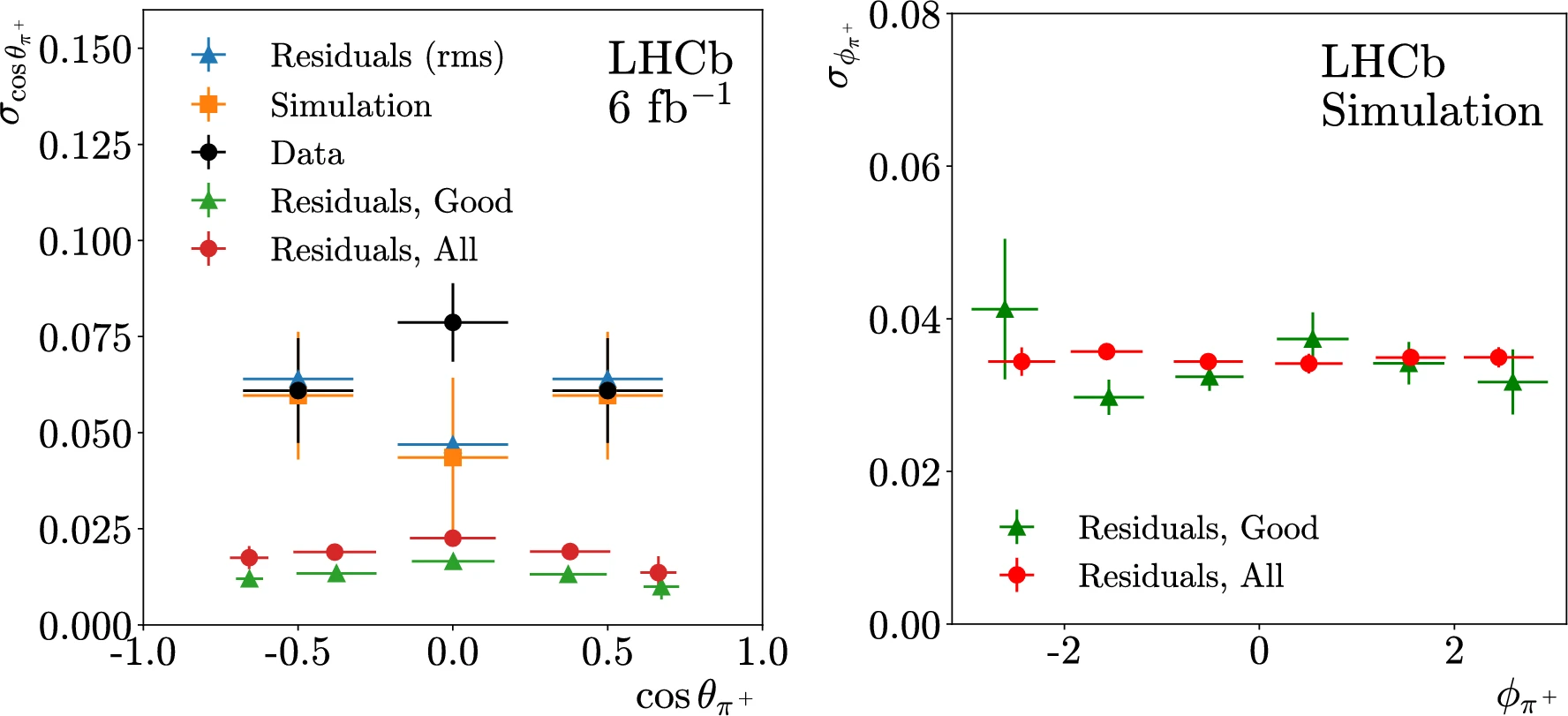}
    \caption{Helicity angle distributions of protons from $\Lambda$ baryons (top) and $\pi^+$ from $K_S^0$ mesons (bottom) reconstructed from decays occurring in the magnet region. As shown in Ref.~\cite{LHCb:2022kwc}.}
    \label{fig:helicity}
\end{figure}

Feasibility studies have been performed with the goal of reconstructing displaced decays, in order to study the electric and magnetic dipole moments of $\Lambda$ baryons~\cite{LHCb:2022kwc}. In these studies, decays occurring between 6~m and 7.6~m from the interaction point were successfully reconstructed. Figure~\ref{fig:invariant_mass} shows the invariant mass distributions obtained for $\Lambda$ baryons and $K_S^0$ mesons, with clearly visible peaks indicating the viability of the reconstruction strategy. Figure~\ref{fig:helicity} shows the estimated helicity angle resolutions.

These studies demonstrate that if $\Lambda$ and $K_S^0$ can be reconstructed and their decay properties studied at such large distances from the interaction point, then BSM LLPs with very high lifetimes ($\tau\sim\mathcal{O}(10$~ns)) become plausible targets. The reconstruction methods developed have been implemented in the second level of the LHCb trigger system (HLT2), and now cover a number of both SM and BSM channels.

\paragraph{$ \boldsymbol{b\to s H'(\to\mu\mu)}$ efficiency}

The expected efficiencies for $ b\to s H'(\to\mu\mu)$ decays have been evaluated using simulated $B^0\to K^*(\to \pi K) H'(\mu\mu)$ decays.

\begin{figure}
    \centering
    \includegraphics[width=0.6\textwidth]{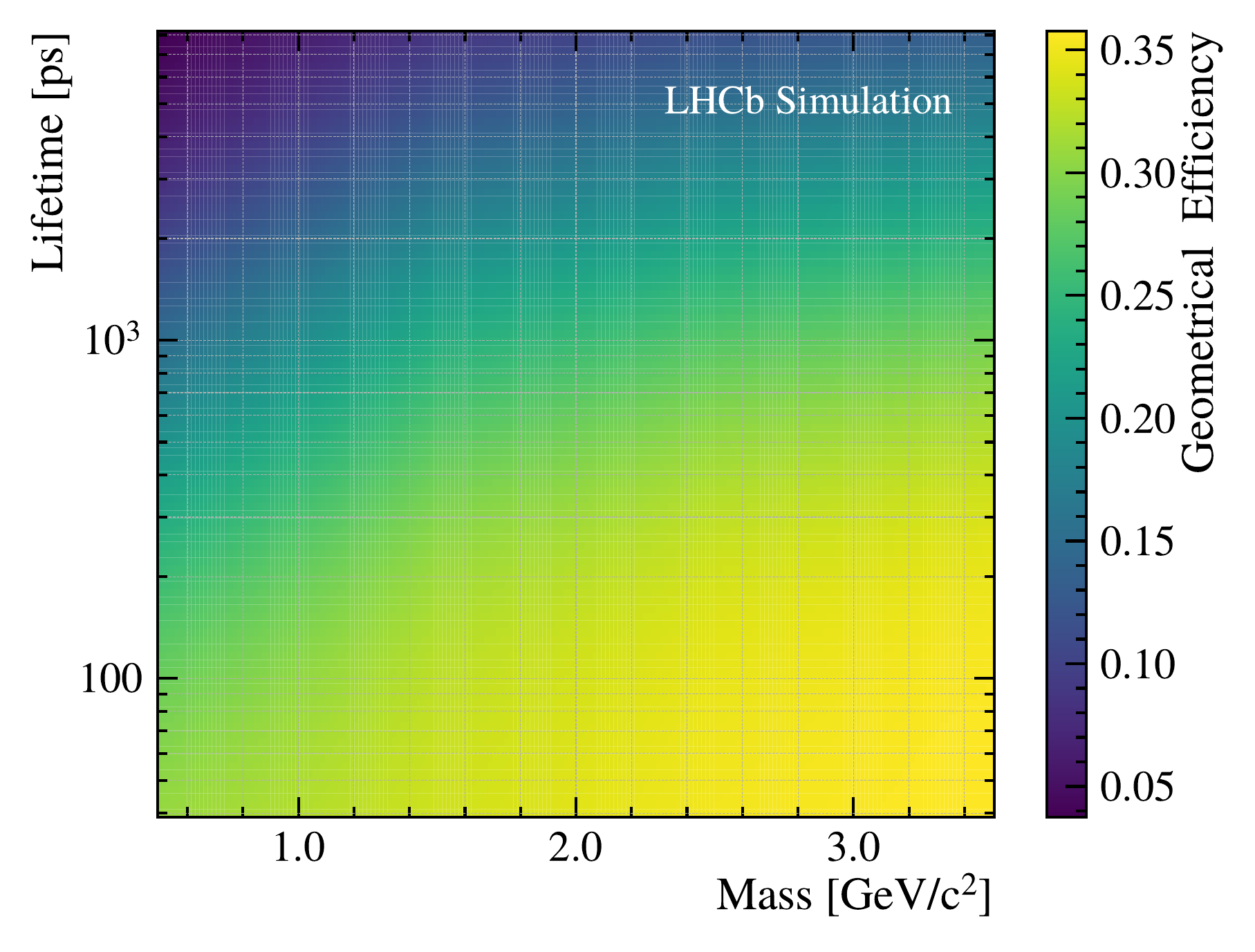}
    \caption{Geometrical acceptance efficiency as a function of LLP lifetime and mass.}
    \label{fig:acceptance_efficiency}
\end{figure}

The geometrical acceptance efficiency of the detector, which requires that both decay products have polar angles $\theta < 500$~mrad and that the reconstructed vertex lies within 8~m of the nominal proton-proton interaction point, is shown in Figure~\ref{fig:acceptance_efficiency}.

The trigger efficiency is influenced by different factors in HLT1 and HLT2:
\begin{itemize}
    \item \textbf{HLT1:} the HLT1 efficiency depends on prompt track signatures, including one-track and two-track topologies. Therefore, in these studies, it is independent of the LLP and depends on the decay of the associated kaon. An estimated average efficiency of approximately 40\% is observed for the studied channel.
    \item \textbf{HLT2:} This trigger level employs kinematic and topological criteria aimed at selecting the displaced LLP vertices in the magnet region. Its design is inclusive, targeting any dimuon signature in the magnet region, with performance ranging between 15\% and 35\%, depending.
\end{itemize}

\begin{figure}
    \centering
    \includegraphics[width=0.6\textwidth]{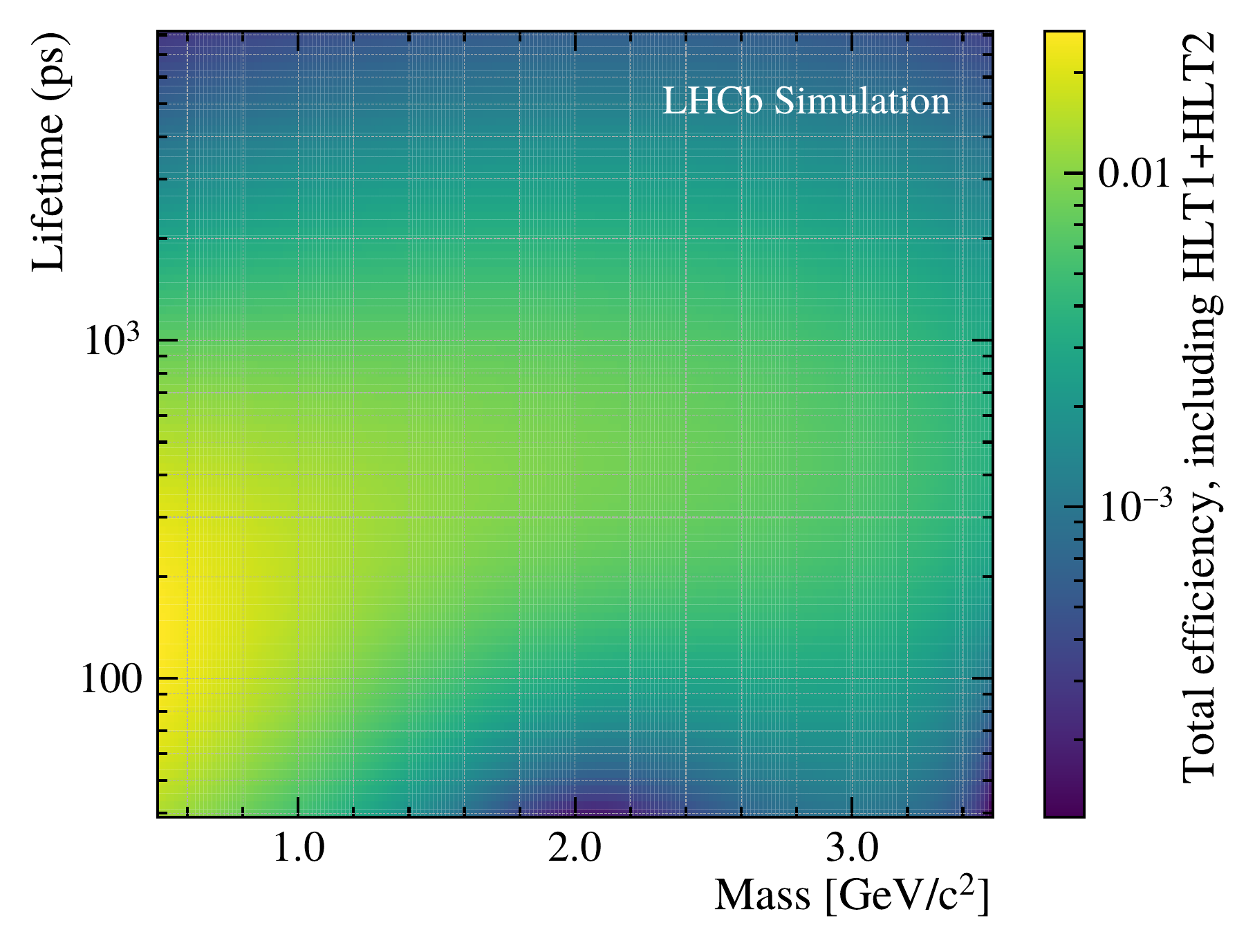}
    \caption{The total trigger efficiency, including acceptance, as a function of LLP lifetime and mass.}
    \label{fig:trigger_efficiency}
\end{figure}

Figure~\ref{fig:trigger_efficiency} illustrates the dependence of the trigger efficiency on the LLP lifetime and mass.

\paragraph{Offline Selection}

Challenges related to mass resolution have been identified, especially as the LLP mass increases. The increase in the $Q$--value of the decay, combined with the reduced magnetic field across the SciFi stations, leads to a degradation in momentum resolution of the daughter tracks. Offline refitting techniques using a primary vertex constraint help to mitigate this issue and improve the mass resolution significantly. However, it is worth noting that the background distribution is expected to be exponentially decreasing, so whilst the width of the search window increases as a function of mass, the expected number of background events decreases as a function of mass.

Robust background rejection is critical when searching for rare LLP decays. Offline selections have been developed based on a boosted decision tree (BDT) classifier, which combines multiple kinematic and topological variables. The BDT was trained using both data and simulated signal events, and a threshold on the BDT score was chosen to reduce the background level to approximately one to two events per pb$^{-1}$.

Figure~\ref{fig:bdt_performance} shows the performance comparison between the BDT output for data and Monte Carlo simulations. Although the BDT has not yet been fully optimised, initial studies indicate that further improvements in signal efficiency can be achieved without compromising the low background levels.

\begin{figure}
    \centering
    \includegraphics[width=0.6\textwidth]{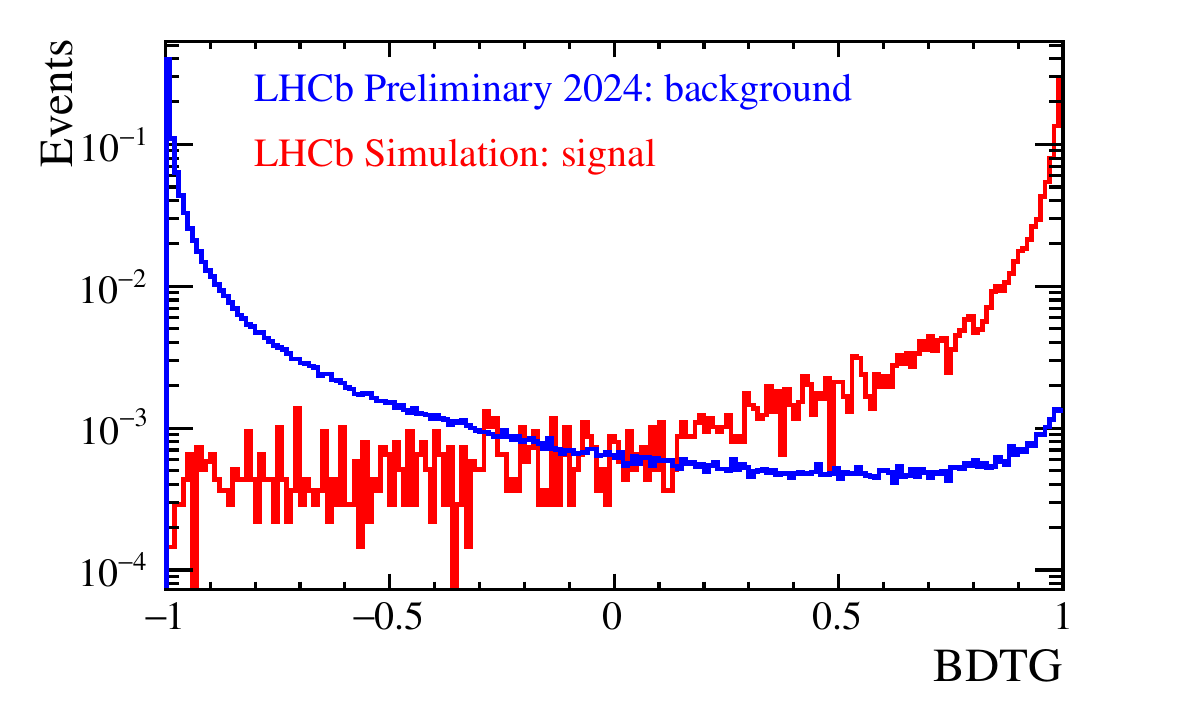}
    \caption{The BDT performance for data and simulated signal events.}
    \label{fig:bdt_performance}
\end{figure}

\begin{figure}
    \centering
    \includegraphics[width=0.49\textwidth]{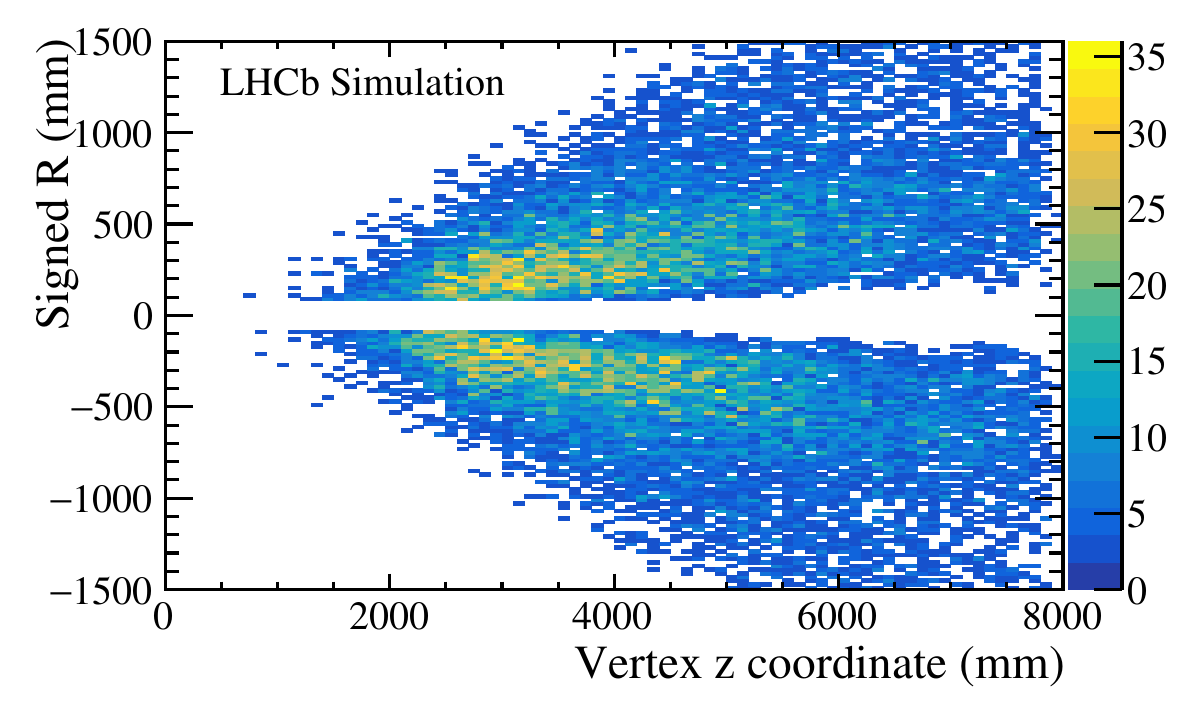}
    \includegraphics[width=0.49\textwidth]{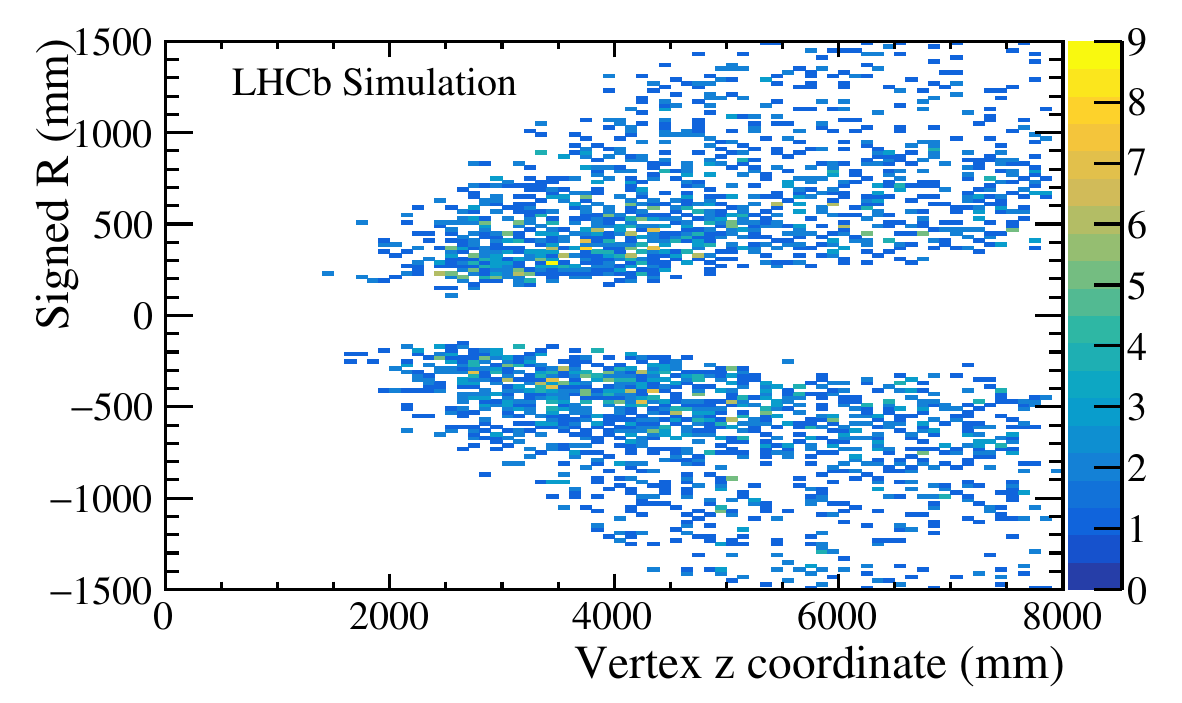}
    \caption{The BDT performance for data and simulated signal events.}
    \label{fig:bdt_performance}
\end{figure}

\paragraph{Preliminary sensitivity estimates}

A preliminary sensitivity estimate has been derived using the current detector and trigger performance, though without offline selection criteria. This estimate, while still subject to further refinement, indicates that the reconstruction of decays in the magnet region enables competitive sensitivity to dedicated LLP experiments, particularly in the region of lower couplings.

\begin{figure}
    \centering
    \includegraphics[width=0.8\textwidth]{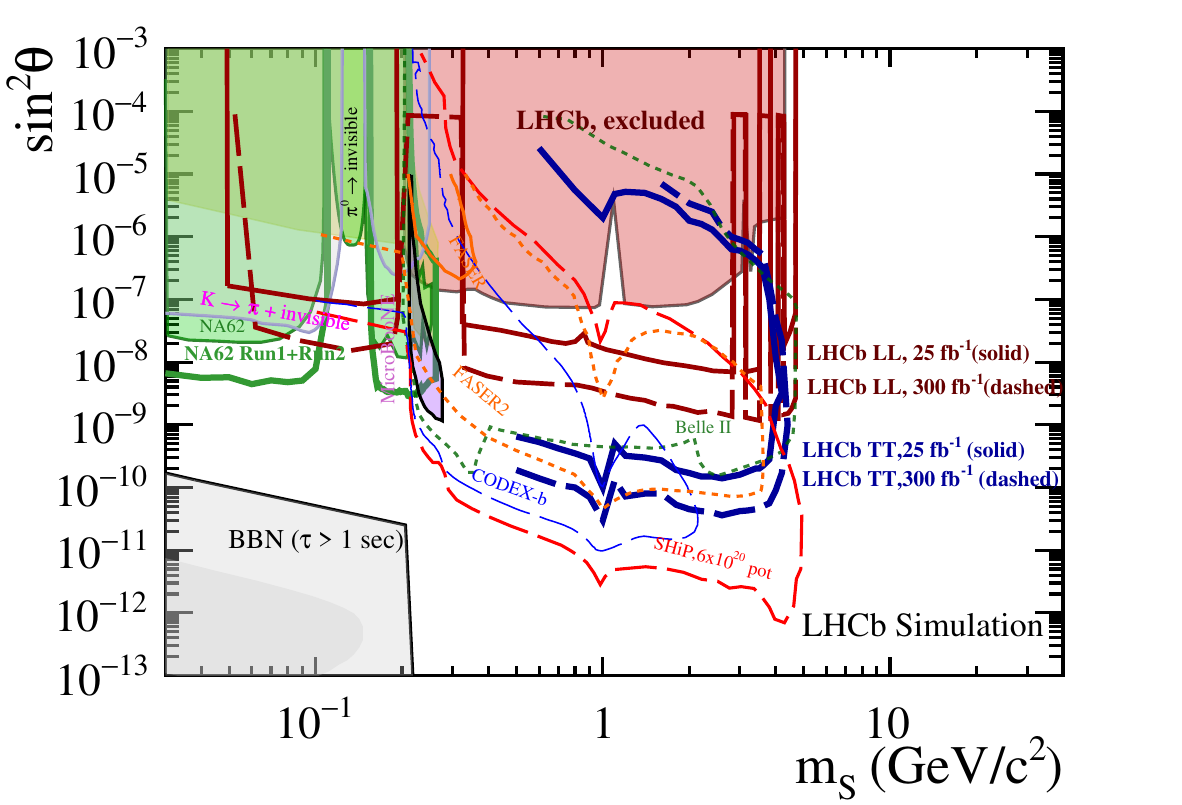}
    \caption{Comparing the sensitivity of the current LHCb approach to other experiments.}
    \label{fig:sensitivity}
\end{figure}

Figure~\ref{fig:sensitivity} illustrates the reach of the current approach compared to dedicated LLP experiments. Continued optimisation of the trigger and offline selections is anticipated to further enhance the sensitivity of the LHCb detector to new physics phenomena.

\paragraph{Future Prospects} \label{sec:prospects}

Several avenues for future improvements have been identified:
\begin{itemize}
    \item \textbf{HLT2 Optimisation:} Enhanced topological selections and improved reconstruction algorithms are being developed to increase the efficiency of the HLT2 trigger whilst maintaining good background rejection.
    \item \textbf{Calorimeter Integration:} The inclusion of calorimeter information for T tracks is expected to improve the particle identification, especially helpful to distinguish muons from pion and kaon decays.
    \item \textbf{Expanded Decay Modes:} The current strategy is being extended to include additional decay modes and incorporate advanced hadron identification techniques.
\end{itemize}

\paragraph{Conclusion}

A study of trigger and reconstruction strategies for LLPs with very displaced vertices has been presented. The unique forward geometry and discrete tracking system of the LHCb detector facilitate the reconstruction of decays in the magnet region, thereby extending sensitivity to long-lived particles decaying far from the interaction point. Feasibility studies based on $\Lambda$ and $K_S^0$ reconstructions have validated the approach, and preliminary sensitivity estimates indicate that the LHCb detector is competitive with dedicated LLP experiments in exploring new physics scenarios.

Future work will focus on further optimisation of the HLT2 trigger, refinement of offline selection algorithms, and expansion to additional decay modes. The integration of additional detector information, such as calorimeter and advanced hadron identification, is expected to further improve the sensitivity and overall performance.

\subsubsection{LHCb Upgrade 1: results and prospects for triggers with very displaced vertices with focus on HNLs --- \textit{S.~Collaviti}}
\label{sssec:collaviti}
\textit{Author: Spencer Collaviti, \email{spencer.collaviti@cern.ch}}  \\
\textit{Author: Andrea Merli, \email{andrea.merli@cern.ch}}  \\
\textit{Author: Lesya Shchutska, \email{lesya.shchutska@cern.ch}}
\paragraph{Introduction}

The Standard Model of particle physics (SM) \cite{Goldstone1961,Goldstone1962,Higgs1964a,Higgs1964b,Higgs1966,Weinberg1967,Weinberg1973,Fritzsch1973,Gross1973a,Gross1973b} encapsulates the best human understanding of all fundamental particles as well as three of the four known fundamental forces. However, despite this success, several deficiencies in the SM necessitate the existence of new physics Beyond the Standard Model (BSM physics). Some of the more promising BSM models are Heavy Neutral Leptons (HNL) models, with the capacity to concurrently explain the nature of dark matter, the matter-antimatter asymmetry produced in the early universe (baryogenesis) and a non-zero neutrino mass (required to explain neutrino flavour oscillations)~\cite{Asaka:2005pn,Asaka:2005an}.

In this subsection we present an analysis strategy and sensitivity study for an ongoing LHCb analysis which will use Run 3 (2022-2026) data from the LHCb Upgrade 1 detector to search for HNLs. We will begin by collecting the experimentally relevant phenomenolgy of HNLs. We will then discuss the challenges of this phenomenology in an LHCb search and how our novel analysis strategy with very displaced vertex reconstruction allows us to navigate around these challenges. Finally, we will present our methodology and results for forecasting the sensitivity of our analysis. This will demonstrate that, for our sensitive mass range, our analysis method and the LHCb detector has potentially world-leading sensitivity.

\paragraph{Experimentally Relevant Phenomenology}

In HNL models, one or more HNLs (typically three, as described in the neutrino-minimal Standard Model; $\nu$MSM \cite{Asaka:2005pn,Asaka:2005an}) are added to the SM as the right-handed counterpart to the SM neutrino. Provided that the masses of any HNLs are not degenerate or approximately degenerate \cite{Anamiati2018,Antusch:2017ebe}, each HNL couples to the SM analogously to SM neutrinos. No mixing between HNL flavours in production and decay occurs and four parameters, a mass and three mixing angles (between HNLs and SM neutrinos) uniquely specify the properties of each HNL. Mathematically, the coupling is as follows~\cite{Bondarenko:2018ptm}, with $U_\alpha$ the mixing angle between a HNL and a SM neutrino of flavour $\alpha$:

\begin{equation}
    \mathcal{L}_{\rm int} = \frac{g}{2\sqrt{2}}W_\mu^+ \overline{N^c}\sum_\alpha U_\alpha^* \gamma^\mu (1-\gamma_5) l^-_\alpha + \frac{g}{4 \cos \theta_W} Z_\mu \overline{N^c}\sum_\alpha U_\alpha^* \gamma^\mu (1-\gamma_5) \nu_\alpha + {\rm h.c.}
\end{equation}

All except the lightest HNL mass eigenstates (at least in the $\nu$MSM) naturally appear between the GeV and TeV energy scales. This, coupled with the current non-observation of HNLs (and thus necessarily small $U_\alpha$) makes HNLs feebly interacting particles (FIPs) interesting for collider searches.

\begin{figure}[h]
    \centering
    \includegraphics[width=0.52\linewidth]{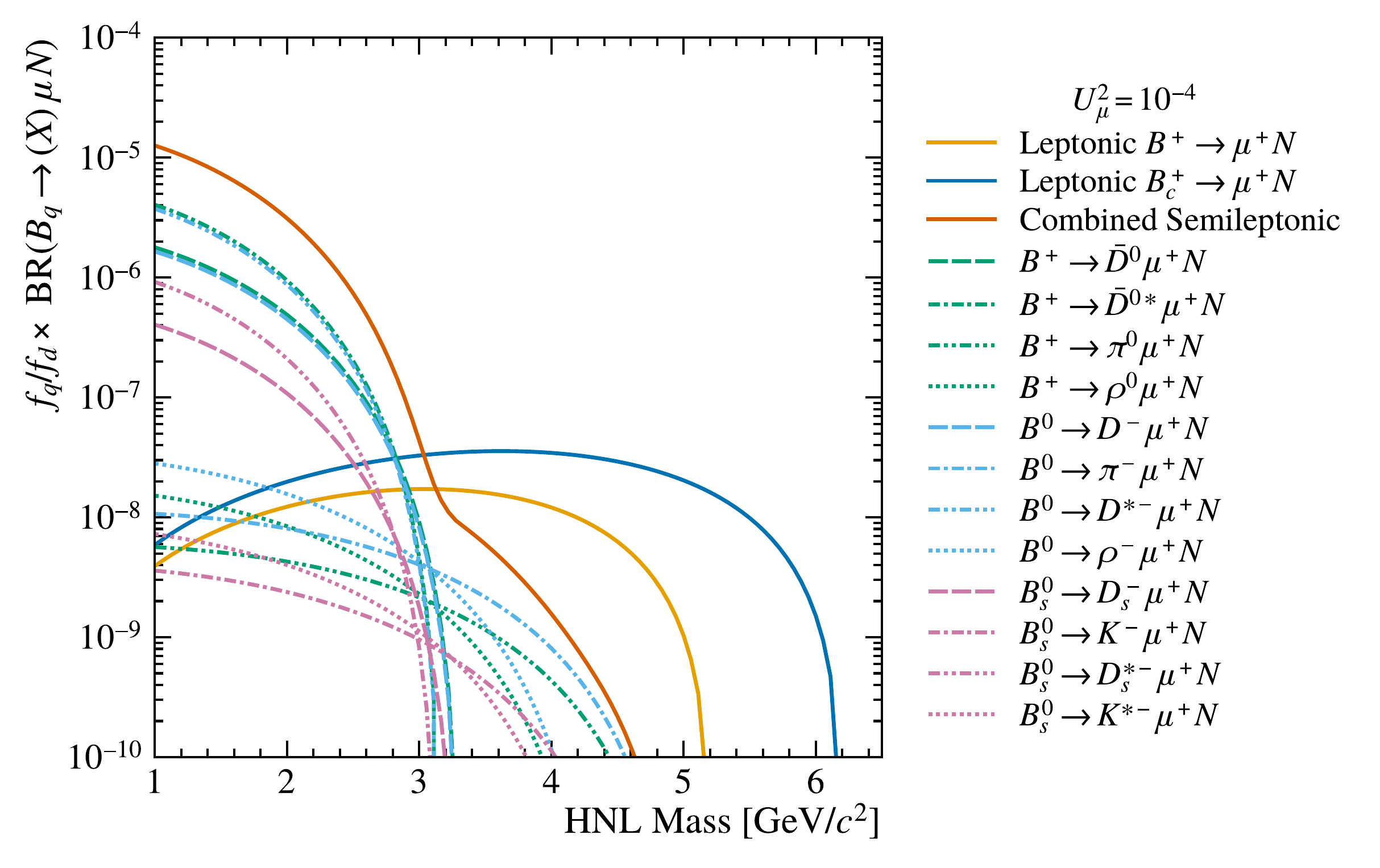}
    \includegraphics[width=0.42\linewidth]{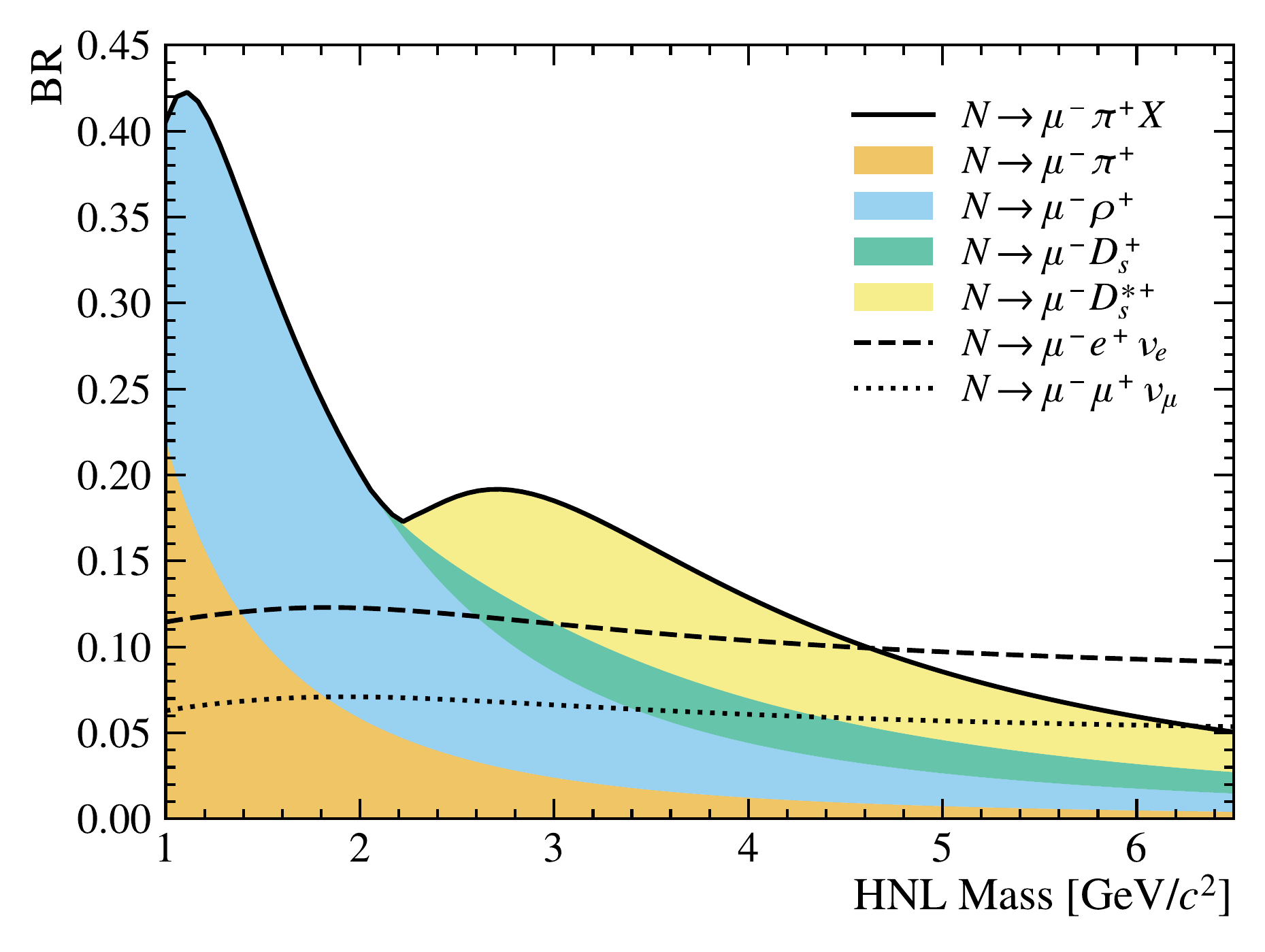}
    \caption{Left: Branching ratios to HNLs from B mesons. Right: Branching ratios in the decay of HNLs. B mesons of quark content $\bar{b}q$ are normalised by the production fraction $f_q$ relative to $f_d$, i.e. the number of $\bar{b}q$ mesons relative to $B^0$. The `Combined Semileptonic' and $N \rightarrow\mu^- \pi^+ X$ categories are combinations of the broken (e.g., dashed) lines, and the shaded regions, respectively~\cite{Bondarenko:2018ptm}.}
    \label{fig:LHCb HNL BRs}
\end{figure}

From the LHCb perspective, we are particularly interested in HNLs with a mass between the D- and B-meson production thresholds (i.e between approximately 1.6 and 6.0 GeV/$c^2$). In this mass range, HNLs are predominantly produced through the decay of B mesons. This gives LHCb with its specialised acceptance for $b$-phyiscs, the potential to be particularly competitive. The dominant production modes and observationally interesting decay modes of HNLs are shown in the left and right of Figure~\ref{fig:LHCb HNL BRs}, respectively~\cite{Bondarenko:2018ptm}. Note that here we assume strictly muonic coupling, i.e. $U_e=U_\tau=0$.

\paragraph{Challenges and Analysis Strategy}

The challenges inherent in an LHCb HNL analyses are twofold and are associated with the nature of HNLs as FIPs. The first challenge is that HNLs (if they exist in the 1.6 to 6.0 GeV/$c^2$ mass range), must be reasonably long lived, i.e. have a lifetime on the order of 1 ns or longer. With this long lifetime we would expect HNLs produced at LHCb to have a mean flight distance larger than the 500 mm covered by our first tracking detector, the Vertex Locator (VELO). The second challenge is that HNLs must have a reasonably small branching fraction from B mesons. This limits how many HNL events we would expect, especially at a lower instantaneous luminosity than either CMS or ATLAS. To get around these challenges, we adopt a novel analysis strategy for our Run 3 analysis where we, one, reconstruct HNLs as very displaced vertices and, two, consider an inclusive selection for HNL production and decay. We expound upon these two points in the following paragraphs.

In the context of LHCb, reconstructing HNLs as very displaced vertices means constructing HNL vertices out of tracks which miss our first tracking detector, VELO. This means tracks with displacement greater than 500 mm from the proton-proton (pp) collision point. These very displaced tracks are termed downstream tracks if only VELO is missed (displacement between 500 and 2,500 mm) and T tracks if both VELO and the Upstream Tracker (UT) is missed (displacement between 2,500 mm and 8,000 mm). The utilisation of these tracks allows a decay length approximately 16-times longer than just using long tracks. Equivalently, this gives us sensitivity to particle lifetimes which are 16-times larger.

By considering an inclusive selection, we mean to consider all of the production and decay modes shown in Figure~\ref{fig:LHCb HNL BRs}. This includes leptonic ($B_{(c)}^+ \rightarrow\mu^+ N$) and semileptonic ($B_q\rightarrow \mu^+ X N$) production and HNL decay with ($N \rightarrow \mu \mu \nu$, $N\rightarrow\mu e \nu$, $N\rightarrow \mu \pi X$) and without ($N\rightarrow\mu \pi$) an unreconstructed particle. To trigger on these events, we consider two approaches. When  all particles are reconstructed (i.e. $B_{(c)}^+ \rightarrow\mu^+ (N \rightarrow\mu^\pm \pi^\mp)$), we trigger on a $\mu\mu\pi$ final state. When some particles are unreconstructed (i.e. everything else), we aim to trigger on just a very displaced two-prong vertex with appropriate PID (e.g. $e \mu$) and save the whole event. We then have the option of either directly using the corrected mass of the HNL as our search variable (increasing our fiducial volume as no long lepton is required) or combining our vertex with other missing charged particles. By considering this inclusive selection, we find up to a 40-fold increase in the number of expected events as opposed to considering just a $N\rightarrow \mu \pi$ final state.

For the moment we restrict ourselves to consider only HNLs with a purely muonic coupling, i.e. $U_e=U_\tau=0$. However we note that our analysis strategy is equally applicable to an HNL of arbitrary SM coupling (i.e. arbitrary mixing angles $U_\alpha$).

\paragraph{Methodology and Results of Sensitivity Study}

As mentioned in previous paragraphs, in our analysis we plan to consider an inclusive selection of HNLs reconstructed as downstream and T tracks. However, at the time of our sensitivity study, T track selections were not complete in the LHCb software configuration for all decay modes (e.g. for $N \rightarrow \mu e \nu$). Further, downstream reconstruction of heavy (i.e. BSM) particles was under development. This precluded the simple evaluation of our HNL sensitivity using an inclusive Monte Carlo (MC) simulation and the combination of all HNL selections. As a result, we adopted an alternative methodology.

To estimate the T track sensitivity we begin by computing the predicted T track online selection efficiency of leptonic $B_c$ and combined semileptonic production with the $N\rightarrow\mu \pi$ decay (for which we had functional selections). To do so, we generated MC for each of these production methods and used this to compute the acceptance efficiency of our HNL events. We then ran our accepted MC through the two online High Level Trigger selection phases (HLT1 and HLT2). To extrapolate to different lifetimes than the generated MC, we used a lifetime reweighting approach. The result was then extrapolated to the alternative production mode, leptonic $B_u$, and to the alternative decay modes (for which we did not have functional selections) using the assumptions detailed later.

As the aforementioned work on the HLT2 downstream reconstruction in Run 3 affects all HNL modes, we cannot compute online selection efficiencies with downstream tracks in the same manner. Instead we make some `sensible' assumptions (detailed later) for the downstream efficiency based on Run 3's HLT1 (already mature) and the performance of the LHCb detector in Run 2.

To move from our T track and downstream online selection efficiencies to our projected sensitivity, the T-track and downstream online trigger efficiencies are multiplied by the number of produced HNL events determined using the branching ratios from Ref~\cite{Bondarenko:2018ptm} and assuming an integrated luminosity of 10 ${\rm fb}^{-1}$. We do not at present consider offline selection or the effect of background.

More details on how we generate our efficiencies and sensitivities are given below, alongside the relevant results. This is split into three distinct sections: 1) evaluating T track sensitivity for $N\rightarrow \mu \pi$, 2) extrapolating T-track sensitivity to alternative decay modes and 3) predicting downstream sensitivity

\textit{Evaluating T track sensitivity for $N\rightarrow \mu \pi$:} We trigger, reconstruct and select events with very displaced vertices made of two T tracks where the particles are identified as $\mu \pi$.

\begin{figure}[h]
     \centering
     \includegraphics[width=0.47\linewidth]{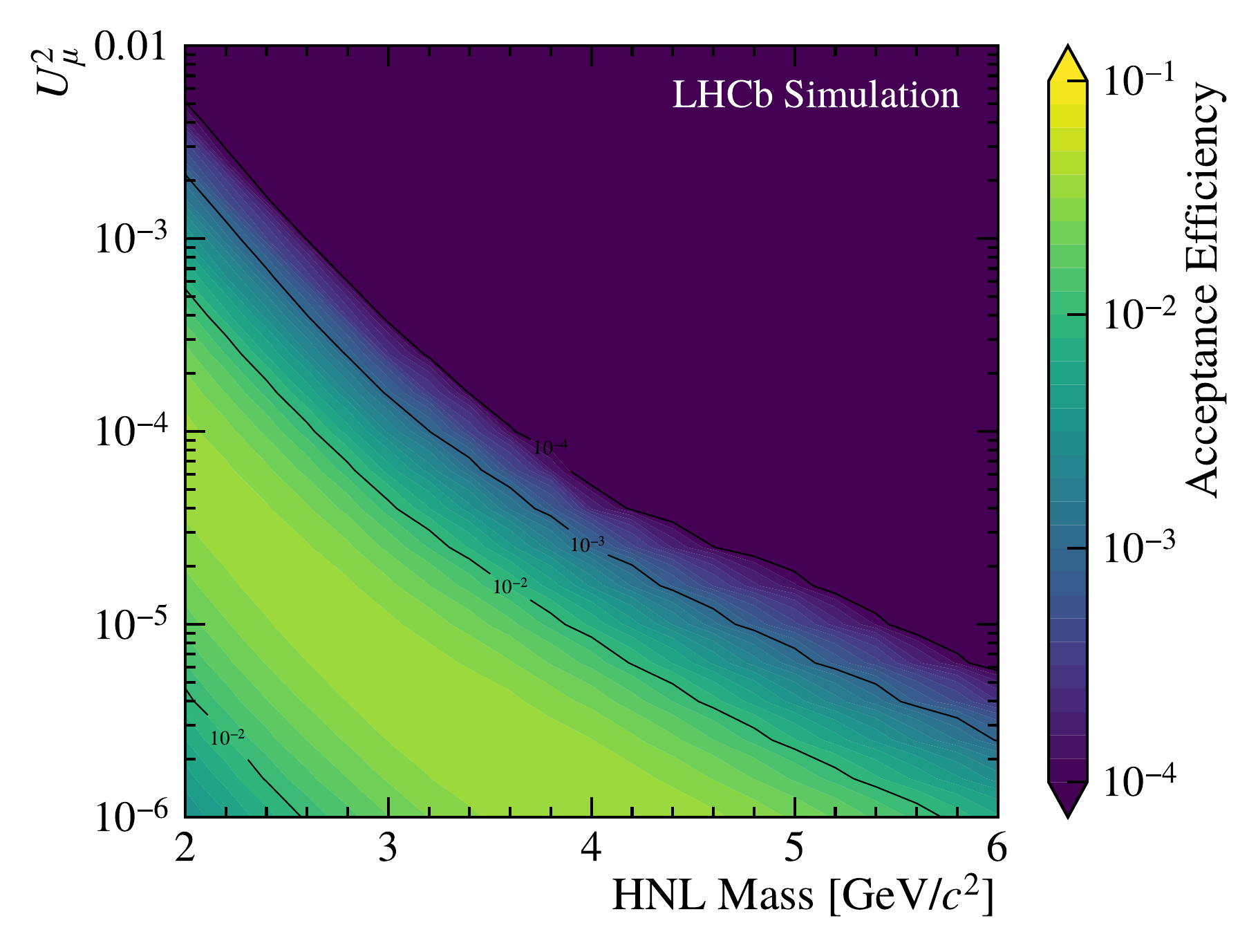}
     \includegraphics[width=0.47\linewidth]{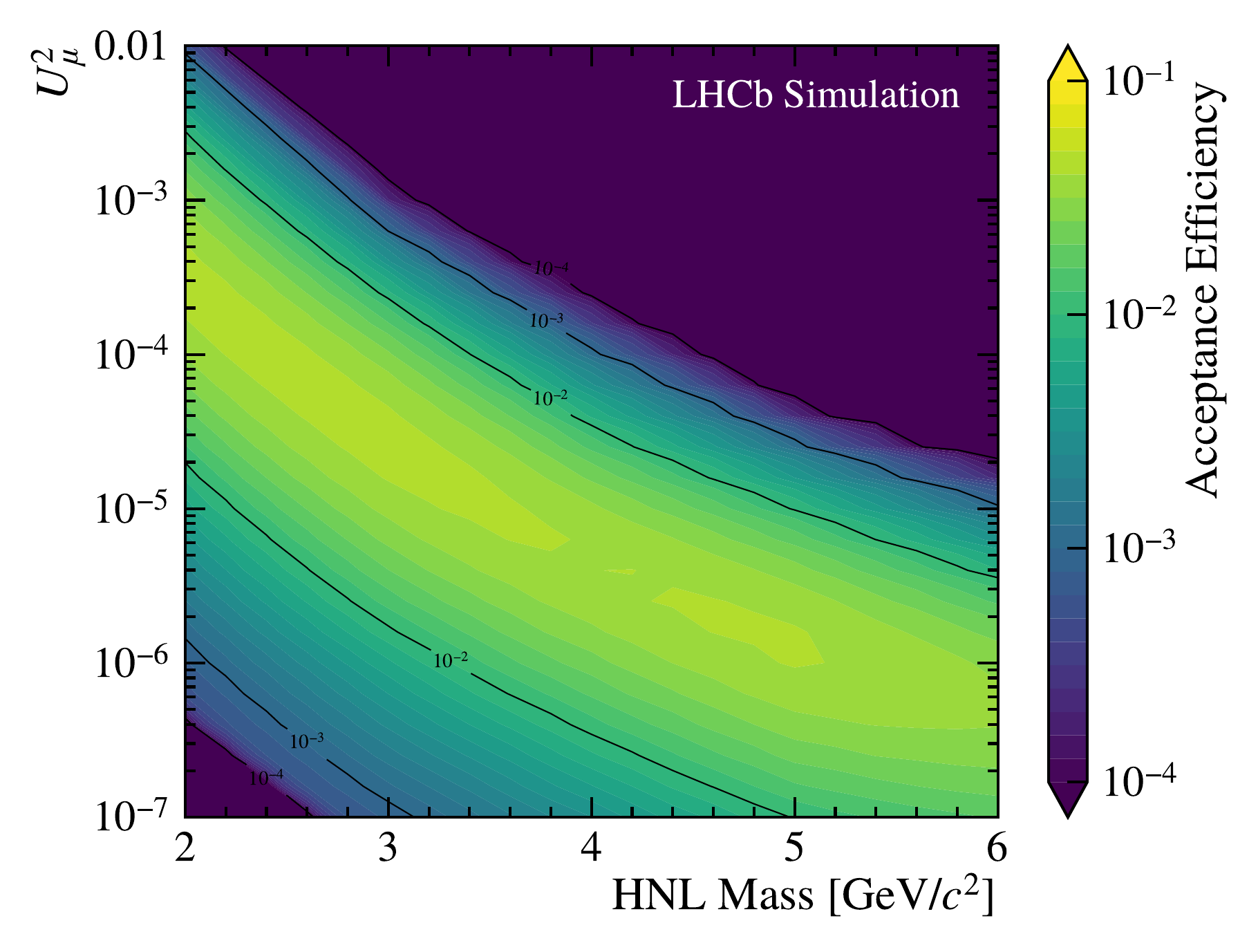}
     \caption{Acceptance efficiency of leptonic $B_c$ production with $N\rightarrow\mu \pi$ decay. This is the proportion of generated HNL events at all solid angles which pass the generator-level cuts, have the $\mu$ from the $B_c$ flagged as reconstructible as long, and have the $\mu$ and $\pi$ from the HNL flagged as either reconstructible as T track (left) or downstream (right).}
     \label{fig:HNL Acceptance}
 \end{figure}

The combination of generator-level efficiency and the efficiency for an event to be flagged as  T-track reconstructible constitutes our acceptance efficiency. This is shown on the left of Figure~\ref{fig:HNL Acceptance} for the $B_{(c)}^+ \rightarrow\mu^+ (N \rightarrow\mu^\pm \pi^\mp)$ mode.

The events are triggered at HLT1 using the Allen (HLT1 software) configuration which ran during data collection at the end of 2024. We do not have a dedicated HLT1 trigger line for HNLs, so we use the combination of all events passing HLT1. Order 20\% efficiency on events with HNLs flagged as reconstructible as T-track vertices, i.e. within acceptance, is achieved with this approach.

At HLT2, the Moore (HLT2 software) configuration which ran during data collection at the end of 2024 was used. To this, a new HLT2 selection on a $N \rightarrow\mu^\pm \pi^\mp$ vertex was added (now included in the LHCb code). For $B_c^+ \rightarrow\mu^+ (N \rightarrow\mu^\pm \pi^\mp)$, the trigger was made on an exclusive $\mu\mu\pi$ final state using a HLT2 selection already in the LHCb software configuration. For $B_q \rightarrow\mu^+ X (N \rightarrow\mu^\pm \pi^\mp)$, the new selection was used.

\begin{figure}
    \centering
    \includegraphics[width=0.47\linewidth]{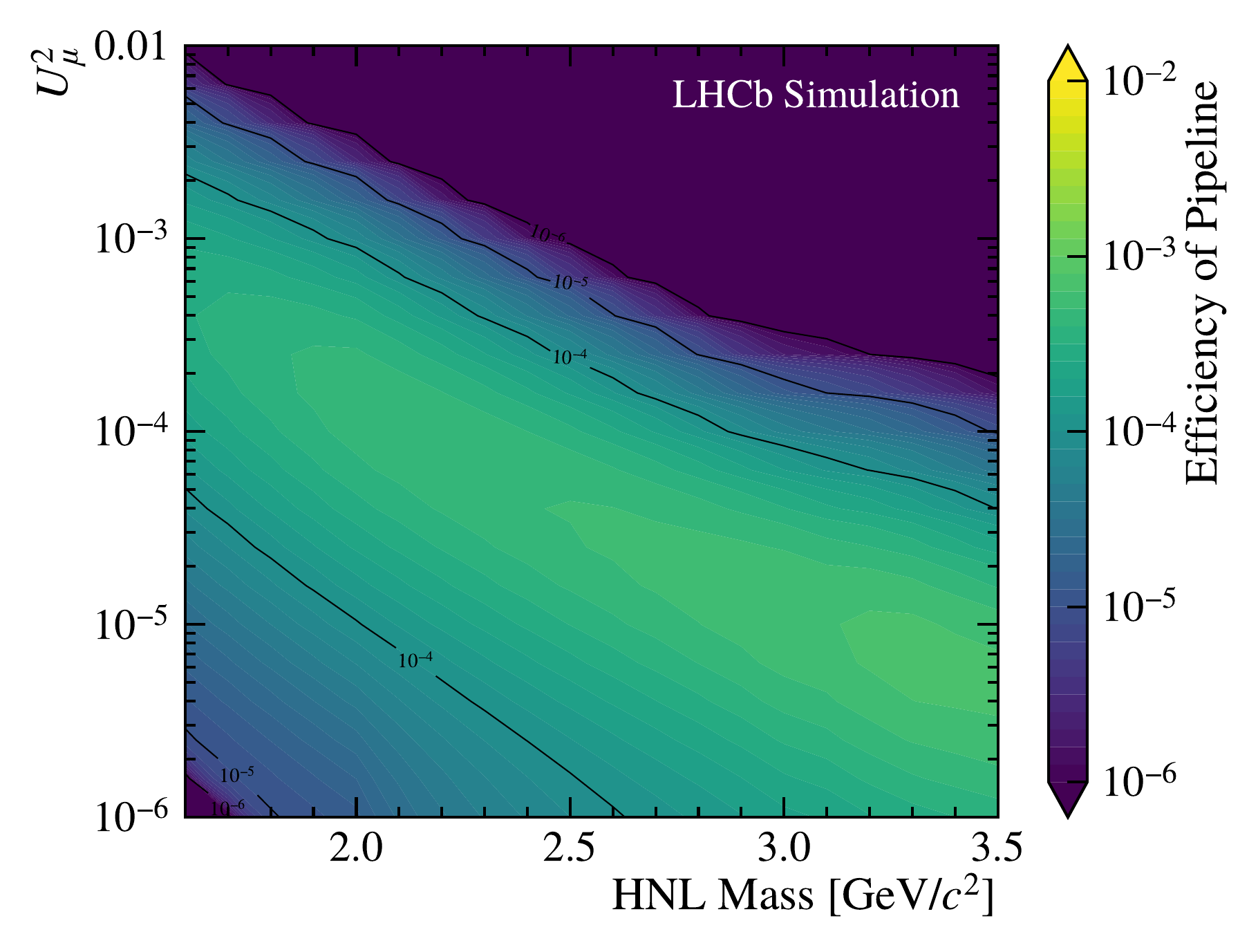}
    \includegraphics[width=0.47\linewidth]{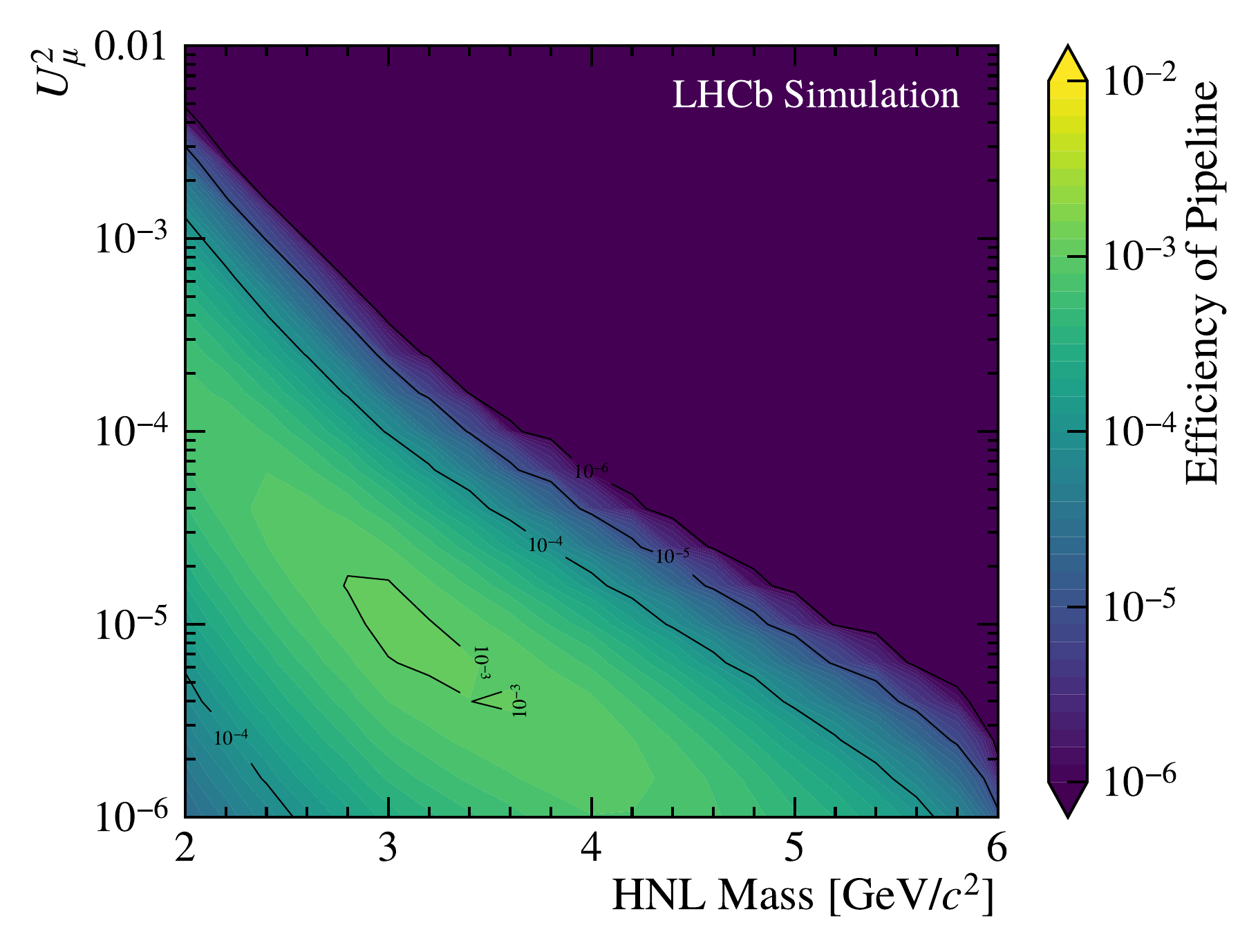}
    \caption{Efficiency of online selection pipeline up to and including HLT2 for T track-reconstructed HNLs. Left: Semileptonic production with HNLs decaying as $N\rightarrow\mu \pi$. Right: Leptonic $B_c$ production with HNLs decaying as $N\rightarrow\mu \pi$.}
    \label{fig:HNL T track efficiency}
\end{figure}

Acceptance efficiency along with the tracking efficiency, HLT1 efficiency and HLT2 efficiency is combined to generate the efficiency of our online selection pipeline for a given HNL mass and lifetime. This is the efficiency for a HNL event generated at a random angle at the proton-proton collision point to pass through to HLT2, and is shown in Figure~\ref{fig:HNL T track efficiency}.

\begin{figure}
    \centering
    \includegraphics[width=0.47\linewidth]{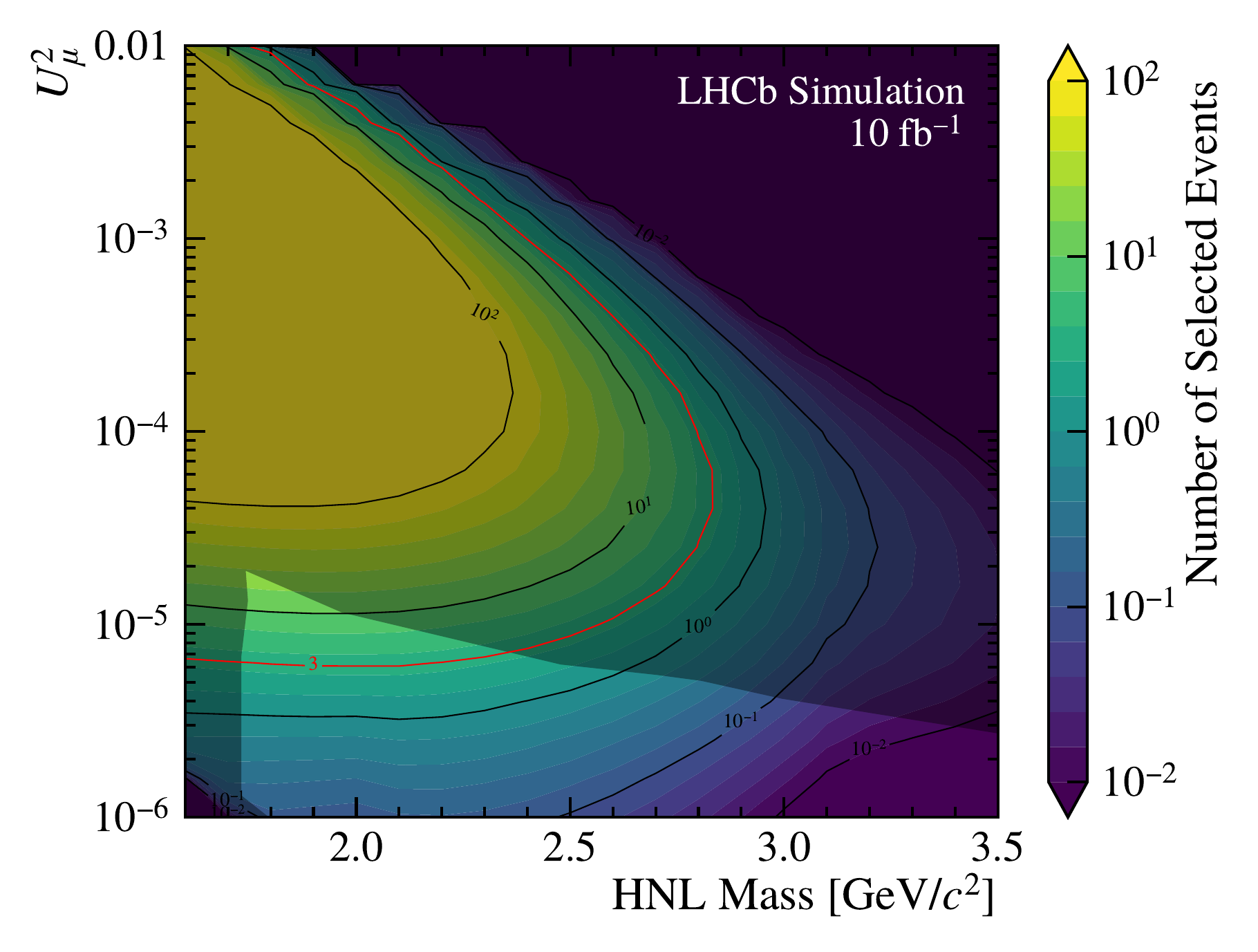}
    \includegraphics[width=0.47\linewidth]{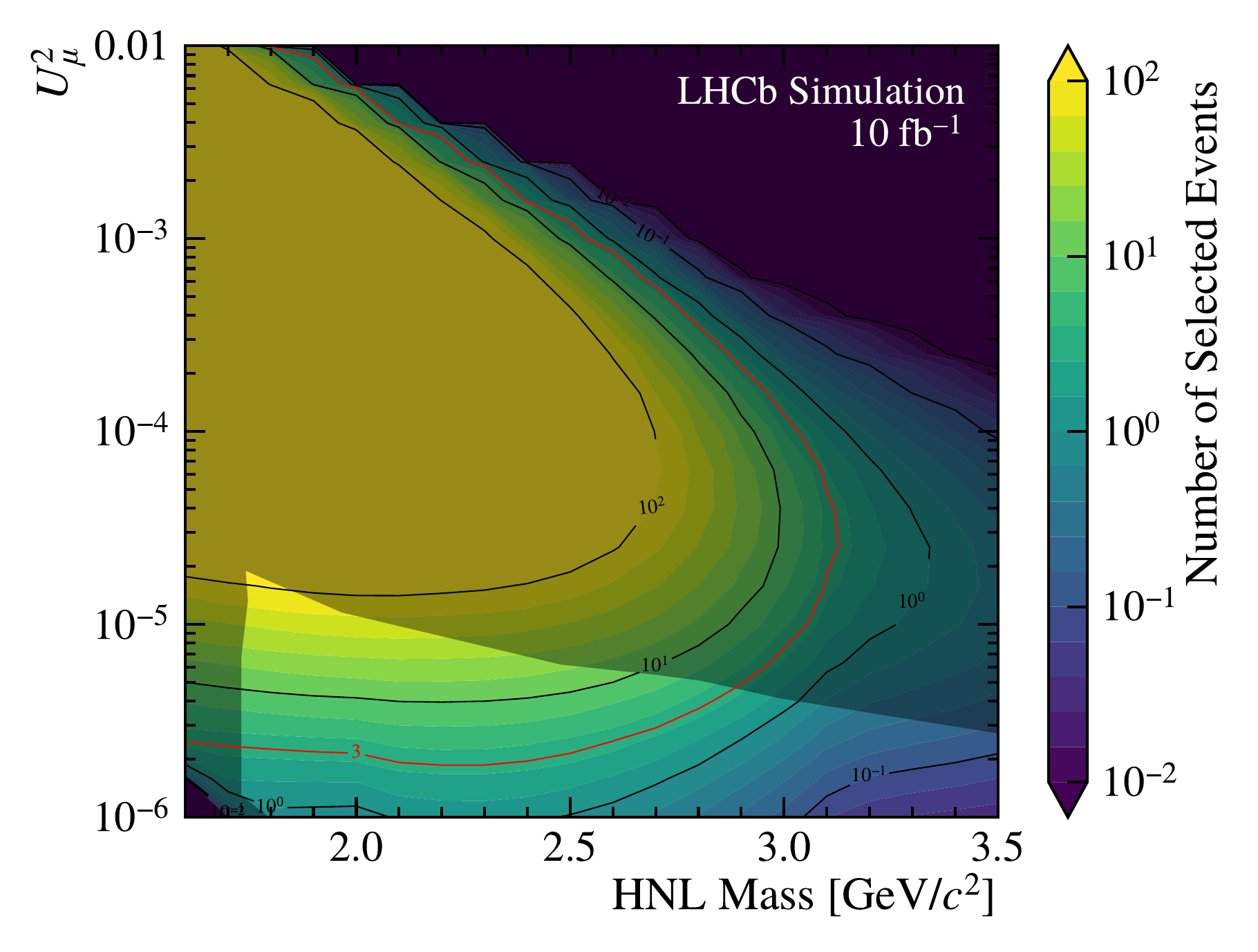}

    \caption{LHCb sensitivity to semileptonically produced HNLs using T track reconstruction and 10 $\rm fb^{-1}$ integrated luminosity. The contour where 3 HNL events are expected is marked in red and corresponds to the 95\% rejection limit assuming no background. The 95\% rejection limit placed by past experiments is overlaid as a transparent black region. Left: HNLs decaying only as $N\rightarrow\mu \pi$. Right: HNLs decaying as $N\rightarrow\mu \pi [X]$, $N\rightarrow\mu e \nu_e$ or $N\rightarrow\mu \mu \nu_\mu$.
    It can be seen that, for T tracks, combined semileptonic production modes can exceed existing limits, especially if multiple HNL decay modes are considered.}
    \label{fig:HNL T track}
\end{figure}

The resultant sensitivity (the online efficiency multiplied by the number of expected HNL events) is shown on the left of Figure~\ref{fig:HNL T track} for semileptonic production. It can be seen that the red contour where 3 HNL events are expected (the 95\% rejection limit we could place in the absence of background) extends outside the black shadowed region representing the 95\% rejection limit placed by past analyses. This indicates that, even with just the $N \rightarrow\mu \pi$ decay mode, we can exceed previous limits. The case of leptonic production (not shown) is less promising.

\textit{Extrapolating T-track sensitivity to alternative decay modes:} When speculating on alternative T track modes, we also consider: production of HNLs through leptonic $B_u$ decay and HNL decays through $N \rightarrow \mu \pi X$, $N \rightarrow \mu e \nu_e$  and $N \rightarrow \mu \mu \nu_\mu$.

We assume acceptance and HLT1 efficiency is only affected by the HNL production mode, and is therefore the same for events with $N \rightarrow \mu \pi X$, $N \rightarrow \mu e \nu_e$  and $N \rightarrow \mu \mu \nu_\mu$ decays as for events with $N\rightarrow \mu \pi$ decay. We assume the same acceptance and HLT1 efficiency for leptonic $B_u$ as for leptonic $B_c$.

For HLT2 efficiency, we consider the use of an inclusive selection on only the T track vertex. For that reason, we assume the efficiency of our selection on the $N\rightarrow \mu \pi$ vertex. For semileptonic production, this is assumed to be the same as for $B_q \rightarrow\mu^+ X (N \rightarrow\mu^\pm \pi^\mp)$. For leptonic $B_u$ or $B_c$ production, we assume the same efficiency as our $N\rightarrow \mu \pi$ vertex selection on the $B_c^+ \rightarrow\mu^+ (N \rightarrow\mu^\pm \pi^\mp)$ state.

The sensitivity projections are shown on the right of Figure~\ref{fig:HNL T track} for semileptonic production. With the additional decay modes included, the reach from semileptonic production modes is extended by a factor $\sim4$, showing further gains over past limits. Leptonic production (not shown) sees similar improvements, but even with all HNL decay modes, we find the limits are not competitive with past analyses. This is attributed to the reduced lifetime of heavier HNLs and resultantly lower T track acceptance. For this reason we must consider downstream as opposed to T track reconstruction to achieve competitive sensitivity with leptonic production (and thus higher mass HNLs).

\textit{Predicting downstream sensitivity:} When speculating on the sensitivity possible for downstream reconstruction, we evaluate  the downstream acceptance efficiency with our MC sample (right of Figure~\ref{fig:HNL Acceptance}). We then assume:
\begin{itemize}
    \item 50\% HLT1 trigger efficiency (trigger efficiency of HLT1 in Run 2),
    \item 80\% HLT2 tracking efficiency of reconstructible downstream-events (reconstruction efficiency achieved at HLT1 in Run 3),
    \item 80\% HLT2 trigger efficiency on reconstructible events (projection based off Run 2).
\end{itemize}

This gives a 32\% chance of a reconstructible downstream event within acceptance passing through HLT2.

\begin{figure}
    \centering
    \includegraphics[width=0.47\linewidth]{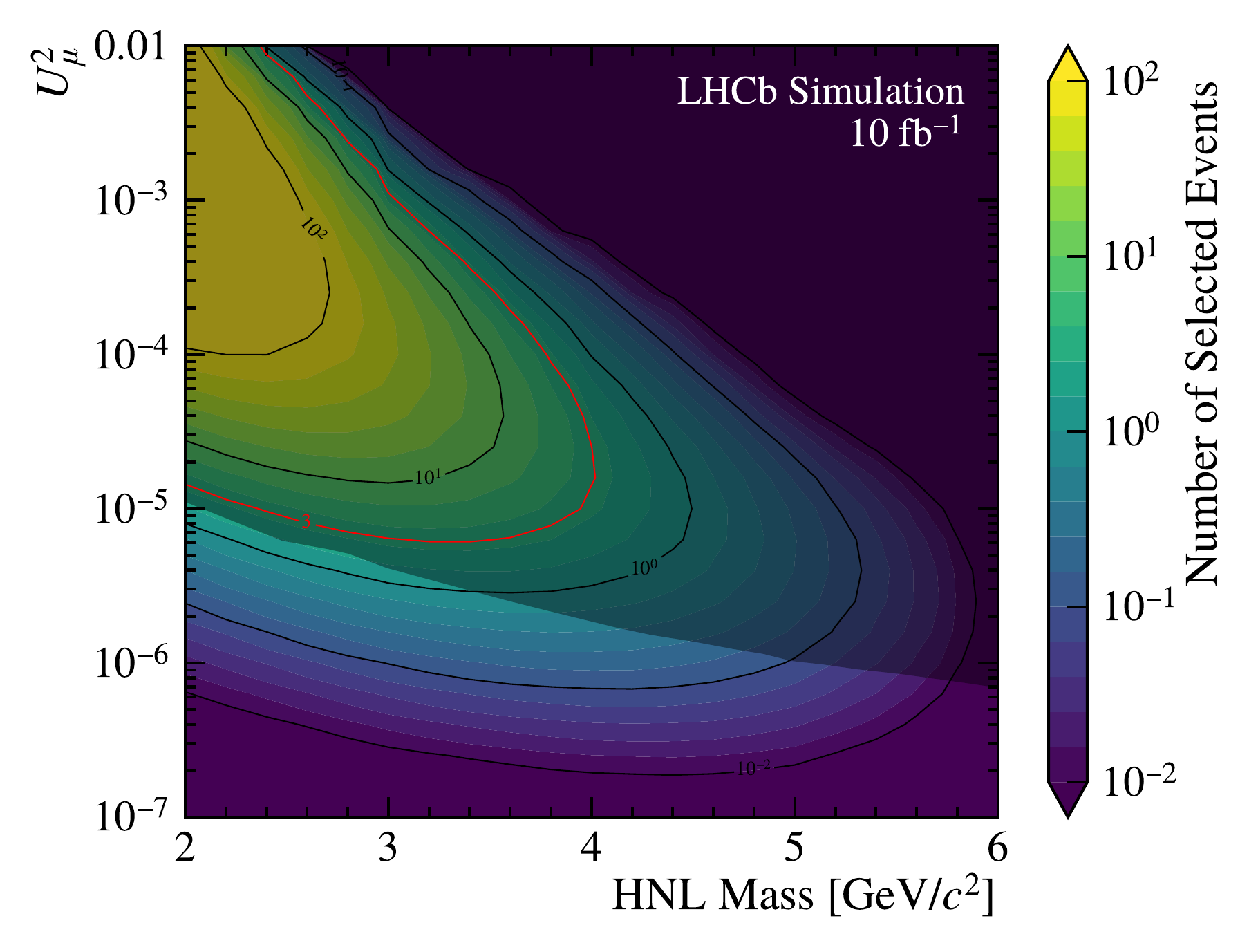}
    \includegraphics[width=0.47\linewidth]{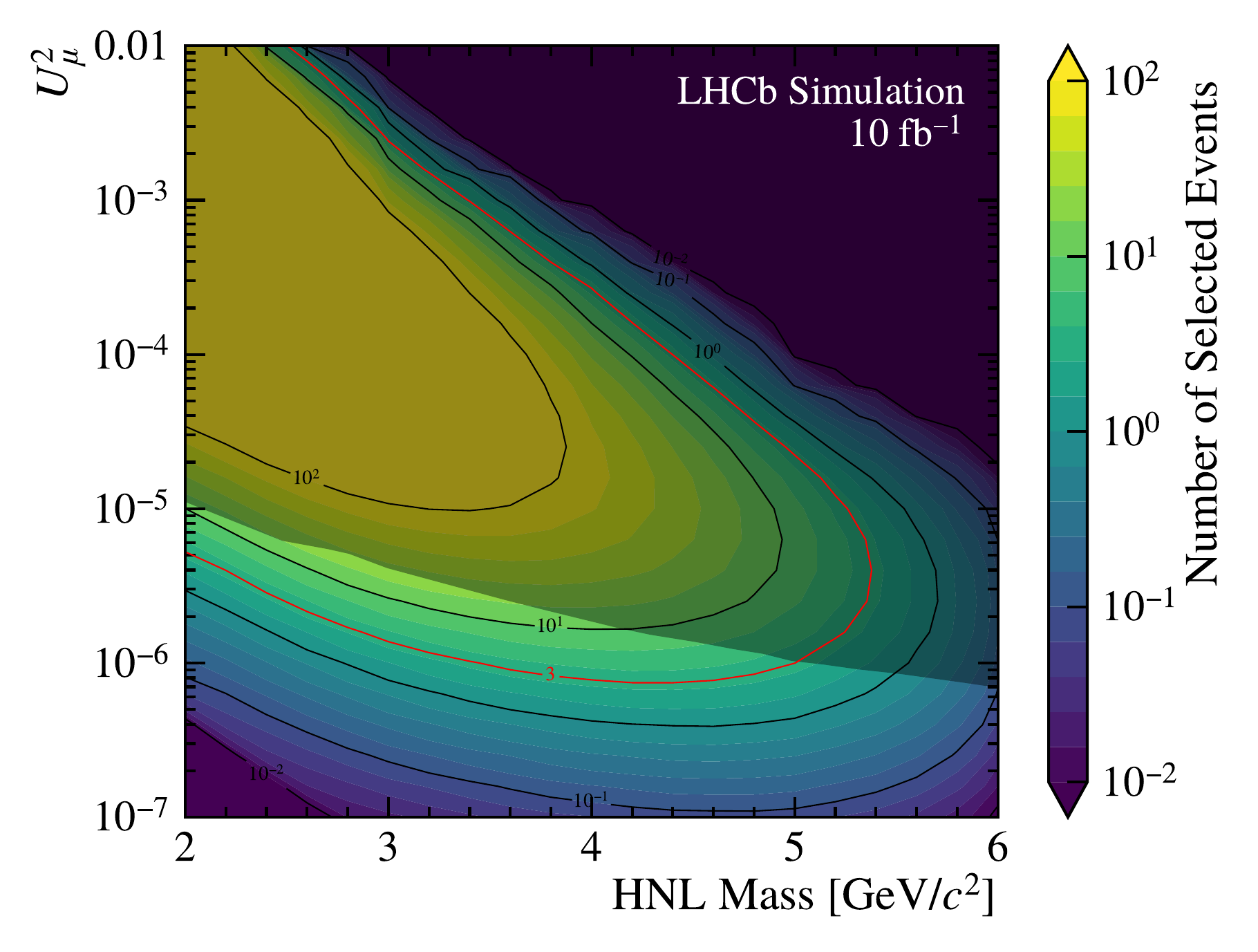}
    \caption{Projected LHCb sensitivity to leptonically produced HNLs using downstream reconstruction and 10 $\rm fb^{-1}$ integrated luminosity. The contour where 3 HNL events are expected is marked in red and corresponds to the 95\% rejection limit assuming no background. The 95\% rejection limit placed by past experiments is overlaid as a transparent black region. Left: HNLs decaying as $N\rightarrow\mu \pi$. Right: HNLs decaying as $N\rightarrow\mu \pi [X]$, $N\rightarrow\mu e \nu_e$ or $N\rightarrow\mu \mu \nu_\mu$.}
    \label{fig:HNL Exclusive Downstream Projection}
\end{figure}

Translating this downstream online efficiency to a sensitivity for leptonic production with first only $N\rightarrow \mu \pi$, and then all HNL decay modes, we find the results on the left and right of Figure~\ref{fig:HNL Exclusive Downstream Projection}, respectively. We find that considering only the $N\rightarrow \mu \pi$ is insufficient to extend our sensitivity beyond that of past analyses. However, if we consider the full inclusive selection, then we are able to achieve improved sensitivity out to an HNL mass of approximately 5 GeV$/c^2$.

\paragraph{Conclusion}

HNLs are a promising BSM theory and, when occurring with a mass between 1.6 and 6.0 GeV/$c^2$, predominantly originate from the decay of B mesons. This makes HNLs a prime target for collider searches with LHCb. However, existing strong bounds on HNLs imply that, if HNLs exist in this mass range, they must have both a small branching ration from B mesons and a long lifetime. This makes them observationally challenging for a typical LHCb analysis searching for vertices originating near the proton-proton collision point. As a result, in our ongoing HNL analysis, we adopt a novel strategy of reconstructing an inclusive selection of very displaced vertices.  This yields an approximately 40-fold increase in the total considered branching ratio from B mesons and a 16-fold increase in the lifetime of HNLs to which we are sensitive. Projecting the sensitivity of the LHCb detector with our analysis strategy, we demonstrate that LHCb has potentially world-leading sensitivity and the capacity to probe and constrain new muon-coupled HNL parameter space outside of existing limits.

\subsubsection{LHCb Upgrade 2: prospects for reconstruction of very displaced vertices --- \textit{C.~Langenbruch}}
\label{sssec:langenbruch}
\textit{Author: Christoph Langenbruch, \email{christoph.langenbruch@cern.ch}}  \\
\paragraph*{Introduction}
In searches for long-lived particles the LHCb experiment profits from the enormous $b\bar{b}$ production cross-section at the LHC of around $500\,\mu{\rm b}$. This is by a factor of ${\cal O}(10^5)$ larger than the $b\bar{b}$ production cross-section available at fixed-target experiments at the SPS. 
LHCb can therefore be competitive in searches for long-lived particles (LLPs) produced in $b$-hadron decays. 
Searches for LLPs require the reconstruction of particles that decay far downstream of their production, 
these searches therefore require typically large detector volumes. 

\begin{figure}
    \centering
    \includegraphics[height=0.3\linewidth]{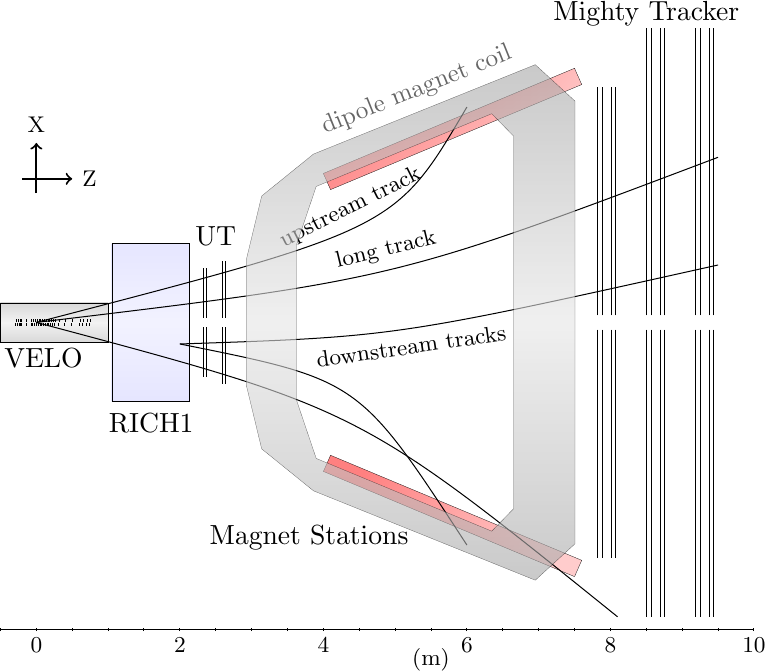}
    \hspace{0.05\linewidth}
    \includegraphics[height=0.32\linewidth]{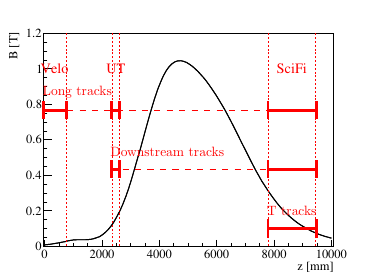}    
    \caption{(Left) The LHCb tracking system together with the different resulting track types reconstructed by LHCb. 
    (Right) The magnetic field provided by the LHCb dipole magnet, overlayed with the positions of the tracking detectors in $z$.}
    \label{fig:lhcbtrackingsystem}
\end{figure}

\begin{figure}
    \centering
    \includegraphics[height=0.3\linewidth]{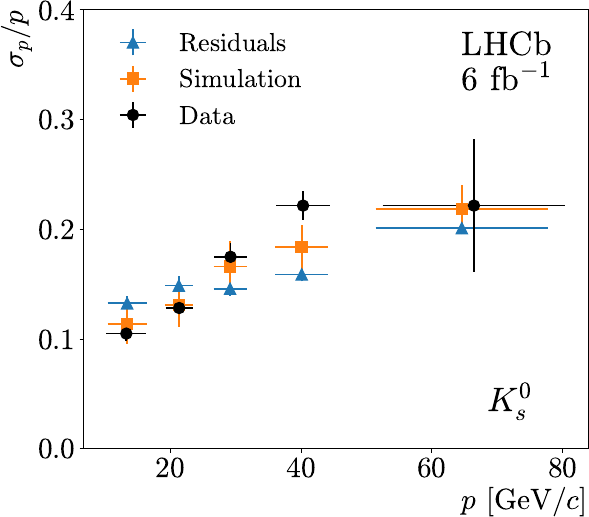}
    \hspace{0.05\linewidth}
    \includegraphics[height=0.3\linewidth]{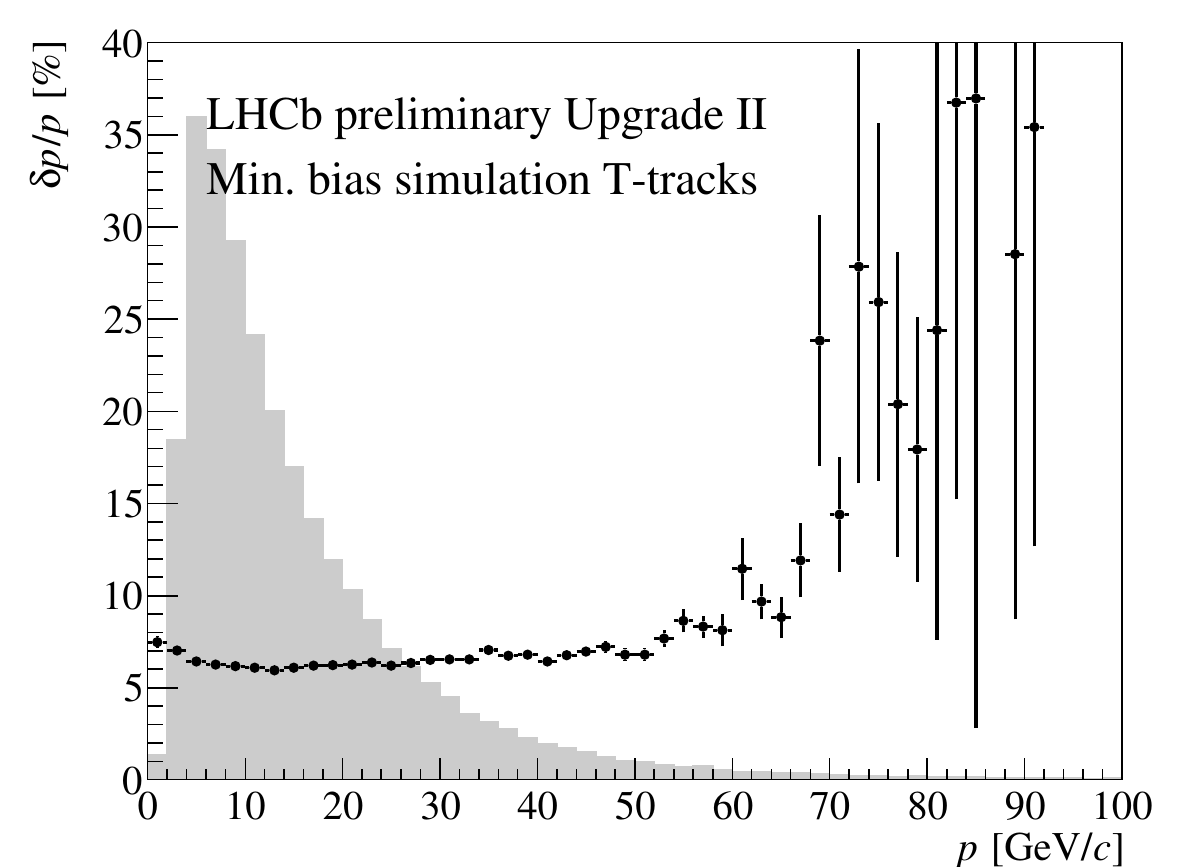}
    \caption{(Left) Momentum resolution for T-tracks in Run~3 using the SciFi tracker. Figure taken from Ref.~\cite{LHCb:2022kwc}.
    (Right) Expected momentum resolution for T-tracks reconstructed in the inner part of the MightyTracker, instrumented by silicon pixel sensors. Figure from Ref.~\cite{LHCB-FIGURE-2025-001}.}
    \label{fig:ttrackresolution}
\end{figure}

The tracking system at LHCb is shown in Fig.~\ref{fig:lhcbtrackingsystem} (left). 
It consists of several subdetectors that are used to reconstruct tracks resulting from decays of long-lived particles. 
Depending on the $z$-coordinate of the decay vertex of the long-lived particle different subdetectors can contribute to the reconstruction of the tracks, resulting in different track types:
\begin{itemize}
    \item \textit{Long tracks:} Long tracks are built from particle tracks that traverse the vertex detector (VELO), the upstream tracker (UT), and the downstream tracking station (SciFi in Run~3). Since these tracks traverse the maximum integrated magnetic field of around $4\,{\rm Tm}$ provided by the LHCb dipole magnet, as shown in Fig.~\ref{fig:lhcbtrackingsystem} (right), and have the longest lever arm, their momentum resolution is best. 
    In Run~3 long tracks have a relative momentum resolution $\delta p/p$ of $0.45\text{--}1\%$, depending on the track momentum~\cite{LHCB-FIGURE-2024-040}. 
    Since the tracks need to traverse the full LHCb tracking system, LLPs need to decay inside of the approximately $1\,{\rm m}$ long VELO detector. 
    \item \textit{Downstream tracks:} Downstream tracks only traverse the UT and the downstream tracker, therefore the LLPs need to decay before the UT at around $2.5\,{\rm m}$ to result in downstream tracks. Since measurements before the magnet can only be obtained from the four layers of the UT, the momentum resolution of downstream tracks is $\delta p/p\approx 2\,\%$~\cite{Davis:2240723}. The momentum determination for downstream tracks is therefore less precise than for long tracks. 
    \item \textit{T-tracks:} T-tracks can be reconstructed from LLPs decaying before the downstream tracking stations, located at around $7.8\,{\rm m}$. This significantly increases the available decay volume, however the momentum measurement relies solely on the fringe magnetic field in the SciFi detector, corresponding to around $0.3\,{\rm Tm}$. The resulting momentum resolution in Run~3 is around $\delta p/p =10\text{--}25\%$~\cite{LHCb:2022kwc}, as shown in Fig.~\ref{fig:ttrackresolution} (left).
\end{itemize}
While using T-tracks or downstream tracks exhibit thus a larger acceptance for LLP decays, 
particular care needs to be taken with respect to the control of backgrounds due to the deteriorated momentum resolution. 

\begin{figure}
    \centering
\includegraphics[height=0.2\linewidth]{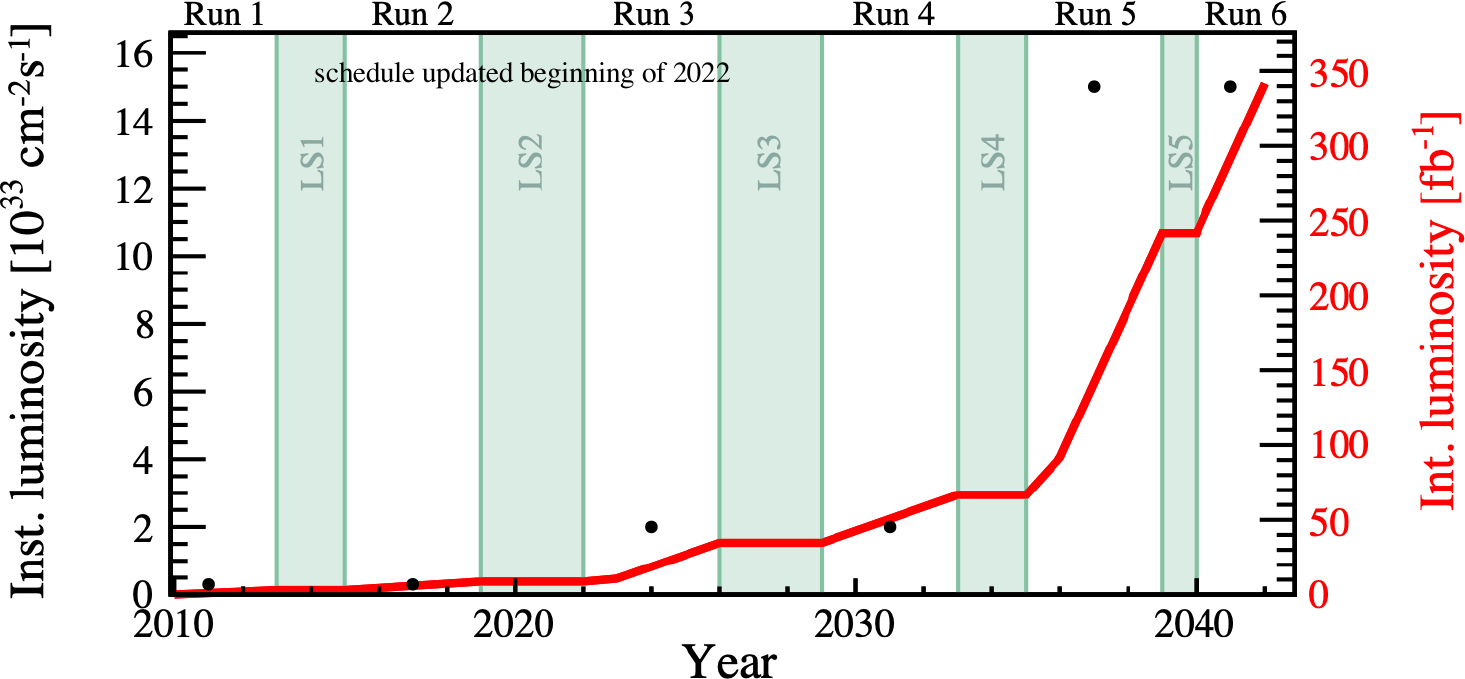}
\hspace{0.05\linewidth}
    \includegraphics[height=0.22\linewidth]{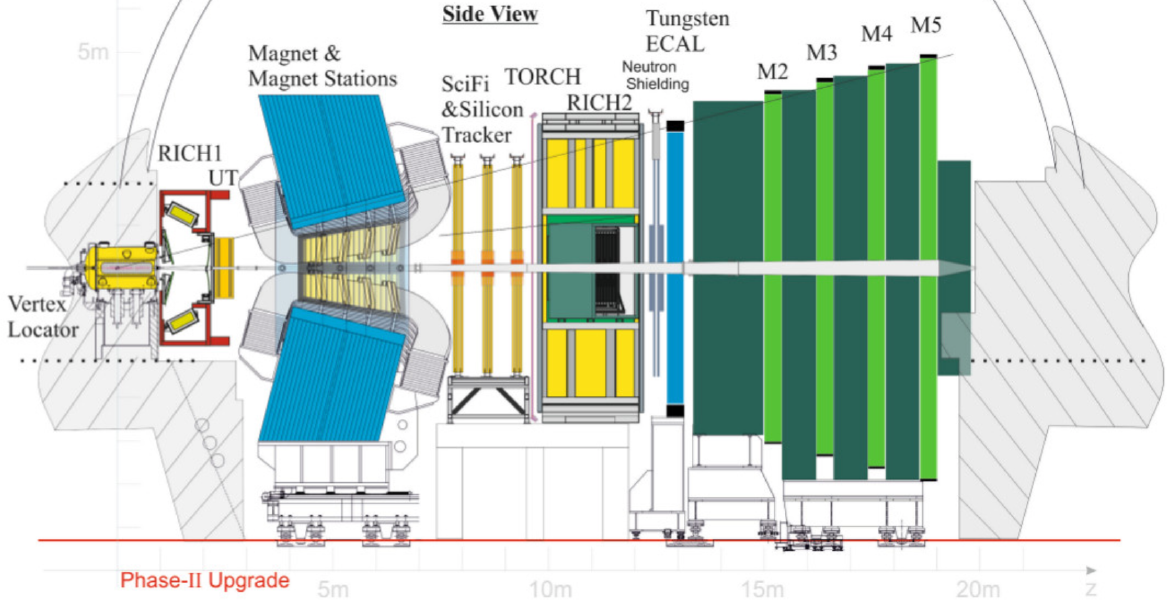}
    \caption{(Left) Expected integrated luminosity in the LHCb Upgrade~I and Upgrade~II. (Right) Side view of the LHCb Upgrade~II detector.}
    \label{fig:upgradeIIdetector}
\end{figure}

\paragraph*{The LHCb Upgrade~II}
The currently operating LHCb detector, the so-called Upgrade~I, is planned to be operated until the end of LHC Run~4 and collect data corresponding to an integrated luminosity of around $50\,{\rm fb}^{-1}$, as illustrated in Fig.~\ref{fig:upgradeIIdetector} (left). Following Run~4, the LHCb Upgrade~II detector, described in detail in Refs.~\cite{LHCb:2021glh,LHCbcollaboration:2903094}, is planned to be installed during Long Shutdown~4. 
During the LHC Runs~5 and 6 the LHCb collaboration intends to collect data corresponding to more than $300\,{\rm fb}^{-1}$ with the Upgrade~II detector, allowing for unprecedented reach also for searches for LLPs. 
The major challenge in the Upgrade~II will be the increase of the instantaneous luminosity to up to $1.5\times 10^{34}\,{\rm cm}^{-2}{\rm s}^{-1}$, which will result in a pile-up of around 40. 
Compared to the pile-up of around 6 during the LHC Run~3 this constitutes an increase by around a factor 8, 
and poses extreme requirements on the radiation hardness and granularity of the detectors. 
A schematic side-view of the LHCb Upgrade~II detector is shown in Fig.~\ref{fig:upgradeIIdetector} (right). 

The tracking system in the LHCb Upgrade~II consists of the Timing VELO~(TV) vertex detector which encloses the primary interaction vertices, the Upstream Pixel~(UP) upstream the LHCb dipole magnet, and the MightyTracker (MT) consisting of 3 tracking stations downstream the magnet. 
In addition, the magnet sidewalls are planned to be instrumented with the magnet stations to increase tracking efficiency for low momentum particles. 
The TV addresses the challenge of reconstructing ${\cal O}(40)$ primary vertices through the use of timing. 
To this end the TV will provide a time stamp per hit with a resolution of $50\,{\rm ps}$, resulting in a time resolution per track of a few tens of ${\rm ps}$. 
Since the primary vertices are distributed in a time interval of around $180\,{\rm ps}$ this allows to significantly reduce the effective pile-up. 
The UP and the inner area of the MightyTracker, called the MightyPixel (MP), will use high-granularity silicon pixel sensors. 
It is planned to use radiation hard Depleted Monolithic Active Pixel sensors (DMAPs), implemented in cost-effective HV-CMOS technology. 
While the high occupancy inner part of the MightyTracker will consist of 6 layers of silicon pixel sensors, 
the large outer area will be covered by scintillating fibres. 
This component of the MT, the MightySciFi (FT), will consist of 12 layers of scintillating fibres, similar to the technology currently used in the LHCb SciFi tracker. 
Relative improvements in performance are expected from cryogenic cooling of the SiPMs, which significantly reduces the dark count rate, and from increased photon detection efficiency through the use of micro-lenses. 
In addition, improved fibre materials are under study. 
The occupancy in the FT in the Upgrade~II is expected to be similar to the highest occupancy areas during Run~3. 

Particle identification in the Upgrade~II will be provided by two Ring Imaging CHerenkov~(RICH) detectors, RICH~1 and RICH~2. 
To handle the high occupancy in the Upgrade~II, highly granular photo-detectors and timing information will be used. 
The TORCH detector will provide time-of-flight information with a resolution of around $15\,{\rm ps}$ for charged tracks, 
which will give important particle identification information in the momentum range below $\sim 15\,{\rm GeV}/c$. 
The detector concept is based on internally reflected Cherenkov photons produced in the passage of charged particles through quartz plates. 
The PicoCal constitutes the electromagnetic calorimenter in the LHCb Upgrade~II. 
The PicoCal will be implemented in high granularity Spaghetti Calorimeter~(SpaCal) technology for the inner region and Shashlik modules in the outer regions. The PicoCal will be able to also perform timing measurements with a hit resolution of a few tens of ${\rm ps}$. 
The muon chambers will provide muon identification in the high occupancy environment of the Upgrade~II. 
The inner regions will be instrumented using high granularity $\mu$-RWELL chambers, the outer areas will be using MWPCs with varying  granularity. 

\paragraph*{Expected tracking performance for LLPs in the Upgrade~II}
A major improvement in the Upgrade~II will be the excellent resolution in $y$ provided by the pixel sensors in the UP and the MP, which will have a hit resolution of ${\cal O}(50\,\mu m\times 150\,\mu m)$~\cite{LHCb:2021glh}. The silicon strips in the UT and the fibres in the SciFi in the current Run~3 detector in comparison rely on stereo layers with a small stereo angle of $\pm 5^\circ$ to determine the $y$-coordinate. For downstream and T-tracks reconstructed in the MP, the LHCb Upgrade~II detector will therefore provide a significantly improved spatial resolution for the reconstruction of LLP decay vertices. This will help reduce combinatorial background from the random combination of unrelated tracks. 

In addition, the Upgrade~II detector will be able to profit from the improved position resolution in the bending plane of the magnet. 
The improvement is particularly prominent for the MP covering the inner part of the MightyTracker. 
Compared to the Run~3 momentum resolution of $10\text{--}25\%$ shown in Fig.~\ref{fig:ttrackresolution}~(left), the LHCb Upgrade~II will provide a momentum resolution of $6\text{--}8\%$ for T-tracks, as shown in Fig.~\ref{fig:ttrackresolution}~(right).  
The preliminary performance numbers for the Upgrade~II were estimated using minimum bias simulation with an instantaneous luminosity of $1.3\times 10^{34}\,{\rm cm}^{-2}{\rm s}^{-1}$, and using hits from the pattern recognition to reconstruct a Kalman-filtered T-track. 
The improved momentum resolution in the Upgrade~II will significantly improve background suppression and simplify the use of T-tracks in searches for LLPs.  

\paragraph*{Conclusions}
In its searches for LLPs LHCb is able to profit from the enormous $b\bar{b}$ production cross-section at the LHC. 
The data-taking during Run~3 was very successful, 
and the data sample taken by LHCb during 2024, which corresponds to an integrated luminosity of more than $9\,{\rm fb}^{-1}$, 
will allow to perform searches for LLPs with excellent sensitivity. 
The LHCb Upgrade~II will be installed during the LS~4, operate during the LHC Runs~5 and~6, 
and will collect a data sample corresponding to an integrated luminosity of $300\,{\rm fb}^{-1}$. 
The challenges of the high multiplicity environment with a pile-up of around $40$ can be overcome by using 
highly granular and radiation hard detectors and by exploiting precision timing information. 
The Upgrade~II tracking system will be well suited for the reconstruction of LLPs  
due to the improved vertexing and momentum resolution for downstream and T-tracks. 
The LHCb Upgrade~II data sample will therefore allow for searches for LLPs with unprecedented reach.

\subsection{FIPs at ATLAS and CMS: Challenges and opportunities}
\label{ssec:fips_at_ATLAS_CMS}

\subsubsection{ATLAS: FIPs results and prospects --- \textit{E.~Torro~Pastor}}
\label{sssec:torro-pastor}
\textit{Author: Emma Torro Pastor, \email{emma.torro.pastor@cern.ch}}  \\

Despite many years of data collection at the Large Hadron Collider (LHC), no evidence for physics beyond the Standard Model (SM) has yet been observed. Among the various potential reasons for this absence of discovery, this contribution discusses two possibilities, both of which involve scenarios characterized by low sensitivity to conventional LHC search strategies.

First, new particles may be very light or have very low production cross sections, making them challenging to detect due to the difficulty to trigger on them and the overwhelming SM backgrounds. Addressing such cases requires specialized techniques, including the use of specific triggers running on dedicated data streams, which permit significantly higher trigger rates while minimizing data volume.

Second, the possibility of new particles with extended lifetimes is considered. Long-lived particles (LLPs) have been a central component of the ATLAS physics program since Run 1, with numerous searches performed over the years. 
Within ATLAS, the Exotics Physics group includes the Long-Lived and Unconventional Physics subgroup, which is responsible for conducting a broad range of searches targeting unconventional signatures. These efforts encompass several tens of distinct analyses, collectively probing a wide array of theoretical models.

The specific experimental signature to be searched for depends on the properties of the LLP, including its electric charge, decay modes, and lifetime, which in turn influence the likelihood of its decay within a particular sub-detector of ATLAS. 
Each search focuses on a distinct unconventional signature and aims to maximize sensitivity across diverse theoretical scenarios. Detailed results are provided to facilitate reinterpretation by the theoretical physics community in the context of alternative models.

A broad spectrum of Beyond the Standard Model (BSM) theories predicts the existence of feebly interacting particles (FIPs), which may decay promptly or with measurable displacement. A summary of selected results is presented below, with particular emphasis on the dedicated experimental strategies required to make such searches possible.

\paragraph{Scalar portal}
A well-motivated class of benchmark models explored by the ATLAS experiment involves the presence of a Hidden Sector (HS), which extends the Standard Model (SM) through the introduction of a new non-Abelian gauge group, $G_{\mathrm{HS}}$. In these scenarios, particles charged under $G_{\mathrm{HS}}$ are neutral with respect to the SM gauge group, and conversely, SM particles carry no charges under the hidden gauge group. The two sectors interact via a heavy mediator particle, enabling suppressed but non-negligible communication between them.

A representative example frequently used in ATLAS studies is a simplified Hidden Sector model~\cite{Chang:2005ht}, in which the interaction between the SM and HS proceeds via a heavy neutral scalar boson $\Phi$. This mediator may be identified with the SM Higgs boson or a similar scalar particle of a different mass. In this model, the mediator decays into a pair of neutral long-lived particles, each of which subsequently decays into SM fermion–antifermion pairs, with branching fractions governed by the Yukawa couplings of $\Phi$.

Such models predict distinctive experimental signatures characterized by displaced activity in the ATLAS detector, with different signatures depending on the proper lifetime of the LLPs, making these scenarios an important target for dedicated searches at the LHC.

The following searches in ATLAS have explored different lifetime ranges of the resulting neutral long-lived scalar by looking at different parts of the detector:
\begin{itemize}
    \item Searches for displaced jets in the tracking system, making use of the Large Radius Tracking (LRT), a dedicated tracking technique developed in ATLAS for the reconstruction of secondary vertices well displaced from the interaction point~\cite{ATLAS:2021jig}. 
    \item Searches for pairs of neutral LLPs decaying hadronically in the hadronic calorimeter and leading to displaced jets~\cite{ATLAS:2019qrr}, \cite{ATLAS:2022zhj}. These have a number of unique characteristics that were used to train an adversarial Neural Network for their identification. 
    \item Searches for pairs of neutral LLPs decaying hadronically in the muon spectrometer~\cite{ATLAS:2018tup}, \cite{ATLAS:2022gbw}.
    \item Searches for pairs of neutral LLPs decaying hadronically, one in the inner tracking and the other in the muon spectrometer~\cite{ATLAS:2019jcm}.
    \item In addition, searches for H$\rightarrow$invisible and monojet signatures have been re-interpreted in the context of this model~\cite{ATLAS:2021kxv}.
    
\end{itemize}

The following plot summarises the results from all these searches at 13 TeV, showing a good complementarity among them.

\begin{figure}[hbt!]
    \centering
    \includegraphics[width=0.95\textwidth]{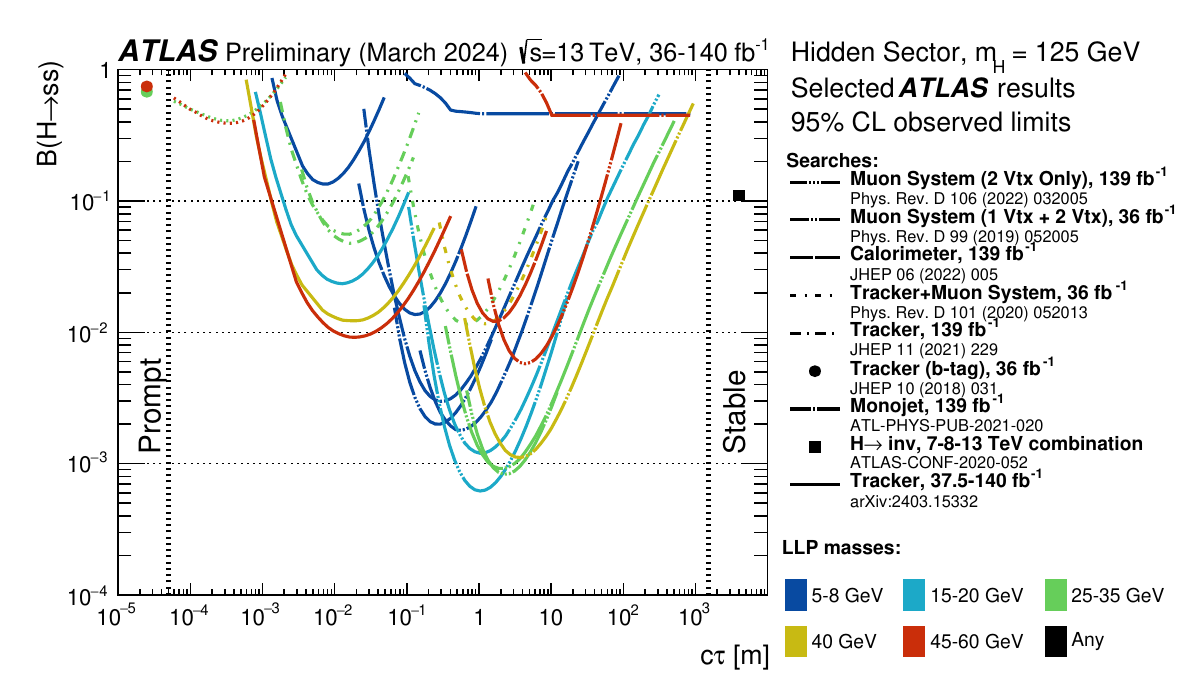}
\caption{The $95\%$ confidence level exclusion limits on the Higgs boson branching to a pair of long-lived neutral
spin-0 bosons ($s$) as a function of the proper decay length ($c\tau$) of $s$. The limits are scaled assuming Yukawa-ordered
branching ratios of $s$. The colored lines represent the regions excluded by the analyses listed in the legend, and
the colors of the lines refer to the long-lived particle mass (or mass range) under consideration. Also shown are
exclusions for models where $s$ is prompt or detector-stable. The plot shows a selection of the most sensitive individual ATLAS 13 TeV results. For clarity, parts of the exclusion curves outside the most sensitive region are omitted~\cite{ATLAS:summaryPlots}.}
    \label{fig:models}
\end{figure}

\subsubsection{CMS: FIPs results and prospects --- \textit{J.~Alimena}}
\label{sssec:alimena}
\textit{Author: Juliette Alimena, \email{juliette.alimena@cern.ch}}  \\
\newcommand{\ptmiss}{\ensuremath{p_{\text{T}}^{\text{miss}}}\xspace}
\newcommand{\PA}{\ensuremath{\text{A}}\xspace}
\newcommand{\PAprime}{\ensuremath{\PA^{\prime}}\xspace}
\newcommand{\mAprime}{\ensuremath{m_{\PAprime}}\xspace}
\newcommand{\mixpar}[1]{\ensuremath{V_{#1}}\xspace}
\newcommand{\mixparsq}[1]{\ensuremath{|\mixpar{#1}|^2}\xspace}
\newcommand{\mixparsqlN}{\mixparsq{lN}}
\newcommand{\mhnl}{\ensuremath{m_{N}}\xspace}
\newcommand{\hinv}{\ensuremath{H \to \text{inv}}\xspace}

\paragraph{Introduction to CMS results and prospects}
The \emph{dark sector} is a collection of yet-unobserved quantum fields and their corresponding new particles. The dark sector particles interact weakly with standard model (SM) particles, and these interactions are typically mediated through other new particles, such as the dark photon, sterile neutrino, and axion. The primary motivation for a dark sector is to explain the source of dark matter (DM). The dark sector may also solve other fundamental issues in the SM, such as providing a natural value for the Higgs vacuum expectation value and a mechanism for baryogenesis~\cite{Foot:2014uba}.

The CMS Collaboration has recently written a report~\cite{CMS:2024zqs} about the latest searches with the CMS Experiment~\cite{CMS:2008xjf} in the dark sector. This report is a comprehensive review of dark sector searches with the CMS experiment at the LHC, using proton-proton and heavy ion collision data collected in Run~2, from 2016 to 2018, or, in some cases, from Run~1 (2011--2012) or Run~3 (2022). This section summarizes some of the models, techniques, and results presented in this report.

\paragraph{Signatures}
The broad CMS dark sector search program spans many different signatures, including those with invisible particles, those with particles promptly decaying into fully visible final states, and those with long-lived particles (LLPs).

Direct DM signatures, in their simplest form, consist of the production of the mediator particle, which subsequently decays into DM. Final states from such processes feature the presence of \ptmiss because the DM particles interact sufficiently weakly to be invisible in the detector; here, we call the DM particle a weakly interacting massive particle (WIMP)~\cite{Jungman:1995df}. For these invisible particles to be detectable, the DM particle must be accompanied by at least one visible object, such as a jet (which is a collimated spray of energetic particles produced by the hadronization of a high-energy quark or gluon), lepton, photon, or the decay products of a heavy SM boson, such as the Higgs (H), W, or Z boson. These characteristic signatures are the mainstay of ``mono-X'' searches, where X denotes the visible, radiated object that recoils off the system that directly produces the DM. Moreover, we focus on DM with masses ranging from the MeV to the TeV range.

Any mediator between the dark sector and the SM that is produced at colliders by the interaction of SM particles must also be able to decay back to those SM particles, such as the process $q\overline{q} \to Z \to q'\overline{q}'$. Correspondingly, we can also search for the DM indirectly via fully visible resonances arising from the mediator production. This approach is only sensitive to the SM interactions of the mediator and therefore makes no additional assumptions about the portal.

Different signatures appear from rich dark sector dynamics that can produce more mediators, additional unstable particles, or new interactions. These extended DM models give rise to a number of signatures that can be probed at the LHC. Moreover, these added signatures enhance the sensitivity of the LHC to the dark sector with additional visible particles and energy in the final state, compared to mono-X searches. One such signature is particles with long lifetimes, which often appear in BSM scenarios, notably in models that describe the elementary particle nature of DM. These LLPs can be produced as a result of small couplings, little available phase space for the particle’s decay, or high mass scales~\cite{Curtin:2018mvb}. These mechanisms generically appear in BSM scenarios, including those in the dark sector. For example, small portal couplings between the SM and the dark sector give rise to LLPs in hidden valley models~\cite{Strassler:2006im,Strassler:2006ri}.

\paragraph{Dark sector models and results}

\begin{figure}[hbt!]
    \centering
    \includegraphics[width=0.95\textwidth]{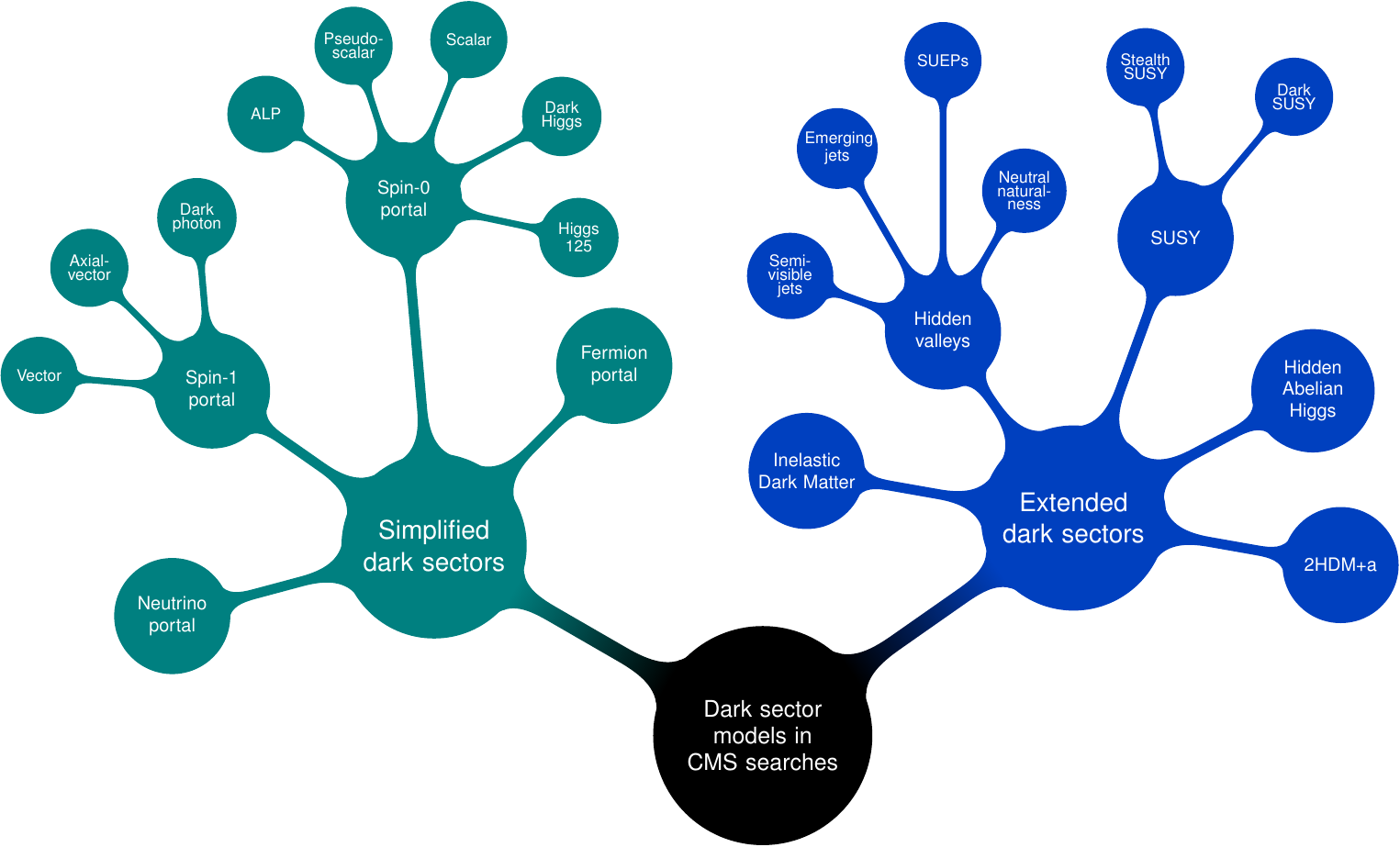}
\caption{Map of the models probed in CMS searches for dark sectors. Nodes are separated by emphasis of the interpretation and may overlap in physics origin. From Ref.~\cite{CMS:2024zqs}.}
    \label{fig:models}
\end{figure}

Searches making use of these three different types of signatures have been interpreted in two categories of models, shown in Fig.~\ref{fig:models}: models that consist of a single mediator particle and DM are denoted \emph{simplified dark sectors}, and models with more complicated dark sector dynamics are denoted \emph{extended dark sectors}. In simplified dark sectors, the DM particle is a WIMP~\cite{Jungman:1995df}, whereas in extended dark sectors, the nature of the DM particle or particles may vary. We note that simplified dark sector models are not complete models, but proxies for more complete models~\cite{Morgante:2018tiq,Kahlhoefer:2015bea,Abdallah:2015ter}.

We will discuss four representative dark sector models, namely, dark photons in the minimal dark photon model (Section~\ref{sec:darkPhotons}), heavy neutral leptons in the neutrino portal (Section~\ref{sec:HNLs}), Higgs boson decays into LLPs in models of neutral naturalness (Section~\ref{sec:HiggsToLLPs}), and dark Higgs bosons in the dark Higgs portal (Section~\ref{sec:darkHiggs}). We will show interpretations of CMS search results within these model frameworks.

\paragraph{Dark photons} \label{sec:darkPhotons}

\begin{figure}[hbt!]
    \centering
    \includegraphics[width=0.4\textwidth]{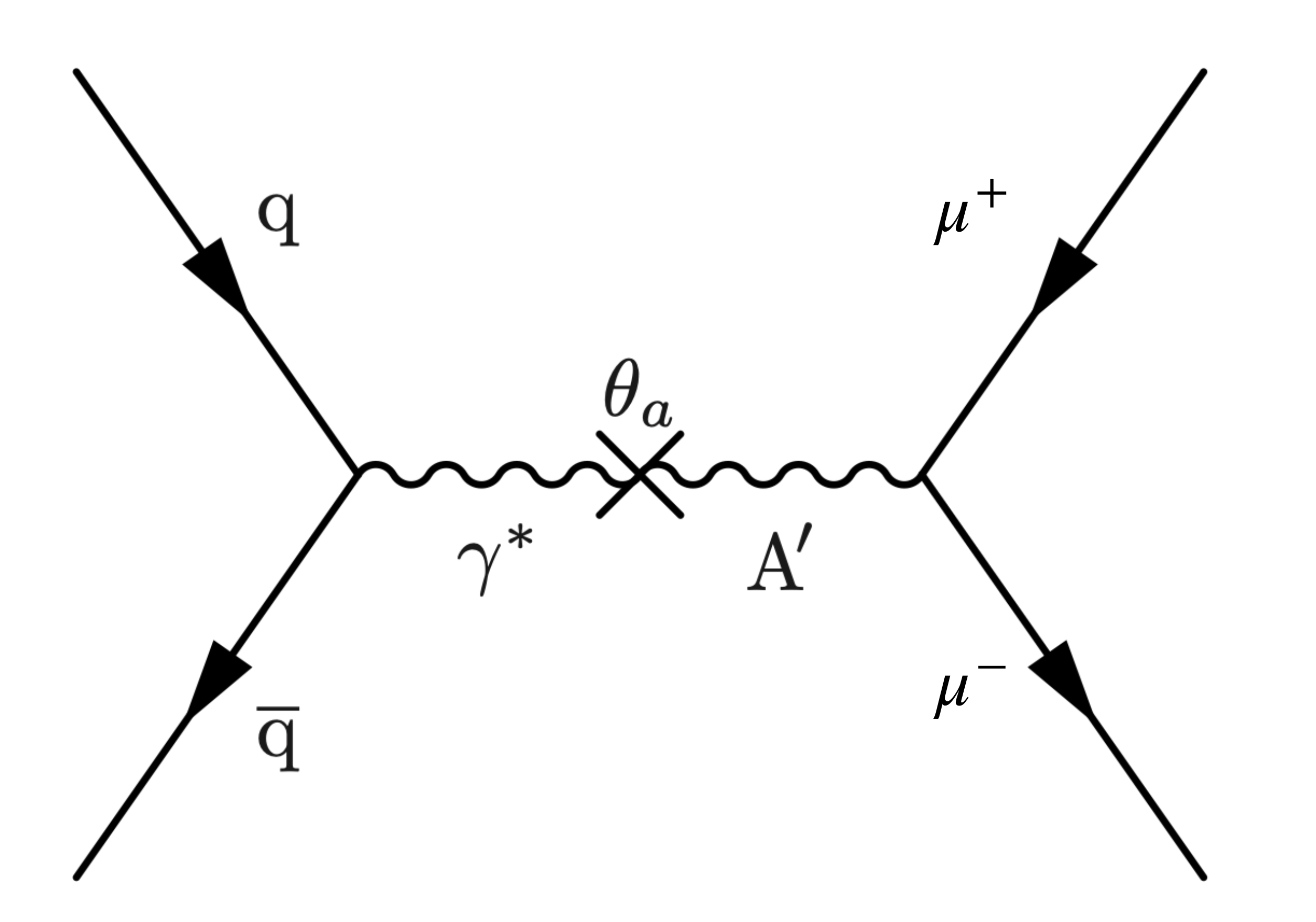}
\caption{Feynman diagram for a dark-photon mediator \PAprime, via mixing with the SM photon.}
    \label{fig:darkPhotonFeynmanDiagram}
\end{figure}

\begin{figure}[hbt!]
\centering
\includegraphics[width=0.97\linewidth]{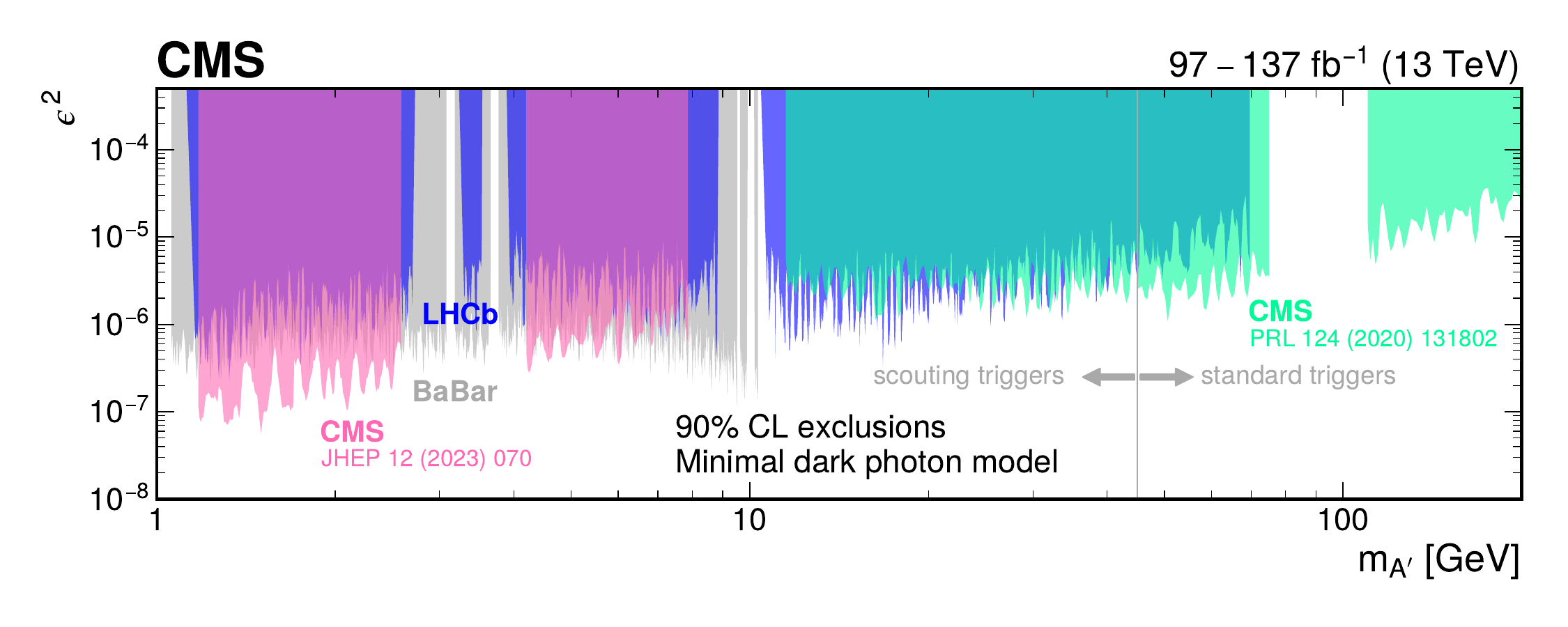}
\caption{Observed upper limits at 90\% CL on the square of the kinetic mixing coefficient $\epsilon$ in the minimal model of a dark photon from a CMS dimuon search~\cite{CMS:2023hwl} in the mass ranges of 1.1--2.6 GeV and 4.2--7.9 GeV (pink) and from another CMS dimuon search~\cite{CMS:2019buh} at larger masses (green). The limits are compared with the existing limits at 90\% CL provided by LHCb (blue)~\cite{LHCb:2017trq,LHCb:2019vmc} and BaBar (gray)~\cite{BaBar:2014zli}. From Ref.~\cite{CMS:2024zqs}.}
\label{fig:darkphotonscouting}
\end{figure}

An example of a simplified dark sector model is the minimal dark photon model. A dark photon (\PAprime) is a spin-1 mediator with a pure vector coupling that mixes with the SM Z boson. Here, we explore fully visible final states, such as those where the dark photon decays to dimuons, as illustrated in the Feynman diagram presented in Fig.~\ref{fig:darkPhotonFeynmanDiagram}. Figure~\ref{fig:darkphotonscouting} presents the 90\% CL limits on the squared kinetic mixing coefficient from the CMS prompt dimuon searches with and without data scouting~\cite{CMS:2023hwl,CMS:2019buh} as a function of \mAprime, along with the LHCb~\cite{LHCb:2017trq,LHCb:2019vmc} and BaBar~\cite{BaBar:2014zli} limits. \emph{Data scouting}~\cite{CMS:2024zhe} is a technique often used in CMS analyses that records events using a lighter data format instead of the full raw data, allowing for an increase in the data acquisition rate while keeping the bandwidth low. In particular, data scouting and dedicated selections have allowed CMS to probe dark photon masses down to about 1 GeV. Data scouting, together with and \emph{data parking}, which is another CMS technique used to collect data that is not reconstructed promptly after acquisition but instead is ``parked'' to be reconstructed later when the Tier-0 computing resources are idle, are the subject of a dedicated report~\cite{CMS:2024zhe}. Figure~\ref{fig:darkphotonscouting} shows that values of the squared kinetic mixing coefficient in the dark-photon model above are excluded at the $10^{-6}$ level for most of the dark-photon mass range of the searches.

\begin{figure}[hbt!]
    \centering
    \includegraphics[width=0.7\textwidth]{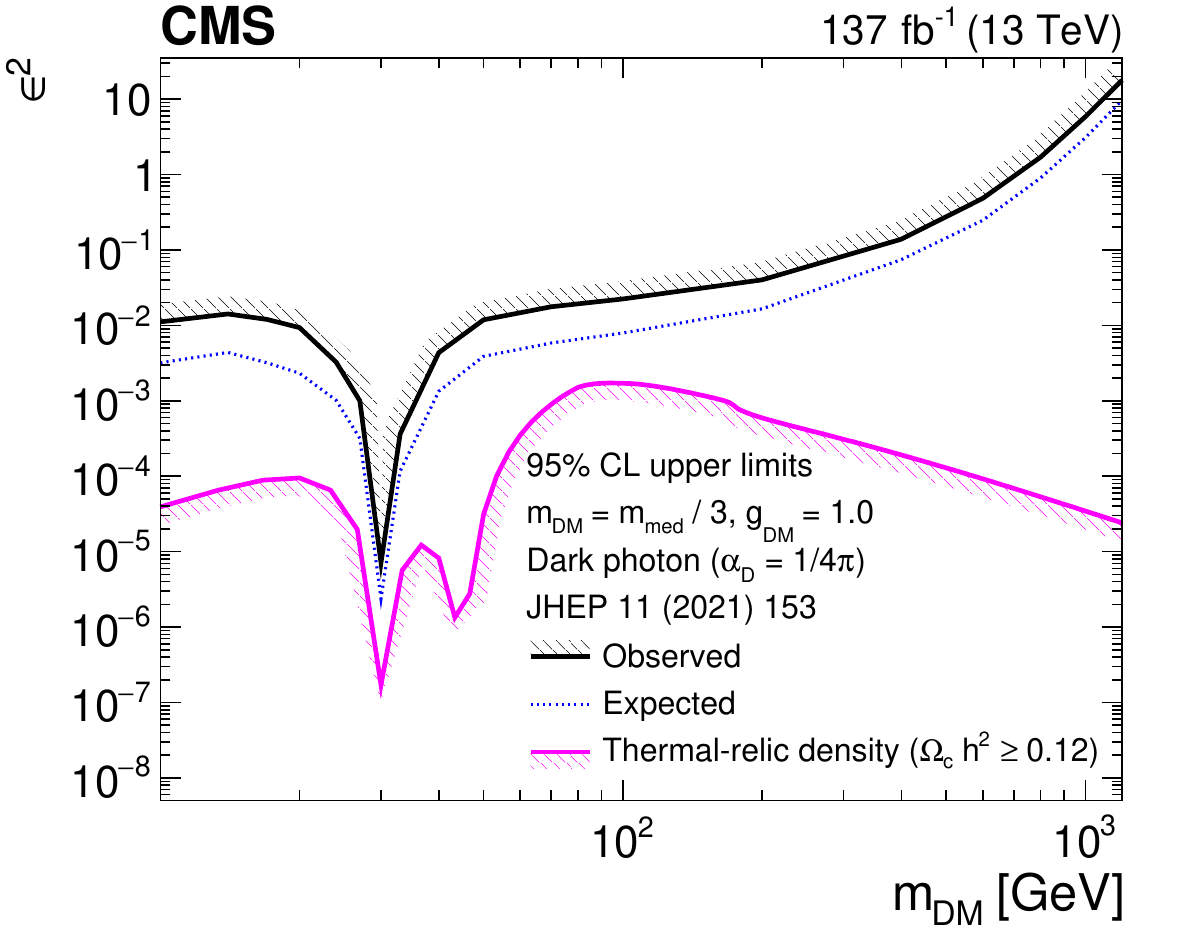}
    \caption{Limits at 95\% CL from a CMS monojet search~\cite{CMS:2021far} interpreted via \textsc{MadAnalysis}~\cite{DVN/IRF7ZL_2021} for a dark-photon model with a DM coupling. The limits are presented in terms of the mixing parameter $\epsilon^{2}$ with DM coupling $g_{\text{DM}} = 1.0$ and $\alpha_{\text{dark}} = g_{\text{DM}}^{2}/(4\pi)$. The constraint from the thermal-relic density ($\Omega_{c} \mathrm{h}^2 \geq 0.12$), obtained from WMAP~\cite{WMAP:2012fli} and Planck~\cite{Planck:2015fie}, is plotted in magenta. From Ref.~\cite{CMS:2024zqs}.}
    \label{fig:eps_darkpho}
\end{figure}

Figure~\ref{fig:eps_darkpho} presents the 95\% CL limits from a search for monojets~\cite{CMS:2021far} in a dark-photon model with a DM coupling. The exclusion is presented in terms of the mixing parameter $\epsilon^{2}$ as a function of the DM mass $m_{\text{DM}}$. The interference with the Z boson can be observed for $m_{\text{DM}}$ near $1/3$ of the Z boson mass $m_Z$, which leads to a more stringent limit in that region by up to three orders of magnitude. At small $m_{\text{DM}}$ values, ${<}1$ GeV, low-energy experiments are more sensitive~\cite{Antel:2023hkf}. The thermal-relic density additionally constrains $\epsilon^{2}$ to lower values for $m_{\text{DM}}$ near $m_Z/2$, when thermal Z boson production becomes resonant. The thermal-relic density constraint also tightens at large $m_{\text{DM}}$, which corresponds to large mediator masses. This occurs when the dark-photon width is dominated by kinetic mixing decays and is therefore proportional to $\epsilon^{2}$; when the mediator DM and SM couplings are of a similar order, the $\epsilon$ dependence in the DM annihilation cross section nearly cancels~\cite{Izaguirre:2015yja}.

\paragraph{Heavy neutral leptons} \label{sec:HNLs}
The neutrino portal is another example of a simplified dark sector model. In this portal, heavy neutral leptons (HNLs, denoted $N$)~\cite{Abdullahi:2022jlv,Dasgupta:2021ies} are sterile neutrinos with very small mixing with active neutrinos. HNLs often take the form of a right-handed neutrino, either paired with the three existing neutrino flavors or with a fourth, sterile neutrino. The HNL model is capable of producing the observed DM thermal-relic density when the HNL masses and flavors are adjusted appropriately. HNLs may exhibit different lifetimes based on their mass and mixing parameters. A detailed review of the CMS searches using HNL models is provided in Ref.~\cite{CMS:2024bni}.

\begin{figure}[hbt!]
\centering
\includegraphics[width=0.48\textwidth]{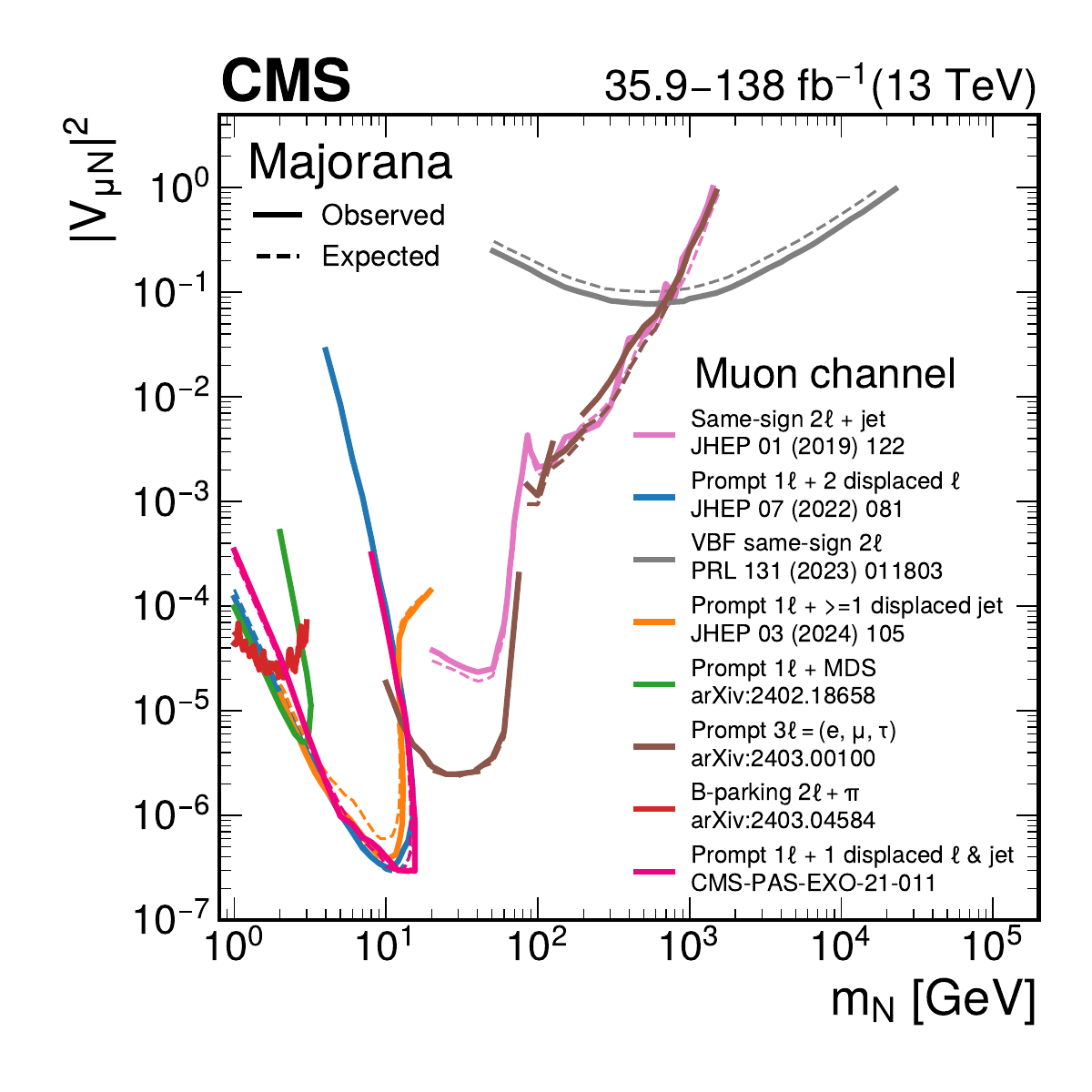}
\hfill
\includegraphics[width=0.48\textwidth]{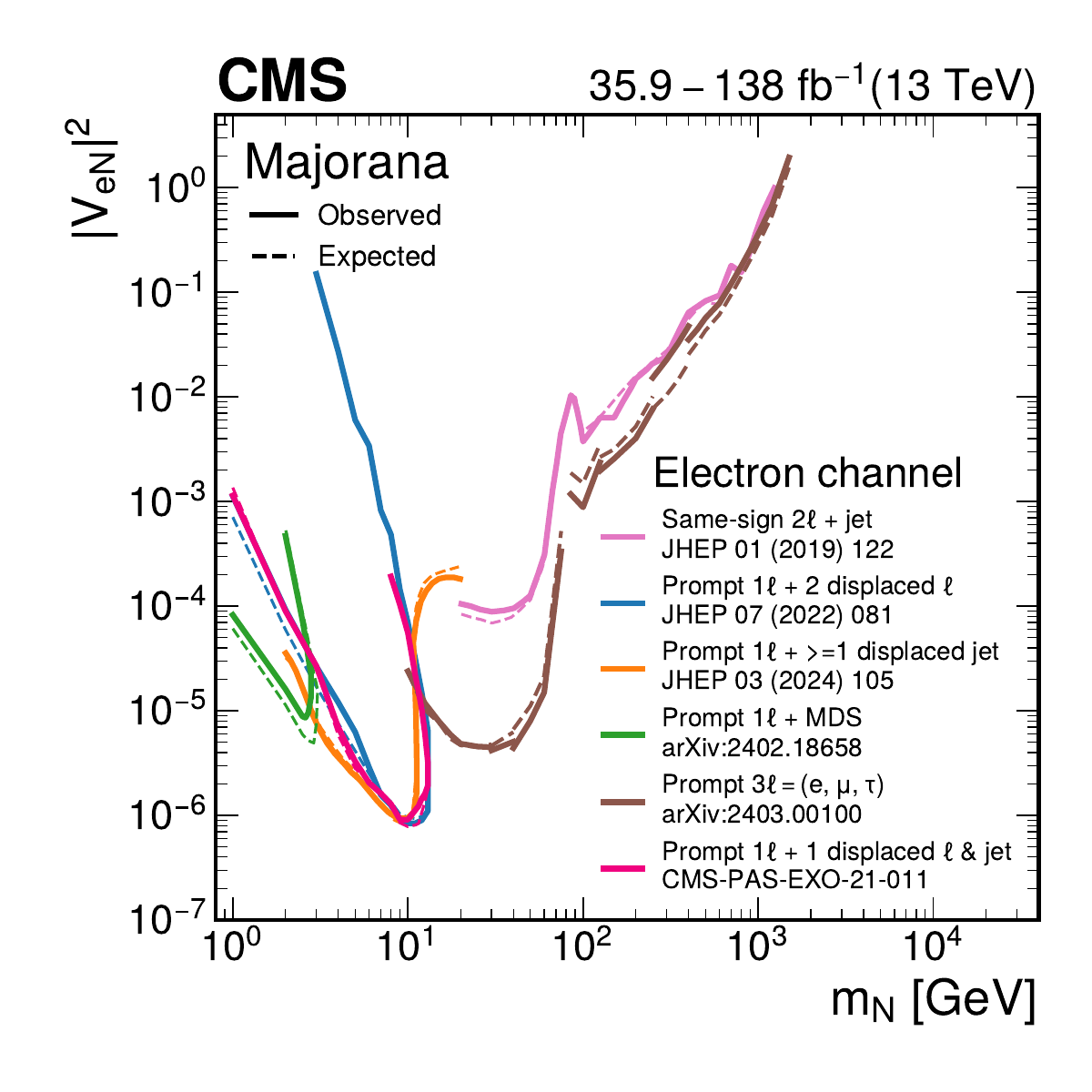} \\
\includegraphics[width=0.48\textwidth]{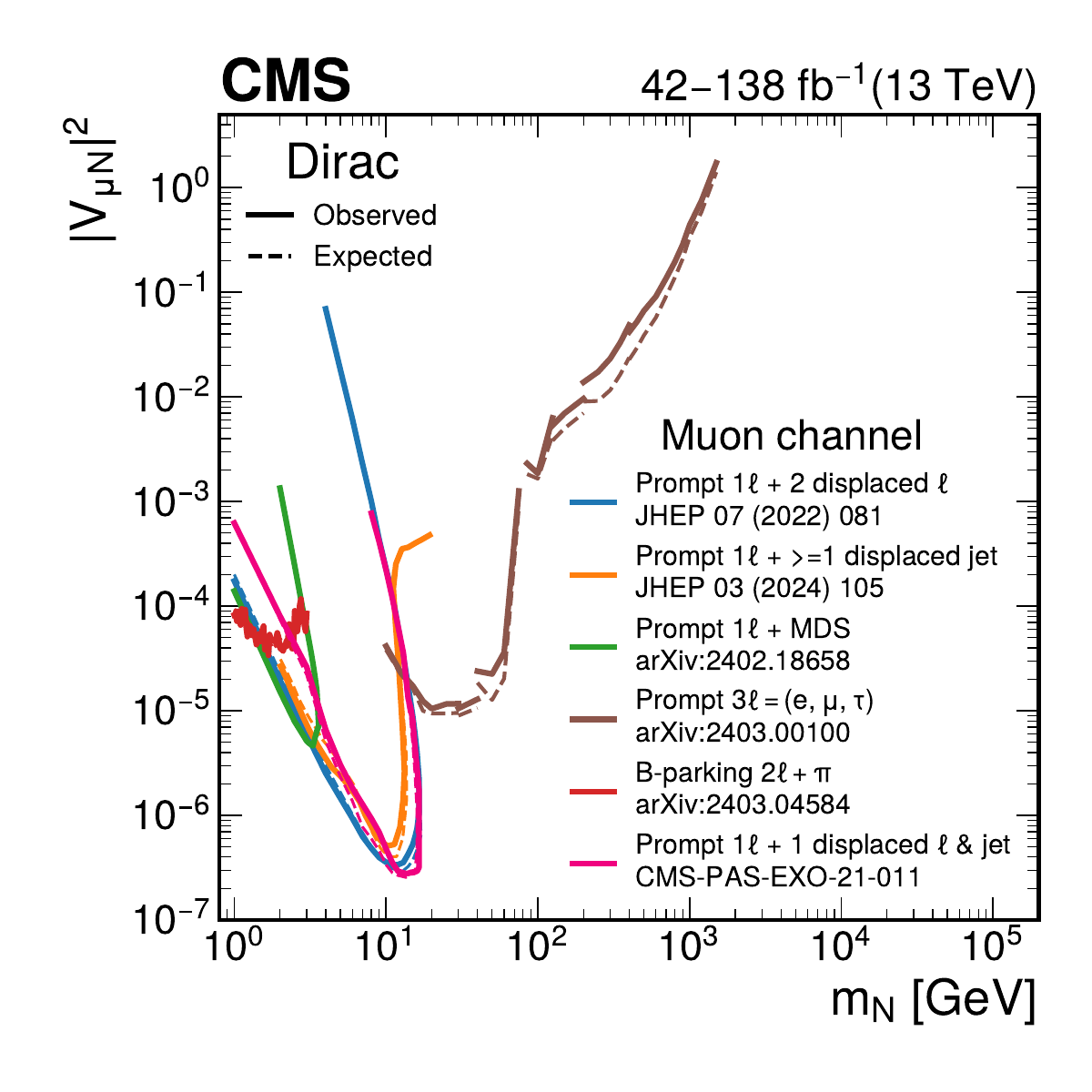}
\hfill
\includegraphics[width=0.48\textwidth]{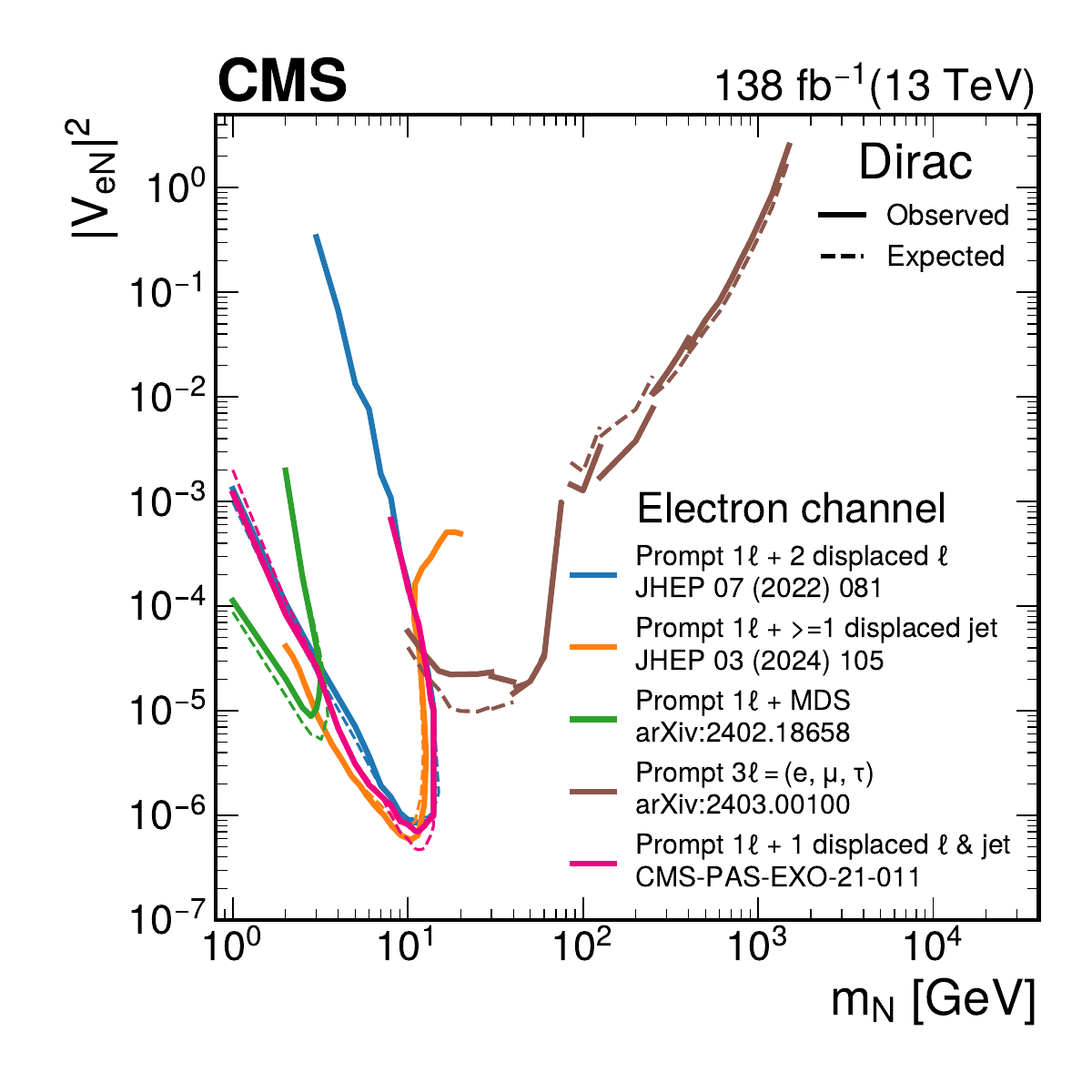}
\caption{Summary of searches at the CMS experiment for HNLs in the Type I seesaw model~\cite{CMS:2018jxx,CMS:2022fut,CMS:2022hvh,CMS:2023jqi,CMS:2024ake,CMS:2024xdq,CMS:2024ita,CMS:2024hik}. The expected and observed upper limits at 95\% CL on the mixing parameter \mixparsqlN as a function of the HNL mass \mhnl are shown, for Majorana and Dirac HNLs (upper and lower row, respectively), and in the muon and electron channel (left and right column, respectively). From Ref.~\cite{CMS:2024bni}.}
\label{fig:HNL}
\end{figure}

\begin{figure}[hbt!]
\centering
\includegraphics[width=0.48\textwidth]{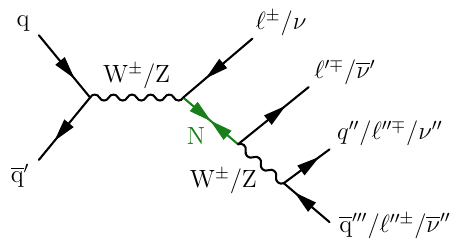}
\caption{Representative Feynman diagram of a Majorana HNL, labelled as N, produced through the decay of a W or Z boson.}
\label{fig:FD_HNLs}
\end{figure}

Figure~\ref{fig:HNL} presents a summary of the 95\% CL limits on the mixing parameter \mixparsqlN from prompt and long-lived Type I seesaw HNL searches, covering a broad mass range from 1 GeV to 10 TeV for the pure muon and electron mixing scenarios, for both Dirac and Majorana HNLs. A typical Feynman diagram for this case is shown in Fig.~\ref{fig:FD_HNLs}. For the high-mass regime, typically above $\approx$20 GeV, HNLs are expected to have a relatively short lifetime. Searches for short-lived (prompt) HNLs benefit from standard detector reconstruction techniques to capture their signatures in the CMS detector. In this mass region, when $N$ exclusively mixes with muons, the search for prompt $N \to ll\nu$~\cite{CMS:2024xdq} provides the most stringent limits in the \mhnl range from $\approx$20 GeV up to $\approx$100 GeV for both Dirac and Majorana HNLs. In the higher mass range, the search for prompt Majorana $N \to l q\overline{q}$~\cite{CMS:2018jxx} exhibits comparable sensitivity despite relying solely on the 2016 data set. This suggests that expanding this search using the full Run 2 data set and probing Dirac HNLs could enhance the reach in this parameter space region. In addition, the vector boson fusion (VBF) search~\cite{CMS:2022hvh} proves valuable in covering the very high mass region, where it attains the strongest constraints up to $\mhnl\approx10$ TeV.

In the low-mass regime, typically below $\approx$20 GeV, HNLs are anticipated to be long-lived. Searches for HNLs with long lifetimes rely on innovative techniques to reconstruct the displaced decays of the HNLs. Techniques targeting displaced $N \to ll\nu$ and $N \to l q\overline{q}$ decays within the tracker volume, such as displaced-vertex reconstruction and displaced jet tagging, are used, leading to the most stringent limits in the mass range of 3 to 20 GeV~\cite{CMS:2022fut,CMS:2023jqi,CMS:2024hik}. Notably, for lower masses between 1 and 3 GeV, the search utilizing a muon detector shower (MDS) signature~\cite{CMS:2024ake} results in the strongest bounds. An MDS could arise from the decay of a neutral LLP with a large lifetime ($c\tau>1~m$) that decays and produces a high-multiplicity shower (hundreds of hits per cluster) in the CMS muon system. Between 1 to 2 GeV, the search using the B-parking data set~\cite{CMS:2024ita} provides the most stringent limits for muon-type HNLs, due to the large cross section of B meson production. These searches are complementary, with each dominating in a specific mass region. The successful implementation of challenging reconstruction methods for the detection of long-lived HNLs has significantly increased their discovery potential at the LHC.

\paragraph{Higgs boson decays into LLPs}\label{sec:HiggsToLLPs}
To solve the electroweak hierarchy problem, one can introduce new discrete symmetries. In this case, the new particles need not carry SM charges and thus can naturally arise from a dark sector or hidden valley. Consequently, the top quark partner is not abundantly produced at the LHC and evades the stringent constraints on colored top quark partners. This scenario is referred to as ``neutral naturalness'', realizations of which include the twin Higgs~\cite{Chacko:2005pe}, folded SUSY~\cite{Burdman:2006tz}, and quirky little Higgs~\cite{Cai:2008au} models. A common benchmark signature is the exotic decay of the 125 GeV Higgs boson to two long-lived scalars, each of which further decays to a pair of displaced particles~\cite{Craig:2015pha,Curtin:2015fna}.

\begin{figure}[hbt!]
    \centering
    \includegraphics[width=0.3\linewidth]{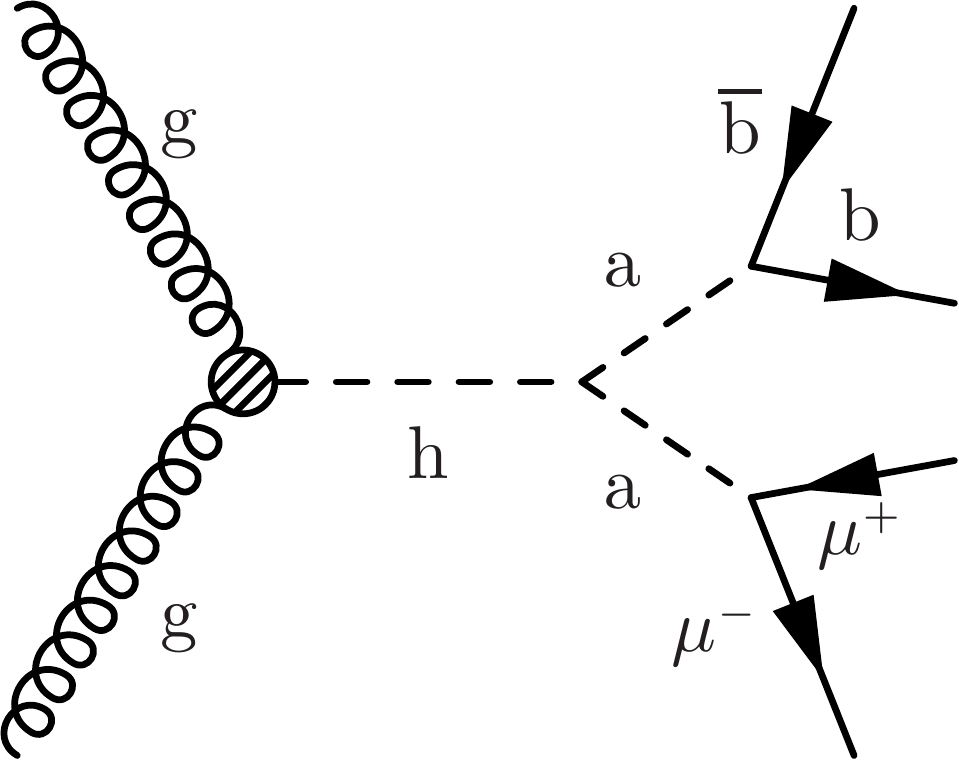}
    \caption{Representative Feynman diagram for an exotic decay of the SM-like Higgs boson ($h$) decaying to two LLPs ($a$).
    }
    \label{fig:exoticHiggsFeynmanDiagram}
\end{figure}

\begin{figure}[hbt!]
\centering
\includegraphics[width=0.65\linewidth]{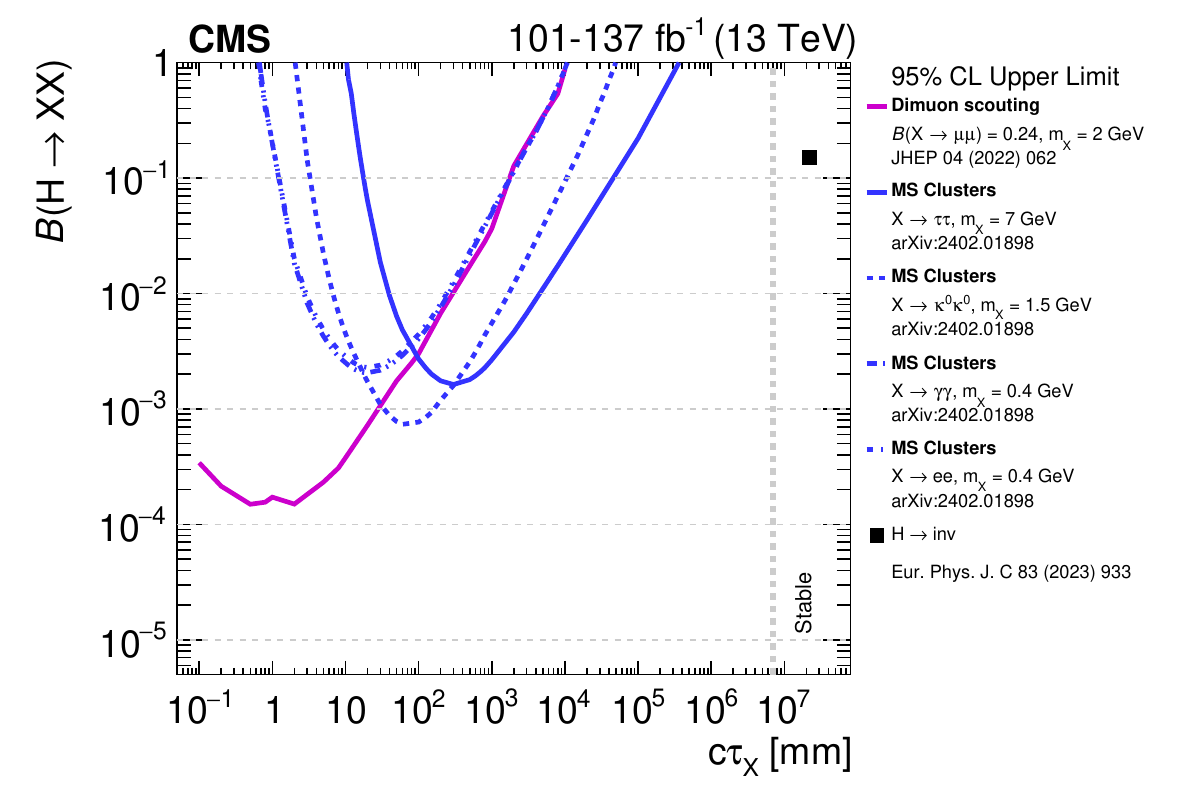}
\includegraphics[width=0.65\linewidth]{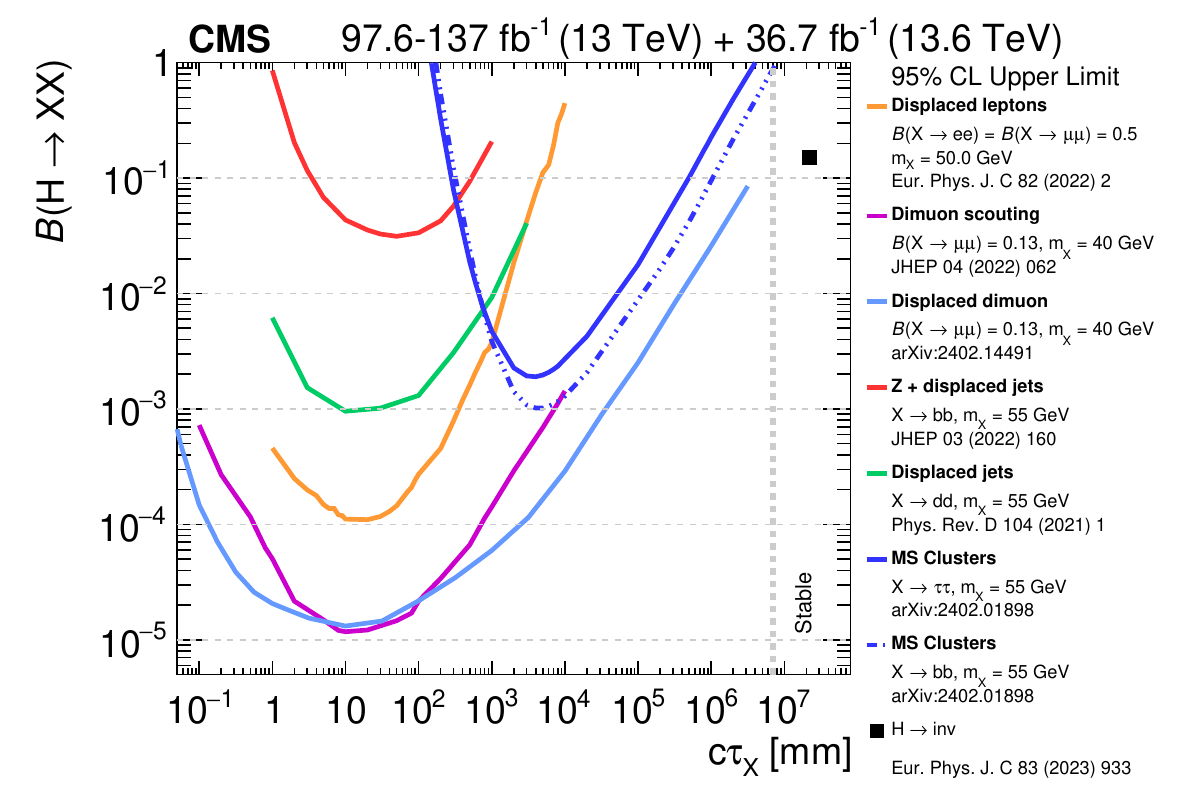}
\includegraphics[width=0.65\linewidth]{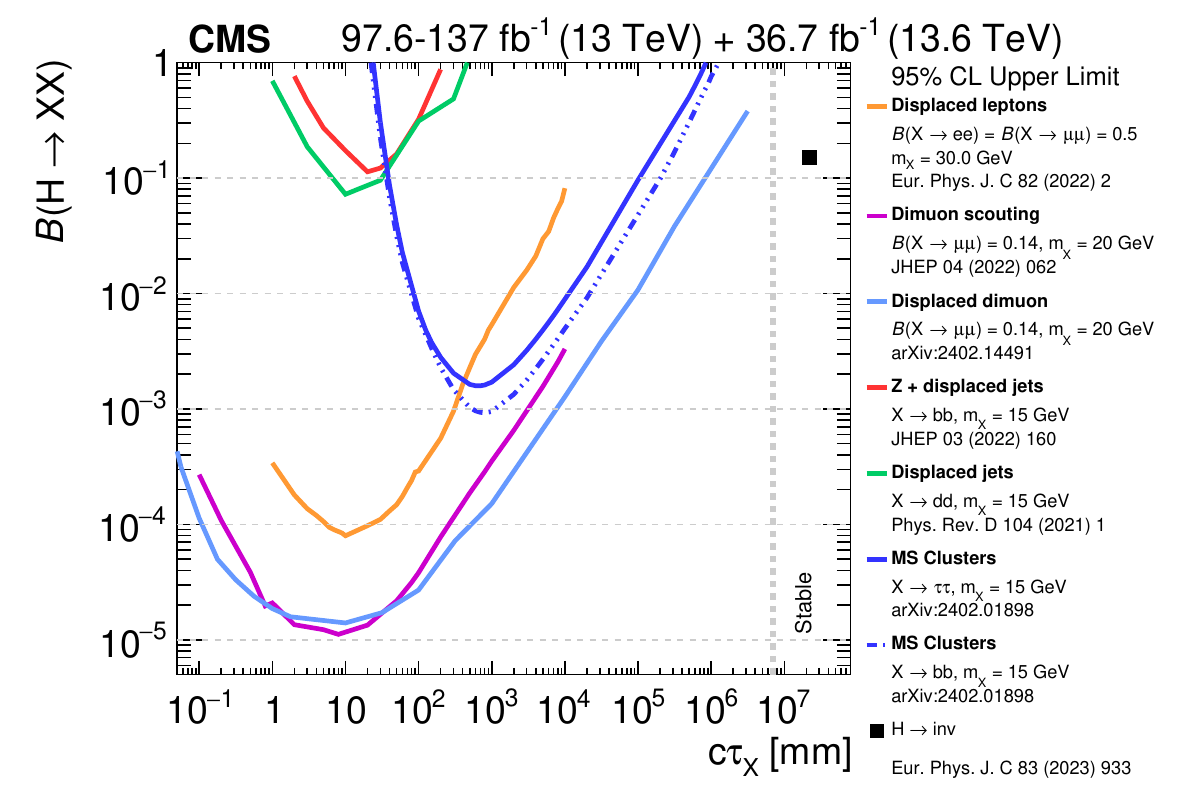}
\caption{Observed 95\% CL upper limits on the branching fraction of Higgs bosons decaying into LLPs with masses between 0.4 and 7 GeV (upper), 15 and 30 GeV (middle), and 40 and 55 GeV (lower)~\cite{CMS:2021kdm,CMS:2021sch,CMS:2024qxz,CMS:2021yhb,CMS:2020iwv,CMS:2023arc,CMS:2023sdw}, for various CMS LLP searches. The LLP mass and decay assumptions are given in the legend. From Ref.~\cite{CMS:2024zqs}.}
\label{fig:higgsllp}
\end{figure}

Exotic decays of the Higgs boson into LLPs are well motivated in a variety of models, such as those motivated by neutral naturalness. Several CMS searches have been reinterpreted in a scenario in which an exotic Higgs boson is produced in $pp$ collisions and then decays into two LLPs, here denoted $a$, as shown in Fig.~\ref{fig:exoticHiggsFeynmanDiagram}. These interpretations are shown in Fig.~\ref{fig:higgsllp}. The 95\% CL upper limits on the branching fraction of Higgs bosons decaying into LLPs with hadronic and leptonic decays and different masses (specified in the legends) are shown as functions of the LLP proper decay length. The dedicated LLP searches presented in these figures probe a variety of final states, and LLP masses and lifetimes. These are shown to provide complementary sensitivity for lifetimes between 0.1 and $10^6$~mm that far exceeds the inclusive limit achieved by the \hinv analysis~\cite{CMS:2023sdw}. For example, a search employing the MDS signature~\cite{CMS:2023arc} is highly sensitive at small masses and large lifetimes, and a search that uses dimuon scouting triggers~\cite{CMS:2021sch} is powerful at small masses and small lifetimes. In addition, the displaced dimuon search~\cite{CMS:2024qxz} is highly inclusive and sets stringent limits for a wide range of lifetimes. The sensitivity to models with smaller lifetimes is limited by the background from B meson decays and the resolution of the tracker. The CMS reach in this regime could be extended further by considering the sensitivity of prompt searches to b quark final states.

\paragraph{Dark Higgs bosons}\label{sec:darkHiggs}

\begin{figure}
    \centering
    \includegraphics[width=0.33\textwidth]{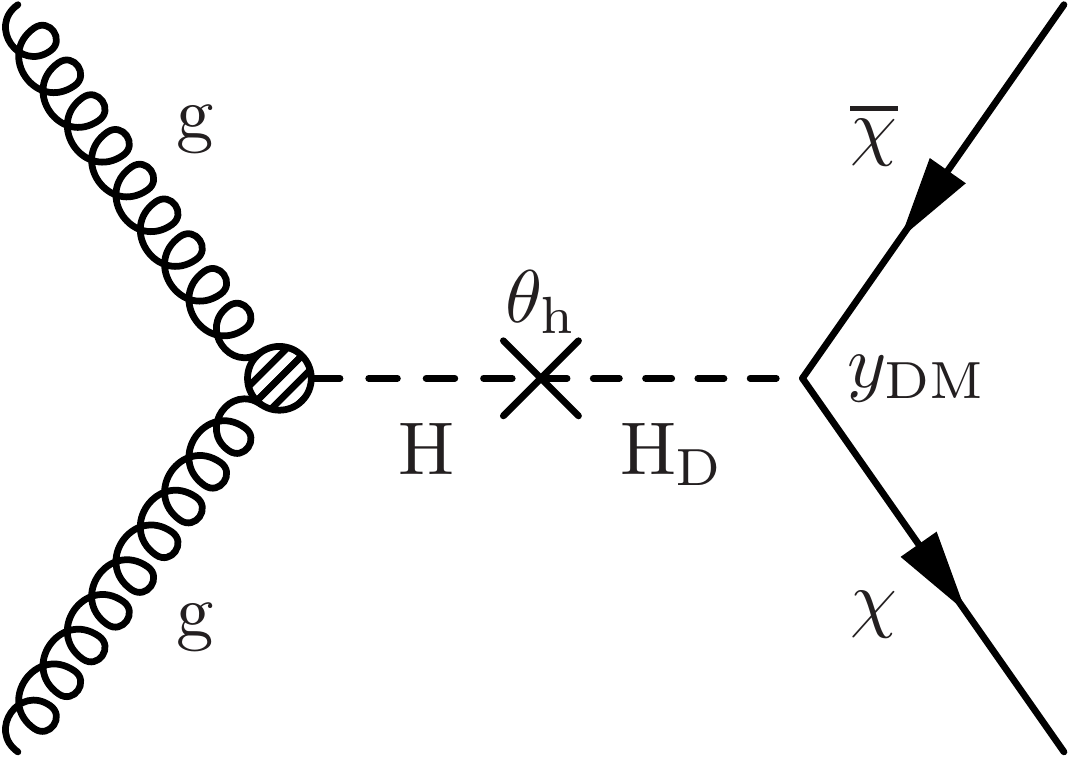}
\caption{Feynman diagram for a dark Higgs mediator $H_{\text{dark}}$ produced via mixing $\theta_{\text{h}}$ with the SM Higgs boson $H$.}
    \label{fig:darkHiggsFeynmanDiagram}
\end{figure}

The spin-0 portal consists of a scalar or pseudoscalar mediator that couples to DM. In a dark Higgs boson model, it is assumed that this scalar mediator acts as the portal to the dark sector and mixes with the SM Higgs boson, giving rise to the Higgs boson decay into invisible final states ($H \to \text{inv}$) signature that is the cornerstone of many DM searches. The mixing of the dark Higgs boson with the SM Higgs boson is typically written in terms of the angle $\theta_{\text{h}}$ between these two bosons. This scenario is shown in the Feynman diagram in Fig.~\ref{fig:darkHiggsFeynmanDiagram}.

\begin{figure}[thb!]
    \centering
    \includegraphics[width=0.6\textwidth]{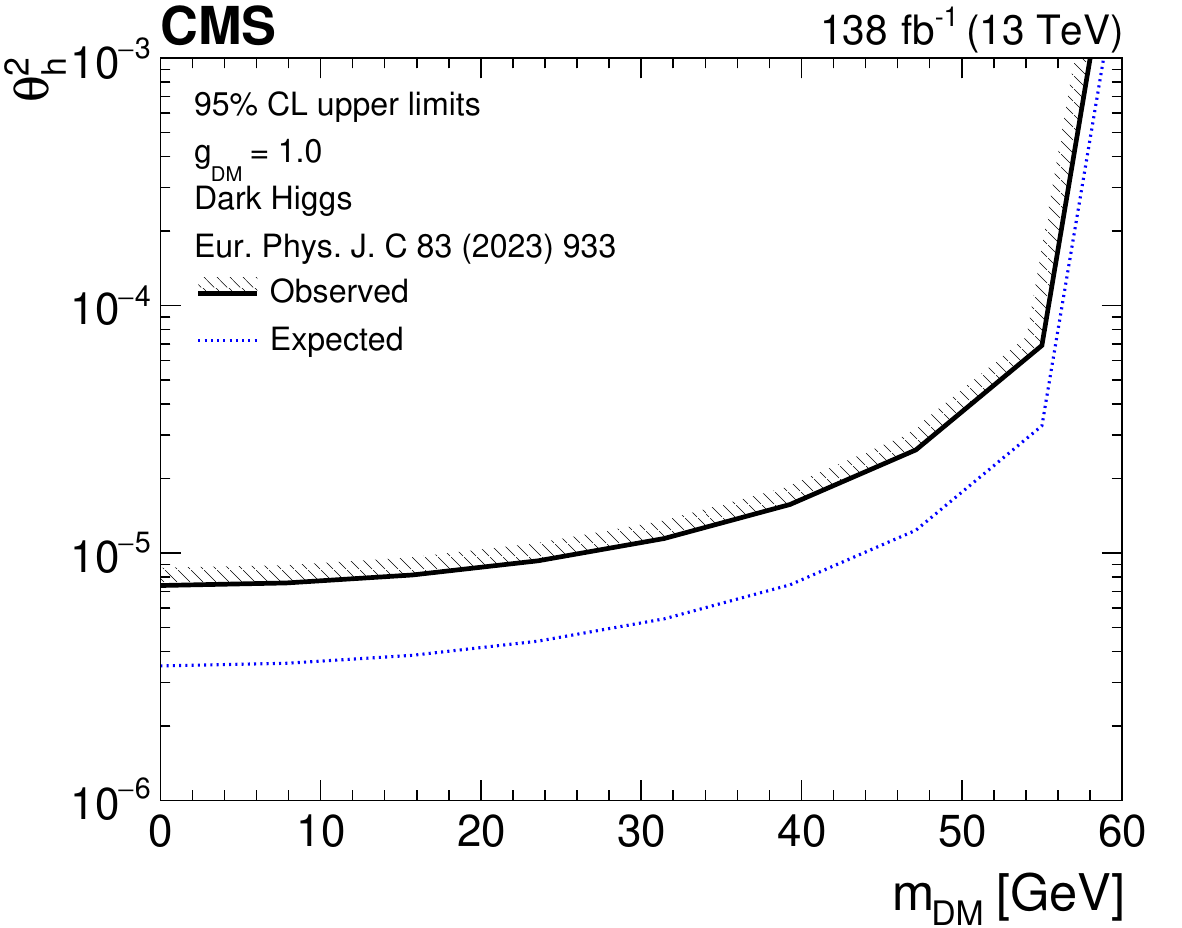}
    \caption{95\% CL upper limits on the mixing parameter $\theta_{\text{h}}^{2}$ from a CMS $H \to \text{inv}$ analysis~\cite{CMS:2023sdw} interpreted with a dark-Higgs boson model. From Ref.~\cite{CMS:2024zqs}.}
    \label{fig:darkhiggsthetaH}
\end{figure}

The limits on $\mathcal{B}(H \to \text{inv})$ can be reinterpreted to place limits on the mixing parameter $\theta_{\text{h}}$, as shown in Fig.~\ref{fig:darkhiggsthetaH}. The exclusion worsens as $m_{\text{DM}}$ approaches $m_H/2$. These results are largely independent of the dark-Higgs boson mass. In these limits, the predicted thermal-relic density for the prescribed couplings and fermionic DM yields an overabundance of DM by several orders of magnitude~\cite{Krnjaic:2015mbs}.

\begin{figure}[hbt!]
    \centering
    \includegraphics[width=1.0\textwidth]{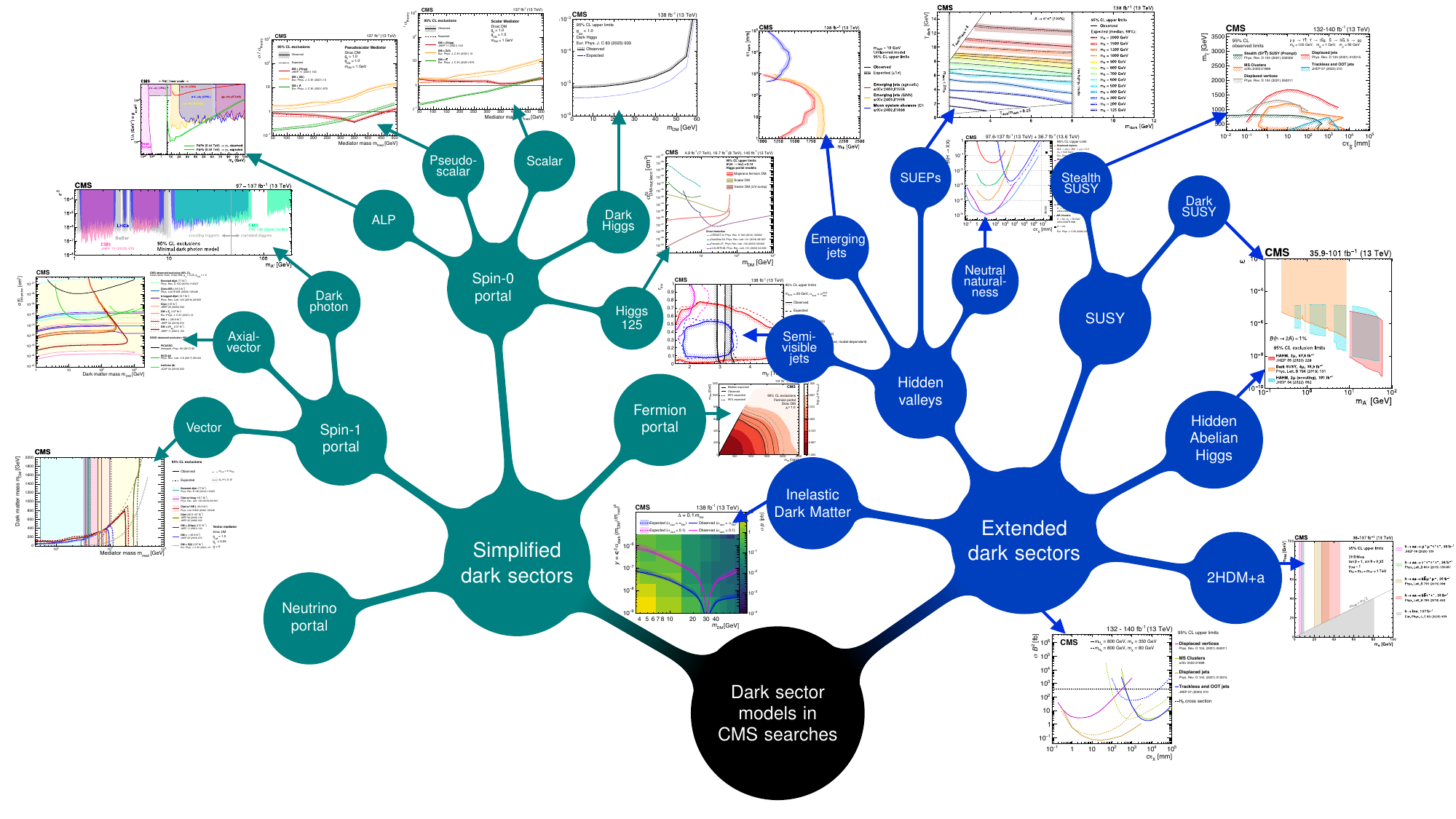}
    \caption{A qualitative depiction of how the results map onto the models probed in CMS searches for dark sectors. From Ref.~\cite{CMS:2024zqs}.}
    \label{fig:summarySketch}
\end{figure}

\paragraph{Summary of CMS results and prospects}
A comprehensive review of dark sector searches with the CMS experiment at the LHC has been presented in Ref.~\cite{CMS:2024zqs}, using proton-proton and heavy ion collision data collected in Run~2, from 2016 to 2018, or, in some cases, from Run~1 (2011--2012) or Run~3 (2022). These searches have been interpreted in simplified and extended dark sector models. Figure~\ref{fig:summarySketch} qualitatively illustrates how the CMS results map into this theoretical framework. The broad CMS dark sector search program spans many different signatures, including those with invisible particles, those with particles promptly decaying into fully visible final states, and those with LLPs. The broad variety of searches provides sensitivity across a wide range of models and parameter space, and the results represent the most complete set of constraints on dark sector models obtained by the CMS Collaboration to date.

Several unique techniques of data collection, reconstruction, and analysis were developed during Run~2 to increase the sensitivity of these searches. For example, data scouting has been exploited for displaced muons for the first time, significantly expanding sensitivity to low-mass resonances. In addition, entirely new LLP reconstruction techniques have been deployed to significantly expand the sensitivity to displacements beyond the tracker, including the first uses of delayed calorimetry for hadronic final states, the development of muon detector shower reconstruction for both hadronic and electromagnetic final states, and the development of new reconstruction algorithms for highly displaced muons. Together, these and other developments have greatly extended the ranges of lifetimes, masses, and even types of signatures that can be probed by the CMS detector. This expansion in accessibility can be seen in the sensitivity across up to seven orders of magnitude in signal lifetime, from 0.1 to $10^6$~mm, for the wide array of displaced signatures.

Future improvements will further increase the sensitivity of searches for LLPs and other novel final states. For Run~3 of the LHC~\cite{CMS:2023gfb}, a range of new displaced trigger algorithms have been developed and deployed for both the Level 1 and the high-level trigger. This includes entirely new delayed calorimetry, using both the electromagnetic and the hadronic calorimeter, and muon detector shower algorithms to increase acceptance for many models by over an order of magnitude~\cite{LLPRun3TriggerDPNote}. New machine-learning-based anomaly detection triggers, using calorimeter or global information, are also being deployed to access final states and kinematic ranges that are not covered by conventional triggers~\cite{DP-2023-079,DP-2023-086}. The performance of data scouting for displaced muons is improved in Run~3 by removing the requirement of a hit in the pixel tracker, thus extending the sensitivity to larger lifetimes. There are also opportunities to employ data parking for several displaced signatures. CMS will continue to seek out opportunities to improve the performance of the full range of prompt and displaced reconstruction algorithms.

The High-Luminosity LHC will include multiple detector upgrades that will substantially enhance the performance of the techniques that have been explored with the current calorimetry, tracking, and trigger capabilities~\cite{Contardo:2015bmq,CMS:2017lum,CERN-LHCC-2017-011,CMS:2017jpq,Hebbeker:2017bix,Zabi:2020gjd,CMS:2667167,CMS-DP-2022-025}. Timing resolution of order 10~ps will be available across multiple upgraded and new subsystems, and there will be a new calorimeter to provide high-granularity energy and position information in the forward region. For the Level 1 trigger, tracking will be implemented, and sophisticated machine-learning algorithms will be deployed. Finally, increasing the integrated luminosity by an order of magnitude will allow probing rarer processes with smaller cross sections, exceeding existing limits on mediator particles. Together, these upgrades will substantially improve the sensitivity to a wide range of dark sector models, as shown in several studies of the physics performance at the High-Luminosity LHC~\cite{CMS:2022cju,Dainese:2019rgk,CMS:2022sfl}.

\subsection{ATLAS, CMS, LHCb: Discussion on displaced objects}
\label{sssec:displaced_objects}

The LHCb experiment, although originally designed to study the decays and properties of heavy-flavor~\cite{LHCb:2008vvz} particles, has significantly extended its physics program in the past years and becoming a general-purpose detector (GPD), featuring Higgs and electroweak measurements~\cite{LHCb:2025nob}, heavy-ions measurements~\cite{LHCb:2025elk}, and searches for direct new physics, both prompt~\cite{LHCb:2020ysn} and displaced~\cite{LHCb:2021dyu}. A comparative discussion of the capabilities of LHCb to ATLAS and CMS on searches for displaced objects is presented in the following paragraphs, in the context of the production vertex, mass, and lifetime of the new physics displaced object. 
This discussion aims to serve as a guide not only for the theory community but also for data analysts, to help understand which benchmark models could be of better interest for each detector, given their search capabilities.

\subsubsection{Advantages and disadvantages}
Concerning displaced new physics objects, the LHCb experiment could be considered as a natural search apparatus for moderately {\bf{long-lived}} and {\bf{low-mass particles}}, given the typical mass and displacement ranges of heavy-flavor\footnote{Containing a $b$ or a $c$ quark.} particles: masses from 1 to 10 GeV with lifetimes down to 1 picosecond. 
Within the quoted ranges, very precise mass and momentum resolutions are mandatory to achieve the required, and highly discriminant, background separation. 
These capabilities are achieved thanks to the combination of a series of features, in particular,
\begin{itemize} 
\item the capability to select {\bf{very low-momentum particles}} and reconstruct displaced vertices in the forward region, thanks to the formidable Vertex Locator (VELO); a fast software-based first-level trigger as discussed in Sec.~\ref{sssec:casais-vidal}, and,
\item an {\bf{unique particle identification}} and hadron separation capabilities with ML-based algorithms trained with information from the muon stations, the calorimeter system, and the Cherenkov detectors.
\end{itemize}

However, the LHCb experiment has a series of disadvantages in comparison with ATLAS and CMS, concerning the instantaneous luminosity delivered to the detector, together with its hermeticity and acceptance. More in detail:
\begin{itemize}
    \item The LHCb detector is a forward single-arm spectrometer, with reduced acceptance in a pseudorapidity range of $2<\eta<5$, this is, covering approximately up to 4~\% of the solid angle of that range. This design choice is motivated by the collimated angular production of heavy quarks in proton-proton collisions, favoring forward rather than transverse regions. However, this {\bf{greatly reduces the detector capabilities to reconstruct}} particles produced with a high degree of transversality, that is, {\bf{high-mass particles}} up to a few GeV above the Higgs boson mass, becoming a more pronounced effect when the final state particles have lighter masses.
    \item This reduced acceptance also significantly {\bf{decreases its ability to study signatures with an invisible component}} by using missing transverse momentum or energy techniques, as ATLAS and CMS collaborations do in their many results on searches involving invisible particles.
    \item Finally, an instantaneous luminosity reduction at the proton-proton interaction point, to ensure a precise separation between displaced vertices reconstructed by the vertex detectors, is a practice not exclusive to, but mainly used by, the LHCb experiment. Reducing to an average number of few collisions per bunch crossing, to allow the VELO to achieve an outstanding vertex resolution, it is done via a technique known as {\it{luminosity leveling}}, {\bf{having a sizable impact in the integrated luminosity recorded by LHCb}}, around one order of magnitude less than the dataset recorded by ATLAS and CMS.
\end{itemize}
In summary, the main advantages of LHCb to ATLAS and CMS are its excellent low mass and momentum resolution, allowing it to trigger on and reconstruct very low momenta objects, resolving low masses from MeV to GeV, having a very high precision as well for moderately low displacements down to a picosecond and even hundreds of femtoseconds. Its main disadvantages, on the other hand, are its reduced acceptance and luminosity, and non-hermeticity, reducing the sensitivity of the detector to masses above the electroweak scale, very transverse objects, or signatures featuring invisible particles. 
This allows LHCb to focus its scouting efforts in a region of the parameter space complementary to ATLAS and CMS: lower FIP masses, especially those produced directly from the proton-proton collision, deemed inaccessible to ATLAS and CMS due to trigger bandwidth restrictions, which become even more critical for hadronic final states; and moderately-low FIP displacements, benefiting from the outstanding capabilities of the LHCb VELO. Therefore, ATLAS and CMS would focus their efforts on moderately higher masses and larger displacements, including final states with invisible particles and very transverse objects.
\subsubsection{Latest technological improvements}
However, this complementarity depends on the production mode considered and on the continuously evolving technological improvements to the detectors, creating overlap regions in the parameter space among the three experiments. ATLAS and CMS have been able to reach even lower masses during the past years, while LHCb can now achieve a high-precision track reconstruction for higher displacements:
\begin{itemize}
\item For instance, while LHCb can reach a higher precision for low-mass objects directly produced from proton-proton collisions, the detector can also study FIPs produced from B-meson (Higgs) decays since this is simply translated into a more displaced (transversal) object, and where ATLAS and CMS have a competitive (better) precision with respect to LHCb: the ATLAS collaboration has been developing dedicated data-handling {\bf{trigger techniques to access low masses}}, as shown in Sec.~\ref{sssec:torro-pastor}; while the CMS experiment can now access masses as low as 1 GeV on HAHM-produced dark photon searches thanks to their improved {\bf{data scouting and data parking techniques}}, as presented in Sec.~\ref{sssec:alimena}. 
\item While the BSM displaced searches in LHCb using Run 1 and 2 data used {\it{long}} tracks, this is, tracks reconstructed using hits in all LHCb substations (VELO, UT and SciFi), therefore, produced from decay vertices in the VELO region; {\bf{new tracking algorithms for higher displacements}} using other track types {\bf{have been developed and implemented in the LHCb trigger and reconstruction system during Run 3}}. These "other" track types considered for displaced BSM searches are two: {\it{downstream}} tracks, where hits from the UT and SciFi are used; and {\it{T-tracks}}, where only hits from SciFi stations are used. The track types are not used only for BSM searches but also for SM analyses such as measurements of $\Lambda$ baryons, with lifetimes of the order of 200--300 ps, and have been used also during previous LHCb runs, but not with such higher precision as with the new algorithms implemented in Run 3 \cite{Kholoimov:2025cqe}. More details on these new algorithms and studies are presented in Sec.~\ref{sssec:zhuo} for {\it{downstream}} tracks, and in Secs.~\ref{sssec:sanderswood} and \ref{sssec:collaviti} for {\it{T-tracks}}.
\end{itemize}
\subsubsection{Summary}
In conclusion, there are regions of the parameter space where LHCb achieves better precision than ATLAS and CMS, and vice versa, due to the intrinsic design of the detectors. 
There are also regions where complementarity among GPDs can be exploited, and where there is still room for future improvements during Run 4 and beyond. A summary is presented in the following paragraph.
Search capabilities regarding the mass of the displaced object are essentially related to: (i) the momentum resolution and vertex reconstruction capability of the detector towards discriminating against backgrounds, and to (ii) the ability of the detector trigger system to record and process large data volumes.
\begin{itemize}
    \item {\bf{Higher masses:}} The higher the mass, the higher the probability of the displaced object to decay outside the LHCb fiducial volume. {\bf{Displaced objects with masses above the electroweak scale are a search region mostly specific to ATLAS and CMS}}, which have a bigger detector fiducial volume with full coverage.
    \item {\bf{Lower masses:}} The lower the mass, the harder it is to suppress against low-mass QCD backgrounds. However, there is also a dependence on the production of the displaced object:
    \begin{itemize}
        \item On the one hand, if produced directly from the proton-proton collision, LHCb has a better precision for electron/muon signatures up to roughly 10 GeV, even higher for hadronic final states and hadronic $\tau$-lepton decays. This happens mainly due to the LHCb fully software-based trigger system, which has no low-$p_T$ threshold. {\bf{This mass region could be considered, for now, highly specific to the LHCb detector.}}
        \item On the other hand, if the displaced object is produced from another high mass object, the higher the mass, the better ATLAS and CMS precisions will be. This happens thanks to the ATLAS and CMS ability to overcome their trigger bandwidth nominal restrictions with smart dedicated techniques, such as having dedicated streams or using data parking and scouting. This case is where the three experiments are competitive, still having ATLAS and CMS room for improvement.
    \end{itemize}
\end{itemize}
The ability to reconstruct lower displacements relies on the precision of the vertex detectors of each experiment, while for higher displacements, both the detector design and the precision of the tracking systems become highly relevant.
\begin{itemize}
    \item {\bf{Lower displacements:}} LHCb can resolve very low displacements thanks to the excellent vertex resolution of the LHCb VELO, required to resolve $B_s^0$ meson oscillations, which becomes crucial to study $\mathcal{CP}$-violation in the heavy flavor sector. For this purpose, the LHCb VELO resolution goes way beyond the picosecond, a region specific to LHCb where ATLAS and CMS may categorize particles already as prompt.
    \item {\bf{Higher displacements:}} ATLAS and CMS naturally have a better precision on high displacements due to a larger detector size with full coverage. Hermeticity also allows ATLAS and CMS to apply missing momentum/energy techniques to study signatures with invisibles in the final state. ATLAS has developed a dedicated algorithm, LRT or Large Radius Tracking, for even higher displacements, as already mentioned. However, the LHCb experiment is extending its displacement reach thanks to the development and implementation of novel tracking algorithms in Run 3, using both {\it{downstream}} tracks and {\it{T-tracks}}, extending the reach of LHCb up to the nanosecond. From Run 3 onwards, the region of higher displacements has become an area of interest for LHCb, ATLAS, and CMS. LHCb has room for improvement, especially during the Upgrade 2, with new tracking algorithms based on {\it{Upstream}} tracks, as discussed in Sec.~\ref{sssec:langenbruch}.
\end{itemize}
A final comment can be made concerning displaced objects with low masses, produced from displaced objects, {\it{e.g.}} heavy neutral leptons. In this case, both ATLAS and CMS improvements to search for lower mass objects, and the LHCb new tracking algorithms to reconstruct more displaced tracks, enter into direct competition, depending on the mass and displacement ranges considered. 

A conclusion from this comparative discussion may be the fact that all these improvements are highly beneficial for both the theory and experimental community, allowing each experiment to provide complementary results in regions where all of them are competitive, while, the existing room for improvement becomes a motivation for all the experiment to enhance their capabilities and reach for BSM displaced objects, helping to unveil the mass-lifetime parameter space as much as possible.

\subsection{FIPs at fixed-target experiments: NA62 and kaon experiments --- \textit{E.~Goudzovski}}
\label{ssec:fips_at_NA62}
\textit{Author: Evgueni Goudzovski, \email{evgueni.goudzovski@cern.ch}} \\
Rare kaon decays represent sensitive probes of both heavy and light new physics. In terms of searches for FIPs, a comprehensive summary of the relevant phenomenological models and experimental results, and prospects is provided in Ref.~\cite{Goudzovski:2022vbt}. Most of the recent results on FIP searches in kaon experiments come from NA62 at CERN~\cite{NA62:2017rwk}, a multi-purpose $K^+$ and $\pi^0$ decay experiment with the main goal of measuring the ultra-rare $K^+\to\pi^+\nu\bar\nu$ decay. The high-intensity 75~GeV/$c$ secondary $K^+$ is produced by the 400~GeV/$c$ primary protons slowly extracted from the SPS impinging of a beryllium target, delivered in 4.8-second spills at a typical intensity of $2\times 10^{12}$ protons per spill. The experiment has collected 850k good SPS spills in 420~days of operation in 2016--2018 (the Run~1 dataset), and 1.5M good SPS spills in 650~days of operation in 2021--2024. The data collection is continuing in 2025 and is expected to be completed in 2026. Furthermore, the experiment has collected so far a dataset of $6\times 10^{17}$ protons on target (pot) in 30~days of operation in the beam-dump mode.

A measurement of the  $K^+\to\pi^+\nu\bar\nu$ decay with the NA62 Run~1 data~\cite{NA62:2021zjw}, now superseded with an analysis based on the 2016--2022 dataset~\cite{NA62:2024pjp}, has been used to search for peaks in the decay spectrum corresponding to production of an invisible particle in the $K^+\to\pi^+X_{\rm inv}$. Invisible particle mass ranges of 0--110~MeV/$c^2$ and 154--260~MeV/$c^2$, on the two sides of the $\pi^0$ peak, are covered by the search. Interpretation of the Run~1 data in terms of limits on the BC4 (dark scalar) scenario is presented in Ref.~\cite{NA62:2021zjw}. Interpretation of a partial Run~1 dataset in terms of limits on the BC10 (fermion-coupled ALP) phase space is given in Ref.~\cite{NA62:2020xlg}. Interpretation of the Run~1 dataset, including also the results of a $K^+\to\pi^+\gamma\gamma$ decay measurement, in terms of limits on the BC11 (gluon-coupled ALP) phase space is provided in Ref.~\cite{NA62:2023olg}. Finally, BC4 and BC10 interpretations of a dedicated search for the fully invisible $\pi^0$ decay in the region 
$m_X\approx m_{\pi^0}$, based on 10\% of the NA62 Run~1 dataset containing $4\times 10^9$ tagged $\pi^0$ mesons produced in the $K^+\to\pi^+\pi^0$ decay and limited by background, are presented in Ref.~\cite{NA62:2020pwi}. An upcoming dedicated paper will report a systematic interpretation of all relevant NA62 results published to date, including $K^+\to\pi^+\nu\bar\nu$~\cite{NA62:2024pjp}, $\pi^0\to X_{\rm inv}$~\cite{NA62:2020pwi}, $K^+\to\pi^+\gamma\gamma$~\cite{NA62:2023olg} and $K^+\to\pi^+\mu^+\mu^-$~\cite{NA62:2022qes} measurements, in terms of the above scenarios (BC4, BC10, BC11) and additionally BC2 (invisible dark photon). The KOTO experiment at J-PARC has recently published a new upper limit of the $K_L\to\pi^0\nu\bar\nu$ decay rate using the 2021 dataset, ${\cal B}(K_L\to\pi^0\nu\bar\nu)<2.2\times 10^{-9}$ at 90\%~CL, and its interpretation in terms of upper limits of the $K_L\to\pi^0X_{\rm inv}$ decay in the $X$ mass range 0--260~MeV/$c^2$~\cite{KOTO:2024zbl}.

Large samples of tagged $\pi^0$ mesons produced in the $K^+\to\pi^+\pi^0$ decays have been used to place stringent upper limits on the dark photon coupling using the $\pi^0\to\gamma A'$ decay. The NA48/2 experiment at CERN, the predecessor of NA62, has established the world's strongest upper limits on the BC1 scenario involving a promptly decaying dark photon ($A'\to e^+e^-$) in the mass range 20--100~MeV/$c^2$ using its full dataset collected in 2003--2004~\cite{NA482:2015wmo}. The NA62 experiment has placed limits on the BC2 scenario involving an invisible dark photon, using about 1\% of the Run~1 dataset collected in dedicated conditions~\cite{NA62:2019meo}.

\begin{figure}[t]
\begin{center}
\resizebox{0.6\textwidth}{!}{\includegraphics{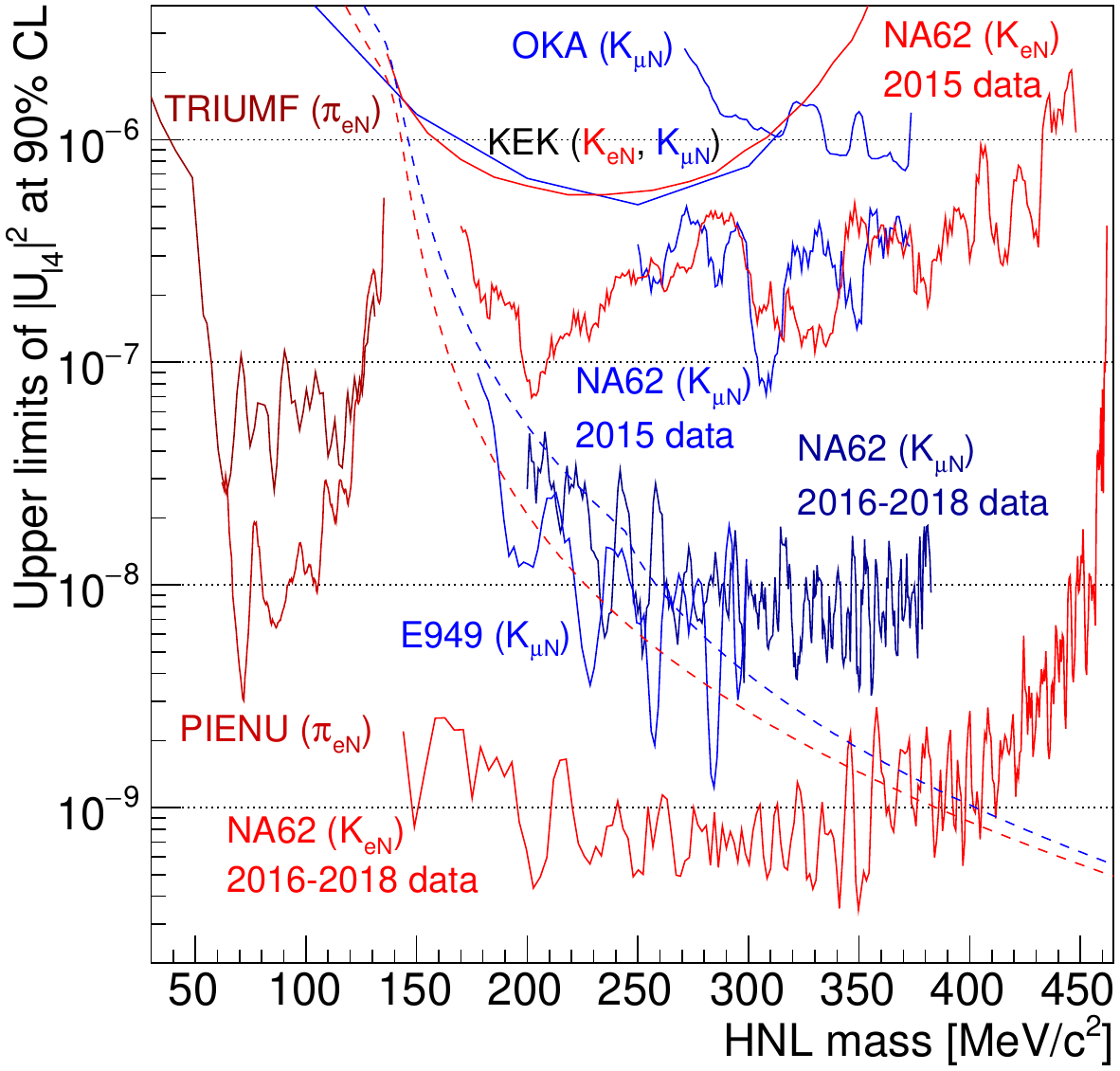}}
\end{center}
\caption{Summary of upper limits at 90\% CL of $|U_{e4}|^2$ (red solid lines) and $|U_{\mu4}|^2$ (blue solid lined) obtained from HNL production searches in $K^+$ decays: NA62~\cite{NA62:2020mcv,NA62:2021bji}, BNL-E949, OKA, KEK; and in $\pi^+$ decays: TRIUMF, PIENU~\cite{PIENU:2017wbj}. The lower bounds of $|U_{e4}|^2$ and $|U_{\mu4}|^2$ imposed by the BBN constraint are shown by the lower and upper dashed lines, respectively. The plot is reproduced from Ref.~\cite{NA62:2021bji}.}
\label{fig:hnl-world-data}
\end{figure}

The NA62 experiment has conducted a search for HNL production in the $K^+\to e^+N$ and $K^+\to\mu^+N$ decays with the Run~1 dataset~\cite{NA62:2020mcv,NA62:2021bji}. The former search uses the main trigger line designed for the $K^+\to\pi^+\nu\bar\nu$ decay, while the latter search is based on the control trigger downscaled by a factor of 400. The analysis represents a peak search in the reconstructed squared missing mass, $m^2_{\rm miss} = (P_K-P_{\ell})^2$, where $P_K$ and $P_\ell$ are the measured kaon and lepton four-momenta. The HNL mass ranges covered are 144--462~MeV/$c^2$ in the electron case, and 200--384~MeV/$c^2$ in the muon case. Upper limits of the decay branching fractions and the mixing parameters $|U_{\ell 4}|^2$ have been established. The sensitivity is limited by backgrounds in both cases; in particular, the $K^+\to\mu^+\nu$ decay followed by the $\mu^+\to e^+\nu\bar\nu$ decay in flight represents an irreducible background to the $K^+\to e^+N$ process. The obtained exclusion limits at 90\% CL of the mixing parameters $|U_{e4}|^2$ and $|U_{\mu4}|^2$ are shown in Fig.~\ref{fig:hnl-world-data}. With the full NA62 dataset, the sensitivity to $|U_{\ell4}|^2$ is expected to be improved by a factor of about 2. Recently, NA62 has presented a search for the $\pi^+\to e^+N$ decay of the beam pions with the 2016--2024 dataset. The results, to be published soon, improve on the PIENU upper limits~\cite{PIENU:2017wbj} shown in Fig.~\ref{fig:hnl-world-data} in the mass range 95--126~MeV/$c^2$. With additional datasets included, the sensitivity to the HNL mixing parameters $|U_{\ell 4}|^2$ is expected to improve as the inverse square root of the integrated kaon flux. In the electron case, the gap at $m_N\approx m_\pi$ between searches in $\pi^+$ and $K^+$ decays can be covered by the search for $K^+\to\pi^0e^+N$ decay, although with a limited sensitivity~\cite{Tastet:2020tzh}.

Further recent searches for FIP production in kaon decays, not interpreted within any specific benchmark scenarios, include: a search for the prompt $K^+\to\pi^+aa$, $a\to e^+e^-$ and $K^+\to\pi^+S$, $S\to A'A'$, $A'\to e^+e^-$ decay chains with the NA62 Run~1 dataset~\cite{NA62:2023rvm};  a search for the $K^+\to\mu^+\nu X_{\rm inv}$ and $K^+\to\mu^+\nu\nu\bar\nu$ decays with the NA62 Run~1 dataset~\cite{NA62:2021bji}; a search for the $K^+\to\pi^+\pi^0 X_{\rm inv}$ decay at the OKA experiment at IHEP-Protvino~\cite{Sadovsky:2023cxu}; and a search for the prompt $K_L\to XX$, $X\to\gamma\gamma$ decay chain at the KOTO experiment~\cite{KOTO:2022lxx}.

The NA62 beam-dump sample collected in 10~days of operation in 2021, corresponding to $1.4\times 10^{17}$ pot, has been analysed to search for FIP decays into di-electron~\cite{NA62:2023nhs}, di-muon~\cite{NA62:2023qyn} and hadronic~\cite{NA62:2025yzs} final states. The interpretation of the combined results in terms of the benchmark scenarios BC1 (dark photon), BC4 (dark scalar), BC10 (fermion-coupled ALP), and BC11 (gluon-coupled ALP) is presented in Ref.~\cite{NA62:2025yzs}: the exclusion limits improve on the previous experimental results in each scenario considered. The analysis has demonstrated no background limitations up to the total expected NA62 beam-dump statistics of $10^{18}$~pot.

Both NA62 and KOTO experiments are currently collecting and analysing data. New FIP results from these experiments at improved sensitivities, from both kaon and beam-dump samples, are expected in the coming years.


\changelocaltocdepth{2}
\subsection{FIPs at the LHC in the on-axis direction: FASER --- \textit{J.~Boyd}}
\label{ssec:fips_at_FASER}
\textit{Author: Jamie Boyd, \email{jamie.boyd@cern.ch}}  \\
\subsubsection{Introduction to FASER}
\begin{figure}[hbt!]
    \centering
        \includegraphics[trim=0cm 4.5cm 1cm 4.5cm, clip=true, width=\textwidth]{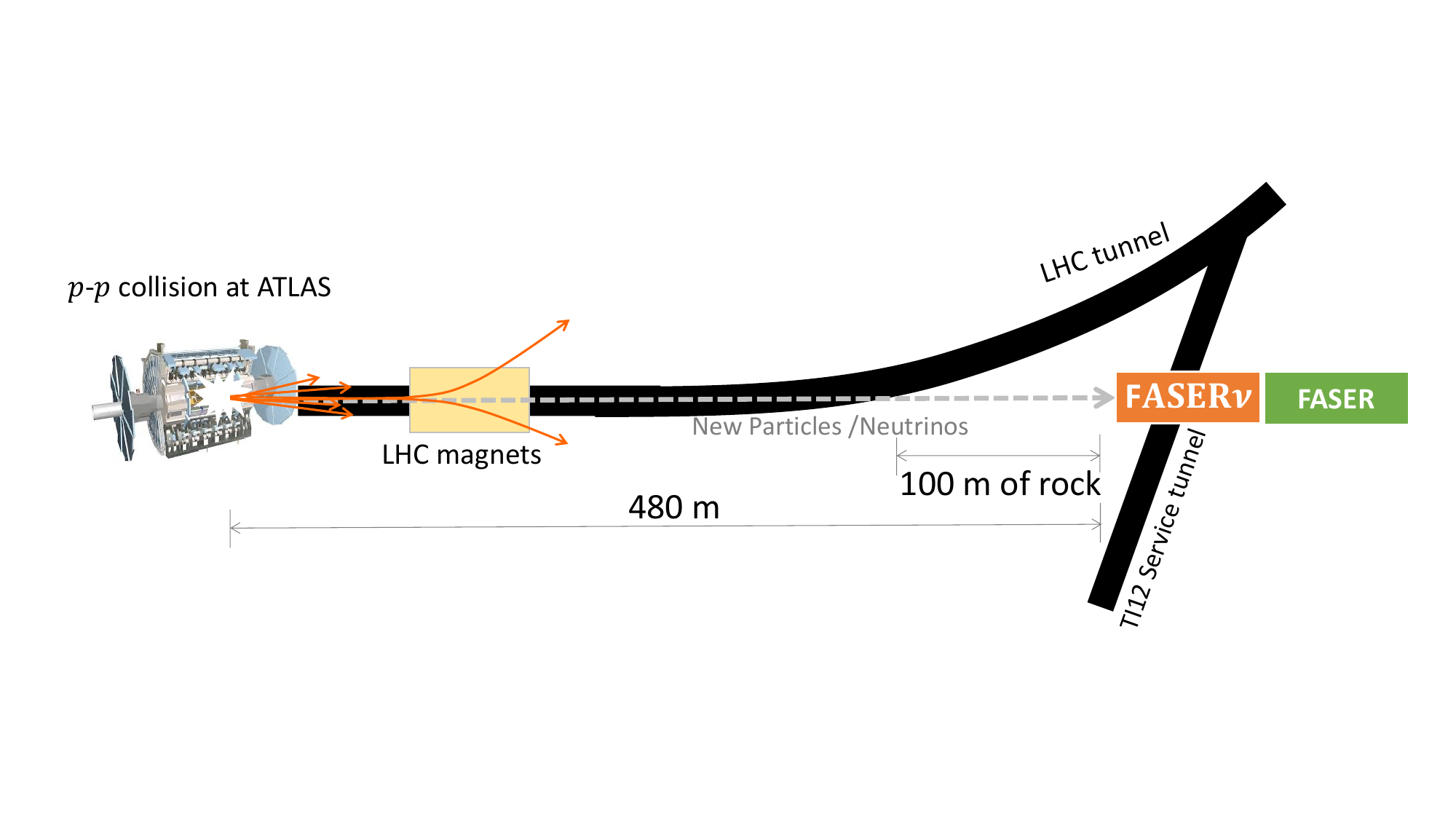}
        \caption{A sketch of the FASER location, situated on the collision axis LOS in the TI12 tunnel in the LHC complex. }
    \label{fig:FASERlocationSketch}
\end{figure}

The ForwArd Search ExpeRiment, or FASER~\cite{Feng:2017uoz,FASER:2018bac}, is a small experiment that was installed into the LHC tunnel during the LHC Long Shutdown 2 (between 2019 and 2021). It is designed to search for new long-lived and weakly interacting particles which could be produced in the LHC collisions at the  ATLAS collision point (IP1), and with the addition of the FASER$\nu$ sub-detector~\cite{FASER:2019dxq}, also to study high-energy neutrinos produced in the LHC collisions. The detector is placed on the collision-axis line-of-sight (LOS), 480~m away from the collision point, after the LHC has bent away from the LOS, as shown in Figure~\ref{fig:FASERlocationSketch}. As can be seen in the figure, the detector is placed in an unused service tunnel TI12, formerly used as an injection beamline for the old LEP collider. The TI12 tunnel is an excellent location for such an experiment, since backgrounds from particles produced in the IP1 collisions are largely suppressed by the location: such particles would need to travel through strong LHC magnets (sweeping away charged particles) and 100~m of rock before they reach FASER.

The FASER active detector is only 20~cm across in the transverse plane; thus, it covers an extremely small angular region of 0.1~mrad around the LOS (and only $10^{-8}$ of the full solid angle). However, given that light particles are predominantly produced in the LHC collisions at very small angles to the beam collision axis, even such a small detector has excellent prospects for searching for new particles in interesting regions of parameter space.

\subsubsection{The FASER Detector}
Since the active area of the detector is very small, the experiment was able to make use of spare detector modules from existing experiments, where silicon microstrip detector modules from the ATLAS experiment are used in the FASER tracking detector, and spare electromagnetic calorimeter modules from the LHCb experiment are used for the FASER calorimeter. The use of these spare modules allowed the detector to be constructed and tested in the short time available before installation, and also significantly reduced the overall cost of the experiment.

\begin{figure}[hbt!]
    \centering
        \includegraphics[trim=0cm 0.8cm 1cm 4.5cm, clip=true, width=\textwidth]{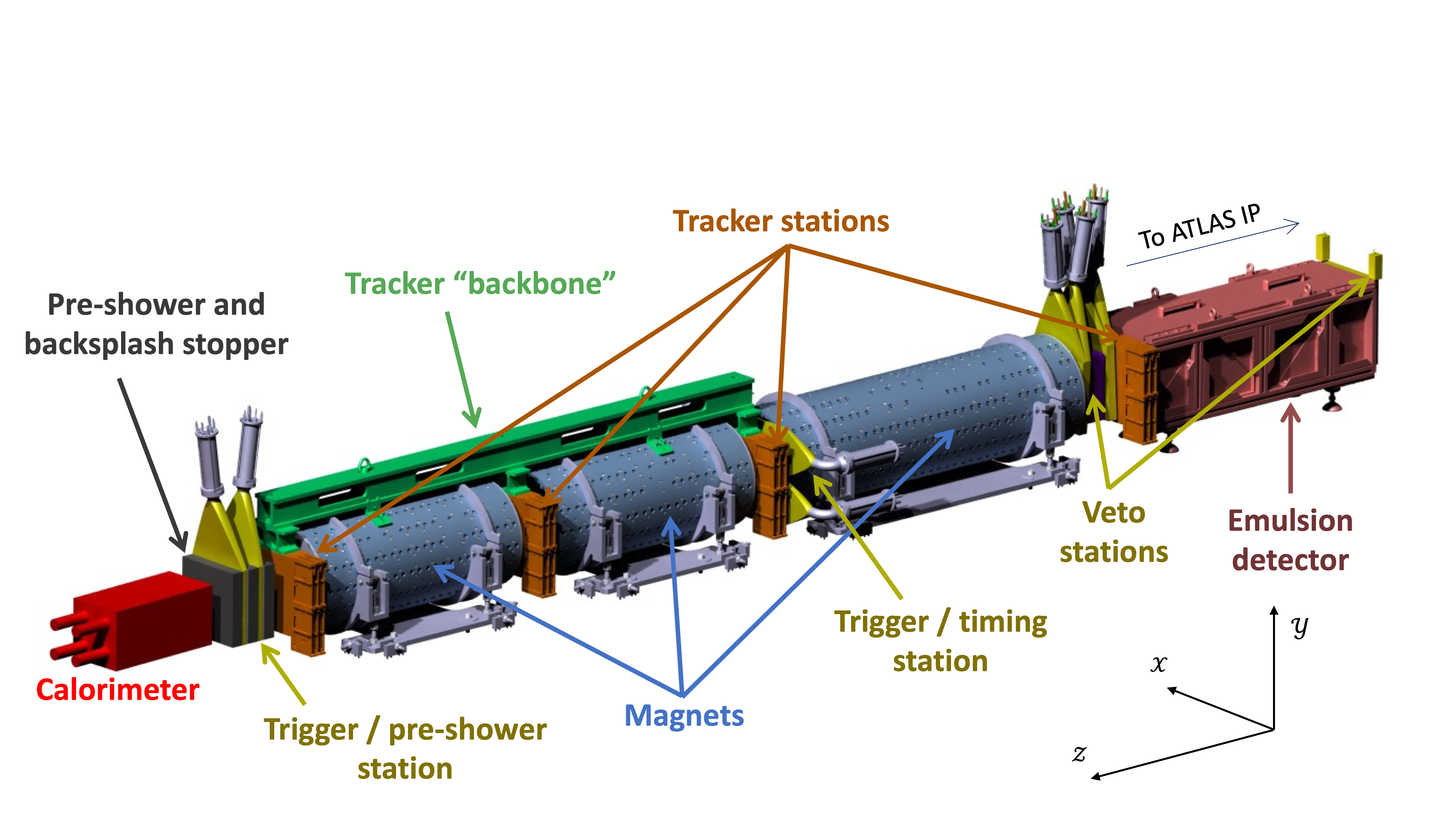}
        \caption{A sketch of the FASER  detector, showing the different sub-detector systems. Particles from the IP enter the detector from the right (upstream) end.}
    \label{fig:FASER_labels}
\end{figure}

A sketch showing the different components of the FASER detector is shown in Figure~\ref{fig:FASER_labels}. The detector, which is described in detail in Ref.~\cite{FASER:detectorPaper}, is made up of a number of sub-detectors as detailed below:
\begin{itemize}
\item Scintillator detector stations. These are used to be able to veto the presence of charged particles entering the detector in the physics analysis. They are also used to trigger events of interest and for precise time measurements of particles in the detector.
\item A decay volume, immersed in a 0.6~T dipole magnetic field. This is 1.5~m long, and is the region in which the decay of new physics particles would be able to be detected by the experiment.
\item A tracking spectrometer, made up of three tracking stations placed before, in the middle, and after two 1-m long dipole magnets (each with a magnetic field strength of 0.6~T). The tracking spectrometer measures the position and momentum of charged particles arising from new particles decaying in the decay volume. The tracking detector is described in detail in Ref.~\cite{FASER:2021ljd}.
\item An electromagnetic calorimeter, placed at the back of the detector to measure the energy of particles traversing the detector.
\item The FASER$\nu$ neutrino detector. Situated in front of the main FASER detector, this is made up of interleaved tungsten plates and nuclear-emulsion films, with 730 of each forming a detector that is 1.1~m long and weighs 1.1~tonnes.
\end{itemize}

In order to be able to place the detector on the collision axis LOS within a few cm, a small trench was excavated from the floor of the TI12 tunnel, in which the detector is installed in. All the detector components except for the emulsion-based detector have their signals read-out following a trigger which fires when a charged particle traverses one of the scintillators, or when a particle leaves significant energy in the calorimeter. The trigger rate is around 1.5~kHz
from background muons produced in IP1 passing through the detector. More details on the FASER trigger and data acquisition system are given in Ref.~\cite{FASER:2021cpr}.

The detector was installed into the TI12 tunnel in March 2021, and has taken physics data since July 2022 when LHC Run 3 started. Up to the end of 2024, the detector had recorded 190~fb$^{-1}$ of data with a data taking efficiency of more than 97\%.

\subsubsection{Physics Results and Prospects}

\paragraph{Dark Photon search}

In 2023, FASER released its first search for decaying dark photons~\cite{FASER:2023tle}.
The model searched for has long-lived dark photons decaying to Standard Model particles, and in the region relevant for FASER, the decay is exclusively to electron-positron pairs.
The search used the 2022 dataset (the first year of FASER operations) corresponding to 27.0~fb$^{-1}$ of proton collision data.
Since this was the first analysis with FASER data, a simple and robust event selection was defined with the aim of having a background-free search. A blind analysis was carried out to avoid any unintentional bias.
The analysis selected events with no signal in any of the five veto scintillators, two reconstructed charged particle tracks, and a large energy deposit in the electromagnetic calorimeter, corresponding to greater than 500~GeV of energy. In order to ensure that background muons do not miss the veto system and fake the signal topology, the reconstructed tracks are required to have momentum greater than 20~GeV and when extrapolated to the front veto system are required to be within the central part of the scintillators (at least 5~cm from the edge of the scintillators). The event selection is about 50\% efficient for typical dark photons decaying in the FASER decay volume.
Backgrounds from several sources were studied:
\begin{itemize}
    \item Muons passing through the veto system due to inefficiencies in the scintillators. This was assessed by measuring the per scintillator inefficiency with data, selecting events without any requirements on the scintillator under test, while requiring signals in the other scintillators. These measurements showed individual scintillator inefficiencies at the $10^{-5}$ level, giving an overall inefficiency of the veto system of smaller than $\mathcal{O}(10^{-20})$. Given that $\mathcal{O}(10^{8})$ muons are expected to pass through FASER in this dataset, the background from the veto inefficiency is negligible.
    \item Neutral hadrons created upstream of the detector by muon interactions in the rock. This background is suppressed by the fact that i) the parent muon must scatter to miss the veto scintillators and ii) the neutral hadron must pass through the 8 interaction lengths of tungsten in FASER$\nu$ to be able to fake the signal. The background was estimated by using a data control samples with three reconstructed tracks (the third being the parent muon) and then extrapolating the yield to that in the signal region. This yields a background estimate of $(0.8 \pm 1.2) \times 10^{-3}$ events.
    \item Background from cosmic rays or beam-induced backgrounds was studied using events taken without colliding bunches at IP1. These studies showed that this background is negligible.
    \item Background from high energy neutrinos interacting in the detector volume downstream of the veto system was studied using high-statistics Monte Carlo simulation samples. The background is mitigated by having little material inside the active volume of the detector (between the veto system and the tracker). The simulations predict a background of $(1.5 \pm 2.0) \times 10^{-3}$ events.
\end{itemize}
The overall expected background is $(2.3 \pm 2.3) \times 10^{-3}$ events, dominated by neutrino interactions in the detector material. Thus, a near-background free search was achieved.

\begin{figure}[hbt!]
    \centering
        \includegraphics[width=0.49\textwidth]{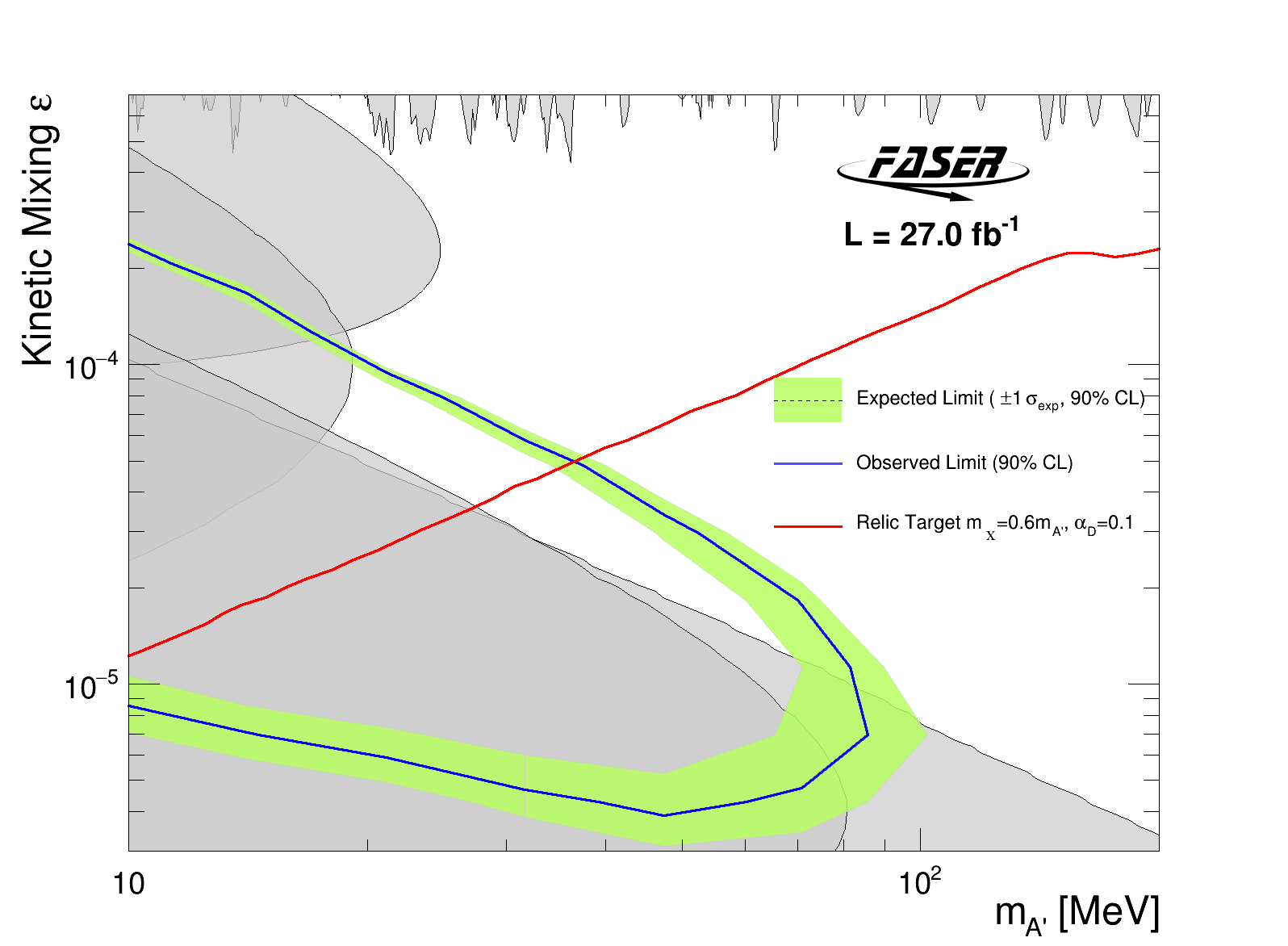}
        \caption{The 90\% exclusion limit for dark photons at FASER.}
    \label{fig:FASER-darkPhoton}
\end{figure}

After unblinding, zero events were observed in the signal region, and exclusion limits on the dark photon model were set at 90\% CL, as shown in Figure~\ref{fig:FASER-darkPhoton}.
\newline

\paragraph{Axion-like-particle (ALP) search}

\begin{figure}[hbt!]
    \centering
        \includegraphics[width=0.49\textwidth]{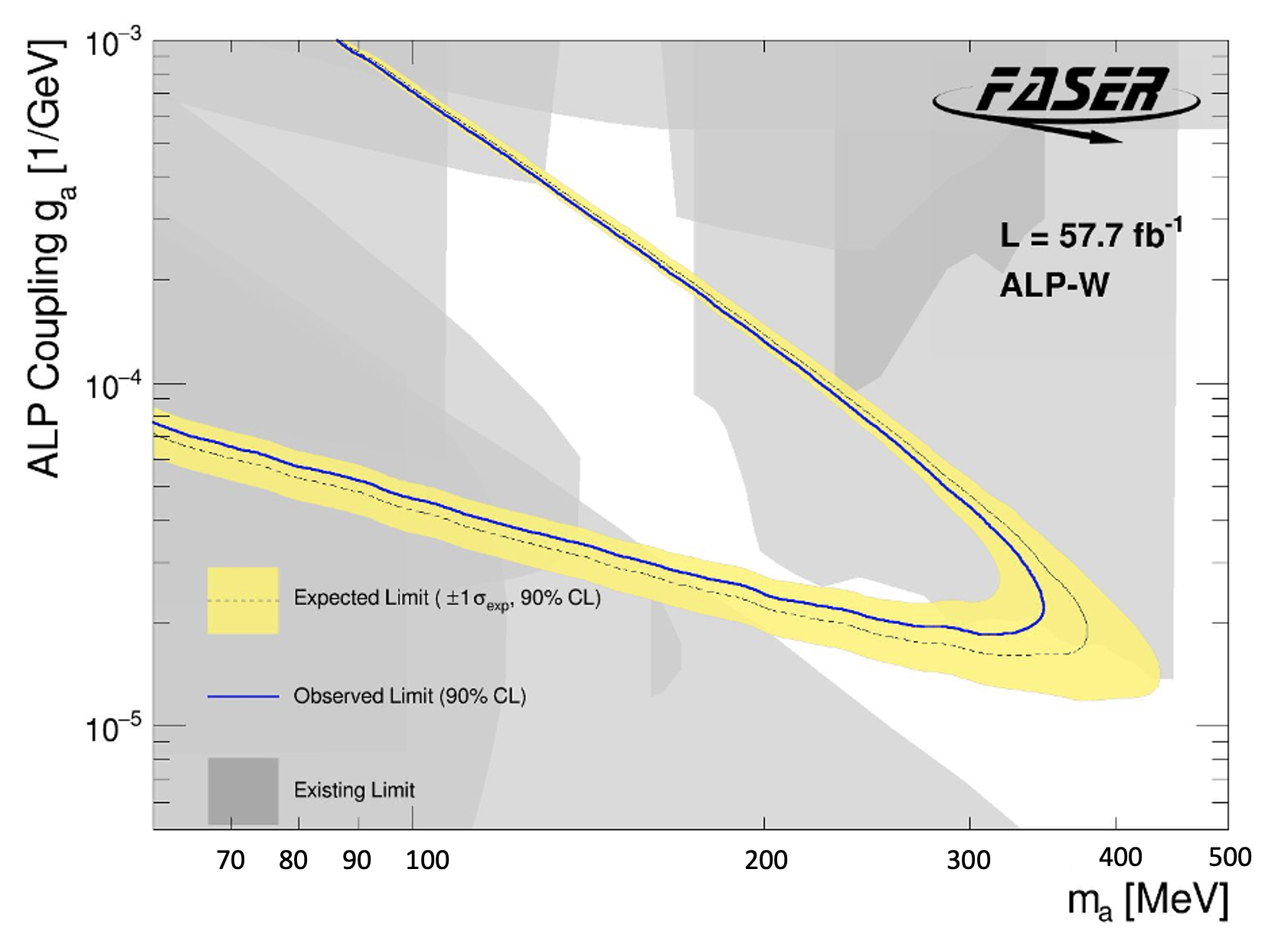}
                \includegraphics[width=0.49\textwidth]{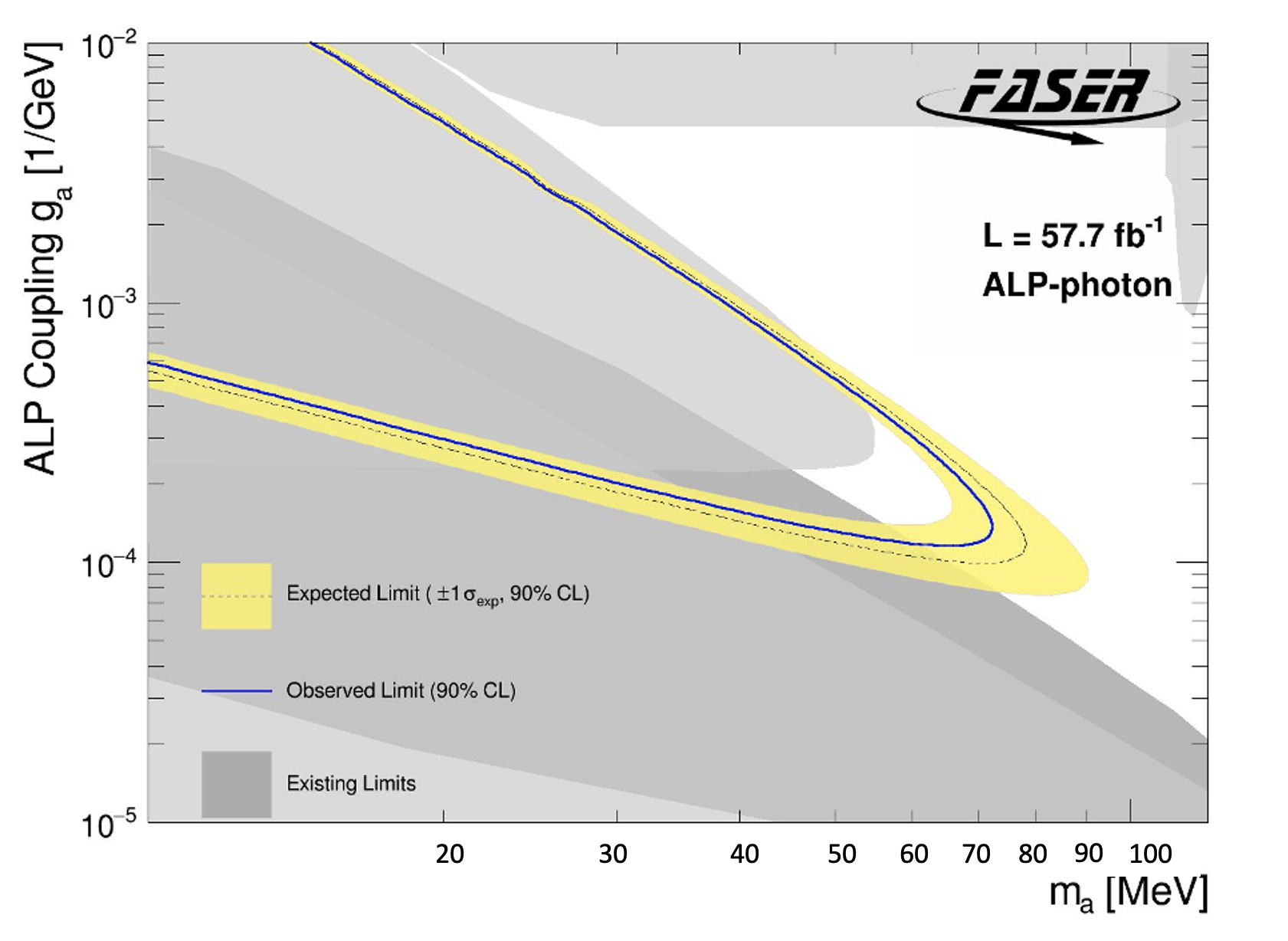}
        \caption{The 90\% exclusion limit from FASER for ALPs. (left) for the ALP-W model; (right) for the ALP-photon model}
    \label{fig:FASER-ALP-limits}
\end{figure}

The FASER ALP search~\cite{FASER:2024bbl} was released in 2024, using the combined 2022 and 2023 dataset for a total integrated luminosity of 57.7~fb$^{-1}$. Since the ALP decays to two photons, the analysis does not use the tracker, and the ALP can decay anywhere within the detector between the veto system and the calorimeter. The current detector can not distinguish the two closely separated photons from the ALP decay, and so the signal topology searched for is just a single large energy deposit in the calorimeter. In the future, a high-granularity silicon/tungsten preshower~\cite{Boyd:2803084} will allow the two photons from the ALP decay to be individually reconstructed.  In a similar way to the dark photon search, incoming charged particles are vetoed using the scintillator veto system, this time also including the trigger/timing scintillator after the decay volume which covers a larger transverse area than the other scintillators allowing to remove background from large angle muons, despite the fact that no track requirements are applied. The main background arises from neutrino interactions in the calorimeter, these can be suppressed by requiring a signature in the preshower detector of an incoming electromagnetic shower. In order to minimize as much as possible the neutrino background while still keeping good sensitivity to the signal, a very large calorimeter energy threshold of 1.5~TeV was required. With this selection, different backgrounds were studied, but these were all found to be negligible except for the neutrino background. This was estimated from high-statistics Monte Carlo simulation samples and validated in data control regions. The final background estimate was 0.4 $\pm$ 0.4 events dominated by electron neutrino interactions. After unblinding, one event was observed, compatible with the background estimate. The result was interpreted in several different ALP models, which all lead to the same detector signature. Example 90\% exclusion curves are shown in Figure~\ref{fig:FASER-ALP-limits} for two of the models considered. The left is for ALPs with couplings to gauge bosons, predominately produced in heavy flavour decays, whereas the right is for ALPs with couplings to photons, produced in high energy photon interactions in the TAN absorber in the LHC.
\newline

\paragraph{Neutrino results}

As well as searching for new particles, FASER is well suited to studying high energy neutrinos of all flavours produced in the LHC collisions. These are the highest-energy neutrinos from an artificial source, and their measurements have important implications for QCD, neutrino physics, and astroparticle physics. FASER neutrino results so far include the first interation candidates at the LHC~\cite{FASER:2021mtu}, the first observation of collider neutrinos~\cite{FASER:2023zcr}, the first measurement of the neutrino interaction cross section in the TeV energy range for both muon and electron neutrinos~\cite{FASER:2024hoe}, and the first differential measurement of the muon neutrino cross section and flux~\cite{FASER:2024ref}.
\newline

\paragraph{Future prospects}

FASER has sensitivity for several additional dark sector
models~\cite{FASER:2018eoc,Feng:2024zgp}, and searches for these will be carried out with the Run 3 data. In addition, the experiment has recently been approved to continue data taking in LHC Run 4, which is expected to add an additional 680~fb$^{-1}$ on top of the expected Run 3 dataset of 350~fb$^{-1}$. Sensitivity projections with the combined Run 3 and Run 4 data are shown in~\cite{Boyd:2882503}.

The upgraded high-granularity silicon/tungsten preshower will enable the neutrino background in ALP searches to be mitigated with much lower calorimeter energy thresholds required, thereby improving the sensitivity to lower energy ALPs decaying in FASER (that is ALPs with lower coupling values).

Finally, a strong physics case is building up for future larger versions of the FASER and FASER$\nu$ detectors, to operate at the HL-LHC, as part of the proposed Forward Physics Facility~\cite{Feng:2022inv}. These detectors would have much stronger sensitivity for different dark sector models, and in addition would allow much more precise studies of LHC neutrinos with important implications for QCD, neutrino physics, and astroparticle physics.

\subsection{FIPs at the LHC in the  off-axis direction: CODEX-b --- \textit{J. Pfaller}}
\label{ssec:fips_at_CODEXb}
\textit{Author: Jake Pfaller, \email{jacob.pfaller@cern.ch}}  \\
\subsubsection{Motivation for a Transverse LLP Detector}

\begin{wrapfigure}{l}{0.5\textwidth}
  \centering
  \includegraphics[width=0.5\textwidth]{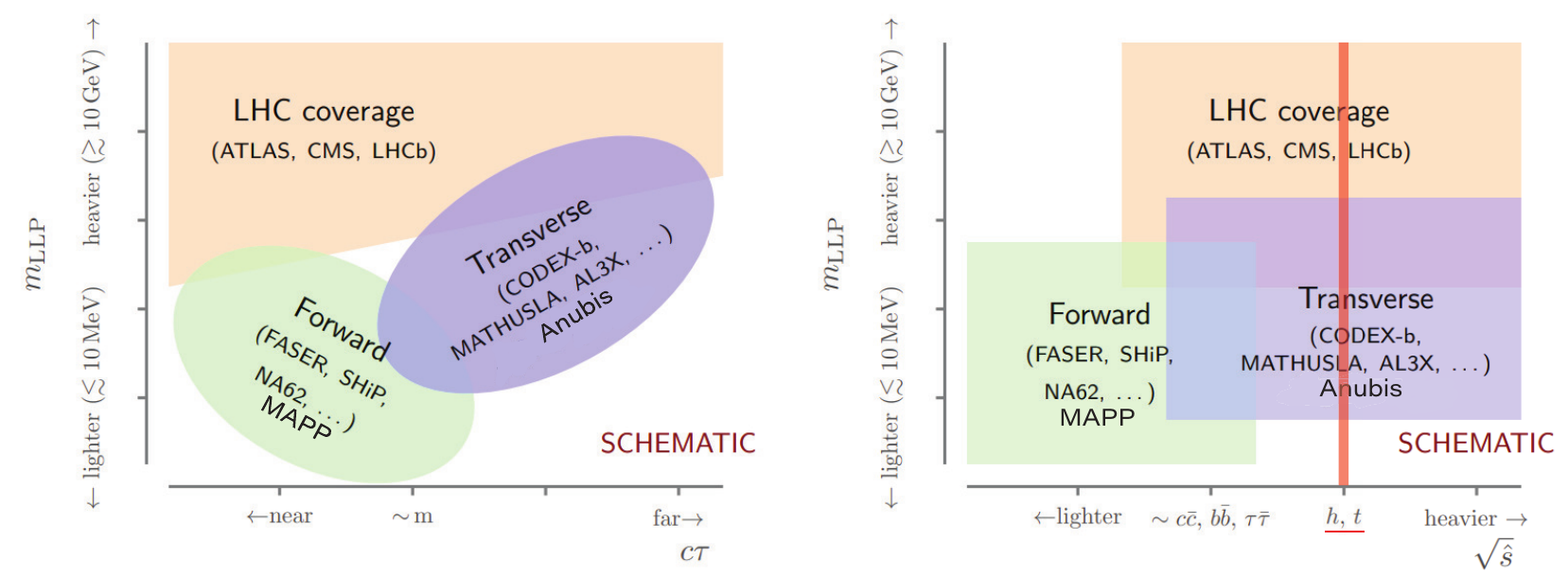}
  \caption{Schematic illustrating the approximate coverage in parameter space of main LHC, forward, and transverse experiments. Figure adapted from Ref.~\cite{Aielli:2019ivi}.}
  \label{fig:Pfaller:schematic}
\end{wrapfigure}

Particle physics experiments at colliders have been very successful over the years at confirming the theoretical predictions of the standard model. Today, however, we are still left with unanswered questions about the universe, so numerous extensions to the standard model have been made to remedy this discrepancy. These beyond standard model (BSM) theories often predict new particles, mediators, or portals between the standard model (SM) and the dark sector. Loose in their constraints of these particles, many of these theories predict particles with relatively long decay times, compared to the majority of particles commonly observed at collider experiments. The long lifetimes predicted for these particles can arise from several reasons, such as suppressed couplings, heavy mediators, or limited phase space. As shown in Fig.~\ref{fig:Pfaller:schematic}, at the main LHC experiments (ATLAS, CMS, LHCb, and ALICE), the large amount of backgrounds in the low-mass region masks much of the potential signal from feebly-interacting BSM particles. This is less of a challenge for experiments in the forward region since these experiments detect only heavily boosted particles, which limits the $c\tau$ of potential BSM particles to which they would have sensitivity. Furthermore, since the center-of-mass energy of the collisions in these experiments is well below the Higgs mass, forward experiments are not sensitive to exotic Higgs boson decays. Therefore, there is a strong case for a displaced, transverse detector, which can provide complementary coverage to existing detectors to fully probe the BSM landscape at the LHC. \textbf{CODEX-b (COmpact Detector for EXotics at LHCb)}~\cite{Aielli:2019ivi,Aielli:2022awh,CODEX-b:2024tdl,CODEX-b:2025rck} is a proposed transverse detector with the ability to provide a comprehensive expansion to the BSM physics program at the LHC.

\pagedepth\maxdimen

\subsubsection{Baseline Detector Design}

\begin{figure}[hbt!]
  \centering
  \begin{subfigure}[b]{0.35\textwidth}
    \centering
    \includegraphics[width=\textwidth]{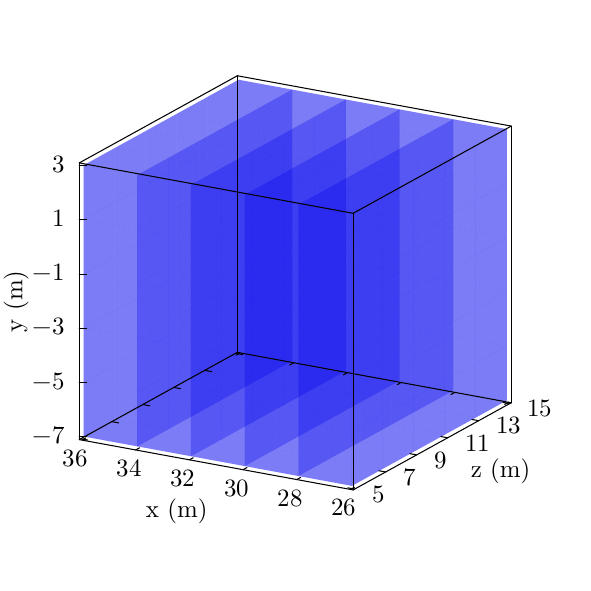}
    \caption{}
    \label{fig:Pfaller:baseline}
  \end{subfigure}
  \hfill
  \begin{subfigure}[b]{0.55\textwidth}
    \centering
    \includegraphics[width=\textwidth]{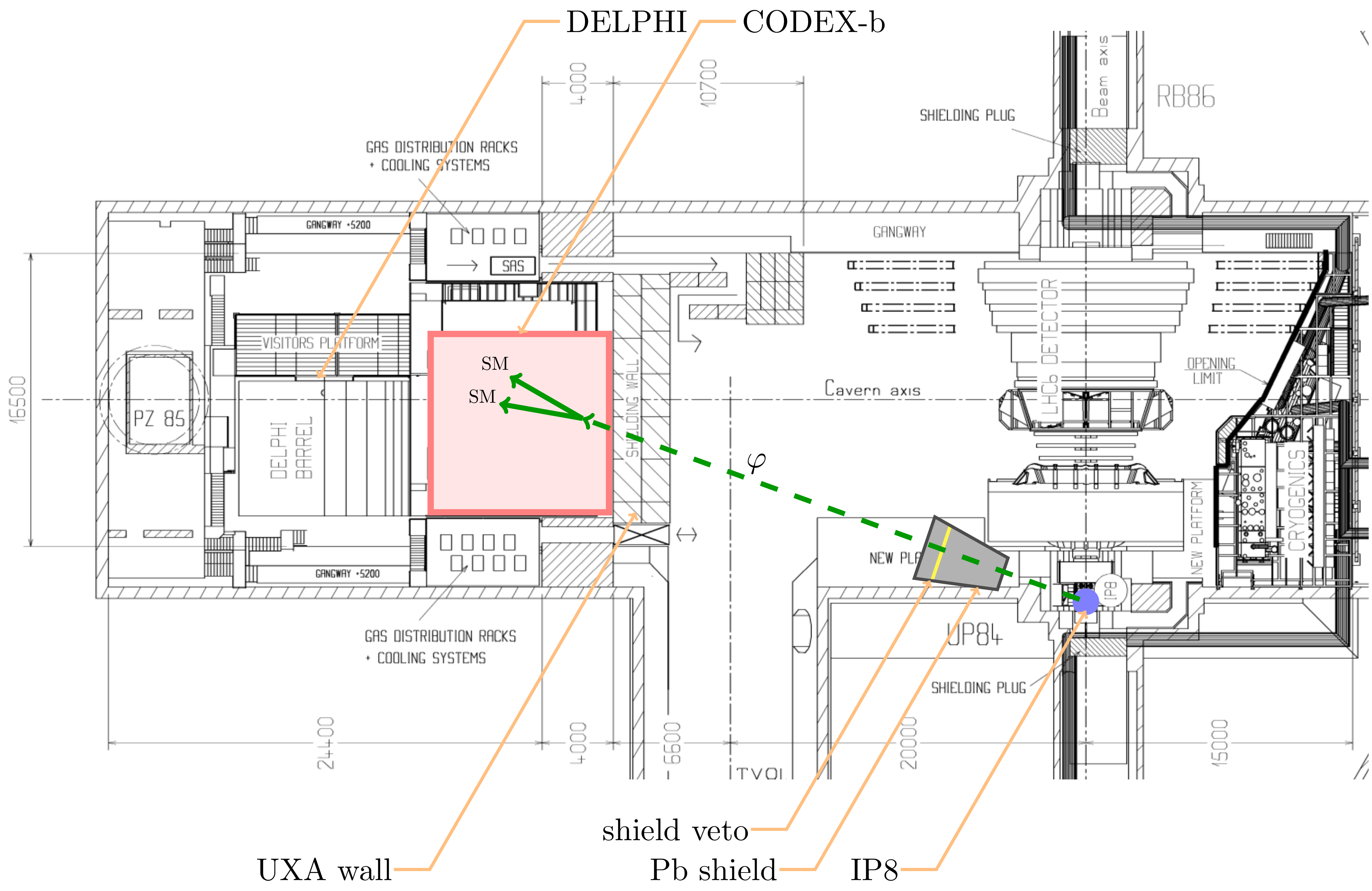}
    \caption{}
    \label{fig:Pfaller:cavern}
  \end{subfigure}
  \caption{The baseline design of the CODEX-b detector (a)~\cite{CODEX-b:2025rck}, and a drawing of the UX85 cavern (b), showing the LHCb detector and the proposed location of CODEX-b in red~\cite{Aielli:2019ivi}.}
  \label{fig:Pfaller:baselineAndCavern}
\end{figure}

The baseline design for the CODEX-b detector (Fig.~\ref{fig:Pfaller:baseline}) is a $10\times10\times10$ meter cube of resistive plate chambers (RPCs)~\cite{Aielli:2019ivi}. The detector would be instrumented about 25 meters from the LHC interaction point 8 (IP8), in the volume currently occupied by the LHCb D1 barracks (Fig.~\ref{fig:Pfaller:cavern}). A key factor in making CODEX-b sensitive to new physics is achieving a near-zero background environment, which is dependent on active and passive shielding. Thus, an important factor in the design of this experiment is using a combination of shielding and vetoes to eliminate SM backgrounds. The concrete UXA wall separates the CODEX-b side of the cavern from the LHCb side of the cavern, and is already in place. Particles created at the interaction point can produce secondary interactions in the concrete wall and background events in CODEX-b. To eliminate such events, additional passive shielding must be introduced adjacent to IP8, accompanied by an active scintillating veto.

\begin{figure}[htb!]
  \centering
  \begin{subfigure}[b]{0.45\textwidth}
    \centering
    \includegraphics[width=\textwidth]{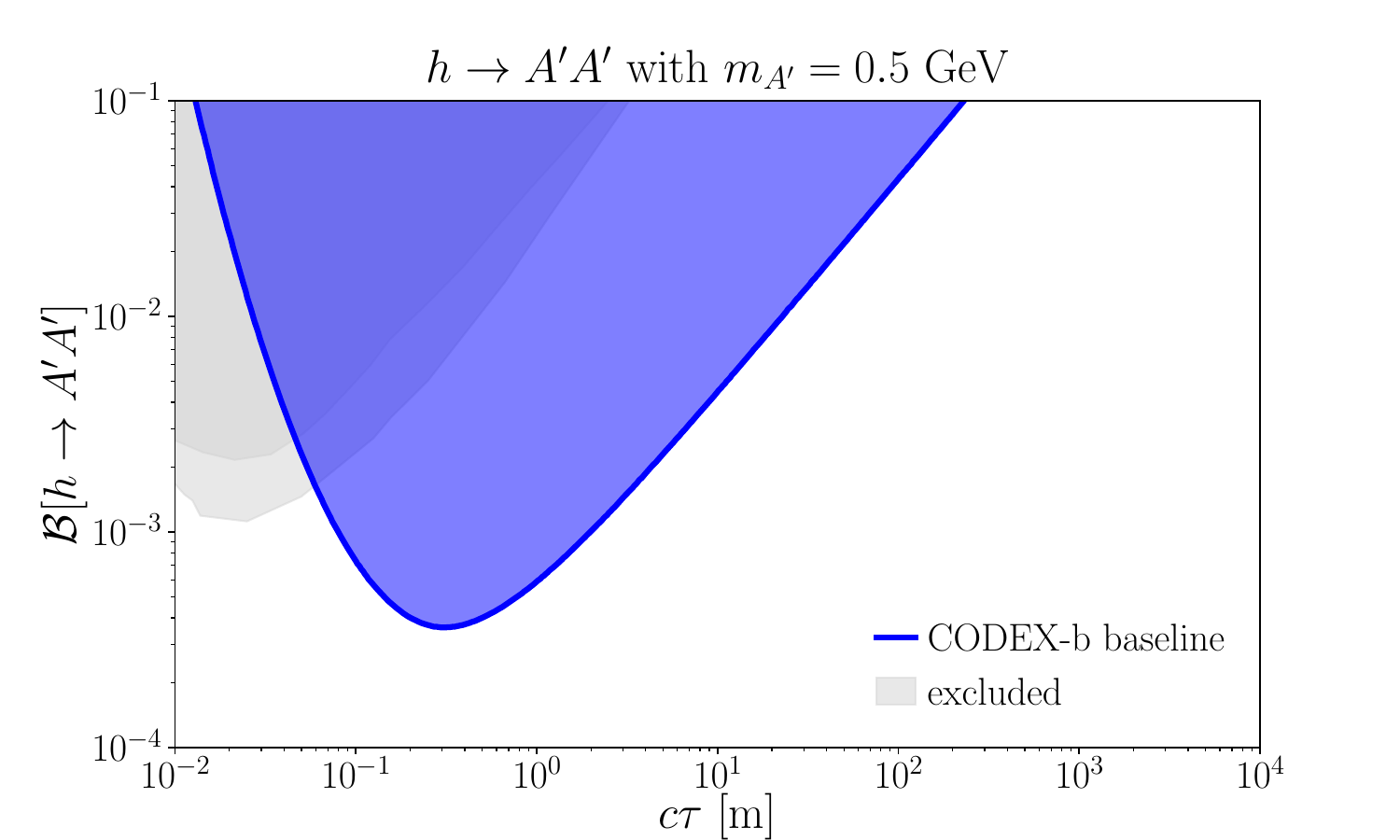}
    \caption{}
    \label{fig:Pfaller:haa_5}
  \end{subfigure}
  \begin{subfigure}[b]{0.45\textwidth}
    \centering
    \includegraphics[width=\textwidth]{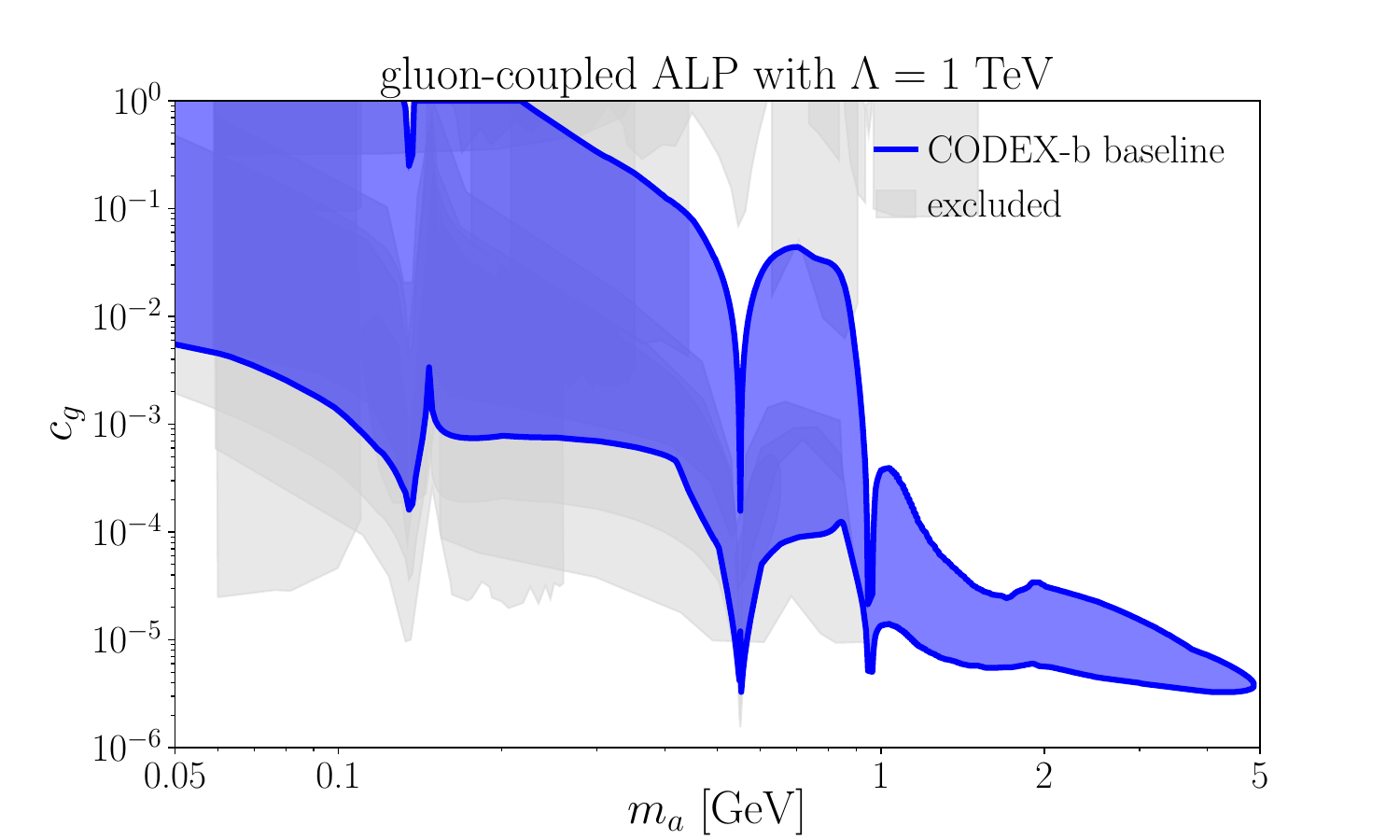}
    \caption{}
    \label{fig:Pfaller:alp_g}
  \end{subfigure}
  \begin{subfigure}[b]{0.45\textwidth}
    \centering
    \includegraphics[width=\textwidth]{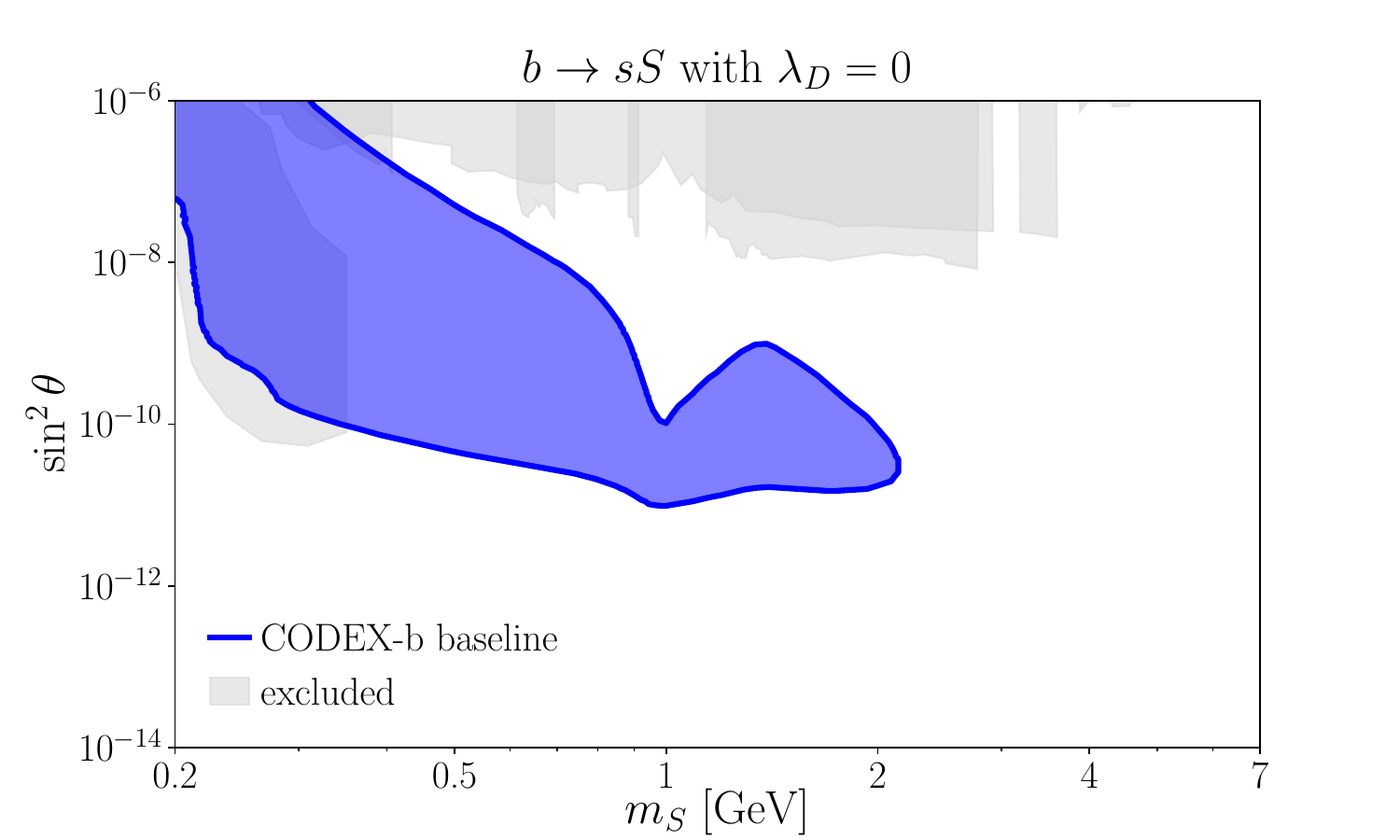}
    \caption{}
    \label{fig:Pfaller:bh_0}
  \end{subfigure}
  \begin{subfigure}[b]{0.45\textwidth}
    \centering
    \includegraphics[width=\textwidth]{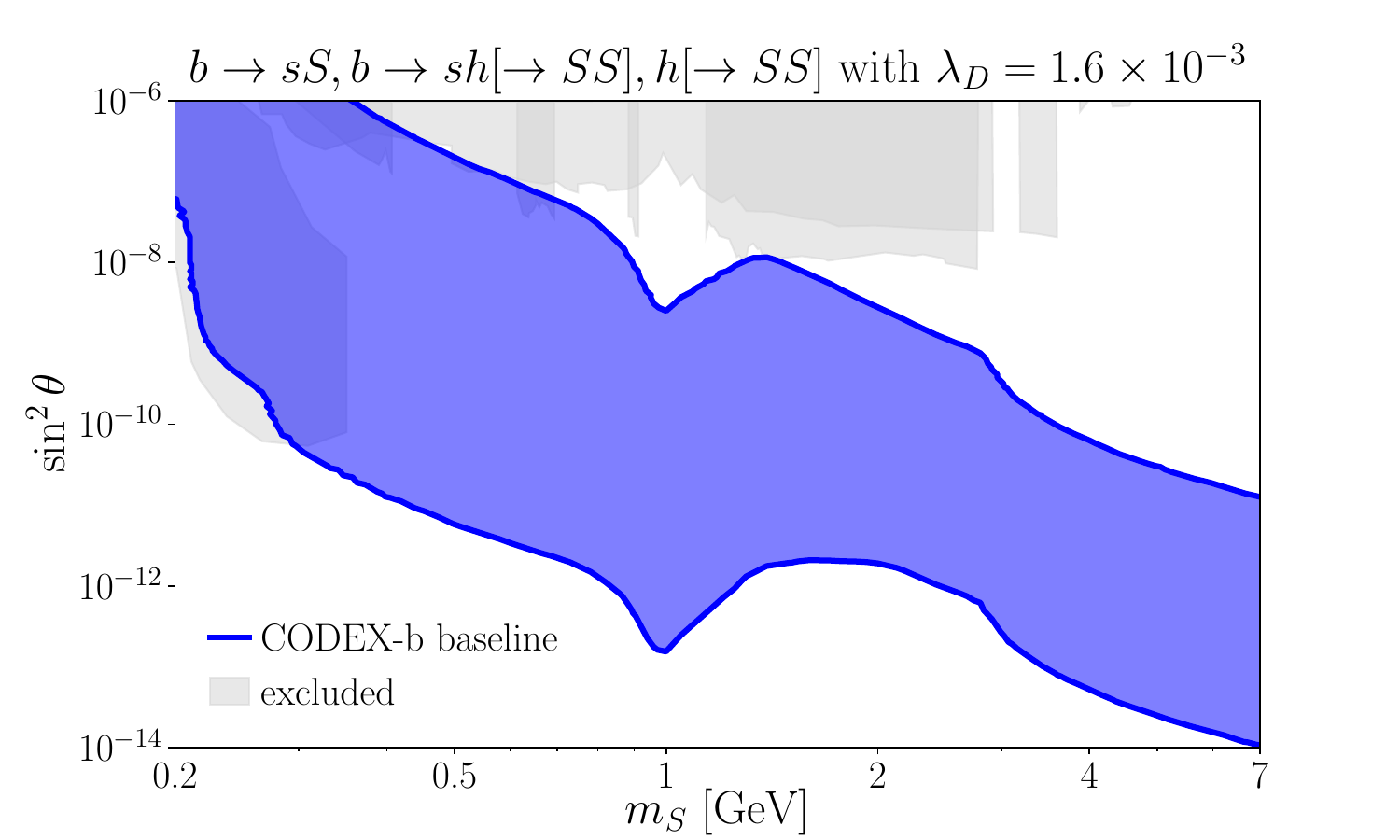}
    \caption{}
    \label{fig:Pfaller:bh_1}
  \end{subfigure}
  \caption{Projected sensitivity~\cite{CODEX-b:2025rck} of the baseline CODEX-b experiment to Higgs boson decays to dark photons (a), gluon-coupled ALPs (b), and dark scalars produced in b-decays (c,d). The model parameters chosen in each case are indicated on the figures. The regions previously excluded by running experiments are shown in gray.}
  \label{fig:Pfaller:sensitivity}
\end{figure}

Figure~\ref{fig:Pfaller:sensitivity} shows the projected sensitivity of CODEX-b to various BSM signal scenarios, with the baseline CODEX-b detector design.

\subsubsection{Backgrounds}

The active and passive shielding planned to accompany the CODEX-b detector is essential to the experimental design~\cite{Dey:2019vyo}. If a feebly-interacting BSM particle is produced in the hard interaction, it will travel through the shielding unimpeded. Then, with some probability, it will decay into charged SM tracks, and the decay vertex can be reconstructed with some displacement from the interaction point and the radiation wall. This process filters out (almost) all SM particles, as they would be stopped by the shielding.

\subsubsection{Alternative Designs}

The ability to fully instrument the baseline design of CODEX-b is not without question, and communication between the CODEX-b and LHCb collaborations is still ongoing. Furthermore, even in the event of full permission from the LHCb collaboration and CERN management, the detector is still subject to logistical constraints of the space in providing the necessary services and mechanical support to the detector modules. Therefore, contingency plans must be in place if the baseline design cannot be used.

\begin{wrapfigure}{r}{0.45\textwidth}
  \centering
  \includegraphics[width=0.4\textwidth]{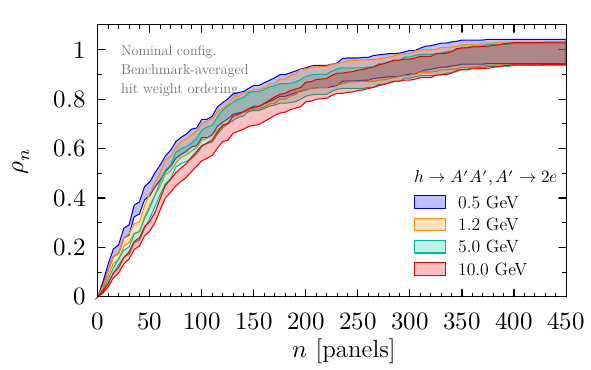}
  \caption{Vertex reconstruction efficiency as a function of the number of RPC panels ($1\sigma$ CL bands)~\cite{Gorordo:2022rro}.}
  \label{fig:Pfaller:recEff}
\end{wrapfigure}

The challenge in determining alternative detector designs is to find a compromise between technical feasibility and maintaining signal sensitivity. To aid in this task, simulations were performed to study exactly how much of the baseline design could be stripped away without sacrificing detector performance~\cite{Gorordo:2022rro}. Figure~\ref{fig:Pfaller:recEff} shows the vertex reconstruction efficiency as a function of the number of RPC panels. This demonstrates that removing a third of the detector panels corresponds to an efficiency loss of just about $10\%$, which means alternative designs for CODEX-b can be quite malleable to logistical constraints without needing to worry about the loss in sensitivity.

\begin{figure}[htb!]
  \centering
  \begin{subfigure}[b]{0.35\textwidth}
    \centering
    \includegraphics[width=\textwidth]{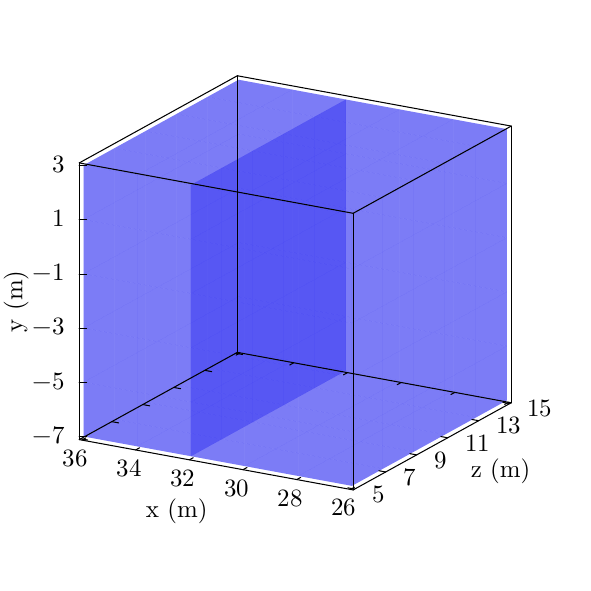}
    \caption{}
    \label{fig:Pfaller:design1}
  \end{subfigure}
  \hfill
  \begin{subfigure}[b]{0.35\textwidth}
    \centering
    \includegraphics[width=\textwidth]{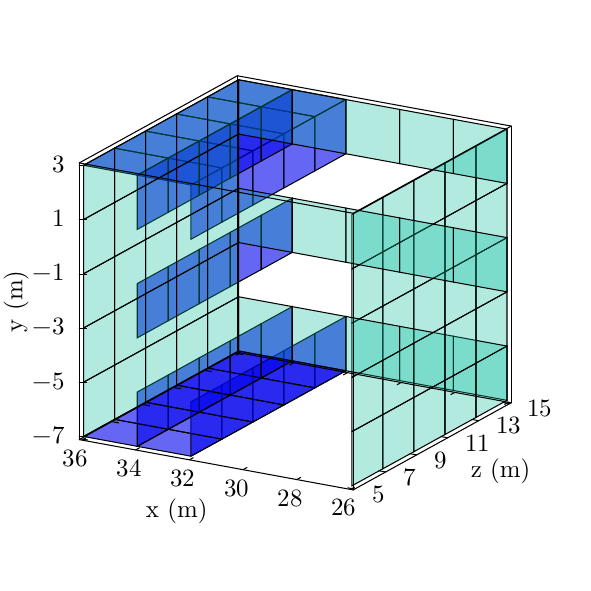}
    \caption{}
    \label{fig:Pfaller:design2}
  \end{subfigure}
  \caption{Two alternative designs for CODEX-b, with RPC triplets (sextets) drawn in blue (green)~\cite{CODEX-b:2025rck}. (a) is a fully hermetic design, with all internal tracking layers removed except one, and (b) is a non-hermetic design more optimized to cavern constraints.}
  \label{fig:Pfaller:altDesigns}
\end{figure}

Figure~\ref{fig:Pfaller:design1} shows a possible alternative detector design, which features a single internal tracking layer~\cite{CODEX-b:2025rck}. Figure~\ref{fig:Pfaller:design2}, on the other hand, demonstrates a more optimal scenario, where existing cavern infrastructure is used to minimize the need for additional support structure. Furthermore, it utilizes RPC sextet modules in addition to the baseline triplet modules to maintain high signal sensitivity and vertex resolution.

\subsubsection{The CODEX-$\beta$ Demonstrator}

A prototype detector is needed to validate many aspects that contribute to the potential success, performance, and feasibility of CODEX-b. Specifically, the goals of the CODEX-$\beta$ demonstrator~\cite{CODEX-b:2024tdl} are:

\begin{itemize}
    \item Demonstrate the scalability of RPC mechanical support structures to the larger-scale detector.
    \item Validate the background estimations and simulations at the experiment site.
    \item Integrate with LHCb data acquisition, so full event information can be known in both detectors simultaneously.
    \item Test the suitability of RPCs as the primary detector technology, in terms of spatial resolution, timing resolution, and overall efficiency.
\end{itemize}

\begin{wrapfigure}{r}{0.45\textwidth}
  \centering
  \includegraphics[width=0.4\textwidth]{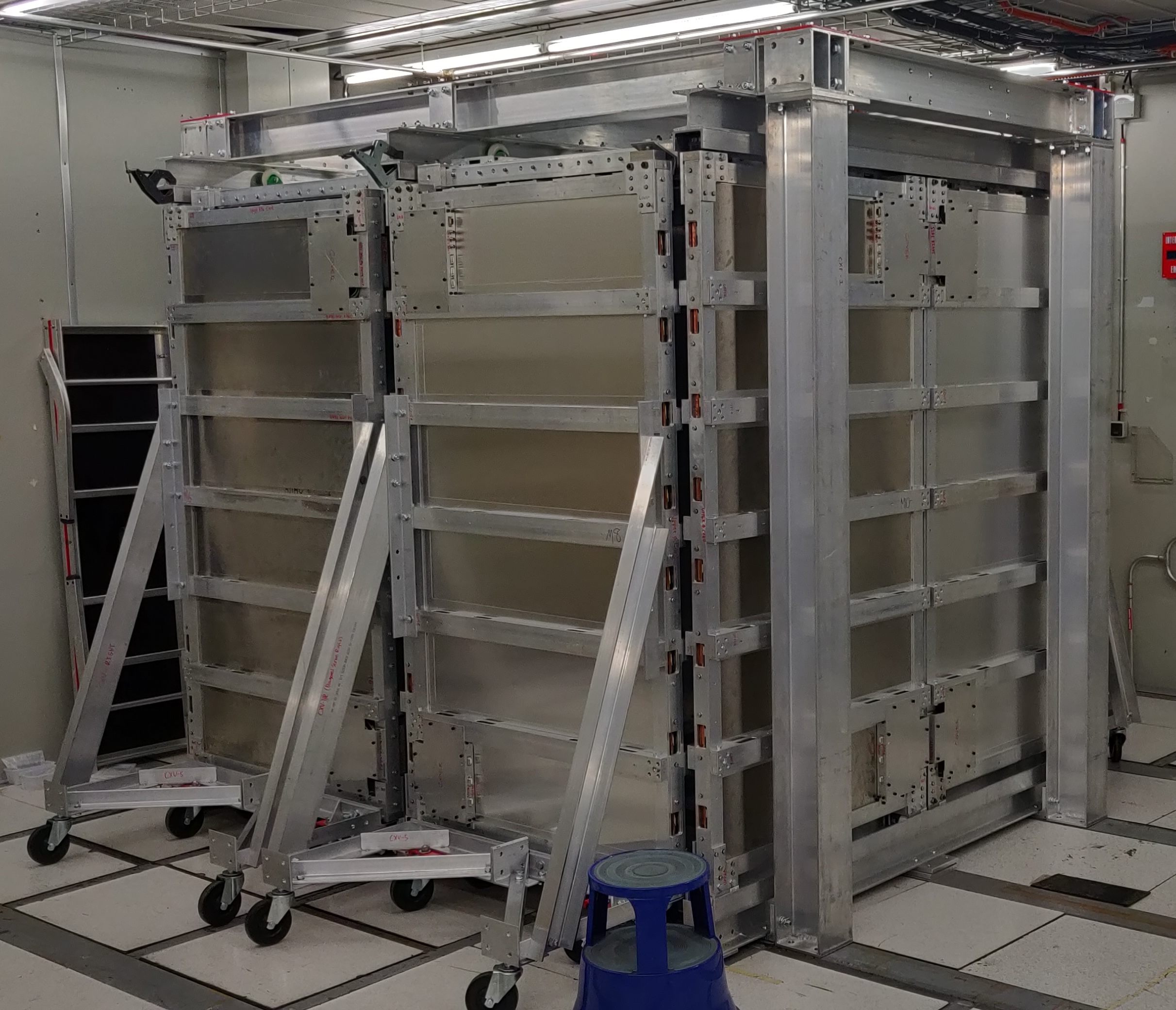}
  \caption{Photograph of the CODEX-$\beta$ detector, installed in the D1 barracks at LHC IP8~\cite{CODEX-b:2025rck}.}
  \label{fig:Pfaller:codexbeta}
\end{wrapfigure}

The design of CODEX-$\beta$ is a scaled-down version of the baseline CODEX-b, namely, a $2\times 2\times 2$ meter cube with one internal tracking layer. A photograph of the CODEX-$\beta$ demonstrator is shown in Fig.~\ref{fig:Pfaller:codexbeta}. The RPCs follow the design of the BIS78 modules from the ATLAS muon upgrade. Ultra-fast front-end electronics provide a timing resolution on the order of $350$ ps, and when three RPCs are layered to form a triplet, the module offers a spatial resolution on the order of 1 mm. The mechanical frame used to support the triplet modules is also adapted from the ATLAS BIS78 design, but suited more specifically to the rigidity needs of CODEX-$\beta$ and for seamless integration into the detector superstructure. These mechanical frames sit at about $2\mathrm{m} \times 1\mathrm{m}\times10\mathrm{cm}$, so two modules side-by-side form one face of CODEX-$\beta$. This brings the full detector count to 14 triplet modules, or 42 RPC singlets, for the demonstrator.

CODEX-$\beta$ will operate with a gas mixture of $R134a$, iso-$C_4H_{10}$, $SF_6$, and $CO_2$. Minus the $CO_2$, this is the standard RPC gas mixture, with the fractional mixtures varying slightly across multiple experiments. This mixture does, however, carry with it a high global warming potential, making the construction of a large-scale RPC detector environmentally worrisome. Research has been done to explore the feasibility of alternative gas mixtures that have a lower environmental impact without sacrificing RPC efficiency. One such mixture has been named the ECO65 mixture, which is a member of the HFO family of gases and features a low global warming potential, but in return, is about 7 times more expensive than the standard mixture. Due to the relatively small size and short runtime of CODEX-$\beta$, both mixtures can be tested without much sacrifice to the environment or budget.

The biggest challenge in reaching a fully-operational CODEX-$\beta$ is the data acquisition, as the ATLAS-designed detector front-end needs to communicate with the LHCb back-end. Data collection and transmission (DCT) modules, recently designed for the ATLAS upgrade, will serve as the interface for this task. One DCT will be mounted on each detector module, and will collect the raw signals, process them with an FPGA, and then send the signals via fiber-optic cables to LHCb SOL40 and TEL40 data acquisition servers. Once all this is in place, CODEX-$\beta$ can be used as a trigger for LHCb events, and event information can be processed in both detectors simultaneously. CODEX-$\beta$ will take data through the end of LHC Run 3, which will end in the summer of 2026.

Though CODEX-$\beta$ is not designed to have new physics sensitivity that is competitive with existing experiments, it can still be used for such analyses as a calibration for the detector and as a proof-of-concept for the larger CODEX-b. Since the dominant background for CODEX-$\beta$ is $K^0_s$ produced in the UXA shielding wall, a $K^0_s$ lifetime measurement can be performed, serving as a calibration for the detector. Furthermore, since all of the potential SM backgrounds in the experimental area decay to at most two tracks, an example new-physics analysis one could do with CODEX-$\beta$ is a search for BSM particles with high-multiplicity decays. These are just a few examples included in the physics program for CODEX-$\beta$.

\clearpage
\changelocaltocdepth{3}
\section{FIPs: Current status and prospects}
\label{sec:current-status}
The concept of feebly interacting particles (FIPs) has (re)generated a lot of interest in the recent years, both on the theoretical and experimental ends.
This chapter represents a tentative summary of the state of the art for FIPs searches at accelerator-based experiments.

\vskip 2mm
Defining a set of crisps criteria for defining FIPs is difficult as their theory motivation is more subtle than more well known theoretical paradigms such as weak scale supersymmetry or WIMPs.
Still, without loosing in generality we can define
FIPs, following Ref.~\cite{Lanfranchi:2020crw}, as any  new  (massive  or  massless)  particle  coupled  to  SM particles  via  small couplings.  The small strength of these couplings can naturally be due to the presence of an approximate symmetry only slightly broken and/or to the presence of a large mass hierarchy between particles.  FIPs are neutral or feebly charged under the SM gauge interactions.

{\it Small couplings} are not necessarily weaker than SM couplings (for example, millicharged particles can interact much more strongly with our detectors than the SM neutrinos), but for most cases they are small enough to translate in lifetimes that range from cm to several km at typical accelerator energies and for masses in the MeV-GeV range. Hence a FIP is also a Long-Lived Particle (LLP), while it is not necessarily true the opposite\footnote{For example, a neutron or a $K_{S,L}$ are long-lived particles but not feebly-interacting.}.

\vskip 2mm
At the present day, the vast majority of the FIPs-related activities are focused on \textit{ dark photons, axion-like particles (ALPs), dark Higgs bosons, and heavy neutral leptons (HNLs)}, see Tab.~\ref{tab:portals} \cite{Beacham:2019nyx}.

\begin{table}[h!]
\centering
\caption{The four main portals for FIPs~\cite{Beacham:2019nyx}. 
}
\begin{tabular}{rcl} \toprule
Portal & & Coupling \\ \midrule
Dark Photon & $A'$ &  $-\frac{\varepsilon}{2 \cos\theta_W} F'_{\mu\nu} B^{\mu\nu}$ \\\addlinespace
Axion-like particles & $a$ &  $\frac{a}{f_a} F_{\mu\nu}  \tilde{F}^{\mu\nu}$, $\frac{a}{f_a}  G_{i, \mu\nu}  \tilde{G}^{\mu\nu}_{i}$,
$\frac{\partial_{\mu} a}{f_a} \overline{\psi} \gamma^{\mu} \gamma^5 \psi$
\\\addlinespace
Dark Higgs & $S$  & $(\mu S + \lambda_\text{HS} S^2) H^{\dagger} H$  \\\addlinespace
Heavy Neutral Lepton & $N$ & $y_N L H N$ \\ \bottomrule
\end{tabular}

\label{tab:portals}
\end{table}

The popularity of this formalism is due to its theoretical simplicity, which in turn is dictated by the cardinal principle in effective field theory, stating that in perturbative models operators with higher operator dimensions should be parametrically suppressed. Indeed, the four operators in Table~\ref{tab:portals} merely represent the leading interactions for the four lowest dimensional representations of the Lorentz group: vector (dark photon), pseudoscalar (ALP), the scalar (dark Higgs), and spinor (HNL).

\vskip 2mm
While these portals represent good tools to investigate the complementarity between various experimental approaches, 
they should be considered as \emph{guest characters in more UV complete models which do address important problems}, such as the hierarchy problem, the strong CP problem, the origin of dark matter, the neutrino masses and the baryon asymmetry of the Universe.
Only in this much broader context these portals can be really predictive, but in a broader context the phenomenology can also be much more rich and diverse from what is presented in this report.

\vskip 2mm
In particular, it is important to emphasize that the benchmark models considered here were chosen because of their simplicity. They typically consist of one new particle with their mass and one coupling as a free parameters.
This permits to plot constraints and projected sensitivities in a two-dimensional parameter space spanned by the mass and coupling, and allow for a comparison of many experiments in one figure.

\vskip 2mm
We stress however, that this simplicity also imposes strong restrictions on the phenomenology and resulting experimental sensitivities. For example, models in which the same coupling controls both the production rate and lifetime of the new state result in a characteristic shape of the sensitivity curves. As a result, the relative position of sensitivities from different experiments are similar in different models.  The most prominent examples are dark Higgs and ALPs with universal couplings with fermions or the dark photon and B-L gauge boson, which look quite similar. In all these scenarios, heavy FIPs with masses above 10 GeV typically require very small couplings in order to be long-lived, leading to strongly suppressed production rates, negligible event rates, and ultimately vanishing sensitivity at far detector experiments.

\vskip 2mm
In contrast, many more complete models are described by more than one relevant coupling or multiple new states. A prominent example for the former scenario are ALPs at the GeV scale, where multiple couplings are expected to contribute to their phenomenology at the same time. Another example is a model of light dark sector containing a dark matter candidate, a dark photon as a mediator, and dark Higgs to generate their mass. In these well motivated cases the production rates, lifetimes, and kinematics and, therefore, sensitivity bounds would be very different from what is assumed here.

\vskip 2mm
New states also permit additional mechanisms to generate a long lifetime. For example, in the case of inelastic dark matter, a small mass splitting between two dark sector states generates a long lifetime for the heavier one while permitting large couplings and hence production rates even for heavy FIPs. Finally, the presence of multiple states could also lead to completely new effects, for example oscillations between the states as in the case of nearly mass-degenerate HNLs.

\vskip 2mm
All these examples highlight that the simplified models discussed below cannot represent the diverse phenomenology that models of feebly interacting particles can offer. Conclusions drawn from these presented results should therefore be taken with the necessary caution.

\vskip 2mm
Another important remark should be made upfront is that all the cosmological bounds present in the plots shown in this chapter (eg: Big Bang Nucleosynthesis (BBN), Cosmic Microwave Background (CMB), dark matter relic target) are valid only in the context of the standard $\Lambda_\text{CDM}$ cosmology, hence also these bounds should be taken with a lot of caution.



\vskip 2mm
Keeping this in mind, in the next Sections we will show the state-of-the-art of the searches performed at accelerator-based experiments for each of the four portals and the related open problems from a theoretical view point. The experimental results and sensitivities are shown in the large majority of cases as exclusion limits at 90\% CL.

\vskip 2mm
Following the prescriptions shown in Ref.~\cite{Antel:2023hkf},
all the plots have been made using the following graphical conventions: filled coloured areas indicate existing bounds,  filled shaded gray areas are interpretations of astrophysical/cosmological measurements and/or reinterpretation of old data sets performed by people not belonging to the original collaboration, solid lines are extrapolations from existing data sets, dashed lines are projections obtained using a full Monte Carlo with background simulated (at different levels); Dotted lines are projections obtained using toy Monte Carlo;

\vskip 2mm
For reader's convenience we have isolated results coming from past and existing experiments (including their possible upgrades if already approved) from those coming from projects still in the  design phase. The reason for this choice is that during the design phase, projects can still undergo to sizeable changes that make the projected sensitivities not fully reliable. For some of the benchmarks we also provide a short summary of open theoretical issues related to computations of cross-sections, lifetimes and branching fractions that are discussed in specific contributions across this document.

\subsection{Dark Photon}

The dark photon \cite{Holdom:1985ag} is perhaps the model builder's primary workhorse, and often the first ingredient one reaches for when constructing a dark sector with some coupling to the SM. Its main appeal is perhaps due to the following salient features: \emph{(i)} all its parameters can be made technically natural, \emph{(ii)} one can easily couple it to the SM fermions without fearing gauge anomalies, anomalous flavor violation or Yukawa suppression factors for the lower generations and \emph{(iii)} its phenomenology is reassuringly familiar from our experience with the SM photon and $Z$ bosons. The dark photon is particularly useful in dark matter models, where it can act as the mediator in the freeze-out process.

\vskip 2mm
Another handy feature of the dark photon is that it is very natural to have an $\mathcal{O}(1)$ coupling of the dark photon ($A'$) with dark sector matter while maintaining a very small coupling with the SM. This is achieved simply by setting its gauge coupling to be $\mathcal{O}(1)$ but its mixing with the photon to be tiny ($\epsilon \ll1$), which is technically natural. This is particularly convenient
to deplete the relic abundance of a light relic dark matter particle $\chi$ that could overclose the Universe.
This is conveniently done just switching on $\chi \chi\to A'A'$ annihilation or $\chi\to A'A'$ decay channels that deplete the relic density of $\chi$ by
increasing the abundance of $A'$. As long as $m_{A'}\gtrsim 10$ MeV, the $A'$ can then harmlessly decay back to the SM before the onset of Big Bang Nucleosynthesis (BBN).

\subsubsection{Dark photon in visible final states (BC1)}
If $m_{A'} < 2 m_{\chi}$ the dark photon once produced decays back to SM states and the parameter space is defined by its mass and its coupling to SM particles, $(m_{A'} , \epsilon)$.

\begin{figure}[H]
    \centering
    \includegraphics[width=\textwidth]{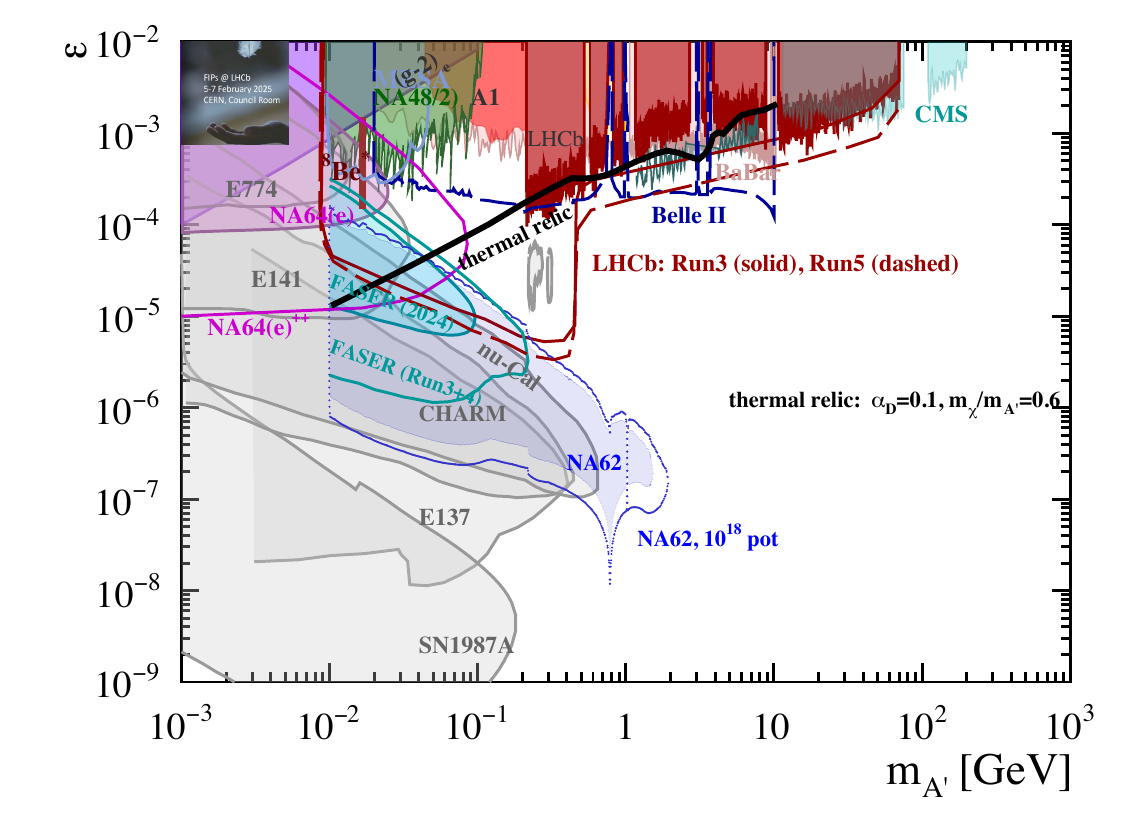}
    \caption{{\bf Dark photon into visible final states (BC1). Past and running experiments}. The parameter space is defined by $\varepsilon$ versus $m_{A'}$.  Current bounds and future projections for 90\% CL exclusion limits. {\it Filled gray areas} bounds coming from interpretation of old data sets or astrophysical data: E774~\cite{Bross:1989mp}, E141~\cite{Riordan:1987aw}, E137~\cite{Bjorken:1988as,Batell:2014mga,Marsicano:2018krp}), $\nu$-Cal~\cite{Blumlein:2011mv,Blumlein:2013cua}, CHARM~\cite{Gninenko:2012eq}, supernovae SN1987A~\cite{Chang:2016ntp} and $(g-2)_e$~\cite{Pospelov:2008zw}. {\it  Filled coloured areas } are bounds set by experimental collaborations: A1~\cite{Merkel:2014avp}, LHCb~\cite{LHCb:2019vmc}, CMS~\cite{CMS:2023hwl},
    BaBar~\cite{BaBar:2014zli}, KLOE~\cite{KLOE-2:2011hhj,KLOE-2:2012lii,KLOE-2:2014qxg,KLOE-2:2016ydq}, NA64(e)~\cite{Andreev:2021fzd}, NA48/2~\cite{NA482:2015wmo}, NA62 in dump mode~\cite{NA62:2025yzs}), FASER~\cite{FASER:2024bbl}
    The $^8 Be$ line is the interpretation in the dark photon model of the ATOMKI anomaly~\cite{Krasznahorkay:2019lyl}: A recent MEG result~\cite{MEGII:2024urz} excludes at 90\% CL this interpretation, while a recent PADME analysis~\cite{Bossi:2025ptv} shows a $\sim 2 \sigma$ excess at around a mass of 16.9~MeV.
    {\it Solid coloured lines} are projections based on existing data sets: Belle II~\cite{Belle-II:2018jsg}; LHCb~\cite{LHCb:upgrade}; NA62 in dump mode with $10^{18}$~\cite{NA62-1018pot}; NA64(e)~\cite{Gninenko:2013rka, Andreas:2013lya}, FASER for Run 3 (250/fb) and Run 4 (680/fb)~\cite{Kling:2020mch}. The relic density curve~\cite{Kling:2021fwx} has been computed for a specific value of the dark matter-dark photon coupling $\alpha_D = 0.1$ and mass ratio  $m_{\chi}/m_{A'} = 0.6$.
}
    \label{fig:DP-BC1-running}
\end{figure}

\begin{figure}[htb]
    \centering
    \includegraphics[width=\textwidth]{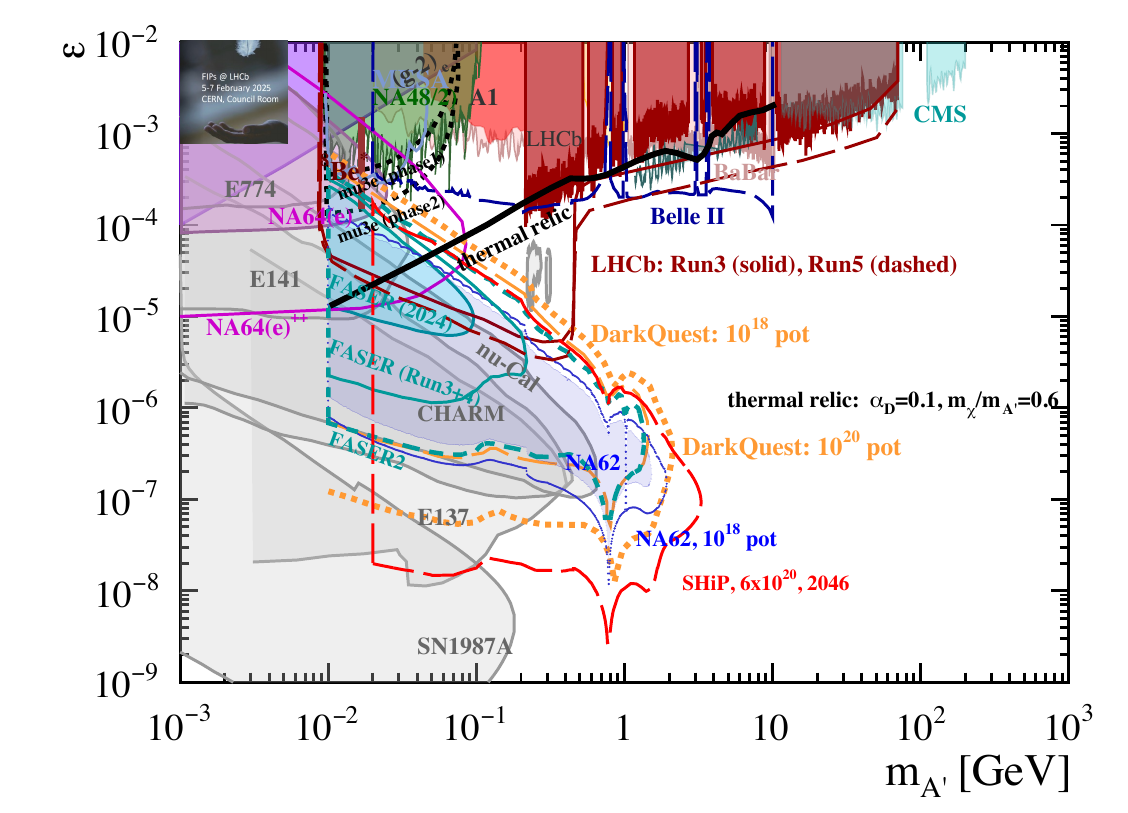}
    \hspace{4mm}
    \caption{ {\bf Dark photon into visible final states (BC1). Past, running, and future experiments}. The parameter space is defined by $\varepsilon$ versus $m_{A'}$.  Current bounds and future projections for 90\% CL exclusion limits. In addition to the filled gray areas, filled coloured areas, and solid curves described in Figure~\ref{fig:DP-BC1-running}, we add here {\it dashed and dotted coloured lines } that are projections from experiments in a design phase:
    FASER2~\cite{Anchordoqui:2021ghd, Feng:2022inv};
    DarkQUEST~\cite{Apyan:2022tsd},
    SHiP~\cite{Ahdida:2023okr}.
    \label{fig:DP-BC1-all}}.
\end{figure}

The results from past and running experiments are shown in Figure~\ref{fig:DP-BC1-running}. Projections from experiment still in the design phase are added in Figure~\ref{fig:DP-BC1-all}.

\vskip 2mm
The two plots show also the parameter space compatible with the dark matter relic density, for the dark matter-dark photon coupling $\alpha_D = 0.1$ and  the very specific mass ratio  $m_{\chi}/m_{A'} = 0.6$~\cite{Kling:2021fwx}. This choice of the mass ratio guarantees from the one hand that the dark photon decays back to SM particles and on the other hand that the DM
relic density is set by annihilations into the SM fermions, $\chi \chi \to A^{*'} \to f^+ f^-$  and it is not suppressed by the otherwise
(if kinematically allowed) very efficient annihilations in the dark sector, $\chi \chi \to A' A'$~\cite{DAgnolo:2015ujb}, the dark photon initial
state radiation $\chi \chi \to A' A^{'*} \to A' \ell^+ \ell^-$~\cite{Rizzo:2020jsm}, and the resonant annihilations into the SM particles,
$\chi \chi \to A' \to \ell^+ \ell^-$
~\cite{Feng:2017drg,Berlin:2020uwy, Bernreuther:2020koj}.

\vskip 2mm
To this respect is important to emphasize two points, that are valid also for Section~\ref{sssec:BC2}:
\begin{itemize}
    \item[-] The region below the relic target line would lead to an overabundance of DM, and is hence excluded.
    \item[-] Within reasonable variations, the relic target line moves a bit up/down. But unless one fine-tunes some mass ratio, it does not move a lot, and hence provides some nice target region for searches.
 \end{itemize}

\clearpage
\subsubsection{Dark Photon in invisible final states (BC2)}
\label{sssec:BC2}

\begin{figure}[htb]
    \centering
    \includegraphics[width=0.9\textwidth]{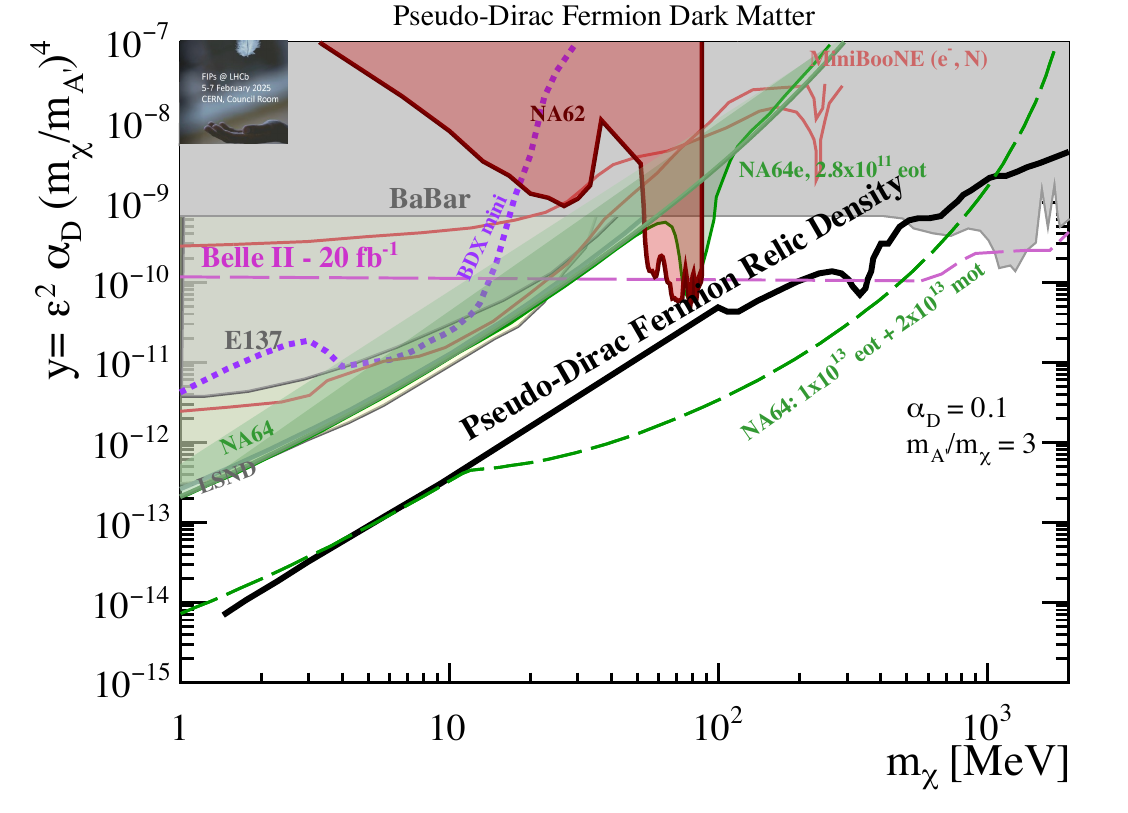}
        \includegraphics[width=0.85\textwidth]{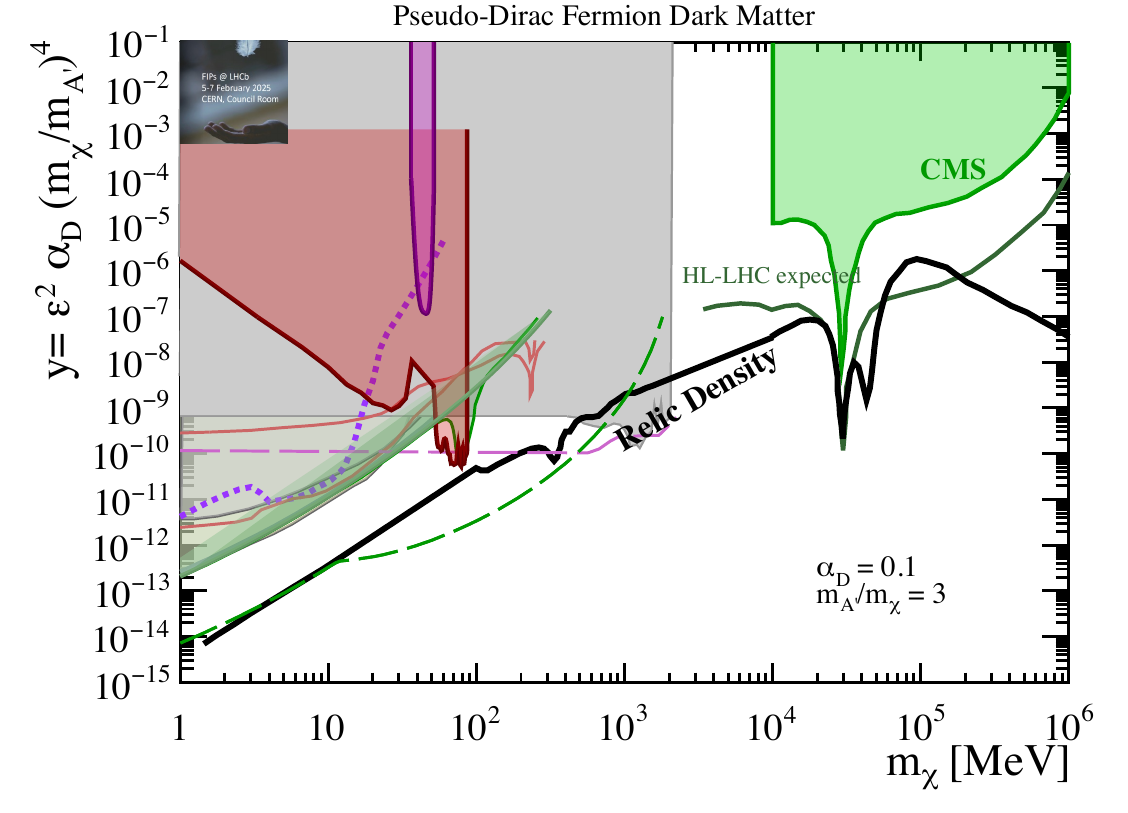}
    \hspace{4mm}
    \caption{\small {\bf Dark Photon into invisible final states (BC2). Past and running experiments}.
  Current bounds and future projections for 90\% CL exclusion limits
 for light dark matter production through a dark photon in the plane defined by the "yield" variable $y$ as a function of DM mass $m_{\chi}$ for a specific choice of $\alpha_D = 0.1$ and $m_{A'}/m_{\chi} = 3$. The DM candidate is assumed to be a pseudo-Dirac fermion.  Top plot shows the DM mass range up to a few GeV, bottom plot up to 1 TeV.
{\it filled coloured areas} are bounds set by experimental collaborations: BaBar~\cite{BaBar:2017tiz}; CMS~\cite{CMS:2021far} with reinterpretation done in \cite{EXO-20-004}.
; NA64$_e$~\cite{Andreev:2021fzd}; reinterpretation of the data from E137~\cite{Batell:2014mga} and LSND~\cite{deNiverville:2011it};  result from MiniBooNE~\cite{MiniBooNEDM:2018cxm}; NA62~\cite{NA62:2025upx}.
{\it Solid coloured lines} are projections based on existing data sets:  NA64~\cite{Gninenko:2019qiv},   BDX-mini~\cite{Battaglieri:2020lds},
 Belle-II~\cite{Belle-II:2018jsg}. The "HL-LHC expected" sensitivities come from \cite{Boveia:2022adi}. Preliminary results obtained by the NA64 collaboration with the muon beam are discussed in Refs.~\cite{NA64:2024klw} and ~\cite{NA64:2024nwj}.
}
    \label{fig:DP-BC2-running}
\end{figure}

\begin{figure}[htb]
    \centering
    \includegraphics[width=0.9\textwidth]{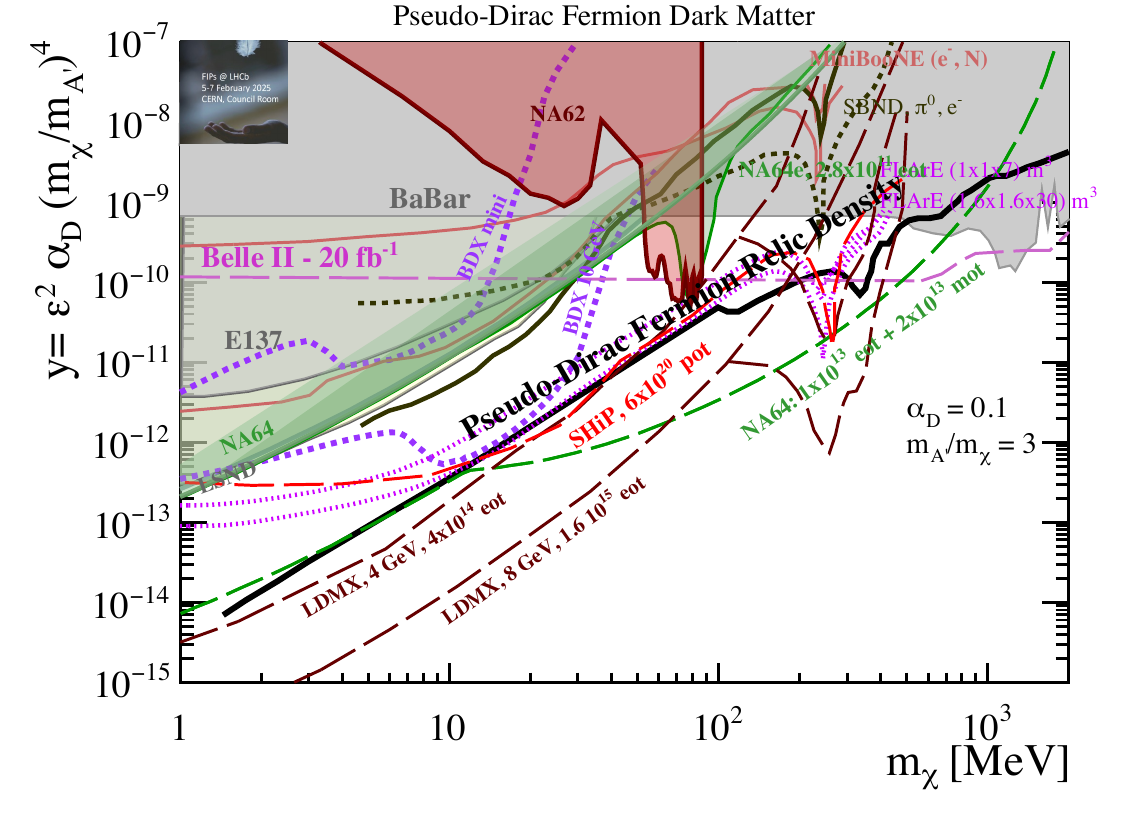}
       \includegraphics[width=0.85\textwidth]{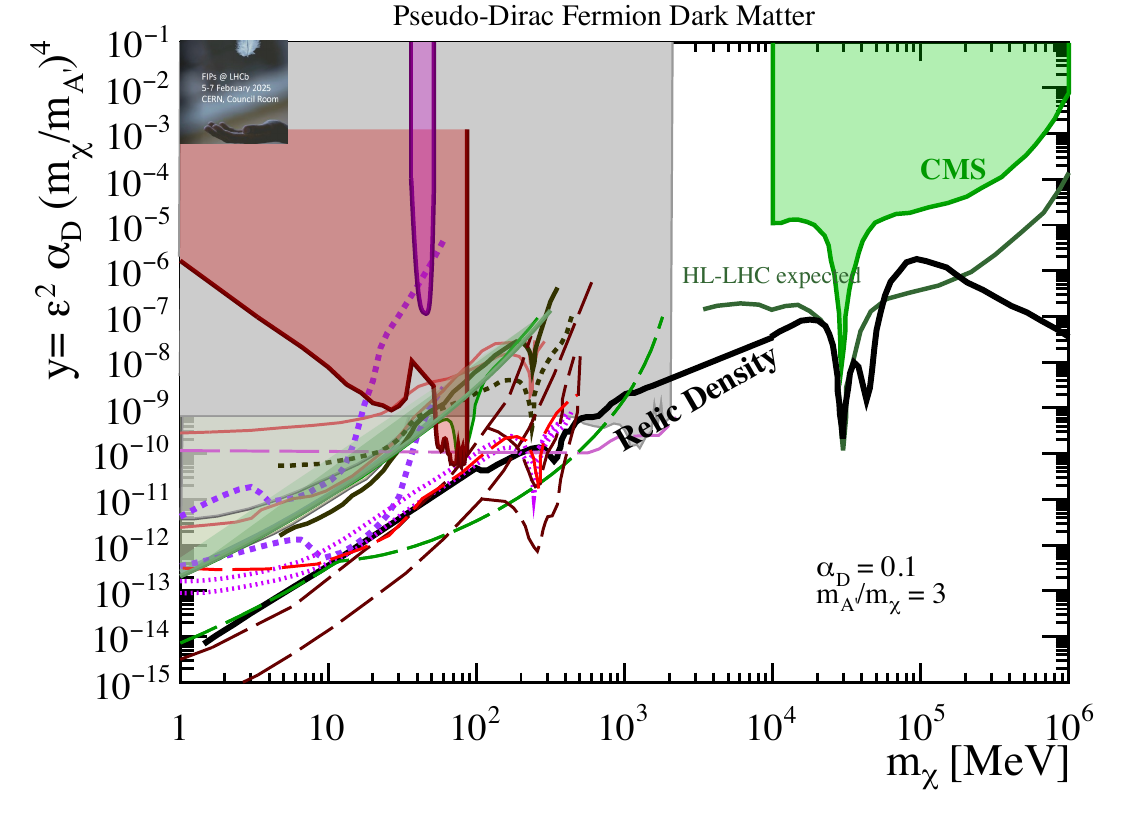}
    \hspace{4mm}
       \caption{\small {\bf Dark Photon into invisible final states (BC2). Past, running, and future experiments}.
         Current bounds and future projections for 90\% CL exclusion limits
 for light dark matter production through a dark photon in the plane defined by the "yield" variable $y$ as a function of DM mass $m_{\chi}$ for a specific choice of $\alpha_D = 0.1$ and $m_{A'}/m_{\chi} = 3$. The DM candidate is assumed to be a pseudo-Dirac fermion.  Top plot shows the DM mass range up to a few GeV, bottom plot up to 1 TeV.
In addition to the filled gray areas, filled coloured areas, and solid curves described in Figure~\ref{fig:DP-BC2-running},
we add here
{\it dashed and dotted coloured lines} as projections for experiments still in a design phase: SHiP~\cite{SHiP-ECN3-LoI},  BDX~\cite{BDX:2016akw}, SBND~\cite{MicroBooNE:2015bmn}, FLArE~\cite{Anchordoqui:2021ghd},
LDMX~\cite{LDMX:2018cma,Akesson:2022vza}.}
    \label{fig:DP-BC2-all}
\end{figure}

This benchmark represents the model of minimal dark  matter thermal freeze-out. In fact thermal WIMPs interacting solely through the electroweak force must be heavier than 1-2 GeV~\cite{Lee:1977ua}, hence  sub-GeV thermal DM $\chi$ requires new forces with light mediators~\cite{Boehm:2003hm}, as for example, a dark photon.

\vskip 2mm
The benchmark values of dark coupling constant $\alpha_{D} = g_D^2/(4\pi)$ are such that the decay of $A'$ occurs predominantly into the invisible $\chi\chi^*$ final state. The parameter space for this model is $\{ m_{A'}, \epsilon, m_\chi, \alpha_D \}$ with further model-dependence associated with properties of $\chi$ (boson or fermion). The suggested choices for the parameter space are (i) $\epsilon$ vs $m_{A'}$ with $\alpha_D \gg \epsilon^2 \alpha$ and $2m_\chi <m_{A'}$, (ii) $y$ vs. $m_\chi$ plot where the {\it yield} variable $y$,  $y = \alpha_D \epsilon^2 (m_\chi/m_{A'})^4$, contains a combination of parameters relevant for the freeze-out and DM-SM particles scattering cross section. One possible choice is $\alpha_D = 0.1$ and $m_{A'}/m_\chi = 3$.

\vskip 2mm
Our choice is to report results in the $y$ versus $m_{\chi}$ plane for a specific choice of the DM candidate, that in our case is a pseudo-Dirac fermion. The results are reported in Figure~\ref{fig:DP-BC2-running} (for past and running experiments) and in Figure~\ref{fig:DP-BC2-all} (including experiments in the design phase).

\vskip 2mm
The choice of DM as a pseudo-Dirac fermion is essentially due to the CMB bounds. In facts, if DM annihilates during CMB era, strong constraints exist on the energy injected in the photon plasma during recombination.  The CMB bounds are based on visible energy injection at T$\sim$~eV, which reionizes the newly recombined hydrogen and thereby modifies the ionized fraction of the early universe. Hence: for s-wave annihilating DM, CMB bounds rule out $m_\text{DM} < 10$~GeV.

\vskip 2mm
In order to escape this bound, for thermal DM in the MeV-GeV range there are viable options.
\begin{enumerate}
\item DM annihilates in $p-$wave ($\sigma v$ is $v^2$ suppressed, hence smaller at low temperature): assuming a vector mediator, DM can be a scalar particle.
\item Presence of a mechanism that cuts off late time annihilation, as e.g.\ mass splitting in the $\chi - \overline{\chi}$ system: DM can be a pseudo-Dirac fermion, as considered here.
Similar considerations apply to DM as Majorana particle.
\end{enumerate}

In case the DM is a pseudo-Dirac fermion  the $U(1)_{A'}$ breaking mass term $\Delta m^2\chi^2$ generates a mass splitting $\Delta m$ between the real components of $\chi = 2^{-1/2}(\chi_1+i\chi_2)$ that is easily larger than the kinetic energy of DM today ($E_\text{kin} \sim \frac12 m_\chi c^2 (v_\text{SM}/c)^2 \sim 10^{-5}{\rm eV}\times m_\chi/20\,{\rm MeV}$). In this case, $s$-wave elastic direct detection scattering is completely quenched, while having a negligible effect on the primordial abundance.

\vskip 2mm
The quenching effect is also the reason why the direct detection experiments cannot not see any signals  and that is why their results are not reported in Figures~\ref{fig:DP-BC2-running} and~\ref{fig:DP-BC2-all}. Hence this particular model is important for accelerator-based experiments that are the only ones able to detect a signal due to higher energy of the involved particles.

\clearpage
\subsubsection{Open theoretical issues}
\label{sssec:DP-theory}

The main theoretical open issues are related to dark photon production at proton accelerators (colliders and beam dump) where the uncertainty in the production rate can reach up to 1-2 orders of magnitude in the 1~GeV dark photon mass range.
This is clearly shown in Figure~\ref{fig:dark-photon-uncertainty} and Figure~\ref{fig:LDM-uncertainty} of Section~\ref{ssec:fips-pheno} of this document.

As explained in Section~\ref{sssec:DP-proton-bremss} these uncertainties mainly come from the proton bremsstrahlung (where the elastic proton form-factor includes the contributions from vector meson excitations) and deep inelastic scattering (which probes the domain of low parton's energy fractions $x$ and scales $Q \simeq m_{\text{FIP}}$).

For dark photon masses around 1 GeV, the most popular production mechanisms are neutral meson
decays ($m_{A'} <$ 0.4~GeV), neutral meson mixing (in the mass range of $\rho^{0},\omega,\phi$ mesons and their excitations), proton bremsstrahlung (0.4 GeV $< m_{A'} <$ 1.8 GeV), and the Drell-Yan process ($m_{A'} >$ 1.8 GeV). The neutral meson mixing and the proton bremsstrahlung have been discussed in details in Sections~\ref{sssec:DP-proton-bremss} and~\ref{ssec:kyselov}, while a thorough computation of dark photon productions in elastic and inelastic proton bremsstrahlung is performed in Section~\ref{sssec:DP-proton-bremss}. Here we just summarize the outcome of these chapters.

\begin{itemize}

\item {\bf Proton elastic bremsstrahlung}\\
Several methods have been used in literature to compute the fully elastic proton brems\-strah\-lung cross section for dark photons. An extensive discussion is performed in Section~\ref{sssec:DP-proton-bremss}. The results obtained in
\cite{Foroughi-Abari:2021zbm,Gorbunov:2023jnx,Kim:1973he,Blumlein:2013cua} for the fully elastic bremsstrahlung cross section are compared in Figure~\ref{fig:comp-elastic} of this document.
This discussion shows that all of them, except the Blumlein-Brunner (BB) approximation~\cite{Blumlein:2013cua}, are consistent with each other, in particular the Weizsacker-Williams (WW) approximation that, therefore, can be safely used to calculate the elastic proton bremsstrahlung with good accuracy.

\item {\bf Proton inelastic bremsstrahlung}\\
For the inelastic proton
bremsstrahlung, the role of the accuracy in determining the proton electromagnetic form factors in the unphysical region has been carefully presented in Section~\ref{sssec:DP-proton-bremss}. In this Section, it was also shown that the inelastic bremsstrahlung cross section always depends on three auxiliary splitting functions and each of them should be taken into
account in the answer.  Possible sources of uncertainties in the bremsstrahlung cross section and the benchmarks of quasi-real approximation are also discussed.
Here the main issue is the validity of the assumed approximations, like for example, threshold in $p_\text{T}$
or form factor values.
Further work will be needed to assess these uncertainties and to fold these computations in some widely-used generator (eg: PYTHIA).

\item {\bf Mixing with neutral mesons}\\
 Dark photons have mixing with $\rho^{0},\omega,\phi$ mesons and their excitations. Mixing plays a crucial role in high-energy accelerator experiments, as it substantially contributes
to the flux of LLPs in the GeV mass range through production processes such as meson decays, parton hadronization, and proton bremsstrahlung.
A thorough discussion about the  dark photon production channels via mixing  and their incorporation in event generator \texttt{PYTHIA8}~\cite{Bierlich:2022pfr} is done in Section~\ref{ssec:kyselov}. The current treatment carries a large uncertainty for cases where the dark photon is nearly degenerate with the SM mesons~\cite{LoChiatto:2024guj, Fuchs:2016swt} and additional work is required to determine reliably the production cross sections and decay widths.

\item {\bf Spin correlations}\\
Many simulations assume that the dark photos beam is unpolarized  and that the dark photon decays isotropically in its restframe. In reality, however, the decay
$\pi^0 \to \gamma A'$ to a longitudinally-polarized $A'$ is forbidden by angular momentum conservation leading to a polarized beam and spin correlations between production and decay. The authors of Ref.~\cite{Feng:2025gji} studied these spin correlations and found that the error made in them is typically small, leading to a reduction of event rate of up to a few percent at FASER and up to 10\% at SHiP. 

\end{itemize}

\clearpage
\subsection{Dark Higgs}

\begin{figure}[htb]
    \centering
    \includegraphics[width=0.9\textwidth]{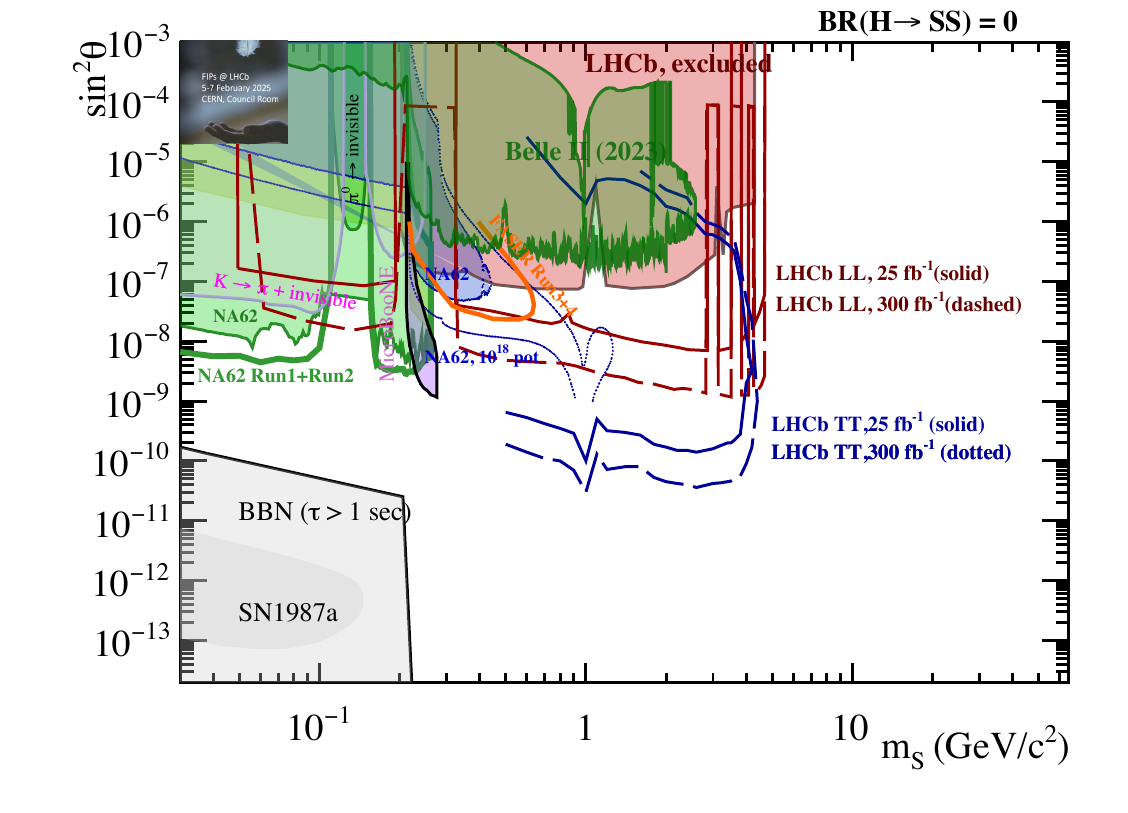}
    \includegraphics[width=0.9\textwidth]{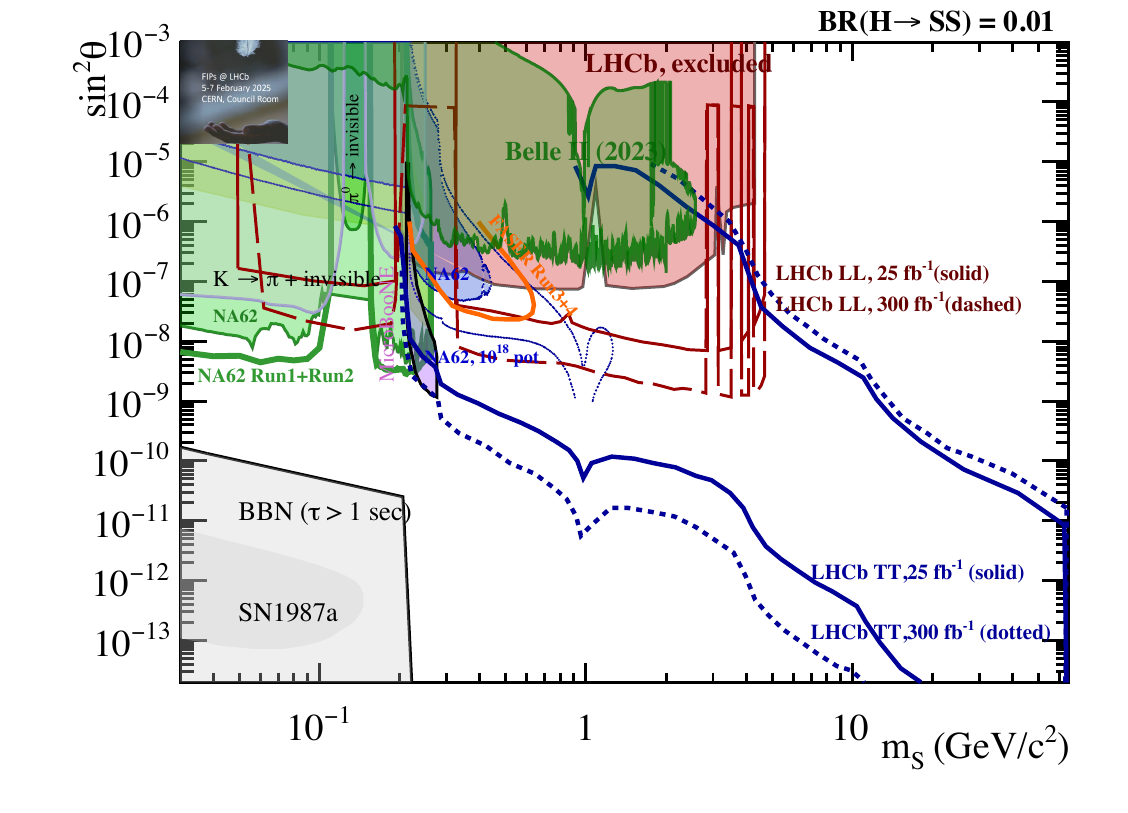}
    \caption{{\bf Sensitivity to light dark scalar with BR(H$\to$SS) = 0 (top) and BR(H$\to$SS)=0.01 (bottom). Past and running experiments.} Current bounds and future projections for 90\% CL exclusion limits.
{\it Filled areas} come from: reinterpretation~\cite{Winkler:2018qyg} of results from CHARM experiment~\cite{Bergsma:1985qz};
NA62 in kaon~\cite{CortinaGil:2020fcx, NA62:2021zjw, NA62:2020pwi,NA62:2025upx} and dump mode~\cite{NA62:2025yzs};
NA62 from the analysis of the $\pi^0 \to$ invisible decays~\cite{NA62:2020pwi};
MicroBooNE from NuMI data~\cite{MicroBooNE:2022ctm}; LHCb~\cite{Aaij:2016qsm, Aaij:2015tna} and Belle II~\cite{Belle-II:2023ueh}.
{\it Coloured lines} are projections of existing 
experiments: LHCb with 25/fb and 300/fb using long tracks~\cite{Craik:2022riw} and T-tracks~\cite{LHCb-DarkHiggs}; NA62 Run1+Run2 is the projection for data collected until 2026~\cite{NA62-proj};
FASER for Run 3 (250/fb) and Run 4 (680/fb)~\cite{Kling:2020mch}.
BBN and SN1987A are from ~\cite{Fradette:2017sdd} and ~\cite{Dev:2020eam}. A private reinterpretation of published CMS results within this framework is shown in Ref.~\cite{delValle:2025zvo}.}
    \label{fig:DS-running}
\end{figure}

\begin{figure}[htb]
    \centering
    \includegraphics[width=0.95\textwidth]{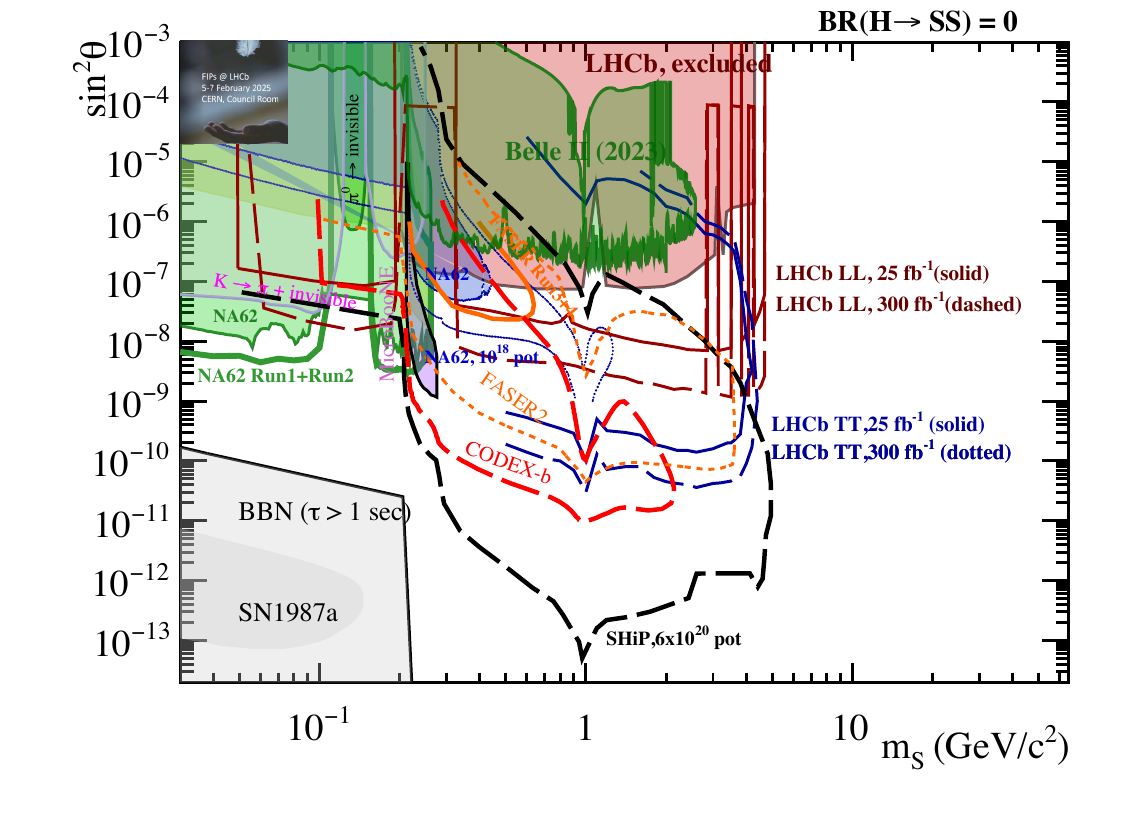}
    \includegraphics[width=0.95\textwidth]{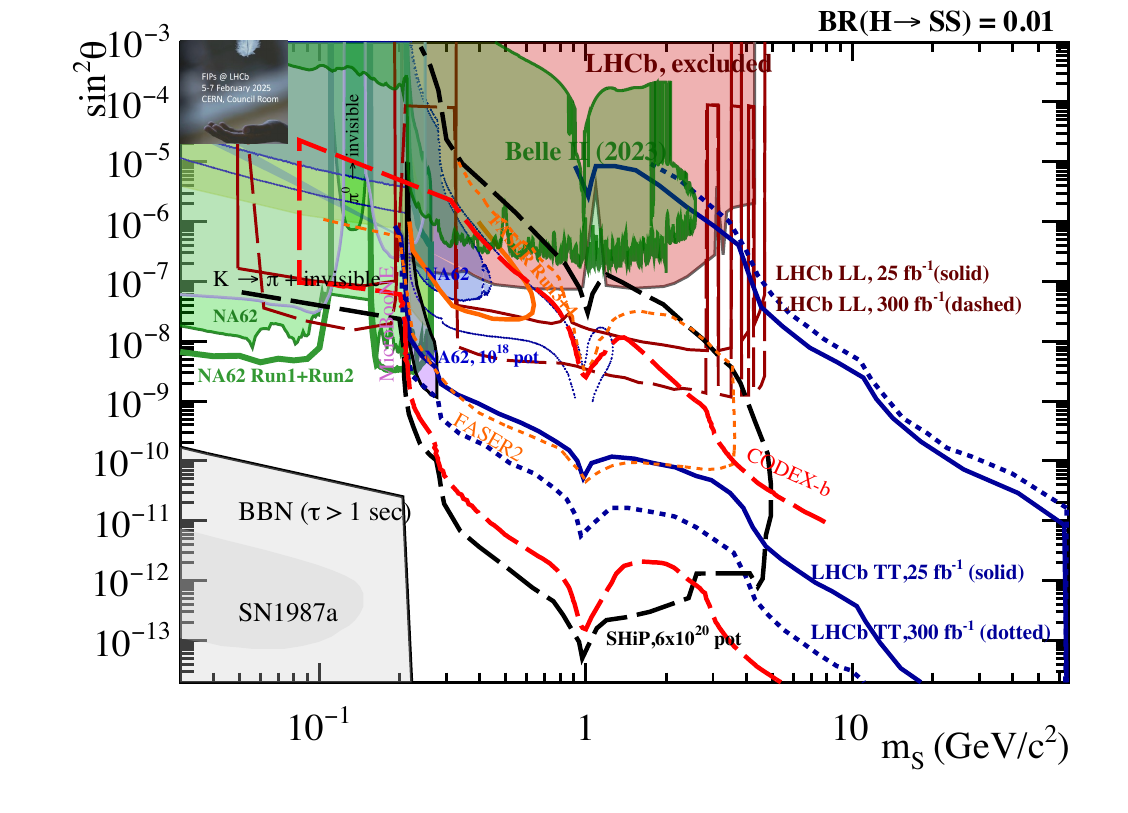}
    \hspace{4mm}
        \caption{{\bf Light dark scalar with BR(H$\to$SS) = 0 (top) and BR(H$\to$SS)=0.01 (bottom). Past, running, and future experiments.} Current bounds and future projections for 90\% CL exclusion limits.
        {\it Gray/coloured filled areas} and {\it coloured solid lines} are for past and current bounds from existing experiments, as in Figure~\ref{fig:DS-running}.
{\it Dashed and dotted coloured lines} are for experiments still in the design phase: SHiP~\cite{Ahdida:2023okr};
FASER2~\cite{Feng:2022inv}, and CODEX-b~\cite{Aielli:2019ivi,Aielli:2022awh,CODEX-b:2025rck}
.}
    \label{fig:DS-all}
\end{figure}

\subsubsection{Description of the benchmarks BC4 and BC5}
The minimal scalar portal model operates with one extra singlet field $S$ and two types of couplings, $\mu$ and $\lambda$~\cite{OConnell:2006rsp},
\begin{equation}
\label{scalar}
{\cal L}_\text{scalar} = {\cal L}_\text{SM} + {\cal L}_\text{DS} - (\mu S+ \lambda S^2)H^\dagger H \, .
\end{equation}

At low energy, the Higgs field can be substituted for $H = (v + h)/\sqrt{2}$, where $v = 246$\,GeV
is the the EW vacuum expectation value, and  $h$ is the field corresponding to the physical 125\,GeV Higgs boson.
The non-zero $\mu$ leads to the mixing of $h$ and $S$ states. In the limit of small mixing, it can be written as
\begin{equation}
\theta = \frac{\mu v}{m_h^2- m_S^2} \, .
\end{equation}

The dark sector Lagrangian ${\cal L}_\text{DS}$ may include the interaction with dark matter $\chi$, ${\cal L}_\text{DS}= S\bar\chi \chi+...$, however
most viable dark matter models in the sub-EW scale range imply $2 m_\chi > m_S$ \cite{Krnjaic:2015mbs} and hence the dark scalar, once produced, decays back to SM particles.

The established PBC benchmarks $BC4$ and $BC5$~\cite{Beacham:2019nyx} that are based on this Langrangian are repeated here below for convenience.

\begin{itemize}

\item {\em BC4, Higgs-mixed scalar without large pair-production channel} \\
This benchmark represents the limit for $BR(H \to SS) = 0$ (equivalent to $\lambda = 0$) of the subsequent $BC5$ benchmark.

\item {\em BC5, Higgs-mixed scalar with large pair-production channel}\\
  In this benchmark the parameter space is $\{\lambda, \theta, m_S\}$, and $\lambda$ is assumed to dominate
  the production via {\em e.g.} $h\to SS$, $B \to K^{(*)}SS$, $B^0 \to SS$ etc.
  In the sensitivity plots a value  of the branching fraction $BR({h \to SS})$
  close to $10^{-2}$ is assumed in order to be complementary to the LHC searches for the Higgs to invisible channels.

 \end{itemize}

The current status of experimental searches and projections for running accelerator-based experiments for the minimal scalar portal model is shown in Figure~\ref{fig:DS-running}, for $BC4$ (top) and $BC5$ (bottom), respectively. Figure~\ref{fig:DS-all} includes also the projections from experiments still in the design phase.

\clearpage
\subsubsection{Open theoretical issues}
Most of the still open theoretical issues of the dark scalar benchmark are discussed in Sections~\ref{ssec:fips-pheno},~\ref{sssec:ovchynnikov},~\ref{sssec:gorbunov}. Here we summarize the main points.

\vskip 2mm
The main production mode for a light dark scalar $S$ are the FCNC decays
$B \to X_s + S$, $B \to X_s + S S$,  and $H(125) \to S S$. Then, the main channel becomes the proton bremsstrahlung~\cite{Boiarska:2019jym}.  While the decay
$B \to  K S S$ is theoretically clean, for the $B \to  X_s S S $ there are several descriptions in literature (see for example Ref.~\cite{Buonocore:2023kna}, Figure~7): In order to make uniform the description across the experiments, we recommend to use the {\it spectator model}.
For dark scalar masses near the SM scalar resonances the main production channel becomes the proton bremsstrahlung due to the mixing. The current understanding of these scalar resonances is however limited. Moreover, similar issues described in Section~\ref{sssec:DP-theory} apply here for determining precisely the production cross sections.

\vskip 2mm
All couplings of the light scalar to the SM fields are due to the mixing with the SM Higgs bosons, hence all the scalar decay modes and the corresponding branching ratios are those exhibited by the SM Higgs boson, would it be light, of the same mass as the light scalar. Likewise, the production modes of the light scalar are the same as those for the light SM Higgs boson.
However, the scalar decay rates into hadrons  suffer of uncertainties and generically prevent one from placing precise direct limits on the model parameters. This is because the hadronic modes dominate the scalar decay and shape the hadronic partial decay widths and, hence, the scalar lifetime.

\vskip 2mm
In fact, contrary to the dark photon case, the scalar decay widths cannot be directly extracted from the experimental data. In the literature, they are calculated using the information from the scattering processes $\pi \pi \to \pi \pi/ KK$,  which provide the input in the framework using the method of dispersion relations~\cite{Monin:2018lee,Blackstone:2024ouf}. The
decay uncertainty on the decays into a pair of mesons is within an order of magnitude, whereas the widths of more complicated decays (potentially dominating the decay width of GeV-scale scalars), such as scalar $\to 4\pi$ have not even been computed.


\vskip 2mm
For example, a very recent study~\cite{Blackstone:2024ouf} presents not only the rates but also the estimate of the
uncertainty in the prediction for the scalar decay rate into pions, obtained within the dispersion relation
technique. The results are shown in Figure~\ref{fig:unc} by grey colours. One observes that the uncertainty factor
varies from several to several hundreds indeed. The very important note concerns the blue line, which is
typically adopted by experimentalists to use in evaluations of the project sensitivities: At large mass it deviates from the QCD estimates by up two orders of magnitude.
All these uncertainties should be taken into account when translating the searches in exclusion limits of the parameters of the model.

\clearpage
\subsection{Heavy neutral leptons}

As discussed in Section~\ref{ssec:HNL}, the commonly studied simplified phenomenological model introduces one HNL field $N$ that interacts with the SM $W$, $Z$ and Higgs ($h$) fields through the \emph{neutrino portal} according to the Lagrangian shown in~\eqref{PhenoModelLagrangian}.

\begin{figure}[htb]
    \centering
    \includegraphics[width=\textwidth]{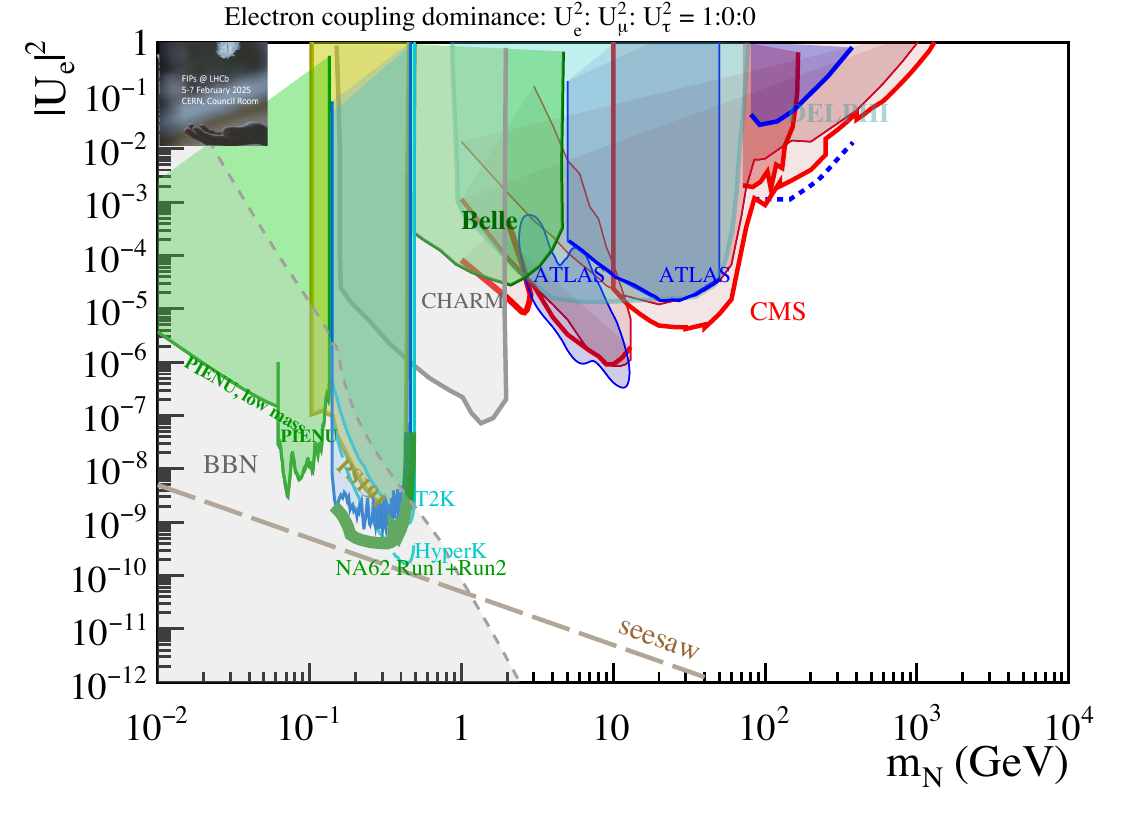}
    \hspace{4mm}
\caption{{\bf HNLs with electron coupling (BC6). Past and running experiments.} Current bounds and future projections for 90\% CL exclusion limits. {\it Filled gray areas} are bounds coming from interpretation of old datasets or astrophysical data: PS191~\cite{Bernardi:1987ek}, CHARM~\cite{Bergsma:1985qz}, PIENU~\cite{Aguilar-Arevalo:2017vlf},  BBN \cite{Boyarsky:2020dzc}.
{\it Filled coloured areas} are bounds set by experimental collaborations: NA62 ($K_{eN}$)~\cite{NA62:2020xlg},
T2K~\cite{extracted_beams:T2K_HNL_results},
Belle~\cite{Liventsev:2013zz}, DELPHI~\cite{Abreu:1996pa}, ATLAS~\cite{ATLAS:2019kpx, ATLAS:2022atq}, and CMS~\cite{CMS:2018iaf, CMS:2022fut, CMS:2024hik, CMS:2024ake, CMS:2024xdq}.
\underline{\it Solid coloured lines} are projections based on existing data sets:
NA62 with the full dataset collected in kaon mode up to 2026~\cite{Ahdida:2023okr}.
The dashed seesaw  line is given by $|U_\alpha|^2 =\sqrt{\Delta m^2_{atm}}/m_N$ corresponding to the naive seesaw scaling and should be considered only as indicative, as it depends on the still unknown value of the mass of the lightest active neutrino. The seesaw and BBN bounds depend stronger on the choice of model parameters than the direct search bounds, see~e.g.~\cite{Drewes:2019mhg,Boyarsky:2020dzc}.
}
    \label{fig:HNL-BC6-running}
\end{figure}

\begin{figure}[htb]
    \centering
        \includegraphics[width=\textwidth]{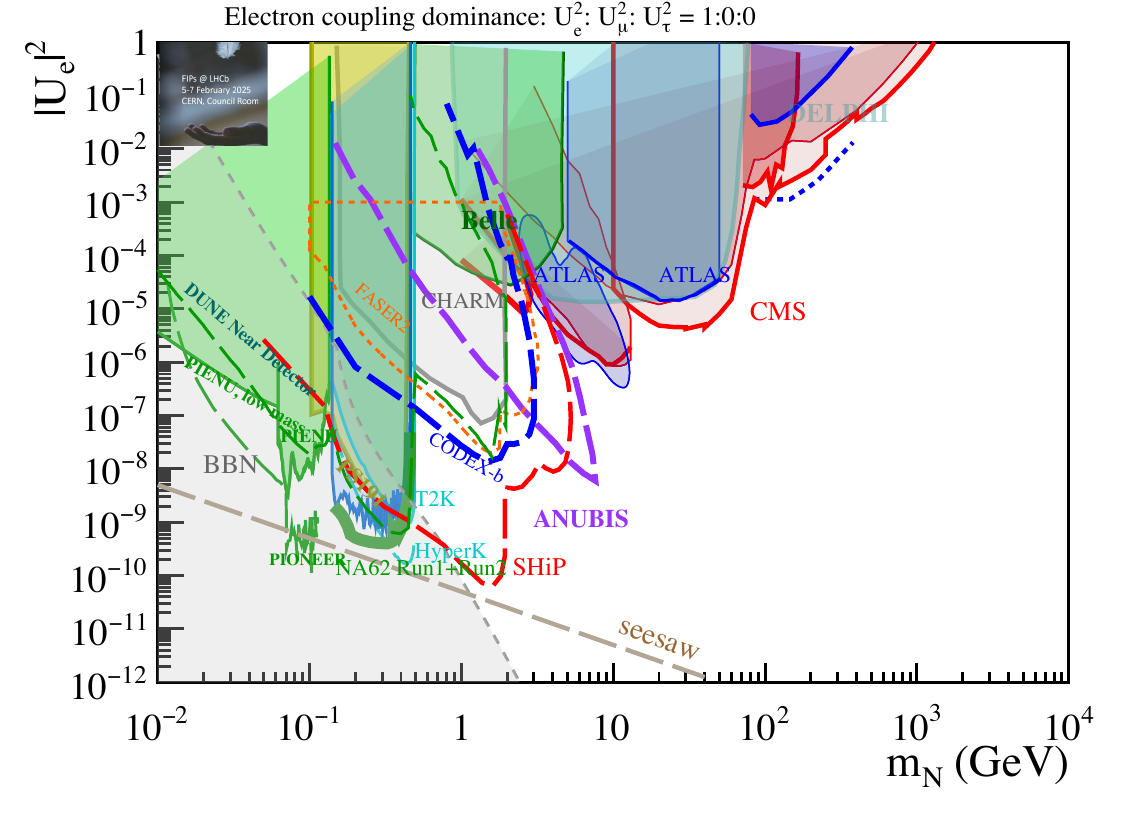}
    \hspace{4mm}
    \caption{{\bf Sensitivity to HNLs with electron coupling (BC6). Past, running, and future experiments.} Current bounds and future projections for 90\% CL exclusion limits. Past and running experiments are represented as filled gray and coloured areas,  and solid coloured lines, as in Figure~\ref{fig:HNL-BC6-running}. {\it Dashed and dotted lines} are for experiments in the design phase:
PIONEER~\cite{PIONEER:2022yag},
DarkQuest~\cite{Blinov:2021say},
Belle II~\cite{Dib:2019tuj},
FASER2~\cite{Ariga:2018uku};
DUNE near detector~\cite{Abdullahi:2022jlv},
Hyper-K (projections based on ~\cite{T2K:2019jwa}),
CODEX-b~\cite{CODEX-b:2025rck},
SHiP~\cite{SHiP-ECN3-LoI},
ANUBIS~\cite{PBC:2025sny}.
}
\label{fig:HNL-BC6-all}
\end{figure}

\begin{figure}[htb]
    \centering
    \includegraphics[width=\textwidth]{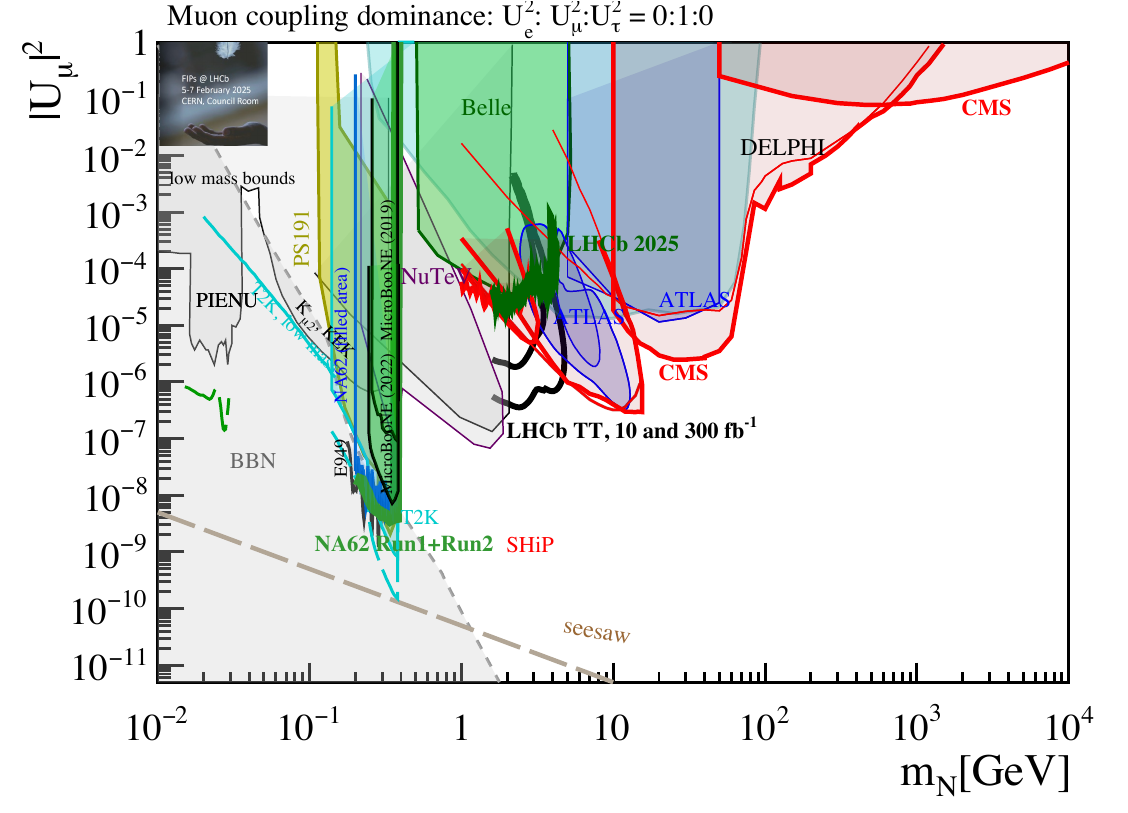}
    \hspace{4mm}
\caption{{\bf HNLs with muon couplings (BC7). Past and running experiments.} Current bounds and future projections for 90\% CL exclusion limits. {\it Filled gray areas} are bounds coming from interpretation of old datasets or astrophysical data:
PS191~\cite{Bernardi:1987ek}, CHARM~\cite{Bergsma:1985qz}, DELPHI~\cite{Abreu:1996pa}, and
BBN~\cite{Boyarsky:2020dzc}.
{\it Filled coloured areas} are bounds set by existing experiments:
NA62 ($K_{\mu N}$)~\cite{NA62:2021bji},
T2K~\cite{extracted_beams:T2K_HNL_results},
Belle~\cite{Liventsev:2013zz},
ATLAS~\cite{ATLAS:2019kpx, ATLAS:2022atq}, CMS~\cite{CMS:2018iaf, CMS:2022fut, CMS:2024hik, CMS:2024ake, CMS:2024xdq}, and LHCb~\cite{LHCb:2020wxx}.
{\it Coloured curves} are projections from existing experiments:  LHCb with 10/fb and 300/fb  using very displaced vertices~\cite{LHCb-DarkHiggs}, NA62 with the full dataset collected in kaon mode up to 2026~\cite{Ahdida:2023okr}.
The dashed seesaw  line is given by $|U_\alpha|^2 =\sqrt{\Delta m^2_{atm}}/m_N$ corresponding to the naive seesaw scaling and should be considered only as indicative.
The seesaw and BBN bounds depend stronger on the choice of model parameters than the direct search bounds, see~e.g.~\cite{Drewes:2019mhg,Boyarsky:2020dzc}
}
    \label{fig:HNL-BC7-running}
\end{figure}

\begin{figure}[htb]
\centering\includegraphics[width=0.9\textwidth]{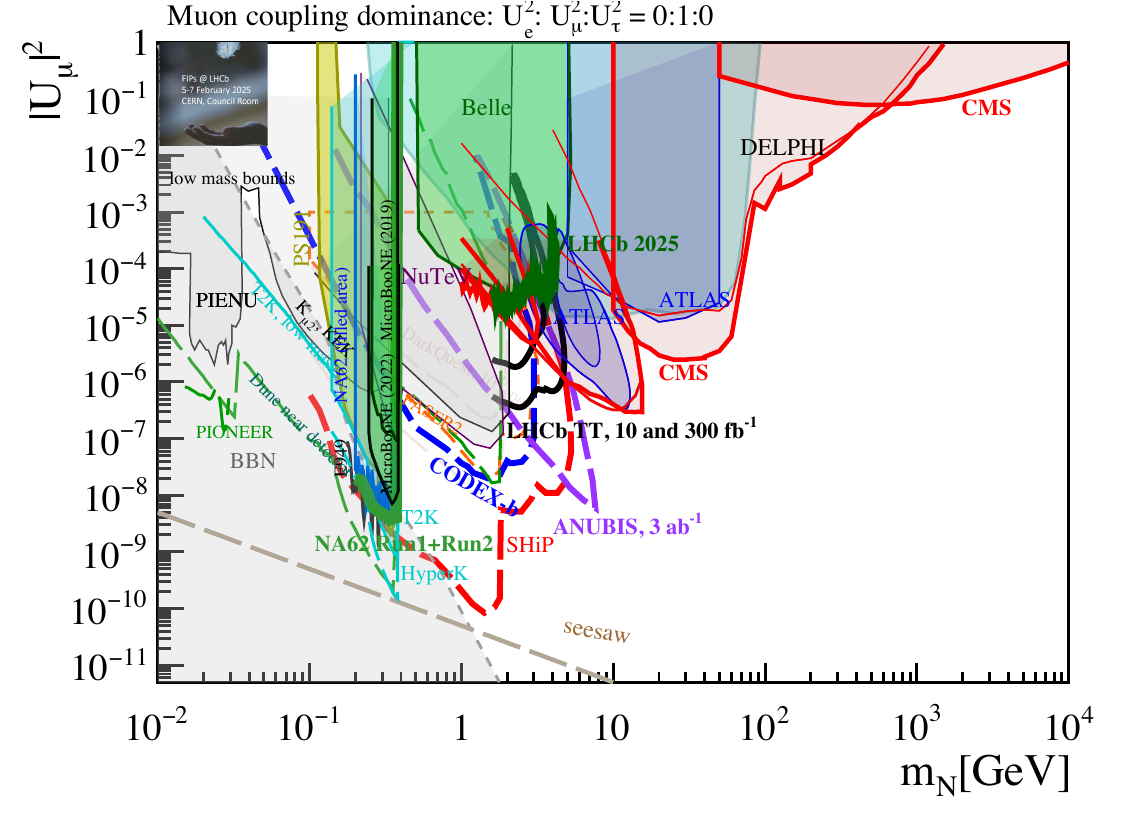}
\caption{{\bf Sensitivity to HNLs with muon coupling (BC7). Past, running, and future experiments.} Current bounds and future projections for 90\% CL exclusion limits. Past and running experiments are represented as filled gray/coloured areas and solid coloured lines, as in Figure~\ref{fig:HNL-BC7-running}. {\it Dashed and dotted lines} are for experiments in the design phase:
PIONEER~\cite{PIONEER:2022yag},
FASER2~\cite{Ariga:2018uku};
DUNE near detector~\cite{Abdullahi:2022jlv},
Hyper-K (projections based on ~\cite{T2K:2019jwa}),
CODEX-b~\cite{CODEX-b:2025rck},
SHiP~\cite{SHiP-ECN3-LoI}, and ANUBIS~\cite{PBC:2025sny}.
} \label{fig:HNL-BC7-all}
\end{figure}

Most of the searches performed at accelerator based experiemnts focus on specific HNL benchmarks --- assuming either exact Dirac or Majorana HNLs that only couple to a single lepton flavour.
These assumptions corresponds to the three benchmark models \ref{BC6}--\ref{BC8} defined in~\cite{Beacham:2019nyx},
\begin{subequations}
\begin{align}
\label{BC6}
U_e^2 : U_\mu^2 : U_\tau^2 = 1:0:0, \tag{BC6} \\
\label{BC7}
U_e^2 : U_\mu^2 : U_\tau^2 = 0:1:0, \tag{BC7} \\
\label{BC8}
U_e^2 : U_\mu^2 : U_\tau^2 = 0:0:1, \tag{BC8}
\end{align}
\end{subequations}
Figures~\ref{fig:HNL-BC6-running} and~\ref{fig:HNL-BC7-running}, show
the current bounds for HNLs under the approximations reported above, with pure electron-coupling and muon-coupling, respectively.
Figures~\ref{fig:HNL-BC6-all} and~\ref{fig:HNL-BC7-all} include the future projections for existing experiments (and their upgrades).








\subsubsection{Open theoretical issues}

\paragraph{Theoretical issues related to model building}

Realistic HNL models based on the type-I seesaw mechanism differ from the simplified phenomenological framework of Eq.~\eqref{PhenoModelLagrangian} in several essential respects.
The most fundamental distinction concerns the number of HNLs required by the theory.
In the type-I seesaw, each massive light neutrino demands the presence of at least one corresponding HNL.
Since neutrino oscillation data indicate the existence of only two massive light neutrinos, the minimal type-I seesaw predicts one massless neutrino ($m_\text{lightest} = 0$), implying $n_\text{HNL} \geq 2$.

To naturally account for the smallness of the light neutrino masses without fine-tuning, the two heavy Majorana states must form an approximate Dirac pair.
This motivates the two benchmark scenarios typically probed at accelerators: a single Majorana HNL or a Dirac HNL.
However, both options face theoretical tensions.
A Dirac HNL preserves lepton number and therefore cannot account for the observed neutrino masses, while a lone Majorana HNL tends to overproduce light neutrino masses due to excessive lepton-number violation.

Finally, if the lightest neutrino were experimentally established to be massive ($m_\text{lightest} \neq 0$), the seesaw framework would necessarily require the presence of a third HNL.

A key difference and a major source of difficulty, compared to the phenomenological model arises from the proliferation of parameters once multiple HNLs are introduced.
After removing unphysical phases, the type-I seesaw with $n_\text{HNL}$ heavy states contains $7n-3$ physical parameters.
The central challenge in defining suitable benchmark models is therefore to establish a consistent mapping between the realistic type-I seesaw and the simplified framework of Eq.~\eqref{PhenoModelLagrangian}.

Fortunately, many of these parameters are already constrained by the requirement that the seesaw reproduce the observed light neutrino masses and mixings.
Specifically, once the measured mass-squared differences, the three mixing angles, and the Dirac $CP$-violating phase of the Pontecorvo-Maki-Nakagawa-Sakata (PMNS) matrix are taken into account, six of the free parameters become fixed.

Figure~\ref{fig:ternary} illustrates the resulting parameter space for $U^2_{e,\mu,\tau}$ in the case $n_\text{HNL}=2$, incorporating both present data and projected sensitivities after DUNE.
The figure highlights that the commonly used {\it single-flavor dominance} assumption, often adopted in experimental analyses, is unphysical in this minimal scenario.
However, for $n_\text{HNL}>2$, the accessible parameter space broadens substantially, and the {\it single-flavor dominance} approximation becomes far more reliable.

To make a closer connection to realistic neutrino mass models, two additional benchmarks were proposed in \cite{Drewes:2022akb}:
\begin{subequations}
\label{NewBenchmarks}
\begin{align}
\label{NObenchmark}
U_e^2 : U_\mu^2 : U_\tau^2 &= 0:1:1,\\
\label{IObenchmark}
U_e^2 : U_\mu^2 : U_\tau^2 &= 1:1:1,
\end{align}
\end{subequations}
Together with the benchmarks in~\ref{BC6}--\ref{BC8}, these capture a more complete physical picture of realistic models in accelerator-based experiments (see Figure~\ref{fig:ternary}).

\paragraph{Theoretical issues related to the branching fraction computations}

The main theoretical issues related to the HNLs decays modes are related to the computation of the HNL branching fractions into hadronic modes, particularly relevant at high masses. More precisely, while the HNL partonic decay widths are known, we also know that the parton model is not reliable in a mass region where hadronic resonances can appear. Moreover the hadronization process for low mass systems is relatively badly described in MC event generators like Pythia, leading to a major uncertainty in the computation of rates of specific hadronic decays. On the other hand, computation of rates of specific hadronic decays is mandatory as different hadronic modes are affected differently from trigger, selection, reconstruction efficiencies and background contaminations.
Hence: the uncertainty in the computation of specific branching fractions containing hadronic modes translates directly into an uncertainty in the exclusion bounds on the model parameters.

\clearpage
\subsection{Axion-Like Particles}

Taking a single pseudoscalar field $a$ one can write a set of its couplings
to photons, quarks, leptons and other fields of the SM. This is shown in Eq.~(\ref{Leff_a}).
\begin{equation}
\begin{aligned}
   {\cal L}_\text{eff}
   &= \frac12 \left( \partial_\mu a\right)\!\left( \partial^\mu a\right)  - \frac{m_{a,0}^2}{2}\,a^2
    +  \frac{\partial^\mu a}{f_a}\,
\sum_F \,\bar\psi_F \gamma_\mu C_F \psi_F
\\
   &\quad\mbox{}  -C_{aGG}\,\frac{\alpha_s}{4\pi}\,\frac{a}{f_a}\,G_{\mu\nu}^a\,\tilde G^{\mu\nu,a}
     -C_{aWW}\,\frac{\alpha_2}{4\pi}\,\frac{a}{f_a}\,W_{\mu\nu}^A\,\tilde W^{\mu\nu,A}
    - C_{aBB}\,\frac{\alpha_1}{4\pi}\,\frac{a}{f_a}\,B_{\mu\nu}\,\tilde B^{\mu\nu}  .
\end{aligned}
\label{Leff_a}
\end{equation}
Here, $\alpha_s=g_s^2/(4\pi)$, $\alpha_2=g^2/(4\pi)$ and $\alpha_1=g^{\prime\,2}/(4\pi)$ are the respective SM gauge coupling parameters, and
$F$ denotes the left-handed fermion multiplets in the SM. $C_F$ is a
Hermitian matrix in generation space with dimensionless entries which, together with the dimensionless coefficients $C_{aGG}$, $C_{aWW}$ and
$C_{aBB}$, depend on the specific UV completion featuring the Abelian global symmetry. Since all the interactions with the SM in Eq.~\eqref{Leff_a} are inversely proportional to $f_a$,  axions and ALPs with $f_a\gg v$ are indeed feebly-interacting pseudoscalar particles.

\vskip 2mm
For axions that solve the strong CP problem, experimental bounds require decay constants $f_a$ well above the PeV, with details depending on the specific model. This is also the order of magnitude of $f_a$ used in many ALP studies. However, for ALPs arising from other dynamical mechanisms, such as composite Higgs models, a decay constant as low as a few TeVs is still allowed, as long as the ALP mass is sufficiently large to avoid bounds from meson decay. This latter type of ALPs is promptly decaying, has gluon fusion as leading production mode at hadron colliders, and is also an attractive target for LHCb, as discussed in Section~\ref{sssec:ferretti} and \cite{Ferretti:2025zsq}.

\vskip 2mm
In principle, the set of possible couplings is very large and only the flavour-diagonal subset is considered in this study.
Following the PBC proposal~\cite{Beacham:2019nyx}, they are
{\it ALPs with photon-coupling dominance (BC9)}, {\it  ALPs with fermion-coupling dominance (BC10)}, {\it ALPs with gluon-coupling dominance (BC11)} and we add {\it ALPs with W-coupling dominance (BC12)} not included in the first PBC list but included in the Eq.~\eqref{Leff_a}.

\vskip 2mm
The ALP portals are {\em effective} interactions, and would typically require UV completion at or below $f_i$ scales. This is fundamentally different from vector, scalar, and neutrino portals that do not require external UV completion.   Moreover, the renormalization group evolution is capable of inducing new couplings. All the sensitivity plots in this Section assume a cut-off scale\footnote{ We use $\Lambda =1$\, TeV for ALPs with fermion couplings and $\Lambda = 4\pi$\, TeV for ALPs with gluon couplings.}.

\vskip 2mm
While the current searches assume a phenomenology where the ALPs have only one coupling switched on at the time, including all the possible interactions in the calculation of experimental predictions also paves the way for the next milestone in the study of ALP phenomenology: scenarios with more than one type of interactions (so-called \emph{codominance}). An interesting example, explored recently in Refs.~\cite{Ertas:2020xcc,Kelly:2020dda}, is the case $C_{BB} = C_{WW} = C_{GG}$ (and negligible couplings to fermions), which leads to an accidental cancellation in the effective ALP-photon coupling.

\vskip 2mm
While a detailed prescription about how to practically implement the co-dominance scenario in future experimental searches is still being worked out, we emphasize that they naturally occur in many UV completions and that it will be essential to consider them in order to obtain a comprehensive understanding of the parameter space of ALP models.

\vskip 2mm
In the following sections we will show the current experimental results and projections for the searches for ALPs in the MeV-TeV range with only one coupling switched on at the time.

\subsubsection{ALPs with photon couplings (BC9)}

\begin{figure}[htb]
    \centering
    \includegraphics[width=0.8\textwidth]{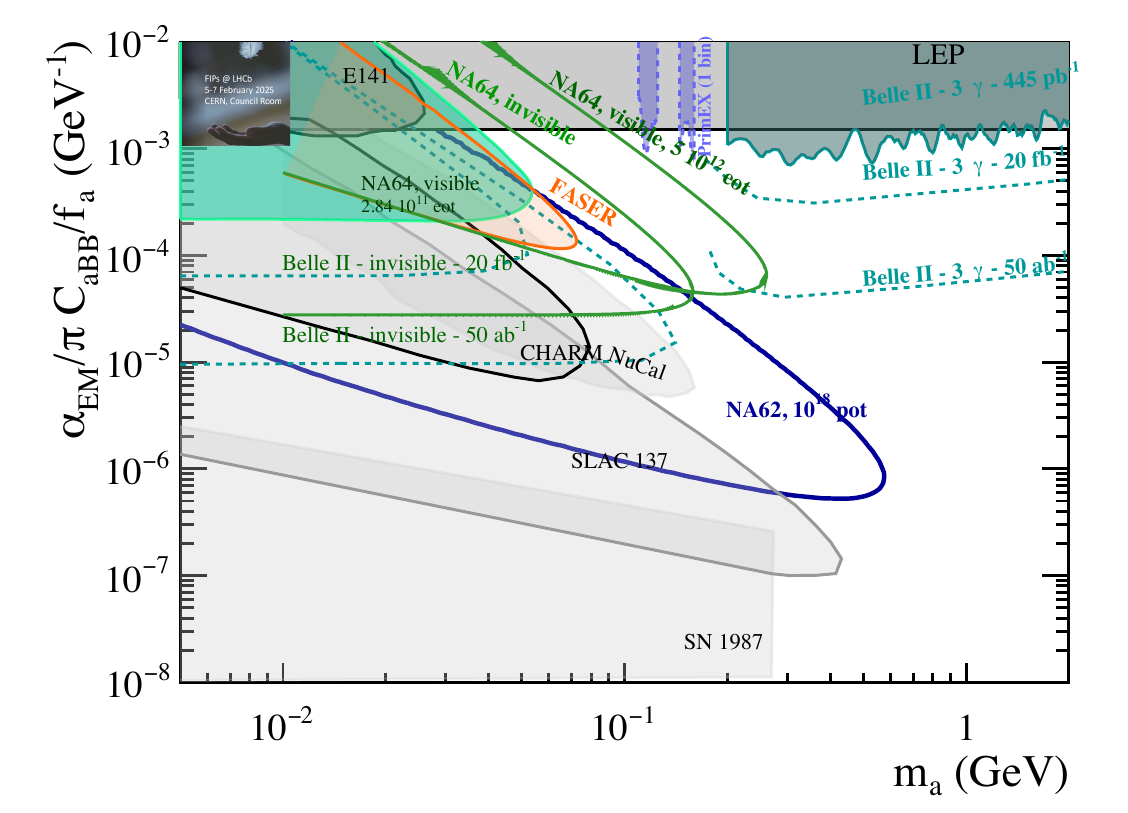}
        \includegraphics[width=0.8\textwidth]{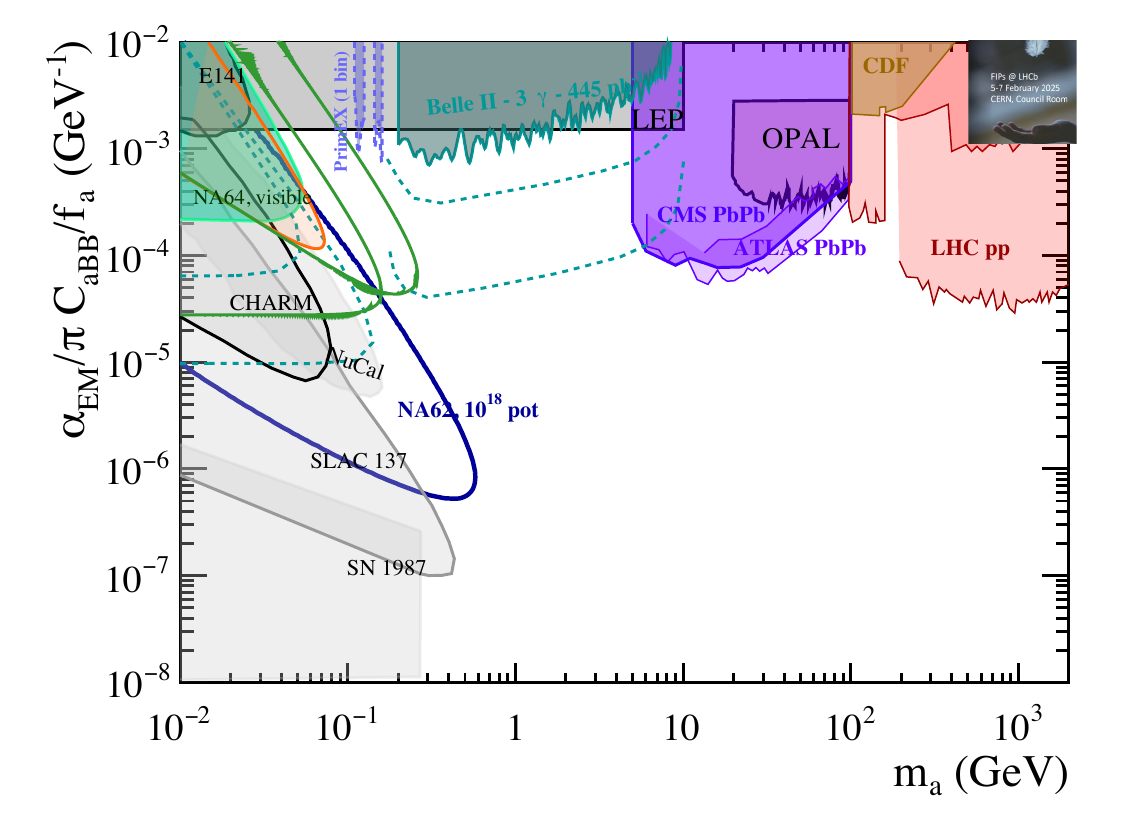}
        \caption{{\bf ALPs with photon couplings (BC9). Past and running experiments. ALP mass range up to 2 GeV (top) and up to 2~TeV (bottom)} Current bounds and future projections for 90\% CL exclusion limits.
{\it Filled gray areas} are bounds coming from interpretation of old datasets or astrophysical data:
LEP (data:~\cite{L3:1994shn, DELPHI:1991emv, DELPHI:1994mra, L3:1995nbq}; interpretation:~\cite{Knapen:2016moh} above 100 MeV and ~\cite{Jaeckel:2015jla} below 100 MeV. Caveat: the LEP line above 100 MeV is likely extendable also in the region below 100 MeV, down to the current bound from NA64).
SLAC 137~\cite{Bjorken:1988as};
CHARM~\cite{Gninenko:2012eq};
NuCal~\cite{Blumlein:1990ay}.
{\it Filled coloured areas} are bounds set by existing experiments: Belle II~\cite{Belle-II:2020jti};
NA64~\cite{NA64:2020qwq};
FASER~\cite{FASER:2024bbl};
PrimEx~\cite{Aloni:2019ruo} based on~\cite{PrimEx:2010fvg}.
Above a few GeV, the current limits come from analyses of data from CDF~\cite{Aaltonen:2013mfa}, and ATLAS and CMS in PbPb~\cite{CMS:2018erd,ATLAS:2020hii} and pp collisions ((details on the latter can be found in Figure~31 of Ref.~\cite{Agrawal:2021dbo}).
{\it Coloured curves} are projections from existing experiments:  Belle II~\cite{Dolan:2017osp};
NA64$_e^{++}$~\cite{NA64:eplus} in visible and invisible modes.
}
\label{fig:ALP-photon-running}
\end{figure}

\begin{figure}[htb]
    \centering
    \includegraphics[width=0.9\textwidth]{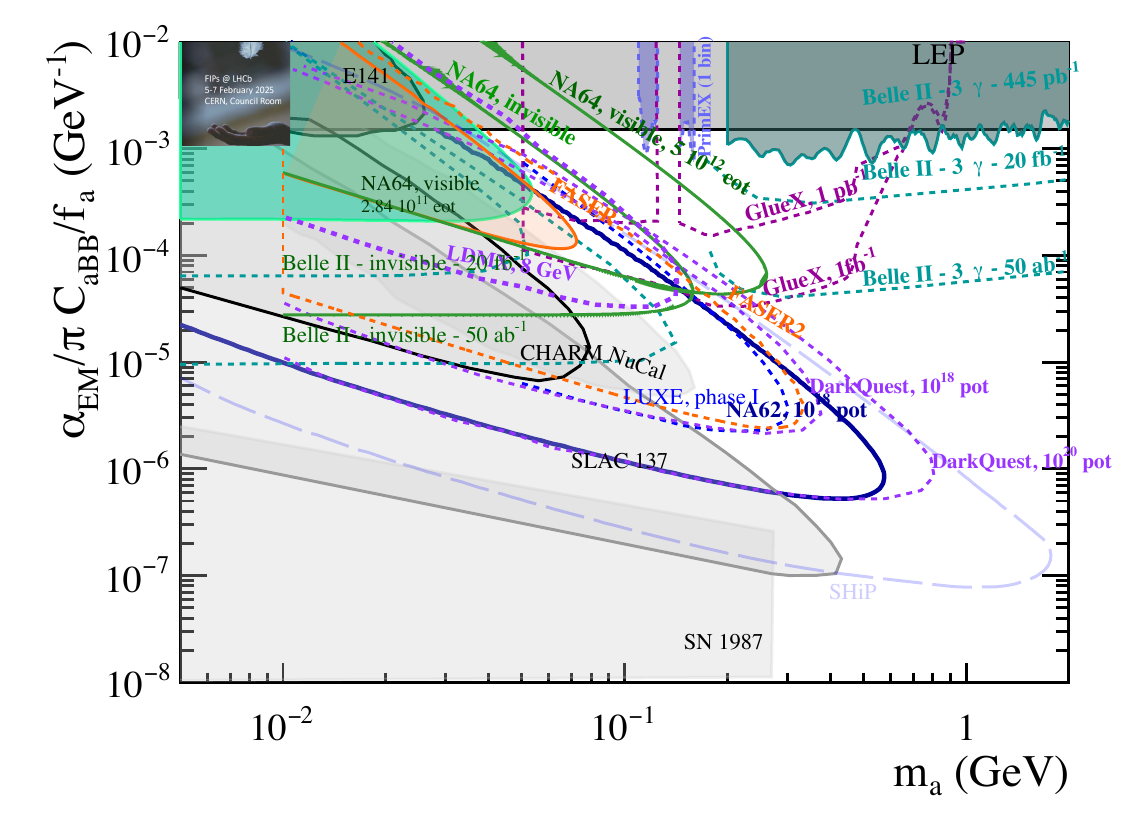}
    \includegraphics[width=0.9\textwidth]{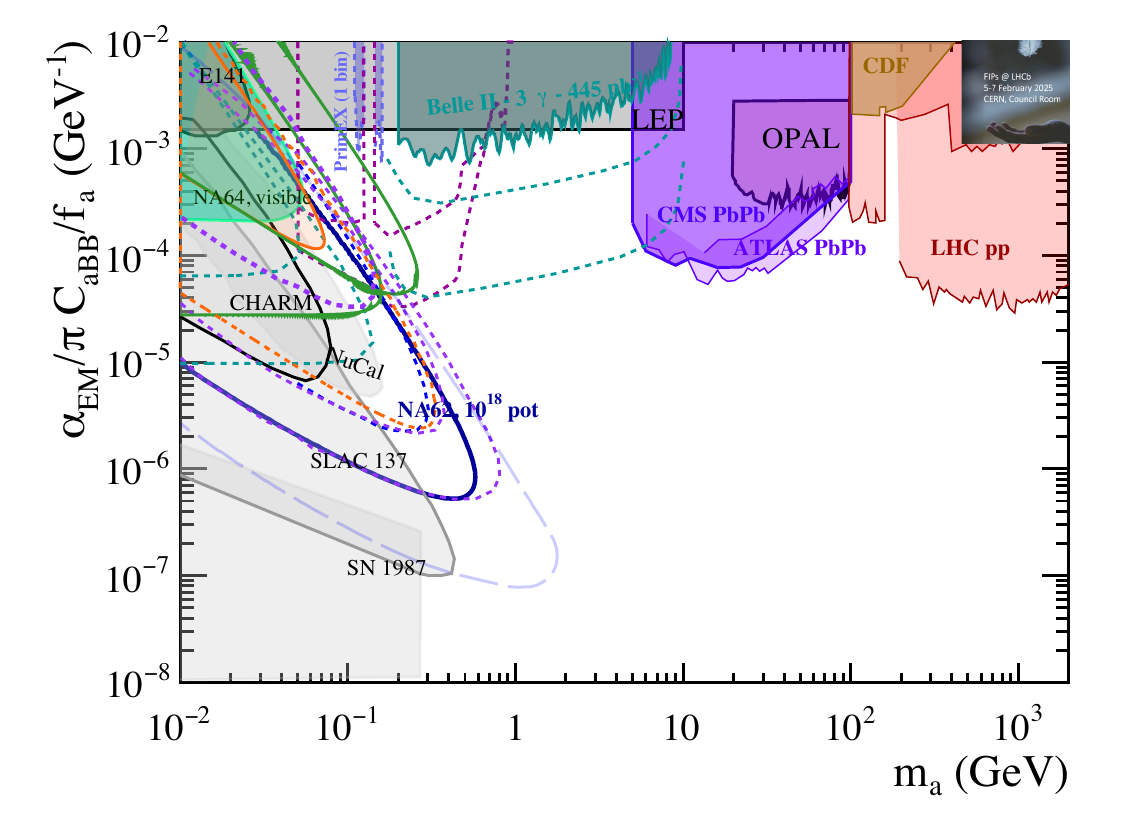}
            \caption{{\bf ALPs with photon couplings (BC9). Past, running, and future experiments.} ALP mass range up to 2 GeV (top) and up to 2~TeV (bottom). Current bounds and future projections for 90\% CL exclusion limits. Past and running experiments are represented as filled gray/coloured areas and solid coloured lines, as in Figure~\ref{fig:ALP-photon-running}. {\it Dashed and dotted lines} are for experiments in the design phase:
            SHiP~\cite{SHiP-ECN3-LoI};
            DarkQuest~\cite{Blinov:2021say};
FASER2~\cite{Feng:2022inv};
LUXE-phase 1~\cite{Bai:2021gbm};
Interpretation of the physics reach~\cite{Aloni:2019ruo}
of PrimEx~\cite{PrimEx:2010fvg} and GlueX experiments at JLab. There also interesting results from ATLAS \cite{ATLAS:2022abz} and LHCb \cite{LHCb:2025gbn} that probe the photon and gluon coupling in $pp$ collisions that are not shown in this plot because the assumptions do not hold exactly.}
    \label{fig:ALP-photon-all}
\end{figure}

Assuming a single ALP state $a$, and its predominant coupling to photons, all phenomenology (production, decay, oscillation in the magnetic field) can be determined as functions on  $\{m_a, g_{a\gamma} \}$ parameter space, where the $g_{a\gamma}=f^{-1}_\gamma$ notation is used.

The current status of experimental bounds and projections for past and current  experiments are shown in Figure~\ref{fig:ALP-photon-running}. Projections from experiments still in the design phase are added in Figure~\ref{fig:ALP-photon-all}.

\subsubsection{ALPs with fermion couplings (BC10)}

\begin{figure}[htb]
    \centering
    \includegraphics[width=\textwidth]{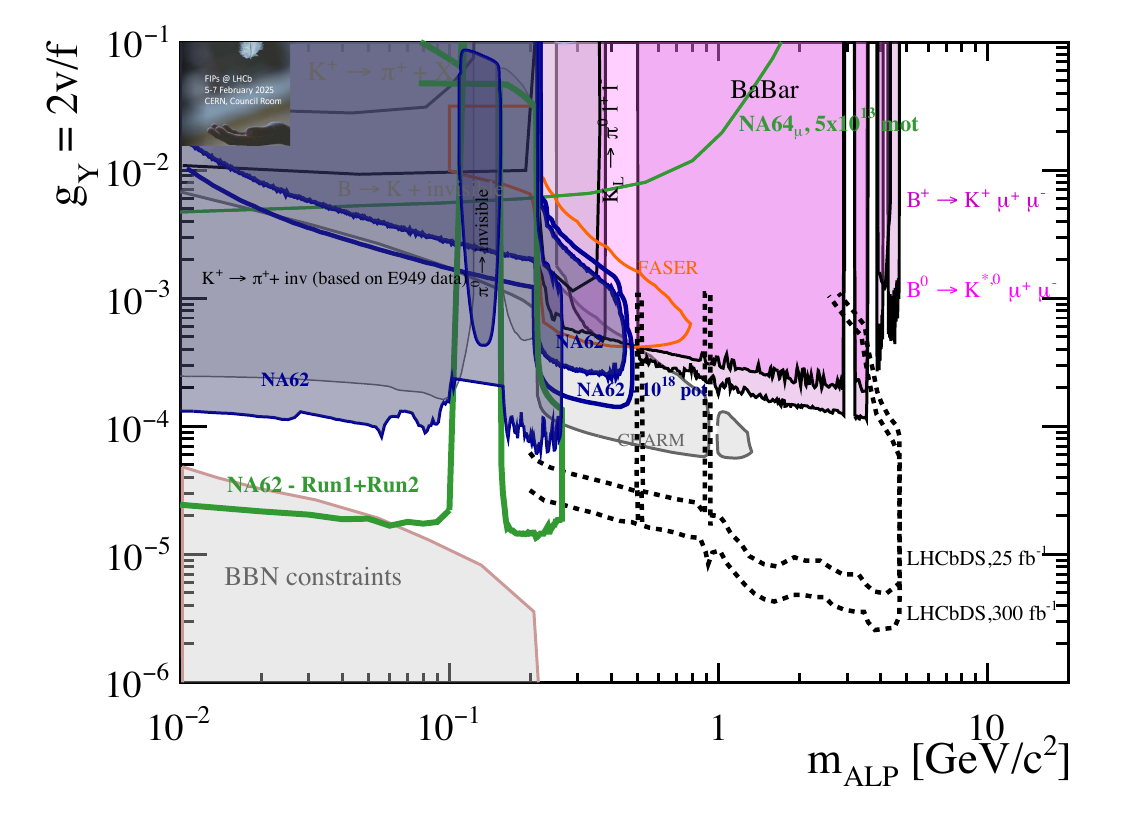}
    \caption{ {\bf ALPs with fermion couplings (BC10). Past and running experiments.} Current bounds and future projections for 90\% CL exclusion limits.
    We assume for the coupling $C_F/f_a$ given in Eq.~\ref{Leff_a} $C_F = 1$ and $f_a = f_l = f_q = f$.
{\it Filled gray areas} are bounds coming from interpretation of old datasets or astrophysical data.
{\it Filled coloured areas} are bounds set by existing experiments.
{\it Coloured curves} are projections from existing experiments.
NA62 results~\cite{NA62:2025yzs} and ~\cite{NA62:2020pwi}, NA62 projections with $10^{18}$~pot in beam dump mode~\cite{NA62-proj}, NA62 Run1+Run2  projections for data collected until 2026 in kaon mode~\cite{Ahdida:2023okr}. LHCb searches for displaced dimuon vertices in $B^+ \to K^+ \mu^+ \mu^-$~\cite{LHCb:2016awg} and $B^0 \to K^* \mu^+ \mu^-$ decays~\cite{LHCb:2015nkv}.  CHARM~\cite{CHARM:1985anb}. 
CHARM and LHCb filled areas have been adapted
by F.~Kahlhoefer, following Ref.~\cite{Dobrich:2018jyi}.
LHCb with very displaced vertices and 25/fb and 300/fb~\cite{Gorkavenko:2023nbk}.
}
    \label{fig:ALP-fermions-running}
\end{figure}

\begin{figure}[htb]
    \centering
    \includegraphics[width=\textwidth]{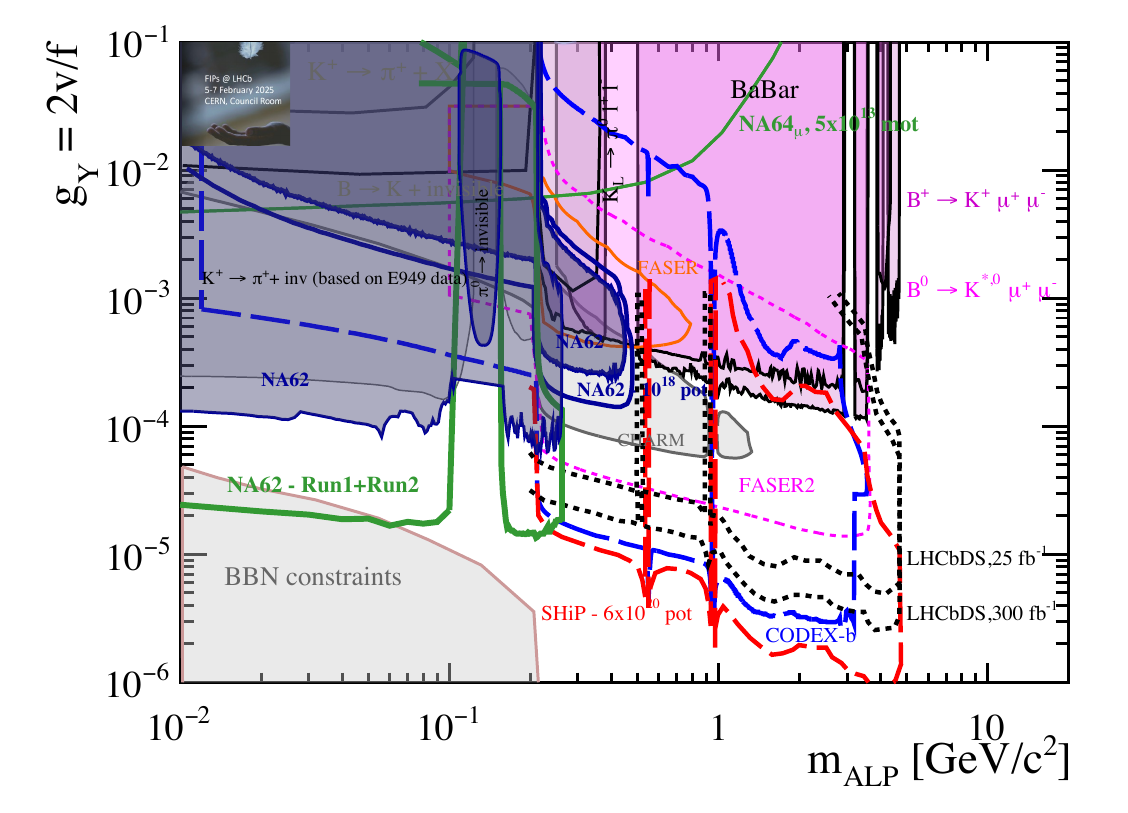}
              \caption{{\bf ALPs with fermion couplings (BC10). Past, running, and future experiments.} Current bounds and future projections for 90\% CL exclusion limits. Past and running experiments are represented as filled gray/coloured areas and solid coloured lines, as in Figure~\ref{fig:ALP-fermions-running}. {\it Dashed and dotted lines} are for experiments in the design phase:
               FASER2~\cite{Ariga:2018uku},
               CODEX-b~\cite{CODEX-b:2025rck};
               SHiP~\cite{SHiP-ECN3-LoI}.}
    \label{fig:ALP-fermions-all}
\end{figure}

Assuming a single ALP state $a$, and its predominant coupling to fermions, all phenomenology (production and decay) can be determined as functions of  $\{m_a, f^{-1}_l, f^{-1}_q \}$.
Furthermore, for the sake of simplicity, we take the assumption of universal fermion coupling, hence $f_q=f_l$.  Hence,  from Eq.~\ref{Leff_a}, we assume for the coupling $C_F/f_a$: $C_F = 1$ and $f_a = f_l = f_q = f$.

The current status of experimental bounds and projections for past and current  experiments are shown in Figure~\ref{fig:ALP-fermions-running}.
Projections from experiments still in the design phase are added in Figure~\ref{fig:ALP-fermions-all}.
For historical reasons the we plot on the $y-$axis the quantity $g_Y = 2 v/f$, where $v$ = 246 GeV is the Higgs vev and $f$, defined above, is the universal ALP couling to fermions.

\vskip 2mm
The ALPs coupled to fermions 
are mostly produced by flavor-changing neutral current (FCNC) decays $K/B \to X_\text{d/s}$ + ALP, where $X_\text{s/d}$ is a hadronic state including an $s/d$ quark.
The ALP decay width is still driven mostly by the di-lepton widths as the contribution of hadronic widths is relatively small\footnote{The LHCb bound has been revisited adding the hadronic width contributions in Ref.~\cite{DallaValleGarcia:2023xhh}.}

\subsubsection{ALPs with gluon couplings (BC11)}

\begin{figure}[htb]
    \centering
    \includegraphics[width=\textwidth]{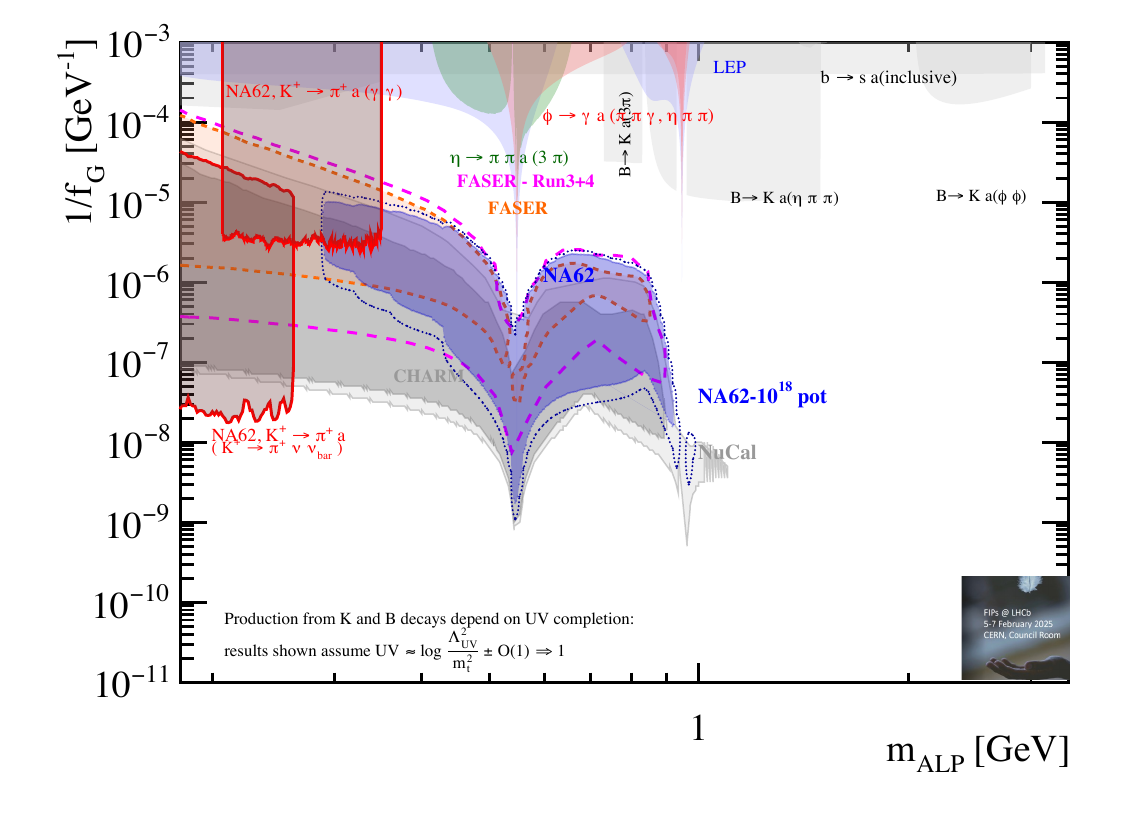}
       \caption{ {\bf ALPs with gluon couplings (BC11). Past and running experiments.} Current bounds and future projections for 90\% CL exclusion limits.
       We assume $1/f_G = 1/{16 \pi^2} C_{aGG}/f_a$ of Eq.~\ref{Leff_a}.
{\it Filled gray and coloured areas} are bounds coming from interpretation of old datasets or astrophysical data or from existing experiments. CHARM and NuCal gray filled area has been computed by reinterpreting using the ALPINIST code~\cite{Jerhot:2022chi} old results from CHARM~\cite{Bergsma:1985qz} and NuCal~\cite{Blumlein:2013cua}. Other coloured filled areas from prompt decays are kindly provided by Mike Williams and revisited from Ref.~\cite{Aloni:2018vki} (eg: the results from "prompt searches" have been rescaled by $(1/8)$ in order to make them compatible with the notation used in Eq.~\eqref{Leff_a}). These bounds depend on UV completion and the results shown assume $\approx [{\rm log \Lambda^2_{UV}}/m^2_t \pm {\mathcal{O}}(1)] \Rightarrow 1$. NA62 from $K^+ \to \pi^+ a, a \to {\rm invisible}$ ~\cite{NA62:2025upx}, NA62 $K^+ \to \pi^+ a, a \to \gamma \gamma$ ~\cite{NA62:2023olg}
NA62 in dump mode~\cite{NA62:2025yzs} and FASER~\cite{FASER:2024bbl}.
{\it Coloured curves} are projections from existing experiments:
FASER projections for Run3 (250/fb)+Run4 (680/fb)~\cite{Boyd:2882503}.}
    \label{fig:ALP-gluon-running}
\end{figure}

\begin{figure}[htb]
    \centering
    \includegraphics[width=\textwidth]{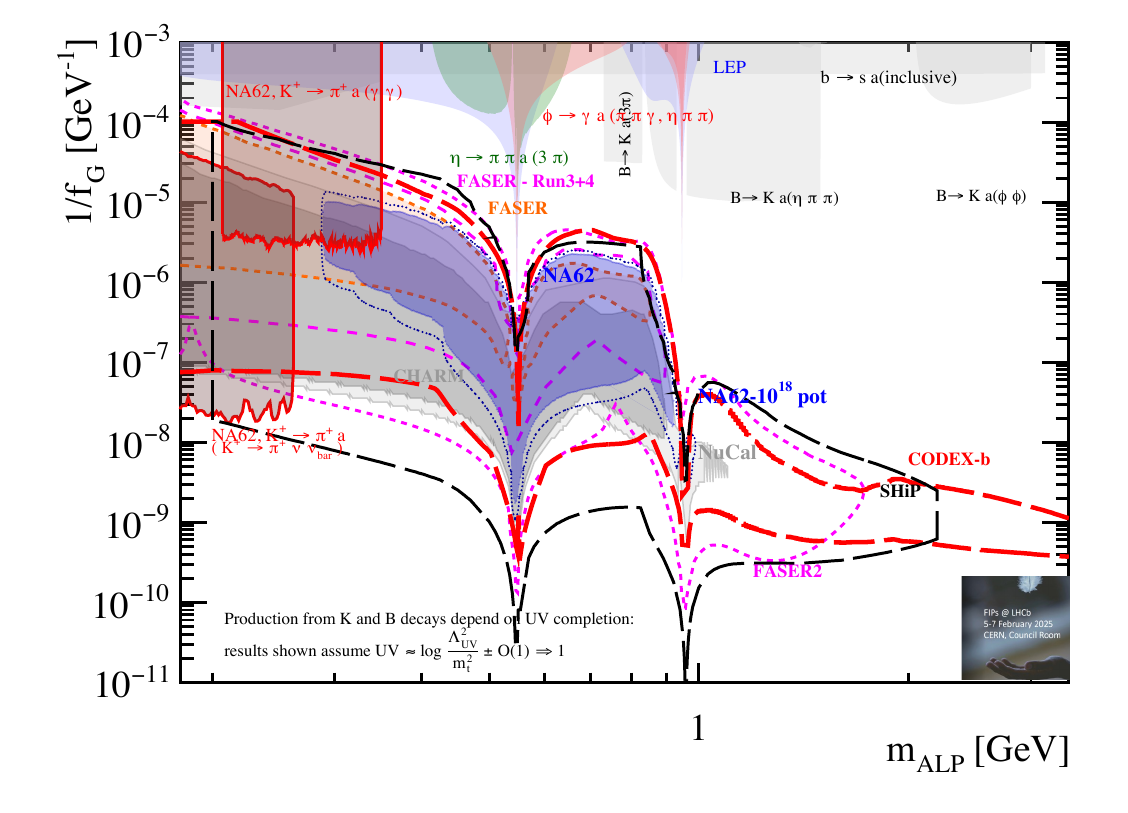}
 \caption{{\bf ALPs with gluon couplings (BC11). Past, running, and future experiments.} Current bounds and future projections for 90\% CL exclusion limits. Past and running experiments are represented as filled gray/coloured areas and solid coloured lines, as in Figure~\ref{fig:ALP-gluon-running}. {\it Dashed and dotted lines} are for experiments in the design phase:
 CODEX-b~\cite{Aielli:2022awh});
 SHiP~\cite{SHiP-ECN3-LoI};
 FASER2~\cite{Feng:2022inv}.
 }
    \label{fig:ALP-gluon-all}
\end{figure}

This case assumes an ALP coupled to gluons. The parameter space is $\{m_a, f^{-1}_G \}$ where   $1/f_G = 1/16 \pi^2 C_{aGG}/f_a$ of Eq.~\ref{Leff_a}. Notice that in this case, the limit of $m_a < m_{a,QCD}|_{f_a=f_G}$ is unnatural as it requires fine tuning and therefore is less motivated.

The current status of experimental bounds and projections for past and current  experiments are shown in Figure~\ref{fig:ALP-gluon-running}. Projections from experiments still in the design phase are added in Figure~\ref{fig:ALP-gluon-all}.



\subsubsection{ALPs with $W$ couplings }

\begin{figure}[htb]
    \centering
    \includegraphics[width=\textwidth]{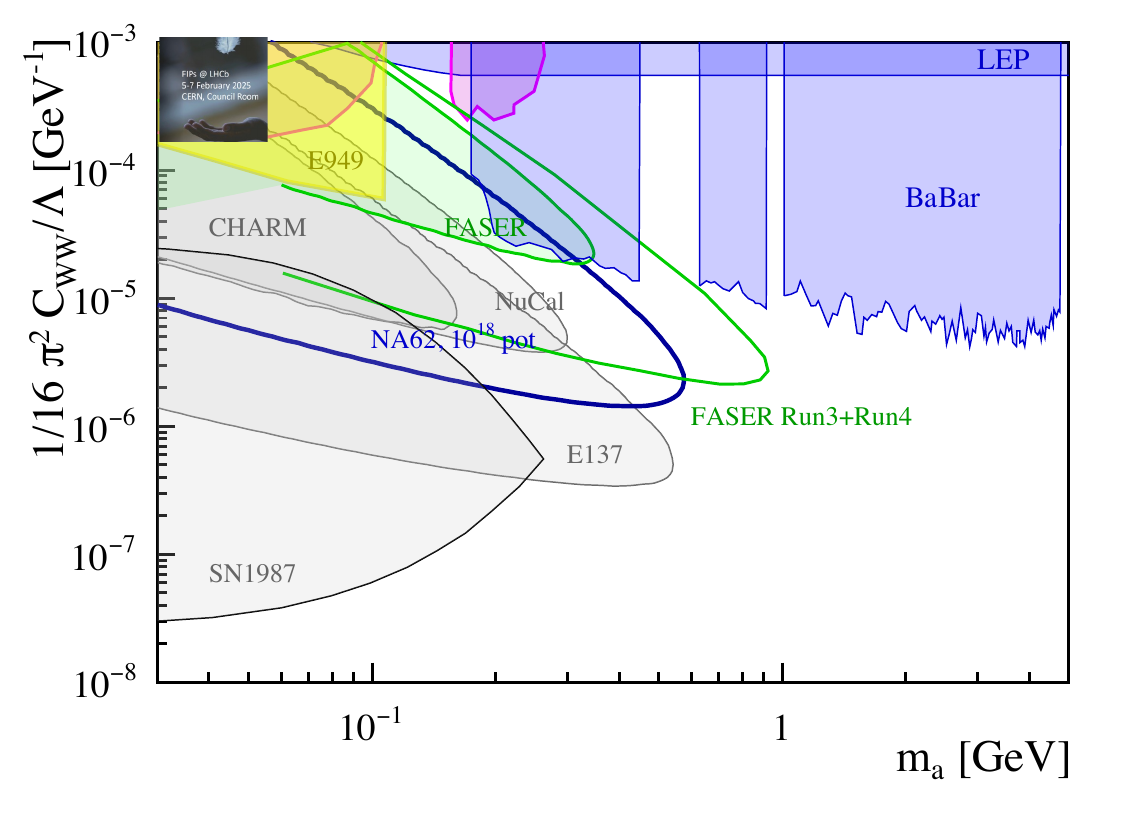}
    \caption{FASER~\cite{FASER:2024bbl},
$K$ and $B$ results from Ref.~\cite{Bauer:2021mvw} (and references therein)
FASER Run3+Run4~\cite{Kling:2020mch},
CHARM, NuCAL, E137 are reinterpretation of old results performed with ALPINIST code~\cite{Jerhot:2022chi}.
.}
    \label{fig:ALP-W-running}
\end{figure}

This scenario, which was first proposed in Ref.~\cite{Izaguirre:2016dfi}, differs from the three existing benchmark scenarios in that the production of ALPs is dominated by rare decays (as in the case of \emph{fermion dominance}),\footnote{We note that the theoretical predictions for rare meson decays involving ALPs with couplings to $W$ bosons are theoretically cleaner than the ones for \emph{fermion dominance} (which suffer from a sensitivity to unknown UV physics) and the ones for \emph{gluon dominance} (which are affected by large hadronic uncertainties).} but the decays are dominated by the effective ALP-photon coupling (as in the case of \emph{photon dominance}). The relevant formulas for the various rare meson decays can be found in Ref.~\cite{Izaguirre:2016dfi}.


 \noindent
 The ALP decay width into photons is given by

\begin{equation}
 \Gamma_{a\gamma\gamma} = \frac{4\pi\alpha^2 m_a^3}{\Lambda^2} \left|C_{WW}\right|^2 \; ,
\end{equation}
where $\alpha$ is the electromagnetic fine-structure constant.
The interplay between these effects leads to a different relation between production cross section and decay length than in any of the currently studied benchmarks. Furthermore, to constrain this scenario it will be crucial to carry out new searches, for example of $B \to K + a(\to \gamma \gamma)$.

\subsubsection{Open theoretical issues for ALPs with fermion and gluon couplings}

Both fermion and gluon-coupled ALPs can mix with the SM pseudoscalar mesons, such as $\pi^0, \eta, \eta'$ and their excitations which affects both the ALP production cross sections and their decay widths. The ALP production cross section uncertainty suffers from the same issues as for the dark photon and dark scalar due to the mixing.

\vskip 2mm
The ALP decay widths, unlike the dark photons and other vector mediators, and similarly to the scalar case, cannot be extracted from the measured $e^+e^- \to {\rm hadrons}$ data and one has to build the effective ALP interaction Lagrangian with mesons~\cite{Aloni:2018vki,DallaValleGarcia:2023xhh}.
The resulting phenomenology description is very sensitive to the addition of various interaction operators of heavy pseudoscalar mesons such as $\pi^{0}(1300), \eta(1295), \pi^{0}(1800)$, which also have mixing with the ALPs. The decay widths of such mesons are typically of the order of the $\rho$ meson width or smaller~\cite{Giacosa:2024epf,ParticleDataGroup:2024cfk}; as a result, their contributions to the ALP observables are non-overlapping and have to be carefully added. This has to be done in a self-consistent way to reproduce the observables, such as partial decay widths of the mesons~\cite{Parganlija:2016yxq}. Hence the computations of the hadronic widths in ALP decays are subject to large uncertainties~\cite{Ovchynnikov:2025gpx} (see also Section~\ref{ssec:fips-pheno}).

\vskip 2mm
An important challenge for the near future will be also to improve the calculations to obtain predictions for the \emph{gluon dominance} benchmark, in particular regarding the different ways in which ALPs with couplings to gluons can be produced (including production in partonic showers~\cite{Aielli:2019ivi,Kyselov:2025uez} and production from gluon fusion~\cite{Kelly:2020dda}).

\clearpage
\section{Conclusions and Outlook}
\label{sec:conclusions}
Feebly-interacting particles (FIPs) are currently one of the most discussed topics in fundamental physics, reaching from Particle Physics to Astrophysics and Cosmology. Breaking with the traditional approach, which favoured the existence of new particles with relatively large couplings to the Standard Model (SM) and masses commensurate to the electroweak scale, an ever-increasing effort has been devoted to the low energy frontier, at lower masses and much more feeble interaction strengths.

The LHCb experiment, thanks to the new software trigger, enabling triggering on low-momentum and long-lived signatures, is starting its journey in this fascinating field, joining the already established actors producing results at CERN and elsewhere.

\vskip 2mm
In the coming years, a wealth of experimental results for FIPs is expected from all
the major laboratories in the world. These include accelerator-based experiments (ATLAS,
CMS, LHCb, FASER, NA64, NA62 at CERN; MEG-II and Mu3e at PSI; Belle II at KEK; BDX and
HPS at JLab; MiniBooNE at FNAL; T2K ND280 and KOTO at J-Park) as well as dark
matter direct detection experiments and dedicated experiments searching for axions/ALPs
in Europe and the US. In addition to established experimental efforts, a multitude of new
initiatives has recently emerged both at CERN (SHiP, MATHUSLA,
FASER-2, CODEX-b) and elsewhere (LDMX at SLAC, long and short baseline neutrino near
detectors at Fermilab, Beam Dump experiment at Mainz, and others) aiming at covering in
the coming decade uncharted regions of FIPs parameter space inaccessible to traditional
experiments.

\vskip 2mm
No single experiment or experimental approach is sufficient alone to cover the large
parameter space in terms of masses and couplings that FIPs models can suggest: hence
synergy and complementarity among a great variety of experimental facilities are paramount.
For example, in the accelerator-based experiments, the interplay between prompt and displaced signatures is paramount, as FIPs can be not necessarily only long-lived.
In addition, a deep collaboration and cross-fertilisation across different communities is mandatory. An example is the understanding of the interplay between active neutrino mixing parameters
from ongoing and future neutrino experiments and the favoured ranges of HNL parameters, as discussed in this report, or more comprehensive comparisons of sub-GeV direct
detection searches and accelerator-based searches. 
Likewise, the wealth of data coming from
accelerator-based experiments should be compared in greater detail to existing and future
astrophysics and cosmology-driven sources of insight that can affect the current bounds coming from the study of the BBN onset or the CMB data.

\vskip 2mm
The multitude of experimental efforts today calls for a reassessment of the theoretical framework for FIPs that emphasizes the most promising and relevant models going beyond the minimal portal framework. Theoretical studies should continue to map out how the known portal operators are embedded in complete models addressing the varied puzzles of the SM, and this will be important for identifying additional promising sensitivity milestones around which to focus experimental efforts. Simultaneously, there are several theoretical open issues related to the computation of decay widths, cross-sections, and lifetimes that require further study and should be addressed in the coming years.

\vskip 2mm
The fast theory progress could lead in the coming years to the emergence of a new (and more realistic) FIP framework with a different phenomenology from what is currently studied.
Hence, experiments should keep a broad view and keep exploring FIP signatures beyond what is known today.

\clearpage
\section{Acknowledgements}
\label{sec:acknowledgements}

\noindent
{\it  G. Dalla Valle Garcia } thanks the Doctoral School  ``Karlsruhe School of Elementary and Astroparticle Physics: Science and Technology (KSETA)” for financial support through the GSSP program of the German Academic Exchange Service (DAAD).

\noindent 
{\it J. Alimena} acknowledges support from DESY (Hamburg, Germany), a member of the Helmholtz Association HGF, and support by the Deutsche Forschungsgemeinschaft under Germany's Excellence Strategy - EXC 2121 Quantum Universe - 390833306.

\noindent
{\it Felix Kling} is supported in part by Heising-Simons Foundation Grant 2020-1840 and in part by U.S. National Science Foundation Grant PHY-2210283 and PHY-2514888.

\noindent
{\it Jose Zurita} is supported by the {\it Generalitat Valenciana} (Spain) through the {\it plan GenT} program (CIDEGENT/2019/068), by the Spanish Government (Agencia Estatal de
Investigaci\'on), ERDF funds from European Commission (MCIN/AEI/10.13039/501100011033, Grant No. PID2020-114473GB-I00 and No. PID2023-146220NB-I00), and by the Spanish Research Agency (Agencia Estatal
de Investigaci\'on, MCIU/AEI) through the grant IFIC Centro de Excelencia Severo Ochoa No. CEX2023-001292-S.

\noindent 
{\it Gabriele Ferretti} is supported by the Swedish Research Council (grant nr. 2024-04347), as well as by grants from the Adlerbert Research Foundation via the KVVS foundation, the Carl Tryggers Foundation (CTS 24:3453) and the Kungl. Vetenskapsakademien (PH2024-0076).

\noindent 
{\it Xabier Cid Vidal} is supported by the Spanish Research State Agency under projects PID2022-139514NA-C33 and PCI2023-145984-2; by the ``María de Maeztu'' grant CEX2023-001318-M, funded by MICIU/AEI /10.13039/501100011033; and by the Xunta de Galicia (CIGUS Network of Research Centres).

\noindent
{\it Luiz Vale Silva} is supported by the Spanish Government (Agencia Estatal de Investigación MCIN/AEI/ 10.13039/501100011033) Grants No. PID2020–114473GB-I00 and No. PID2023-146220NB-I00, and CEX2023-001292-S (Agencia Estatal de Investigación MCIU/AEI (Spain) under grant IFIC Centro de Excelencia Severo Ochoa).

\noindent
{\it Carlos Vázquez Sierra} is supported by Agencia Estatal de Investigación (Spain) through the Ramón y Cajal program RYC2023-043804-I, and through the Universidade da Coruña-InTalent program.

\noindent
{\it P. Reimitz} acknowledges support by Funda\c{c}\~{a}o de Amparo \`a Pesquisa do Estado de S\~{a}o Paulo (FAPESP) under the contract 2020/10004-7 and in part by NSERC, Canada.

\noindent
The work of {\it Ekaterina Kriukova} on inelastic bremsstrahlung is supported by the Russian Science Foundation RSF grant No. 25-12-00309. {\it Ekaterina Kriukova} also acknowledges the Foundation for the Advancement of Theoretical Physics and Mathematics “BASIS” for the PhD fellowship No. 21-2-10-37-1 that supported her work on elastic bremsstrahlung.

\noindent
{\it Andrea Merli} and {\it Spencer Collaviti} acknowledge the support from the Swiss National Science Foundation under the Ambizione grant n. 216324

\clearpage

\bibliographystyle{JHEP}
\bibliography{biblio/biblio.bib,biblio/biblio2.bib,biblio/biblio_plots,biblio/biblio_fips_at_lhcb, biblio/ferretti.bib, biblio/LVS_mybib}

\end{document}